\documentclass[12pt,a4paper]{book}

\usepackage{ngerman}
\usepackage[T1]{fontenc}
\usepackage{fancyhdr}
\usepackage{graphicx}
\usepackage{amsmath}
\usepackage{amssymb}
\usepackage{caption}
\usepackage{mystyle}

\def\thebibliography#1{\markboth
{}{}\list
 {[\arabic{enumi}]}{\settowidth\labelwidth{[#1]}\leftmargin
\labelwidth
 \advance\leftmargin\labelsep
 \usecounter{enumi}}
 \def\newblock{\hskip .11em plus .33em minus -.07em}
 \sloppy
 \sfcode`\.=1000\relax}

\newenvironment{npb}{\interlinepenalty9999
                     \postdisplaypenalty9999
                     \interdisplaylinepenalty9999
                     \csname@beginparpenalty\endcsname9999
                     \csname@endparpenalty\endcsname9999
                     \csname@itempenalty\endcsname9999
                     \csname@secpenalty\endcsname9999}{}

\newcommand{\Int}[2]{\int\limits_{#1}^{#2}}
\newcommand{\Sum}[2]{\sum\limits_{#1}^{#2}}
\newcommand{\Prod}[2]{\prod\limits_{#1}^{#2}}

\newcommand{\D}{\displaystyle}
\newcommand{\T}{\textstyle}
\newcommand{\winkel}{<\!\!\!)}
\newcommand{\CdoT}{\!\cdot\!}
\newcommand{\Cdot}{\!\cdot}
\newcommand{\cdoT}{\cdot\!}

\newcommand{\Left}{\!\left}
\newcommand{\up}[2]{\raisebox{#1ex}{$\scriptscriptstyle #2$}}
\newcommand{\uP}[2]{\raisebox{#1ex}{$\scriptstyle #2$}}

\newcommand{\HH}{{\text{H}}}
\newcommand{\TT}{{\!\text{T}}}
\newcommand{\Hh}{\raisebox{0.3ex}{$\scriptstyle\text{H}$}}
\newcommand{\Kk}{\raisebox{0.3ex}{$\scriptstyle *$}}
\newcommand{\Tt}{\raisebox{0.3ex}{$\scriptstyle\!\text{T}$}}
\newcommand{\hH}{\raisebox{-0.4ex}{$\scriptstyle\text{H}$}}
\newcommand{\kK}{\raisebox{-0.4ex}{$\scriptstyle *$}}
\newcommand{\tT}{\raisebox{-0.4ex}{$\scriptstyle\!\text{T}$}}
\newcommand{\snorm}[1]{\|#1\|_{\raisebox{-0.4ex}{$\scriptstyle\text{2}$}}}
\newcommand{\Snorm}[1]{\big\|#1\big\|_{\raisebox{-0.4ex}{$\scriptstyle\text{2}$}}}
\newcommand{\fnorm}[1]{\|#1\|_{\raisebox{-0.4ex}{$\scriptstyle\text{F}$}}}

\newcommand{\zu}{\mbox{$\stackrel{\raisebox{-0.3ex}{\tiny?}}{z}_0$}}

\newcommand{\dftran}[0]{
{\setlength{\unitlength}{0.5pt}
\begin{picture}(60,15)(5,-2)
\put(5,5){\circle{10}}
\put(55,5){\circle*{10}}
\put(10,5){\line(1,0){10}}
\put(40,5){\line(-2,1){20}}
\put(40,5){\line(1,0){10}}
\end{picture}}}

\allowdisplaybreaks[1]

\renewcommand{\chaptermark}[1]{\markboth{\small\sf\thechapter.\ #1}{\small\sf\thechapter.\ #1}}
\renewcommand{\sectionmark}[1]{\markright{\small\sf\thesection.\ #1}}

\renewcommand{\headrulewidth}{0pt}
\begin{document}
\tracingpages=1
\thispagestyle{empty}
\begin{titlepage}
\begin{center}
{\sf {\large
\vspace*{2cm}
{\LARGE \bf Ein Fenster zur gleichzeitigen Messung der "Ubertragungsfunktion
eines realen Systems und des Leistungsdichtespektrums des "uberlagerten
Rauschens am Systemausgang (Teil 1)}\\
\vspace*{8cm}
{\bf Helmut Repp}\\
\vspace*{1cm}
Erlangen - 2025
}}
\newpage
\rule{0pt}{0pt}
\vfill
Dipl.-Ing. Helmut Repp\\
Liegnitzer Stra"se 1\\
D-91058 Erlangen\\
Germany\\\vspace{6pt}
Telefon: +49-9131-3\,66\,41\\
Mobile: +49-173-57\,19\,350\\
E-mail: Helmut.Repp@gmx.de
\end{center}

\end{titlepage}
\thispagestyle{empty}
\chapter*{"Ubersicht}

Mit Hilfe des aus der Literatur bekannten Rauschklirrmessverfahrens
gelingt es in einer Messung sowohl die "Ubertragungsfunktion als auch
das Rauschleistungsdichtespektrum eines gest"orten zeitinvarianten
Systems zu messen. Das Verfahren wird hier auf komplexwertige Systeme
erweitert. Durch die Einf"uhrung einer Fensterung wird es unter anderem
m"oglich, die spektralen Eigenschaften der St"orung wesentlich
frequenzselektiver und dabei dennoch leistungsrichtig zu ermitteln.

Da bei der Messung mit dem Rauschklirrmessverfahren bisher
implizit eine Rechteckfensterung verwendet wurde, war es bisher
bei Systemen, die durch Prozesse mit "uber der Frequenz stark
schwankender Leistungsdichte gest"ort werden, nur bedingt m"oglich
aussagekr"aftige Messwerte f"ur das Leistungsdichtespektrum
der St"orung zu erhalten. 
Dies kann durch den Einsatz eines anderen Fensters als des
Rechteckfensters deutlich verbessert werden. Es kann jedoch
nicht jede beliebige Fensterfolge verwendet werden. Es werden
zum einen exakte Bedingungen und zum anderen gew"unschte
Eigenschaften angegeben, die das Spektrum einer Fensterfolge
erf"ullen sollte, wenn man sie beim Rauschklirrmessverfahren
einsetzen will. Weil die Berechnung der Messwerte der
"Ubertragungsfunktion und des Leistungsdichtespektrums von der Wahl
der Fensterfolge abh"angt, ist bei Verwendung einer ungeeigneten
Fensterfolge entweder ein extrem erh"ohter Speicherbedarf und
Rechenaufwand n"otig oder es ist mit unbrauchbaren Messergebnissen
zu rechnen. Es wird ein Algorithmus vorgestellt, mit dem
eine geeignete Fensterfolge numerisch berechnet werden kann.
Die Bedingungen, die es erm"oglichen, die Fensterfolge beim
Rauschklirrmessverfahren einzusetzen, werden von der mit diesem 
Algorithmus berechneten Fensterfolge mit Abweichungen erf"ullt,
die in der Gr"o"senordnung der minimal m"oglichen Rechenfehler
liegen, also der Fehler, die auf Grund der endlichen Wortl"ange
der Zahlendarstellung des zur Berechnung verwendeten Rechners
unvermeidbar sind. Dadurch, und durch die Tatsache, dass der
hier vorgestellte Algorithmus nicht iterativ arbeitet, unterscheidet
er sich von anderen in der Literatur bekannten Verfahren zum Entwurf von
Fensterfolgen, die die gew"unschten Eigenschaften nur n"aherungsweise
erf"ullen. Bei dem hier angegebenen Algorithmus kann die L"ange der Fensterfolge
als ein ganzzahliges Vielfaches der L"ange des Rechteckfensters gew"ahlt
werden. Dadurch kann die Frequenzselektivit"at der Messung des
Leistungsdichtespektrums der St"orung nahezu beliebig erh"oht werden,
wobei die Gesamtrauschleistung der St"orung immer erwartungstreu
abgesch"atzt werden kann. Mit der hier vorgestellten Fensterfolge
ist eine erwartungstreue Messung der "Ubertragungsfunktion m"oglich,
wobei der durch die Fensterung bedingte zus"atzliche Rechenaufwand
als gering zu bewerten ist. Dabei ergibt sich bei Systemen, die durch
Prozesse mit "uber der Frequenz stark schwankender Leistungsdichte
gest"ort werden, eine zum Teil deutlich verringerte Varianz f"ur die
Messwerte der "Ubertragungsfunktion.

Die zwei wichtigsten Eigenschaften der hier vorgestellten Fensterfolge
seien kurz erw"ahnt. Zum einen l"asst sich die Fensterfolge als eine
Periode eines periodischen Signals darstellen, das nur niederfrequente
Anteile aufweist, und zum anderen handelt es sich bei dem Betragsquadrat
des Spektrums der Fensterfolge um das Spektrum eines M-tel-Band-Filters.
Diese Eigenschaften machen die Fensterfolge auch f"ur andere Anwendungen,
wie z.~B. den Einsatz in einer Filterbank, oder als Sendeimpuls bei der
"Ubertragung digitaler Signale interessant. Auf diese Einsatzgebiete der
Fensterfolge wird im Rahmen dieser Abhandlung abgesehen von einem
beispielhaften Augendigramm im zweiten Teil nicht n"aher eingegangen.

\tableofcontents
\cleardoublepage

\pagenumbering{arabic} 
\setcounter{page}{1}
\frenchspacing
\renewcommand{\headrulewidth}{0.4pt}
\lhead[\fancyplain{\small\sf\thepage}{\small\sf\thepage}]{\fancyplain{}{{\rightmark}}}
\rhead[\fancyplain{}{{\leftmark}}]{\fancyplain{\small\sf\thepage}{\small\sf\thepage}}
\chapter{Einleitung}\vspace{30pt}

Von vielen Systemen in der Nachrichten"ubertragung w"unscht man ein lineares,
zeitinvariantes und stabiles Verhalten. Reale Systeme k"onnen solches Verhalten
nur n"aherungsweise erf"ullen. So ist z.~B. in der Realit"at der Bereich
der zul"assigen Aussteuerung immer begrenzt, bei digitalen Systemen kann nur
mit endlicher Wortl"ange gearbeitet werden, und bei kontinuierlichen
Systemen zeigen nichtlineare Bauteilkennlinien ihre Wirkung. Auch externe 
St"orungen, die von au"sen in das System einstreuen, verursachen Abweichungen 
des realen Systemverhaltens vom gew"unschten Systemverhalten. Um beurteilen 
zu k"onnen, wie gut das in Bild~\ref{b1h}a
\begin{figure}[btp]
\rule{\textwidth}{0.5pt}\vspace{-5pt}
\begin{center}
{
\begin{picture}(250,197)
\put(-40,185){\makebox(0,0)[l]{a) Reales System:}}
\put(30,160){\circle{6}}
\put(90,140){\circle{6}}
\put(210,150){\circle{6}}
\put(33,159){\vector(1,0){96}}
\put(33,161){\vector(1,0){96}}
\put(93,139){\vector(1,0){36}}
\put(93,141){\vector(1,0){36}}
\put(171,149){\vector(1,0){36}}
\put(171,151){\vector(1,0){36}}
\put(131,131){\framebox(38,38){\LARGE ${\cal S}$}}
\put(129,129){\framebox(42,42){}}
\put(22,160){\makebox(0,0)[r]{$\boldsymbol{v}(k)$}}
\put(82,140){\makebox(0,0)[r]{$\boldsymbol{n}_{ext}(k)$}}
\put(218,150){\makebox(0,0)[l]{$\boldsymbol{y}(k)$}}
\put(-40,90){\makebox(0,0)[l]{b) Systemmodell:}}
\put(30,60){\circle{6}}
\put(180,30){\circle{6}}
\put(210,60){\circle{6}}
\put(175,60){\line(1,0){10}}
\put(180,55){\line(0,1){10}}
\put(175,55){\framebox(10,10){}}
\put(33,59){\vector(1,0){37}}
\put(33,61){\vector(1,0){37}}
\put(110,59){\vector(1,0){65}}
\put(110,61){\vector(1,0){65}}
\put(185,59){\vector(1,0){22}}
\put(185,61){\vector(1,0){22}}
\put(179,33){\vector(0,1){22}}
\put(181,33){\vector(0,1){22}}
\put(70,40){\framebox(40,40){\LARGE ${\cal S}_{lin}$}}
\put(22,60){\makebox(0,0)[r]{$\boldsymbol{v}(k)$}}
\put(180,22){\makebox(0,0)[t]{$\boldsymbol{n}(k)$}}
\put(132,65){\makebox(0,0)[b]{$\boldsymbol{x}(k)$}}
\put(218,60){\makebox(0,0)[l]{$\boldsymbol{y}(k)$}}
\end{picture}}
\end{center}\vspace{-20pt}
\setlength{\belowcaptionskip}{-6pt}
\caption{Modell eines gest"orten nichtlinearen realen Systems}
\label{b1h}
\end{figure}
dargestellte reale
System ${\cal S}$ das gew"unschte lineare Systemverhalten erf"ullt, modellieren 
wir in Bild~\ref{b1h}b die Abweichungen vom linearen Systemverhalten als eine 
St"orung \mbox{$\boldsymbol{n}(k)$}, die einem ideal linearen Modellsystem 
${\cal S}_{lin}$ ausgangsseitig "uberlagert ist. Alle St"orungen, die 
in das reale System einstreuen, sind in Bild~\ref{b1h}a zu dem St"orer 
\mbox{$\boldsymbol{n}_{ext}(k)$} zusammengefasst. Im weiteren werde ich das
lineare System ${\cal S}_{lin}$ als Modellsystem bezeichnen, w"ahrend das
gesamte System, also das lineare Modellsystem, das um die Modellst"orung
\mbox{$\boldsymbol{n}(k)$} erweitert ist, als Systemmodell bezeichnet wird.

In dieser Abhandlung werde ich mich auf die Untersuchung zeitdiskreter 
Systeme beschr"anken. Bei zeitkontinuierlichen Systemen\label{f.1} 
sind in der Regel alle an dem System auftretenden Signale ---
Nutz- als auch St"orsignale --- auf ein bestimmtes Frequenzband
begrenzt. Daher lassen sich diese kontinuierlichen Signale theoretisch
ohne Informationsverlust anhand ihrer Abtastwerte analysieren
und synthetisieren, wenn die Abtastung in geeigneter Weise erfolgt. 
Bei Bandpasssignalen ist gegebenenfalls die bei der Abtastung inh"arente
Abmischung ins Basisband zu ber"ucksichtigen. Zusammen mit der Synthese
des Eingangssignals des realen kontinuierlichen Systems aus einer
digitalen Signalfolge und mit der Abtastung des Ausgangssignals
ergibt sich aus dem kontinuierlichen realen System ein neues reales
System, das digital arbeitet. Bei der Synthese des Eingangssignals
mit einer realen Digital-Analog-Wandlung mit anschlie"sender
Tiefpassfilterung und eventueller Modulation, entstehen ebenso 
Fehler, wie auch bei einer realen Tief- bzw. Bandpassfilterung mit
anschlie"sender Abtastung und Analog-Digital-Wandlung auf der
Analyseseite. Will man Messungen an kontinuierlichen Systemen
durchf"uhren, so sollten diese unvermeidbaren  Fehler klein
gegen"uber den St"orungen im zu vermessenden kontinuierlichen System sein.
 Nur dann wird man erwarten k"onnen, dass die Messergebnisse,
die sich auf das digitale Gesamtsystem beziehen,  das reale
kontinuierliche System brauchbar beschreiben. Im weiteren wird 
unter dem realen System immer das digitale Gesamtsystem verstanden. 

In aller Regel sind reale Systeme nicht nur f"ur die Erregung mit einer 
konkreten Signalfolge entworfen. Man wird daher nicht nur die St"orung, 
sondern auch die Erregung des Systems als einen Zufallsprozess 
\mbox{$\boldsymbol{v}(k)$} betrachten, von dem die wichtigsten 
stochastischen Eigenschaften --- wie z.~B. die Varianz --- bekannt sind. 

Den ersten Teil der Einleitung bilden einige Vorbemerkungen zu den am System 
auftretenden Zufallsprozesse. Im folgenden Unterkapitel, stelle ich die 
Struktur dieser Abhandlung vor. Es beginnt mit einigen Aspekten, die bei der 
theoretischen Aufteilung des realen Systems in das lineare Modellsystem und 
die Modellst"orung zu beachten sind. Es folgt ein kurze Einleitung zu der hier 
vorgestellten Erweiterung eines Messverfahrens zur empirischen Bestimmung der 
das Systemmodell beschreibenden Gr"o"sen. Die Erweiterung besteht dabei in der 
Einf"uhrung einer Fensterung. Die Konstruktion geeigneter Fensterfolgen bildet 
den folgenden Schwerpunkt der Abhandlung. Beispiele f"ur Messergebnisse, die sich 
mit dem modifizierten Messverfahren erzielen lassen, runden das Ganze ab. Im letzten 
Unterkapitel der Einleitung gehe ich auf die Systematik der in dieser Abhandlung 
verwendeten Schreibweisen ein. 

\section{Die Zufallsprozesse des Systems}

Um das reale System zu modellieren, ist man nun daran interessiert, den Prozess 
\mbox{$\boldsymbol{y}(k)$} am Ausgang des realen Systems ${\cal S}$ in zwei 
Anteile \mbox{$\boldsymbol{x}(k)$} und \mbox{$\boldsymbol{n}(k)$} 
aufzuspalten, wie dies in Bild~\ref{b1h} im unteren Teilbild dargestellt ist.
Der eine Anteil \mbox{$\boldsymbol{x}(k)$} ist die Reaktion eines idealen
linearen zeitinvarianten Modellsystems
${\cal S}_{lin}$, das durch seine "Ubertragungsfunktion \mbox{$H(\Omega)$}
beschrieben wird, auf den erregenden Prozess \mbox{$\boldsymbol{v}(k)$}.
Der zweite Anteil ist der Modellrauschprozess
\begin{equation}
\boldsymbol{n}(k)\,=\,\boldsymbol{y}(k)-\boldsymbol{x}(k),
\label{1.1}
\end{equation}
der die Abweichungen vom linearen Systemverhalten modelliert.
Es sei darauf hingewiesen, dass durch diese Wahl des
Modellrauschprozesses \mbox{$\boldsymbol{n}(k)$} die Zufallsprozesse
an den Ausg"angen des realen Systems und des Systemmodells dieselben sind.

Im weiteren werde ich mich in dieser Abhandlung auf den Fall beschr"anken,
dass der Verbundprozess aus \mbox{$\boldsymbol{v}(k)$} und 
\mbox{$\boldsymbol{y}(k)$} station"ar und mittelwertfrei ist.
Wie man das Systemmodell f"ur reale Systeme, die von instation"aren
Prozessen gest"ort oder erregt werden, erweitern kann, und dass man auch
bei solchen Systemen das Rauschklirrmessverfahren anwenden kann, wird in
\cite{Erg} gezeigt.

Eine Ensemblemittelung --- d.~h. eine Mittelung "uber mehrere
Messungen im gleichen Zeit"-intervall, bei denen das System mit 
unterschiedlichen Musterfolgen des eingangsseitigen
Zufallsprozesses erregt wird --- kann zur Gewinnung der das System
beschreibenden Gr"o"sen an einem einzelnen System nicht vorgenommen 
werden. Wenn wir jedoch annehmen k"onnen, dass die Ergodenhypothese 
erf"ullt sei, was wir im Weiteren tun wollen, kann die Messung der das 
System beschreibenden Gr"o"sen mit Hilfe einer Zeitmittelung erfolgen. 
Deshalb werden bei der Messung zeitlich begrenzte Musterfolgen verwendet, 
die in unterschiedlichen Zeitintervallen das zu messende System erregen. 
Am Ausgang des Systems werden dann die entsprechenden zeitlich begrenzten 
Musterfolgen des Ausgangsprozesses gemessen. 

In aller Regel l"asst sich der Prozess \mbox{$\boldsymbol{n}(k)$}
niemals vollst"andig beschreiben. Dazu m"usste man n"amlich die in der
Dimension unbegrenzte Verbundverteilungsfunktion des Prozesses f"ur alle
Zeitpunkte $k$ angeben. Im Rahmen dieser Abhandlung werde ich mich 
daher bei der theoretischen Beschreibung des Modellrauschprozesses
\mbox{$\boldsymbol{n}(k)$} auf die Momente zweiter Ordnung beschr"anken,
die sich aus der zweidimensionalen Verbundverteilungsfunktion des 
Rauschprozesses f"ur zwei beliebige Zeitpunkte berechnen lassen.
Ein Teil der Momente zweiter Ordnung bildet die Autokorrelationsfolge (\,AKF\,) 
dieses Prozesses. Sie beinhaltet die Korrelationen der Zufallsgr"o"sen
des Prozesses zu einem Zeitpunkt mit den Zufallsgr"o"sen zu einem weiteren
Zeitpunkt. Da wir im weiteren immer komplexe Zufallsprozesse betrachten, 
ist die Korrelation der beiden zeitversetzten Zufallsgr"o"sen des Prozesses
als der Erwartungswert des Produkts der einen Zufallsgr"o"se mit dem 
{\em Konjugierten} der anderen Zufallsgr"o"se definiert.
\begin{equation}
\phi_{\boldsymbol{n}}(\kappa)\;=
\text{E}\big\{\boldsymbol{n}(k)^{\Kk}\!\CdoT
\boldsymbol{n}(k\!+\!\kappa)\big\}
\label{1.2}
\end{equation}
Weil wir angenommen haben, dass es sich um einen mittelwertfreien und somit 
zentralen Prozess handelt, ist die Auto{\em korrelations}\/folge identisch mit der 
Auto{\em kovarianz}\/folge, die aus den entsprechenden zweiten {\em zentralen}\/ 
Momenten gebildet wird. Da wir davon ausgehen, dass ein im weiten Sinne 
station"arer Prozess vorliegt, h"angt dessen AKF nur von der Zeitdifferenz 
$\kappa$, nicht aber von der absoluten Lage $k$ der beiden am Produkt der 
Kovarianz beteiligten Zufallsgr"o"sen des Prozesses ab.
"Uber eine eindimensionale Fouriertransformation bez"uglich dieser
Zeitdifferenz erh"alt man das Leistungsdichtespektrum~(\,LDS\,)
\begin{equation}
\Phi_{\boldsymbol{n}}(\Omega)\;=
\Sum{\kappa=-\infty}{\infty}
\phi_{\boldsymbol{n}}(\kappa)\cdot
e^{\!-j\cdot\Omega\cdot\kappa}
\qquad\qquad\forall\qquad\Omega\in\mathbb{R}.
\label{1.3}
\end{equation}
Bei komplexen Rauschprozessen ist die Beschreibung der zweiten Momente 
nur vollst"andig, wenn neben der AKF auch noch die diskrete Folge der 
Kovarianzen der Zufallsgr"o"sen des Prozesses zu einem Zeitpunkt mit 
den konjugierten Zufallsgr"o"sen zu einem weiteren Zeitpunkt angegeben wird. 
Weil bei der Definition der Kovarianz nach Gleichung (\ref{1.2}) die eine 
daran beteiligte Zufallsgr"o"se bereits konjugiert wird, ist also auch noch 
die Folge der Erwartungswerte der Produkte zweier {\em nicht}\/ konjugierter 
Zufallsgr"o"sen anzugeben.
\begin{equation}
\psi_{\boldsymbol{n}}(\kappa)\;=
\text{E}\big\{\boldsymbol{n}(k)\CdoT
\boldsymbol{n}(k\!+\!\kappa)\big\}
\label{1.4}
\end{equation}
Diese sei im weiteren als modifizierte Autokorrelationsfolge (MAKF) bezeichnet.
Bei einem im weiten Sinne station"aren Prozess h"angt auch diese Folge
nur von der Zeitdifferenz $\kappa$ der daran beteiligten Zufallsgr"o"sen ab.
Durch eine diskrete Fouriertransformation gewinnt man daraus ein Spektrum
\begin{equation}
\Psi_{\boldsymbol{n}}(\Omega)\;=
\Sum{\kappa=-\infty}{\infty}
\psi_{\boldsymbol{n}}(\kappa)\cdot
e^{\!-j\cdot\Omega\cdot\kappa}
\qquad\qquad\forall\qquad\Omega\in\mathbb{R},\quad{}
\label{1.5}
\end{equation}
das im weiteren als modifiziertes Leistungsdichtespektrum (\,MLDS\,) 
bezeichnet werden soll. Im Rahmen dieser Abhandlung will ich mich bei der 
Beschreibung der stochastischen Eigenschaften des Rauschprozesses mit der 
Angabe des LDS und MLDS bzw. der Messwerte dieser Funktionen begn"ugen. 

Die Einschr"ankung auf diese i. Allg. unvollst"andige Beschreibung des 
Rauschprozesses ist deshalb sinnvoll, weil es oft nicht notwendig
ist, eine vollst"andige Beschreibung zu kennen. Bei vielen Berechnungen, wie
z.~B. bei der Berechnung von Signal-Ger"auschverh"altnissen, gehen nur diese
Momente in die Rechnung ein. Eine weitere wichtige Motivation ist durch
den zentralen Grenzwertsatz gegeben, der besagt, dass bei der
additiven "Uberlagerung hinreichend vieler unabh"angiger Rauschquellen,
die alle einen verschwindenden Beitrag zur gesamten St"orung liefern,
ein Rauschprozess entsteht, der n"aherungsweise normalverteilt ist.
Bei einem normalverteilten Rauschprozess gen"ugt es, die ersten beiden
Momente zu kennen, um diesen vollst"andig zu beschreiben, weil sich
alle weiteren Momente aus den ersten beiden Momenten berechnen lassen.
Bei vielen realen Systemen kann von dem eingestreuten, externen Rauschprozess
angenommen werden, dass er durch eine derartige additive "Uberlagerung entstanden ist.
Es sei noch darauf hingewiesen, dass die Signale in realen Systemen in ihrem
Wertebereich begrenzt sind. Einerseits bedeutet das, dass der Rauschanteil
eines gemessenen Signals niemals exakt normalverteilt sein kann, so dass
die Beschreibung des Rauschprozesses niemals vollst"andig sein kann.
Andererseits garantiert diese Tatsache die Existenz der Momente
zweiter Ordnung f"ur alle in realen Systemen gemessenen
Rauschprozesse.

\section{Strukturierung der Abhandlung}

Es wird in Kapitel \ref{theo} gezeigt, welche "Ubertragungsfunktion
\mbox{$H(\Omega)$} sich im Systemmodell theoretisch ergibt, wenn das
Modellsystem das lineare Verhalten des realen Systems so approximieren
soll, dass das zweite Moment des Modellrauschprozesses
\mbox{$\boldsymbol{n}(k)$} m"oglichst klein wird. Die "Uberf"uhrung des
realen Systems in das Systemmodell wird somit als die L"osung einer
linearen Regression dargestellt. Der Modellrauschprozess ergibt sich
daher als der Prozess des Restfehlers, der bei der optimalen Approximation des
realen Systems durch das lineare zeitinvariante System verbleibt.
Er wird daher im weiteren auch oft als Approximationsfehlerprozess bezeichnet.

Da die Approximation des realen Systems durch das Systemmodell nach
Bild \ref{b1h} von den i.~Allg. unbekannten statistischen Eigenschaften
der St"orungen am realen System abh"angt, kann man die zu bestimmenden, in 
$\Omega$ kontinuierlichen Funktionen \mbox{$H(\Omega)$},
\mbox{$\Phi_{\boldsymbol{n}}(\Omega)$} und \mbox{$\Psi_{\boldsymbol{n}}(\Omega)$}
in der Regel nicht theoretisch in geschlossener Form angeben.
Man muss daher zun"achst endlich viele Werte finden, mit deren
Hilfe die das Systemmodell beschreibenden Funktionen m"oglichst
aussagekr"aftig charakterisiert werden k"onnen. Desweiteren wird
ein Verfahren ben"otigt, das es erlaubt, mit Hilfe einer Messung
am realen System diese endlich vielen Werte abzusch"atzen.

Bei der "Ubertragungsfunktion \mbox{$H(\Omega)$} wird man sich daher mit
der Angabe endlich vieler "aquidistanter Abtastwerte begn"ugen.
Nach dem Abtasttheorem ist diese Beschreibung der "Ubertragungsfunktion
immer dann vollst"andig, wenn das theoretische, lineare Modellsystem
eine zeitlich begrenzte Im"-puls"-ant"-wort besitzt, und die Anzahl der
Abtastwerte der "Ubertragungsfunktion hinreichend gro"s gew"ahlt wird.
Bei vielen realen Systemen kann man davon ausgehen, dass diese
Art der Beschreibung der "Ubertragungsfunktion wenigstens in
guter N"aherung ausreichend ist. Durch die Verwendung eines
bereichsweise periodischen Eingangsprozesses gelingt es, das
Regressionsproblem in der Art zu modifizieren, dass darin nur
mehr die endlich vielen "aquidistanten Abtastwerte der
"Ubertragungsfunktion auftreten.

Da man bei einigen Systemen nicht mit einer hinreichend engen
zeitlichen Begrenzung der AKF rechnen kann, wird das LDS durch 
die Sch"atzung endlich vieler Abtastwerte oft
nur unzureichend beschrieben. Desweiteren kann man zeigen
(~z.~B. in~\cite{Dittrich}~), dass es keine konsistenten Messwerte
f"ur die Abtastwerte des LDS eines Zufallsprozesses gibt,
selbst wenn man die Messdauer theoretisch "uber alle Grenzen
wachsen l"asst. Daher muss man sich "uberlegen, welche andere
dem LDS eng verwandte spektrale Folge man zur Beschreibung
des Modellrauschprozesses verwenden kann. Diese "Uberlegung f"uhrt
in Kapitel \ref{W} zu der Angabe des Erwartungswertes des Periodogramms
eines gefensterten Ausschnitts des Modellzufallsprozesses. Damit
diese Beschreibung des Rauschprozesses eine hohe Aussagekraft
erh"alt, muss man gewisse Forderungen an die dabei verwendete
Fensterfolge stellen. Eine der Forderungen ist, dass die "uber den 
Erwartungswert des Periodogramms berechnete Gesamtrauschleistung 
gleich der tats"achlichen Rauschleistung sein soll, 
die gleich der AKF an der Stelle Null ist. Es wird gezeigt, 
dass dies erf"ullt wird, wenn die Fensterfolge ein 
leistungskomplement"ares Spektrum aufweist. Bei einem solchen 
Spektrum "uberlagern sich die Betragsquadrate mehrerer verschobener 
Versionen des Spektrums zu einer Konstanten. Die Fenster-AKF 
\mbox{$f(k)\ast f(-k)/M$} ist dann die Impulsantwort eines 
\mbox{$M$-tel-Band-}Tiefpassfilters. Es wird weiterhin gezeigt, 
dass ein rechteckf"ormiger Verlauf f"ur das Betragsquadrat des 
Spektrums der Fensterfolge w"unschenswert w"are. Da dieser 
Wunschverlauf von einer endlich langen Fensterfolge niemals 
exakt realisiert werden kann, wird man versuchen, diesen Wunschverlauf 
des Betragsquadrats des Spektrums m"oglichst gut zu approximieren. 
Dabei zeigt sich, dass es w"unschenswert ist, m"oglichst einen 
D"ampfungsfrequenzgang  zu erzielen, bei dem die Sperrd"ampfung 
f"ur gro"se Frequenzen mit einer frei vorgebbaren Potenz ansteigt. 
Im Fall einer zeitlich hinreichend begrenzten Autokorrelationsfolge 
soll es die verwendete Fensterfolge erm"oglichen, die AKF aus den 
endlich vielen Erwartungswerten des Periodogramms perfekt rekonstruieren 
zu k"onnen.

In Kapitel \ref{W} wird noch eine weitere Bedingung f"ur das Spektrum 
der Fensterfolge hergeleitet. Wird diese Bedingung eingehalten, so l"asst 
sich zeigen, dass bei der L"osung des Regressionsproblems die Minimierung 
des zweiten Moments des Approximationsfehlers "aquivalent zur Minimierung 
des Erwartungswertes des gefensterten Periodogramms ist. Somit ergeben sich 
bei beiden Minimierungen dieselben Parameter des linearen Modellsystems also 
dieselben Werte der "Ubertragungsfunktion.

Fr"uher wurde f"ur die Absch"atzung der das Systemmodell
beschreibenden endlich vielen Frequenzwerte ein Verfahren
--- das Rauschklirrmessverfahren (\,RKM\,) --- angewandt,
das sich aus vielen Einzelmessungen f"ur jede zu messende
Frequenz zusammensetzte \cite{Sch}. In \cite{Sch/D} und
\cite{Dong} wurde ein Verfahren vorgestellt, bei dem die
Messung f"ur den gesamten zu messenden Frequenzbereich auf
einmal erfolgt. Dabei wurde der Name "`Rauschklirrmessverfahren"'
"ubernommen, obwohl jetzt nicht mehr mit Sinuseintonsignalen
gemessen wird, und auch keine Klirrfaktoren mehr bestimmt werden.
In \cite{Sch/H} wurde dieses Verfahren modifiziert, um den
Rechenaufwand des RKM deutlich zu verringern. In Kapitel \ref{RKM}
wird nun eine weitere Modifikation des Verfahrens nach \cite{Dong}
und \cite{Sch/H} vorgestellt. Das RKM wird dabei als die empirische
Bestimmung der Regressionskoeffizienten dargestellt. Bei St"orungen
mit "uber der Frequenz stark schwankendem LDS wird durch die
Einf"uhrung einer Fensterung bei nur geringf"ugig erh"ohtem
Aufwand erreicht, dass die Messwerte das LDS deutlich besser
beschreiben. Von Beginn an wird dabei die Verallgemeinerung
auf komplexwertige Systeme und Signale vorgenommen.
Mit diesem Messverfahren kann daher sowohl das LDS des
Approximationsfehlers als auch dessen MLDS abgesch"atzt werden.

Sinnvoll ist ein Messverfahren nur dann, wenn die damit
gewonnenen Messwerte die zu messenden Gr"o"sen erwartungstreu
und konsistent absch"atzen. Dies ist der Fall,
wenn sich die Messwerte mit zunehmender Messdauer in
Wahrscheinlichkeit an die zu messenden Gr"o"sen ann"ahern.
Um das beurteilen zu k"onnen, werden in Kapitel \ref{RKM} im
Anschluss die Erwartungswerte und die Varianzen der Messwerte
untersucht. Wenn man die theoretischen Varianzen und
Kovarianzen der Real- und Imagin"arteile der Messwerte kennt,
und wenn man annimmt, dass die  Messwerte normalverteilt sind,
kann man Konfidenzgebiete um die Messwerte herum angeben, deren Gr"o"se
von einer vorgebbaren Wahrscheinlichkeit abh"angt. Mit dieser
Wahrscheinlichkeit beinhaltet dann das Konfidenzgebiet den wahren Wert
der abzusch"atzenden theoretischen Gr"o"se. Da man die wahren theoretischen
(Ko)varianzen der Messwerte nicht kennt, kann man die Konfidenzgebiete
nur anhand der Messung absch"atzen. In Kapitel \ref{RKM} wird
beschrieben, wie man diese Sch"atzwerte f"ur die Konfidenzgebiete aus
den Messwerten berechnen kann. Desweiteren wird dort der Vorteil der 
Verwendung zuf"alliger Mehrton- und Chirpsignale untersucht. 

Die Vereinfachungen, die sich bei der Messung reellwertiger Systeme ergeben, 
werden in Kapitel \ref{Resys} diskutiert. Dabei wird ein neuartiges zuf"alliges 
reelles Chirpsignal vorgestellt. 

Auf den Spezialfall der klassischen Spektralsch"atzung eines reellen 
mittelwertfreien Zufallsprozesses, bei dem kein Modellsystem 
angesetzt wird, wird in Kapitel \ref{LDS} eingegangen. 

Eine implementierungsnahe Auflistung der bei diesem Messverfahren
durchzuf"uhrenden Einzelschritte, sowie eine Absch"atzung des
ben"otigten Speicherbedarfs findet sich in \cite{Erg}. Au"serdem wird
dort untersucht, wie sich das Modellsystem erweitern l"asst, wenn
sich das reale System nicht oder nur unzureichend durch ein lineares
zeitinvariantes System modellieren l"asst, und wie das Messverfahren
abzu"andern ist, wenn sich ein Approximationsfehlerprozess
mit einem zeitabh"angigen ersten Moment ergibt, oder wenn weitere
Kreuzkorrelationen des Ein- und Ausgangsprozesses ber"ucksichtigt
werden sollen. Auch die theoretische Behandlung eines zyklostation"aren
Approximationsfehlerprozesses, sowie die erweiterte Messung der
spektralen Eigenschaften eines solchen Fehlerprozesses wird in \cite{Erg}
diskutiert. Weitere Messwerte, die man aus den bis dahin beschriebenen
Messergebnissen ableiten kann, sind die ggf. vom Zeitpunkt der Erregung 
abh"angige Impulsantwort sowie die AKF und die MAKF des komplexen 
Approximationsfehlerprozesses. Deren Berechnung wird ebenfalls
in \cite{Erg} angegeben.

Die Angabe eines Verfahrens zur Berechnung einer f"ur das RKM
geeigneten Fensterfolge erfolgt in Kapitel \ref{Algo}. Dieses
ist von den davorliegenden Kapiteln nur insofern abh"angig,
dass die damit berechenbaren Fensterfolgen die Forderungen
erf"ullen, die in Kapitel \ref{W} hergeleitet wurden.
Wer nicht an der Verwendung der Fensterfolge beim RKM
interessiert ist, kann nun gleich zu Kapitel \ref{Algo}
weiterspringen, und muss lediglich die an die Fensterfolgen
gestellten Forderungen anhand der angegebenen Gleichungsnummern in 
Kapitel~\ref{W} nachschlagen. Zun"achst wird in Kapitel \ref{Fen} das 
Prinzip der Berechnung der Fensterfolge vorgestellt, und gezeigt, dass 
eine so konstruierte Fensterfolge die gew"unschten Eigenschaften aufweist.
Ein Algorithmus wird im folgenden angegeben, mit dessen Hilfe die
Fensterfolge numerisch berechnet werden kann. Dieser ist bez"uglich
der Genauigkeit der Berechnung der Fensterfolge mit Hilfe eines
Prozessors optimiert, der mit Gleitkommaarithmetik arbeitet.
Bei solchen Prozessoren ist die Gr"o"senordnung des wegen der
endlichen Wortl"ange unvermeidbaren relativen Fehlers der
Zahlendarstellung "uber viele Dekaden hinweg nahezu konstant.
Die Gr"o"senordnung der bei der Berechnung des Fensters
auftretenden Fehler wird abgesch"atzt. Die hier vorgestellte
Methode eine geeignete Fensterfolge zu berechnen, ist ein Sonderfall
eines allgemeineren Algorithmus, der in \cite{Erg} zu finden ist.
Dort wird auch eine Variante behandelt, mit deren Hilfe man eine
kontinuierliche Fensterfunktion berechnen kann, die "ahnliche
Eigenschaften wie die diskrete Fensterfolge aufweist, und die
in einem gewissen Sinne als Grenzwertl"osung anzusehen ist.

In Kapitel \ref{FenBeisp} folgen einige graphische Darstellungen,
die die Eigenschaften der mit dem angegebenen Algorithmus konstruierten
Fensterfolge f"ur unterschiedliche Parametereinstellungen demonstrieren.
Anschlie"send werden andere aus der Literatur bekannten Fensterfolgen
auf ihre Einsetzbarkeit beim RKM "uberpr"uft.

Anhand einiger am Rechner simulierter Beispielsysteme wird in Kapitel
\ref{RKMBeisp} die Verbesserung der Messgenauigkeit des RKM durch den
Einsatz der Fensterung verdeutlicht. Auch f"ur den Sonderfall
der Spektralsch"atzung des LDS eines Rauschprozesses wird an
einem Beispiel die Verwendbarkeit der so konstruierten Fensterfolge
gezeigt. Die in diesem Kapitel angef"uhrten Beispiele
dienen auch der Verdeutlichung einiger im theoretischen Teil der
Abhandlung beschriebener Sachverhalte.

Neben einigen Beweisen und Herleitungen, von denen manche an mehreren
Stellen des Hauptteils gebraucht werden, befindet sich im Anhang
auch der Rumpf eines Programms in der Interpretersprache
{\tt MATLAB} zur Berechnung der Fensterfolge.

\section{Konventionen und Formelzeichen }

Mit einigen Worten zu der Systematik der in dieser Abhandlung 
verwendeten Schreibweisen m"ochte ich die Einleitung beenden.
Eine ausf"uhrliche "Ubersicht "uber die verwendeten Abk"urzungen und
Formelzeichen befindet sich im Anhang.

Unter einer reellen Zufallsgr"o"se soll im weiteren das verstanden werden,
was in der Literatur auch oft als zuf"allige Ver"anderliche oder
Zufallsvariable bezeichnet wird. Dabei handelt es sich um
eine Abbildung des Ergebnisraums eines Zufallsexperimentes auf
den K"orper der reellen Zahlen, die gewissen Einschr"ankungen unterliegt.
Unter einer komplexen Zufallsgr"o"se soll die Linearkombination zweier
reeller Zufallsgr"o"sen verstanden werden, bei der die erste reelle
Zufallsgr"o"se den Koeffizienten $1$ und die zweite den Koeffizienten $j$
aufweist. Die beiden daran beteiligten reellen Zufallsgr"o"sen unterliegen
keinen Einschr"ankungen, und k"onnen daher sowohl stochastisch abh"angig,
als auch korreliert sein. Wenn nicht explizit angegeben wird, dass
eine reelle Zufallsgr"o"se gemeint ist, ist immer davon auszugehen,
dass es sich um eine komplexe Zufallsgr"o"se handelt.

Ein Zufallsvektor ist ein geordnetes $n$-Tupel von Zufallsgr"o"sen. 
Diese werden in der Literatur oft auch als $n$-dimensionale Zufallsgr"o"sen 
bezeichnet. Zufallsprozesse sind Zufallsgr"o"sen, die von einem Parameter 
abh"angen. F"ur jeden Parameterwert aus dem Definitionsbereich des Parameters 
kann man diese als eine eigene Zufallsgr"o"se interpretieren, die von den 
anderen Zufallsgr"o"sen des Prozesses f"ur andere Parameterwerte
abh"angen kann. Zufallsprozesse mit einem Parameter, der
einer endlichen und abz"ahlbaren Menge entnommen wird, kann man als
Zufallsvektoren betrachten, wobei man jedem m"oglichen Parameterwert
ein Element des Zufallsvektors eineindeutig zuordnet. Daher werde ich solche
Zufallsprozesse auch als Zufallsvektoren bezeichnen und ich werde den Begriff
des Zufallsprozesses haupts"achlich auf solche Zufallsprozesse anwenden,
deren Parameter einer unendlichen Menge entstammt.

Es sei noch auf eine Besonderheit komplexer Zufallsgr"o"sen und Vektoren
hingewiesen. Da \label{komplrel}man f"ur komplexe Zahlen keine
"`gr"o"ser als"' oder "`kleiner als"' Relation angeben kann
---~die Schreibweise \mbox{$\boldsymbol{x}\!<\!x$} also sinnlos ist ---
gibt es f"ur komplexe Zufallsgr"o"sen auch keine Verteilung
\mbox{$\text{P}(\boldsymbol{x}\!<\!x)$},
welche die Wahrscheinlichkeit ist, dass die Zufallsgr"o"se
\mbox{$\boldsymbol{x}$} kleiner als der freie Parameter $x$ der Verteilung
ist. Da aber andererseits auch bei einer reellen
Zufallsgr"o"se die Schreibweise \mbox{$\boldsymbol{x}\!<\!x$} in dem Ausdruck
\mbox{$\text{P}(\boldsymbol{x}\!<\!x)$} selbst nur eine symbolische
Schreibweise ist, wird diese symbolische Schreibweise im weiteren
auch f"ur komplexe Zufallsgr"o"sen verwendet. Bei einer reellen
Zufallsgr"o"se steht die Schreibweise \mbox{$\text{P}(\boldsymbol{x}\!<\!x)$}
f"ur die Wahrscheinlichkeit, dass das Ergebnis (\,Elementarereignis\,)
eines Zufallsexperimentes in dem Unterraum der Ergebnismenge des
Experimentes liegt, der nur die Ergebnisse enth"alt, denen die Werte
der Zufallsgr"o"se zugeordnet sind, die kleiner als $x$ sind.
Im Fall einer komplexen Zufallsgr"o"se verwende ich im weiteren die
symbolische Schreibweise \mbox{$\text{P}(\boldsymbol{x}\!<\!x)$}
f"ur die Wahrscheinlichkeit, dass ein Zufallsexperiment ein Ergebnis liefert,
das in dem Unterraum der Ergebnismenge liegt, der nur die Ergebnisse
enth"alt, denen die komplexen Werte der Zufallsgr"o"se zugeordnet sind,
deren Realteil kleiner als der Realteil von $x$ ist und deren
Imagin"arteil kleiner als der Imagin"arteil von $x$ ist. Die symbolische
Schreibweise \mbox{$\text{P}(\boldsymbol{x}\!<\!x)$}
sei also gleichbedeutend mit sonst "ublichen, aufwendigeren und
ebenfalls symbolischen Schreibweise
\mbox{$\text{P}\big(\Re\{\boldsymbol{x}\}<\Re\{x\}\wedge\Im\{\boldsymbol{x}\}<\Im\{x\}\big)$}.
Mit anderen Worten: es wird im weiteren unter der Verteilung einer
komplexen Zufallsgr"o"se immer die Verbundverteilung ihres
Real- und Imagin"arteils verstanden. Entsprechendes gilt f"ur die 
Verbundverteilung eines komplexen Zufallsvektors mit $L$ Elementen. Mit
\mbox{$\text{P}(\Vec{\boldsymbol{x}}<\Vec{x})$} sei die Verbundverteilung
\mbox{$\text{P}\big(\Re\{\boldsymbol{x}_1\!\}<\Re\{x_1\!\}\,\wedge\,
\Im\{\boldsymbol{x}_1\!\}<\Im\{x_1\!\}\,\wedge\,\ldots\,\wedge\,
\Re\{\boldsymbol{x}_L\}<\Re\{x_L\}\,\wedge\,\Im\{\boldsymbol{x}_L\}<\Im\{x_L\}\big)$}
der Real- und Imagin"arteile aller Elemente des Zufallsvektors bezeichnet.
Analog wird, wenn von der Dimension des komplexen Raumes, "uber
dem die Verteilung eines komplexen Zufallsvektors definiert ist,
oder wenn von einer Anzahl komplexer Freiheitsgrade die Rede ist,
gegebenfalls die entsprechende doppelte Dimension oder doppelte Anzahl
der Freiheitsgrade der entsprechenden reellen Real- und
Imagin"arteilzufallsgr"o"sen gemeint sein.

Eine Zufallsgr"o"se ist eine skalare Gr"o"se, also eine \mbox{$1\!\times\!1$}
Matrix. $L$ konkrete Realisierungen einer Zufallsgr"o"se bilden eine konkrete
Stichprobe vom Umfang $L$. Diese werde ich im Weiteren immer zu einem
{\em nicht}\/ zuf"alligen \mbox{$1\!\times\!L$} {\em Zeilen}\/vektor zusammenfassen.
Eine konkrete Stichprobe vom Umfang $L$ ist das Ergebnis von $L$
Zufallsexperimenten, n"amlich der $L$ Experimente "`Ziehen einer konkreten
Realisierung einer Zufallsgr"o"se"'. Nur wenn die Methode der Stichprobenentnahme
geeignet gew"ahlt wurde, sind die $L$ Zufallsgr"o"sen aller $L$
Zufallsexperimente unabh"angig und besitzen alle die gleiche\label{MathStich}
Verteilung, n"amlich die Verteilung der Zufallsgr"o"se, aus der
die Stichprobe entnommen wurde. Nur wenn diese beiden Voraussetzungen 
f"ur die Entnahme der Stichprobe erf"ullt sind, spricht man bei dem zuf"alligen
\mbox{$1\!\times\!L$} Zeilenvektor, der diese $L$ Zufallsgr"o"sen enth"alt,
von einer mathematischen Stichprobe vom Umfang $L$. In Gegensatz zur konkreten 
Stichprobe, dessen Elemente $L$ {\em nicht}\/zuf"allige, konkret gezogene Werte sind, 
sind die Elemente des mathematischen Stichprobenvektors unabh"angige Zufallsgr"o"sen 
mit identischer Verteilung.

Werden dagegen andere $M$ Zufallsgr"o"sen zu einem Zufallsvektor
zusammengefasst, dessen Elemente nicht die Zufallsgr"o"sen einer
mathematischen Stichprobe vom Umfang $L$ sind, so werden diese im Weiteren
zu einen zuf"alligen \mbox{$M\!\times\!1$} {\em Spalten}\/vektor zusammengefasst. 
Eine konkrete Realisierung eines solchen Zufallsspaltenvektors ist ebenfalls 
ein \mbox{$M\!\times\!1$} Spaltenvektor, der aber nicht zuf"allig ist. Eine 
konkrete Stichprobe vom Umfang $L$ eines \mbox{$M\!\times\!1$} Zufallsvektors 
ist eine nicht zuf"allige \mbox{$M\!\times\!L$} Matrix, die eine konkrete
Re"-ali"-sie"-rung der zuf"alligen \mbox{$M\!\times\!L$} Matrix der mathematischen 
Stichprobe des Zufallsvektors vom Umfang $L$ ist. Zufallsgr"o"sen, 
Zufallsvektoren, Zufallsmatrizen und Zufallsprozesse werden
in Fettdruck dargestellt. Konkrete Realisierungen, Stichproben
und daraus gewonnene Messwerte werden mit demselben Symbol aber
{\em nicht}\/ fettgedruckt dargestellt. Sch"atzwerte sind mit einem
Dach $\Hat{\phantom{x}}$ gekennzeichnet\footnote{Umgekehrt muss ein
$\Hat{\phantom{x}}$ {\em nicht}\/ bedeuten, dass es sich um
einen konkreten Sch"atzwert von etwas Zuf"alligem handelt.
Diese Kennzeichnung wird vor allen auch bei Mehrfachsummen
zur Unterscheidung unterschiedlicher Lauf"|indizes verwendet.}.

Bei Vektoren, die dadurch entstanden sind, dass mehrere Werte einer Folge
zusammengefasst worden sind oder, dass  Zufallsgr"o"sen unterschiedlicher
Werte des Parameters eines Prozesses zusammengefasst worden sind, wird zur
Unterscheidung der Elemente des Vektors das Argument der Folge bzw. der 
Parameter des Prozesses in runden Klammern verwendet. Steht hingegen die 
Vektorcharakteristik im Vordergrund, so wird die Schreibweise mit einem 
tiefgestellten Index f"ur den Zugriff auf einzelne Vektorelemente verwendet. 
Entsprechendes gilt f"ur Matrizen. 

Am Beispiel des Zufallsspaltenvektors $\Vec{\boldsymbol{V}}$ sei erl"autert, welche 
Elemente in Zeilen- und welche Elemente in Spaltenvektoren zusammengefasst 
werden, und es wird klargestellt, wann es sich bei diesen Elementen um konkrete, 
nichtzuf"allige Werte und wann um Zufallswerte handelt. Der Zufallsspaltenvektor 
$\Vec{\boldsymbol{V}}$ enth"alt die $M$ Elemente \mbox{$\boldsymbol{V}(\mu)$}, 
also die $M$ Zufallsgr"o"sen, die sich f"ur die $M$ ganzzahligen Werte $\mu$ 
von $0$ bis \mbox{$M\!-\!1$} des Parameters des Zufallsprozesses 
\mbox{$\boldsymbol{V}(\mu)$} des Spektrums der Erregung ergeben. 
Eine konkrete, Stichprobe vom Umfang $L$ eines Elementes --- des $\mu$-ten
Elementes --- ist der nicht zuf"allige Zeilenvektor \mbox{$\Vec{V}(\mu)$}
mit den $L$ Elementen \mbox{$V_{\lambda}(\mu)$}. Eine konkrete
Stichprobe vom Umfang $L$ des Zufallsvektors $\Vec{\boldsymbol{V}}$
ist demnach die nicht zuf"allige \mbox{$M\!\times\!L$} Matrix
$\underline{V}$. Jeder Spaltenvektor $\Vec{V}_{\lambda}$
ist ein Element der Stichprobe des Zufallsvektors $\Vec{\boldsymbol{V}}$
und somit eine konkrete Realisierung dieses Zufallsvektors.
Die \mbox{$M\!\times\!L$} Matrix $\underline{\boldsymbol{V}}$
ist die zuf"allige Matrix der {\em mathematischen}\/ Stichprobe vom
Umfang $L$ des Zufallsvektors $\Vec{\boldsymbol{V}}$. Sie ber"ucksichtigt,
dass das Entnehmen einer Stichprobe selbst ein Zufallsexperiment ist.
Die nicht zuf"allige Matrix $\underline{V}$ ist eine konkrete
Realisierung der zuf"alligen Matrix $\underline{\boldsymbol{V}}$.
Der zuf"allige Zeilenvektor \mbox{$\Vec{\boldsymbol{V}}(\mu)$}
ist die mathematische Stichprobe vom Umfang $L$ der Zufallsgr"o"se
\mbox{$\boldsymbol{V}(\mu)$} und somit zugleich eine Zeile der
Zufallsmatrix $\underline{\boldsymbol{V}}$.

Wenn von Normen die Rede ist, so ist bei Vektoren immer die
euklidische Norm gemeint, und bei Matrizen --- wenn
nichts anders gesagt wird --- die mit der euklidischen Vektornorm kompatible
Spektralnorm, die gleich dem gr"o"sten Singul"arwert
der Matrix ist. Abh"angig vom Kontext kann mit der L"ange eines Vektors
entweder dessen euklidische Norm, oder die Anzahl seiner Elemente --- also
seine von Eins verschiedene Dimension --- gemeint sein. 

Die Spaltenvektoren der Matrizen, die mit dem Index $\vphantom{x}_{\bot}$ 
gekennzeichnet sind, stehen auf einigen Vektoren orthogonal und spannen 
somit den Nullraum dieser Vektoren auf. Welche Vektoren dies im einzelnen 
sind, ist im Text detailliert beschrieben. 

$\lambda$ ist die Laufvariable der beim RKM auftretenden Mittelung.
Gr"o"sen, die den Index $f$ tragen, sind gefensterte Signale oder Musterfolgen,
deren Spektren oder die entsprechenden Zufallsprozesse, -gr"o"sen, -vektoren
und -matrizen. Gr"o"sen im Zeitbereich sind klein geschrieben und haben
als ver"anderliche Variablen $k$ oder $\kappa$.
Gr"o"sen im Frequenzbereich sind gro"s geschrieben und es werden als
ver"anderliche Variablen wahlweise die komplexen Variablen
$z$ bzw. $\Tilde{z}$ oder die reellen Frequenzen $\Omega$ bzw.
$\widetilde{\Omega}$ verwendet. Wenn f"ur die
reellen Frequenzen nur die "aquidistanten diskreten Werte
\mbox{$\Omega= \mu\CdoT2\pi/M$},
\mbox{$\Omega= \nu\CdoT2\pi/F$} bzw.
\mbox{$\widetilde{\Omega}=\eta\CdoT2\pi/\widetilde{M}$} mit
\mbox{$F,\,M,\,\widetilde{M}\,\in\,\mathbb{N}$} auftreten, werden die
Gr"o"sen im Frequenzbereich auch in Abh"angigkeit von $\mu$, $\nu$ bzw.
$\eta$ dargestellt. \mbox{$\mu,\,\nu,\,\eta\in{}\mathbb{Z}$}
werden dann als diskrete Frequenzvariablen bezeichnet. Bei den
kontinuierlichen Variablen $z$ und $\tilde{z}$ bzw.
$\Omega$ und $\widetilde{\Omega}$ entscheidet der "uberwiegende
Gebrauch, ob jeweils die reelle oder die komplexe Frequenzvariable
als ver"anderliche Variable verwendet wird. Wenn in Einzelf"allen
doch die jeweils andere Frequenzvariable zu verwenden ist, wird
an diesen Stellen \mbox{$z=e^{j\Omega}$} bzw.
\mbox{$\Tilde{z}=e^{j\widetilde{\Omega}}$}
f"ur die komplexe Frequenzvariable und \mbox{$\Omega=-j\Cdot\ln(z)$} bzw.
\mbox{$\widetilde{\Omega}=-j\Cdot\ln(\Tilde{z})$} f"ur die reelle Frequenzvariable
als Argument der Frequenzfunktion eingesetzt. Man h"atte auch
extra ein neues Formelzeichen f"ur eine Funktion einf"uhren k"onnen,
die nur in Ausnahmef"allen mit der jeweils nicht gew"ahlten
Frequenzvariable auftritt. Da dies eher verwirrend wirken d"urfte,
wurde darauf jedoch verzichtet, um insgesamt eine kompakte und gut
lesbare Darstellung der teils umfangreichen Formeln zu erreichen.

Einige Konstanten und Funktionen oder Folgen tragen trotz unterschiedlicher
Bedeutung gleiche Formelzeichen. Man kann diese dann dadurch unterscheiden,
dass das Formelzeichen der einen Bedeutung nie mit einem Index oder einem
Argument auftritt, w"ahrend das Formelzeichen der anderen Bedeutung
immer durch eine solche Kennzeichnung spezifiziert ist.
Eine Verwechslung ist dadurch ausgeschlossen.
Beispielsweise tritt die konstante L"ange $F$ der Fensterfolge \mbox{$f(k)$}
niemals mit einen Argument auf, w"ahrend die Fouriertransformierte
\mbox{$F(\Omega)$} der Fensterfolge \mbox{$f(k)$} immer als von
$\Omega$ abh"angig dargestellt wird. Gleiche Formelzeichen wurden nur
dann verwendet, wenn sich auf andere Art der enge Zusammenhang
mit einer anderen Gr"o"se --- im Beispiel ist das \mbox{$f(k)$} ---
nicht oder nur schlecht deutlich machen l"asst.

Bei einigen Formeln ist der G"ultigkeitsbereich der darin auftretenden
Parameter in Form einer Angabe der Menge, aus der sie entnommen
werden k"onnen, mit angegeben (~\,z.~B. \mbox{$\forall\;\mu\in{}\mathbb{Z}$}\,).
Wenn die zugrundeliegende Menge (\,z.~B. die ganzen
Zahlen\,) eines Parameters unzweifelhaft ist, wird diese Angabe auch
weggelassen. Ist eine Formel nur f"ur Parameter aus einer Teilmenge 
g"ultig, so wird meist nur der zul"assige Bereich des Parameters
mit Hilfe einer Ungleichung eingeschr"ankt. Beispielsweises
wird statt der etwas umst"andlichen korrekten Schreibweise
\mbox{$\forall\;\mu\in\{\;x\;\pmb{|}\;\;0\le x<M\;\wedge\;x\in\mathbb{Z}\,\}$}
die einfachere Angabe \mbox{$0\le\mu<M$} verwendet. Bei einigen Gleichungen
handelt es sich nicht um Formeln, bei denen ein Parameter einer bestimmten
Definitionsmenge entnommen werden muss, sondern es wird ein Parameter
verwendet, um eine kompakte Schreibweise f"ur eine Vielzahl von Gleichungen
zu erzielen. Dann wird f"ur diesen Parameter die Inkrementschreibweise
benutzt. Beispielsweise bedeutet \mbox{$\forall\;\;\;\mu=0\;(1)\;M\!-\!1$}
dass es sich um $M$ Gleichungen f"ur die unterschiedlichen Werte des
Parameters $\mu$ handelt. Diese Gleichungen k"onnen dann dazu dienen,
bereits definierte Gr"o"sen festzulegen oder noch nicht definierte
Gr"o"sen implizit oder explizit zu definieren. Je nachdem ob nur ein Teil
der Gr"o"sen oder alle festgelegt werden sollen, muss deren 
Gesamtanzahl nicht mit der Anzahl der Gleichungen "ubereinstimmen.
Dieselbe Schreibweise wird auch f"ur Wertzuweisungen verwendet,
wenn also mehrere Gr"o"sen zu berechnen sind.
Bei allen drei Schreibweisen (\,Mengenschreibweise, Ungleichung
oder Inkrementschreibweise\,) ist bei von Natur aus periodischen Folgen und
Funktionen gegebenfalls in den Argumenten mit Moduloarithmetik
zu rechnen, wenn zur Definition oder Berechnung neuer Gr"o"sen Werte verwendet
werden, die mit dem nicht modulo berechneten Argument bisher noch nicht
definiert oder berechnet wurden, mit dem modulo-reduzierten Argument aber
bereitstehen. Beispielsweise kann die Schreibweise
\[A(\mu)=\text{Funktion}(\mu)
\qquad\qquad\forall\qquad\mu=0\;(1)\;M\!-\!1
\]
bedeuten, dass die $M$ Variablen \mbox{$A(\mu)$} als $M$ spezielle
Funktionswerte definiert werden, oder dass die $M$ Gleichungen f"ur
alle Werte von $\mu$ gleichzeitig von den Werten \mbox{$A(\mu)$} und
der Funktion erf"ullt sein m"ussen, aber auch, dass man die $M$
Werte \mbox{$A(\mu)$} mit Hilfe der angegebenen Funktion zu berechnen hat.
Welche konkrete Interpretation zu w"ahlen ist wird aus dem Zusammenhang
ersichtlich. Wenn die folgende Gleichung als Definition oder als
Wertzuweisung interpretiert wird, so ist mit
\[
B(\mu)= A(\mu)\cdot A(-\mu)
\qquad\qquad\forall\qquad\mu=0\;(1)\;M\!-\!1
\]
bei einer mit $M$ periodischen Folge \mbox{$A(\mu)$} gemeint, dass in korrekter
Schreibweise ohne implizite Moduloarithmetik im Argument die Werte
\[B(\mu)= A(\mu)\cdot A(M-\mu)\quad\forall\quad\mu=1\;(1)\;M\!-\!1
\quad\text{ und }\quad B(0)= A(0)^2.
\]
definiert werden oder zu berechnen sind. Die vereinfachte Schreibweise
f"uhrt zu einer knapperen und besser lesbaren
Darstellung und f"uhrt an keiner Stelle zu Unklarheiten und wird vor
allem bei Fouriertransformierten von diskreten Folgen verwendet. Wenn 
---~abweichend vom letzten Beispiel~--- der zu definierende oder zu
berechnende Wert f"ur einen speziellen Wert von $\mu$ echt abweichend
definiert oder berechnet werden soll, wird dies nat"urlich explizit angegeben.

\chapter{Das Systemmodell}\label{SysMod}

In diesem Kapitel m"ochte ich erl"autern, wie die Aufspaltung des realen 
Systems in das in Bild~\ref{b1h} dargestellte, lineare Modellsystem 
und die "uberlagerte Modellst"orung erfolgt. Die Aufspaltung erfolgt dabei 
theoretisch, also auf Grund der im Allgemeinen unbekannten stochastischen 
Eigenschaften der am realen System anliegenden Zufallsprozesse. Hier 
auftretende Erwartungswerte sind also immer die Integrale "uber Produkte 
einer Funktion und einer Verteilungsdichtefunktion. Erst im n"achsten Kapitel, 
bei der Vorstellung eines Messverfahren, spielen empirisch gewonnene Mittelwerte 
eine Rolle. Dort wird von dem Messverfahren gezeigt, dass man damit in der Lage ist, 
die Aufspaltung des realen Systems in der Art empirisch vorzunehmen, dass man damit 
die in diesem Kapitel hergeleitete Aufspaltung erwartungstreu und konsistent absch"atzt. 

Im ersten Abschnitt wird hergeleitet, welches lineare Modellsystem sich bei einer 
optimalen Aufspaltung ergibt. Der zweite Abschnitt besch"aftigt sich mit der Frage, 
wie man den bei der Systemaufspaltung verbleibenden Fehlerprozess sinnvoll beschreiben 
kann. Dabei wird die Fensterung der am System anliegenden Prozesse eingef"uhrt. Im 
letzten Abschnitt wird gezeigt, dass sich bei geeigneter Wahl der Fensterfolge 
dieselbe optimale Systemaufspaltung ergibt, wie bei den ungefensterten, urspr"unglichen 
Prozessen.

\section{Das optimale lineare Modellsystem}\label{theo}

Zun"achst soll aufgezeigt werden, wie sich die Aufgabe der Systemaufspaltung im 
Zeitbereich exakt formulieren und l"osen l"asst. Es wird dann gezeigt, wie sich 
die Aufgabe der Minimierung des Approximationsfehlers in den Frequenzbereich 
"ubertragen l"asst, wenn man den erregenden Zufallsprozess geeignet substituiert. 
Es zeigt sich, dass es sich bei der dabei auftretenden Minimierungsaufgabe 
um eine lineare Regression der zweiten Art handelt. Zum Vergleich wird in 
diesem Unterabschnitt auch die Regression der ersten Art diskutiert. Dass
die Substitution des erregenden Zufallsprozesses problematisch sein kann, 
wird anschlie"send erl"autert. Welche L"osung der Minimierungsaufgabe sich im 
Frequenzbereich ergibt, wird abschlie"send hergeleitet.

\subsection{Systemaufspaltung im Zeitbereich}\label{SysZeit}

Das Modellsystem ist ein lineares, stabiles, zeitdiskretes
und i.~Allg. komplexwertiges System ${\cal S}$ dessen Antwort
\mbox{$h_{\kappa}(k)$} auf einen Impuls \mbox{$v(k)=\gamma_0(k\!-\!\kappa)$}
zum Zeitpunkt $\kappa$ wir zun"achst als vom Zeitpunkt des erregenden 
Impulses abh"angig ansetzen. 
Da reale Systeme immer kausal sind, ist es ausreichend ein 
Modellsystem anzusetzen, dessen Impulsantwort f"ur
\mbox{$k\!<\!\kappa$} Null ist.
Desweiteren kann man bei realen stabilen Systemen davon ausgehen, 
dass die Impulsantwort nach einer hinreichend gro"s gew"ahlten
Einschwingzeit $E$ soweit abgeklungen ist, dass die Impulsantwort
in guter N"aherung als zeitlich begrenzt anzusehen ist.
Man kann somit ein Modellsystem ansetzen, dessen Impulsantwort
nur f"ur \mbox{$k\le \kappa\!+\!E$} von Null verschieden sein kann.
Erregt wird das Modellsystem von dem Zufallsprozess
\mbox{$\boldsymbol{v}(k)$}. Am Ausgang des Modellsystems
erhalten wir den Prozess\vspace{-4pt}
\begin{equation}
\boldsymbol{x}(k)\;=
\Sum{\kappa=k-E}{k}\boldsymbol{v}(\kappa)\CdoT h_{\kappa}(k)
\qquad\qquad\forall\qquad k\in\mathbb{Z}.
\label{2.1}
\end{equation}
Der Prozess des Fehlers der Approximation des realen Systems
durch das Modellsystem wird mit \mbox{$\boldsymbol{n}(k)$} bezeichnet.
"Uberlagert man den Ausgangsprozess des Modellsystems und den
Approximationsfehlerprozess, so erh"alt man am Ausgang des gesamten
Systemmodells denselben Prozess \mbox{$\boldsymbol{y}(k)$}, wie am Ausgang
des realen Systems.
\begin{equation}
\boldsymbol{n}(k)\,=\,\boldsymbol{y}(k)-\boldsymbol{x}(k)
\label{2.2}
\end{equation}
Bei der Approximation des realen Systems durch das Modellsystem
w"ahlt man die Impuls"-antwort \mbox{$h_{\kappa}(k)$} in der Art,
dass das zweite Moment des Approximationsfehlerprozesses
\mbox{$\boldsymbol{n}(k)$} m"oglichst klein wird. In dem hier 
betrachteten Fall der Mittelwertfreiheit des 
Verbundzufallsprozesses aus \mbox{$\boldsymbol{v}(k)$} und 
\mbox{$\boldsymbol{y}(k)$} ist dies zugleich die Varianz, die das zweite
{\em zentrale}\/ Moment ist. Bei realen Systemen kann man davon ausgehen, 
dass dieses Moment immer existiert, da der Zufallsprozess 
\mbox{$\boldsymbol{n}(k)$} in der Amplitude
begrenzt ist, so dass dessen Verteilungsfunktion unter- bzw. oberhalb
bestimmter Werte immer $0$ bzw. $1$ ist. Das bei der Berechnung
des zweiten Moments "uber \mbox{$|n(k)|^2$} gebildete uneigentliche
Integral\footnote{Gegebenenfalls ist diese Integration im Stieltjesschen
Sinne durchzuf"uhren.} existiert daher immer. 

F"ur die Zeitpunkte $k$ erhalten wir folgende Minimierungsaufgaben 
f"ur das zweite Moment des Prozesses \mbox{$\boldsymbol{n}(k)$}:
\begin{equation}
\text{E}\big\{|\boldsymbol{n}(k)|^2\big\}\,=\,
\text{E}\Bigg\{\,\bigg|\,\boldsymbol{y}(k)\,-\!
\Sum{\kappa=k-E}{k}\!\!\boldsymbol{v}(\kappa)\CdoT h_{\kappa}(k)
\,\bigg|^2\Bigg\}\;\stackrel{!}{=}\;\text{minimal}\qquad
\forall\quad k\in\mathbb{Z}.
\label{2.3}
\end{equation}
Die L"osung dieser Minimierungsaufgabe erh"alt man f"ur jeden
Zeitpunkt $k$, indem man den zu minimierenden Term nach den zu
bestimmenden Gr"o"sen --- den Werten der Impulsantwort ---
partiell ableitet\footnote{Die Ableitung nach den komplexen Werten 
der Impulsantwort ist hier im klassischen Sinne nicht m"oglich, da 
die Betragsquadratbildung keine analytische (holomorphe) Funktion ist. 
Dies stellt jedoch kein Problem dar, weil der zu minimierende Term 
ebenso reell ist, wie die Realteile und die Imagin"arteile der 
Impulsantwort. Es handelt sich also um eine reelle Funktion mehrerer 
reeller Variablen. Von diesem reellen Term berechnet ganz konventionell 
die partiellen Ableitungen indem man zun"achst nach allen Realteilen,
und dann nach allen Imagin"arteilen der Impulsantwort partiell ableitet.
Man erh"alt so zwei Gleichungen f"ur jeden Wert der Impulsantwort. 
Die Gleichung, die man bei der partiellen Ableitung nach dem Imagin"arteil 
erh"alt, multipliziert man anschie"send mit $j$ und fasst sie 
mit der Gleichung, die man bei der partiellen Ableitung nach dem Realteil
erh"alt, zu einer komplexen Gleichung zusammen. Alternativ ist es m"oglich
die Ableitung im Sinne des Wirtinger Kalk"uls zu berechnen. Dies liefert 
dasselbe Ergebnis.} und zu Null setzt. Man erh"alt so f"ur jeden Zeitpunkt $k$ 
ein lineares Gleichungssystem zur Bestimmung der Werte der Impulsantwort, 
das sich in Matrixschreibweise folgenderma"sen darstellen l"asst:
\begin{equation}
\Vec{h}\cdot\underline{C}_{\boldsymbol{v},\boldsymbol{v}}=
\Vec{C}_{\boldsymbol{y},\boldsymbol{v}}.
\label{2.4}
\end{equation}
Dabei ist das Element in der $i$-ten Zeile und der $j$-ten Spalte
der \mbox{$(E\!+\!1)\times(E\!+\!1)$} Matrix $\underline{C}_{\boldsymbol{v},\boldsymbol{v}}$ 
die Kovarianz \mbox{$\text{E}\big\{\boldsymbol{v}(k\!+\!1\!-\!i)\CdoT
\boldsymbol{v}(k\!+\!1\!-\!j)^{\Kk}\big\}$}. Es sei noch einmal darauf hingewiesen, 
dass es sich hierbei um die Kovarianz handelt, die sich mit Hilfe der Verbundverteilung 
der beiden daran beteiligten Zufallsgr"o"sen berechnet. Um diese Kovarianz besser 
von der empirischen Kovarianz, die man messtechnisch mit Hilfe einer Mittelung "uber eine 
Stichprobe gewinnen w"urde, abzugrenzen, werde ich die erstgenannte Kovarianz im Weiteren 
auch als theoretische Kovarianz bezeichnen. Das $j$-te Element des Zeilenvektors 
$\Vec{C}_{\boldsymbol{y},\boldsymbol{v}}$ ist die Kovarianz 
\mbox{$\text{E}\big\{\boldsymbol{y}(k)\CdoT\boldsymbol{v}(k\!+\!1\!-\!j)^{\Kk}\big\}$}. 
Die zu bestimmenden Werte der Impulsantwort sind in dem 
Zeilenvektor $\Vec{h}$ zusammengefasst, wobei das $i$-te Element des Vektors 
\mbox{$h_{k+1-i}(k)$} ist. Da wir hier nur den Fall eines station"aren 
Verbundprozesses aus \mbox{$\boldsymbol{v}(k)$} und  \mbox{$\boldsymbol{y}(k)$} 
behandeln, ist sowohl die Matrix $\underline{C}_{\boldsymbol{v},\boldsymbol{v}}$ 
als auch der Vektor $\Vec{C}_{\boldsymbol{y},\boldsymbol{v}}$ f"ur alle Zeitpunkte 
$k$ identisch. Somit ergibt sich f"ur den Vektor der zu bestimmenden Werte der 
Impulsantwort eine L"osung bzw. ein L"osungsraum\footnote{Auch wenn die Matrix 
$\underline{C}_{\boldsymbol{v},\boldsymbol{v}}$ singul"ar ist, existiert hier immer
eine L"osung, weil der zu minimierende Term in Gleichung (\ref{2.3}) stets positiv ist 
und die Minimierungsparameter --- die Real- und Imagin"arteile der Impulsantwort --- nur 
linear oder quadratisch auftreten.}, der nicht von der Zeit $k$ 
abh"angt. Man kann daher ein zeitinvariantes Modellsystem mit einer zeitlich 
begrenzten Impulsantwort
\begin{equation}
h_{\kappa}(k)\;=\;
\begin{cases}
\quad h(k\!-\!\kappa)\qquad\qquad&\forall\qquad k\!-\!\kappa=0\;(1)\;E\\
\quad 0&\text{sonst.}
\end{cases}
\label{2.5}
\end{equation}
ansetzen, und es gen"ugt in diesem Fall die Minimierungsaufgabe (\ref{2.3}) 
f"ur {\em einen} beliebig w"ahlbaren Zeitpunkt $k$ zu l"osen.

\subsection{Die Minimierungsaufgabe im Frequenzbereich}

Die Anzahl \mbox{$E\!+\!1$} der maximal von Null verschiedenen Werte 
der gesuchten Impulsantwort kann sehr gro"s sein und ist i. Allg. a priori 
nicht bekannt. Deshalb, und weil man oft nur an der "Ubertragungsfunktion 
\begin{equation}
H(\Omega)\,=\,\Sum{\kappa=0}{E}\,h(\kappa)\cdot e^{\!-j\cdot\Omega\cdot \kappa}
\qquad\qquad\forall\qquad\Omega\in\mathbb{R}
\label{2.6}
\end{equation}
des Modellsystems f"ur die $M$ "aquidistanten Frequenzen
\mbox{$\Omega= \mu\CdoT2\pi/M$} mit \mbox{$\mu=0\;(1)\;M\!-\!1$}
interessiert ist, erscheint es jedoch sinnvoll, die Anzahl der freien 
Parameter in der Minimierungsaufgabe auf den frei w"ahlbaren, positiven, 
endlichen Wert $M$ in sinnvoller Gr"o"senordnung zu begrenzen. Man erreicht
dies, indem man den Zufallsprozess am Eingang des Systems 
geeignet festlegt. 

Zun"achst gibt man sich einen Zufallsvektor 
$\Vec{\boldsymbol{v}}$ der L"ange $M$ vor. Dieser Zufallsvektor bildet 
den Teil des Zufallsprozesses am Eingang des Systems f"ur den Zeitbereich 
\mbox{$0\le k<M$}. Den restlichen Zufallsprozess erh"alt man, indem man 
diesen Zufallsvektor periodisch wiederholt. Jede Musterfolge des 
Zufallsprozesses ist dann ein Signal, das mit $M$ periodisch ist.
Wegen der Periodizit"at des Zufallsprozesses am Eingang des
Systems sind dessen stochastische Eigenschaften durch die
stochastischen Eigenschaften des Ausschnitts \mbox{$k=0\;(1)\;M\!-\!1$}
und somit durch die stochastischen Eigenschaften des Zufallsvektor 
$\Vec{\boldsymbol{v}}$ bereits vollst"andig festgelegt. 

Durch die eineindeutige diskrete Fouriertransformation (\,DFT\,)
\begin{equation}
\boldsymbol{V}(\mu)\;=
\Sum{k=0}{M-1}\boldsymbol{v}(k)\cdot e^{\!-j\cdot\frac{2\pi}{M}\cdot\mu\cdot k}
\qquad\qquad\forall\qquad\mu=1\;(1)\;M\!-\!1
\label{2.7}
\end{equation}
des Zufallsvektors $\Vec{\boldsymbol{v}}$ erh"alt man $M$
Zufallsgr"o"sen \mbox{$\boldsymbol{V}(\mu)$}, die zu
einem neuen Zufallsvektor $\Vec{\boldsymbol{V}}$ der L"ange $M$
zusammengefasst werden, der zur Beschreibung des Zufallsprozesses am
Eingang des Systems genausogut geeignet ist, wie der Zufallsvektor
$\Vec{\boldsymbol{v}}$.

F"ur den Approximationsfehlerprozess nach Gleichung (\ref{2.2}) 
erh"alt man bei Verwendung des periodischen Eingangsprozesses
\begin{equation}
\boldsymbol{v}(k\!+\!\Tilde{k}\CdoT M) = \boldsymbol{v}(k)
\qquad\qquad\forall\qquad \Tilde{k}\in\mathbb{Z}
\label{2.8}
\end{equation}
mit Gleichung (\ref{2.1}) und dem Ansatz (\ref{2.5}) 
den Zufallsprozess mit den Elementen

\begin{gather}
\boldsymbol{n}(k)\;=\;\boldsymbol{y}(k)\,-\!
\Sum{\kappa=0}{E}\boldsymbol{v}(k\!-\!\kappa)\CdoT h(\kappa)\;=
\label{2.9}\\*[10pt]
=\;\boldsymbol{y}(k)\,-\!\Sum{\kappa=0}{M-1}\boldsymbol{v}(k\!-\!\kappa)\cdot
\!\Sum{\Tilde{k}=0}{E\,/\,M}h(\Tilde{k}\CdoT M+\kappa)\;=
\notag\\[14pt]
=\;\boldsymbol{y}(k)-
\frac{1}{M}\cdoT\Sum{\kappa=0}{M-1}\boldsymbol{v}(k\!-\!\kappa)\,\cdot
\Sum{\mu=0}{M-1}H\big({\T\mu\CdoT\frac{2\pi}{M}}\big)\cdot 
e^{j\cdot\frac{2\pi}{M}\cdot\mu\cdot\kappa}\;=
\notag\\*[12pt]
=\;\boldsymbol{y}(k)-\frac{1}{M}\cdoT\Sum{\mu=0}{M-1}
H\big({\T\mu\CdoT\frac{2\pi}{M}}\big)\CdoT
\boldsymbol{V}(\mu)\cdot
e^{j\cdot\frac{2\pi}{M}\cdot\mu\cdot k}.
\notag
\end{gather}
Damit ergibt sich aus der Minimierungsaufgabe (\ref{2.3}) die
modifizierte Minimierungsaufgabe\vspace{-6pt}
\begin{equation}
\text{E}\big\{\,|\boldsymbol{n}(k)|^2\big\}\,=\,
\text{E}\Bigg\{\,\bigg|\,\boldsymbol{y}(k)-
\frac{1}{M}\cdoT\Sum{\mu=0}{M-1}H\big({\T\mu\CdoT\frac{2\pi}{M}}\big)\CdoT 
\boldsymbol{V}(\mu)\CdoT e^{j\cdot\frac{2\pi}{M}\cdot\mu\cdot k}\,
\bigg|^2\Bigg\}\,\stackrel{!}{=}\,\text{minimal}\,,
\label{2.10}
\end{equation}
die nur mehr die $M$ neuen Optimierungsparameter \mbox{$H(\mu\CdoT2\pi/M)$}
enth"alt.\vspace{6pt}

Da es sich hier f"ur alle Werte von $k$ um ein und dieselbe Minimierungsaufgabe 
handelt, gibt es immer wenigstens eine L"osung f"ur die Optimierungsparameter 
\mbox{$H(\mu\CdoT2\pi/M)$}. Allenfalls kann sich ein L"osungsraum f"ur die 
$M$ Werte \mbox{$H(\mu\CdoT2\pi/M)$} ergeben. Der Fall, dass es keine L"osung 
gibt, kann nicht eintreten. Man beachte, dass auf Grund des Abtasttheorems durch die 
L"osung f"ur die $M$ Optimierungsparameter  \mbox{$H(\mu\CdoT2\pi/M)$}
der Systemapproximation nur die $M$ Werte der periodisch fortgesetzten Impulsantwort
\mbox{$\sum_{\Tilde{k}=-\infty}^{\infty}h(\kappa\!+\!\Tilde{k}\CdoT M)$}
festgelegt sind, weil der komplexe Drehfaktor bei der diskreten
Fouriertransformation nach Gleichung (\ref{2.6}) bei diesen Frequenzen in 
$\kappa$ mit $M$ periodisch ist. Nur wenn weitere Eigenschaften der Impulsantwort 
bekannt sind\footnote{ Z.~B. weil heuristische "Uberlegungen vermuten lassen, 
dass die Impulsantwort des realen Systems auf ein Intervall der L"ange 
$M$ beschr"ankt ist.}, ist \mbox{$h(\kappa)$} durch die $M$ 
Optimalwerte der "Ubertragungsfunktion eindeutig bestimmt.\vspace{6pt}

\subsection{Systemaufspaltung als lineare Regression}

F"ur einen beliebig gew"ahlten, festen Zeitpunkt $k$ ist der Term 
am Ausgang des Modellsystems 
\mbox{$x(k)\,=\,\frac{1}{M}\cdoT\sum_{\mu=0}^{M-1}H(\mu\CdoT2\pi/M)\CdoT
V(\mu)\cdot e^{j\cdot\frac{2\pi}{M}\cdot\mu\cdot k}$}
eine lineare Funktion der $M$ Variablen \mbox{$V(\mu)$}
des Vektors $\Vec{V}$. Bei dieser linearen "Uberlagerung sind die 
Koeffizienten \mbox{$H(\mu\CdoT2\pi/M)\cdot
e^{j\cdot\frac{2\pi}{M}\cdot\mu\cdot k}\,/M$}. Wenn nun die unabh"angigen
Variablen des Vektors \mbox{$\Vec{\boldsymbol{V}}$} zuf"allig sind, 
ist die lineare Funktion \mbox{$\boldsymbol{x}(k)$} dieser
Zufallsvariablen selbst eine Zufallsgr"o"se. Gleichung (\ref{2.10})
beschreibt, wie man die Koeffizienten zu w"ahlen hat, wenn man 
erreichen will, dass die lineare Funktion \mbox{$\boldsymbol{x}(k)$}
die Zufallsgr"o"se \mbox{$\boldsymbol{y}(k)$} in Sinne des kleinsten 
quadratischen Fehlers optimal ann"ahert. 

Allgemein bezeichnet man die Aufgabe eine Zufallsgr"o"se, die von ein oder 
mehreren Zufallsgr"o"sen abh"angt, durch eine lineare "Uberlagerung deterministischer 
Funktionen dieser Zufallsgr"o"sen mit einem m"oglichst kleinen Fehler zu approximieren, 
als lineare Regressionsaufgabe. Die lineare N"aherung \mbox{$x(k)$}, 
die man mit den so gew"ahlten Koeffizienten erh"alt, kann man f"ur den festen 
Zeitpunkt $k$ in einem \mbox{$M\!+\!1$}-dimensionalen Raum "uber dem 
\mbox{$M$-dimensionalen} Vektor $\Vec{V}$ auftragen, und man erh"alt eine 
$M$-dimensionale Hyperebene, die als Regressionshyperebene der zweiten Art 
bez"uglich \mbox{$\Vec{\boldsymbol{V}}$} bezeichnet wird. Die Koeffizienten 
werden Regressionskoeffizienten genannt. Im weiteren werde ich auch 
die Optimierungsparameter \mbox{$H(\mu\CdoT2\pi/M)$} ohne die 
Drehterme \mbox{$e^{j\cdot\frac{2\pi}{M}\cdot\mu\cdot k}\,/M$}, 
die nicht zur Optimierung der Systemapproximation verwendet werden k"onnen, 
als Regressionskoeffizienten bezeichnen. Diese Drehterme modifizieren 
f"ur den festen Zeitpunkt $k$ lediglich die Elemente des Vektors $\Vec{V}$,
so dass man sich die Regressionshyperebene in einem Raum mit einer anderen
Basis dargestellt denken kann. Der nach Einsetzen der Optimall"osung in 
(\ref{2.10}) verbleibende Wert hei"st Restdispersion.

In derselben Art kann man auch den bedingten Erwartungswert
\mbox{$\text{E}\big\{\,\boldsymbol{y}(k)\;\pmb{\big|}\;\Vec{V}\big\}$} 
der Zufallsgr"o"se \mbox{$\boldsymbol{y}(k)$} am Ausgang des Systems 
f"ur denselben festen Zeitpunkt $k$ in demselben 
\mbox{$M\!+\!1$}-dimensionalen Raum
"uber demselben $M$-dimensionalen Vektor $\Vec{V}$ auftragen.
Man erh"alt so einen i. Allg nichtlinearen $M$-dimensionalen Unterraum, 
der Regressionsfl"ache erster Art der zuf"alligen Ver"anderlichen
\mbox{$\boldsymbol{y}(k)$} bez"uglich $\Vec{\boldsymbol{V}}$ genannt wird.

In Bild \ref{b1e}
\begin{figure}[btp]
\begin{center}
{ \setlength{\unitlength}{1.3pt}
\begin{picture}(340,150)(0,0)
\input{mbild1e}
\put(123,135){\makebox(0,0)[r]{$y(k)$}}
\put(340,45){\makebox(0,0)[rt]{$V(\mu)\cdot
e^{j\cdot\frac{2\pi}{M}\cdot\mu\cdot k}$}}
\put(127,60){\makebox(0,0)[rt]{$0$}}
\put(145,20){\makebox(0,0)[l]{Restdispersion}}
\put(145,10){\makebox(0,0)[l]{in Gleichung (\ref{2.10})}}
\put(155,90){\makebox(0,0)[r]{II}}
\put(205,60){\makebox(0,0)[l]{I}}
\put(295,10){\makebox(0,0)[t]{1. Art}}
\put(290,81){\rotatebox[origin=lt]{15}{2. Art}}
\put(260,88){\rotatebox[origin=lt]{15}{Steigung:
$H\big({\T\mu\CdoT\frac{2\pi}{M}}\big)$}}
\end{picture}}
\end{center}\vspace{-18pt}
\setlength{\belowcaptionskip}{-2pt}
\caption{Regressionsfl"ache und Regressionshyperebene}
\label{b1e}
\rule{\textwidth}{0.5pt}\vspace{-4pt}
\end{figure}
werden die Regressionsfl"ache der ersten Art, die 
Regressionshyperebene der zweiten Art und die Bedeutung der Regressionskoeffizienten 
f"ur einen festen Zeitpunkt $k$ prinzipiell veranschaulicht. Um diese Zusammenh"ange 
"uberhaupt als zweidimensionale Graphik darstellen zu k"onnen, wird nur die Ebene 
des \mbox{$M\!+\!1$}-dimensionalen Raums dargestellt, die 
durch die Variable \mbox{$y(k)$} und {\em eine}\/ der 
Variablen \mbox{$V(\mu)$} aufgespannt wird. Au"serdem wird 
von den Zufallsgr"o"sen \mbox{$\boldsymbol{V}(\mu)\cdot 
e^{j\cdot\frac{2\pi}{M}\cdot\mu\cdot k}$} und 
\mbox{$\boldsymbol{y}(k)$} angenommen, dass sie reell sind, 
so dass die Imagin"arteile nicht graphisch dargestellt werden m"ussen. 
Die "`krumme"' Kurve ist die Regressionsfl"ache erster Art. 
Die Gerade, also die Regressionshyperebene zweiter Art, 
hat die Steigung \mbox{$H(\mu\CdoT2\pi/M)$}.
Desweiteren wurde eine konkrete Stichprobe vom Umfang $200$
der Zufallsgr"o"senpaare \mbox{$\boldsymbol{V}(\mu)\cdot
e^{j\cdot\frac{2\pi}{M}\cdot\mu\cdot k}$} und
\mbox{$\boldsymbol{y}(k)$} als Messpunkte eingetragen. Auch wenn
die Gerade nach Gleichung (\ref{2.10}) {\em nicht}\/ den quadratischen
Abstand zu den konkreten Messpunkten minimiert, sondern das zweite
Moment der zuf"alligen Abweichung, so wurde doch zur Interpretation
der Restdispersion der Minimierung nach Gleichung (\ref{2.10})
der entsprechende Abstand zu einem Messpunkt eingezeichnet.

Die lineare Systemapproximation mit den Regressionskoeffizienten, die
die Minimierungsaufgabe (\ref{2.10}) l"osen, liefert zugleich diejenige
Hyperebene, bei der das zweite Moment des Abstands der Hyperebene zur
Regressionsfl"ache der ersten Art minimal wird. Dieser Abstand ist 
in Bild \ref{b1e} mit {\bf I} gekennzeichnet. In Kapitel \ref{Regress} 
des Anhangs wird die Restdispersion in zwei Anteile aufgeteilt. Dessen 
erster Anteil ist genau das zweite Moment des ebengenannten Abstandes.
Dieses zweite Moment ist der erste Summand auf der rechten Seite der 
Gleichung (\ref{A.1.9}) des Anhangs \ref{Regress}. Er l"asst sich mit Hilfe 
der Regressionskoeffizienten minimieren. Der zweite Anteil der 
Restdispersion auf der rechten Seite der Gleichung (\ref{A.1.9}) ist 
das zweite Moment des Abstands der Zufallsgr"o"se \mbox{$\boldsymbol{y}$} 
zu der Regressionsfl"ache der ersten Art. Von diesem Anteil wird gezeigt, 
dass er konstant --- also von den Regressionskoeffizienten unabh"angig --- ist. 
Somit wird das zweite Moment des mit {\bf I} gekennzeichneten Abstands minimiert, 
wenn die Restdispersion minimiert wird. Der Abstand, dessen Varianz sich 
nicht minimieren l"asst, ist anhand eines Messpunktes in Bild \ref{b1e} 
eingetragen und mit {\bf II} gekennzeichnet.

Es sei noch einmal explizit darauf hingewiesen, dass weder
 die Regressionsfl"ache erster Art, noch die Lage der 
Regressionshyperebene zweiter Art zuf"allig sind. Im Gegensatz zur 
Regressionsfl"ache erster Art ist die Regressionshyperebene der zweiten 
Art jedoch von den stochastischen Eigenschaften der Erregung abh"angig. 
Man kann sich bei dem Beispiel in Bild \ref{b1e} leicht vorstellen, 
dass eine gr"o"sere Streuung bei der Zufallsgr"o"se 
\mbox{$\boldsymbol{V}(\mu)\cdot e^{j\cdot\frac{2\pi}{M}\cdot\mu\cdot k}$} 
zu einer geringeren Steigung der Geraden f"uhrt.

\subsection{Zur Wahl des Eingangsprozesses}

Als Eingangsprozess hatten wir einen periodischen Prozess gew"ahlt, 
um so eine fest vorgebbare Anzahl $M$ an Optimierungsparametern 
\mbox{$H(\mu\CdoT2\pi/M)$} zu erhalten. Bei Systemen, die empfindlich
auf eine Ver"anderung der stochastischen Eigenschaften des Eingangsprozesses 
reagieren, muss man bei der Wahl des Zufallsvektors $\Vec{\boldsymbol{V}}$, 
darauf achten, dass die wesentlichen stochastischen Eigenschaften des 
periodischen Erregungsprozesses mit denjenigen "ubereinstimmen, 
die der Prozess aufweist, der das System in normalen Betrieb erregt.
Nur dann kann man davon ausgehen, dass die L"osung der Regressionsaufgabe 
eine Re"-gres"-sions"-hyper"-ebene liefert, die mit der Hyperebene
"ubereinstimmt, die sich ergeben w"urde, wenn man das System mit dem
normalerweise im Betrieb vorliegenden Prozess erregen w"urde. Die 
optimalen Werte f"ur die "Ubertragungsfunktion, die man mit Hilfe des
periodischen Prozesses gewinnt, beschreiben dann ein Modellsystem, das 
das lineare Verhalten des realen Systems im normalen Betrieb zutreffend
charakterisiert. Da ein Messverfahren, wie z.~B. das im n"achsten 
Kapitel vorgestellte, mit dessen Hilfe man die optimalen Werte der 
"Ubertragungsfunktion empirisch bestimmt, bestenfalls die Werte
der optimalen theoretischen Systemmodellierung erwartungstreu 
absch"atzen kann, muss man sich bereits vor der Messung "uberlegen,
wie der erregende periodische Prozess zu w"ahlen ist, um aussagekr"aftige 
Messwerte zu erhalten. Im Allgemeinen werden heuristische "Uberlegungen
und an realen Systemen gemessene stochastische Eigenschaften bei der Wahl
eines geeigneten periodischen Eingangsprozesses hilfreich sein.
Bei vielen Systemen wird man darauf zu achten haben, dass man 
zumindest die Varianz sowie die Kovarianzen eng benachbarter
Signalwerte und u. U. auch die Verbundverteilungsfunktionen der
 Eingangsgr"o"sen \mbox{$\boldsymbol{v}(k_1)$} und 
\mbox{$\boldsymbol{v}(k_2)$} f"ur nahe benachbarte Zeitpunkte 
$k_1$ und $k_2$ dem realen Eingangsprozess entsprechend w"ahlt.
Beispiele f"ur Systeme, bei denen zu erwarten ist, dass sie 
empfindlich auf eine Ver"anderung der stochastischen Eigenschaften 
des Eingangssignals reagieren, sind Systeme, die auf diese hin 
optimiert worden sind.

Als Beispiel betrachten wir ein
System zur Nachrichten"ubertragung, das eine Vektorquantisierung
enth"alt. Man unterteilt hier zun"achst das zu "ubertragende
Signal in Abschnitte, die man zu Vektoren zusammenfasst. 
Den Vektorraum unterteilt man nun in eine 
begrenzte Anzahl durchnummerierter Unterr"aume. 
Man "ubertr"agt nun statt des Vektors nur die Nummer des Unterraums, 
in dem sich der Vektor befindet. Der Empf"anger generiert dann f"ur
jede Unterraumnummer einen repr"asentativen Vektor. Die 
aufeinanderfolgenden Repr"asentantenvektoren bilden das Ausgangssignal.
Die Unterr"aume und die Repr"asentanten werden dabei so gew"ahlt, dass 
der Erwartungswert des quadratischen Fehlers zwischen Ein- und Ausgangssignal
bei Verwendung des im normalen Betrieb des realen Systems vorliegenden 
Eingangsprozesses mit seinen spezifischen stochastischen Eigenschaften 
minimal wird. Bei dieser Vektorquantisierung wird man erwarten 
k"onnen, dass bei der Messung des Systems mit einem Eingangsprozess
mit stark abweichenden stochastischen Eigenschaften eine gr"o"sere Rauschleistung 
gemessen wird, als die, die im normalen Betrieb des Systems auftritt.

Wie dieses Beispiel zeigt, ist im Einzelfall immer zu pr"ufen, ob und in 
welcher Weise der i.~Allg. nichtperiodische, f"ur einen Betriebszustand 
typische Eingangsprozess durch einen periodischen Prozess ersetzt werden kann. 
Bei dem Beispiel der Vektorquantisierung w"are z.~B. die Periode des 
Zufallsvektors mehr als doppelt so gro"s zu w"ahlen als die Vektorl"ange 
der Vektorquantisierung und die Verbundverteilung der Vektorelemente
sollte dieselbe sein wie bei dem nichtperiodischen Eingangsprozess. 
An diesem Beispiel sieht man, dass man bei der periodischen 
Fortsetzung des Zufallsvektors $\Vec{\boldsymbol{v}}$ darauf
achten muss, dass auch die stochastischen Eigenschaften eines
Ausschnittes des Zufallsvektors, der einen Zeitpunkt enth"alt,
am dem die periodisch fortgesetzten Zufallsvektoren $\Vec{\boldsymbol{v}}$
aneinandersto"sen, mit den entsprechenden Eigenschaften des
typischen Eingangsprozesses "ubereinstimmen m"ussen.

Andererseits gibt es auch Systeme, bei
denen sich die L"osung f"ur \mbox{$H(\mu\CdoT2\pi/M)$} bei einer
Variation der Varianz des Prozesses $\boldsymbol{v}(k)$ nicht oder nur
unwesentlich "andert, wenn die Varianz innerhalb einer sinnvollen
Gr"o"senordnung bleibt (\,keine Unter- oder "Ubersteuerung des realen
Systems\,). Solche Systeme wurden in \cite{Dong} und \cite{Sch/D} als schwach
nichtlineare Systeme bezeichnet und f"ur solche Systeme wurde das RKM
zun"achst eingesetzt. 

Allgemein kann das RKM aber auch zur Vermessung anderer als schwach 
nichtlinearer Systeme verwendet werden. Dann sollten jedoch von den
stochastischen Eigenschaften der bei der Messung verwendeten Erregung
immer diejenigen angegeben werden, die einen wesentlichen
Einfluss auf die Ergebnisse haben. Desweiteren gibt es eine Vielzahl von
Systemen, bei denen die L"osung f"ur \mbox{$H(\mu\CdoT2\pi/M)$} nicht oder nur
unwesentlich von den stochastischen Eigenschaften des Prozesses
$\boldsymbol{v}(k)$ abh"angt. Solche Systeme lassen sich mit Zufallsprozessen
erregen, die besonders gut zur Messung mit dem RKM geeignet sind.
Man kann dann z.~B. einen normalverteilten unkorrelierten Zufallsvektor
verwenden, um den bereichsweise periodischen Prozess f"ur die Erregung
des zu vermessenden Systems zu generieren. Auch die Verwendung von
Mehrton- oder Chirpsignalen --- diese werden in Kapitel \ref{Spesig} vorgestellt ---
ist beim RKM vorteilhaft.

\subsection{L"osung der Systemapproximation im Frequenzbereich}

Im Weiteren wollen wir nun annehmen, dass die Substitution des f"ur den 
Betriebszustand typischen erregenden Zufallsprozesses durch den periodischen 
Zufallsprozess in der Art vorgenommen wurde, dass es f"ur die periodische
Erregung dieselbe L"osung bzw. denselben L"osungsraum  f"ur \mbox{$H(\mu\CdoT2\pi/M)$} 
gibt, wie im Fall der Erregung durch den f"ur den Betriebszustand typischen Zufallsprozess.
Daraus ergibt sich auch eine Forderung bez"uglich der Varianzen der Zufallsgr"o"sen
des Zufallsvektors $\Vec{\boldsymbol{V}}$, die weiter unten vorgestellt wird.

Der zu minimierende Term (\ref{2.10}) h"angt von den $M$
komplexen Parametern der "Ubertragungsfunktion \mbox{$H(\mu\CdoT2\pi/M)$} ab.
Zerlegt man diese komplexen Gr"o"sen in ihre Real- und Imagin"arteile,
so stehen genau \mbox{$2\CdoT M$} unabh"angige reelle Minimierungsparameter 
zur Verf"ugung. Sie k"onnen unabh"angig voneinander eingestellt werden. 
Der zu minimierende Ausdruck (\ref{2.10}) ist ein zweites Moment, 
also eine reelle, nichtnegative, und in den \mbox{$2\CdoT M$} reellen 
Minimierungsparametern stetige und stetig differenzierbare 
Funktion\footnote{Die zu minimierende Varianz 
\mbox{$\text{E}\big\{\,|\boldsymbol{n}(k)|^2\big\}$} l"asst sich in 
einer Form darstellen, die neben einer Konstanten nur lineare Terme
und eine quadratische Form in \mbox{$H(\mu\CdoT2\pi/M)$} enth"alt.
Die Matrix der quadratischen Form ist eine Kovarianzmatrix und
weist daher nur nichtnegative Eigenwerte auf. Die quadratische Form 
ist somit positiv semidefinit. Tr"agt man die zu minimierende Varianz
"uber den Minimierungsparametern auf, ergibt sich ein nach oben
ge"offnetes Paraboloid. Daher ist das Minimum der Varianz 
\mbox{$\text{E}\big\{\,|\boldsymbol{n}(k)|^2\big\}$} immer ein globales Minimum.}, 
die nur f"ur die Einstellung der Minimierungsparameter ein Minimum
aufweisen kann, bei der alle partiellen Ableitungen nach allen 
Minimierungsparametern Null werden. Wir leiten nun den Ausdruck 
(\ref{2.10}) jeweils nach dem Real- und Imagin"arteil von 
\mbox{$H(\Tilde{\mu}\CdoT2\pi/M)$}\footnote{$\mu$ ist der Lauf"|index  
der Summe in (\ref{2.10}). Um klarzustellen, dass man nur nach einem 
Minimierungsparameter, der lediglich bei einem Summanden der Summe 
auftritt, ableitet, wurde hier zu Unterscheidung $\Tilde{\mu}$ im 
Argument des Minimierungsparameters \mbox{$H(\Tilde{\mu}\CdoT2\pi/M)$}, 
nach dem partiell abgeleitet wird, gew"ahlt.} getrennt partiell 
ab\footnote{Die Ableitung nach den komplexen Werten ist hier im 
klassischen Sinne nicht m"oglich, da die Betragsquadratbildung 
keine analytische (holomorphe) Funktion ist. Die Ableitung im Sinne 
des Wirtinger Kalk"uls ist m"oglich und liefert dasselbe Ergebnis.}. 
Man erh"alt so \mbox{$2\CdoT M$} reelle Gleichungen, von denen man 
jeweils zwei Gleichungen als Real- und Imagin"arteilgleichung einer 
komplexen Gleichung zusammenfasst. Man erh"alt nach kurzer Rechnung 
$M$ komplexe Gleichungen zur Bestimmung der $M$ gesuchten Werte der 
"Ubertragungsfunktion. 
\begin{gather}
\frac{1}{M}\cdoT\Sum{\mu=0}{M-1}\text{E}\big\{
\boldsymbol{V}(\Tilde{\mu})^{\!*}\!\CdoT
\boldsymbol{V}(\mu)\big\}\cdot
e^{j\cdot\frac{2\pi}{M}\cdot\mu\cdot k}\cdot
H\big({\T\mu\CdoT\frac{2\pi}{M}}\big)\;=\;
\text{E}\big\{
\boldsymbol{V}(\Tilde{\mu})^{\!*}\!\CdoT
\boldsymbol{y}(k)\big\}
\notag\\*[4pt]
\forall\qquad \Tilde{\mu}=0\;(1)\;M\!-\!1
\label{2.11}
\end{gather}
Jede einzelne dieser $M$ Gleichungen unterwirft man nun einer 
Fouriertransformation. Dazu setzt man in die Gleichung f"ur 
einen festen Wert $\Tilde{\mu}$ die $M$ Werte 
\mbox{$k=0\;(1)\;M\!-\!1$} f"ur die diskrete Zeit $k$ ein, 
so dass man f"ur ein und denselben Wert $\Tilde{\mu}$ 
$M$ Gleichungen erh"alt, die aufgrund der angenommenen 
Stationarit"at der beteiligten Zufallprozesse alle dieselbe 
L"osung besitzen. Jede dieser $M$ Gleichungen multipliziert 
man nun auf beiden Seiten mit dem Drehfaktor 
\mbox{$e^{\!-j\cdot\frac{2\pi}{M}\cdot\Tilde{\mu}\cdot k}$}.
Abschlie"send summiert man die $M$ Gleichungen mit demselben
$\Tilde{\mu}$ auf, so dass man f"ur jedes $\Tilde{\mu}$ wieder
eine Gleichung erh"alt. F"ur alle $M$ m"oglichen Werte von 
$\Tilde{\mu}$ ergeben sich somit insgesamt die $M$ Gleichungen.
\begin{gather}
\text{E}\Big\{\big|\boldsymbol{V}(\Tilde{\mu})\big|^2\Big\}\cdot
H\big({\T\Tilde{\mu}\CdoT\frac{2\pi}{M}}\big)\;=\;
\text{E}\big\{\boldsymbol{V}(\Tilde{\mu})^{\!*}
\Cdot\boldsymbol{Y}(\Tilde{\mu})\big\}
\notag\\*[4pt]
\forall\qquad \Tilde{\mu}=0\;(1)\;M\!-\!1
\label{2.12}
\end{gather}

Wie man an dieser fast expliziten Form der $M$ Gleichungen sieht, 
ergibt sich f"ur die $M$ Werte der "Ubertragungsfunktion nur dann 
ein wenigstens eindimensionaler L"osungsraum, wenn eine oder mehrere 
zweite Momente der $M$ Zufallsgr"o"sen \mbox{$\boldsymbol{V}(\mu)$} 
Null sind. Hat man es mit einem System zu tun, bei dem dieser Fall 
eintritt, so kann man die durch die letzten Gleichungen nicht 
festgelegten Werte der "Ubertragungsfunktion auf beliebige Werte 
---~ z.~B. Null~--- setzen und so die Anzahl der Optimierungsparameter 
verringern. F"ur die restlichen Minimierungsparameter erh"alt man wieder 
eine eindeutige L"osung, die nicht von den willk"urlich festgelegten 
Werten der "Ubertragungsfunktion abh"angt. Damit erh"alt man nat"urlich nur 
eine noch weiter eingeschr"ankte Aussage "uber die "Uber"-tra"-gungs"-funk"-tion des 
realen Systems, als die Einschr"ankung auf die diskreten Frequenzen ohnehin 
schon bewirkt. Selbst bei Systemen, von denen bekannt ist, dass ihre Impulsantwort 
auf $M$ aufeinanderfolgende Werte zeitlich begrenzt ist, kann die Impulsantwort 
dann nicht mehr eindeutig berechnet werden. Da es nicht allzu anspruchsvoll, 
aber doch relativ umfangreich ist, den Fall verschwindender Varianzen getrennt 
zu behandeln, wird im weiteren angenommen, dass der Zufallsvektor $\Vec{\boldsymbol{V}}$
so gew"ahlt worden sei, dass dessen Varianz bei allen $M$ Frequenzen
{\em nicht}\/ Null wird, und somit eine eindeutige L"osung f"ur die
Werte der "Ubertragungsfunktion existiert.

\section[Leistungsdichtespektrum, Periodogramm und Fensterung]
{Leistungsdichtespektrum, Periodogramm und\\Fensterung}\label{W}

In diesem Kapitel werden die unterschiedlichen M"oglichkeiten aufgezeigt, 
das in $\Omega$ kontinuierliche LDS mit Hilfe eine endlichen Anzahl von 
Werten sinnvoll zu charakterisieren. Desweiteren wird gezeigt, dass man 
bei der Aufspaltung des realen Systems dasselbe Modellsystem erh"alt, wenn 
man statt der zeitlich unbegrenzten Prozesse gefensterte Ein- und 
Ausgangsprozesse verwendet, sofern man dabei eine geeignete Fensterfolge 
einsetzt. Es werden drei Forderungen aufgestellt, die eine Fensterfolge 
erf"ullen sollte, um beim RKM sinnvoll eingesetzt werden zu k"onnen.

In der Einleitung wurde festgelegt, dass in dieser Abhandlung 
nur der Fall mittelwertfreier station"arer Prozesse 
\mbox{$\boldsymbol{v}(k)$} und \mbox{$\boldsymbol{y}(k)$} behandelt 
wird. Der Prozess \mbox{$\boldsymbol{x}(k)$} am Ausgang des linearen 
stabilen Modellsystems ist dann ebenfalls station"ar und mittelwertfrei, 
und somit auch der nach Gleichung (\ref{2.2}) definierte 
Approximationsfehlerprozess \mbox{$\boldsymbol{n}(k)$}.

Die nach Gleichung (\ref{1.2}) definierte AKF h"angt nur 
von der Zeitdifferenz $\kappa$, nicht aber von der absoluten 
Lage $k$ der beiden am Produkt der Kovarianz beteiligten 
Zufallsgr"o"sen des Prozesses ab. Desweiteren besitzt die AKF 
einen geradesymmetrischen Real- und einen schiefsymmetrischen 
Imagin"arteil. Durch eine eindimensionale diskrete 
Fouriertransformation haben wir in Gleichung (\ref{1.3}) das
Leistungsdichtespektrum (LDS) erhalten, das reell und mit $2\pi$ periodisch, 
aber bei einem komplexen Prozess i. Allg. {\em nicht}\/ geradesymmetrisch ist, 
da die AKF einen von Null verschiedenen Imagin"arteil enthalten kann. 

\subsection{Abtastwerte des LDS}

I.~Allg. ist die AKF zeitlich nicht begrenzt. 
Ein Beispiel, das als N"aherung f"ur eine St"orung betrachtet 
werden kann, wie sie an realen Systemen durchaus auftreten kann, 
ist ein Sinuseintonst"orer mit zuf"alliger, gleichverteilter Phasenlage. 
Solch eine St"orung besitzt eine kosinusf"ormige AKF. 
Dementsprechend findet sich im LDS dieses Prozesses bei der 
positiven und der negativen Frequenz der periodischen St"orung
jeweils ein Dirac-Impuls. 

Selbst wenn man das LDS geschlossen angeben k"onnte, 
m"usste man sich bei der Messung immer auf die Angabe des 
LDS f"ur endlich viele Frequenzen beschr"anken. W"are die Frequenz 
der periodischen St"orung keine der ausgew"ahlten Frequenzen, so 
w"urde diese St"orung aufgrund der Impulseigenschaft ihres Spektrums 
mit den endlich vielen Werten des LDS nicht erfasst werden k"onnen. 
W"are dagegen die Frequenz der periodischen St"orung eine der ausgew"ahlten
Frequenzen so w"urde das LDS --- also die unendliche Summe in der 
Gleichung (\ref{1.3}) --- bei dieser Frequenz divergieren, und eine 
Angabe, die einen R"uckschluss auf die Leistung der periodischen 
St"orung zul"asst, w"are nicht m"oglich.

Betrachten wir uns dazu das Beispiel in 
\begin{figure}[btp]
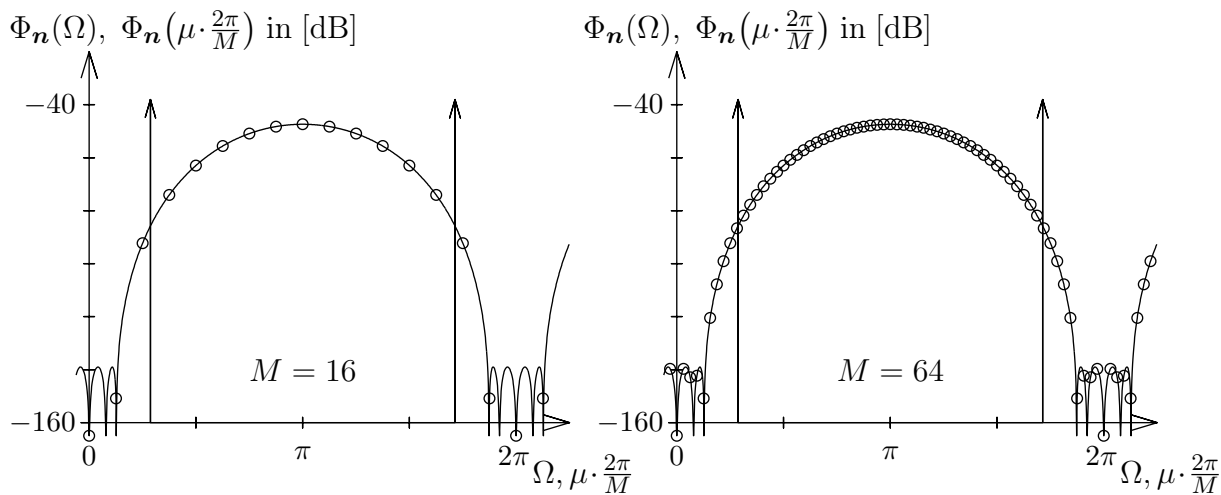

\begin{center}
{ 
\begin{picture}(454,190)
\input{mbild1c1}
\input{mbild1c2}
\put(0,175){$\Phi_{\boldsymbol{n}}(\Omega),\;
\Phi_{\boldsymbol{n}}\big(\mu\CdoT\frac{2\pi}{M}\big)\;\text{in}\;[\text{dB}]$}
\put(215,175){$\Phi_{\boldsymbol{n}}(\Omega),\;
\Phi_{\boldsymbol{n}}\big(\mu\CdoT\frac{2\pi}{M}\big)\;\text{in}\;[\text{dB}]$}
\put(25,150){\makebox(0,0)[r]{\small$-40$}}
\put(25,30){\makebox(0,0)[r]{\small$-160$}}
\put(245,150){\makebox(0,0)[r]{\small$-40$}}
\put(245,30){\makebox(0,0)[r]{\small$-160$}}
\put(30,21){\makebox(0,0)[t]{\small$0$}}
\put(110,21){\makebox(0,0)[t]{\small$\pi$}}
\put(189,21){\makebox(0,0)[t]{\small$2\pi$}}
\put(250,21){\makebox(0,0)[t]{\small$0$}}
\put(330,21){\makebox(0,0)[t]{\small$\pi$}}
\put(409,21){\makebox(0,0)[t]{\small$2\pi$}}
\put(215,18){\makebox(0,0)[t]{$\Omega,\mu\CdoT\frac{2\pi}{M}$}}
\put(435,18){\makebox(0,0)[t]{$\Omega,\mu\CdoT\frac{2\pi}{M}$}}
\put(110,50){\makebox(0,0){$M=16$}}
\put(330,50){\makebox(0,0){$M=64$}}
\end{picture}}
\end{center}\vspace{-25pt}
\setlength{\belowcaptionskip}{-4pt}
\caption[Abtastwerte des Leistungsdichtespektrums eines Prozesses mit
einem periodischen St"oranteil]{Abtastwerte des
Leistungsdichtespektrums eines Prozesses mit einem periodischen St"oranteil.
Das LDS des Prozesses ist mit der durchgezogenen Linie dargestellt,
w"ahrend die Abtastwerte des LDS  mit "`o"' gekennzeichnet sind.}
\label{b1c}
\rule{\textwidth}{0.5pt}\vspace{-6pt}
\end{figure}
Bild \ref{b1c}, bei 
dem ein Sinuseintonst"orer einer weiteren St"orung "uberlagert ist, die durch 
Filterung eines reellen, wei"sen Prozesses entstanden ist und dessen LDS 
"uber der Frequenz stark schwankt. Der Dirac-Impuls, der sich als Fouriertransformierte 
des periodischen Anteils in der AKF ergibt, ist als Pfeil symbolisch dargestellt, 
wobei die H"ohe des Pfeils der St"arke des Dirac-Impulses --- also $\pi$ mal dem
halben Quadrat der Amplitude des Sinusst"orers mit der Zufallsphase ---
entspricht. Man erkennt, dass die durch kleine Kreise 
markierten Abtastwerte die Sinuseintonst"orung nicht erfassen. Dies gilt sowohl 
f"ur \mbox{$M=16\,$} als auch f"ur \mbox{$M=64$} Abtastwerte des LDS. Andererseits kann 
der Verlauf des kontinuierlichen Anteils im LDS exakt rekonstruiert werden, 
wenn man wei"s, dass die AKF zeitlich hinreichend begrenzt ist.

\subsection{Stufenapproximation des LDS}

\begin{figure}[btp]
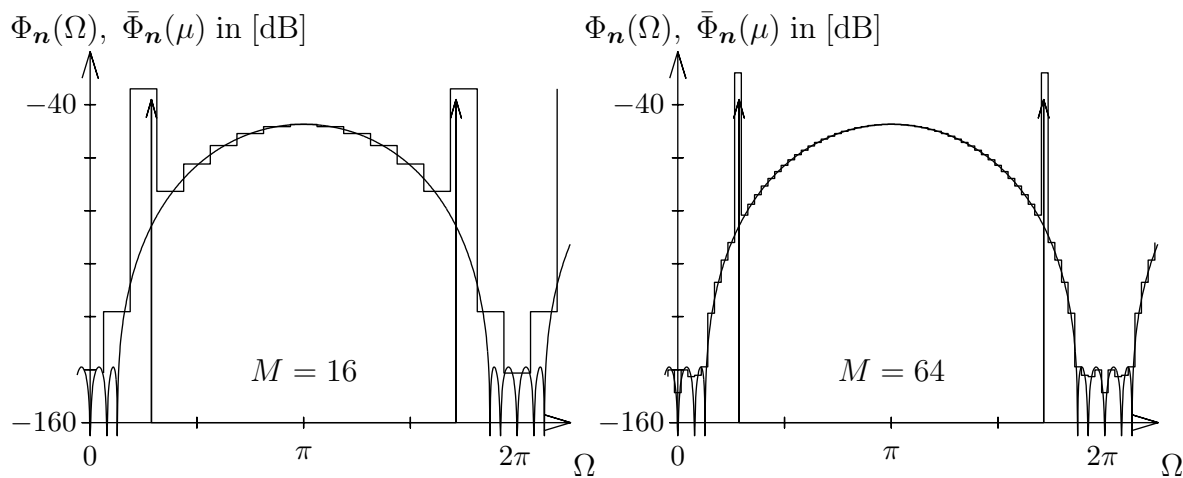

\begin{center}
{ 
\begin{picture}(454,190)
\input{mbild1b1}
\input{mbild1b2}
\put(0,175){$\Phi_{\boldsymbol{n}}(\Omega),\;
\Bar{\Phi}_{\boldsymbol{n}}(\mu)\;\text{in}\;[\text{dB}]$}
\put(215,175){$\Phi_{\boldsymbol{n}}(\Omega),\;
\Bar{\Phi}_{\boldsymbol{n}}(\mu)\;\text{in}\;[\text{dB}]$}
\put(25,150){\makebox(0,0)[r]{\small$-40$}}
\put(25,30){\makebox(0,0)[r]{\small$-160$}}
\put(245,150){\makebox(0,0)[r]{\small$-40$}}
\put(245,30){\makebox(0,0)[r]{\small$-160$}}
\put(30,21){\makebox(0,0)[t]{\small$0$}}
\put(110,21){\makebox(0,0)[t]{\small$\pi$}}
\put(189,21){\makebox(0,0)[t]{\small$2\pi$}}
\put(250,21){\makebox(0,0)[t]{\small$0$}}
\put(330,21){\makebox(0,0)[t]{\small$\pi$}}
\put(409,21){\makebox(0,0)[t]{\small$2\pi$}}
\put(215,18){\makebox(0,0)[t]{$\Omega$}}
\put(435,18){\makebox(0,0)[t]{$\Omega$}}
\put(110,50){\makebox(0,0){$M=16$}}
\put(330,50){\makebox(0,0){$M=64$}}
\end{picture}}
\end{center}\vspace{-25pt}
\setlength{\belowcaptionskip}{-4pt}
\caption[Stufenapproximation des Leistungsdichtespektrums eines Prozesses mit
einem periodischen St"oranteil]{Stufenapproximation des
Leistungsdichtespektrums eines Prozesses mit einem periodischen St"oranteil.}
\label{b1b}
\rule{\textwidth}{0.5pt}\vspace{-6pt}
\end{figure}

Will man die gesamte St"orung bei einem Prozess erfassen, dessen AKF nicht 
hinreichend begrenzt ist, so erscheint es nicht sinnvoll, die Abtastwerte des 
LDS f"ur bestimmte Frequenzen anzugeben. Wir werden daher versuchen einen 
Satz von $M$ Werten zu finden, der zur Beschreibung der spektralen 
Eigenschaften des Approximationsfehlerprozesses besser geeignet ist. 
Eine m"ogliche und sinnvolle Wahl ist es, f"ur die $M$ Frequenzpunkte 
\mbox{$\Omega=\mu\CdoT2\pi/M$} jeweils das Integral "uber das LDS 
in einer symmetrischen Umgebung der Breite \mbox{$2\pi/M$} zu bilden, 
und die auf diese Breite normierten Werte anzugeben. 
\begin{gather}
\Bar{\Phi}_{\boldsymbol{n}}(\mu)\;=\;
\frac{M}{2\pi}\cdoT\!\Int{\mu\cdot\frac{2\pi}{M}-\frac{\pi}{M}}
{\mu\cdot\frac{2\pi}{M}+\frac{\pi}{M}}\!\!
\Phi_{\boldsymbol{n}}(\Omega)\cdot\, d\Omega\;=\;
\frac{M}{2\pi}\cdoT\!\Int{-\frac{\pi}{M}}{\frac{\pi}{M}}
\Phi_{\boldsymbol{n}}\big({\T\mu\CdoT\frac{2\pi}{M}\!-\!\Omega}\big)
\cdot\, d\Omega
\notag\\*[6pt]
\forall\qquad \mu=0\;(1)\;M\!-\!1
\label{2.13}
\end{gather}
Mit Hilfe dieser Werte l"asst sich das zu messende LDS durch eine
Treppenfunktion approximieren, die in jedem Intervall der Breite
\mbox{$2\pi/M$} die gleiche Fl"ache aufweist, wie das LDS. F"ur die Werte
\mbox{$M=16$} und \mbox{$M=64$} ist diese Approximation an demselben Beispiel
mit dem periodischen St"oranteil in Bild \ref{b1b}
dargestellt. 
Man erkennt hier, dass die Treppenkurve, deren Plateaus durch die $M$ Werte 
\mbox{$\Bar{\Phi}_{\boldsymbol{n}}(\mu)$} festgelegt sind, die sich aus 
dem theoretischen LDS mit Hilfe der letzten Gleichung berechnen lassen,
eine Aussage "uber die ungef"ahre Lage der Frequenz des Sinusst"orers zul"asst. 
Man erkennt au"serdem, dass der Wert
\mbox{$\Bar{\Phi}_{\boldsymbol{n}}(\mu)$} bei dem Frequenzintervall, das
den periodischen St"oranteil beinhaltet, mit steigendem $M$ zunimmt.
Das liegt daran, dass bei gleichbleibender Leistung des periodischen
St"oranteils\footnote{Bei der Berechnung der Leistung des periodischen 
St"oranteils ist das Integral "uber die Dirac Impulse in Gleichung 
(\ref{2.13}) in Sinne der Distributionentheorie auszuwerten.} die 
Intervallbreite $2\pi/M$ mit steigendem $M$ abnimmt. Auf die Intervallbreite 
wurde das Integral in der letzten Gleichung aber gerade normiert. 
Da $M$ vorgegeben --- also bekannt --- ist, kann somit anhand des Wertes 
\mbox{$\Bar{\Phi}_{\boldsymbol{n}}(\mu)$} die Leistung des periodischen 
St"oranteils abgelesen werden. 

Die Wahl diese $M$ Werte \mbox{$\Bar{\Phi}_{\boldsymbol{n}}(\mu)$} statt 
der $M$ Abtastwerte \mbox{$\Phi_{\boldsymbol{n}}(\mu\CdoT2\pi/M)$} des LDS 
anzugeben, bietet drei wesentliche Vorteile. Der erste Vorteil besteht 
darin, dass periodische St"oranteile leistungsrichtig im LDS erscheinen. 
Zum Zweiten l"asst sich die Gesamtleistung des Approximationsfehlerprozesses, 
die sich als das Integral "uber das LDS berechnen l"asst, aus 
den $M$ Werten \mbox{$\Bar{\Phi}_{\boldsymbol{n}}(\mu)$} exakt und einfach 
mit Hilfe einer Summation berechnen, wie die folgende Umformung zeigt:
\begin{gather}
\sigma_{\boldsymbol{n}}^2\;=\;
\text{E}\big\{\,|\boldsymbol{n}(k)|^2\big\}\;=\;
\frac{1}{2\pi}\cdoT\!\Int{-\pi}{\pi}
\Phi_{\boldsymbol{n}}(\Omega)\cdot d\Omega\;=
\label{2.14}\\*[6pt]
=\;\frac{1}{2\pi}\cdoT\Sum{\mu=0}{M-1}
\Int{-\frac{\pi}{M}}{\frac{\pi}{M}}
\!\Phi_{\boldsymbol{n}}\big({\T\mu\CdoT\frac{2\pi}{M}\!-\!\Omega}\big)\cdot
d\Omega\;=\;
\frac{1}{M}\cdoT\Sum{\mu=0}{M-1}\Bar{\Phi}_{\boldsymbol{n}}(\mu).\notag
\end{gather}
Der dritte Vorteil besteht darin, dass die Frequenzunsch"arfe auf ein
Frequenzintervall der Breite \mbox{$2\pi/M$} begrenzt ist. D.~h. Nur der
Anteil des LDS geht in den Wert bei einer diskreten Frequenz $\mu$ ein, der
in der auf \mbox{$\pm\pi/M$} begrenzten unmittelbaren Umgebung der Frequenz 
\mbox{$\mu\CdoT2\pi/M$} liegt.

\subsection{Periodogramm}

Obwohl die $M$ Werte \mbox{$\Bar{\Phi}_{\boldsymbol{n}}(\mu)$} theoretisch 
geeignet erscheinen, das LDS mit endlich vielen Werten zu beschreiben, 
kann man diese Werte in der Praxis nicht verwenden. Wie wir sp"ater noch 
sehen werden, ben"otigt man zur Berechnung dieser $M$ Werte die gesamte 
AKF f"ur alle Zeitpunkte \mbox{$\kappa\in{}\mathbb{Z}$}.
Man kann also die $M$ Werte \mbox{$\Bar{\Phi}_{\boldsymbol{n}}(\mu)$}
nur angeben, wenn die  AKF a priori bekannt ist. In diesem Fall 
ben"otigt man kein Messverfahren. Wenn es jedoch darum geht,
messtechnisch eine Aussage "uber das LDS zu gewinnen, wird man aufgrund
der endlichen Messdauer nicht die komplette AKF, und somit 
auch nicht die Werte \mbox{$\Bar{\Phi}_{\boldsymbol{n}}(\mu)$} absch"atzen 
k"onnen. Hier tritt nun das bekannte Problem auf, wie man anhand eines 
endlich langen Ausschnitts des Approximationsfehlerprozesses $M$ Werte 
erhalten kann, die eine "ahnlich gute Beschreibung des LDS liefern, wie 
die $M$ Werte \mbox{$\Bar{\Phi}_{\boldsymbol{n}}(\mu)$}. Im weiteren sei
durch \mbox{$k\in[0;F)$} der betrachtete Ausschnitt des 
Approximationsfehlerprozesses festgelegt. Die Aufgabe besteht nun darin,
die $F$ Zufallsgr"o"sen dieses Ausschnitts in geeigneter Weise zu
verkn"upfen, um so $M$ Zufallsgr"o"sen zu erhalten, deren 
Erwartungswerte wenigstens n"aherungsweise mit den $M$ Werten
\mbox{$\Bar{\Phi}_{\boldsymbol{n}}(\mu)$} "ubereinstimmen. Die Messung 
dieser $M$ N"aherungswerte f"ur \mbox{$\Bar{\Phi}_{\boldsymbol{n}}(\mu)$}
besteht dann darin, deren Erwartungswerte durch ihre empirischen Mittelwerte,
die man aus einer konkreten Stichprobe gewinnt, abzusch"atzen. 

Es ist bekannt, dass die $M$ Werte des Periodogramms
solche Zufallsgr"o"sen sind. Das Periodogramm berechnet 
sich in zwei Schritten aus dem zeitlich auf das Intervall
\mbox{$[0;F)$} begrenzten Approximationsfehlerprozess. Zun"achst w"ahlt
man die L"ange $F$ des Messintervalls gleich der Anzahl $M$ der abzusch"atzenden
Werte und bildet die DFT des zeitbegrenzten Prozesses f"ur die $M$ Frequenzen
\mbox{$\Omega=\mu\CdoT2\pi/M$} mit \mbox{$\mu=0\;(1)\;M\!-\!1$}.
Nun bildet man von diesen Zufallsgr"o"sen das
Betragsquadrat und normiert sie auf $M$, so erh"alt man die $M$ Werte des
Periodogramms:
\[
\Tilde{\Phi}_{\boldsymbol{n}}(\mu)\;=\;
\text{E}\Big\{\,\frac{1}{M}\cdot|\boldsymbol{N}(\mu)|^2\Big\}\;=\;
\frac{1}{M}\cdot\text{E}\Bigg\{\,\bigg|\Sum{k=0}{M-1}
\!\boldsymbol{n}(k)\cdot
e^{\!-j\cdot\frac{2\pi}{M}\cdot\mu\cdot k}\bigg|^2\,\Bigg\}
\]
Wie Bild 
\begin{figure}[t!]
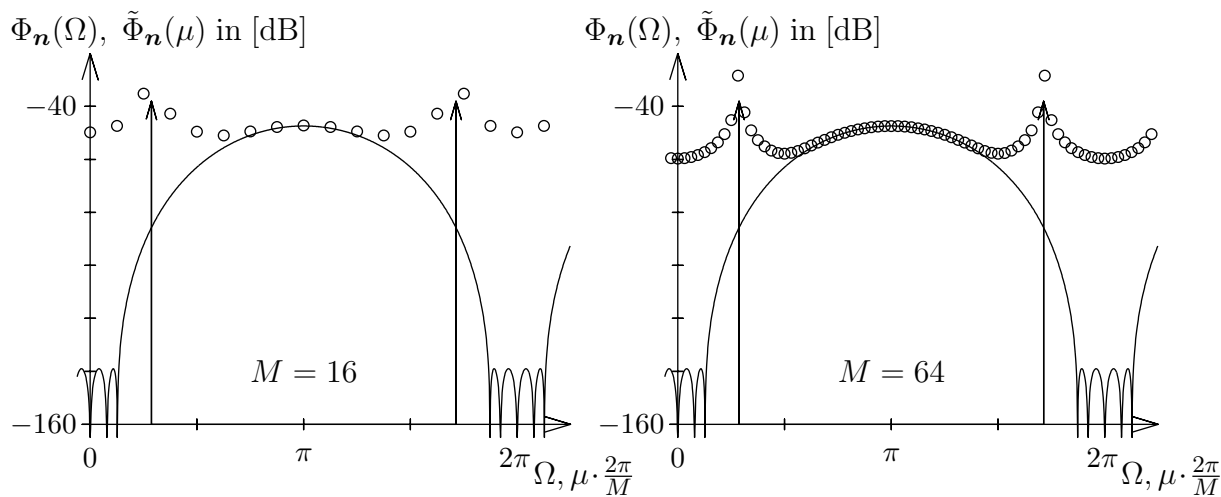

\begin{center}
{ 
\begin{picture}(454,188)
\input{mbild1d1}
\input{mbild1d2}
\put(0,175){$\Phi_{\boldsymbol{n}}(\Omega),\;
\Tilde{\Phi}_{\boldsymbol{n}}(\mu)\;\text{in}\;[\text{dB}]$}
\put(215,175){$\Phi_{\boldsymbol{n}}(\Omega),\;
\Tilde{\Phi}_{\boldsymbol{n}}(\mu)\;\text{in}\;[\text{dB}]$}
\put(25,150){\makebox(0,0)[r]{\small$-40$}}
\put(25,30){\makebox(0,0)[r]{\small$-160$}}
\put(245,150){\makebox(0,0)[r]{\small$-40$}}
\put(245,30){\makebox(0,0)[r]{\small$-160$}}
\put(30,21){\makebox(0,0)[t]{\small$0$}}
\put(110,21){\makebox(0,0)[t]{\small$\pi$}}
\put(189,21){\makebox(0,0)[t]{\small$2\pi$}}
\put(250,21){\makebox(0,0)[t]{\small$0$}}
\put(330,21){\makebox(0,0)[t]{\small$\pi$}}
\put(409,21){\makebox(0,0)[t]{\small$2\pi$}}
\put(215,18){\makebox(0,0)[t]{$\Omega,\mu\CdoT\frac{2\pi}{M}$}}
\put(435,18){\makebox(0,0)[t]{$\Omega,\mu\CdoT\frac{2\pi}{M}$}}
\put(110,50){\makebox(0,0){$M=16$}}
\put(330,50){\makebox(0,0){$M=64$}}
\end{picture}}
\end{center}\vspace{-21pt}
\setlength{\belowcaptionskip}{-6pt}
\caption[Erwartungswert des Periodogramms des Prozesses
mit einem periodischen St"oranteil]{Erwartungswert des Periodogramms des
Prozesses mit einem periodischen St"oranteil.
Das LDS des Prozesses ist mit der durchgezogenen Linie dargestellt,
w"ahrend die N"aherung mit Hilfe des Erwartungswertes des Periodogramms
des Prozesses mit "`o"' gekennzeichnet ist.}
\label{b1d}
\rule{\textwidth}{0.5pt}\vspace{5pt}
\end{figure}
\ref{b1d} zeigt, gelingt es mit den $M$ Werten des Periodogramms zwar 
die Lage der beiden Impulse ungef"ahr abzusch"atzen, aber das LDS des gefilterten 
Prozesses ist in deren Umgebung "uberhaupt nicht messbar. Daher wird das Periodogramm 
eigentlich immer nur dann eingesetzt, wenn man wei"s, dass das LDS "uber der Frequenz 
nur unwesentlich schwankt.

\subsection{Periodogramm des gefensterten Prozesses}

Die Tatsache, dass die Erwartungswerte der so erhaltenen Zufallsgr"o"sen 
nur eine sehr schlechte N"aherung f"ur die $M$ abzusch"atzenden 
Gr"o"sen \mbox{$\Bar{\Phi}_{\boldsymbol{n}}(\mu)$} sind, wenn das LDS 
einen "uber der Frequenz stark schwankenden Verlauf aufweist, ist lange bekannt. 
Aus diesem Grund haben Welch und Bartlett eine Methode entwickelt, 
mit deren Hilfe sich die Gr"o"sen \mbox{$\Bar{\Phi}_{\boldsymbol{n}}(\mu)$} 
auch in diesem Fall besser absch"atzen lassen. Sie ist beispielweise in 
\cite{Kam} beschrieben. Bei dieser Methode wird statt des im Intervall 
\mbox{$[0;M)$} gleich stark --- n"amlich mit eins --- gewichteten Zufallsprozesses, 
der mit unterschiedlichen Gewichten \mbox{$f(k)$} gewichtete Zufallsprozess 
im Intervall \mbox{$[0;F)$} zur Fouriertransformation verwendet. 
Durch die anschlie"sende Betragsquadratbildung und geeignete Normierung 
erh"alt man so $M$ Zufallsgr"o"sen, deren Erwartungswerte eine bessere 
N"aherung der $M$ Gr"o"sen \mbox{$\Bar{\Phi}_{\boldsymbol{n}}(\mu)$} sind.

Um aus einer zeitlich unbegrenzten Folge einen Ausschnitt 
herauszuschneiden, wobei unterschiedliche Werte des Ausschnitts 
unterschiedlich gewichtet werden, multipliziert man die unbegrenzte 
Folge mit einer Folge, die au"serhalb des herauszuschneidenden 
Zeitbereichs immer Null ist:
\begin{equation}
f(k)=0\qquad\qquad\forall\qquad k<0\quad\vee\quad k\ge F.
\label{2.15}
\end{equation}
Diesen Vorgang bezeichnet man als Fensterung und dementsprechend wird 
die Folge \mbox{$f(k)$} als Fensterfolge oder kurz als Fenster bezeichnet. 

\begin{figure}[t!]
\begin{center}
{ \setlength{\unitlength}{0.8425pt}
\begin{picture}(535,347)(-6,0)
\input{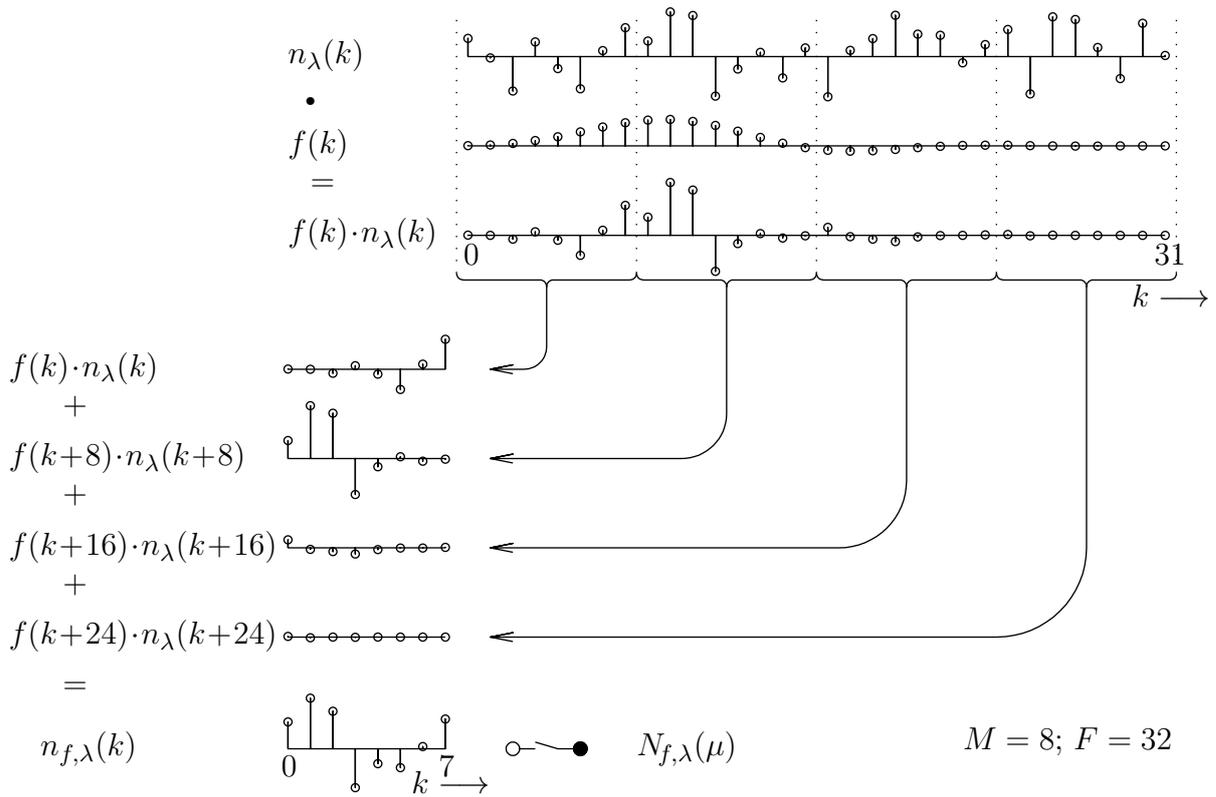}
\put(120,327){$n_{\lambda}(k)$}
\put(130,310){\circle*{3}}
\put(120,287){$f(k)$}
\put(130,270){$=$}
\put(120,247){$f(k)\CdoT n_{\lambda}(k)$}
\put(-4,187){$f(k)\CdoT n_{\lambda}(k)$}
\put(20,170){$+$}
\put(-4,147){$f(k\!+\!8)\CdoT n_{\lambda}(k\!+\!8)$}
\put(20,130){$+$}
\put(-4,107){$f(k\!+\!16)\CdoT n_{\lambda}(k\!+\!16)$}
\put(20,90){$+$}
\put(-4,67){$f(k\!+\!24)\CdoT n_{\lambda}(k\!+\!24)$}
\put(20,45){$=$}
\put(10,17){$n_{f,\lambda}(k)$}
\put(275,17){$N_{\!f,\lambda}(\mu)$}
\put(198,237){$0$}
\put(505,237){$31$}
\put(495,218){$k\longrightarrow$}
\put(117,8){$0$}
\put(187,8){$7$}
\put(175,0){$k\longrightarrow$}
\put(420,20){$M=8$; $F=32$}
\end{picture}}
\end{center}\vspace{-15pt}
\setlength{\belowcaptionskip}{-6pt}
\caption{Fensterung des Approximationsfehlerprozesses mit blockweiser
"Uberlagerung und anschlie"sender DFT}
\label{b1a}
\rule{\textwidth}{0.5pt}\vspace{-13pt}
\end{figure}

Wie bei der Gewinnung des Periodogramms, das den Spezialfall
der Fensterung mit einem Rechteckfenster der L"ange $M$ darstellt,
werden nun mit der diskreten Fouriertransformation die $M$
Spektralwerte des gefensterten Rauschprozesses gebildet.

\begin{gather}
\boldsymbol{N}_{\!\!f}(\mu)=\!\Sum{k=-\infty}{\infty}\!\!
f(k)\CdoT\boldsymbol{n}(k)\CdoT
e^{\uP{0.4}{\!\!-j\cdot\frac{2\pi}{M}\cdot\mu\cdot k}}=\!\!
\Sum{\kappa=-\infty}{\infty}\,\Sum{k=0}{M-1}
f(k\!+\!\kappa\CdoT M)\CdoT\boldsymbol{n}(k\!+\!\kappa\CdoT M)\CdoT
e^{\uP{0.4}{\!\!-j\cdot\frac{2\pi}{M}\cdot\mu\cdot(k+\kappa\cdot M)}}\!=\!
\notag\\*[10pt]
=\,\Sum{k=0}{M-1}\;\underbrace{\Sum{\kappa=-\infty}{\infty}
f(k\!+\!\kappa\CdoT M)\CdoT\boldsymbol{n}(k\!+\!\kappa\CdoT M)}
_{\T =\;\boldsymbol{n}_f(k)}\cdot\,
e^{\!-j\cdot\frac{2\pi}{M}\cdot\mu\cdot k}\;=\;
\Sum{k=0}{M-1}\boldsymbol{n}_f(k)\cdot
e^{\!-j\cdot\frac{2\pi}{M}\cdot\mu\cdot k}
\label{2.16}\\*[-2pt]
\qquad\qquad\qquad\forall\qquad\mu=0\;(1)\;M\!-\!1\notag
\end{gather}
Wie diese Gleichung zeigt, l"asst sich die Fouriertransformierte 
des diskreten, gefensterten Approximationsfehlerprozesses 
in vier Schritten berechnen:
\begin{itemize}
\item Der Approximationsfehlerprozess wird mit der Fensterfolge 
      gewichtet: \mbox{$f(k)\CdoT\boldsymbol{n}(k)$}.
\item Der gefensterte Approximationsfehlerprozess wird in Bl"ocke der L"ange 
      $M$ zerlegt. F"ur jeden Wert $\kappa$ bildet der Prozessausschnitt 
      \mbox{$f(k\!+\!\kappa\CdoT M)\CdoT\boldsymbol{n}(k\!+\!\kappa\CdoT M)$}
      f"ur die $M$ Werte \mbox{$k=0\;(1)\;M\!-\!1$} einen Block der L"ange $M$.
\item Die Bl"ocke f"ur alle Werte von $\kappa$ werden additiv "uberlagert:\\[12pt]
      \mbox{$\boldsymbol{n}_f(k)\;= \Sum{\kappa=-\infty}{\infty}
      f(k\!+\!\kappa\CdoT M)\CdoT\boldsymbol{n}(k\!+\!\kappa\CdoT M)$}
\item Die so entstandene Folge der $M$ Zufallsgr"o"sen
      \mbox{$\boldsymbol{n}_f(k)$} wird einer DFT der L"ange $M$ unterworfen:\\[12pt]
      \mbox{$\boldsymbol{N}_{\!\!f}(\mu)\;=
      \Sum{k=0}{M-1}\boldsymbol{n}_f(k)\cdot e^{\!-j\cdot\frac{2\pi}{M}\cdot\mu\cdot k}$}.
\end{itemize}
In Bild \ref{b1a} wird diese Vorgehensweise beispielhaft anhand einer Musterfolge 
\mbox{$n_{\lambda}(k)$} des Prozesses \mbox{$\boldsymbol{n}(k)$} 
symbolisch dargestellt. Dabei steht das Symbol \dftran{} f"ur die DFT einer 
diskreten Folge. Die Zufallsgr"o"sen des Prozesses "uberlagern sich in der gleichen Weise, 
wie die Abtastwerte der Musterfolge.

Anschlie"send 
bildet man von den so entstandenen 
$M$ zuf"alligen Spektralwerten die Betragsquadrate und normiert
diese auf $M$. Man erh"alt so die $M$ neuen, reellen Zufallsgr"o"sen
\mbox{$|\boldsymbol{N}_{\!\!f}(\mu)|^2/M$}, deren Erwartungswerte

\begin{gather}
\Tilde{\Phi}_{\boldsymbol{n}}(\mu)\;=\;
\text{E}\Big\{\,\frac{1}{M}\cdot|\boldsymbol{N}_{\!\!f}(\mu)|^2\Big\}\;=\;
\label{2.17}\\*[8pt]
=\;\frac{1}{M}\cdot\text{E}\bigg\{\,\Sum{k_1=-\infty}{\infty}
\!\!\boldsymbol{n}(k_1)\CdoT f(k_1)\cdot
e^{\!-j\cdot\frac{2\pi}{M}\cdot\mu\cdot k_1}\;\cdoT
\Sum{k_2=-\infty}{\infty}\!\!
\boldsymbol{n}(k_2)^{\Kk}\!\CdoT f(k_2)^{\Kk}\!\CdoT
e^{j\cdot\frac{2\pi}{M}\cdot\mu\cdot k_2}\,\bigg\}\;=
\notag\\[4pt]
=\;\frac{1}{M}\cdoT\!\Sum{k_1=-\infty}{\infty}\:\Sum{k_2=-\infty}{\infty}
\text{E}\big\{\boldsymbol{n}(k_1)\CdoT\boldsymbol{n}(k_2)^{\Kk}\big\}\cdot
f(k_1)\CdoT f(k_2)^{\Kk}\Cdot
e^{\!-j\cdot\frac{2\pi}{M}\cdot\mu\cdot(k_1-k_2)}\;=
\notag\\[4pt]
=\;\frac{1}{M}\cdoT\!\Sum{\kappa=-\infty}{\infty}\:\Sum{k_2=-\infty}{\infty}
\text{E}\big\{\boldsymbol{n}(k_2\!+\!\kappa)\CdoT
\boldsymbol{n}(k_2)^{\Kk}\big\}\cdot
f(k_2\!+\!\kappa)\CdoT f(k_2)^{\Kk}\Cdot
e^{\!-j\cdot\frac{2\pi}{M}\cdot\mu\cdot\kappa}\;=
\notag\\[6pt]
=\;\frac{1}{M}\cdoT\!\Sum{\kappa=-\infty}{\infty}\:\Sum{k_2=-\infty}{\infty}
\frac{1}{2\pi}\cdoT\!\Int{-\pi}{\pi}\Phi_{\boldsymbol{n}}(\Omega)\cdot
e^{j\cdot\Omega\cdot\kappa}\Cdot d\Omega\,\cdot\,
f(k_2\!+\!\kappa)\CdoT f(k_2)^{\Kk}\Cdot
e^{\!-j\cdot\frac{2\pi}{M}\cdot\mu\cdot\kappa}\;=
\notag\\[2pt]
=\;\frac{1}{2\pi\CdoT M}\cdoT\Int{-\pi}{\pi}\Phi_{\boldsymbol{n}}(\Omega)
\cdot\big|\,F\big({\T\mu\CdoT\frac{2\pi}{M}\!-\!\Omega}\big)\,\big|^2
\cdot\:d\Omega
\notag\\[-2pt]
\qquad\qquad\forall\qquad\mu=0\;(1)\;M\!-\!1\notag
\end{gather}
bei geeigneter Wahl der Fensterfolge eine bessere N"aherung f"ur 
die zu bestimmenden Gr"o"sen \mbox{$\Bar{\Phi}_{\boldsymbol{n}}(\mu)$} sind, 
als dies beim Periodogramm der Fall ist. Bei der Umformung der letzten 
Gleichung wurde ber"ucksichtigt, dass auf Grund der Stationarit"at von 
\mbox{$\boldsymbol{n}(k)$} der Erwartungswert des Produktes 
\mbox{$\boldsymbol{n}(k_1)\CdoT\boldsymbol{n}(k_2)^*$} nur von der 
Differenz \mbox{$k_1\!-\!k_2$} abh"angt, und er sich daher nach der 
Substitution \mbox{$\kappa=k_1\!-\!k_2$} vor die Summe "uber $k_2$ 
ziehen l"asst. Au"serdem wurde der Satz verwendet, der besagt, 
dass die diskrete Fouriertransformierte des Produktes zweier 
Folgen der Faltung der beiden Spektren der Folgen entspricht. 
Desweiteren wurde die Fouriertransformierte \mbox{$F(\Omega)$}
der Fensterfolge \mbox{$f(k)$} eingef"uhrt\footnote{Die 
Fouriertransformierte \mbox{$F(\Omega)$} der Fensterfolge hat 
dasselbe Formelzeichen wie die L"ange $F$ der Fensterfolge. Die
L"ange unterscheidet sich jedoch eindeutig von der Fouriertransformierten, 
weil diese immer mit dem eingeklammerten Argument $F(\ldots)$ auftritt, 
w"ahrend die Fensterl"ange $F$ als Konstante niemals ein Argument tr"agt. 
Eine Verwechslung sollte dadurch ausgeschlossen sein.}.
\begin{equation}
F(\Omega)\;=\,\Sum{k=0}{F-1}f(k)\cdot
e^{\!-j\cdot\Omega\cdot k}
\qquad\qquad\forall\qquad\Omega\in\mathbb{R}
\label{2.18}
\end{equation}
W"urde man f"ur das Betragsquadrat des Spektrums der Fensterfolge
eine mit $2\pi$ periodische Rechteckfunktion w"ahlen, die im
Intervall \mbox{$(-\pi;\pi]$} f"ur \mbox{$|\Omega|<\pi/M$} den
Wert $M^2$ annimmt, und sonst Null ist, so w"aren die Erwartungswerte
\mbox{$\Tilde{\Phi}_{\boldsymbol{n}}(\mu)$} der Zufallsgr"o"sen
\mbox{$|\boldsymbol{N}_{\!\!f}(\mu)|^2/M$} gleich den gew"unschten Gr"o"sen
\mbox{$\Bar{\Phi}_{\boldsymbol{n}}(\mu)$}. Ein Vergleich mit Gleichung 
(\ref{2.13}) zeigt dies. Die Fensterfolge mit solch einem 
Betragsquadratspektrum ist die abgetastete si-Funktion
\mbox{$\text{si}(\pi\CdoT k/M)$}, die zeitlich nicht auf ein endliches 
Intervall begrenzt ist. Mit dieser Fensterfolge gelingt es daher nicht, 
mit einer Messung endlicher Dauer Stichproben der Zufallsgr"o"sen 
\mbox{$|\boldsymbol{N}_{\!\!f}(\mu)|^2/M$} zu erhalten, deren 
empirische Mittelwerte als Sch"atzwert f"ur die gesuchten Erwartungswerte
\mbox{$\Tilde{\Phi}_{\boldsymbol{n}}(\mu)$} dienen k"onnen. Man wird 
daher versuchen, eine Fensterfolge zu finden, die zeitlich begrenzt ist,
und deren Betragsquadratspektrum n"aherungsweise rechteckf"ormig ist.
In Kapitel \ref{Algo} wird ein Algorithmus angegeben, mit dessen
Hilfe man eine Fensterfolge berechnen kann, die diesem Ziel sehr nahe
kommt.

\subsection{Fenster mit leistungskomplement"arem Spektrum}

Wie Gleichung (\ref{2.14}) zeigt, l"asst sich die Leistung des Prozesses
\mbox{$\boldsymbol{n}(k)$} als die auf $M$ normierte Summe der Werte
\mbox{$\Bar{\Phi}_{\boldsymbol{n}}(\mu)$} angeben. Es w"are nun sehr 
w"unschenswert, wenn sich auch bei Verwendung einer endlich langen
Fensterfolge die Gesamtleistung als die auf $M$ normierte Summe 
der Erwartungswerte \mbox{$\Tilde{\Phi}_{\boldsymbol{n}}(\mu)$} ergibt. 
Diese erste wichtige Forderung l"asst sich mit Gleichung (\ref{2.17}) 
wie folgt exakt formulieren:

\begin{gather}
\frac{1}{M}\cdoT\Sum{\mu=0}{M-1}\Tilde{\Phi}_{\boldsymbol{n}}(\mu)\;=\;
\frac{1}{2\pi\CdoT M^2}\cdoT\Sum{\mu=0}{M-1}\;
\Int{-\pi}{\pi}\Phi_{\boldsymbol{n}}(\Omega)\cdot
\big|\,F\big({\T\mu\CdoT\frac{2\pi}{M}\!-\!\Omega}\big)\,\big|^2
\Cdot\:d\Omega\;=
\label{2.19}\\*[6pt]
=\;\frac{1}{2\pi\CdoT M^2}\cdoT\Int{-\pi}{\pi}\Phi_{\boldsymbol{n}}(\Omega)
\cdoT\Sum{\mu=0}{M-1}\:
\big|\,F\big({\T\mu\CdoT\frac{2\pi}{M}\!-\!\Omega}\big)\,\big|^2
\Cdot\:d\Omega\;\stackrel{!}{=}\;\frac{1}{2\pi}\cdoT\Int{-\pi}{\pi}
\Phi_{\boldsymbol{n}}(\Omega)\cdot d\Omega\;=\;\sigma_{\boldsymbol{n}}^2
\notag
\end{gather}
Man erkennt, dass diese Forderung f"ur beliebige
\mbox{$\Phi_{\boldsymbol{n}}(\Omega)$} erf"ullt wird, wenn das
Betragsquadrat des Spektrums der Fensterfolge die Forderung\vspace{0pt}
\begin{equation}
\Sum{\mu=0}{M-1}\:
\big|\,F\big({\T\Omega\!-\!\mu\CdoT\frac{2\pi}{M}}\big)\,\big|^2\;\stackrel{!}{=}\;M^2
\qquad\qquad\forall\qquad\Omega\in\mathbb{R}
\label{2.20}
\end{equation}
erf"ullt. Diese exakt formulierbare Forderung (\ref{2.20}) besagt, 
dass sich das Betragsquadrat des Spektrums des Fensters --- also 
dessen spektrale Leistungsdichte --- zu einer Konstanten erg"anzt, 
wenn man diese mit sich selbst verschoben "uberlagert. Solche Fensterfolgen 
werde ich daher im Weiteren als leistungskomplement"ar bezeichnen.
Fensterfolgen, die mit dem in Kapitel \ref{Algo} angegebenen 
Algorithmus berechnet worden sind, erf"ullen diese Forderung 
mit einem Fehler, der in der Gr"o"senordnung der unvermeidbaren 
Fehler liegt, also der Fehler, die an einem Rechner wegen der 
begrenzten Wortl"ange der Zahlendarstellung immer auftreten. 
F"ur die Fensterautokorrelationsfolge 
\begin{gather}
d(\kappa)\;=\;\frac{1}{M}\cdot\big(\,f(\kappa)\ast f(\!-\kappa)^{\Kk}\big)\;=
\label{2.21}\\*[6pt]
=\;\frac{1}{M}\cdoT\!\Sum{\Tilde{k}=-\infty}{\infty}\!
f(\Tilde{k})^{\Kk}\Cdot f(\Tilde{k}+\kappa)\;=\;
\frac{1}{2\pi\CdoT M}\cdoT\Int{-\pi}{\pi}
\big|F(\Omega)\big|^2\Cdot e^{j\cdot\Omega\cdot\kappa}\cdot\:d\Omega,
\notag
\end{gather}
deren Fouriertransformierte gerade \mbox{$|F(\Omega)|^2/M$} ist, bedeutet das, 
dass die Fouriertransformierte der mit der Schrittweite $M$ unterabgetasteten 
Fenster-AKF
\begin{equation}
\Sum{\Tilde{k}=-\infty}{\infty}\!d(\Tilde{k}\CdoT M)\cdot
e^{\!-j\cdot\Omega\cdot\Tilde{k}\cdot M}\;=\;
\frac{1}{M^2}\cdoT\Sum{\mu=0}{M-1}\:
\big|F\big({\T\Omega\!-\!\mu\CdoT\frac{2\pi}{M}}\big)\big|^2\,\stackrel{!}{=}\;1
\qquad\quad\forall\qquad\Omega\in\mathbb{R}\quad{}
\label{2.22}
\end{equation}
konstant Eins sein muss, also
\begin{equation}
d(\kappa)\;\stackrel{!}{=}\;\begin{cases}
\quad 0&\quad\text{ f"ur }\quad\kappa=\Tilde{k}\CdoT M\quad
\text{ mit }\quad\Tilde{k}\in\mathbb{Z}\backslash\{0\}\\
\quad 1&\quad\text{ f"ur }\quad\kappa=0\\
\quad\text{beliebig }&\quad\text{ sonst}
\end{cases}
\label{2.23}
\end{equation}
gelten muss. Betrachtet man die Folge \mbox{$d(\kappa)$} als Koeffizienten
eines FIR-Filters, so handelt es sich aufgrund der eben genannten
Eigenschaften von \mbox{$d(\kappa)$} um ein sogenanntes M-tel Band-Filter.
W"are $\kappa$ eine kontinuierliche Gr"o"se, so w"urde man sagen, dass
\mbox{$d(\kappa)$} das erste Nyquist-Kriterium erf"ullt.
Filter mit solchen Koeffizienten \mbox{$d(\kappa)$}, als auch Filter,
mit den Koeffizienten \mbox{$f(k)$}, deren Spektren die Bedingung (\ref{2.20})
erf"ullen, spielen in vielen Bereichen der digitalen Signalverarbeitung
eine wichtige Rolle. Au"serdem sei noch angemerkt, dass aus Gleichung
(\ref{2.17}) folgt, dass die Werte \mbox{$\Tilde{\Phi}_{\boldsymbol{n}}(\mu)$}
sich auch aus der AKF \mbox{$\phi_{\boldsymbol{n}}(\kappa)$} 
durch eine Fensterung mit \mbox{$d(\kappa)$} und eine anschlie"sende
diskrete Fouriertransformation berechnen lassen. 

\subsection{Aspekte zur Wahl der Fensterfolge}

Durch die letzte Forderung wird noch nichts dar"uber ausgesagt,
in welcher Weise das Betragsquadrat des Spektrums der Fensterfolge
die gew"unschte Rechteckfunktion ann"ahert. Wie gut die Erwartungswerte
\mbox{$\Tilde{\Phi}_{\boldsymbol{n}}(\mu)$} mit den gew"unschten
Werten \mbox{$\Bar{\Phi}_{\boldsymbol{n}}(\mu)$} "ubereinstimmen,
h"angt von dem Produkt des Integranden in Gleichung (\ref{2.17}), also
zum einen von dem Betragsquadrat des Spektrums der Fensterfolge und zum
anderen von dem LDS des Prozesses \mbox{$\boldsymbol{n}(k)$} ab.
Die Frequenzbereiche, innerhalb derer das Produkt des Integranden 
besonders gro"s ist, liefern die gr"o"sten Beitr"age zum Erwartungswert 
\mbox{$\Tilde{\Phi}_{\boldsymbol{n}}(\mu)$}. Da das Betragsquadrat des 
Spektrums der Fensterfolge im Integranden in Gleichung (\ref{2.17})
mit einem Frequenzversatz von \mbox{$\mu\cdot2\pi/M$} auftritt,
wird der Frequenzgang des Produktes im Integral vom diskreten 
Frequenzpunkt $\mu$ abh"angen. Die Fensterung soll bewirken, 
dass aus dem LDS nur der Bereich in unmittelbarer Umgebung der 
Frequenz \mbox{$\mu\cdot2\pi/M$} herausgeschnitten wird, indem dort
das Betragsquadrat des Spektrums der Fensterfolge gro"s ist, w"ahrend
es im gesamten anderen Frequenzbereich m"oglichst klein sein sollte,
um dort ein kleines Produkt als Integranden zu gew"ahrleisten. 
Man ben"otigt also eine Fensterfolge, dessen Spektrum \mbox{$F(\Omega)$}
f"ur \mbox{$|\Omega|>\pi/M$} eine hohe Sperrd"ampfung aufweist.

Betrachten wir nun den Fall, der durch die drei folgenden Aussagen 
charakterisiert sei. Erstens werde wie bei der Berechnung des 
Periodogramms in Bild \ref{b1d} ein Rechteckfenster der L"ange $M$ verwendet, 
zweitens sei die Leistung des Prozesses \mbox{$\boldsymbol{n}(k)$} 
auf wenige Frequenzbereiche konzentriert, so dass der Verlauf des 
LDS \mbox{$\Phi_{\boldsymbol{n}}(\Omega)$} "uber der Frequenz stark
schwankt und drittens werde \mbox{$\Tilde{\Phi}_{\boldsymbol{n}}(\mu)$} 
f"ur eine diskrete Frequenz $\mu$ berechnet, bei der \mbox{$\mu\CdoT2\pi/M$} 
in einem Bereich von $\Omega$ liegt, bei dem das LDS besonders klein ist.
In diesem Fall wird der Verlauf des Produkts im Integral sein Maximum 
{\em nicht} im Bereich des Maximums des Betragsquadrats des Spektrums der 
Fensterfolge haben, weil die Sperrd"ampfung dieses Fensters\footnote{Die 
Nebenmaxima von \mbox{$F(\Omega)$} fallen hier nur mit \mbox{$\sin(\Omega/2)^{-1}$} 
ab.} in dem Frequenzbereich, in dem das LDS gro"s ist, zu gering 
ist, um das LDS hinreichend zu unterdr"ucken. Der integrale Wert 
\mbox{$\Tilde{\Phi}_{\boldsymbol{n}}(\mu)$}  wird somit keine
Aussage "uber das LDS im Bereich um \mbox{$\mu\CdoT2\pi/M$}
zulassen. Man setzt daher "ublicherweise Fenster mit einer h"oheren 
Sperrd"ampfung ein. 

\begin{figure}[btp]
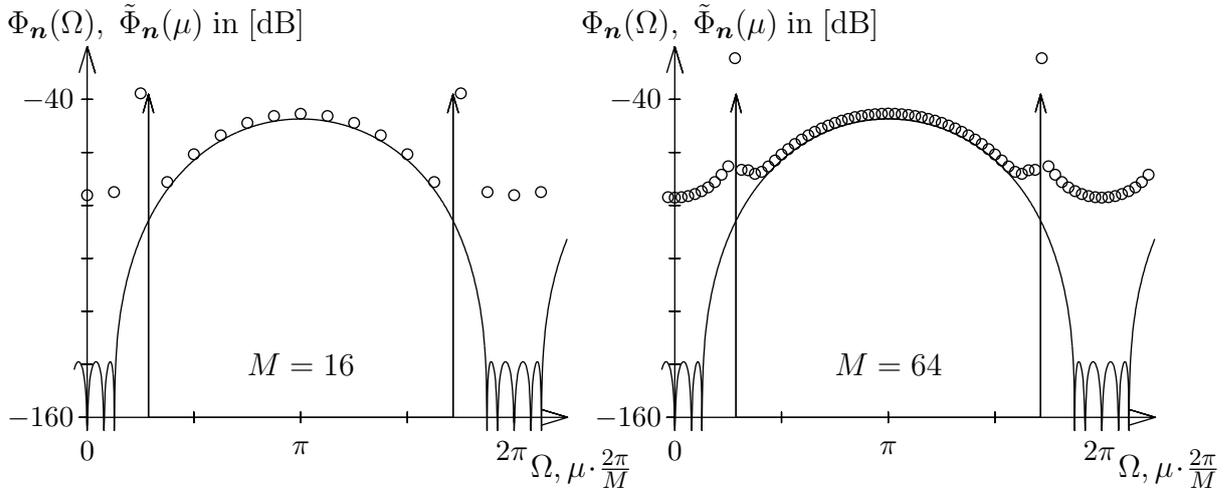

\begin{center}
{ 
\begin{picture}(454,188)
\input{mbild1f1}
\input{mbild1f2}
\put(0,175){$\Phi_{\boldsymbol{n}}(\Omega),\;
\Tilde{\Phi}_{\boldsymbol{n}}(\mu)\;\text{in}\;[\text{dB}]$}
\put(215,175){$\Phi_{\boldsymbol{n}}(\Omega),\;
\Tilde{\Phi}_{\boldsymbol{n}}(\mu)\;\text{in}\;[\text{dB}]$}
\put(25,150){\makebox(0,0)[r]{\small$-40$}}
\put(25,30){\makebox(0,0)[r]{\small$-160$}}
\put(245,150){\makebox(0,0)[r]{\small$-40$}}
\put(245,30){\makebox(0,0)[r]{\small$-160$}}
\put(30,21){\makebox(0,0)[t]{\small$0$}}
\put(110,21){\makebox(0,0)[t]{\small$\pi$}}
\put(189,21){\makebox(0,0)[t]{\small$2\pi$}}
\put(250,21){\makebox(0,0)[t]{\small$0$}}
\put(330,21){\makebox(0,0)[t]{\small$\pi$}}
\put(409,21){\makebox(0,0)[t]{\small$2\pi$}}
\put(215,18){\makebox(0,0)[t]{$\Omega,\mu\CdoT\frac{2\pi}{M}$}}
\put(435,18){\makebox(0,0)[t]{$\Omega,\mu\CdoT\frac{2\pi}{M}$}}
\put(110,50){\makebox(0,0){$M=16$}}
\put(330,50){\makebox(0,0){$M=64$}}
\end{picture}}
\end{center}\vspace{-21pt}
\setlength{\belowcaptionskip}{-6pt}
\caption[Erwartungswert des Periodogramms eines Hamming-gefensterten Prozesses
mit einem periodischen St"oranteil]{Erwartungswert des Periodogramms eines
Hamming-gefensterten Prozesses mit einem periodischen St"oranteil.
Das LDS des Prozesses ist mit der durchgezogenen Linie dargestellt,
w"ahrend die N"aherung mit Hilfe des Erwartungswertes des Periodogramms
des gefensterten Prozesses mit "`o"' gekennzeichnet ist.}
\label{b1f}
\rule{\textwidth}{0.5pt}\vspace{-10pt}
\end{figure}

Eines der ersten Fenster, das hierf"ur vorgeschlagen wurde, ist 
das nach seinem Entwickler benannte Hamming-Fenster \cite{Kam}. 
In Bild  \ref{b1f}
 sind bei demselben Beispielprozess, 
der auch in den Bildern \ref{b1c}, \ref{b1b} und \ref{b1d} verwendet wurde, 
die Erwartungswerte des Periodogramms dargestellt, die sich bei einer 
Fensterung des Prozesses mit dem Hamming-Fenster der L"ange \mbox{$F=4\CdoT M$} 
ergeben. Obwohl nun die Werte um die Frequenz \mbox{$\Omega=0$} schon fast 20~dB 
niedriger sind, zeigt sich, dass auch dieses Fenster hier eine zu geringe 
Sperrd"ampfung besitzt. Die Faltung des Betragsquadrats des Spektrums der 
Hamming-Fensterfolge mit den Dirac-Impulsen der periodischen St"orung bewirkt, 
dass die Erwartungswerte des Periodogramms des gefensterten Prozesses im 
Frequenzbereich um \mbox{$\Omega=0$} im wesentlichen durch die um 
die Sperrd"ampfung des Hamming-Fensters abged"ampften Impulsst"arken 
bestimmt werden, und nicht mehr durch das LDS in diesem Frequenzbereich. 
Es sei noch angemerkt, dass das Hamming-Fenster die Forderung (\ref{2.20}) 
nicht erf"ullt, so dass eine einfache Berechnung der Gesamtrauschleistung mit 
Gleichung (\ref{2.19}) aus den Werten \mbox{$\Tilde{\Phi}_{\boldsymbol{n}}(\mu)$} 
hier nicht m"oglich ist.

Jede zeitlich begrenzte Fensterfolge weist im Sperrbereich des Frequenzgangs 
eine endliche und im Durchgangsbereich eine nicht exakt konstante D"ampfung auf. 
Unter der Randbedingung der zeitlichen Begrenzung der Fensterfolge hat man 
jedoch die M"oglichkeit f"ur den Verlauf der D"ampfung einen Kompromiss 
in der Art zu finden, dass die mit Gleichung (\ref{2.17}) berechneten 
Werte \mbox{$\Tilde{\Phi}_{\boldsymbol{n}}(\mu)$} f"ur alle $\mu$ 
die gew"unschten Werte \mbox{$\Bar{\Phi}_{\boldsymbol{n}}(\mu)$} 
gut ann"ahern. Dieser Kompromiss h"angt jedoch vom tats"achlich vorhanden 
LDS des Prozesses \mbox{$\boldsymbol{n}(k)$} ab, da dieses in Gleichung (\ref{2.17}) 
mit dem Frequenzgang der Fensterfolge multiplikativ verkn"upft ist. 
Da man jedoch das LDS i.~Allg. nicht a priori kennt, wird man sich damit 
begn"ugen m"ussen, eine Fensterfolge zu finden deren D"ampfungsfrequenzgang 
f"ur viele \mbox{real} vorkommende Rauschprozesse geeignet erscheint.

Um eine weitere Aussage "uber den gew"unschten Verlauf der Sperrd"ampfung 
zu erhalten wollen wir nun annehmen, dass viele reale Rauschprozesse 
ein LDS aufweisen, das dem LDS eines gefilterten wei"sen Rauschprozesses 
"ahnelt. Es l"asst sich dann als eine gebrochen rationale Funktion in $z$ 
mit \mbox{$z=e^{j\cdot\Omega}$} schreiben (\,AR-MA-Modell \cite{Kam}\,). 
Dabei wird die gebrochen rationale Funktion endlich viele Pole und 
Nullstellen aufweisen, die alle symmetrisch zum Einheitskreis liegen. 
Bei solchen Rauschprozessen "andert sich das LDS nicht sprungartig. 
Daher braucht die D"ampfung des Fensters nicht gleich in unmittelbarer  
Umgebung des gew"unschten rechteckf"ormigen Durchlassbereichs auf den zu 
fordernden hohen Wert der Sperrd"ampfung anzusteigen. Solche Systeme weisen 
typischerweise immer dann starke Schwankungen in Frequenzverlauf des LDS auf, 
wenn Pol- oder Nullstellen des AR"~MA-Systems besonders nahe am Einheitskreis 
liegen. Gewisserma"sen als den schlimmsten Fall eines solchen LDS wollen wir nun 
ein wei"ses Rauschsignal betrachten, das man $B$ mal durch ein n"aherungsweise 
differenzierendes System schickt, das durch die Differenzengleichung 
\mbox{$y(k)=x(k)-x(k\!-\!1)$} beschrieben wird. Die Z-Transformierte 
der Impulsantwort des mehrfach differenzierenden Systems weist eine 
$B$-fache Nullstelle am Einheitskreis der z-Ebene bei \mbox{$z=1$} auf. 
Das LDS des Rauschens ist dann proportional zu \mbox{$\sin(\Omega/2)^{2\cdot B}$} 
und hat eine \mbox{$2\cdoT B$}-fache Nullstelle bei \mbox{$\Omega=0$}. 
Um bei diesem LDS daf"ur zu sorgen, dass der Erwartungswert 
\mbox{$\Tilde{\Phi}_{\boldsymbol{n}}(0)$} einigerma"sen gut mit 
\mbox{$\Bar{\Phi}_{\boldsymbol{n}}(0)$} "ubereinstimmt, muss 
der Anteil des Integrals in Gleichung (\ref{2.17}) im Durchlassbereichs 
des Fensterfrequenzgangs immer deutlich h"oher sein als der Anteil 
im Sperrbereich. Deshalb sollte der Integrand ein Maximum im Bereich 
der Frequenzen \mbox{$|\Omega|<\pi/M$} aufweisen. Dies wird nur dann 
der Fall sein, wenn \mbox{$\left|F(\Omega)\right|^2$} mit $\Omega$ 
steiler abf"allt als \mbox{$\Phi_{\boldsymbol{n}}(\Omega)$} ansteigt. 
Daher sollte der Betragsfrequenzgang der Fensterfolge schneller als mit 
\mbox{$\sin(\Omega/2)^{-B}$} abfallen.  

Aus dieser "Uberlegung heraus l"asst sich nun eine weitere 
Forderung f"ur das Betragsquadrat des Spektrums der Fensterfolge 
verallgemeinern. Wenn man ein LDS eines Rauschprozesses am Ausgang 
eines gebrochen rationalen Systems, das eine Anh"aufung von maximal 
$2\cdoT B$ Polstellen nahe bei einem Punkt am Einheitskreis aufweist, 
durch die Erwartungswerte \mbox{$\Tilde{\Phi}_{\boldsymbol{n}}(\mu)$} 
beschreiben will, sollte man eine Fensterfolge verwenden, deren 
Betragsfrequenzgang schneller als mit \mbox{$\sin(\Omega/2)^{-B}$} 
abf"allt. Diese zweite Forderung, die wir an eine Fensterfolge stellen, 
die beim RKM eingesetzt werden soll, ist nicht so pr"azise formulierbar 
wie die erste Forderung (\ref{2.20}).

In Bild \ref{b1g} wird bei demselben Beispielprozess wie in den 
Bildern \ref{b1c}, \ref{b1b}, \ref{b1d} und \ref{b1f} eine Fensterfolge verwendet,
deren Betragsfrequenzgang asymptotisch mit \mbox{$\sin(\Omega/2)^{-4}$}
abf"allt. Diese Fensterfolge wurde mit dem in Kapitel \ref{Algo} angegebenen
Algorithmus berechnet, bei dem die Potenz des Anstiegs der Sperrd"ampfung 
in weiten Grenzen frei eingestellt werden kann. Es ergibt sich hier eine 
Fensterl"ange von \mbox{$F=4\CdoT (M\!-\!1)$}. Man erkennt, dass nun f"ur 
\mbox{$M=64$} auch im 
\begin{figure}[t!]
\begin{center}
{ 
\begin{picture}(454,206)
\input{mbild1g1}
\input{mbild1g2}
\put(0,175){$\Phi_{\boldsymbol{n}}(\Omega),\;
\Tilde{\Phi}_{\boldsymbol{n}}(\mu)\;\text{in}\;[\text{dB}]$}
\put(215,175){$\Phi_{\boldsymbol{n}}(\Omega),\;
\Tilde{\Phi}_{\boldsymbol{n}}(\mu)\;\text{in}\;[\text{dB}]$}
\put(25,150){\makebox(0,0)[r]{\small$-40$}}
\put(25,30){\makebox(0,0)[r]{\small$-160$}}
\put(245,150){\makebox(0,0)[r]{\small$-40$}}
\put(245,30){\makebox(0,0)[r]{\small$-160$}}
\put(30,21){\makebox(0,0)[t]{\small$0$}}
\put(110,21){\makebox(0,0)[t]{\small$\pi$}}
\put(189,21){\makebox(0,0)[t]{\small$2\pi$}}
\put(250,21){\makebox(0,0)[t]{\small$0$}}
\put(330,21){\makebox(0,0)[t]{\small$\pi$}}
\put(409,21){\makebox(0,0)[t]{\small$2\pi$}}
\put(215,18){\makebox(0,0)[t]{$\Omega,\mu\CdoT\frac{2\pi}{M}$}}
\put(435,18){\makebox(0,0)[t]{$\Omega,\mu\CdoT\frac{2\pi}{M}$}}
\put(110,50){\makebox(0,0){$M=16$}}
\put(330,50){\makebox(0,0){$M=64$}}
\end{picture}}
\end{center}\vspace{-18pt}
\setlength{\belowcaptionskip}{-2pt}
\caption[Erwartungswert des Periodogramms eines gefensterten Prozesses mit einem periodischen
St"oranteil]{Erwartungswert des Periodogramms eines gefensterten Prozesses mit einem periodischen
St"oranteil. Das LDS des Prozesses ist mit der durchgezogenen Linie
dargestellt, w"ahrend die N"aherung mit Hilfe des Erwartungswertes des
Periodogramms des gefensterten Prozesses mit "`o"' gekennzeichnet ist.}
\label{b1g}
\rule{\textwidth}{0.5pt}\vspace{-5pt}
\end{figure}
Bereich um die Frequenz \mbox{$\Omega=0$} 
das LDS, bzw. dessen Stufenapproximation gut angen"ahert werden. Bei diesem 
Beispielprozess ist f"ur \mbox{$M=16$} die Frequenzselektivit"at der
Fensterung auch bei dem nach Kapitel \ref{Algo} berechneten Fenster 
offensichtlich noch nicht hoch genug.

\subsection{Systemapproximation mit Hilfe der gefensterten Prozesse}\label{AppPhi}

Die $M$ Werte der "Ubertragungsfunktion \mbox{$H(\mu\CdoT2\pi/M)$} 
des linearen Modellsystems haben wir nach Gleichung (\ref{2.10}) 
so gew"ahlt, dass der Approximationsfehler \mbox{$\boldsymbol{n}(k)$} 
eine minimale Varianz auf aufweist. Nach Gleichung (\ref{2.9}) 
ergibt sich der Approximationsfehlerprozess \mbox{$\boldsymbol{n}(k)$}, 
und somit auch dessen Varianz, aus den Prozessen \mbox{$\boldsymbol{V}(\mu)$} 
und \mbox{$\boldsymbol{y}(k)$} in Abh"angigkeit von den zu optimierenden 
Werten der "Ubertragungsfunktion. Somit sind auch die $M$ Erwartungswerte 
\mbox{$\Tilde{\Phi}_{\boldsymbol{n}}(\mu)$} eine Funktion
der Werte der "Ubertragungsfunktion.

\begin{gather}
\Tilde{\Phi}_{\boldsymbol{n}}(\mu)\;=\;
\frac{1}{M}\cdot\text{E}\big\{\,|\boldsymbol{N}_{\!\!f}(\mu)|^2\big\}\;=\;
\frac{1}{M}\cdot\text{E}\Bigg\{\,
\bigg|\,\Sum{k=-\infty}{\infty}\!\boldsymbol{n}(k)\CdoT
f(k)\cdot e^{\!-j\cdot\frac{2\pi}{M}\cdot\mu\cdot k}\,\bigg|^2\Bigg\}\;=
\notag\\[8pt]
=\;\frac{1}{M}\cdot\text{E}\Bigg\{\,\bigg|\,\Sum{k=-\infty}{\infty}\!
\big(\,\boldsymbol{y}(k)-\boldsymbol{x}(k)\,\big)\cdot f(k)\cdot
e^{\!-j\cdot\frac{2\pi}{M}\cdot\mu\cdot k}\,\bigg|^2\Bigg\}\;=
\notag\displaybreak[2]\\[6pt]
=\frac{1}{M}\cdot\text{E}\Bigg\{
\bigg|\,\boldsymbol{Y}_{\!\!\!f}(\mu)-\frac{1}{M}\cdoT\!
\Sum{\Breve{\mu}=0}{M-1}\!\boldsymbol{V}(\Breve{\mu})\CdoT
H\big({\T\Breve{\mu}\CdoT\frac{2\pi}{M}}\big)\CdoT
F\big({\T(\mu\!-\!\Breve{\mu})\CdoT\frac{2\pi}{M}}\big)\,\bigg|^2\Bigg\}
\notag\\[2pt]
\forall\qquad \mu=0\;(1)\;M\!-\!1
\label{2.24}
\end{gather}
Dabei sind \mbox{$\boldsymbol{Y}_{\!\!\!f}(\mu)$} die $M$ Elemente
des Zufallsvektors \mbox{$\Vec{\boldsymbol{Y}}_{\!\!\!f}$},
der durch Fensterung und anschlie"sende DFT aus dem
Zufallsprozess \mbox{$\boldsymbol{y}(k)$} entsteht. Die Fensterung und
DFT erfolgt dabei analog zur Berechnung von \mbox{$\boldsymbol{N}_{\!\!f}(\mu)$}
aus \mbox{$\boldsymbol{n}(k)$} nach Gleichung (\ref{2.16}).
\begin{gather}
\boldsymbol{Y}_{\!\!\!f}(\mu)=\!\!\Sum{k=-\infty}{\infty}\!\!
f(k)\CdoT\boldsymbol{y}(k)\CdoT e^{\!-j\cdot\frac{2\pi}{M}\cdot\mu\cdot k}=\!
\Sum{\kappa=-\infty}{\infty}\Sum{k=0}{M-1}
f(k\!+\!\kappa\CdoT M)\CdoT\boldsymbol{y}(k\!+\!\kappa\CdoT M)\CdoT
e^{\!-j\cdot\frac{2\pi}{M}\cdot\mu\cdot(k+\kappa\cdot M)}=\!
\notag\\[6pt]
=\,\Sum{k=0}{M-1}\,\underbrace{\Sum{\kappa=-\infty}{\infty}\!
f(k\!+\!\kappa\CdoT M)\CdoT\boldsymbol{y}(k\!+\!\kappa\CdoT M)}
_{\T =\;\boldsymbol{y}_{\!f}(k)}\cdot\,
e^{\!-j\cdot\frac{2\pi}{M}\cdot\mu\cdot k}\;=\;
\Sum{k=0}{M-1}\boldsymbol{y}_{\!f}(k)\CdoT
e^{\!-j\cdot\frac{2\pi}{M}\cdot\mu\cdot k}
\label{2.25}\\*[-6pt]
\qquad\qquad\forall\qquad\mu=0\;(1)\;M\!-\!1\notag
\end{gather}
Wir wollen nun untersuchen, welche Optimall"osungen sich f"ur die 
$M$ Werte \mbox{$H(\mu\CdoT2\pi/M)$} ergeben, wenn wir nicht die 
Varianz des Approximationsfehlers \mbox{$\boldsymbol{n}(k)$}, sondern 
stattdessen die $M$ Erwartungswerte \mbox{$\Tilde{\Phi}_{\boldsymbol{n}}(\mu)$}
minimieren. Dazu leitet man diese $M$ Terme --- genauso wie bei der 
Berechnung des Minimums der Varianz von \mbox{$\boldsymbol{n}(k)$} 
--- partiell nach Real- und Imagin"arteil der $M$ Werte der 
"Ubertragungsfunktion \mbox{$H(\Tilde{\mu}\CdoT2\pi/M)$} getrennt ab, 
und erh"alt so \mbox{$2\CdoT M^2$} reelle Gleichungen, von denen man jeweils 
zwei Gleichungen als Real- und Imagin"arteil einer komplexen 
Gleichung zusammenfasst.  Man erh"alt nach kurzer Rechnung $M^2$ komplexe
Gleichungen zur Bestimmung der $M$ gesuchten Werte der 
"Uber"-tra"-gungs"-funk"-tion.\vspace{-10pt}
\begin{gather}
F\big({\T(\mu\!-\!\Tilde{\mu})\CdoT\frac{2\pi}{M}}\big)^{\Kk}\Cdot
\text{E}\big\{
\boldsymbol{V}(\Tilde{\mu})^{\Kk}
\Cdot\boldsymbol{Y}_{\!\!\!f}(\mu)\big\}\;=
\label{2.26}\\*[6pt]
=F\big({\T(\mu\!\!-\!\!\Tilde{\mu})\CdoT\frac{2\pi}{M}}\!\big)^{\Kk}\Cdot
\frac{1}{M}\cdoT\Sum{\Breve{\mu}=0}{M-1}
\text{E}\big\{\boldsymbol{V}(\Tilde{\mu})^{\Kk}\Cdot
\boldsymbol{V}(\Breve{\mu})\,\big\}\cdot
F\big({\T(\mu\!\!-\!\!\Breve{\mu})\CdoT\frac{2\pi}{M}}\!\big)\cdot
H\big({\T\Breve{\mu}\CdoT\frac{2\pi}{M}}\!\big)
\notag\\*[6pt]
\forall\qquad\mu=0\;(1)\;M\!-\!1\quad\text{ und }
\quad\Tilde{\mu}=0\;(1)\;M\!-\!1\notag
\end{gather}
Dieselben $M^2$ Gleichungen ergeben sich auch, wenn man 
die Gleichung (\ref{2.11}) zur Bestimmung der $M$
optimalen Werte der "Ubertragungsfunktion bei der Minimierung
der Varianz des Approximationsfehlers \mbox{$\boldsymbol{n}(k)$}
mit \mbox{$f(k)$} gewichtet, anschlie"send diskret fouriertransformiert 
und abschlie"send auf beiden Seiten mit den Spektralwerten der
Fensterfolge multipliziert. Man schreibt dazu zun"achst die auf beiden Seiten 
mit \mbox{$f(k)$} multiplizierte Gleichung (\ref{2.11}) $M$ mal untereinander.
Jede dieser identischen Gleichungen multipliziert man auf beiden Seiten mit 
den Drehfaktoren \mbox{$e^{\!-j\cdot\frac{2\pi}{M}\cdot\Breve{\mu}\cdot k}$}, 
wobei man f"ur $\Breve{\mu}$ die $M$ Werte \mbox{$0\;(1)\;M\!-\!1$} einsetzt, 
so dass die Drehfrequenz bei allen $M$ Gleichungen unterschiedlich ist.
Anschlie"send wird "uber alle \mbox{$k\,=\,0\;(1)\;F\!-\!1$} aufsummiert. 

Da die mit Gleichung (\ref{2.12}) berechneten Werte 
\mbox{$H(\mu\CdoT2\pi/M)$} aufgrund der angenommenen Stationarit"at des 
Verbundprozesses aus \mbox{$\boldsymbol{v}(k)$} und \mbox{$\boldsymbol{y}(k)$} 
die Gleichung (\ref{2.11}) l"osen, sind diese Optimalwerte zugleich
eine L"osung des Systems der $M^2$ Gleichungen (\ref{2.26}). Somit wird also
nicht nur die Varianz von \mbox{$\boldsymbol{n}(k)$} --- also eine einzelne 
Gr"o"se --- minimiert, sondern zugleich jeder der $M$  Erwartungswerte 
\mbox{$\Tilde{\Phi}_{\boldsymbol{n}}(\mu)$}, deren Summe gleich der Varianz 
ist, sofern die Fensterfolge der Bedingung (\ref{2.20}) gen"ugt.

Leider l"asst sich keine notwendige und hinreichende Bedingung f"ur die Wahl
der Fensterfolge angeben, die sicherstellt, dass auch die $M^2$
Gleichungen (\ref{2.26}) keinen m"achtigeren L"osungsraum besitzen, wie 
Gleichung (\ref{2.11}). Es ist jedoch relativ einfach eine hinreichende 
Bedingung f"ur das Spektrum der Fensterfolge zu finden, die die Existenz 
einer eindeutigen L"osung f"ur den Fall garantiert, dass alle Elemente des 
Zufallsvektors $\Vec{\boldsymbol{V}}$ eine von Null verschiedene Varianz 
besitzen. Diese Bedingung besagt, dass das Spektrum der Fensterfolge bei ganzzahligen
Vielfachen der Frequenz \mbox{$2\pi/M$} au"ser bei ganzzahligen Vielfachen
von \mbox{$2\pi$} Nullstellen aufweist. Desweiteren empfiehlt es sich
die Fensterfolge so zu normieren, dass sich bei der Frequenz \mbox{$\Omega=0$}
im Spektrum der Wert $M$ ergibt.
\begin{equation}
F\big({\T\mu\CdoT\frac{2\pi}{M}}\big)\;=\;\begin{cases}
\quad M\quad&\text{ f"ur }\quad\mu=0\\
\quad 0&\text{ f"ur }\quad\mu=1\;(1)\;M\!-\!1
\end{cases}
\label{2.27}
\end{equation}
Die Forderung (\ref{2.27}) ist die dritte Forderung, die wir an eine 
Fensterfolge stellen, wenn wir sie beim RKM einsetzen wollen. Wie auch 
die erste Forderung (\ref{2.20}) ist auch diese Forderung exakt formulierbar. 
Die mit dem in Kapitel \ref{Algo} angegebenen Algorithmus berechneten 
Fensterfolgen erf"ullen die Bedingung (\ref{2.27}). 
 
Von den $M^2$ Gleichungen (\ref{2.26}) sind bei Verwendung einer
Fensterfolge, die dieser Bedingung gen"ugt, alle Gleichungen mit
\mbox{$\mu\!\neq\!\Tilde{\mu}$} sowieso erf"ullt, da in diesen Gleichungen
auf beiden Seiten mit Null multipliziert wird. Bei den Summentermen
in den verbleibenden $M$ Gleichungen bleibt jeweils nur der Summand mit
\mbox{$\Breve{\mu}\!=\!\mu$} "ubrig. Die L"osung 
\begin{equation}
H\big({\T\mu\CdoT\frac{2\pi}{M}}\big)\;=\;
\frac{\D \text{E}\big\{
\boldsymbol{V}(\mu)^{\Kk}\Cdot
\boldsymbol{Y}_{\!\!\!f}(\mu)\big\}}
{\D \text{E}\big\{|\boldsymbol{V}(\mu)|^2\big\}}
\qquad\qquad\forall\qquad\mu=0\;(1)\;M\!-\!1\quad{}
\label{2.28}
\end{equation}
dieser $M$ Gleichungen ist eindeutig festgelegt, wenn man voraussetzt, dass 
die Varianzen aller $M$ Zufallsgr"o"sen \mbox{$\boldsymbol{V}(\Tilde{\mu})$} 
von Null verschieden sind. Der L"osungsraum der  $M^2$ Gleichungen (\ref{2.26}) 
kann somit nicht m"achtiger sein als der L"osungsraum der $M$ urspr"unglichen 
Gleichungen (\ref{2.12}).

Die nach dem Einsetzen der Forderung (\ref{2.27}) verbleibenden $M$ Gleichungen 
und die $M$ Gleichungen (\ref{2.12}) unterscheiden sich, abgesehen von dem 
konstanten von Null verschiedenen Faktor $M$ und davon, dass der Parameter der 
Gleichungsnummer dort $\Tilde{\mu}$ und hier $\mu$ hei"st, nur dadurch, dass 
hier nun das Spektrum des gefensterten Prozesses auftritt. Somit stellt die 
Forderung (\ref{2.27}) sicher, dass man bei der Systemapproximation auch 
dann dasselbe Modellsystem erh"alt, wenn man daf"ur den gefensterten 
Ausgangsprozess heranzieht.

\subsection{Orthogonalit"at der Systemapproximation mit Fensterung}

Die zuf"alligen Spektralwerte \mbox{$\boldsymbol{N}_{\!\!f}(\mu)$} 
lassen sich bei Verwendung einer Fensterfolge, deren Spektrum der 
Bedingung (\ref{2.27}) gen"ugt, in der einfachen Form
\begin{equation}
\boldsymbol{N}_{\!\!f}(\mu)\;=\;\boldsymbol{Y}_{\!\!\!f}(\mu)-
H\big({\T\mu\CdoT\frac{2\pi}{M}}\big)\CdoT\boldsymbol{V}(\mu)
\label{2.29}
\end{equation}
darstellen. Dazu wurde zun"achst in die Definitionsgleichung (\ref{2.16}) 
der zuf"alligen Spektralwerte \mbox{$\boldsymbol{N}_{\!\!f}(\mu)$} 
der Approximationsfehlerprozess nach Gleichung (\ref{2.9}) eingesetzt. 
Anschlie"send wurde in der sich ergebenden Gleichung f"ur eine Summe 
die Fouriertransformierte \mbox{$\boldsymbol{Y}_{\!\!\!f}(\mu)$} des 
diskreten, gefensterten Ausgangsprozesses \mbox{$\boldsymbol{y}(k)$}
nach Gleichung (\ref{2.25}) substituiert. In der anderen Summe wurde die 
Nullstelleneigenschaft (\ref{2.27}) des Spektrums der Fensterfolge 
ber"ucksichtigt, so dass von dieser Summe nur ein Summand "ubrig bleibt. 

Da wir uns hier auf den Fall mittelwertfreier Ein- und Ausgangsprozesse 
\mbox{$\boldsymbol{v}(k)$} und \mbox{$\boldsymbol{y}(k)$} beschr"anken, 
sind auch bei Verwendung einer Fensterung deren Fouriertransformierte
\mbox{$\boldsymbol{V}(\mu)$} und \mbox{$\boldsymbol{Y}_{\!\!\!f}(\mu)$}
mittelwertfrei, da es sich hierbei um eine lineare Super"-po"-si"-tion 
mittelwertfreier  Zufallsgr"o"sen handelt. Daher sind auch die zuf"alligen 
Spektralwerte \mbox{$\boldsymbol{N}_{\!\!f}(\mu)$} mittelwertfrei.
Deren zweite Momente \mbox{$M\cdot\Tilde{\Phi}_{\boldsymbol{n}}(\mu)$}
sind somit zugleich deren Varianzen ($=$~zweite {\em zentrale} Momente).
Desweiteren l"asst sich die G"ultigkeit der Aussage
\begin{equation}
\text{E}\big\{\boldsymbol{V}(\mu)^{\Kk}\!\CdoT
\boldsymbol{N}_{\!\!f}(\mu)\big\}\;=\;0
\label{2.30}
\end{equation}
"uber die Kovarianz der Zufallsgr"o"sen \mbox{$\boldsymbol{V}(\mu)$} 
und \mbox{$\boldsymbol{N}_{\!\!f}(\mu)$} durch Einsetzen der Schreibweise 
(\ref{2.29}) und der Optimall"osung (\ref{2.28}) zeigen. Die Approximation 
des realen Systems durch das Modellsystem wird also so vorgenommen, dass 
die sich ergebenden zuf"alligen Spektralwerte des gefensterten 
Approximationsfehlers zu den Spektralwerten des erregenden Zufallsvektors 
orthogonal sind.

\subsection{Auswirkung der Fensterung auf das modifizierte LDS}

Die zweiten Momente --- also die Korrelationen des Real- und des 
Imagin"arteils --- des Prozesses \mbox{$\boldsymbol{n}(k)$} sind 
bei einem allgemeinen komplexen Rauschprozess durch die AKF
\mbox{$\text{E}\big\{\,\boldsymbol{n}(k)^{\Kk}\Cdot
\boldsymbol{n}(k\!+\!\kappa)\,\big\}$}
nicht vollst"andig beschrieben. Vollst"andig wird die Beschreibung der
zweiten Momente durch die zus"atzliche Angabe der modifizierten 
Autokorrelationsfolge (\,MAKF\,), die als \mbox{$\text{E}\big\{\,\boldsymbol{n}(k)\cdot
\boldsymbol{n}(k\!+\!\kappa)\,\big\}$} definiert sei.
Aufgrund der geforderten Stationarit"at h"angt auch diese
Korrelationsfolge nur von der Zeitdifferenz $\kappa$ und nicht von $k$ ab.
Sowohl der Real- als auch der Imagin"arteil dieser Korrelationsfolge
sind gerade. Diese Korrelationsfolge ist durch ihre Fouriertransformierte
\begin{equation}
\Psi_{\boldsymbol{n}}(\Omega)\;=
\Sum{\kappa=-\infty}{\infty}
\text{E}\big\{\boldsymbol{n}(k)\CdoT
\boldsymbol{n}(k\!+\!\kappa)\big\}\cdot
e^{\!-j\cdot\Omega\cdot\kappa}
\qquad\qquad\forall\qquad\Omega\in\mathbb{R},\quad{}
\label{2.31}
\end{equation}
die eine mit $2\pi$ periodische und in Real- und Imagin"arteil
gerade Funktion ist, vollst"andig beschrieben. Der Realteil von 
\mbox{$\Psi_{\boldsymbol{n}}(\Omega)$} ist bei einem mittelwertfreien 
Approximationsfehlerprozess die Differenz der 
Leistungsdichtespektren des Real- und des Imagin"arteilprozesses. 
Den Imagin"arteil von \mbox{$\Psi_{\boldsymbol{n}}(\Omega)$} erh"alt man
durch Fouriertransformation aus der doppelten Kreuzkorrelationsfolge der
Real- und Imagin"arteilprozesse. \mbox{$\Psi_{\boldsymbol{n}}(\Omega)$}
wird im Weiteren als modifiziertes Leistungsdichtespektrum (\,MLDS\,) 
des Prozesses \mbox{$\boldsymbol{n}(k)$} bezeichnet.

Aus den oben genannten Gr"unden ist --- wie bei dem LDS
\mbox{$\Phi_{\boldsymbol{n}}(\Omega)$} --- die Angabe von Abtastwerten dieser
Spektralfunktion nicht m"oglich und sinnvoll. Stattdessen kann man
analog zum LDS die $M$ Erwartungswerte \mbox{$\Tilde{\Psi}_{\boldsymbol{n}}(\mu)$}
der $M$ Zufallsgr"o"sen
\mbox{$\boldsymbol{N}_{\!\!f}(\mu)\CdoT\boldsymbol{N}_{\!\!f}(\!-\mu)/M$}
angeben. Diese berechnen sich mit Hilfe der Fensterung entsprechend zu:
\begin{gather}
\Tilde{\Psi}_{\boldsymbol{n}}(\mu)\;=\;
\text{E}\Big\{\,\frac{1}{M}\cdot
\boldsymbol{N}_{\!\!f}(\mu)\CdoT\boldsymbol{N}_{\!\!f}(\!-\mu)\,\Big\}\;=
\label{2.32}\\*[8pt]
=\;\frac{1}{M}\cdot\text{E}\bigg\{\,\Sum{k_1=-\infty}{\infty}\!\!
\boldsymbol{n}(k_1)\CdoT f(k_1)\cdot
e^{\!-j\cdot\frac{2\pi}{M}\cdot\mu\cdot k_1}\:\cdoT
\Sum{k_2=-\infty}{\infty}\!\!\boldsymbol{n}(k_2)\CdoT f(k_2)\cdot
e^{j\cdot\frac{2\pi}{M}\cdot\mu\cdot k_2}\,\bigg\}\;=
\notag\\[8pt]
=\;\frac{1}{M}\cdoT\Sum{k_2=-\infty}{\infty}\,\Sum{k_1=-\infty}{\infty}\!
\text{E}\big\{\boldsymbol{n}(k_2)\CdoT\boldsymbol{n}(k_1)\big\}\cdot
f(k_2)\CdoT f(k_1)\cdot
e^{j\cdot\frac{2\pi}{M}\cdot\mu\cdot(k_2-k_1)}\;=
\notag\\[10pt]
=\;\frac{1}{M}\cdoT\Sum{\Tilde{k}=-\infty}{\infty}\,
\Sum{\kappa=-\infty}{\infty}\!
\text{E}\big\{\boldsymbol{n}(\Tilde{k})\CdoT
\boldsymbol{n}(\Tilde{k}\!+\!\kappa)\big\}\cdot
f(\Tilde{k})\CdoT f(\Tilde{k}\!+\!\kappa)\cdot
e^{\!-j\cdot\frac{2\pi}{M}\cdot\mu\cdot\kappa}\;=
\notag\\[8pt]
=\;\frac{1}{M}\cdoT\Sum{\Tilde{k}=-\infty}{\infty}\,
\Sum{\kappa=-\infty}{\infty}
\frac{1}{2\pi}\cdoT\!\Int{-\pi}{\pi}\Psi_{\boldsymbol{n}}(\Omega)\cdot
e^{j\cdot\Omega\cdot\kappa}\Cdot d\Omega\,\cdot\,
f(\Tilde{k})\CdoT f(\Tilde{k}\!+\!\kappa)\cdot
e^{\!-j\cdot\frac{2\pi}{M}\cdot\mu\cdot\kappa}\;=
\notag\\[10pt]
=\;\frac{1}{2\pi\CdoT M}\cdoT\Int{-\pi}{\pi}\Psi_{\boldsymbol{n}}(\Omega)\cdot
F\big({\T\mu\CdoT\frac{2\pi}{M}\!-\!\Omega}\big)\,\cdoT\!
\Sum{\Tilde{k}=-\infty}{\infty}\!f(\Tilde{k})\cdot
e^{\!-j\cdot(\Omega-\frac{2\pi}{M}\cdot\mu)\cdot\Tilde{k}}\cdot\:d\Omega\;=
\notag\\[8pt]
=\;\frac{1}{2\pi\CdoT M}\cdoT\Int{-\pi}{\pi}\Psi_{\boldsymbol{n}}(\Omega)\cdot
F\big({\T\mu\CdoT\frac{2\pi}{M}\!-\!\Omega}\big)\cdot
F\big({\T\Omega\!-\!\mu\CdoT\frac{2\pi}{M}}\big)\,\cdot\:d\Omega
\notag\\*[4pt]
\qquad\qquad\forall\qquad\mu=0\;(1)\;M\!-\!1.\notag
\end{gather}

Bei Verwendung einer reellen Fensterfolge
kann man f"ur diese Werte 
\begin{equation}
\Tilde{\Psi}_{\boldsymbol{n}}(\mu)\;=\;
\frac{1}{2\pi\CdoT M}\cdoT\Int{-\pi}{\pi}\Psi_{\boldsymbol{n}}(\Omega)\cdot
\big|F\big({\T\mu\CdoT\frac{2\pi}{M}\!-\!\Omega}\big)\big|^2
\Cdot\,d\Omega
\quad\qquad\forall\qquad \mu=0\;(1)\;M\!-\!1\quad{}
\label{2.33}
\end{equation}
schreiben. Wie bei den Werten \mbox{$\Tilde{\Phi}_{\boldsymbol{n}}(\mu)$}
wird auch bei den Werten \mbox{$\Tilde{\Psi}_{\boldsymbol{n}}(\mu)$}
bei jeder diskreten Frequenz $\mu$ das Integral "uber die durch das
Betragsquadratspektrum der Fensterfolge ausgeblendete unmittelbare
Umgebung der Frequenz \mbox{$\Omega=\mu\CdoT2\pi/M$} der Funktion
\mbox{$\Psi_{\boldsymbol{n}}(\Omega)$} angegeben. Die Werte
\mbox{$\Tilde{\Psi}_{\boldsymbol{n}}(\mu)$} lassen sich analog auch wieder
aus der Korrelationsfolge \mbox{$\text{E}\big\{\boldsymbol{n}(k)\CdoT
\boldsymbol{n}(k\!+\!\kappa)\big\}$} berechnen, indem man diese mit der
Folge \mbox{$f(\kappa)\!\ast\!f(\!-\kappa)/M$}, die bei Verwendung einer reellen
Fensterfolge gleich der Fenster-AKF \mbox{$d(\kappa)$} ist,
fenstert und anschlie"send diskret fouriertransformiert. 
Bei geeigneter Wahl der Fensterfolge hat man mit
\mbox{$\Tilde{\Psi}_{\boldsymbol{n}}(\mu)$} eine gute N"aherung
f"ur die Gr"o"sen
\begin{equation}
\Bar{\Psi}_{\boldsymbol{n}}(\mu)\;=\;
\frac{M}{2\pi}\cdoT\Int{-\frac{\pi}{M}}{\frac{\pi}{M}}
\Psi_{\boldsymbol{n}}\big({\T\mu\CdoT\frac{2\pi}{M}\!-\!\Omega}\big)
\cdot\, d\Omega.
\label{2.34}
\end{equation}
Auch hier gilt bei Verwendung einer Fensterfolge, die der Bedingung
(\ref{2.20}) gen"ugt:
\begin{gather}
\frac{1}{M}\cdoT\Sum{\mu=0}{M-1}\Tilde{\Psi}_{\boldsymbol{n}}(\mu)\;=\;
\frac{1}{2\pi\CdoT M^2}\cdoT\Sum{\mu=0}{M-1}\:
\Int{-\pi}{\pi}\Psi_{\boldsymbol{n}}(\Omega)\cdot
\big|F\big({\T\mu\CdoT\frac{2\pi}{M}\!-\!\Omega}\big)\big|^2
\Cdot\:d\Omega\;=
\label{2.35}\\*[6pt]
=\;\frac{1}{2\pi}\cdoT\!\Int{-\pi}{\pi}\Psi_{\boldsymbol{n}}(\Omega)
\cdot\frac{1}{M^2}\cdoT\Sum{\mu=0}{M-1}\:
\big|F\big({\T\mu\CdoT\frac{2\pi}{M}\!-\!\Omega}\big)\big|^2
\Cdot\:d\Omega\;=\;
\frac{1}{2\pi}\cdoT\!\Int{-\pi}{\pi}
\Psi_{\boldsymbol{n}}(\Omega)\cdot d\Omega\;=\;
\text{E}\big\{\boldsymbol{n}(k)^2\big\}.\notag
\end{gather}
Gleichung (\ref{2.16}) definiert die $M$ Zufallsgr"o"sen
\mbox{$\boldsymbol{N}_{\!\!f}(\mu)$} als endliche 
Linearkombination der einzelnen Zufallsgr"o"sen, die f"ur die
unterschiedlichen Werte $k$ aus dem Zufallsprozess
\mbox{$\boldsymbol{n}(k)$} entnommen sind. Die Koeffizienten dieser
Linearkombination sind die Produkte aus den Werten der Fensterfolge
und den Drehfaktoren der DFT. Wenn die Erwartungswerte
\mbox{$\text{E}\big\{|\boldsymbol{N}_{\!\!f}(\mu)|^2\big\}$}
und \mbox{$\text{E}\big\{\boldsymbol{N}_{\!\!f}(\mu)\CdoT
\boldsymbol{N}_{\!\!f}(\!-\mu)\big\}$} existieren, kann man diese nach
Gleichung (\ref{2.17}) und (\ref{2.32}) in die Ungleichung
\begin{equation}
\big|\Tilde{\Psi}_{\boldsymbol{n}}(\mu)\big|^2 \,\le\;
\Tilde{\Phi}_{\boldsymbol{n}}(\mu) \cdot \Tilde{\Phi}_{\boldsymbol{n}}(\!-\mu)
\label{2.36}
\end{equation}
einsetzen, und deren G"ultigkeit mit der im Anhang \ref{Cauchy}
hergeleiteten Ungleichung zeigen.

\section[Zusammenfassung der Systemapproximation mit Fensterung]{Zusammenfassung der Systemapproximation mit \\Fensterung}

In diesem Kapitel wurde gezeigt, wie sich ein reales gest"ortes 
System modellieren l"asst, wenn die an dem System anliegenden 
Ein- und Ausgangsprozesse station"ar und mittelwertfrei sind. 
Die sich bei einer periodischen zuf"alligen Erregung ergebenden 
{\em theoretisch} optimalen Werte der "Ubertragungsfunktion des 
Modellsystems wurden in der Art bestimmt, dass der verbleibende 
Approximationsfehlerprozess eine minimale Varianz aufweist. 
Es wurde zur Beschreibung des LDS des Fehlerprozesses eine 
Spektralfolge endlich vieler Werte angegeben, die sich mit Hilfe 
einer Fensterung, einer DFT und einer Erwartungswertbildung aus den 
Ein- und Ausgangsprozessen berechnen l"asst. Es wurde gezeigt, dass 
die Fensterung keinen Einfluss auf die optimale Systemaufspaltung hat.
Es wurden die folgenden drei Bedingungen f"ur die dabei verwendete 
Fensterfolge angegeben, die sicherstellen, dass die Systemaufspaltung 
nicht verf"alscht wird und die Spektralfolge in der Lage ist das LDS 
des Approximationsfehlerprozesses aussagekr"aftig zu beschreiben.
\begin{enumerate}
\item Die Fensterfolge soll ein leistungskomplement"ares Spektrum gem"a"s
      Gleichungen (\ref{2.20}) besitzen. Dies stellt die korrekte Erfassung 
      der Gesamtrauschleistung sicher.
\item Das Betragsquadrat des Spektrums der Fensterfolge sollte einen 
      potenzm"a"sigen Anstieg der Sperrd"ampfung besitzen, wobei die 
      Potenz h"oher sein sollte als die schlimmstenfalls zu erwartende 
      Potenz des Abfalls des Leistungsdichtespektrums. Dies erm"oglicht 
      die ad"aquate Beschreibung des LDS eines Prozesses mit stark 
      schwankendem LDS. 
\item Das Spektrum der Fensterfolge soll die Nullstellen gem"a"s
      Gleichungen (\ref{2.20}) besitzen. Dies stellt die korrekte 
      Systemaufspaltung sicher.
\end{enumerate}


\chapter{Das Rauschklirrmessverfahren mit Fensterung}\label{RKM}

Nun wollen wir uns der Frage widmen, wie man mit Hilfe
einer Messung Sch"atzwerte \mbox{$\Hat{H}(\mu)$} f"ur die 
theoretischen Optimall"osungen der "Ubertragungsfunktion 
\mbox{$H\big({\T\mu\CdoT\frac{2\pi}{M}}\big)$} sowie Sch"atzwerte 
\mbox{$\Hat{\Phi}_{\boldsymbol{n}}(\mu)$} und 
\mbox{$\Hat{\Psi}_{\boldsymbol{n}}(\mu)$} f"ur die N"aherungen
\mbox{$\Tilde{\Phi}_{\boldsymbol{n}}(\mu)$} und 
\mbox{$\Tilde{\Psi}_{\boldsymbol{n}}(\mu)$}
der Stufenapproximationen des LDS und des MLDS des Approximationsfehlers
gewinnen kann. Die mit der Messung f"ur diese Gr"o"sen gewonnenen Sch"atzwerte
werden im weiteren meist als Messwerte bezeichnet.

Wir beginnen damit, das prinzipielle Verfahren zur Bestimmung der Messwerte 
der "Ubertragungsfunktion anzugeben. Bei diesem Verfahren wird eine Fensterung
vorgenommen. Es wird dann gezeigt, wie sich der Aufwand des Verfahrens 
deutlich reduzieren l"asst, wenn man die dabei eingesetzte Fensterfolge 
geeignet w"ahlt. Die Erwartungstreue der Messwerte der "Ubertragungsfunktion 
wird anschlie"send gezeigt. 

Wie sich bei diesem Messverfahren zugleich Sch"atzwerte f"ur das LDS und 
das MLDS des Approximationsfehlers gewinnen lassen, wird im Folgenden gezeigt. 
Auch hier wird zun"achst das prinzipielle Vorgehen geschildert. Es wird dann 
gezeigt, wie sich das Verfahren vereinfachen l"asst, wenn man eine dabei 
auftretende Matrix geschickt konstruiert. 

Es folgt die Berechnung der Messwertvarianzen und Kovarianzen. Damit 
kann anschlie"send die Konsistenz der Messwerte nachgewiesen werden. 
In diesen Unterkapiteln werden zudem Sch"atzwerte f"ur die Messwertvarianzen 
und Kovarianzen angegeben. Nachdem einige Anmerkungen zu Besonderheiten 
dieses Messverfahrens durchgef"uhrt worden sind, wird gezeigt, wie man mit 
Hilfe der Sch"atzwerte der Messwertvarianzen und Kovarianzen Konfidenzgebiete 
f"ur die Messwerte angeben kann.

Abschlie"send werden zwei Arten von Signalen angegeben, die f"ur den Einsatz 
als Erregung beim Rauschklirrmessverfahren besonders geeignet sind.

\section{Prinzip der Messung der "Ubertragungsfunktion}

Um die Sch"atzwerte mit Hilfe einer Messung bestimmen zu k"onnen,
ben"otigen wir eine Stichprobe vom Umfang $L$ des Zufallsvektors
\mbox{$\Vec{\boldsymbol{v}}$}, also $L$ konkrete Realisierungen 
--- die Vektoren \mbox{$\Vec{v}_{\lambda}$} mit \mbox{$\lambda=1\;(1)\;L$} --- 
des Zufallsvektors \mbox{$\Vec{\boldsymbol{v}}$}, die voneinander 
unabh"angig gewonnen werden. Die $L$ Vektoren \mbox{$\Vec{V}_{\lambda}$},
die man mit Hilfe einer DFT aus den $L$ Vektoren \mbox{$\Vec{v}_{\lambda}$}
berechnet, sind somit eine Stichprobe vom Umfang $L$ des Zufallsvektors 
\mbox{$\Vec{\boldsymbol{V}}$}. Wahlweise kann man daher auch eine 
Stichprobe vom Umfang $L$ des Zufallsvektors \mbox{$\Vec{\boldsymbol{V}}$} 
entnehmen und daraus die $L$ Vektoren \mbox{$\Vec{v}_{\lambda}$} berechnen. 
Jeweils einer dieser $L$ Vektoren bildet das Eingangssignal \mbox{$v_{\lambda}(k)$} 
im Zeitintervall \mbox{$[0;M\!-\!1]$} bei der Einzelmessung mit der Nummer
$\lambda$. Das gesamte Eingangssignal bei der Einzelmessung erh"alt man durch
periodische Fortsetzung des Signals mit der Periode $M$. Da wir in Gleichung 
(\ref{2.6}) angenommen haben, dass die Impulsantwort \mbox{$h(k)$} des Systems 
nach $E$ Takten soweit abgeklungen ist, dass die weiteren Abtastwerte mit 
\mbox{$k>E$} keinen Einfluss auf die Systemmodellierung mehr haben, gen"ugt es, 
wenn wir die Erregung bei den Einzelmessungen ab einem Zeitpunkt anlegen, der 
um $E$ Takte vor dem Zeitpunkt liegt, an dem wir mit der Messung der Signale am 
Systemausgang beginnen. Das System befindet sich dann im eingeschwungenen Zustand 
und am Systemausgang liegen dann dieselben Signale an, die sich ergeben w"urden, 
wenn man das System mit einer zeitlich unbegrenzten periodischen Folge erregt h"atte.
Die Messung am Ausgang des realen Systems soll die $F$ aufeinanderfolgenden 
Abtastwerte der Musterfolge \mbox{$y_{\lambda}(k)$} mit \mbox{$k=0\;(1)\;F\!-\!1$}
enthalten. Die in Gleichung (\ref{2.6}) angenommene Kausalit"at des Modellsystems
f"uhrt dazu, dass das reale System f"ur \mbox{$k\ge F$} nicht weiter erregt 
zu werden braucht. Die periodisch fortgesetzte Musterfolge der Erregung muss 
also bei jeder Einzelmessung f"ur \mbox{$k=-E\;(1)\;F\!-\!1$} anliegen.

Die Elemente der Vektoren \mbox{$\Vec{v}_{\lambda}$} sind aufgrund
der begrenzten Wortl"ange des zur Si"-gnal"-er"-zeu"-gung verwendeten Rechners
fehlerbehaftet. Diese Fehler entstehen einerseits durch die Quantisierung
der Elemente der Vektoren \mbox{$\Vec{V}_{\lambda}$} und andererseits
durch die inverse DFT dieser Vektoren, die auch nur mit einer endlichen
Wortl"ange durchgef"uhrt wird. Die Quantisierung der Vektoren
\mbox{$\Vec{V}_{\lambda}$} hat keinen direkten Einfluss auf die
Genauigkeit der Berechnung der gesuchten Messwerte, weil in der
gesamten Messung --- also auch bei der Erzeugung der Erregung ---
nur die quantisierten Vektoren eingehen. Man kann daher diese
Quantisierung als Teil des Zufallsgenerators betrachten, mit dessen
Hilfe die Stichprobe des Zufallsvektors $\Vec{\boldsymbol{V}}$ gewonnen wird.
Der Zufallsvektor $\Vec{\boldsymbol{V}}$ ist also diskret, und besitzt als
Menge der Ergebnisvektoren eine Untermenge aller am Rechner mit
endlicher Wortl"ange darstellbaren quantisierten Vektoren. Die
Quantisierungsfehler, die bei der Berechnung der inversen DFT bei der
Erzeugung der Systemerregung entstehen, werden als additive St"orung
am Eingang des realen Systems betrachtet, und diesem zugeschlagen.

Am Ausgang des Systems messen wir innerhalb des Intervalls
\mbox{$[0;F\!-\!1]$} die $L$ Ausgangssignalfolgen
\mbox{$y_{\lambda}(k)$} mit \mbox{$\lambda=1\;(1)\;L$}, die eine
konkrete Stichprobe vom Umfang $L$ des Zufallsvektors sind, 
der aus dem Zufallsprozess \mbox{$\boldsymbol{y}(k)$}
durch zeitliche Begrenzung auf das Zeitintervall \mbox{$[0;F\!-\!1]$}
entsteht. Die Messung muss dabei in der Art erfolgen, 
dass die $L$ Einzelmessungen voneinander unabh"angig 
sind und alle $L$ Stichprobenelemente dieselben 
stochastischen Eigenschaften aufweisen, wie der 
Zufallsprozess \mbox{$\boldsymbol{y}(k)$} in diesem 
Zeitintervall. Das bedeutet, dass sich die gemeinsame 
Verbundverteilung aller $L$ Ausgangssignalabschnitte 
in die $L$ identischen Verbundverteilungen der einzelnen 
Ausgangssignalabschnitte faktorisieren l"asst. Dass
es sich bei den $L$ Einzelmessungen um eine zul"assige
Stichprobe vom Umfang $L$ mit den eben erw"ahnten 
Eigenschaften handelt, ist im Einzelfall z.~B.
anhand heuristischer "Uberlegungen sicherzustellen. 
Beispielsweise ist bei einer Simulation eines nichtlinearen 
Systems ohne externe St"orungen an einem Computer nicht damit 
zu rechnen, dass Abh"angigkeiten der Einzelmessungen vorliegen, 
wenn das System bei jeder Einzelmessung neu initialisiert wird, und
wenn man einen Zufallsgenerator zur Generierung der Erregung 
verwendet, bei dem die Unabh"angigkeit der Stichprobenelemente 
"uberpr"uft worden ist. Weitere "Uberlegungen zur geeigneten Wahl 
der Stichprobenerhebung sind in \cite{Erg} f"ur den Fall 
periodisch zeitvarianter Systeme, die von zyklostation"aren 
Prozessen gest"ort werden, angegeben.

Durch eine Fensterung, eine blockweise "Uberlagerung und eine 
anschlie"sende DFT der $L$ Ausgangssignalausschnitte \mbox{$y_{\lambda}(k)$}
werden dann die $L$ Vektoren \mbox{$\Vec{Y}_{\!f,\lambda}$} 
gebildet, die eine konkrete Stichprobe vom Umfang $L$ des Zufallsvektors
\mbox{$\Vec{\boldsymbol{Y}}_{\!\!\!f}$} mit den Elementen
\mbox{$\boldsymbol{Y}_{\!\!\!f}(\mu)$} darstellen. Die Fensterung 
erfolgt dabei mit den gemessenen Musterfolgen in derselben Art wie 
beim Ausgangsprozess in Gleichung (\ref{2.25}) in vier Schritten:
\begin{itemize}\label{yFen}
\item Die Musterfolge wird mit der Fensterfolge multipliziert:
      \mbox{$f(k)\CdoT y_{\lambda}(k)$}.
\item Die gefensterte Musterfolge wird in Bl"ocke der L"ange 
      $M$ zerlegt.\\ F"ur jeden Wert $\kappa$ bildet der Musterfolgenabschnitt 
      \mbox{$f(k\!+\!\kappa\CdoT M)\CdoT y_{\lambda}(k\!+\!\kappa\CdoT M)$}
      f"ur die $M$ Werte \mbox{$k=0\;(1)\;M\!-\!1$} einen Block der L"ange $M$.
\item Die Bl"ocke f"ur alle Werte von $\kappa$ werden additiv "uberlagert:\\
      \mbox{$y_{f,\lambda}(k)\;= \Sum{\kappa=-\infty}{\infty}
      f(k\!+\!\kappa\CdoT M)\CdoT y_{\lambda}(k\!+\!\kappa\CdoT M)$}
\item Die so entstandene Musterfolge der L"ange $M$ wird einer DFT unterworfen 
      und liefert die $M$ Elemente des Vektors \mbox{$\Vec{Y}_{\!f,\lambda}$}:\\
      \mbox{$Y_{\!f,\lambda}(\mu)\;=
      \Sum{k=0}{M-1}y_{f,\lambda}(k)\cdot e^{\!-j\cdot\frac{2\pi}{M}\cdot\mu\cdot k}$}.
\end{itemize}

Auch die Vektoren \mbox{$\Vec{Y}_{\!f,\lambda}$} sind aufgrund
der begrenzten Wortl"ange des zur Messung verwendeten Rechners
fehlerbehaftet. Diese Fehler entstehen einerseits durch die Quantisierung
der Werte \mbox{$y_{\lambda}(k)$} der $L$ Folgen am Systemausgang
und andererseits durch die Fensterung und die anschlie"sende DFT dieser 
Folgen. Wie bei den Fehlern bei der Berechnung der inversen DFT bei der
Erzeugung der Systemerregung, wollen wir auch diese Fehler als eine
additive St"orung des realen Systems --- diesmal aber am Systemausgang ---
interpretieren. Wir wollen daher die Vektoren \mbox{$\Vec{V}_{\lambda}$}
und \mbox{$\Vec{Y}_{\!f,\lambda}$} als fehlerfrei betrachten, m"ussen uns
aber dar"uber im Klaren sein, dass wir ein modifiziertes reales System
messen. Bei den meisten realen Systemen, wird der Einfluss dieser
zus"atzlichen St"orungen verschwindend gering sein. In Kapitel
\ref{Mess0} wird an einem Verz"ogerungsglied, das keine weiteren
Quantisierungsfehler hervorruft, der Einfluss der Quantisierungen
bei der Berechnung des Spektrums der Erregung und der Messwerte 
exemplarisch dargestellt.

Bei der theoretischen Modellierung des Systems, also der Berechnung der 
Regressionshyperebene der zweiten Art, bestand in Kapitel \ref{AppPhi} 
die Aufgabe darin die {\em theoretischen}\/ Regressionskoeffizienten so zu
w"ahlen, dass die Erwartungswerte \mbox{$\Tilde{\Phi}_{\boldsymbol{n}}(\mu)$}
minimal werden. Gleichung (\ref{2.24}) zeigt, dass dies gleichbedeutend damit ist,
die theoretischen Regressionskoeffizienten so zu w"ahlen,
dass der Zufallsvektor mit den Elementen
\begin{equation}
\frac{1}{M}\cdoT\Sum{\Tilde{\mu}=0}{M-1}\boldsymbol{V}(\Tilde{\mu})\CdoT
H\big({\T\Tilde{\mu}\CdoT\frac{2\pi}{M}}\big)\CdoT
F\big({\T(\mu\!-\!\Tilde{\mu})\CdoT\frac{2\pi}{M}}\big)
\label{3.1}
\end{equation}
im Sinne des kleinsten {\em Erwartungswertes}\/ des quadratischen Fehlers
m"oglichst gut mit dem Zufallsvektor mit den Elementen
\mbox{$\boldsymbol{Y}_{\!\!\!f}(\mu)$} "ubereinstimmt. Bei der
{\em empirischen}\/ Bestimmung der Regressionskoeffizienten 
mit dem Rauschklirrmessverfahren wird man die 
Regressionskoeffizienten so w"ahlen, dass der {\em mittlere}\/
quadratische Fehler der Abweichung der entsprechenden Sichprobe
\begin{equation}
\frac{1}{M}\cdoT\Sum{\Tilde{\mu}=0}{M-1}V_{\lambda}(\Tilde{\mu})\CdoT
\Hat{H}(\Tilde{\mu})\CdoT
F\big({\T(\mu\!-\!\Tilde{\mu})\CdoT\frac{2\pi}{M}}\big),
\label{3.2}
\end{equation}
der aus der Stichprobe des Eingangssignals abgeleitet wird, 
von der gemessenen, gefensterten und diskret fouriertransformierten
Stichprobe des Ausgangssignals \mbox{$Y_{\!f,\lambda}(\mu)$}
m"oglichst klein wird. Die Mittelung des quadratischen Fehlers 
geschieht dabei "uber alle $L$ Einzelmessungen mit dem Index 
$\lambda$, also "uber alle $L$ Stichprobenelemente.

Die Aufgabe besteht somit darin, eine Ausgleichsl"osung f"ur
das "uberbestimmte Gleichungssystem 
\begin{gather}
\frac{1}{M}\cdoT\Sum{\Tilde{\mu}=0}{M-1}\Hat{H}(\Tilde{\mu})\CdoT
V_{\lambda}(\Tilde{\mu})\CdoT
F\big({\T(\mu\!-\!\Tilde{\mu})\CdoT\frac{2\pi}{M}}\big)
\;=\;Y_{\!f,\lambda}(\mu)\notag\\
\forall\qquad \mu=0\;(1)\;M\!-\!1\quad\text{ und }
\quad\lambda=1\;(1)\;L
\label{3.3}
\end{gather}
zu finden. Diese Gleichungssystem besteht aus \mbox{$L\CdoT M$} Gleichungen, 
da sich f"ur jede m"ogliche Kombinationen von $\mu$ und $\lambda$ jeweils eine 
Gleichung ergibt. Die Ausgleichsl"osung ist dabei derjenige Satz von Werten 
\mbox{$\Hat{H}(\Tilde{\mu})$} bei dem das Betragsquadrat der Differenz der 
rechten und der linken Seite des Gleichungssystems m"oglichst klein 
wird\footnote{Als Ausgleichsprinzip verwenden wir somit die Methode der 
kleinsten Quadrate, die auf C. F. Gau"s zur"uckgeht.}. Das Gleichungssystem 
(\ref{3.3}) kann man auch in Matrixdarstellung als
\begin{equation}
\Hat{\Vec{H}}^{\,\Tt}\Cdot\underline{M}\;=\;\Tilde{\Vec{Y}}_{\!f}
\label{3.4}
\end{equation}
schreiben. Der \mbox{$M\!\times\!1$} Spaltenvektor \mbox{$\Hat{\Vec{H}}$}
enth"alt die zu bestimmenden Messwerte, wobei das \mbox{$\mu\!+\!1$}-te 
Element der Spektralwert bei der diskreten Frequenz \mbox{$\mu\CdoT2\pi/M$} ist.
\begin{equation}
\Hat{\Vec{H}}\;=\;\big[\,\Hat{H}(0),\,\ldots\,,\Hat{H}(\mu),\,\ldots\,,
\Hat{H}(M\!-\!1)\,\big]^{\TT}
\label{3.5}
\end{equation}
Der \mbox{$1\!\times(L\CdoT M)$} Zeilenvektor \mbox{$\Tilde{\Vec{Y}}_{\!f}$}
setzt sich aus den $L$ Zeilenvektoren \mbox{$\Vec{Y}_{\!f,\lambda}^{\,\Tt}$}
der Dimension \mbox{$1\!\times\!M$} zusammen, die jeweils die $M$ Werte des
Spektrums des gefensterten Ausgangssignals des zu messenden Systems bei
der Einzelmessung mit der Nummer $\lambda$ enthalten, wobei auch hier
wieder der Spektralwert bei der diskreten Frequenz \mbox{$\mu\CdoT2\pi/M$}
das \mbox{$\mu\!+\!1$}-te Element ist.
\begin{subequations}
\begin{align}
\Tilde{\Vec{Y}}_{\!f}\;&=\;\big[\,\Vec{Y}_{\!f,1}^{\,\TT},\,\ldots\,,
\Vec{Y}_{\!f,\lambda}^{\,\TT},\,\ldots\,,\Vec{Y}_{\!f,L}^{\,\TT}\,\big]
&&\text{mit}\label{3.6.a}\\
\Vec{Y}_{\!f,\lambda}\;&=\;\big[\:Y_{\!f,\lambda}(0),\,\ldots\,,
Y_{\!f,\lambda}(\mu),\,\ldots\,,Y_{\!f,\lambda}(M\!-\!1)\,\big]^{\,\TT}
&\qquad&\forall\qquad\lambda=1\,(1)\,L
\label{3.6.b}
\end{align}
\end{subequations}
Die \mbox{$M\times(L\CdoT M)$} Matrix $\underline{M}$ hat
folgenden Aufbau.
\begin{equation}
\underline{M}\;=\;
\bigg[\underline{V}_1\CdoT\frac{1}{M}\CdoT\underline{F},\,\ldots\,,
\underline{V}_{\lambda}\CdoT\frac{1}{M}\CdoT\underline{F},\,\ldots\,,
\underline{V}_L\CdoT\frac{1}{M}\CdoT\underline{F}\bigg]
\label{3.7}
\end{equation}
Dabei ist $\underline{F}$ die \mbox{$M\!\times\!M$} Matrix, deren 
Element in der \mbox{$\Tilde{\mu}\!+\!1$-ten} Zeile und in der 
\mbox{$\mu\!+\!1$}-ten Spalte der Wert 
\mbox{$F\big((\mu\!-\!\Tilde{\mu})\CdoT2\pi/M\big)$} 
des Spektrums der Fensterfolge ist.
\begin{equation}
\underline{F}\;=\;\begin{bmatrix}
F(0)&\cdots&
F\big({\T\mu\CdoT\frac{2\pi}{M}}\big)&\cdots&
F\big({\T(M\!-\!1)\CdoT\frac{2\pi}{M}}\big)\\[-2pt]
\vdots&&\vdots&&\vdots\\[-2pt]
F\big({\T-\Tilde{\mu}\CdoT\frac{2\pi}{M}}\big)&\cdots&
F\big({\T(\mu\!-\!\Tilde{\mu})\CdoT\frac{2\pi}{M}}\big)&\cdots&
F\big({\T(M\!-\!1\!-\!\Tilde{\mu})\CdoT\frac{2\pi}{M}}\big)\\[-2pt]
\vdots&&\vdots&&\vdots\\[-2pt]
F\big({\T(1\!-\!M)\CdoT\frac{2\pi}{M}}\big)&\cdots&
F\big({\T(\mu\!-\!M\!+\!1)\CdoT\frac{2\pi}{M}}\big)&\cdots&
F(0)
\end{bmatrix}
\label{3.8}
\end{equation}
$\underline{F}$ ist eine zirkulante Matrix, d.~h. jeder Zeilenvektor
dieser Matrix ist gerade die um ein Element rotierte Version des
dar"uberliegenden Zeilenvektors. $\underline{F}$ l"asst sich daher
mit der Matrixmultiplikation 
\mbox{$\underline{T}^{\HH}\Cdot\underline{F}\cdoT\underline{T}$} 
unit"ar auf Diagonalform transformieren. Die dazu ben"otigte
Transformationsmatrix $\underline{T}$ ist die Matrix, die als 
Elemente die auf $\sqrt{\!M\,}$ normierten Drehfaktoren der DFT 
enth"alt\footnote{Das Element \mbox{$T_{\mu,k}$} in der 
\mbox{$\mu\!+\!1$}-ten Zeile und in der \mbox{$k\!+\!1$-ten} 
Spalte der Transformationsmatrix enth"alt den Drehfaktor 
\mbox{$\D \frac{1}{\sqrt{\!M\,}}\cdot e^{j\cdot\frac{2\pi}{M}\cdot\mu\cdot k}$}.}.
Die sich bei Transformation ergebende Diagonalmatrix 
hat als \mbox{$k\!+\!1$}-tes Diagonalelement $M$ mal die Summe
\mbox{$\sum_{\kappa=0}^{F/M}f(k\!+\!\kappa\CdoT M)$} der Werte der
Fensterfolge. Die Matrix $\underline{F}$ ist also regul"ar,
wenn diese Summe der Werte der Fensterfolge f"ur alle \mbox{$\:0\le k<M\:$}
von Null verschieden ist.

Die in der Gleichung (\ref{3.7}) auftretende
Matrix $\underline{V}_{\lambda}$ ist eine \mbox{$M\!\times\!M$}
Diagonalmatrix, deren Hauptdiagonalenelemente gerade die Elemente des
Vektors \mbox{$\Vec{V}_{\lambda}$} sind, der bei der Einzelmessung mit
der Nummer $\lambda$ zur Berechnung der Erregung verwendet wurde.
Das \mbox{$\mu\!+\!1$}-te Diagonalelement von $\underline{V}_{\lambda}$
ist also der Wert \mbox{$V_{\lambda}(\mu)$} des Spektrums der Erregung
bei der $\lambda$-ten Einzelmessung und bei der diskreten Frequenz
\mbox{$\mu\CdoT2\pi/M$}.

Als Ausgleichsl"osung f"ur das "uberbestimmte Gleichungssystem (\ref{3.4})
erh"alt man
\begin{equation}
\Hat{\Vec{H}}^{\,\Tt}\,=\;
\Tilde{\Vec{Y}}_{\!f}\cdot\underline{M}^{\HH}\Cdot
\big(\underline{M}\cdot\underline{M}^{\HH}\big)^{\!-1}\!\!.
\label{3.9}
\end{equation}
Voraussetzung f"ur die Existenz einer eindeutigen
Ausgleichsl"osung ist die Regularit"at der Matrix
\mbox{$\underline{M}\cdot\underline{M}^{\HH}$}.
Selbst wenn die Matrix $\underline{F}$ regul"ar ist und die empirischen
zweiten Momente
\begin{equation}
\Hat{C}_{\boldsymbol{V}(\mu),\boldsymbol{V}(\mu)}\;=\;
\frac{1}{L}\cdoT\Sum{\lambda=1}{L}\,\big|V_{\lambda}(\mu)\big|^2
\label{3.10}
\end{equation}
f"ur alle Werte $\mu$ von Null verschieden sind,
kann man die Regularit"at der Matrix
\mbox{$\underline{M}\cdot\underline{M}^{\hH}$}
nicht zeigen. Au"serdem ist zur Berechnung des Messwertvektors 
\mbox{$\Hat{\Vec{H}}$} die Inversion einer \mbox{$M\!\times\!M$} 
Matrix erforderlich, "uber deren Konditionierung i.~Allg. keine 
Aussage gemacht werden kann. Es ist jedoch aufgrund der Struktur 
der Matrix \mbox{$\underline{M}$} anzunehmen, dass die zu 
invertierende Matrix mit hoher Wahrscheinlichkeit gut konditioniert ist, 
wenn die Matrix $\underline{F}\,$ gut konditioniert ist, wenn das empirische
zweite Moment der Spektralwerte des erregenden Signals bei allen Frequenzen
etwa gleich ist und wenn die Spektralwerte der Erregung unkorreliert sind.
Dennoch stellt die Invertierung der \mbox{$M\!\times\!M$} Matrix einen 
erheblichen Rechenaufwand dar. Au"serdem ist zur Berechnung
der zu invertierenden Matrix bei jeder Einzelmessung das Matrixprodukt
\mbox{$\underline{V}_{\lambda}\Cdot\underline{F}\cdot\underline{F}^{\HH}
\Cdot\underline{V}_{\lambda}^{\hH}$} zu berechnen und "uber alle $L$ 
Einzelmessungen zu akkumulieren. Neben dem Rechenaufwand der 
Matrixmultiplikation ist daher ein gro"ser Speicher f"ur die 
Akkumulation notwendig. 

\section{Messung der "Ubertragungsfunktion bei geeigneter Fensterung}

Der bei Anwendung einer nicht n"aher spezifizierten Fensterfolge immense 
Rechenaufwand l"asst sich jedoch drastisch reduzieren, wenn man eine Fensterfolge 
verwendet, die der Bedingung (\ref{2.27}) gen"ugt. Dann wird die 
Matrix $\underline{F}\,$ gerade $M$ mal die Einheitsmatrix, so dass 
sich das Gleichungssystem, f"ur das die Ausgleichsl"osung gefunden werden 
soll, in $M$ Gleichungssysteme zu je $L$ Gleichungen aufspalten l"asst, 
wobei in jedem der $M$ Gleichungssysteme jeweils nur die eine Unbekannte 
\mbox{$\Hat{H}(\mu)$} bei {\em einer}\/ Frequenz vorkommt.
Man erh"alt so in Vektorschreibweise $M$ Gleichungssysteme:
\begin{equation}
\Hat{H}(\mu)\CdoT\Vec{V}(\mu)\;=\;\Vec{Y}_{\!f}(\mu)
\qquad\qquad\forall\qquad\mu=0\;(1)\;M\!-\!1.
\label{3.11}
\end{equation}
Die $M$ dabei auftretenden \mbox{$1\!\times\!L$} Zeilenvektoren
\mbox{$\Vec{Y}_{\!f}(\mu)$} setzten sich jeweils aus den $L$ Elementen
\mbox{$Y_{\!f,\lambda}(\mu)$} zusammen.
\begin{equation}
\Vec{Y}_{\!f}(\mu)\;=\;\big[\,Y_{f,1}(\mu),\,\ldots\,,
Y_{f,\lambda}(\mu),\,\ldots\,,Y_{f,L}(\mu)\,\big]
\qquad\qquad\forall\qquad\mu=0\;(1)\;M\!-\!1\quad{}
\label{3.12}
\end{equation}
Jeder dieser Vektoren ist also eine Stichprobe vom Umfang
$L$ der Zufallsgr"o"se \mbox{$\boldsymbol{Y}_{\!\!\!f}(\mu)$}.
Der Zeilenvektor \mbox{$\Vec{V}(\mu)$} ist eine Stichprobe
vom Umfang $L$ der Zufallsgr"o"se \mbox{$\boldsymbol{V}(\mu)$}.
\begin{equation}
\Vec{V}(\mu)\;=\;\big[\,V_1(\mu),\,\ldots\,,
V_{\lambda}(\mu),\,\ldots\,,V_L(\mu)\,\big]
\qquad\qquad\forall\qquad\mu=0\;(1)\;M\!-\!1
\label{3.13}
\end{equation}
Als Ausgleichsl"osung jedes der Gleichungssysteme (\ref{3.11}) erhalten wir
\begin{align}
\Hat{H}(\mu)&\;=\;
\Vec{Y}_{\!f}(\mu)\CdoT\Vec{V}(\mu)^{\Hh}\Cdot
\Big(\Vec{V}(\mu)\CdoT\Vec{V}(\mu)^{\Hh}\,\Big)^{\!-1}\;=\;{}
\label{3.14}\\[6pt]
&\;=\;\Hat{C}_{\boldsymbol{Y}_{\!\!\!f}(\mu),\boldsymbol{V}(\mu)}\Cdot
\Hat{C}_{\boldsymbol{V}(\mu),\boldsymbol{V}(\mu)}^{\,-1}
\qquad\qquad\forall\qquad\mu=0\;(1)\;M\!-\!1.\notag{}
\end{align}

Dabei ist das Skalarprodukt
\begin{gather}
\Hat{C}_{\boldsymbol{V}(\mu_1),\boldsymbol{V}(\mu_2)}\;=\;
\frac{1}{L}\cdot
\Vec{V}(\mu_1)\cdot\Vec{V}(\mu_2)^{\Hh}\;=\;
\frac{1}{L}\cdoT\Sum{\lambda=1}{L}\,
V_{\lambda}(\mu_1)\CdoT V_{\lambda}(\mu_2)^{\Kk}
\label{3.15}\\
\forall\qquad\mu_1=0\;(1)\;M\!-\!1\quad\text{und}
\quad\mu_2=0\;(1)\;M\!-\!1,\notag
\end{gather}
das mit \mbox{$\mu_1=\mu_2=\mu$} in Gleichung (\ref{3.14}) auftritt, 
ein erwartungstreuer und konsistenter Sch"atzwert f"ur die Kovarianz 
der beiden mittelwertfreien Zufallsgr"o"sen
\mbox{$\boldsymbol{V}(\mu_1)$} und \mbox{$\boldsymbol{V}(\mu_2)$}. 
Dieser Sch"atzwert wird im Weiteren auch als die {\em empirische}\/ 
Kovarianz dieser beiden Zufallsgr"o"sen bezeichnet. 
Man beachte, dass bei der Berechnung der empirischen Kovarianz
das Konjugieren des zweiten beteiligten Stichprobenvektors
in dieser De\-fi\-ni\-tion bereits mit eingeschlossen ist.
Dies erfolgt konform mit der Definition der Kovarianz zweier
komplexer, mittelwertfreier Zufallsgr"o"sen, bei der ebenfalls die 
zweite der daran beteiligten Zufallsgr"o"sen konjugiert wird
\mbox{(\,$C_{\boldsymbol{X},\boldsymbol{Y}}=
\text{E}\{\boldsymbol{X}\CdoT\boldsymbol{Y}^*\}$\,)}.
Die empirische Kovarianz l"asst sich durch Akkumulation aus den 
bei den Einzelmessungen $\lambda$ verwendeten Spektralwerten der 
Erregung berechnen, ohne dass dazu die Spektralwerte aller 
Einzelmessungen bis zum Ende der gesamten Messung abgespeichert 
werden m"ussen. F"ur die Messwerte der "Ubertragungsfunktion
wird hier mit \mbox{$\mu_1=\mu_2=\mu$} die empirische Varianz
\mbox{$\Hat{C}_{\boldsymbol{V}(\mu),\boldsymbol{V}(\mu)}$}
des Spektrums der Erregung ben"otigt. Ist sie von Null verschieden,
so sind die L"osungen f"ur die Werte der "Uber\-tra\-gungs\-funk\-tion 
eindeutig. Dass man ein solch einfaches Kriterium f"ur die 
Eindeutigkeit der L"osung erh"alt, ist ein weiterer wesentlicher 
Vorteil der Verwendung einer Fensterfolge, deren Spektrum der 
Bedingung (\ref{2.27}) gen"ugt.

Nach Gleichung (\ref{3.11}) ben"otigt man zur Berechnung der 
Messwerte der "Ubertragungsfunktion auch noch die empirischen Kreuzkovarianzen
\begin{gather}
\Hat{C}_{\boldsymbol{Y}_{\!\!\!f}(\mu_1),\boldsymbol{V}(\mu_2)}\;=\;
\frac{1}{L}\cdot\Vec{Y}_{\!f}(\mu_1)\cdot\Vec{V}(\mu_2)^{\Hh}\;=\;
\frac{1}{L}\cdoT\Sum{\lambda=1}{L}\,
Y_{\!f,\lambda}(\mu_1)\CdoT V_{\lambda}(\mu_2)^{\Kk}
\notag\\
\forall\qquad\mu_1=0\;(1)\;M\!-\!1\quad\text{ und }
\quad\mu_2=0\;(1)\;M\!-\!1
\label{3.16}
\end{gather}
der Spektralwerte der Signale am Ein- und Ausgang des Systems
f"ur \mbox{$\mu_1=\mu_2=\mu$}, die sich ebenfalls durch Akkumulation
aus den bei den Einzelmessungen $\lambda$ gemessenen Signalen
berechnen lassen.

\section[Erwartungstreue der Messwerte der 
"Ubertragungsfunktion]{Erwartungstreue der Messwerte 
der\\"Ubertragungsfunktion}

Bevor wir uns der Berechnung der Erwartungswerte der Messwerte der 
"Ubertragungsfunktion zuwenden, bedarf es einer Vor"uberlegung.

Gleichung (\ref{2.29}) beschreibt den Zusammenhang zwischen den $M$
Zufallsgr"o"sen \mbox{$\boldsymbol{V}(\mu)$} am Eingang und den $M$
Zufallsgr"o"sen \mbox{$\boldsymbol{Y}_{\!\!\!f}(\mu)$} am Ausgang des 
realen Systems und zeigt, wie sich die $M$ Zufallsgr"o"sen 
\mbox{$\boldsymbol{N}_{\!\!f}(\mu)$} des Spektrums des gefensterten 
Approximationsfehlers daraus ergeben. $M$ konkrete Stichprobenvektoren
\begin{equation}
\Vec{N}_{\!f}(\mu)\;=\;\Big[N_{f,1}(\mu),\,\ldots\,,
N_{\!f,\lambda\!}(\mu),\,\ldots\,,N_{f,L}(\mu)\Big]
\qquad\qquad\forall\qquad \mu=0\;(1)\;M\!-\!1\quad{}
\label{3.17}
\end{equation}
vom Umfang $L$ dieser $M$ Zufallsgr"o"sen erhalten wir, wenn in
Gleichung (\ref{2.29}) statt der Zufallsgr"o"sen \mbox{$\boldsymbol{V}(\mu)$} 
und \mbox{$\boldsymbol{Y}_{\!\!\!f}(\mu)$} deren konkrete Realisierungen 
\mbox{$\Vec{V}(\mu)$} und \mbox{$\Vec{Y}_{\!f}(\mu)$} einsetzen.
\begin{equation}
\Vec{N}_{\!f}(\mu)\;=\;\Vec{Y}_{\!f}(\mu)-
H\big({\T\mu\CdoT\frac{2\pi}{M}}\big)\CdoT\Vec{V}(\mu)
\qquad\qquad\forall\qquad \mu=0\;(1)\;M\!-\!1\quad{}
\label{3.18}
\end{equation}
Man beachte, dass hierbei die unbekannte Optimall"osung 
\mbox{$H(\mu\CdoT2\pi/M)$} der theoretischen Regression 
auftritt, die durch die Messung abgesch"atzt werden soll. 
Die letzte Gleichung l"asst sich nach \mbox{$\Vec{Y}_{\!f}(\mu)$} 
aufl"osen und in die Ausgleichsl"osung (\ref{3.14}) einsetzen und
wir erhalten die Messwerte der "Ubertragungsfunktion als 
Funktion der Stichprobe des Spektrums der Erregung, der
Stichprobe des Approximationsfehlerspektrums und der theoretisch
optimalen Regressionskoeffizienten:
\begin{gather}
\Hat{H}(\mu)\,=\,
\Big(H\big({\T\mu\CdoT\frac{2\pi}{M}}\big)\CdoT\Vec{V}(\mu)+
\Vec{N}_{\!f}(\mu)\Big)\cdot\Vec{V}(\mu)^{\Hh}\Cdot
\Big(\Vec{V}(\mu)\CdoT\Vec{V}(\mu)^{\Hh}\,\Big)^{\!-1}\,=
\label{3.19}
\\*[3pt]
=\;H\big({\T\mu\CdoT\frac{2\pi}{M}}\big)+
\frac{1}{L}\cdot\Vec{N}_{\!f}(\mu)\cdoT
\Vec{V}(\mu)^{\Hh}\Cdot
\Hat{C}_{\boldsymbol{V}(\mu),\boldsymbol{V}(\mu)}^{-1}
\qquad\qquad\forall\qquad \mu=0\;(1)\;M\!-\!1.\notag
\end{gather}

Nun wollen wir die Erwartungswerte der Messwerte der
"Ubertragungsfunktion bestimmen. Dazu betrachten wir diese
Messwerte jeweils als eine konkrete Realisierung --- also eine 
Stichprobe vom Umfang eins --- der Zufallsgr"o"sen 
\mbox{$\Hat{\boldsymbol{H}}(\mu)$}. Diese Zufallsgr"o"sen 
erh"alt man, wenn man statt der konkreten Stichprobenvektoren 
\mbox{$\Vec{V}(\mu)$} und \mbox{$\Vec{N}_{\!f}(\mu)$} 
die mathematischen Stichprobenvektoren\footnote{Auf Seite \pageref{MathStich} 
hatten wir gesagt, dass wir von einer mathematischen Stichprobe sprechen, 
wenn die Stichprobe in der Art erhoben wird, dass alle Elemente der Stichprobe voneinander 
unabh"angig sind, und dieselbe Verteilung besitzen, wie die Zufallsgr"o"se aus der die Stichprobe 
entnommen wurde. Im weiteren wird davon ausgegangen, das die Messung beim RKM so durchgef"uhrt wurde,
dass die Vektoren \mbox{$\Vec{V}(\mu)$} und \mbox{$\Vec{Y}_{\!f}(\mu)$} konkrete Realisierungen
mathematischer Stichprobenvektoren \mbox{$\Vec{\boldsymbol{V}}(\mu)$} und 
\mbox{$\Vec{\boldsymbol{Y}}_{\!\!\!f}(\mu)$} sind.} 
\mbox{$\Vec{\boldsymbol{V}}(\mu)$} und \mbox{$\Vec{\boldsymbol{N}}_{\!\!f}(\mu)$}, 
die aufgrund der zuf"alligen Stichprobenentnahme selbst Zufallsvektoren sind,
in die letzte Gleichung einsetzt. Wir bilden also den Erwartungswert
"uber alle m"oglichen Messungen, die sich jeweils aus $L$
Einzelmessungen zusammensetzen.

Behandeln wir zun"achst den Sonderfall der Erregung mit einem Mehrtonsignal, 
dessen Verwendung in \cite{Dong} vorgeschlagen wurde, und das in Kapitel
\ref{Spesig} etwas n"aher untersucht wird.
Bei der Generierung des Eingangssignals verwenden wir dabei einen Zufallsvektor 
\mbox{$\Vec{\boldsymbol{V}}$}, bei dem lediglich die Phasen der $M$ komplexen 
Elemente \mbox{$\boldsymbol{V}(\mu)$} zuf"allig sind, w"ahrend die Betr"age
nicht zuf"allige Konstanten sind, die jedoch f"ur unterschiedliche Werte von 
$\mu$ unterschiedlich sein d"urfen. In diesem Fall sind bei jeder m"oglichen
konkreten Realisierung \mbox{$\Vec{V}_{\lambda}$} des Zufallsvektors 
\mbox{$\Vec{\boldsymbol{V}}$} die Betr"age der Elemente gleich
den Betr"agen der Elemente des Zufallsvektors \mbox{$\Vec{\boldsymbol{V}}$}
und somit nicht zuf"allig. Auch die $M$ empirischen Varianzen
\mbox{$\Hat{C}_{\boldsymbol{V}(\mu),\boldsymbol{V}(\mu)}$} sind dann nicht 
zuf"allig. Wenn wir weiterhin die konstanten Betr"age f"ur alle $\mu$ von 
Null verschieden w"ahlen, sind die $M$ empirischen Varianzen alle ungleich 
Null und deren Inverse existieren dann immer. Sie k"onnen bei der 
Berechnung der $M$ Erwartungswerte der "Ubertragungsfunktion vor die 
Erwartungswertbildung gezogen werden:
\begin{gather}
\text{E}\big\{\Hat{\boldsymbol{H}}(\mu)\big\}\;=\;
H\big({\T\mu\CdoT\frac{2\pi}{M}}\big)\,+\,
\frac{1}{L}\cdot\text{E}\Big\{
\Vec{\boldsymbol{N}}_{\!\!f}(\mu)\CdoT
\Vec{\boldsymbol{V}}(\mu)^{\Hh}\Big\}\cdot
\Hat{C}_{\boldsymbol{V}(\mu),\boldsymbol{V}(\mu)}^{\,-1}
\;=\;\notag\\*
\;=\;H\big({\T\mu\CdoT\frac{2\pi}{M}}\big)\,+\,
\frac{1}{L}\cdot\Sum{\lambda=1}{L}
\text{E}\big\{\boldsymbol{N}_{\!\!f,\lambda}(\mu)\CdoT
\boldsymbol{V}_{\!\!\lambda}(\mu)^{\Kk}\big\}\cdot
\Hat{C}_{\boldsymbol{V}(\mu),\boldsymbol{V}(\mu)}^{\,-1}
\;=\;\notag\\*
\;=\;H\big({\T\mu\CdoT\frac{2\pi}{M}}\big)\,+\,
\text{E}\big\{\boldsymbol{N}_{\!\!f}(\mu)\CdoT
\boldsymbol{V}(\mu)^{\Kk}\big\}\cdot
\Hat{C}_{\boldsymbol{V}(\mu),\boldsymbol{V}(\mu)}^{\,-1}
\;=\;H\big({\T\mu\CdoT\frac{2\pi}{M}}\big).
\label{3.20}
\end{gather}
Zuletzt wurde bei der Berechnung ber"ucksichtigt, dass bei einer zul"assigen 
Stichprobenentnahme mit unabh"angigen Einzelmessungen alle Zufallsgr"o"sentupel 
\mbox{$\big[\boldsymbol{V}_{\!\!\lambda}(\mu),\boldsymbol{N}_{\!\!f,\lambda}(\mu)\big]^{\TT}$} 
dieselbe Verbundverteilung besitzen, wie das Zufallsgr"o"sentupel 
\mbox{$\big[\boldsymbol{V}(\mu),\boldsymbol{N}_{\!\!f}(\mu)\big]^{\TT}$} 
aus dem die Stichprobe entnommen wurde. Somit ist auch der Erwartungswert 
des Produkts der Zufallsgr"o"sen --- die Kovarianz --- von $\lambda$ 
unabh"angig. Die Kovarianz ist nach Gleichung (\ref{2.30}) Null. 
F"ur den Fall der Erregung mit Mehrtonsignalen gen"ugt also die 
Tatsache, dass die beiden Zufallsgr"o"sen \mbox{$\boldsymbol{V}(\mu)$} 
und \mbox{$\boldsymbol{N}_{\!\!f}(\mu)$} unkorreliert sind, um zeigen zu 
k"onnen, dass die Messwerte der "Ubertragungsfunktion erwartungstreu sind.

Bei Erregung des Systems mit allgemeineren Zufallsprozessen werden mit den Elementen 
des Vektors \mbox{$\Vec{\boldsymbol{V}}(\mu)$} auch deren empirische Varianzen
\mbox{$\Hat{\boldsymbol{C}}_{\boldsymbol{V}(\mu),\boldsymbol{V}(\mu)}$} zuf"allig. 
Ohne eine genaue Kenntnis der Verbundverteilungen der $M$ Zufallsgr"o"sentupel
\mbox{$\big[\boldsymbol{V}(\mu),\boldsymbol{N}_{\!\!f}(\mu)\big]^{\TT}$} lassen sich 
die Erwartungswerte der Messwerte der "Ubertragungsfunktion nur dann berechnen, 
wenn man annimmt, dass die Spektralwerte \mbox{$\boldsymbol{N}_{\!\!f}(\mu)$} der 
gefensterten St"orung des realen Systems unabh"angig von den Spektralwerten
\mbox{$\boldsymbol{V}(\mu)$} der Erregung bei derselben Frequenz
sind. Die zweidimensionale Verbundverteilung des Zufallsgr"o"sentupels
\mbox{$\big[\boldsymbol{V}(\mu),\boldsymbol{N}_{\!\!f}(\mu)\big]^{\TT}$}
l"asst sich dann f"ur alle Frequenzen $\mu$ als das Produkt der
Verteilungen der Zufallsgr"o"sen \mbox{$\boldsymbol{V}(\mu)$} und
\mbox{$\boldsymbol{N}_{\!\!f}(\mu)$} schreiben. Diese Forderung nach 
Unabh"angigkeit der Spektralwerte scheint auf den ersten Blick die Art der 
Systeme, f"ur die die Ergebnisse der weiteren Untersuchungen gelten, extrem 
stark einzuschr"anken. Dass dem nicht so ist, ja dass sogar Systeme die eben 
aufgestellte Forderung nach Unabh"angigkeit erf"ullen k"onnen, bei denen die
St"orung nur durch Nichtlinearit"aten im System selbst entsteht,
die daher zu jeder bestimmten Erregung genau eine determinierte
Systemantwort besitzen, wird im Kapitel \ref{Unab} des Anhangs erl"autert. 
Eine wichtige notwendige Voraussetzung f"ur die Unabh"angigkeit der Spektralwerte 
\mbox{$\boldsymbol{N}_{\!\!f}(\mu)$} und \mbox{$\boldsymbol{V}(\mu)$} 
ist die Unkorreliertheit dieser Spektralwerte. Sie ist nach Gleichung 
(\ref{2.30}) durch die optimale Approximation des realen Systems 
gegeben, reicht aber nicht aus, um die im weiteren angenommene 
Unabh"angigkeit der Spektralwerte zu garantieren. 

Das RKM liefert eine Stichprobe des Zufallszahlentupels
\mbox{$\big[\boldsymbol{V}(\mu),\boldsymbol{N}_{\!\!f}(\mu)\big]^{\Tt}\!$} 
vom Umfang $L$. Nun fasst man die $L$ Elemente dieser Stichprobe zu
einer \mbox{$2\!\times\!L$} Matrix zusammen, deren Spalten jeweils eine
konkrete Realisierung des Zufallsgr"o"sentupels --- also ein einzelnes
Element der Stichprobe --- enthalten und deren Zeilen die zwei Stichprobenvektoren
\mbox{$\Vec{V}(\mu)$} und \mbox{$\Vec{N}_{\!f}(\mu)$} der
Zufallsgr"o"sen \mbox{$\boldsymbol{V}(\mu)$} und
\mbox{$\boldsymbol{N}_{\!\!f}(\mu)$} sind. Da wir gefordert hatten,
dass die Stichprobenentnahme zuf"allig und f"ur alle Einzelmessungen
--- die Elemente der Stichprobe --- unabh"angig erfolgt sein soll, 
l"asst sich die gemeinsame Verbundverteilung aller Elemente
\mbox{$\big[\boldsymbol{V}_{\!\!\lambda}(\mu),
\boldsymbol{N}_{\!\!f,\lambda}(\mu)\big]^{\TT}$}
der mathematischen Stichprobe, also die gemeinsame Verbundverteilung 
aller Spalten der zuf"alligen Matrix, die sich unter Ber"ucksichtigung 
der Zuf"alligkeit der Stichprobenentnahme ergibt, als das Produkt 
der Verbundverteilungen der einzelnen Elemente der Stichprobe schreiben. 
Diese Verbundverteilungen der einzelnen Elemente der Stichprobe 
sind alle gleich der Verbundverteilung des Zufallsgr"o"sentupels 
\mbox{$\big[\boldsymbol{V}(\mu),\boldsymbol{N}_{\!\!f}(\mu)\big]^{\TT}\!\!$}, 
von der wir annehmen, dass sie sich als das Produkt zweier Verteilungen schreiben 
l"asst. Daher l"asst sich die Verbundverteilung aller Elemente der zuf"alligen 
\mbox{$2\!\times\!L$} Matrix als das Produkt der Verteilungen der einzelnen 
Elemente der zuf"alligen Matrix schreiben. Somit sind zwei beliebige Vektoren, 
die aus den Elementen der zuf"alligen Matrix gebildet werden, voneinander 
unabh"angig, wenn sie keine gemeinsamen Matrixelemente enthalten. Wenn wir nun 
den Erwartungswert des Produkts einer beliebigen Funktion des Zufallsvektors 
\mbox{$\Vec{\boldsymbol{V}}(\mu)$} und einer anderen beliebigen 
Funktion des Zufallsvektors \mbox{$\Vec{\boldsymbol{N}}_{\!\!f}(\mu)$} 
bilden, berechnet sich dieser als das Produkt 
der beiden Erwartungswerte der einzelnen zuf"alligen Faktoren.
Dies gilt auch f"ur nichtlineare Funktionen, wie z.~B. f"ur die
Inverse der empirischen Varianz.

Mit Gleichung (\ref{3.19}) l"asst sich nun der Erwartungswert der
Messwerte f"ur die "Ubertragungsfunktion berechnen, wobei statt der konkreten
Stichprobenvektoren die mathematischen Stichprobenvektoren und die
daraus abgeleiteten zuf"alligen empirischen Varianzen eingesetzt werden.
\begin{gather}
\text{E}\big\{\Hat{\boldsymbol{H}}(\mu)\big\}\;=\;
H\big({\T\mu\CdoT\frac{2\pi}{M}}\big)\,+\,
\frac{1}{L}\cdot\text{E}\Big\{
\Vec{\boldsymbol{N}}_{\!\!f}(\mu)\CdoT
\Vec{\boldsymbol{V}}(\mu)^{\Hh}\cdot
\Hat{\boldsymbol{C}}_{\boldsymbol{V}(\mu),\boldsymbol{V}(\mu)}^{\,-1}\Big\}
\;=\;\notag\\*
\;=\;H\big({\T\mu\CdoT\frac{2\pi}{M}}\big)\,+\,
\frac{1}{L}\cdot\Sum{\lambda=1}{L}
\text{E}\Big\{\boldsymbol{N}_{\!\!f,\lambda}(\mu)\CdoT
\boldsymbol{V}_{\!\!\lambda}(\mu)^{\Kk}\cdot
\Hat{\boldsymbol{C}}_{\boldsymbol{V}(\mu),\boldsymbol{V}(\mu)}^{\,-1}\Big\}
\;=\;\notag\\*
\;=\;H\big({\T\mu\CdoT\frac{2\pi}{M}}\big)\,+\,
\underbrace{\text{E}\big\{\boldsymbol{N}_{\!\!f}(\mu)\big\}}_{\scriptstyle=0}\cdot
\frac{1}{L}\cdoT\Sum{\lambda=1}{L}
\text{E}\Big\{\boldsymbol{V}_{\!\!\lambda}(\mu)^{\Kk}\cdot
\Hat{\boldsymbol{C}}_{\boldsymbol{V}(\mu),\boldsymbol{V}(\mu)}^{\,-1}\Big\}
\label{3.21}
\end{gather}

Da wir uns hier auf den Fall mittelwertfreier Ein- und Ausgangsprozesse 
\mbox{$\boldsymbol{v}(k)$} und \mbox{$\boldsymbol{y}(k)$} beschr"anken, sind auch 
die zuf"alligen Spektralwerte \mbox{$\boldsymbol{N}_{\!\!f}(\mu)$} des gefensterten 
Approxi\-ma\-tions\-fehlers mittelwertfrei, so dass der erste Erwartungswert auf der 
rechten Seite der letzten Gleichung Null ist. Der Erwartungswert der gesamtem 
Messwertabweichung ist jedoch nur dann Null, wenn alle Erwartungswerte innerhalb 
der Summe existieren und endlich sind.

Wenn wir zur Erregung wertdiskrete Zufallsgr"o"sen \mbox{$\boldsymbol{V}(\mu)$} 
verwenden, bei denen \mbox{$V(\mu)=0$} {\em kein}\/ m"ogliches Ereignis ist, so ist 
die empirische Varianz \mbox{$\Hat{C}_{\boldsymbol{V}(\mu),\boldsymbol{V}(\mu)}$} 
{\em immer}\/ von Null verschieden und somit deren Inverse endlich. Auch alle 
Erwartungswerte innerhalb der Summe in der letzten Gleichung existieren dann und 
sind endlich. Die Messwerte der "Ubertragungsfunktion sind in diesem Fall erwartungstreu. 

Wenn jedoch wertdiskrete Zufallsgr"o"sen \mbox{$\boldsymbol{V}(\mu)$} verwendet 
werden, bei denen \mbox{$V(\mu)=0$} {\em ein}\/ m"ogliches Ereignis ist, das mit einer, 
wenn auch beliebig kleinen, positiven Wahrscheinlichkeit auftritt, kann es vorkommen, dass 
die empirische Varianz \mbox{$\Hat{C}_{\boldsymbol{V}(\mu),\boldsymbol{V}(\mu)}$}
Null ist. Da dann deren Inverse nicht existiert, existieren auch alle Erwartungswerte
innerhalb der Summe in der letzten Gleichung nicht. Es ist somit in diesem Fall keine
Aussage "uber die Erwartungstreue der Messwerte der "Ubertragungsfunktion m"oglich.

Man kann jedoch die Berechnung der Messwerte nach Gleichung (\ref{3.14}) 
geringf"ugig modifizieren um auch in diesem Fall praktisch erwartungstreue Messwerte 
zu erhalten. Im Fall \mbox{$\Hat{C}_{\boldsymbol{V}(\mu),\boldsymbol{V}(\mu)}=0$}
wird statt der inversen empirischen Varianz eine beliebige Konstante eingesetzt. 
An der Unabh"angigkeit der Einzelmessungen und der Separierbarkeit der 
Verbundverteilung des Zufallsgr"o"sentupels 
\mbox{$\big[\boldsymbol{V}(\mu),\boldsymbol{N}_{\!\!f}(\mu)\big]^{\TT}\!\!$} 
"andert diese Modifikation nichts. Es wird lediglich eine andere nichtlineare 
Funktion der Zufallsgr"o"sen \mbox{$\boldsymbol{V}_{\!\!\lambda}(\mu)$} als die 
Inverse der empirischen Varianz zur Berechnung der Messwerte verwendet.
Da der Fall \mbox{$\Hat{C}_{\boldsymbol{V}(\mu),\boldsymbol{V}(\mu)}=0$}
nur dann eintritt, wenn alle Werte \mbox{$V_{\lambda}(\mu)$} bei allen Einzelmessungen
(\mbox{$\lambda=1\;(1)\;L$}) Null sind, wird in Gleichung (\ref{3.14}) auch die 
Kovarianz \mbox{$\Hat{C}_{\boldsymbol{Y}_{\!\!\!f}(\mu),\boldsymbol{V}(\mu)}$} Null.
Somit berechnet sich in diesem Fall der Messwert zu \mbox{$\Hat{H}(\mu)=0$}.
Die Messwertabweichung, die mit der Auftrittswahrscheinlichkeit dieses Falles im
Erwartungswert zu ber"ucksichtigen ist, ist dann \mbox{$-H(\mu\CdoT2\pi/M)$} und 
somit nicht Null. Daher sind die Messwerte genau genommen nicht erwartungstreu.
Da jedoch die Auftrittswahrscheinlichkeit dieses Falles in der Praxis
extrem klein ist, ist der Einfluss dieses Falls auf die Erwartungswerte der 
Messwerte der "Ubertragungsfunktion vernachl"assigbar. In \cite{Erg} 
wird die Auftrittswahrscheinlichkeit des Falles, dass die empirische Varianz 
Null wird, an einem realit"atsnahen Beispiel abgesch"atzt. Dort wird die 
empirische Varianz eines im Allgemeinen nicht mittelwertfreien Prozesses untersucht. 
In der Praxis wird man im Fall einer zu kleinen empirischen Varianz die Messung 
wiederholen, bzw. die Anzahl der Einzelmessungen solange erh"ohen, 
bis sich f"ur alle $M$ Frequenzen $\mu$ eine hinreichend gro"se empirische 
Varianz ergibt. Dieses Verfahren bedarf dann streng genommen einer gesonderten 
theoretischen Behandlung, die in dieser Abhandlung nicht durchgef"uhrt wird. 

\section{Prinzip der Messung des Leistungsdichtespektrums}

Zur Beschreibung des Rauschprozesses \mbox{$\boldsymbol{n}(k)$}
im Systemmodell wollten wir die $M$ Werte
\mbox{$\Bar{\Phi}_{\boldsymbol{n}}(\mu)$} durch die $M$ Erwartungswerte
\mbox{$\Tilde{\Phi}_{\boldsymbol{n}}(\mu)$} der $M$ Zufallsgr"o"sen
\mbox{$|\boldsymbol{N}_{\!\!f}(\mu)|^2/M$}, die nach Gleichung (\ref{2.17})
definiert sind, ann"ahern, die wiederum durch die $M$ Sch"atzwerte
\mbox{$\Hat{\Phi}_{\boldsymbol{n}}(\mu)$} abgesch"atzt werden sollen.
Da die theoretischen Gr"o"sen \mbox{$\Tilde{\Phi}_{\boldsymbol{n}}(\mu)$}
die Erwartungswerte der Zufallsgr"o"sen \mbox{$|\boldsymbol{N}_{\!\!f}(\mu)|^2/M$}
sind, und da der empirische Mittelwert einer Zufallsgr"o"se erwartungstreu
ist, m"usste man die $M$ Mittelwerte f"ur \mbox{$\mu=0\;(1)\;M\!-\!1$}
"uber jeweils eine Stichprobe der L"ange $L$ der Zufallsgr"o"sen
\mbox{$|\boldsymbol{N}_{\!\!f}(\mu)|^2/M$} bilden, um erwartungstreue
Sch"atzwerte f"ur \mbox{$\Tilde{\Phi}_{\boldsymbol{n}}(\mu)$}
zu erhalten. Es w"aren also die $M$ Terme
\begin{equation}
\frac{1}{L}\cdoT\Sum{\lambda=1}{L}\,\frac{1}{M}\cdot
\big|N_{\!f,\lambda\!}(\mu)\big|^2\;=\;
\frac{1}{L\CdoT M}\cdot\Vec{N}_{\!f}(\mu)\CdoT\Vec{N}_{\!f}(\mu)^{\Hh}
\qquad\qquad\forall\qquad \mu=0\;(1)\;M\!-\!1
\label{3.22}
\end{equation}
zu berechnen, wobei die Werte \mbox{$N_{\!f,\lambda\!}(\mu)$}
jeweils die $L$ Elemente einer Stichprobe \mbox{$\Vec{N}_{\!f}(\mu)$}
vom Umfang $L$ der Zufallsgr"o"sen \mbox{$\boldsymbol{N}_{\!\!f}(\mu)$} sind. 
Die Zufallsgr"o"sen \mbox{$\boldsymbol{N}_{\!\!f}(\mu)$} sind
nach Gleichung (\ref{2.29}) als die Differenz der diskreten 
Fouriertransformierten \mbox{$\boldsymbol{Y}_{\!\!\!f}(\mu)$}
des gefensterten ausgangsseitigen Zufallsprozesses und der 
diskreten Fouriertransformierten des Ausgangsprozesses 
\mbox{$H\big({\T\mu\CdoT\frac{2\pi}{M}}\big)\CdoT\boldsymbol{V}(\mu)$}
des linearen Modellsystems mit der theoretischen "Ubertragungsfunktion 
definiert. Demnach berechnen sich die Stichproben
\mbox{$\Vec{N}_{\!f}(\mu)$} nach Gleichung (\ref{3.18})
analog "uber die Optimall"osungen der theoretischen
Regression. Somit kann man ohne die Kenntnis der
theoretisch optimalen Re\-gres\-sions\-ko\-ef\-fi\-zi\-enten keine
Stichproben f"ur die Zufallsgr"o"sen \mbox{$\boldsymbol{N}_{\!\!f}(\mu)$}
erhalten. Demnach lassen sich auch die empirischen Mittelwerte (\ref{3.22})
der Zufallsgr"o"sen \mbox{$|\boldsymbol{N}_{\!\!f}(\mu)|^2/M$}
--- also die auf \mbox{$L\CdoT M$} normierten Betragsquadrate
\mbox{$\Vec{N}_{\!f}(\mu)\CdoT\Vec{N}_{\!f}(\mu)^{\Hh}$}
der L"angen (\,euklidischen Normen\,) der Stichprobenvektoren
\mbox{$\Vec{N}_{\!f}(\mu)$} --- nicht berechnen. Da man lediglich 
die Sch"atzwerte \mbox{$\Hat{H}(\mu)$} f"ur die optimalen 
Regressionskoeffizienten \mbox{$H(\mu\CdoT2\pi/M)$} kennt, ben"otigt man 
$M$ andere Zufallsgr"o"sen, deren Erwartungswerte gleich den Erwartungswerten 
der Zufallsgr"o"sen \mbox{$|\boldsymbol{N}_{\!\!f}(\mu)|^2/M$} sind.

Analog k"onnten wir Sch"atzwerte f"ur die Werte 
\mbox{$\Tilde{\Psi}_{\boldsymbol{n}}(\mu)$} zur Beschreibung des 
MLDS erhalten, wenn wir die optimalen Regressionskoeffizienten 
\mbox{$H(\mu\CdoT2\pi/M)$} kennen w"urden. Man m"usste die 
empirischen Mittelwerte jeweils "uber eine Stichprobe
der L"ange $L$ der Zufallsgr"o"sen
\mbox{$\boldsymbol{N}_{\!\!f}(\mu)\CdoT\boldsymbol{N}_{\!\!f}(\!-\mu)/M$}
bilden: 
\begin{equation}
\frac{1}{L}\cdoT\Sum{\lambda=1}{L}\:\frac{1}{M}\cdot
N_{\!f,\lambda\!}(\mu)\CdoT N_{\!f,\lambda\!}(\!-\mu)\;=\;
\frac{1}{L\CdoT M}\cdot\Vec{N}_{\!f}(\mu)\CdoT\Vec{N}_{\!f}(\!-\mu)^{\Tt}
\qquad\forall\quad\mu=0\;(1)\;M\!-\!1.
\label{3.23}
\end{equation}
Da die theoretischen Werte \mbox{$H(\mu\CdoT2\pi/M)$} jedoch unbekannt sind,
ben"otigen wir auch hier $M$ andere Zufallsgr"o"sen, deren Erwartungswerte 
dann gleich den Erwartungswerten der Zufallsgr"o"sen 
\mbox{$\boldsymbol{N}_{\!\!f}(\mu)\CdoT\boldsymbol{N}_{\!\!f}(\!-\mu)/M$} sind.

Um den Anteil des Vektors \mbox{$\Vec{N}_{\!f}(\mu)$} zu eliminieren, der
nach Gleichung (\ref{3.18}) von den unbekannten, theoretischen Optimall"osungen 
f"ur die "Ubertragungsfunktion abh"angt, bilden wir diesen Vektor mit einer Matrix 
\mbox{$\underline{V}_{\bot}\!(\mu)$} ab, die den Vektor \mbox{$\Vec{V}(\mu)$} 
als Eigenvektor zum Eigenwert Null aufweist. Da bei einer solchen Matrix 
\begin{equation}
\Vec{V}(\mu)\CdoT\underline{V}_{\bot}\!(\mu)\;=\;\Vec{0}
\label{3.24}
\end{equation}
gilt, l"asst sich der abgebildete Vektor \mbox{$\Hat{\Vec{N}}_{\!f}(\mu)$} mit
Gleichung (\ref{3.18}) gem"a"s
\begin{gather}
\Hat{\Vec{N}}_{\!f}(\mu)\:=\:
\Vec{N}_{\!f}(\mu)\CdoT\underline{V}_{\bot}\!(\mu)\:=\:
\Big(\Vec{Y}_{\!f}(\mu)-
H\big({\T\mu\CdoT\frac{2\pi}{M}}\big)\CdoT\Vec{V}(\mu)\Big)
\cdot\underline{V}_{\bot}\!(\mu)\:=\:
\Vec{Y}_{\!f}(\mu)\CdoT\underline{V}_{\bot}\!(\mu)
\notag\\*[3pt]
\forall\qquad\mu=0\;(1)\;M\!-\!1.
\label{3.25}
\end{gather}
berechnen. Die $M$ dazu ben"otigten Matrizen \mbox{$\underline{V}_{\bot}\!(\mu)$} 
lassen sich prinzipiell mit Hilfe der $M$ bekannten Vektoren \mbox{$\Vec{V}(\mu)$} 
ohne die Kenntnis der theoretischen Werte \mbox{$H(\mu\CdoT2\pi/M)$} konstruieren. 
Eine besonders g"unstige Art der Konstruktion dieser Matrizen wird weiter unten 
angegeben. Da sich auch die $M$ Vektoren \mbox{$\Vec{Y}_{\!f}(\mu)$} mit einer Fensterung 
und einer anschlie"senden DFT aus den am Ausgang des realen Systems gemessenen
Signalen berechnen lassen, k"onnen auch die $M$ Vektoren \mbox{$\Hat{\Vec{N}}_{\!f}(\mu)$}
ohne die Kenntnis der theoretisch optimalen "Ubertragungsfunktion berechnet werden.

Nun wollen wir untersuchen, ob die Skalarprodukte 
\mbox{$\Hat{\Vec{N}}_{\!f}(\mu)\CdoT\Hat{\Vec{N}}_{\!f}(\mu)^{\Hh}$} und
\mbox{$\Hat{\Vec{N}}_{\!f}(\mu)\CdoT\Hat{\Vec{N}}_{\!f}(\!-\mu)^{\Tt}$}
der mit der Matrix \mbox{$\underline{V}_{\bot}\!(\mu)$} bzw. 
\mbox{$\underline{V}_{\bot}\!(\!-\mu)$} abgebildeten 
Vektoren ebenfalls als konkrete Sch"atzwerte f"ur die Gr"o"sen 
\mbox{$\Tilde{\Phi}_{\boldsymbol{n}}(\mu)$} und 
\mbox{$\Tilde{\Psi}_{\boldsymbol{n}}(\mu)$} dienen k"onnen, wenn
man sie mit noch n"aher zu bestimmenden Normierungsfaktoren \mbox{$c_{\Phi}(\mu)$} 
bzw. \mbox{$c_{\Psi}(\mu)$} multipliziert.
Mit Hilfe der letzten Gleichung lassen sich die normierten Skalarprodukte als 
\begin{align}
c_{\Phi}(\mu)\CdoT\Hat{\Vec{N}}_{\!f}(\mu)\CdoT\Hat{\Vec{N}}_{\!f}(\mu)^{\Hh}&
\;=\;\Vec{N}_{\!f}(\mu)\CdoT\underline{V}_{\Phi}(\mu)\CdoT\Vec{N}_{\!f}(\mu)^{\Hh}
\;=\;\Vec{Y}_{\!f}(\mu)\CdoT\underline{V}_{\Phi}(\mu)\CdoT\Vec{Y}_{\!f}(\mu)^{\Hh}\label{3.26}\\
&\text{mit }\qquad\underline{V}_{\Phi}(\mu)\;=\;
c_{\Phi}(\mu)\CdoT\underline{V}_{\bot}\!(\mu)\CdoT\underline{V}_{\bot}\!(\mu)^{\Hh}\qquad\text{und}\label{3.27}\\
c_{\Psi}(\mu)\CdoT\Hat{\Vec{N}}_{\!f}(\mu)\CdoT\Hat{\Vec{N}}_{\!f}(\!-\mu)^{\Tt}&
\;=\;\Vec{N}_{\!f}(\mu)\CdoT\underline{V}_{\Psi}(\mu)\CdoT\Vec{N}_{\!f}(\!-\mu)^{\Tt}
\;=\;\Vec{Y}_{\!f}(\mu)\CdoT\underline{V}_{\Psi}(\mu)\CdoT\Vec{Y}_{\!f}(\!-\mu)^{\Tt}\label{3.28}\\
&\text{mit }\qquad\underline{V}_{\Psi}(\mu)\;=\;
c_{\Psi}(\mu)\CdoT\underline{V}_{\bot}\!(\mu)\CdoT\underline{V}_{\bot}\!(\!-\mu)^{\Tt}\label{3.29}
\end{align}
darstellen. Nun ber"ucksichtigen wir, dass es sich bei der Messung selbst 
um ein Zufallsexperiment handelt, und berechnen die Erwartungswerte dieser
Skalarprodukte. Die Vektoren \mbox{$\Hat{\Vec{N}}_{\!f}(\mu)$} sind also 
die bei der Messung gewonnenen konkreten Realisierungen der Zufallsvektoren 
\mbox{$\Hat{\Vec{\boldsymbol{N}}}_{\!\!f}(\mu)$}. Ebenso sind die Vektoren 
\mbox{$\Vec{N}_{\!f}(\mu)$} und \mbox{$\Vec{V}(\mu)$} konkrete Realisierungen 
der Zufallsvektoren \mbox{$\Vec{\boldsymbol{N}}_{\!\!f}(\mu)$} und 
\mbox{$\Vec{\boldsymbol{V}}(\mu)$} der mathematischen Stichprobe 
vom Umfang $L$ der Zufallsgr"o"sen \mbox{$\boldsymbol{N}_{\!\!f}(\mu)$} und 
\mbox{$\boldsymbol{V}(\mu)$}. Wenn auch die Normierungsfaktoren \mbox{$c_{\Phi}(\mu)$} 
bzw. \mbox{$c_{\Psi}(\mu)$} konkrete Realisierungen zuf"alliger Faktoren \mbox{$\boldsymbol{c}_{\Phi}(\mu)$} 
und \mbox{$\boldsymbol{c}_{\Psi}(\mu)$} sind, sind auch die Matrizen
\mbox{$\underline{V}_{\bot}\!(\mu)$}, \mbox{$\underline{V}_{\Phi}(\mu)$} 
und \mbox{$\underline{V}_{\Psi}(\mu)$} konkrete Realisierungen 
zuf"alliger Matrizen \mbox{$\underline{\boldsymbol{V}}_{\bot}\!(\mu)$}, 
\mbox{$\underline{\boldsymbol{V}}_{\Phi}(\mu)$} und 
\mbox{$\underline{\boldsymbol{V}}_{\Psi}(\mu)$}.
Um im weiteren die Erwartungswerte der in den letzten beiden Gleichungen
genannten normierten Skalarprodukte berechnen zu k"onnen, m"ussen wir fordern, 
dass sowohl die Normierungsfaktoren \mbox{$\boldsymbol{c}_{\Phi}(\mu)$} 
und \mbox{$\boldsymbol{c}_{\Psi}(\mu)$} als auch die Matrizen 
\mbox{$\underline{\boldsymbol{V}}_{\bot}\!(\mu)$} und 
\mbox{$\underline{\boldsymbol{V}}_{\bot}\!(-\mu)$} in der Art konstruiert wurden, 
dass die Normierungsfaktoren und alle Elemente der Zufallsmatrizen von den Zufallsgr"o"sen 
\mbox{$\boldsymbol{N}_{\!\!f}(\mu)$} und \mbox{$\boldsymbol{N}_{\!\!f}(-\mu)$} 
unabh"angig sind. Somit sind auch alle Elemente der Zufallsmatrizen 
\mbox{$\underline{\boldsymbol{V}}_{\Phi}(\mu)$} und 
\mbox{$\underline{\boldsymbol{V}}_{\Psi}(\mu)$} von den Zufallsgr"o"sen 
\mbox{$\boldsymbol{N}_{\!\!f}(\mu)$} bzw. \mbox{$\boldsymbol{N}_{\!\!f}(-\mu)$} 
unabh"angig. 

Wir beginnen nun mit der Berechnung des Erwartungswertes des ersten normierten Skalarproduktes 
\mbox{$\boldsymbol{c}_{\Phi}(\mu)\CdoT\Hat{\Vec{\boldsymbol{N}}}_{\!\!f}(\mu)\CdoT\Hat{\Vec{\boldsymbol{N}}}_{\!\!f}(\mu)^{\Hh}$}, 
wobei wir hier die Abh"angigkeit des Faktors, der Vektoren und der Matrix von der
diskreten Frequenz $\mu$ nicht mehr explizit hinschreiben und
das Element der Matrix \mbox{$\underline{\boldsymbol{V}}_{\Phi}(\mu)$}
in der $\lambda_1$-ten Zeile und der $\lambda_2$-ten Spalte mit
\mbox{$\underline{\boldsymbol{V}}_{\lambda_1,\lambda_2}$} bezeichnen.
\begin{subequations}\label{3.30}
\begin{equation}
\text{E}\Big\{\boldsymbol{c}_{\Phi}\CdoT\Hat{\Vec{\boldsymbol{N}}}_{\!\!f}\CdoT
\Hat{\Vec{\boldsymbol{N}}}_{\!\!f}^{\hH}\Big\}=
\text{E}\Big\{\Vec{\boldsymbol{N}}_{\!\!f}\cdot
\underline{\boldsymbol{V}}_{\Phi}\Cdot
\Vec{\boldsymbol{N}}_{\!\!f}^{\hH}\Big\}=
\text{E}\bigg\{\,\Sum{\lambda_1=1}{L}\;\Sum{\lambda_2=1}{L}
\boldsymbol{N}_{\!\!f,\lambda_1}\Cdot
\underline{\boldsymbol{V}}_{\lambda_1,\lambda_2}\Cdot
\boldsymbol{N}_{\!\!f,\lambda_2}^*\bigg\}=\ldots
\label{3.30.a}
\end{equation}
Der Erwartungswert wird als Summe der Erwartungswerte berechnet.
\begin{equation}
\ldots\;=\;\Sum{\lambda_1=1}{L}\;\Sum{\lambda_2=1}{L}\text{E}\Big\{\,
\boldsymbol{N}_{\!\!f,\lambda_1}\Cdot
\underline{\boldsymbol{V}}_{\lambda_1,\lambda_2}\Cdot
\boldsymbol{N}_{\!\!f,\lambda_2}^*\,\Big\}\;=\;\ldots
\label{3.30.b}
\end{equation}
Die Doppelsumme wird in zwei Teilsummen aufgespalten. Die erste Teilsumme
enth"alt die Hauptdiagonalelemente, w"ahrend in der zweiten Teilsumme
die Nebendiagonalelemente auftreten.
\begin{equation}
\ldots\;=\;\Sum{\lambda=1}{L}
\text{E}\Big\{\,\big|\,\boldsymbol{N}_{\!\!f,\lambda}\big|^2\Cdot
\underline{\boldsymbol{V}}_{\lambda,\lambda}\,\Big\}\;+
\Sum{\lambda_1=1}{L}\;
\Sum{\substack{\lambda_2=1\;\;\\\lambda_2\neq\lambda_1}}{L}
\text{E}\Big\{\,\boldsymbol{N}_{\!\!f,\lambda_1}\Cdot
\underline{\boldsymbol{V}}_{\lambda_1,\lambda_2}\Cdot
\boldsymbol{N}_{\!\!f,\lambda_2}^*\,\Big\}\;=\;\ldots
\label{3.30.c}
\end{equation}
Nun wird die vorausgesetzte Unabh"angigkeit der Zufallselemente der
Matrix \mbox{$\Hat{\underline{\boldsymbol{V}}}_{\Phi}(\mu)$} 
von den Zufallsgr"o"sen \mbox{$\boldsymbol{N}_{\!\!f}(\mu)$} dazu
verwendet, den Erwartungswert des Produktes als das Produkt der
Erwartungswerte der einzelnen Faktoren zu schreiben.
\begin{equation}
\ldots\,=\Sum{\lambda=1}{L}
\text{E}\big\{|\boldsymbol{N}_{\!\!f,\lambda}|^2\big\}\CdoT
\text{E}\big\{\underline{\boldsymbol{V}}_{\lambda,\lambda}\big\}+\!
\Sum{\lambda_1=1}{L}\;
\Sum{\substack{\lambda_2=1\;\;\\\lambda_2\neq\lambda_1}}{L}\!
\text{E}\Big\{\boldsymbol{N}_{\!\!f,\lambda_1}\CdoT
\boldsymbol{N}_{\!\!f,\lambda_2}^*\Big\}\CdoT
\text{E}\big\{\underline{\boldsymbol{V}}_{\lambda_1,\lambda_2}\big\}
=\ldots
\label{3.30.d}
\end{equation}
Desweiteren ber"ucksichtigen wir nun die Unabh"angigkeit der Stichprobenelemente, 
die in unterschiedlichen Einzelmessungen mit \mbox{$\lambda_1\!\neq\!\lambda_2$} 
gewonnen wurden. Au"serdem ist bei einer mathematischen Stichprobe die Verteilung 
der Stichprobenelemente gleich der Verteilung der Zufallsgr"o"se, aus der die 
Stichprobe gewonnen worden ist. Somit ist auch der Erwartungswert der 
Stichprobenelemente gleich dem Erwartungswert der Zufallsgr"o"se. 
\begin{equation}
\ldots\;=\;\Sum{\lambda=1}{L}
\text{E}\big\{\,|\boldsymbol{N}_{\!\!f}|^2\big\}\cdot
\text{E}\big\{\,\underline{\boldsymbol{V}}_{\lambda,\lambda}\,\big\}+
\Sum{\lambda_1=1}{L}\;
\Sum{\substack{\lambda_2=1\;\;\\\lambda_2\neq\lambda_1}}{L}
\big|\,\text{E}\{\boldsymbol{N}_{\!\!f}\}\big|^2\Cdot
\text{E}\big\{\,\underline{\boldsymbol{V}}_{\lambda_1,\lambda_2}\,\big\}
\;=\;\ldots
\label{3.30.e}
\end{equation}
Die von $\lambda$ bzw. von $\lambda_1$ und $\lambda_2$ unabh"angigen
Erwartungswerte des Spektrums des gefensterten Approximationsfehlerprozesses
werden vor die Summen gezogen.
\begin{equation}
\ldots\;=\;\text{E}\big\{\,|\boldsymbol{N}_{\!\!f}|^2\big\}\cdot
\Sum{\lambda=1}{L}
\text{E}\big\{\,\underline{\boldsymbol{V}}_{\lambda,\lambda}\,\big\}\;+\;
\big|\,\text{E}\{\boldsymbol{N}_{\!\!f}\}\big|^2\CdoT
\Sum{\lambda_1=1}{L}\;
\Sum{\substack{\lambda_2=1\;\;\\\lambda_2\neq\lambda_1}}{L}
\text{E}\big\{\,\underline{\boldsymbol{V}}_{\lambda_1,\lambda_2}\,\big\}
\;=\;\ldots
\label{3.30.f}
\end{equation}
Die Summe der Erwartungswerte der einzelnen Summanden wird jeweils als
Erwartungswert der Summe geschrieben.
\begin{equation}
\ldots\;=\;\text{E}\big\{\,|\boldsymbol{N}_{\!\!f}|^2\big\}\cdot
\text{E}\bigg\{\Sum{\lambda=1}{L}
\underline{\boldsymbol{V}}_{\lambda,\lambda}\bigg\}+
\big|\!\underbrace{\text{E}\{\boldsymbol{N}_{\!\!f}\}}_{\scriptstyle=0}\!\big|^2\Cdot
\text{E}\bigg\{\Sum{\lambda_1=1}{L}\;
\Sum{\substack{\lambda_2=1\;\;\\\lambda_2\neq\lambda_1}}{L}
\underline{\boldsymbol{V}}_{\lambda_1,\lambda_2}\!\bigg\}\;=\;\ldots
\label{3.30.g}
\end{equation}
Wenn wir ber"ucksichtigen, dass der Erwartungswert des Spektrums des 
wahren Approximationsfehlers Null ist, und wenn wir die Abh"angigkeit 
der Vektoren und der Matrix von der diskreten Frequenz $\mu$ wieder 
explizit angeben, erhalten wir
\begin{equation}
\ldots\;=\;\text{E}\big\{|\boldsymbol{N}_{\!\!f}(\mu)|^2\big\}\cdot
\text{E}\Big\{\,\text{spur}\big(\underline{\boldsymbol{V}}_{\Phi}\!(\mu)\big)\Big\}
\label{3.30.h}
\end{equation}
\end{subequations}

Analog erhalten wir f"ur den Erwartungswert des in Gleichung (\ref{3.28}) 
angegebenen zweiten normierten Skalarprodukts
\begin{equation}
\text{E}\Big\{\boldsymbol{c}_{\Psi}(\mu)\CdoT\Hat{\Vec{\boldsymbol{N}}}_{\!\!f}(\mu)\CdoT
\Hat{\Vec{\boldsymbol{N}}}_{\!\!f}(\!-\mu)^{\tT}\Big\}\;=\;
\text{E}\big\{\boldsymbol{N}_{\!\!f}(\mu)\CdoT\boldsymbol{N}_{\!\!f}(\!-\mu)\big\}\cdot
\text{E}\Big\{\,\text{spur}\big(\underline{\boldsymbol{V}}_{\Psi}\!(\mu)\big)\Big\}.
\label{3.31}
\end{equation}

Wenn wir nun f"ur die zuf"alligen Normierungsfaktoren
\begin{align}
\boldsymbol{c}_{\Phi}(\mu)&\;=\;
\frac{1}{M\cdot\text{spur}\big(\underline{\boldsymbol{V}}_{\bot}\!(\mu)\CdoT
\underline{\boldsymbol{V}}_{\bot}\!(\mu)^{\Hh}\big)}\qquad\text{und}\qquad\label{3.32}\\
\boldsymbol{c}_{\Psi}(\mu)&\;=\;
\frac{1}{M\cdot\text{spur}\big(\underline{\boldsymbol{V}}_{\bot}\!(\mu)\CdoT
\underline{\boldsymbol{V}}_{\bot}\!(\!-\mu)^{\Tt}\big)}\label{3.33}
\end{align}
w"ahlen, sind die Spuren der Matrizen 
\mbox{$\underline{\boldsymbol{V}}_{\Phi}(\mu)$} und 
\mbox{$\underline{\boldsymbol{V}}_{\Psi}(\mu)$} immer $1/M$, und die 
zuf"alligen Normierungsfaktoren sind wie gefordert von den Zufallsgr"o"sen 
\mbox{$\boldsymbol{N}_{\!\!f}(\mu)$} und \mbox{$\boldsymbol{N}_{\!\!f}(\!-\mu)$} 
unabh"angig, wenn dies f"ur die Elemente der Matrizen 
\mbox{$\underline{\boldsymbol{V}}_{\bot}\!(\mu)$} und 
\mbox{$\underline{\boldsymbol{V}}_{\bot}\!(\!-\mu)$} ebenfalls gilt.
Dann sch"atzen die in den Gleichungen (\ref{3.26}) und (\ref{3.28}) 
angegebenen normierten Skalarprodukte die Gr"o"sen 
\mbox{$\Tilde{\Phi}_{\boldsymbol{n}}(\mu)$} und 
\mbox{$\Tilde{\Psi}_{\boldsymbol{n}}(\mu)$} erwartungstreu ab.
Als Messwerte \mbox{$\Hat{\Phi}_{\boldsymbol{n}}(\mu)$} und 
\mbox{$\Hat{\Psi}_{\boldsymbol{n}}(\mu)$} verwenden wir daher konkrete 
Realisierungen dieser normierten zuf"alligen Skalarprodukte:
\begin{gather}
\Hat{\Phi}_{\boldsymbol{n}}(\mu)\;=\;
\frac{1}{M\CdoT\text{spur}\big(\underline{V}_{\bot}\!(\mu)\CdoT
\underline{V}_{\bot}\!(\mu)^{\Hh}\big)}\cdot
\Hat{\Vec{N}}_{\!f}(\mu)\CdoT\Hat{\Vec{N}}_{\!f}(\mu)^{\Hh}\,=
\label{3.34}\\[4pt]
=\;\frac{1}{M\CdoT\text{spur}\big(\underline{V}_{\bot}\!(\mu)\CdoT
\underline{V}_{\bot}\!(\mu)^{\Hh}\big)}\cdot
\Vec{N}_{\!f}(\mu)\CdoT
\underline{V}_{\bot}\!(\mu)\CdoT
\underline{V}_{\bot}\!(\mu)^{\Hh}\!\CdoT
\Vec{N}_{\!f}(\mu)^{\Hh}\,=
\notag\\[4pt]
=\;\frac{1}{M\CdoT\text{spur}\big(\underline{V}_{\bot}\!(\mu)\CdoT
\underline{V}_{\bot}\!(\mu)^{\Hh}\big)}\cdot
\Vec{Y}_{\!f}(\mu)\CdoT
\underline{V}_{\bot}\!(\mu)\CdoT
\underline{V}_{\bot}\!(\mu)^{\Hh}\!\CdoT
\Vec{Y}_{\!f}(\mu)^{\Hh}
\notag\\*[4pt]
\qquad\qquad\forall\qquad\mu=0\;(1)\;M\!-\!1\notag\\*[-25pt]\notag
\end{gather}
\begin{gather}
\Hat{\Psi}_{\boldsymbol{n}}(\mu)\;=\;
\frac{1}{M\CdoT\text{spur}\big(\underline{V}_{\bot}\!(\mu)\CdoT
\underline{V}_{\bot}\!(\!-\mu)^{\Tt}\big)}\cdot
\Hat{\Vec{N}}_{\!f}(\mu)\CdoT\Hat{\Vec{N}}_{\!f}(\!-\mu)^{\Tt}\,=
\label{3.35}\\[4pt]
=\;\frac{1}{M\CdoT\text{spur}\big(\underline{V}_{\bot}\!(\mu)\CdoT
\underline{V}_{\bot}\!(\!-\mu)^{\Tt}\big)}\cdot
\Vec{N}_{\!f}(\mu)\CdoT
\underline{V}_{\bot}\!(\mu)\CdoT
\underline{V}_{\bot}\!(\!-\mu)^{\Tt}\!\CdoT
\Vec{N}_{\!f}(\!-\mu)^{\Tt}\,=
\notag\\[4pt]
=\;\frac{1}{M\CdoT\text{spur}\big(\underline{V}_{\bot}\!(\mu)\CdoT
\underline{V}_{\bot}\!(\!-\mu)^{\Tt}\big)}\cdot
\Vec{Y}_{\!f}(\mu)\CdoT
\underline{V}_{\bot}\!(\mu)\CdoT
\underline{V}_{\bot}\!(\!-\mu)^{\Tt}\!\CdoT
\Vec{Y}_{\!f}(\!-\mu)^{\Tt}
\notag\\*[4pt]
\qquad\qquad\forall\qquad\mu=0\;(1)\;M\!-\!1.\notag
\end{gather}

Die bei der Messung verwendete Erregung ist durch die $M$ Vektoren 
\mbox{$\Vec{V}(\mu)$} bestimmt, mit deren Hilfe sich die $M$ Matrizen 
\mbox{$\underline{V}_{\bot}\!(\mu)$} konstruieren lassen. Aus den bei 
der Messung erhaltenen $L$ Signalmusterfolgen \mbox{$y_{\lambda}(k)$} 
lassen sich die $M$ Stichprobenvektoren \mbox{$\Vec{Y}_{\!f}(\mu)$} 
\mbox{"uber} die Fensterung und die DFT berechnen. Daher lassen sich diese 
Messwerte \mbox{prinzipiell} \mbox{ohne} die Kenntnis der theoretisch optimalen 
"Ubertragungsfunktion berechnen. Wenn wir jedoch die Art der Konstruktion 
der Matrizen \mbox{$\underline{V}_{\bot}\!(\mu)$} nicht geschickt w"ahlen, 
wird im Allgemeinen ein erheblicher Speicher- und Rechenaufwand n"otig sein,
um diese Messwerte zu berechnen. Bisher hatten wir f"ur deren Konstruktion
lediglich festgelegt, dass der Vektor \mbox{$\Vec{V}(\mu)$} ein Eigenvektor 
der Matrix \mbox{$\underline{V}_{\bot}\!(\mu)$} zum Eigenwert Null sein 
muss, und dass alle Matrixelemente von den Zufallsgr"o"sen 
\mbox{$\boldsymbol{N}_{\!\!f}(\mu)$} und \mbox{$\boldsymbol{N}_{\!\!f}(\!-\mu)$} 
unabh"angig sein m"ussen. 

\section[Konstruktion einer Matrix zur Messung des Leistungsdichtespektrums]{Konstruktion 
einer Matrix zur Messung des \\Leistungsdichtespektrums}

Bevor nun eine geeignete Methode zur Konstruktion 
der Matrizen \mbox{$\underline{V}_{\bot}\!(\mu)$} angegeben wird, soll zun"achst 
eine im weiteren n"utzliche Eigenschaft, die diese Matrizen erf"ullen sollen,
hergeleitet werden. 

Bei der Berechnung der Messwerte erf"ullen die Skalarprodukte der
konkreten Stichprobenvektoren \mbox{$\Hat{\Vec{N}}_{\!f}(\pm\mu)$}
immer die Cauchy-Schwarzsche Ungleichung, so dass f"ur sie immer
\begin{equation}
\Big|\Hat{\Vec{N}}_{\!f}(\mu)\CdoT
\Hat{\Vec{N}}_{\!f}(\!-\mu)^{\Tt}\Big|^2\,\le\;
\Big(\Hat{\Vec{N}}_{\!f}(\mu)\CdoT
\Hat{\Vec{N}}_{\!f}(\mu)^{\Hh}\Big)\CdoT
\Big(\Hat{\Vec{N}}_{\!f}(\!-\mu)\CdoT
\Hat{\Vec{N}}_{\!f}(\!-\mu)^{\Hh}\Big)
\label{3.36}
\end{equation}
gilt. Wenn man die Matrix \mbox{$\underline{V}_{\bot}\!(\mu)$}
so festlegt, dass f"ur die Spuren der beiden Matrixprodukte bei der
Berechnung beider Messwerte die Ungleichung
\begin{equation}
\Big|\text{spur}\big(\underline{V}_{\bot}\!(\mu)\CdoT
\underline{V}_{\bot}\!(\!-\mu)^{\Tt}\big)\Big|^2\,\ge\;
\text{spur}\big(\underline{V}_{\bot}\!(\mu)\CdoT
\underline{V}_{\bot}\!(\mu)^{\Hh}\big)\cdot
\text{spur}\big(\underline{V}_{\bot}\!(\!-\mu)\CdoT
\underline{V}_{\bot}\!(\!-\mu)^{\Hh}\big)
\label{3.37}
\end{equation}
gilt, so kann man mit Gleichung (\ref{3.36}) zeigen, dass die Bedingung
\begin{equation}
\big|\Hat{\Psi}_{\boldsymbol{n}}(\mu)\big|^2 \,\le\;
\Hat{\Phi}_{\boldsymbol{n}}(\mu) \cdot \Hat{\Phi}_{\boldsymbol{n}}(\!-\mu)
\label{3.38}
\end{equation}
von den Messwerten immer erf"ullt wird. Diese Bedingung wird sp"ater bei der 
Berechnung der Sch"atzwerte f"ur die Konfidenzgebiete der Messwerte von 
Bedeutung sein. Die analoge Bedingung f"ur die durch die Messung abzusch"atzenden
Gr"o"sen \mbox{$\Tilde{\Phi}_{\boldsymbol{n}}(\mu)$} und 
\mbox{$\Tilde{\Psi}_{\boldsymbol{n}}(\mu)$} ist nach Ungleichung (\ref{2.36}) 
immer erf"ullt. 

Wenn man bedenkt, dass die Spur einer Matrix die Summe ihrer Hauptdiagonalelemente 
ist, kann man unter anderem mit der Cauchy-Schwarzschen-Ungleichung
zeigen, dass die Ungleichung (\ref{3.37}) immer mit dem Ungleichheitsoperator
\mbox{"`$\le$"'} erf"ullt ist. Daher kann die Matrix
\mbox{$\underline{V}_{\bot}\!(\mu)$} bestenfalls so gew"ahlt werden,
dass die Bedingung (\ref{3.37}) mit dem Gleichheitszeichen erf"ullt
wird. Wie man durch Einsetzen leicht zeigen kann, erf"ullt jedes Matrizenpaar
\mbox{$\underline{V}_{\bot}\!(\mu)$} und \mbox{$\underline{V}_{\bot}\!(\!-\mu)$} die Bedingung (\ref{3.37})
mit dem Gleichheitszeichen, wenn
\begin{equation}
\underline{V}_{\bot}\!(\mu)^{\Hh}\;=\;
\underline{V}_{\bot}\!(\!-\mu)^{\Tt}
\label{3.39}
\end{equation}\vspace{-34pt}

gilt. 

Eine M"oglichkeit ein Matrizenpaar \mbox{$\underline{V}_{\bot}\!(\mu)$} und 
\mbox{$\underline{V}_{\bot}\!(\!-\mu)$} dieser Bedingung entsprechend festzulegen, 
besteht darin, eine Matrix \mbox{$\underline{V}_{\bot}\!(\mu)$} zu 
w"ahlen, die jeden beliebigen Vektor auf den Orthogonalraum der
Vektoren \mbox{$\Vec{V}(\mu)$} und \mbox{$\Vec{V}(\!-\mu)^{\Kk}$} projiziert. 
Um die Matrix \mbox{$\underline{V}_{\bot}\!(\mu)$} zu konstruieren,
wird zun"achst die Stichprobenmatrix
\begin{equation}
\Hat{\underline{V}}(\mu)\;=\;
\begin{bmatrix}
\Vec{V}(\mu)\\
\Vec{V}(\!-\mu)^{\Kk}
\end{bmatrix}
\qquad\qquad\forall\qquad\mu=0\;(1)\;M\!-\!1
\label{3.40}
\end{equation}
mit der Permutationssymmetrie
\begin{equation}
\Hat{\underline{V}}(\mu)\;=\;
\begin{bmatrix}0&1\\1&0\end{bmatrix}\Cdot
\Hat{\underline{V}}(\!-\mu)^{\Kk}
\label{3.41}
\end{equation}
gebildet, deren Zeilen die bei der Messung verwendeten 
konkreten Stichproben vom Umfang $L$ der Zufallsgr"o"sen 
\mbox{$\boldsymbol{V}(\mu)$} und \mbox{$\boldsymbol{V}(\!-\mu)$}
enthalten, wobei die Elemente der zweiten Zeile konjugiert sind.
Damit bilden wir die empirische, hermitesche Kovarianzmatrix
\begin{equation}
\Hat{\underline{C}}_{\Hat{\Vec{\boldsymbol{V}}}(\mu),\Hat{\Vec{\boldsymbol{V}}}(\mu)}\;=\;
\frac{1}{L}\cdot\Hat{\underline{V}}(\mu)\CdoT\Hat{\underline{V}}(\mu)^{\;\HH}\;=\;
\Hat{\underline{C}}_{\Hat{\Vec{\boldsymbol{V}}}(\mu),\Hat{\Vec{\boldsymbol{V}}}(\mu)}^{\Hh}
\label{3.42}
\end{equation}
der Dimension \mbox{$2\!\times\!2$}. Sie weist die Permutationssymmetrie
\begin{gather}
\Hat{\underline{C}}_{\Hat{\Vec{\boldsymbol{V}}}(\mu),\Hat{\Vec{\boldsymbol{V}}}(\mu)}\;=\;
\begin{bmatrix}
\Hat{C}_{\boldsymbol{V}(\mu),\boldsymbol{V}(\mu)}&
\Hat{C}_{\boldsymbol{V}(\mu),\boldsymbol{V}(-\mu)^{\Kk}}\\
\Hat{C}_{\boldsymbol{V}(-\mu)^{\Kk}\!,\boldsymbol{V}(\mu)}&
\Hat{C}_{\boldsymbol{V}(-\mu)^{\Kk}\!,\boldsymbol{V}(-\mu)^{\Kk}}
\end{bmatrix}\;=\;
\begin{bmatrix}
\Hat{C}_{\boldsymbol{V}(\mu),\boldsymbol{V}(\mu)}&
\Hat{C}_{\boldsymbol{V}(-\mu),\boldsymbol{V}(\mu)^{\Kk}}\\
\Hat{C}_{\boldsymbol{V}(-\mu),\boldsymbol{V}(\mu)^{\Kk}}^{\Kk}&
\Hat{C}_{\boldsymbol{V}(-\mu),\boldsymbol{V}(-\mu)}
\end{bmatrix}\;=
\notag\\[12pt]
=\;\begin{bmatrix}0&1\\1&0\end{bmatrix}\Cdot
\Hat{\underline{C}}_{\Hat{\Vec{\boldsymbol{V}}}(-\mu),\Hat{\Vec{\boldsymbol{V}}}(-\mu)}^{\;\TT}\Cdot
\begin{bmatrix}0&1\\1&0\end{bmatrix}
\qquad\qquad\forall\qquad\mu=0\;(1)\;M\!-\!1\quad{}
\label{3.43}
\end{gather}
auf. Die Hauptdiagonalelemente sind die nach Gleichung (\ref{3.15})
mit \mbox{$\mu_1=\mu_2=\mu$} bzw. mit \mbox{$\mu_1=\mu_2=-\mu$}
definierten empirischen Kovarianzen, die sich durch Akkumulation 
aus den bei den Einzelmessungen $\lambda$ verwendeten Spektralwerten 
der Erregung berechnen lassen, ohne dass dazu die Spektralwerte
aller Einzelmessungen abgespeichert werden m"ussen. 
Die Nebendiagonalelemente sind die empirischen Kovarianzen
\begin{gather}
\Hat{C}_{\boldsymbol{V}(\mu_1),\boldsymbol{V}(\mu_2)^{\Kk}}\;=\;
\frac{1}{L}\cdot
\Vec{V}(\mu_1)\cdot\Vec{V}(\mu_2)^{\Tt}\;=\;
\frac{1}{L}\cdoT\Sum{\lambda=1}{L}\,
V_{\lambda}(\mu_1)\CdoT V_{\lambda}(\mu_2)
\label{3.44}\\
\forall\qquad\mu_1=0\;(1)\;M\!-\!1\quad\text{und}
\quad\mu_2=0\;(1)\;M\!-\!1,\notag
\end{gather}
die hier mit \mbox{$-\mu_1=\mu_2=\mu$} auftreten und die sich in 
derselben Art ohne Speicherung der Spektralwerte
aller Einzelmessungen berechnen lassen. Mit der Stichprobenmatrix
\mbox{$\Hat{\underline{V}}(\mu)$} und der empirischen Kovarianzmatrix 
\mbox{$\Hat{\underline{C}}_{\Hat{\Vec{\boldsymbol{V}}}(\mu),\Hat{\Vec{\boldsymbol{V}}}(\mu)}$} 
k"onnen wir nun die ben"otigte, hermitesche Matrix \mbox{$\underline{V}_{\bot}\!(\!-\mu)$} 
wie folgt konstruieren:
\begin{equation}
\underline{V}_{\bot}\!(\mu)\;=\;
\underline{V}_{\bot}\!(\mu)^{\Hh}\;=\;
\underline{E}-\frac{1}{L}\cdot
\Hat{\underline{V}}(\mu)^{\Hh}\CdoT
\Hat{\underline{C}}_{\Hat{\Vec{\boldsymbol{V}}}(\mu),\Hat{\Vec{\boldsymbol{V}}}(\mu)}^{\uP{0.4}{\!-1}}\!\CdoT
\Hat{\underline{V}}(\mu)
\qquad\forall\qquad\mu=0\;(1)\;M\!-\!1.
\label{3.45}
\end{equation}
Dass diese Matrix die Forderung (\ref{3.24}), dass der Vektor \mbox{$\Vec{V}(\mu)$} 
ein Eigenvektor zum Eigenwert Null ist, erf"ullt wird weiter unten gezeigt. Dort wird auch 
n"aher erl"autert, wie diese Matrix jeden beliebigen Vektor auf den Orthogonalraum der
Vektoren \mbox{$\Vec{V}(\mu)$} und \mbox{$\Vec{V}(\!-\mu)^{\Kk}$} projiziert. 
Durch Einsetzen der Permutationssymmetrien (\ref{3.41}) und 
(\ref{3.43}) kann man verifizieren, dass diese Matrix auch die in Gleichung 
(\ref{3.39}) geforderte Symmetriebedingung erf"ullt.

Zur Berechnung der Matrix \mbox{$\underline{V}_{\bot}\!(\mu)$}
ist es notwendig, die empirische Kovarianzmatrix
\mbox{$\Hat{\underline{C}}_{\Hat{\Vec{\boldsymbol{V}}}(\mu),\Hat{\Vec{\boldsymbol{V}}}(\mu)}$} zu invertierten.
Im Fall, dass der Vektor \mbox{$\Vec{V}(\!-\mu)^{\Kk}$} ein Vielfaches des
Vektors \mbox{$\Vec{V}(\mu)$} ist, wird die empirische Kovarianzmatrix
singul"ar. In diesem Fall ersetzen wir einen der beiden linear abh"angigen 
Vektoren durch einen Vektor, der kein Vielfaches des Vektors 
\mbox{$\Vec{V}\!(\mu)$} ist. Die Konstruktion dieses Vektors muss dabei
in einer Weise erfolgen, dass dessen Elemente --- unter Ber"ucksichtigung der
Zuf"alligkeit der Messung und der Vektoren \mbox{$\Vec{\boldsymbol{V}}(\mu)$} 
und \mbox{$\Vec{\boldsymbol{N}}_{\!\!f}(\mu)$} --- von den Zufallsgr"o"sen 
\mbox{$\boldsymbol{N}_{\!\!f}(\mu)$} und \mbox{$\boldsymbol{N}_{\!\!f}(\!-\mu)$}
unabh"angig sind, um sicherzustellen, dass die oben durchgef"uhrte Berechnung der 
Erwartungswerte der Messwerte auch im singul"aren Fall ihre G"ultigkeit beh"alt. 

Damit sich die Matrix \mbox{$\underline{V}_{\bot}\!(\mu)$} und somit auch die Messwerte 
numerisch gut berechnen lassen, muss die empirische Kovarianzmatrix gut
konditioniert sein. Wenn die theoretische Kovarianzmatrix\footnote{Wie in 
Kapitel \ref{SysZeit} gesagt, berechnet sich diese mit Hilfe der Verbundverteilung der 
daran beteiligten Zufallsgr"o"sen.}
\begin{equation}
\underline{C}_{\Hat{\Vec{\boldsymbol{V}}}(\mu),\Hat{\Vec{\boldsymbol{V}}}(\mu)}\;=\;
\text{E}\big\{\Hat{\Vec{\boldsymbol{V}}}(\mu)\CdoT
\Hat{\Vec{\boldsymbol{V}}}(\mu)^{\Hh}\big\}
\label{3.46}
\end{equation}
des Zufallsvektors
\begin{equation}
\Hat{\Vec{\boldsymbol{V}}}(\mu)\;=\;
\begin{bmatrix}
\boldsymbol{V}(\mu)\\
\boldsymbol{V}(\!-\mu)^{\Kk}
\end{bmatrix}
\qquad\qquad\forall\qquad \mu=0\;(1)\;M\!-\!1
\label{3.47}
\end{equation}
dessen konkrete Stichprobe vom Umfang $L$ die Matrix \mbox{$\Hat{\underline{V}}(\mu)$} 
ist, gut konditioniert ist, ist zu erwarten, dass die Wahrscheinlichkeit,
eine schlecht konditionierte empirische Kovarianzmatrix 
\mbox{$\Hat{\underline{C}}_{\Hat{\Vec{\boldsymbol{V}}}(\mu),\Hat{\Vec{\boldsymbol{V}}}(\mu)}$}
zu erhalten, mit steigender Mittelungsanzahl $L$ bald so klein wird, dass dies f"ur
die Praxis ebensowenig von Bedeutung ist, wie die Behandlung des singul"aren Falles.
Im Anhang \ref{Komat} wird gezeigt, dass eine obere Schranke f"ur die 
Wahrscheinlichkeit, eine schlecht konditionierte empirische Kovarianzmatrix 
mittelwertfreier Zufallsvektoren zu erhalten, indirekt proportional mit 
steigender Mittelungsanzahl $L$ sinkt. Eine entsprechende Abhandlung f"ur empirische 
Kovarianzmatrizen mittelwertbehafteter Zufallsvektoren ist in \cite{Erg} zu
finden.

Aufgrund der in der Gleichung (\ref{3.39}) angegebenen
Eigenschaft der in der hier angegeben Weise konstruierten Matrizen 
\mbox{$\underline{V}_{\bot}\!(\mu)$} braucht man nun zur Berechnung der 
Messwerte gem"a"s Gleichung (\ref{3.34}) und (\ref{3.35}) f"ur die zwei Frequenzen
$\mu$ und $-\mu$ nicht zwei verschiedene Matrizen zu berechnen, sondern nur die eine 
Matrix, was eine erste deutliche Reduktion des Rechenaufwands darstellt.
Die Matrizen \mbox{$\underline{V}_{\bot}\!(\mu)$} nach Gleichung (\ref{3.45}) 
mit den Stichprobenmatrizen \mbox{$\Hat{\underline{V}}(\mu)$} nach Gleichung 
(\ref{3.40}) weisen noch weitere besondere Eigenschaften auf, die bei der 
Berechnung der Messwerte von Vorteil sind. Die Matrizen \mbox{$\underline{V}_{\bot}\!(\mu)$} 
sind n"amlich idempotent. Eine Matrix $\underline{M}$ ist idempotent, wenn f"ur sie 
\mbox{$\underline{M}^n=\underline{M}\;\;\forall\;\;n\in\mathbb{N}$}
gilt. Um die Idempotenz der Matrix \mbox{$\underline{V}_{\bot}\!(\mu)$} 
nachzuweisen, gen"ugt es zu zeigen, dass 
\begin{equation}
\underline{V}_{\bot}\!(\mu)^2=\underline{V}_{\bot}\!(\mu)
\label{3.48}
\end{equation}
erf"ullt ist. Der Rest folgt aus der vollst"andigen Induktion
\begin{equation}
\underline{V}_{\bot}\!(\mu)^n=
\underline{V}_{\bot}\!(\mu)^{n-2}\Cdot\underline{V}_{\bot}\!(\mu)^2=
\underline{V}_{\bot}\!(\mu)^{n-2}\Cdot\underline{V}_{\bot}\!(\mu)=
\underline{V}_{\bot}\!(\mu)^{n-1}.
\label{3.49}
\end{equation}
Dass Gleichung (\ref{3.48}) erf"ullt wird, zeigt man, indem man 
\mbox{$\underline{V}_{\bot}\!(\mu)$} nach Gleichung (\ref{3.45}) einsetzt:\vspace{-10pt}
\begin{gather}
\underline{V}_{\bot}\!(\mu)^2\;=
\label{3.50}\\*[6pt]
=\;\big(\underline{E}-\frac{1}{L}\cdot\Hat{\underline{V}}(\mu)^{\Hh}\CdoT
\Hat{\underline{C}}_{\Hat{\Vec{\boldsymbol{V}}}(\mu),\Hat{\Vec{\boldsymbol{V}}}(\mu)}^{\uP{0.4}{\!-1}}\!\CdoT
\Hat{\underline{V}}(\mu)\big)\cdot\big(\underline{E}-\frac{1}{L}\cdot
\Hat{\underline{V}}(\mu)^{\Hh}\CdoT
\Hat{\underline{C}}_{\Hat{\Vec{\boldsymbol{V}}}(\mu),\Hat{\Vec{\boldsymbol{V}}}(\mu)}^{\uP{0.4}{\!-1}}\!\CdoT
\Hat{\underline{V}}(\mu)\big)\;=
\notag\\*[10pt]\begin{flalign*}
&=\;\underline{E}\cdot\underline{E}-
\underline{E}\cdot\frac{1}{L}\cdot\Hat{\underline{V}}(\mu)^{\Hh}\CdoT
\Hat{\underline{C}}_{\Hat{\Vec{\boldsymbol{V}}}(\mu),\Hat{\Vec{\boldsymbol{V}}}(\mu)}^{\uP{0.4}{\!-1}}\!\CdoT
\Hat{\underline{V}}(\mu)-\frac{1}{L}\cdot\Hat{\underline{V}}(\mu)^{\Hh}\CdoT
\Hat{\underline{C}}_{\Hat{\Vec{\boldsymbol{V}}}(\mu),\Hat{\Vec{\boldsymbol{V}}}(\mu)}^{\uP{0.4}{\!-1}}\!\CdoT
\Hat{\underline{V}}(\mu)\cdot\underline{E}+{}&&
\end{flalign*}\notag\\*\begin{flalign*}
&&{}+\frac{1}{L^2}\cdot\Hat{\underline{V}}(\mu)^{\Hh}\CdoT
\Hat{\underline{C}}_{\Hat{\Vec{\boldsymbol{V}}}(\mu),\Hat{\Vec{\boldsymbol{V}}}(\mu)}^{\uP{0.4}{\!-1}}\CdoT\!\!
\underbrace{\Hat{\underline{V}}(\mu)\cdot\Hat{\underline{V}}(\mu)^{\Hh}}_{\scriptstyle
=L\cdot\Hat{\underline{C}}_{\Hat{\Vec{\boldsymbol{V}}}(\mu),\Hat{\Vec{\boldsymbol{V}}}(\mu)} 
\text{ nach (\ref{3.42})}}\!\!\Cdot
\Hat{\underline{C}}_{\Hat{\Vec{\boldsymbol{V}}}(\mu),\Hat{\Vec{\boldsymbol{V}}}(\mu)}^{\uP{0.4}{\!-1}}\!\CdoT
\Hat{\underline{V}}(\mu)\;=&
\end{flalign*}
\notag\\[-4pt]
=\;\underline{E}-\frac{1}{L}\cdot\Hat{\underline{V}}(\mu)^{\Hh}\CdoT
\Hat{\underline{C}}_{\Hat{\Vec{\boldsymbol{V}}}(\mu),\Hat{\Vec{\boldsymbol{V}}}(\mu)}^{\uP{0.4}{\!-1}}\!\CdoT
\Hat{\underline{V}}(\mu)\;=\;
\underline{V}_{\bot}\!(\mu).\qquad\qquad\qquad\qquad\qquad\qquad
\notag
\end{gather}
Aufgrund der Idempotenz lassen sich bei der Berechnung der 
Messwerte gem"a"s Gleichung (\ref{3.34}) und (\ref{3.35}) 
alle Matrixprodukte durch die Matrizen \mbox{$\underline{V}_{\bot}\!(\mu)$}
ersetzen, was den Rechenaufwand weiter reduziert.

Zur Berechnung der Messwerte gem"a"s Gleichung (\ref{3.34}) und (\ref{3.35}) 
sind auch noch die Spuren der Matrizen \mbox{$\underline{V}_{\bot}\!(\mu)$}
zu berechnen. Da die Spur einer Matrix die Summe ihrer Eigenwerte ist, wollen
wir diese nun bestimmen. F"ur jeden Vektor $\Vec{x}$, der sowohl zum Vektor 
\mbox{$\Vec{V}(\mu)$} als auch zum Vektor \mbox{$\Vec{V}(\!-\mu)^{\Kk}$} 
orthogonal ist, gilt:
\begin{equation}
\Vec{x}\CdoT\Hat{\underline{V}}(\mu)^{\Hh}\;=\;\Vec{0}.
\label{3.51}
\end{equation}
Bildet man einen solchen Vektor mit der Matrix \mbox{$\underline{V}_{\bot}\!(\mu)$}
nach Gleichung (\ref{3.45}) ab, so wird dieser Vektor nicht ver"andert.
\begin{align}
\Vec{x}\CdoT\underline{V}_{\bot}\!(\mu)&\;=\;
\Vec{x}\cdot\big(\underline{E}-\frac{1}{L}\cdot
\Hat{\underline{V}}(\mu)^{\Hh}\CdoT
\Hat{\underline{C}}_{\Hat{\Vec{\boldsymbol{V}}}(\mu),\Hat{\Vec{\boldsymbol{V}}}(\mu)}^{\uP{0.4}{\!-1}}\!\CdoT
\Hat{\underline{V}}(\mu)\big)\;=\;{}\label{3.52}\\[6pt]
&\;=\;\Vec{x}\CdoT\underline{E}-\frac{1}{L}\CdoT
\underbrace{\Vec{x}\CdoT\Hat{\underline{V}}(\mu)^{\Hh}}_{\scriptstyle=\Vec{0}}\CdoT
\Hat{\underline{C}}_{\Hat{\Vec{\boldsymbol{V}}}(\mu),\Hat{\Vec{\boldsymbol{V}}}(\mu)}^{\uP{0.4}{\!-1}}\!\CdoT
\Hat{\underline{V}}(\mu)\;=\;\Vec{x}\notag{}
\end{align}
Solch ein Vektor $\Vec{x}$ ist daher ein Eigenvektor des Eigenwertes Eins.
Da der zu den beiden Vektoren \mbox{$\Vec{V}(\mu)$} und 
\mbox{$\Vec{V}(\!-\mu)^{\Kk}$} orthogonale Raum von der Dimension 
\mbox{$L\!-\!2$} ist, ist Eins ein \mbox{$L\!-\!2$}-facher Eigenwert der
Matrix \mbox{$\underline{V}_{\bot}\!(\mu)$}. Bildet man die beiden Vektoren 
\mbox{$\Vec{V}(\mu)$} und \mbox{$\Vec{V}(\!-\mu)^{\Kk}$},
die die Zeilenvektoren der Stichprobenmatrix \mbox{$\Hat{\underline{V}}(\mu)$}
sind, mit der Matrix \mbox{$\underline{V}_{\bot}\!(\mu)$} ab, so erh"alt man
die Nullvektoren:
\begin{gather}\begin{flalign}
&\Hat{\underline{V}}(\mu)\CdoT\underline{V}_{\bot}\!(\mu)\;=\;
\Hat{\underline{V}}(\mu)\cdot\big(\underline{E}-\frac{1}{L}\cdot
\Hat{\underline{V}}(\mu)^{\Hh}\CdoT
\Hat{\underline{C}}_{\Hat{\Vec{\boldsymbol{V}}}(\mu),\Hat{\Vec{\boldsymbol{V}}}(\mu)}^{\uP{0.4}{\!-1}}\!\CdoT
\Hat{\underline{V}}(\mu)\big)\;=&&
\end{flalign}\label{3.53}\\*[8pt]\begin{flalign*}
&&=\;\Hat{\underline{V}}(\mu)\CdoT\underline{E}-\frac{1}{L}\cdoT\!\!
\underbrace{\Hat{\underline{V}}(\mu)\CdoT\Hat{\underline{V}}(\mu)^{\Hh}}_{\scriptstyle
=L\cdot\Hat{\underline{C}}_{\Hat{\Vec{\boldsymbol{V}}}(\mu),\Hat{\Vec{\boldsymbol{V}}}(\mu)}
\text{ nach (\ref{3.42})}}\!\!\Cdot
\Hat{\underline{C}}_{\Hat{\Vec{\boldsymbol{V}}}(\mu),\Hat{\Vec{\boldsymbol{V}}}(\mu)}^{\uP{0.4}{\!-1}}\!\CdoT
\Hat{\underline{V}}(\mu)\;=\;\Hat{\underline{V}}(\mu)-\Hat{\underline{V}}(\mu)\;=\;
\underline{0}.&
\end{flalign*}\notag
\end{gather}
Die Forderung (\ref{3.24}) ist somit erf"ullt und Null ist ein Eigenwert 
der Matrix \mbox{$\underline{V}_{\bot}\!(\mu)$} der Vielfachheit zwei. 
Damit haben wir nun alle Eigenwerte bestimmt, und die Spur der Matrix 
\mbox{$\underline{V}_{\bot}\!(\mu)$} ergibt sich als die Summe aller 
Eigenwerte zu \mbox{$L\!-\!2$}. Sie ist von der konkret bei der Messung 
verwendeten Erregung unabh"angig und muss daher nicht berechnet werden. 
Auch dies reduziert den Rechenaufwand erheblich.

Die $L$ Spaltenvektoren der Matrix \mbox{$\underline{V}_{\bot}\!(\mu)$} 
spannen den \mbox{$L\!-\!2$}-dimensionalen Nullraum der Stichprobenmatrix 
\mbox{$\Hat{\underline{V}}(\mu)$} auf. Jeder beliebige Vektor 
l"asst sich als Summe zweier zueinander orthogonaler Vektoren darstellen, 
von denen der eine Vektor in dem Raum liegt, der durch die beiden 
Vektoren der Stichprobenmatrix \mbox{$\Hat{\underline{V}}(\mu)$} 
aufgespannt wird, und der andere Vektor im Nullraum der Stichprobenmatrix 
liegt. Bei der Abbildung des beliebigen Vektors mit der Matrix 
\mbox{$\underline{V}_{\bot}\!(\mu)$} verschwindet der erste Anteil w"ahrend 
zweite Anteil unver"andert bleibt. Den Vorgang, den zu einem Raum orthogonalen 
Anteil eines Vektors zu eliminieren, ohne den restlichen Vektor zu ver"andern, 
bezeichnet man auch als Projektion des Vektors auf den Raum. 

Es sei hier noch eine andere M"oglichkeit, die Spur dieser Matrix zu 
berechnen, genannt. Dabei werden die folgenden beiden S"atze, die 
auch weiter unten bei der Berechnung von weiteren Matrixspuren n"utzlich
sein werden, verwendet. Der erste Satz besagt, dass die Spur der Summe zweier
Matrizen gleich der Summe der Spuren der beiden Matrizen ist:
\begin{equation}
\text{spur}\big(\underline{A}+\underline{B}\big)\;=\;
\text{spur}\big(\underline{A}\big)\;+\;\text{spur}\big(\underline{B}\big).
\label{3.54}
\end{equation}
Dies folgt unmittelbar aus der Definition der Spur einer Matrix als die 
Summe ihrer Hauptdiagonalelemente. Der zweite Satz besagt, dass die Spur 
des Produkts zweier Matrizen sich durch die Vertauschung der 
Reihenfolge der Matrizen nicht "andert. 
\begin{equation}
\text{spur}\big(\underline{A}\CdoT\underline{B}\big)\;=\;
\text{spur}\big(\underline{B}\CdoT\underline{A}\big)
\label{3.55}
\end{equation}
Voraussetzung ist hier nat"urlich, dass die Matrizen in beiden Richtungen 
verkettbar sind, so dass beide Matrixprodukte gebildet werden k"onnen, 
und immer eine quadratische Matrix entsteht. Zeigen l"asst sich dieser Satz, 
indem man das Matrixprodukt als Summe dyadischer Produkte\footnote{$i$-ter 
Spaltenvektor der ersten Matrix mal $i$-ter Zeilenvektor der zweiten 
Matrix} schreibt. Die Spur jedes dieser dyadischen Produkte ist gleich 
dem Skalarprodukt der beiden daran beteiligten Vektoren. Die Summe der Spuren 
der dyadischen Produkte ist die Summe der entsprechenden Skalarprodukte. 
Da diese Skalarprodukte gerade die Hauptdiagonalelemente der Matrix des 
Produkts der Matrizen mit vertauschter Reihenfolge sind, ist die Spur von 
der Reihenfolge der Matrixfaktoren unabh"angig. 

Wenden wir diese beiden S"atze auf die Matrix \mbox{$\underline{V}_{\bot}\!(\mu)$} 
nach Gleichung (\ref{3.45}) an, so erhalten wir die Spur \mbox{$L\!-\!2$} dieser 
Matrix als die Differenz der Spuren der Matrizen \mbox{$\underline{E}$}, deren 
Spur $L$ ist, und der Matrix \mbox{$\Hat{\underline{V}}(\mu)^{\Hh}\CdoT
\Hat{\underline{C}}_{\Hat{\Vec{\boldsymbol{V}}}(\mu),\Hat{\Vec{\boldsymbol{V}}}(\mu)}^{\uP{0.4}{\!-1}}\!\CdoT
\Hat{\underline{V}}(\mu)/L$}, die die gleiche Spur besitzt wie die 
\mbox{$2\!\times\!2$} Matrix\vspace{0pt minus 10pt}
\[
\frac{1}{L}\cdot
\Hat{\underline{C}}_{\Hat{\Vec{\boldsymbol{V}}}(\mu),\Hat{\Vec{\boldsymbol{V}}}(\mu)}^{\uP{0.4}{\!-1}}\!\CdoT
\underbrace{\Hat{\underline{V}}(\mu)\CdoT\Hat{\underline{V}}(\mu)^{\Hh}}_{\scriptstyle
=L\cdot\Hat{\underline{C}}_{\Hat{\Vec{\boldsymbol{V}}}(\mu),\Hat{\Vec{\boldsymbol{V}}}(\mu)}
\text{ nach (\ref{3.42})}}\;=\;\underline{E}
\]
deren Spur zwei ist.

Wenn wir nun all die Eigenschaften (\ref{3.39}), (\ref{3.45}) und (\ref{3.48}) 
sowie die eben berechnete Spur der Matrix \mbox{$\underline{V}_{\bot}\!(\mu)$} 
in die Gleichungen (\ref{3.34}) und (\ref{3.35}) einsetzen, erhalten wir
erwartungstreue Messwerte
\begin{gather}
\Hat{\Phi}_{\boldsymbol{n}}(\mu)\;=\;
\frac{1}{M\CdoT(L\!-\!2)}\cdot
\Hat{\Vec{N}}_{\!f}(\mu)\CdoT\Hat{\Vec{N}}_{\!f}(\mu)^{\Hh}\,=
\label{3.56}\\[6pt]
=\;\frac{1}{M\CdoT(L\!-\!2)}\cdot\Vec{N}_{\!f}(\mu)\CdoT
\underline{V}_{\bot}\!(\mu)\CdoT\Vec{N}_{\!f}(\mu)^{\Hh}\,=
\notag\\[6pt]
=\;\frac{1}{M\CdoT(L\!-\!2)}\cdot\Vec{Y}_{\!f}(\mu)\CdoT
\underline{V}_{\bot}\!(\mu)\CdoT\Vec{Y}_{\!f}(\mu)^{\Hh}\,=
\notag\\[6pt]
=\;\frac{1}{M\CdoT(L\!-\!2)}\cdot\Vec{Y}_{\!f}(\mu)\cdot
\Big(\,\underline{E}-\frac{1}{L}\CdoT
\underline{V}(\mu)^{\Hh}\!\Cdot
\Hat{\underline{C}}_{\Hat{\Vec{\boldsymbol{V}}}(\mu),\Hat{\Vec{\boldsymbol{V}}}(\mu)}^{\uP{0.4}{\!-1}}\!\CdoT
\underline{V}(\mu)\,\Big)\cdot\Vec{Y}_{\!f}(\mu)^{\Hh}\;=
\notag\\[6pt]
=\frac{L}{M\CdoT(L\!-\!2)}\CdoT\Bigg(\frac{
\Vec{Y}_{\!f}(\mu)\CdoT
\Vec{Y}_{\!f}(\mu)^{\Hh}\!}{L}-\frac{
\Vec{Y}_{\!f}(\mu)\CdoT
\underline{V}(\mu)^{\Hh}\!}{L}\cdot
\Hat{\underline{C}}_{\Hat{\Vec{\boldsymbol{V}}}(\mu),\Hat{\Vec{\boldsymbol{V}}}(\mu)}^{\uP{0.4}{\!-1}}\Cdot
\frac{\underline{V}(\mu)\CdoT
\Vec{Y}_{\!f}(\mu)^{\Hh}}{L}\Bigg)=
\notag\\[6pt]
=\;\frac{L}{M\CdoT(L\!-\!2)}\cdoT\Bigg(\!
\Hat{C}_{\boldsymbol{Y}_{\!\!\!f}(\mu),\boldsymbol{Y}_{\!\!\!f}(\mu)}-
\Big[\Hat{C}_{\boldsymbol{Y}_{\!\!\!f}(\mu),\boldsymbol{V}(\mu)},\,
\Hat{C}_{\boldsymbol{Y}_{\!\!\!f}(\mu),\boldsymbol{V}(-\mu)^{\Kk}}\Big]\Cdot
\Hat{\underline{C}}_{\Hat{\Vec{\boldsymbol{V}}}(\mu),\Hat{\Vec{\boldsymbol{V}}}(\mu)}^{\uP{0.4}{\!-1}}\CdoT
\begin{bmatrix}
\Hat{C}_{\boldsymbol{Y}_{\!\!\!f}(\mu),\boldsymbol{V}(\mu)}^{\Kk}\\
\Hat{C}_{\boldsymbol{Y}_{\!\!\!f}(\mu),\boldsymbol{V}(-\mu)^{\Kk}}^{\Kk}\!
\end{bmatrix}\!\Bigg)
\notag\\[4pt]
\forall\qquad\mu=0\;(1)\;M\!-\!1\notag
\end{gather}\vspace{0pt minus 10pt}
und\vspace{0pt minus 10pt}
\begin{gather}
\Hat{\Psi}_{\boldsymbol{n}}(\mu)\;=\;
\frac{1}{M\CdoT(L\!-\!2)}\cdot
\Hat{\Vec{N}}_{\!f}(\mu)\CdoT\Hat{\Vec{N}}_{\!f}(\!-\mu)^{\Tt}\,=
\label{3.57}\\[6pt]
=\;\frac{1}{M\CdoT(L\!-\!2)}\cdot\Vec{N}_{\!f}(\mu)\CdoT
\underline{V}_{\bot}\!(\mu)\CdoT\Vec{N}_{\!f}(\!-\mu)^{\Tt}\,=
\notag\\[6pt]
=\;\frac{1}{M\CdoT(L\!-\!2)}\cdot\Vec{Y}_{\!f}(\mu)\CdoT
\underline{V}_{\bot}\!(\mu)\CdoT\Vec{Y}_{\!f}(\!-\mu)^{\Tt}\;=
\notag\\[6pt]
=\,\frac{1}{M\CdoT(L\!-\!2)}\cdot\Vec{Y}_{\!f}(\mu)\CdoT
\Big(\,\underline{E}-\frac{1}{L}\CdoT\underline{V}(\mu)^{\Hh}\!\Cdot
\Hat{\underline{C}}_{\Hat{\Vec{\boldsymbol{V}}}(\mu),\Hat{\Vec{\boldsymbol{V}}}(\mu)}^{\uP{0.4}{\!-1}}\!\CdoT
\underline{V}(\mu)\,\Big)\cdot\Vec{Y}_{\!f}(\!-\mu)^{\Tt}\,=
\notag\\[6pt]
=\frac{L}{M\CdoT(L\!-\!2)}\cdoT\Bigg(\!
\Hat{C}_{\boldsymbol{Y}_{\!\!\!f}(\mu),\boldsymbol{Y}_{\!\!\!f}(-\mu)^{\Kk}}-
\Big[\Hat{C}_{\boldsymbol{Y}_{\!\!\!f}(\mu),\boldsymbol{V}(\mu)},
\Hat{C}_{\boldsymbol{Y}_{\!\!\!f}(\mu),\boldsymbol{V}(-\mu)^{\Kk}}\!\Big]\!\Cdot\!
\Hat{\underline{C}}_{\Hat{\Vec{\boldsymbol{V}}}(\mu),\Hat{\Vec{\boldsymbol{V}}}(\mu)}^{\uP{0.4}{\!-1}}\!\CdoT\!
\begin{bmatrix}\Hat{C}_{\boldsymbol{V}(\mu),\boldsymbol{Y}_{\!\!\!f}(-\mu)^{\Kk}}\\
\Hat{C}_{\boldsymbol{V}(-\mu),\boldsymbol{Y}_{\!\!\!f}(-\mu)}^{\Kk}\end{bmatrix}\!\Bigg)
\notag\\[4pt]
\forall\qquad\mu=0\;(1)\;M\!-\!1,\notag
\end{gather}
f"ur das LDS \mbox{$\Hat{\Phi}_{\boldsymbol{n}}(\mu)$} und 
das MLDS \mbox{$\Hat{\Psi}_{\boldsymbol{n}}(\mu)$}, die die Ungleichung 
(\ref{3.38}) immer erf"ullen. Die jeweils zuletzt dargestellte Form zeigt,
wie man diese Messwerte aus den empirisch gewonnenen Kovarianzen und 
Kovarianzmatrizen berechnet, die man ihrerseits, wie in den Gleichungen 
(\ref{3.10}), (\ref{3.15}), (\ref{3.16}), (\ref{3.44}) und (\ref{3.43}) gezeigt, 
ohne Zwischenspeicherung der Spektralwerte aller Einzelmessungen bestimmen kann.

\section[Varianzen und Kovarianzen der Messwerte der "Ubertragungsfunktion]{Varianzen 
und Kovarianzen der Messwerte der\\"Ubertragungsfunktion}\label{Kova}

Die Varianzen der Messwerte \mbox{$\Hat{\boldsymbol{H}}(\mu)$} sind die 
Erwartungswerte der Betragsquadrate der Abweichungen der zuf"alligen 
Messwerte von ihren Erwartungswerten. Da die Messwerte erwartungstreu sind, 
sind diese zuf"alligen Abweichungen zugleich die Abweichungen von den 
theoretisch optimalen Werten \mbox{$H\big({\T\mu\CdoT\frac{2\pi}{M}}\big)$}.
Setzt man diese Abweichungen gem"a"s Gleichung (\ref{3.19}) ein so erh"alt
man als Varianzen der Messwerte \mbox{$\Hat{\boldsymbol{H}}(\mu)$}:
\begin{gather}
C_{\Hat{\boldsymbol{H}}(\mu),\Hat{\boldsymbol{H}}(\mu)}\;=\;
\text{E}\bigg\{\Big|\Hat{\boldsymbol{H}}(\mu)\!-\!
\text{E}\big\{\Hat{\boldsymbol{H}}(\mu)\big\}\Big|^2\bigg\}\;=\;
\text{E}\bigg\{\Big|\Hat{\boldsymbol{H}}(\mu)\!-\!
H\!\big({\T\mu\CdoT\frac{2\pi}{M}}\big)\Big|^2\bigg\}\;=
\label{3.58}\\[4pt]
=\;\text{E}\bigg\{\frac{1}{L^2}\cdoT
\Vec{\boldsymbol{N}}_{\!\!f}(\mu)\CdoT
\Vec{\boldsymbol{V}}(\mu)^{\Hh}\Cdot
\Hat{\boldsymbol{C}}_{\boldsymbol{V}(\mu),\boldsymbol{V}(\mu)}^{-1}\Cdot
\Hat{\boldsymbol{C}}_{\boldsymbol{V}(\mu),\boldsymbol{V}(\mu)}^{-1}\CdoT
\Vec{\boldsymbol{V}}(\mu)\CdoT
\Vec{\boldsymbol{N}}_{\!\!f}(\mu)^{\Hh}\bigg\}\;=
\notag\\[4pt]
=\;\text{E}\bigg\{\text{spur}\bigg(\frac{1}{L^2}\cdot
\Vec{\boldsymbol{V}}(\mu)^{\Hh}\Cdot
\Hat{\boldsymbol{C}}_{\boldsymbol{V}(\mu),\boldsymbol{V}(\mu)}^{-1}\Cdot
\Hat{\boldsymbol{C}}_{\boldsymbol{V}(\mu),\boldsymbol{V}(\mu)}^{-1}\CdoT
\Vec{\boldsymbol{V}}(\mu)\bigg)\bigg\}\cdot
\text{E}\big\{|\boldsymbol{N}_{\!\!f}(\mu)|^2\big\}\;=
\notag\\[4pt]
=\;\text{E}\bigg\{\text{spur}\bigg(\frac{1}{L}\cdot
\Hat{\boldsymbol{C}}_{\boldsymbol{V}(\mu),\boldsymbol{V}(\mu)}^{-1}\bigg)\!
\bigg\}\cdot\text{E}\big\{|\boldsymbol{N}_{\!\!f}(\mu)|^2\big\}\;=\;
\frac{M}{L}\cdot\text{E}\big\{
\Hat{\boldsymbol{C}}_{\boldsymbol{V}(\mu),\boldsymbol{V}(\mu)}^{-1}
\big\}\cdot\Tilde{\Phi}_{\boldsymbol{n}}(\mu)\notag
\end{gather}
Die Varianz der Messwerte der "Ubertragungsfunktion sinkt mit
\mbox{$1/L$}, wenn der Erwartungswert der Inversen der
empirischen Varianz der Spektralwerte der Erregung existiert und
mit steigendem $L$ gegen einen endlichen, konstanten Wert geht.
Ist \mbox{$\boldsymbol{V}(\mu)$} eine kontinuierliche Zufallsgr"o"se,
mit einer endlichen Verteilungsdichtefunktion, so ist anzunehmen,
dass der Erwartungswert der Inversen der empirischen Varianz f"ur
hinreichend gro"ses $L$ existiert. Ist dagegen die Zufallsgr"o"se
diskret mit einem Ergebnisraum, der endlich und abz"ahlbar ist
--- Zufallsgeneratoren endlicher Wortl"ange sind von dieser Art ---,
so kann es vorkommen, dass der Erwartungswert der Inversen der 
empirischen Varianz nicht existiert. Bei wertdiskreten Zufallsgr"o"sen 
\mbox{$\boldsymbol{V}(\mu)$}, bei denen \mbox{$V(\mu)=0$} {\em ein}\/ 
m"ogliches Ereignis ist, treten bei der Berechnung des Erwartungswertes
der Inversen der empirischen Varianz immer Summanden auf, deren Nenner Null 
ist und deren Z"ahler die zwar extrem kleine aber immer positive Wahrscheinlichkeit
ist, dass alle Elemente der Stichprobe von \mbox{$\boldsymbol{V}(\mu)$}
gleich Null sind. Bei der Berechnung der Messwerte ist in diesem Fall
statt der inversen empirischen Varianz eine beliebige Konstante eingesetzt worden.
Daher ist streng genommen die eben berechnete Messwertvarianz mit
der Wahrscheinlichkeit, dass die inverse empirische Varianz
{\em nicht}\/ durch die Konstante zu ersetzen ist, zu multiplizieren, 
und zu dem Produkt der Messwertvarianz im singul"aren Fall und der 
Wahrscheinlichkeit des Auftretens des singul"aren Falls zu addieren. 
Im singul"aren Fall wird der Messwert \mbox{$\Hat{H}(\mu)=0$} verwendet,
dessen Varianz Null ist. Daher tr"agt dieses Produkt nichts zur Messwertvarianz bei.
In der Praxis ist der singul"are Fall so unwahrscheinlich,
dass die komplement"are Wahrscheinlichkeit, dass die inverse empirische Varianz
{\em nicht}\/ durch die Konstante zu ersetzen ist, praktisch Eins ist, so
dass die Messwertvarianz mit der in der letzten Gleichung angegebenen
Formel berechnet werden kann.

Messwerte minimaler Varianz erhalten wir offensichtlich dadurch, dass 
wir einen Zufallsvektor $\Vec{\boldsymbol{V}}$ zur Gewinnung unserer 
Erregung verwenden, der eine m"oglichst gro"se Varianz aufweist. 
Insbesondere sollte die Varianz in den Frequenzbereichen besonders 
gro"s sein, wo auch das LDS $\Phi_{\boldsymbol{n}}(\mu)$ gro"s ist bzw. 
die "Ubertragungsfunktion besonders genau gemessen werden soll. Dies geht 
nat"urlich nur dann, wenn man sicher sein kann, dass sich die theoretisch 
optimalen Regressionsl"osungen durch eine Erh"ohung oder eine Ver"anderung 
der Aufteilung der Varianz auf unterschiedliche Frequenzbereiche 
nicht "andern. 

Da die Messwertvarianz von den theoretischen Werten 
\mbox{$\Tilde{\Phi}_{\boldsymbol{n}}(\mu)$} abh"angt, die sich nach 
Gleichung (\ref{2.17}) bei einem station"aren Approximationsfehlerprozess 
aus der Faltung des Betragsquadrats des Spektrums der Fensterfolge 
mit dem LDS des Approximationsfehlerprozesses ergibt, kann man die 
"Ubertragungsfunktion im Frequenzbereich mit hoher St"orleistungsdichte 
nur mit sehr hohen Mittelungsanzahlen $L$ messen. Verwendet man 
eine nicht besonders frequenzselektive Fensterfolge ---~z.~B. ein 
Rechteckfenster der L"ange $M$~--- so kann eine St"orung, 
die in einem schmalen Frequenzband sehr stark ist, wie z.~B. 
ein periodisches St"orsignal mit zuf"alliger Phase oder ein 
im Vergleich zum gemessenen Frequenzbereich schmalbandiger 
St"orsender, eine Messung mit sinnvoller Mittelungsanzahl im 
gesamten Frequenzbereich unm"oglich machen, da dann die Sperrd"ampfung 
der Fensterfolge nicht ausreicht, um im Faltungsintegral in 
Gleichung (\ref{2.17}) in den anderen Frequenzbereichen den Einfluss 
der St"orung hinreichend zu unterdr"ucken. Ein Beispiel daf"ur 
wird in Kapitel \ref{Mess05} gezeigt. Dies ist ein weiterer 
Grund, eine hoch frequenzselektive Fensterfolge, wie die im Kapitel 
\ref{Algo} vorgestellte, zu verwenden.

Sch"atzwerte f"ur die Varianz der Messwerte der 
"Ubertragungsfunktion erhalten wir mit
\begin{equation}
\Hat{C}_{\Hat{\boldsymbol{H}}(\mu),\Hat{\boldsymbol{H}}(\mu)}\;=\;
\frac{M}{L}\cdot\Hat{C}_{\boldsymbol{V}(\mu),\boldsymbol{V}(\mu)}^{-1}
\Cdot\Hat{\Phi}_{\boldsymbol{n}}(\mu)
\qquad\qquad\forall\qquad \mu=0\;(1)\;M\!-\!1
\label{3.59}
\end{equation}
Diese Sch"atzwerte sind erwartungstreu, wenn man die erwartungstreuen 
Messwerte \mbox{$\Hat{\Phi}_{\boldsymbol{n}}(\mu)$} nach Gleichung 
(\ref{3.34}) oder (\ref{3.56}) verwendet. Man zeigt dies, indem man die 
Messwerte \mbox{$\Hat{\Phi}_{\boldsymbol{n}}(\mu)$} einsetzt, und dann 
die Erwartungswerte der Sch"atzwerte f"ur die \pagebreak[2]Messwertvarianzen 
--- ganz analog zur Berechnung der Erwartungswerte der Messwerte 
\mbox{$\Hat{\Phi}_{\boldsymbol{n}}(\mu)$} in Gleichung (\ref{3.30}) --- 
berechnet. Dabei ist lediglich der in Gleichung (\ref{3.32}) definierte Vorfaktor 
\mbox{$\boldsymbol{c}_{\Phi}(\mu)$} zu modifizieren, indem er noch mit 
\mbox{$M / \big(L\CdoT\Hat{\boldsymbol{C}}_{\boldsymbol{V}(\mu),\boldsymbol{V}(\mu)}\big)$}
zu multiplizieren ist.

F"ur die Kovarianzen der Messwerte der "Ubertragungsfunktion erhalten
wir analog
\begin{gather}
C_{\Hat{\boldsymbol{H}}(\mu),\Hat{\boldsymbol{H}}(\mu)^{\Kk}}\:=\;
\text{E}\bigg\{\Big(\Hat{\boldsymbol{H}}(\mu)\!-\!
\text{E}\big\{\Hat{\boldsymbol{H}}(\mu)\big\}\Big)^{\!2}\bigg\}\;=\;
\text{E}\bigg\{\Big(\Hat{\boldsymbol{H}}(\mu)\!-\!
H\big({\T\mu\CdoT\frac{2\pi}{M}}\big)\Big)^{\!2}\bigg\}\;=
\label{3.60}\\[6pt]
=\;\text{E}\bigg\{\frac{1}{L^2}\cdoT
\Vec{\boldsymbol{N}}_{\!\!f}(\mu)\CdoT
\Vec{\boldsymbol{V}}(\mu)^{\Hh}\!\Cdot
\Hat{\boldsymbol{C}}_{\boldsymbol{V}(\mu),\boldsymbol{V}(\mu)}^{-1}\Cdot
\Hat{\boldsymbol{C}}_{\boldsymbol{V}(\mu),\boldsymbol{V}(\mu)}^{-1}\CdoT
\Vec{\boldsymbol{V}}(\mu)^{\Kk}\!\CdoT
\Vec{\boldsymbol{N}}_{\!\!f}(\mu)^{\Tt}\bigg\}\;=
\notag\\[6pt]
=\;\text{E}\bigg\{\text{spur}\bigg(\frac{1}{L^2}\cdot
\Vec{\boldsymbol{V}}(\mu)^{\Hh}\!\Cdot
\Hat{\boldsymbol{C}}_{\boldsymbol{V}(\mu),\boldsymbol{V}(\mu)}^{-1}\Cdot
\Hat{\boldsymbol{C}}_{\boldsymbol{V}(\mu),\boldsymbol{V}(\mu)}^{-1}\CdoT
\Vec{\boldsymbol{V}}(\mu)^{\Kk}\bigg)\bigg\}\cdot
\text{E}\big\{\boldsymbol{N}_{\!\!f}(\mu)^2\big\}\;=
\notag\\[6pt]
=\;\text{E}\bigg\{\text{spur}\bigg(\frac{1}{L}\cdot
\Hat{\boldsymbol{C}}_{\boldsymbol{V}(\mu),\boldsymbol{V}(\mu)}^{-2}\Cdot
\Hat{\boldsymbol{C}}_{\boldsymbol{V}(\mu),\boldsymbol{V}(\mu)^{\Kk}}^*\!\bigg)
\,\bigg\}\cdot\text{E}\big\{\boldsymbol{N}_{\!\!f}(\mu)^2\big\}
\notag
\end{gather}
Da vorausgesetzt wurde, dass der Approximationsfehlerprozess station"ar ist, 
berechnet sich der Erwartungswert des Quadrats des Spektrums des gefensterten 
Approximationsfehlers analog zu Gleichung (\ref{2.32}):
\begin{equation}
\text{E}\big\{\boldsymbol{N}_{\!\!f}(\mu)^2\big\}\;=\;
\frac{1}{2\pi}\cdoT\Int{-\pi}{\pi}\Psi_{\boldsymbol{n}}(\Omega)\CdoT
F\big({\T\mu\CdoT\frac{2\pi}{M}\!-\!\Omega}\big)\CdoT
F\big({\T\mu\CdoT\frac{2\pi}{M}\!+\!\Omega}\big)\cdot\,d\Omega.
\label{3.61}
\end{equation}
Wenn wir eine Fensterfolge verwenden, deren Betragsquadratspektrum den im 
letzten Kapitel hergeleiteten Wunschverlauf einer mit $2\pi$ periodischen 
Rechteckfunktion gut ann"ahert, die im Intervall \mbox{$(-\pi;\pi]$} f"ur 
\mbox{$|\Omega|<\pi/M$} den Wert $M^2$ annimmt, und sonst Null ist, sind 
nur die Werte von \mbox{$\text{E}\big\{\boldsymbol{N}_{\!\!f}(\mu)^2\big\}$} 
f"ur \mbox{$\mu=0$} und --- falls M gerade ist --- \mbox{$\mu=M/2$} 
nennenswert von Null verschieden. F"ur alle anderen Werte von $\mu$
sind die Spektren der Fensterfolge im Integranden in der letzten
Gleichung in der Art gegeneinander verschoben, dass die Frequenz
$\Omega$ wenigstens bei einer der beiden Fensterspektren immer im
Sperrbereich \mbox{( $2\pi/M\le|\Omega|\le\pi$ )} liegt.
Es gilt daher:
\begin{equation}
\text{E}\big\{\boldsymbol{N}_{\!\!f}(\mu)^2\big\}\;=\;
\begin{cases}
\;\text{E}\big\{\boldsymbol{N}_{\!\!f}(\mu)\CdoT
\boldsymbol{N}_{\!\!f}(\!-\mu)\big\}=M\CdoT\Tilde{\Psi}_{\boldsymbol{n}}(\mu)
\qquad&
\text{ f"ur }\quad\mu=0\quad\vee\quad\mu=\frac{M}{2}\\
\;{}\approx\;0&\text{ sonst, }
\end{cases}
\label{3.62}
\end{equation}
so dass wir f"ur die Messwertkovarianzen der "Ubertragungsfunktion 
\begin{equation}
C_{\Hat{\boldsymbol{H}}(\mu),\Hat{\boldsymbol{H}}(\mu)^{\Kk}}\:=\;
=\;\begin{cases}
{\D\;\frac{M}{L}\cdot\text{E}\bigg\{\!\frac{\,
\Hat{\boldsymbol{C}}_{\boldsymbol{V}(\mu),\boldsymbol{V}(-\mu)^{\Kk}}^*}
{\Hat{\boldsymbol{C}}_{\boldsymbol{V}(\mu),\boldsymbol{V}(\mu)}^2}\!\bigg\}\cdot
\Tilde{\Psi}_{\boldsymbol{n}}(\mu)\qquad{}}
&\text{ f"ur }\quad\mu=0\quad\vee\quad\mu=\frac{M}{2}\\
{\D\;{}\approx\;0}&\text{ sonst.}
\end{cases}
\raisetag{12pt}\label{3.63}
\end{equation}
erhalten. Die Erwartungstreue der \pagebreak[2]Sch"atzwerte
\begin{equation}
\Hat{C}_{\Hat{\boldsymbol{H}}(\mu),\Hat{\boldsymbol{H}}(\mu)^{\Kk}}=\,
\frac{M}{L}\Cdot
\Hat{C}_{\boldsymbol{V}(\mu),\boldsymbol{V}(\mu)}^{-2}\Cdot
\Hat{C}_{\boldsymbol{V}(\mu),\boldsymbol{V}(-\mu)^{\Kk}}^*\Cdot
\Hat{\Psi}_{\boldsymbol{n}}(\mu)
\quad\text{ f"ur }{\T\;\;\mu\!=\!0\!\;\;\vee\!\;\;\mu\!=\!\frac{M}{2}}
\label{3.64}
\end{equation}
f"ur die Kovarianz der Messwerte der "Ubertragungsfunktion, ist bei Verwendung
der erwartungstreuen Messwerte \mbox{$\Hat{\Psi}_{\boldsymbol{n}}(\mu)$}
nach Gleichung (\ref{3.35}) oder (\ref{3.57}) gegeben. Da die Messwerte 
\mbox{$\Hat{\Phi}_{\boldsymbol{n}}(\mu)$} und 
\mbox{$\Hat{\Psi}_{\boldsymbol{n}}(\mu)$} der Gleichung (\ref{3.38}) 
gen"ugen, l"asst sich zeigen, dass die Sch"atzwerte
\mbox{$\Hat{C}_{\Hat{\boldsymbol{H}}(\mu),\Hat{\boldsymbol{H}}(\mu)}$}
gr"o"ser oder gleich dem Betrag der Sch"atzwerte
\mbox{$\Hat{C}_{\Hat{\boldsymbol{H}}(\mu),\Hat{\boldsymbol{H}}(\mu)^{\Kk}}$}
sind. Dabei ist zu ber"ucksichtigen, dass immer
\mbox{$\Hat{C}_{\boldsymbol{V}(\mu),\boldsymbol{V}(\mu)}\CdoT
\Hat{C}_{\boldsymbol{V}(-\mu),\boldsymbol{V}(-\mu)}\ge
\big|\Hat{C}_{\boldsymbol{V}(\mu),\boldsymbol{V}(-\mu)^{\Kk}}\big|^2$} gilt.

Bei ungeradem $M$ kann die N"aherung (\ref{3.62}) f"ur die beiden zu
$\pi$ benachbarten Frequenzpunkte nicht so gut erf"ullt werden,
da dann im Integranden in Gleichung (\ref{3.61}) die Frequenz $\Omega$ 
auch bei beiden Fensterspektren in den "Ubergangsbereich
\mbox{( $\pi/M\le|\Omega|\le 2\pi/M$ )} der D"ampfung
fallen kann. Wenn es sich nicht vermeiden l"asst, ein ungerades $M$
zu verwenden, w"are dies zu ber"ucksichtigen indem man f"ur
diese beiden Frequenzen die Sch"atzwerte f"ur
\mbox{$\text{E}\big\{\boldsymbol{N}_{\!\!f}(\mu)^2\big\}$} --- analog zu
\mbox{$\Hat{\Phi}_{\boldsymbol{n}}(\mu)$} mit dem Quadrat statt dem
Betragsquadrat --- aus \mbox{$\Hat{\Vec{N}}_{\!f}(\mu)$} berechnet.
Im weiteren wird angenommen, dass $M$ gerade ist.\vspace*{0pt minus 10pt}

\section{Varianzen und Kovarianzen der Messwerte des LDS und des MLDS}\label{KovaLDS}

Die Varianzen und Kovarianzen der Messwerte
\mbox{$\Hat{\boldsymbol{\Phi}}_{\!\boldsymbol{n}}(\mu)$} und
\mbox{$\Hat{\boldsymbol{\Psi}}_{\!\boldsymbol{n}}(\mu)$} sind 
die Erwartungswerte der Betragsquadrate bzw. Quadrate der 
Messwertabweichungen. Mit den Messwerten nach Gleichung 
(\ref{3.34}) und (\ref{3.35}) und unter Beachtung der 
Erwartungstreue der Messwerte, die in den Gleichungen 
(\ref{3.30}) bis (\ref{3.33}) hergeleitet wurde, erhalten 
wir f"ur die Varianzen und Kovarianzen der Messwerte zun"achst
\begin{subequations}\label{3.65}
\begin{gather}
C_{\Hat{\boldsymbol{\Phi}}_{\!\boldsymbol{n}}(\mu),\Hat{\boldsymbol{\Phi}}_{\!\boldsymbol{n}}(\mu)}\,=\,
\text{E}\Big\{\big|\Hat{\boldsymbol{\Phi}}_{\!\boldsymbol{n}}(\mu)\!-\!
\text{E}\big\{\Hat{\boldsymbol{\Phi}}_{\!\boldsymbol{n}}(\mu)\big\}\big|^2\Big\}\,=\,
\text{E}\Big\{\big|\Hat{\boldsymbol{\Phi}}_{\!\boldsymbol{n}}(\mu)\big|^2\Big\}-
\Big|\text{E}\big\{\Hat{\boldsymbol{\Phi}}_{\!\boldsymbol{n}}(\mu)\big\}\Big|^2=
\label{3.65.a}\\[1pt]
=\;\text{E}\bigg\{\,\bigg|\,\Vec{\boldsymbol{N}}_{\!\!f}(\mu)\cdot
\frac{\D\underline{\boldsymbol{V}}_{\bot}\!(\mu)\CdoT\underline{\boldsymbol{V}}_{\bot}\!(\mu)^{\Hh}}
{\D \,M\CdoT\text{spur}\big(
\underline{\boldsymbol{V}}_{\bot}\!(\mu)\CdoT\underline{\boldsymbol{V}}_{\bot}\!(\mu)^{\Hh}\big)}\cdot
\Vec{\boldsymbol{N}}_{\!\!f}(\mu)^{\Hh}\bigg|^2\,\bigg\}\,-\,
\bigg|\,\frac{\D\;\text{E}\big\{|\boldsymbol{N}_{\!\!f}(\mu)|^2\big\}\,}{\D M}\,\bigg|^2,
\notag\\[7pt]
\!C_{\Hat{\boldsymbol{\Psi}}_{\!\boldsymbol{n}}(\mu),\Hat{\boldsymbol{\Psi}}_{\!\boldsymbol{n}}(\mu)}\,=\,
\text{E}\Big\{\big|\Hat{\boldsymbol{\Psi}}_{\!\boldsymbol{n}}(\mu)-
\text{E}\big\{\Hat{\boldsymbol{\Psi}}_{\!\boldsymbol{n}}(\mu)\big\}\big|^2\Big\}\,=\,
\text{E}\Big\{\big|\Hat{\boldsymbol{\Psi}}_{\!\boldsymbol{n}}(\mu)\big|^2\Big\}-
\Big|\text{E}\big\{\Hat{\boldsymbol{\Psi}}_{\!\boldsymbol{n}}(\mu)\big\}\Big|^2=\!
\label{3.65.b}\\[1pt]
=\;\text{E}\bigg\{\,\bigg|\,\Vec{\boldsymbol{N}}_{\!\!f}(\mu)\cdot
\frac{\D\underline{\boldsymbol{V}}_{\bot}\!(\mu)\CdoT\underline{\boldsymbol{V}}_{\bot}\!(\!-\mu)^{\Tt}}
{\D \,M\CdoT\text{spur}\big(
\underline{\boldsymbol{V}}_{\bot}\!(\mu)\CdoT\underline{\boldsymbol{V}}_{\bot}\!(\!-\mu)^{\Tt}\big)}\cdot
\Vec{\boldsymbol{N}}_{\!\!f}(\!-\mu)^{\Tt}\bigg|^2\bigg\}-\,
\bigg|\frac{\D\;
\text{E}\big\{\boldsymbol{N}_{\!\!f}(\mu)\CdoT
\boldsymbol{N}_{\!\!f}(\!-\mu)\big\}\,}{\D M}\bigg|^2\;\text{ und}
\notag\\[7pt]
C_{\Hat{\boldsymbol{\Psi}}_{\!\boldsymbol{n}}(\mu),\Hat{\boldsymbol{\Psi}}_{\!\boldsymbol{n}}(\mu)^{\Kk}}\;=\;
\text{E}\Big\{\big(\Hat{\boldsymbol{\Psi}}_{\!\boldsymbol{n}}(\mu)-
\text{E}\big\{\Hat{\boldsymbol{\Psi}}_{\!\boldsymbol{n}}(\mu)\big\}\big)^{\!2}\Big\}\;=\;
\text{E}\Big\{\Hat{\boldsymbol{\Psi}}_{\!\boldsymbol{n}}(\mu)^2\Big\}-
\text{E}\big\{\Hat{\boldsymbol{\Psi}}_{\!\boldsymbol{n}}(\mu)\big\}^{\!2}\;=
\label{3.65.c}\\[1pt]
=\;\text{E}\bigg\{\bigg(\,\Vec{\boldsymbol{N}}_{\!\!f}(\mu)\cdot
\frac{\D\underline{\boldsymbol{V}}_{\bot}\!(\mu)\CdoT\underline{\boldsymbol{V}}_{\bot}\!(\!-\mu)^{\Tt}}
{\D \,M\CdoT\text{spur}\big(
\underline{\boldsymbol{V}}_{\bot}\!(\mu)\CdoT\underline{\boldsymbol{V}}_{\bot}\!(\!-\mu)^{\Tt}\big)}\cdot
\Vec{\boldsymbol{N}}_{\!\!f}(\!-\mu)^{\Tt}\bigg)^{\!\!2}\,\bigg\}\,-\,
\bigg(\,\frac{\D\;
\text{E}\big\{\boldsymbol{N}_{\!\!f}(\mu)\CdoT
\boldsymbol{N}_{\!\!f}(\!-\mu)\big\}\,}{\D M}\,\bigg)^{\!\!2}.
\notag
\end{gather}
\end{subequations}
\begin{table}[b!]
\rule{\textwidth}{0.5pt}\vspace{0pt}
\[
\begin{array}{l}
\begin{array}{||c||c|c|c|c||}
\hline
\hline
\rule[-7pt]{0pt}{25pt}\text{(Ko)varianz}&
\Vec{\boldsymbol{N}}_{\!1}=&
\Vec{\boldsymbol{N}}_{\!2}=&
\Vec{\boldsymbol{N}}_{\!3}=&
\Vec{\boldsymbol{N}}_{\!4}=\\
\hline
\rule{0pt}{16pt}\!\!C_{\Hat{\boldsymbol{\Phi}}_{\!\boldsymbol{n}}(\mu),\Hat{\boldsymbol{\Phi}}_{\!\boldsymbol{n}}(\mu)}\!\!&
\Vec{\boldsymbol{N}}_{\!\!f}(\mu)&
\Vec{\boldsymbol{N}}_{\!\!f}(\mu)&
\Vec{\boldsymbol{N}}_{\!\!f}(\mu)^{\Kk}&
\Vec{\boldsymbol{N}}_{\!\!f}(\mu)^{\Kk}\\
\!\!C_{\Hat{\boldsymbol{\Psi}}_{\!\boldsymbol{n}}(\mu),\Hat{\boldsymbol{\Psi}}_{\!\boldsymbol{n}}(\mu)}\!\!&
\Vec{\boldsymbol{N}}_{\!\!f}(\mu)&
\Vec{\boldsymbol{N}}_{\!\!f}(\!-\mu)^{\Kk}&
\Vec{\boldsymbol{N}}_{\!\!f}(\mu)^{\Kk}&
\Vec{\boldsymbol{N}}_{\!\!f}(\!-\mu)\\
\rule[-9pt]{0pt}{17pt}\!C_{\Hat{\boldsymbol{\Psi}}_{\!\boldsymbol{n}}(\mu),\Hat{\boldsymbol{\Psi}}_{\!\boldsymbol{n}}(\mu)^{\Kk}}\!&
\Vec{\boldsymbol{N}}_{\!\!f}(\mu)&
\Vec{\boldsymbol{N}}_{\!\!f}(\!-\mu)^{\Kk}&
\Vec{\boldsymbol{N}}_{\!\!f}(\mu)&
\Vec{\boldsymbol{N}}_{\!\!f}(\!-\mu)^{\Kk}\\
\hline
\end{array}\\
\begin{array}{||c||c||c|c||}
\hline
\hline
\rule[-7pt]{0pt}{23pt}\text{(Ko)varianz}&
\boldsymbol{c}=&
\underline{\boldsymbol{A}}=&
\underline{\boldsymbol{B}}=\\
\hline
\rule{0pt}{16pt}\!\!C_{\Hat{\boldsymbol{\Phi}}_{\!\boldsymbol{n}}(\mu),\Hat{\boldsymbol{\Phi}}_{\!\boldsymbol{n}}(\mu)}\!\!&
\big|\,M\cdot\text{spur}\big(
\underline{\boldsymbol{V}}_{\bot}\!(\mu)\CdoT
\underline{\boldsymbol{V}}_{\bot}\!(\mu)^{\Hh}\big)\,\big|^{-2}&
\underline{\boldsymbol{V}}_{\bot}\!(\mu)\CdoT
\underline{\boldsymbol{V}}_{\bot}\!(\mu)^{\Hh}&
\underline{\boldsymbol{V}}_{\bot}\!(\mu)^{\Kk}\!\CdoT
\underline{\boldsymbol{V}}_{\bot}\!(\mu)^{\Tt}\\
\rule{0pt}{15pt}\!\!C_{\Hat{\boldsymbol{\Psi}}_{\!\boldsymbol{n}}(\mu),\Hat{\boldsymbol{\Psi}}_{\!\boldsymbol{n}}(\mu)}\!\!&\!
\big|M\CdoT\text{spur}\big(
\underline{\boldsymbol{V}}_{\bot}\!(\mu)\CdoT
\underline{\boldsymbol{V}}_{\bot}\!(\!-\mu)^{\Tt}\big)\big|^{-2}\!&\!
\underline{\boldsymbol{V}}_{\bot}\!(\mu)\CdoT
\underline{\boldsymbol{V}}_{\bot}\!(\!-\mu)^{\Tt}\!&\!
\underline{\boldsymbol{V}}_{\bot}\!(\mu)^{\Kk}\!\CdoT
\underline{\boldsymbol{V}}_{\bot}\!(\!-\mu)^{\Hh}\!\\
\rule[-9pt]{0pt}{25pt}\!C_{\Hat{\boldsymbol{\Psi}}_{\!\boldsymbol{n}}(\mu),\Hat{\boldsymbol{\Psi}}_{\!\boldsymbol{n}}(\mu)^{\Kk}}\!&\!
\big(M\CdoT\text{spur}\big(
\underline{\boldsymbol{V}}_{\bot}\!(\mu)\CdoT
\underline{\boldsymbol{V}}_{\bot}\!(\!-\mu)^{\Tt}\big)\big)^{\!-2}\!&
\underline{\boldsymbol{V}}_{\bot}\!(\mu)\CdoT
\underline{\boldsymbol{V}}_{\bot}\!(\!-\mu)^{\Tt}\!&
\underline{\boldsymbol{V}}_{\bot}\!(\mu)\CdoT
\underline{\boldsymbol{V}}_{\bot}\!(\!-\mu)^{\Tt}\!\\
\hline
\hline
\end{array}
\end{array}
\]\setlength{\abovecaptionskip}{-9pt}
\caption{Substitutionen in Gleichung (\ref{A.5.1}) bei der Berechnung der
Varianzen und Kovarianzen der Messwerte des LDS und MLDS}
\label{T3.1}
\end{table}
Da sowohl die Messwerte
\mbox{$\Hat{\boldsymbol{\Phi}}_{\!\boldsymbol{n}}(\mu)$} als auch deren Erwartungswerte 
\mbox{$\text{E}\big\{\Hat{\boldsymbol{\Phi}}_{\!\boldsymbol{n}}(\mu)\big\}$}
immer reell sind, brauchen die Kovarianzen
\mbox{$C_{\Hat{\boldsymbol{\Phi}}(\mu),\Hat{\boldsymbol{\Phi}}(\mu)^{\Kk}}$} 
nicht berechnet zu werden, weil diese immer gleich den Varianzen
\mbox{$C_{\Hat{\boldsymbol{\Phi}}(\mu),\Hat{\boldsymbol{\Phi}}(\mu)}$} 
sind. Wie man sieht ist zu weiteren Berechnung der Messwertvarianzen 
und Kovarianzen jeweils der Erwartungswert eines Quadrats
oder Betragsquadrats einer quadratischen oder bilinearen Form
zu berechnen. Im Anhang \ref{4Mom} wird gezeigt, wie sich so
ein Erwartungswert berechnen l"asst. Dabei wird vorausgesetzt,
dass das Zufallsgr"o"sentupel der Spektralwerte
\mbox{$\big[\boldsymbol{N}_{\!\!f}(\mu),\boldsymbol{N}_{\!\!f}(\!-\mu)^{\Kk}\big]^{\Tt}$} 
der gefensterten St"orung des realen Systems unabh"angig von dem Zufallsvektor ist, 
der alle Zufallsgr"o"sen \mbox{$\boldsymbol{V}(\Tilde{\mu})$} der Spektralwerte
der Erregung bei den Frequenzen $\Tilde{\mu}$ enth"alt,
deren Stichprobenelemente in die Elemente der Matrizen
\mbox{$\underline{\boldsymbol{V}}_{\bot}\!(\pm\mu)$} eingehen.
Desweiteren sind die Herleitungen im Anhang nur g"ultig, wenn man von 
den Zufallsgr"o"sentupeln der Spektralwerte
\mbox{$\big[\boldsymbol{N}_{\!\!f}(\mu),\boldsymbol{N}_{\!\!f}(\!-\mu)\big]^{\TT}$}
annehmen kann, dass sie verbundnormalverteilt sind. Anderenfalls sind die
im weiteren hergeleiteten Ergebnisse h"ochstens als eine Absch"atzung der
Messwertvarianzen und Kovarianzen zu betrachten. 
Wenn wir in den Gleichungen (\ref{A.5.1}) bis (\ref{A.5.9}) des Anhangs die 
in Tabelle \ref{T3.1} aufgelisteten Substitutionen vornehmen, k"onnen wir mit 
Gleichung (\ref{A.5.9}) die Erwartungswerte der Quadrate bzw. Betragsquadrate 
der quadratischen und bilinearen Formen in der letzten Gleichung ersetzen, 
so dass sich f"ur die Messwertvarianzen und Kovarianzen folgende Ausdr"ucke ergeben:
\begin{subequations}\label{3.66}
\begin{align}
C_{\Hat{\boldsymbol{\Phi}}_{\!\boldsymbol{n}}(\mu),\Hat{\boldsymbol{\Phi}}_{\!\boldsymbol{n}}(\mu)}&
{}=\text{E}\bigg\{\!\frac{\text{spur}\big(
\underline{\boldsymbol{V}}_{\bot}\!(\mu)\CdoT
\underline{\boldsymbol{V}}_{\bot}\!(\mu)^{\Hh}\!\CdoT
\underline{\boldsymbol{V}}_{\bot}\!(\mu)^{\Kk}\!\CdoT
\underline{\boldsymbol{V}}_{\bot}\!(\mu)^{\Tt}\big)}
{\big|\text{spur}\big(
\underline{\boldsymbol{V}}_{\bot}\!(\mu)\CdoT
\underline{\boldsymbol{V}}_{\bot}\!(\mu)^{\Hh}\big)\,\big|^2}\bigg\}\cdot
\bigg|\,\frac{\,\text{E}\big\{\boldsymbol{N}_{\!\!f}(\mu)^2\big\}}{M}\,\bigg|^2+{}
\raisetag{44pt}\label{3.66.a}\\*[2pt]
&{}+\text{E}\bigg\{\!\frac{\text{spur}\big(
\underline{\boldsymbol{V}}_{\bot}\!(\mu)^{\Kk}\!\CdoT
\underline{\boldsymbol{V}}_{\bot}\!(\mu)^{\Tt}\!\CdoT
\underline{\boldsymbol{V}}_{\bot}\!(\mu)^{\Kk}\!\CdoT
\underline{\boldsymbol{V}}_{\bot}\!(\mu)^{\Tt}\big)}
{\big|\text{spur}\big(
\underline{\boldsymbol{V}}_{\bot}\!(\mu)\CdoT
\underline{\boldsymbol{V}}_{\bot}\!(\mu)^{\Hh}\big)\,\big|^2}\bigg\}\cdot
\bigg(\!\frac{\;\text{E}\big\{|\boldsymbol{N}_{\!\!f}(\mu)|^2\big\}\,}{M}\!\bigg)^{\!\!2}\!,
\rule[-10pt]{0pt}{10pt}\notag\displaybreak\\
\rule{0pt}{42pt}C_{\Hat{\boldsymbol{\Psi}}_{\!\boldsymbol{n}}(\mu),\Hat{\boldsymbol{\Psi}}_{\!\boldsymbol{n}}(\mu)}&{}=
\text{E}\bigg\{\!\frac{\text{spur}\big(
\underline{\boldsymbol{V}}_{\bot}\!(\mu)\CdoT
\underline{\boldsymbol{V}}_{\bot}\!(\!-\mu)^{\Tt}\!\CdoT
\underline{\boldsymbol{V}}_{\bot}\!(\mu)^{\Kk}\!\CdoT
\underline{\boldsymbol{V}}_{\bot}\!(\!-\mu)^{\Hh}\big)}
{\big|\text{spur}\big(\underline{\boldsymbol{V}}_{\bot}\!(\mu)\CdoT
\underline{\boldsymbol{V}}_{\bot}\!(\!-\mu)^{\Tt}\big)\big|^2}\!\bigg\}\CdoT
\bigg|\frac{\text{E}\big\{\boldsymbol{N}_{\!\!f}(\mu)\CdoT
\boldsymbol{N}_{\!\!f}(\!-\mu)^{\Kk}\big\}}{M}\bigg|^2+{}
\raisetag{60pt}\label{3.66.b}\\*[4pt]
&{}\!\!\!\!\!\!\!\!+\text{E}\bigg\{\frac{\text{spur}\big(
\underline{\boldsymbol{V}}_{\bot}\!(\!-\mu)\CdoT
\underline{\boldsymbol{V}}_{\bot}\!(\mu)^{\Tt}\!\CdoT
\underline{\boldsymbol{V}}_{\bot}\!(\mu)^{\Kk}\!\CdoT
\underline{\boldsymbol{V}}_{\bot}\!(\!-\mu)^{\Hh}\big)}
{\big|\text{spur}\big(
\underline{\boldsymbol{V}}_{\bot}\!(\mu)\CdoT
\underline{\boldsymbol{V}}_{\bot}\!(\!-\mu)^{\Tt}\big)\big|^2}\bigg\}\CdoT
\frac{\text{E}\big\{|\boldsymbol{N}_{\!\!f}(\mu)|^2\big\}}{M}\CdoT
\frac{\text{E}\big\{|\boldsymbol{N}_{\!\!f}(\!-\mu)|^2\big\}}{M}
\notag\\[-8pt]
\intertext{und}
C_{\Hat{\boldsymbol{\Psi}}_{\!\boldsymbol{n}}(\mu),\Hat{\boldsymbol{\Psi}}_{\!\boldsymbol{n}}(\mu)^{\Kk}}\!&{}=
\text{E}\bigg\{\!\frac{\text{spur}\big(
\underline{\boldsymbol{V}}_{\bot}\!(\mu)\CdoT
\underline{\boldsymbol{V}}_{\bot}\!(\!-\mu)^{\Tt}\!\CdoT
\underline{\boldsymbol{V}}_{\bot}\!(\mu)\CdoT
\underline{\boldsymbol{V}}_{\bot}\!(\!-\mu)^{\Tt}\big)}
{\text{spur}\big(\underline{\boldsymbol{V}}_{\bot}\!(\mu)\CdoT
\underline{\boldsymbol{V}}_{\bot}\!(\!-\mu)^{\Tt}\big)^{\!2}}\!\bigg\}\CdoT
\bigg(\!\frac{\,\text{E}\big\{\boldsymbol{N}_{\!\!f}(\mu)\CdoT
\boldsymbol{N}_{\!\!f}(\!-\mu)\big\}}{M}\!\bigg)^{\!\!\!2}+{}
\raisetag{60pt}\label{3.66.c}\\*[4pt]
&{}+\text{E}\bigg\{\!\frac{\text{spur}\big(
\underline{\boldsymbol{V}}_{\bot}\!(\!-\mu)\CdoT
\underline{\boldsymbol{V}}_{\bot}\!(\mu)^{\Tt}\!\CdoT
\underline{\boldsymbol{V}}_{\bot}\!(\mu)\CdoT
\underline{\boldsymbol{V}}_{\bot}\!(\!-\mu)^{\Tt}\big)}
{\text{spur}\big(\underline{\boldsymbol{V}}_{\bot}\!(\mu)\CdoT
\underline{\boldsymbol{V}}_{\bot}\!(\!-\mu)^{\Tt}\big)^{\!2}}\!\bigg\}\CdoT
\frac{\text{E}\big\{\boldsymbol{N}_{\!\!f}(\mu)^2\big\}\!}{M}\CdoT
\frac{\text{E}\big\{\boldsymbol{N}_{\!\!f}(\!-\mu)^2\big\}\!}{M}.\notag
\end{align}
\end{subequations}
In diesen Gleichungen treten neben den beiden Erwartungswerten 
\mbox{$\text{E}\big\{|\boldsymbol{N}_{\!\!f}(\mu)|^2\big\}/M$} und 
\mbox{$\text{E}\big\{\boldsymbol{N}_{\!\!f}(\mu)\CdoT\boldsymbol{N}_{\!\!f}(\!-\mu)\big\}/M$},
die nach Gleichung (\ref{2.17}) und (\ref{2.32}) gleich den theoretischen Werten 
des LDS \mbox{$\Tilde{\Phi}_{\boldsymbol{n}}(\mu)$} und des MLDS
\mbox{$\Tilde{\Psi}_{\boldsymbol{n}}(\mu)$} sind, auch noch die drei Erwartungswerte 
\mbox{$\text{E}\big\{\boldsymbol{N}_{\!\!f}(\mu)^2\big\}/M$}, 
\mbox{$\text{E}\big\{\boldsymbol{N}_{\!\!f}(\!-\mu)^2\big\}/M$} und 
\mbox{$\text{E}\big\{\boldsymbol{N}_{\!\!f}(\mu)\CdoT\boldsymbol{N}_{\!\!f}(\!-\mu)^{\Kk}\big\}/M$} 
auf. Die ersten beiden dieser drei Erwartungswerte berechnen sich nach 
Gleichung (\ref{3.61}) und k"onnen nach Gleichung (\ref{3.62}) bei einem 
station"aren Approximationsfehlerprozess in guter N"aherung f"ur
\mbox{$\mu=1\;(1)\;M\!-\!1\;\wedge\;\mu\!\neq\!M/2$}
vernachl"assigt werden, wenn man eine hoch frequenzselektive
Fensterfolge verwendet, deren Betragsquadratspektrum den im letzten Kapitel
hergeleiteten Wunschverlauf einer mit $2\pi$ periodischen Rechteckfunktion
gut ann"ahert, die im Intervall \mbox{$(-\pi;\pi]$} f"ur \mbox{$|\Omega|<\pi/M$} 
den Wert $M^2$ annimmt, und sonst Null ist. F"ur den Erwartungswert 
\mbox{$\text{E}\big\{\boldsymbol{N}_{\!\!f}(\mu)\CdoT\boldsymbol{N}_{\!\!f}(\!-\mu)^{\Kk}\big\}/M$} 
kann man mit der allgemeineren Beziehung 
\begin{equation}
\text{E}\big\{\boldsymbol{N}_{\!\!f}(\mu_1)\CdoT
\boldsymbol{N}_{\!\!f}(\mu_2)^{\Kk}\big\}\;=\;
\frac{1}{2\pi}\cdoT\Int{-\pi}{\pi}\Phi_{\boldsymbol{n}}(\Omega)\CdoT
F\big({\T\mu_1\CdoT\frac{2\pi}{M}\!-\!\Omega}\big)\CdoT
F\big({\T\mu_2\CdoT\frac{2\pi}{M}\!-\!\Omega}\big)^{\!*}\Cdot d\Omega\quad
\label{3.67}
\end{equation}
zeigen, dass dieser f"ur die hier auftretenden diskreten Frequenzen
\mbox{$\mu_1\!=\!-\mu_2\!=\!\mu$} unter denselben Voraussetzungen 
ebenfalls vernachl"assigbar klein wird: 
\begin{equation}
\text{E}\big\{\boldsymbol{N}_{\!\!f}(\mu)\CdoT\boldsymbol{N}_{\!\!f}(\!-\mu)^{\Kk}\big\}\;=\;
\begin{cases}
\;\text{E}\big\{|\boldsymbol{N}_{\!\!f}(\mu)|^2\big\}=M\CdoT\Tilde{\Phi}_{\boldsymbol{n}}(\mu)
\qquad&
\text{ f"ur }\quad\mu=0\quad\vee\quad\mu=\frac{M}{2}\\
\;{}\approx\;0&\text{ sonst. }
\end{cases}
\raisetag{12pt}\label{3.68}
\end{equation}
Nun wollen wir uns auf den Fall beschr"anken, dass wir die
idempotenten, hermiteschen Matrizen
\mbox{$\underline{\boldsymbol{V}}_{\bot}\!(\pm\mu)$} 
verwenden, deren Konstruktion in den Gleichungen (\ref{3.40})
bis (\ref{3.45}) beschrieben ist. Da diese Matrizen au"serdem 
noch die Gleichung (\ref{3.39}) erf"ullen, tritt nun in den 
Gleichungen (\ref{3.66}) f"ur die Messwert(ko)varianzen 
statt der Spuren der unterschiedlichen Matrixprodukte nur mehr die 
Spur der Matrix \mbox{$\underline{\boldsymbol{V}}_{\bot}\!(\mu)$} 
auf, die --- wie wir oben gesehen haben --- immer \mbox{$L\!-\!2$} ist.
Somit erhalten wir f"ur die theoretischen Messwert(ko)varianzen:
\begin{subequations}\label{3.69}
\begin{align}
C_{\Hat{\boldsymbol{\Phi}}_{\!\boldsymbol{n}}(\mu),\Hat{\boldsymbol{\Phi}}_{\!\boldsymbol{n}}(\mu)}&{}\,=\,
\frac{1}{L\!-\!2}\cdot
\Big(\big|\Tilde{\Psi}_{\boldsymbol{n}}(\mu)\big|^2\!+
\Tilde{\Phi}_{\boldsymbol{n}}(\mu)^{\uP{0.4}{\!2}}\Big),
\label{3.69.a}\\[6pt]
C_{\Hat{\boldsymbol{\Psi}}_{\!\boldsymbol{n}}(\mu),\Hat{\boldsymbol{\Psi}}_{\!\boldsymbol{n}}(\mu)}&{}\,=\,
\frac{2}{L\!-\!2}\cdot
\Tilde{\Phi}_{\boldsymbol{n}}(\mu)^{\uP{0.4}{\!2}}\qquad\text{ und}
\label{3.69.b}\\[6pt]
C_{\Hat{\boldsymbol{\Psi}}_{\!\boldsymbol{n}}(\mu),\Hat{\boldsymbol{\Psi}}_{\!\boldsymbol{n}}(\mu)^{\Kk}}\!&{}\,=\,
\frac{2}{L\!-\!2}\cdot
\Tilde{\Psi}_{\boldsymbol{n}}(\mu)^{\uP{0.4}{\!2}}
\label{3.69.c}\\*[3pt]
&{}\;\;\,\forall{\T\qquad\qquad\mu=0\quad\vee\quad\mu=\frac{M}{2}}
\notag\\[-14pt]\intertext{bzw.\vspace{-2pt}}
C_{\Hat{\boldsymbol{\Phi}}_{\!\boldsymbol{n}}(\mu),\Hat{\boldsymbol{\Phi}}_{\!\boldsymbol{n}}(\mu)}&{}\,=\,
\frac{1}{L\!-\!2}\cdot
\Tilde{\Phi}_{\boldsymbol{n}}(\mu)^{\uP{0.4}{\!2}}\!,
\label{3.69.d}\\[6pt]
C_{\Hat{\boldsymbol{\Psi}}_{\!\boldsymbol{n}}(\mu),\Hat{\boldsymbol{\Psi}}_{\!\boldsymbol{n}}(\mu)}&{}\,=\,
\frac{1}{L\!-\!2}\cdot
\Tilde{\Phi}_{\boldsymbol{n}}(\mu)\cdot
\Tilde{\Phi}_{\boldsymbol{n}}(\!-\mu)\qquad\text{ und}
\label{3.69.e}\\[6pt]
C_{\Hat{\boldsymbol{\Psi}}_{\!\boldsymbol{n}}(\mu),\Hat{\boldsymbol{\Psi}}_{\!\boldsymbol{n}}(\mu)^{\Kk}}\!&{}\,=\,
\frac{1}{L\!-\!2}\cdot
\Tilde{\Psi}_{\boldsymbol{n}}(\mu)^{\uP{0.4}{\!2}}
\label{3.69.f}\\*[3pt]
&{}\;\;\,\forall{\T\qquad\qquad\mu=1\;(1)\;M\!-\!1
\quad\wedge\quad\mu\neq\frac{M}{2}}.
\notag
\end{align}
\end{subequations}
Als Sch"atzwerte f"ur die Messwert(ko)varianzen verwenden wir:
\begin{subequations}\label{3.70}
\begin{align}
\Hat{C}_{\Hat{\boldsymbol{\Phi}}_{\!\boldsymbol{n}}(\mu),\Hat{\boldsymbol{\Phi}}_{\!\boldsymbol{n}}(\mu)}&{}\;=\;
\frac{L\!-\!4}{L\CdoT(L\!-\!3)}\cdot
\Hat{\Phi}_{\boldsymbol{n}}(\mu)^{\uP{0.4}{\!2}}+
\frac{L\!-\!2}{L\CdoT(L\!-\!3)}\cdot
\big|\Hat{\Psi}_{\boldsymbol{n}}(\mu)\big|^2
\label{3.70.a}\\*[8pt]
\Hat{C}_{\Hat{\boldsymbol{\Psi}}_{\!\boldsymbol{n}}(\mu),\Hat{\boldsymbol{\Psi}}_{\!\boldsymbol{n}}(\mu)}&{}\;=\;
\frac{2\CdoT(L\!-\!2)}{L\CdoT(L\!-\!3)}\cdot
\Hat{\Phi}_{\boldsymbol{n}}(\mu)^{\uP{0.4}{\!2}}-
\frac{2}{L\CdoT(L\!-\!3)}\cdot
\big|\Hat{\Psi}_{\boldsymbol{n}}(\mu)\big|^2
\label{3.70.b}\\[8pt]
\Hat{C}_{\Hat{\boldsymbol{\Psi}}_{\!\boldsymbol{n}}(\mu),\Hat{\boldsymbol{\Psi}}_{\!\boldsymbol{n}}(\mu)^{\Kk}}\!&{}\;=\;
\frac{2}{L}\cdot\Hat{\Psi}_{\boldsymbol{n}}(\mu)^{\uP{0.4}{\!2}}
\label{3.70.c}\\
&\qquad\qquad\qquad\qquad\qquad
\forall\qquad{\T\mu=0\quad\vee\quad\mu=\frac{M}{2}}
\notag\\[-6pt]
\intertext{bzw.\vspace{-3pt}}
\Hat{C}_{\Hat{\boldsymbol{\Phi}}_{\!\boldsymbol{n}}(\mu),\Hat{\boldsymbol{\Phi}}_{\!\boldsymbol{n}}(\mu)}&{}\;=\;
\frac{1}{L\!-\!1}\cdot
\Hat{\Phi}_{\boldsymbol{n}}(\mu)^{\uP{0.4}{\!2}}
\label{3.70.d}\\*[8pt]
\Hat{C}_{\Hat{\boldsymbol{\Psi}}_{\!\boldsymbol{n}}(\mu),\Hat{\boldsymbol{\Psi}}_{\!\boldsymbol{n}}(\mu)}&{}\;=\;
\frac{L\!-\!2}{(L\!-\!1)\CdoT(L\!-\!3)}\CdoT
\Hat{\Phi}_{\boldsymbol{n}}(\!-\mu)\CdoT\Hat{\Phi}_{\boldsymbol{n}}(\mu)-
\frac{1}{(L\!-\!1)\CdoT(L\!-\!3)}\CdoT
\big|\Hat{\Psi}_{\boldsymbol{n}}(\mu)\big|^2\!\!
\label{3.70.e}\\*[8pt]
\Hat{C}_{\Hat{\boldsymbol{\Psi}}_{\!\boldsymbol{n}}(\mu),\Hat{\boldsymbol{\Psi}}_{\!\boldsymbol{n}}(\mu)^{\Kk}}\!&{}\;=\;
\frac{1}{L\!-\!1}\cdot\Hat{\Psi}_{\boldsymbol{n}}(\mu)^{\uP{0.4}{\!2}}
\label{3.70.f}\\
&\qquad\qquad\qquad\qquad
\forall\qquad{\T\mu=1\;(1)\;M\!-\!1\quad\wedge\quad\mu\neq\frac{M}{2}}
\notag
\end{align}
\end{subequations}
Dass diese Sch"atzwerte erwartungstreu sind, zeigt man, indem man
zun"achst die Messwerte nach Gleichung (\ref{3.56}) und (\ref{3.57})
einsetzt, und dann mit Hilfe der Gleichung (\ref{A.5.9}) des Anhangs 
die Erwartungswerte berechnet. \pagebreak[2]Dabei ist f"ur die Matrizen 
\mbox{$\underline{\boldsymbol{A}}=\underline{\boldsymbol{B}}=
\underline{\boldsymbol{V}}_{\bot}\!(\mu)$}, f"ur alle Matrixspuren 
\mbox{$L\!-\!2$} und f"ur den skalaren Faktor jeweils 
\mbox{$\boldsymbol{c}=\big(M\CdoT(L\!-\!2)\big)^2$} einzusetzen.
Auch hier l"asst sich zeigen, dass die Sch"atzwerte der Messwertkovarianz
\mbox{$\Hat{C}_{\Hat{\boldsymbol{\Psi}}_{\!\boldsymbol{n}}(\mu),\Hat{\boldsymbol{\Psi}}_{\!\boldsymbol{n}}(\mu)^{\Kk}}$}
nie betraglich gr"o"ser sind als die Sch"atzwerte der Messwertvarianz
\mbox{$\Hat{C}_{\Hat{\boldsymbol{\Psi}}_{\!\boldsymbol{n}}(\mu),\Hat{\boldsymbol{\Psi}}_{\!\boldsymbol{n}}(\mu)}$},
da die Bedingung (\ref{3.38}) bei diesen Messwerten immer erf"ullt ist.
Wie man auf diese Messwerte kommt, ist hier nicht angegeben.
In \cite{Erg} wird angegeben, wie man im allgemeineren Fall
eines zyklostation"aren Fehlerprozesses diese Messwerte erh"alt. 

\section{Konsistenz der Messwerte}

Nun haben wir die Varianzen aller Messwerte, die am Anfang dieses Kapitels
aufgelistet sind, als Funktionen der Momente von
\mbox{$\boldsymbol{N}_{\!\!f}(\mu)$} berechnet. Dabei haben wir lediglich
bei der Absch"atzung der Messwert(ko)varianzen des LDS und des MLDS
angenommen, dass die Zufallsgr"o"sentupel
\mbox{$\big[\boldsymbol{N}_{\!\!f}(\mu),
\boldsymbol{N}_{\!\!f}(\!-\mu)\big]^{\TT}$}
der Spektralwerte des gefensterten Modellzufallsvektors
des Approximationsfehlerprozesses verbundnormalverteilt sind.
So konnten die dort auftretenden
vierten Momente durch ihre zweiten Momente ausgedr"uckt werden,
die bereits gemessen worden sind.
Bei allen anderen Messwerten und Messwert(ko)varianzen war die Kenntnis
ihrer Verbundverteilung nicht ben"otigt worden. 

Falls die Momente von \mbox{$\boldsymbol{N}_{\!\!f}(\mu)$} existieren ---
wovon man bei realen St"orprozessen auf Grund einer inh"arenten
Amplitudenbegrenzung ausgehen kann ---, h"angt die Konsistenz
der Messwerte des RKM nur von dem Verhalten der Vorfaktoren der Momente von
\mbox{$\boldsymbol{N}_{\!\!f}(\mu)$} f"ur den Grenz"ubergang 
\mbox{$L\!\to\!\infty$} ab. Ein Messwert wird dann
als konsistent bezeichnet, wenn der Grenzwert \mbox{$L\!\to\!\infty$}
der Wahrscheinlichkeit, dass der Messwert betraglich um mehr als eine
beliebig kleine, positiv reelle Schranke $\varepsilon_{\!\Delta}$ abweicht,
Null ist. Die Tschebyscheffsche Ungleichung
\cite{Fisz} besagt, dass diese Wahrscheinlichkeit kleiner ist als
die durch $\varepsilon_{\!\Delta}^{\;2}$ dividierte Varianz des Messwertes.
Wenn also die Varianz eines Messwertes, die jeweils vom Grenzwert des
Vorfaktors des Momentes von \mbox{$\boldsymbol{N}_{\!\!f}(\mu)$} abh"angt,
mit steigendem $L$ gegen Null sterbt, ist dieser Messwert konsistent.

Bei den Messwerten \mbox{$\Hat{H}(\mu)$} enthalten die Vorfaktoren der 
Messwertvarianzen in Gleichung (\ref{3.58}), den Faktor $L$ im Nenner
sowie den Erwartungswert des Reziprokwertes der empirischen Varianz
im Z"ahler. Wenn man nun zur Erregung des Systems solche
Zufallsvektoren w"ahlt, bei denen der Erwartungswert des Reziprokwertes 
der empirischen Varianz f"ur hinreichend gro"se, endliche Werte $L$ 
existiert und immer unterhalb einer festen oberen 
Schranke bleibt, dann streben die Vorfaktoren der Messwertvarianzen f"ur 
\mbox{$L\!\to\!\infty$} gegen Null und diese Messwerte sind konsistent.
Wird zur Erregung des Systems jedoch ein Zufallsvektor verwendet, bei dem 
die empirische Varianz der erregenden Spektralwerte Null werden kann, so 
ist dieser Fall gesondert zu behandeln, da dann der Messwert der
"Ubertragungsfunktion --- wie oben geschildert --- zu Null gesetzt wird. 
Die Messwertvarianz ist in diesem Fall Null. Da man in der Praxis den 
erregenden Zufallsvektor so w"ahlen wird, dass die Wahrscheinlichkeit 
dieses Falls extrem klein wird, ist dessen Einfluss bei der Berechnung 
der Messwertvarianz vernachl"assigbar klein. Die Konsistenz der Messwerte 
der "Ubertragungsfunktion ist auch bei der eben geschilderten Art der Behandlung des 
Sonderfalls gegeben. 

Die Vorfaktoren der Varianzen der Messwerte
\mbox{$\Hat{\boldsymbol{\Phi}}_{\!\boldsymbol{n}}(\mu)$} und
\mbox{$\Hat{\boldsymbol{\Psi}}_{\!\boldsymbol{n}}(\mu)$} sind nach Gleichung
(\ref{3.69}) von der Erregung unabh"angig und streben f"ur 
\mbox{$L\!\to\!\infty$} ebenfalls gegen Null, so dass auch diese Messwerte 
konsistent sind. 

Die Varianzen und Kovarianzen aller Messwerte kann man f"ur
einen festen Wert von $L$ mit Hilfe der Gleichungen (\ref{3.59}),
(\ref{3.64}) und (\ref{3.70}) absch"atzen. Die Konsistenz
der Sch"atzwerte f"ur die Varianzen und Kovarianzen der Messwerte
wurde nicht untersucht.

\section{Anmerkungen zur Erhebung der Stichproben}

Selbst bei Verwendung eines Zufallsvektors \mbox{$\Vec{\boldsymbol{V}}\!$}, der
f"ur den Betriebszustand typischen ist, kann es vorkommen, dass der f"ur
die konkrete Messung verwendete Stichprobenvektor \mbox{$\Vec{V}\!(\mu)$}
f"ur den Betriebszustand absolut untypisch ist, und dass die Messung
dadurch an Aussagekraft verliert. An einem Beispiel, das nichts mit dem RKM
zu tun hat, sei dies verdeutlicht. Um eine Prognose f"ur ein Wahlergebnis
abzugeben, w"ahlt man aus der Gesamtheit aller Wahlberechtigten eine
hinreichend gro"se Anzahl von Personen gleichwahrscheinlich aus, um diese
nach ihrem Wahlverhalten zu befragen. Bei dieser Art der Auswahl kann es
vorkommen, dass man nur Personen einer Bev"olkerungsgruppe zieht, von der man
a priori annehmen kann, dass deren Wahlverhalten nicht repr"asentativ ist.
Man kann von vornherein sagen, dass die mit dieser konkreten Stichprobe
gewonnene Wahlvorhersage nicht sehr zutreffend sein wird, und man wird
sich daher "uberlegen, wie man die Stichprobenentnahme modifizieren kann,
um solche konkrete Stichproben auszuschlie"sen (\,etwa indem man aus mehreren
Bev"olkerungsgruppen jeweils weniger Personen gleichwahrscheinlich ausw"ahlt\,).
Analoge Strategien zur Stichprobenentnahme aus dem Zufallsvektor
\mbox{$\Vec{\boldsymbol{V}}$} sollte man sich gegebenenfalls auch f"ur das RKM
"uberlegen. Streng genommen gilt dann die theoretische Berechnung der
Erwartungswerte und Varianzen der Messergebnisse nicht mehr, da wir
dabei eine zuf"allige Stichprobenentnahme zugrundegelegt hatten.
Dennoch kann u.~U. auch eine modifizierte
Stichprobenentnahme in Bezug auf die zu messenden Merkmale
\mbox{$H(\mu\CdoT2\pi/M)$}, \mbox{$\Tilde{\Phi}_{\boldsymbol{n}}(\mu)$} und
\mbox{$\Tilde{\Psi}_{\boldsymbol{n}}(\mu)$} als zuf"allig angesehen werden.
Eine m"ogliche Methode besteht darin, dass man, wenn man eine
untypische konkrete Stichprobe erh"alt, diese nicht verwendet,
sondern eine neue konkrete Stichprobe des Zufallsvektors
\mbox{$\Vec{\boldsymbol{V}}$} zieht. \pagebreak[2]Eine andere Variante w"are es
aus einer Vielzahl von als typisch klassifizierten konkreten Stichproben
eine Stichprobe zuf"allig zu ziehen. Ob die modifizierte Art der
Stichprobenentnahme als zuf"allig in Bezug auf die zu messenden Merkmale
angesehen werden kann, w"are streng genommen an jedem zu messenden System
mit Hilfe eines Hypothesentest zu "uberpr"ufen, wobei die Hypothese
aufgestellt wird, dass die Verteilung des mit der modifizierten
Stichprobenentnahme gemessenen Merkmals, mit der Verteilung des
mit der zuf"alligen Stichprobenentnahme gemessenen Merkmals "ubereinstimmt.
In Bezug auf die geeignete Stichprobenentnahme und den Hypothesentest sei
der Leser auf die entsprechende Literatur verwiesen, da dieser
Hypothesentest in der Praxis wegen der begrenzten Messdauer
wohl kaum durchf"uhrbar sein wird.

Da die Varianz der Messwerte f"ur die "Ubertragungsfunktion
dann besonderes klein wird, wenn die empirische Varianz 
\mbox{$\Hat{\boldsymbol C}_{\boldsymbol{V}(\mu),\boldsymbol{V}(\mu)}$} 
der Erregung f"ur alle Frequenzen $\mu$ besonders gro"s wird, kann man 
eine gute Messung auch dadurch erzwingen, indem man die Anzahl $L$ der 
Einzelmessungen in dem Fall erh"oht, wenn die empirischen Varianzen 
nicht f"ur alle $\mu$ innerhalb eines vorgegebenen Intervalls um die 
theoretischen Varianzen liegen. Bei geeigneter Wahl des zul"assigen 
Intervalls, des Zufallsvektors \mbox{$\Vec{\boldsymbol{V}}$} und der 
Stichprobenentnahme kann davon ausgegangen werden, dass die 
Wahrscheinlichkeit f"ur eine extrem lange Messung mit zunehmenden $L$ 
rasch so klein sein wird, dass dies f"ur die praktische Anwendung des 
RKM mit dynamischer Verl"angerung der Mittelungsanzahl nicht mehr von 
Bedeutung ist. Auch bei dieser Art der dynamischen Wahl der Zahl der 
Einzelmessungen verliert die theoretische Berechnung der Erwartungswerte 
und Varianzen der Messergebnisse ihre G"ultigkeit. Diese Art der dynamischen 
Wahl des Umfangs einer Stichprobe ist in der Literatur unter dem Namen
Sequenzialanalyse bekannt. Es w"are viel zu umfangreich (\,und wohl auch nur
f"ur den Spezialfall einer normalverteilten Erregung mit unkorrelierten Real-
und Imagin"arteilen m"oglich\,) die entsprechende Theorie auf das RKM zu
"ubertragen. Da sich die dynamische Vergr"o"serung der Mittelungsanzahl
$L$ beim RKM --- wie in \cite{Erg} gezeigt wird --- mit minimalem Mehraufwand
realisieren l"asst, und da sich bei Rechnersimulationen an Systemen,
deren theoretische Werte bekannt waren, gezeigt hat, dass die oben
hergeleiteten Erwartungswerte und Varianzen der Messergebnisse auch
in diesem Fall gut mit der Theorie "ubereinstimmen, kann ich diese
Art der Bestimmung der Mittelungsanzahl nur empfehlen. 

\section[Sonderfall der Messung eines linearen, unabh"angig gest"orten Systems]{Sonderfall 
der Messung eines linearen, \\unabh"angig gest"orten Systems}\label{linSys}

Wenn man ein reales System vermessen will, von dem man wei"s, dass sich 
dieses nahezu perfekt linear und zeitinvariant verh"alt, bei dem 
aber eine von der Erregung unabh"angige St"orung vorliegt, die so stark 
ist, dass eine direkte Messung der "Ubertragungsfunktion mit einer 
einzigen Messung nicht m"oglich ist, so kann man das RKM auch mit 
einer nicht zuf"alligen Erregung betreiben. Die Testsignalsequenzen 
der Einzelmessungen, bzw. deren Spektralwerte, die im Vektor 
\mbox{$\Vec{V}_{\lambda}$} zusammengefasst sind, wird man dann 
so w"ahlen, dass sich gut konditionierte Gleichungssysteme ergeben. 
Die von der Erregung abh"angigen Erwartungswerte sind dann in der 
bisherigen Herleitung "uberall durch die Werte zu ersetzen, die sich 
bei den bei der Messung verwendeten Testsignalsequenzen konkret ergeben. 
Es wird nun kurz auf die Theorie dieses Falls eingegangen. 

Da sowohl die theoretischen Regressionskoeffizienten als auch die 
Erregung nicht zuf"allig sind, ist das Signal am Ausgang des Modellsystems kein 
Zufallsprozess und vom Ausgangsprozess des realen Systems ist nur der Anteil 
des nach Gleichung (\ref{2.2}) definierten Approximationsfehlerprozesses 
zuf"allig. Weil wir angenommen haben, dass das System ohne St"orung linear 
und zeitinvariant ist, liegen die Ausgangswerte \mbox{$x_{\lambda}(k)$} 
des ungest"orten realen Systems, wenn wir sie in einem \mbox{$M\!+\!1$}-dimensionalen 
Raum, "uber dem $M$-dimensionalen Vektor $\Vec{V}$ auftragen, 
in einer Hyperebene, die den Ursprung enth"alt. Da au"serdem die 
St"orung nicht von der Erregung abh"angt, ist die Regressionsfl"ache 
\mbox{$\text{E}\big\{\boldsymbol{y}(k)\pmb{\big|}\Vec{V}\big\}$} 
der ersten Art keine gekr"ummte Fl"ache sondern genau diese Hyperebene. 
Die bedingten Erwartungswerte \mbox{$\text{E}\big\{\boldsymbol{y}(k)\pmb{\big|} 
\Vec{V}_{\!\!\lambda}\big\}$}, die man bei den verwendeten 
Testsignalsequenzen erh"alt, sind immer Punkte auf der Regressionsfl"ache 
der ersten Art. Da wir die Testsignalsequenzen so gew"ahlt haben, dass sich 
gut konditionierte Gleichungssysteme ergeben, sind die theoretischen 
Regressionskoeffizienten eindeutig festgelegt, da sie ja als L"osung der 
theoretischen Approximationsfehlerminimierung die Parameter der Hyperebene sind, 
die durch die Punkte \mbox{$\text{E}\big\{\boldsymbol{y}(k)\pmb{\big|} 
\Vec{V}_{\!\!\lambda}\big\}$} geht. 

Wie wir im Kapitel \ref{theo} gesehen haben, minimiert auch die 
Optimall"osung der Approximationsfehlerminimierung bei zuf"alliger 
Erregung das zweite Moment des Abstands der Hyperebene der zweiten 
Art von der Regressionsfl"ache der ersten Art. Da bei diesem realen 
System das zweite Moment des Abstands unabh"angig von der Wahl der 
Art der Erregung immer den Minimalwert Null aufweist, sind die 
Regressionskoeffizienten bei zuf"alliger Erregung immer gleich den 
Regressionskoeffizienten, die man mit Hilfe der deterministischen 
Testsignalsequenzen erh"alt. Die L"osung der empirischen Regression 
besteht nun --- wie bei der zuf"alligen Erregung --- darin, eine Hyperebene 
anzugeben, deren Abstand zu den entsprechenden gemessenen Punkten, 
die auch bei der Erregung mit festgelegten Testsignalsequenzen 
wegen der St"orung nun nicht mehr alle auf einer Hyperebene liegen, 
minimal ist. Man berechnet nun die Erwartungswerte, Varianzen und 
Kovarianzen der Messwerte, sowie die Erwartungswerte der Sch"atzwerte 
f"ur die Messwert(ko)varianzen wie bisher, wobei man allerdings 
ber"ucksichtigt, dass die Erwartungswertbildung bei den von der Erregung 
abh"angigen Termen zu unterbleiben hat, da diese ja nun nicht mehr zuf"allig 
sind. Man wird feststellen, dass auch bei einem solchen System die Messwerte 
erwartungstreu und konsistent sind, und dass die Sch"atzwerte der 
Messwertkovarianzen erwartungstreu sind. 

Ein weiterer wesentlicher Aspekt, der sich bei der eben durchgef"uhrten 
"Uberlegung ergibt, ist, dass man auch bei 
zuf"alliger Erregung bei einem realen, linearen und zeitinvarianten 
System, das unabh"angig gest"ort wird, wirklich dessen wahre 
"Ubertragungsfunktion und die zweiten Momente der realen St"orung 
misst. Weil die Erwartungswerte der Hyperebenen 
der empirischen Anpassungsl"osungen mit den Hyperebenen der 
theoretischen Regression "ubereinstimmen, sind die 
Erwartungswerte der gemessenen "Ubertragungsfunktion, die 
die Steigungen der einen Hyperebene sind, gleich den wahren 
Werten der "Ubertragungsfunktion, die die Steigungen der anderen 
Hyperebene sind. Da die beiden Hyperebenen "ubereinstimmen, ist auch 
der Zufallsvektor des Approximationsfehlerprozesses bei der theoretischen 
Regression identisch mit dem Zufallsvektor der real am System anliegenden 
St"orung, und somit sind auch die zweiten Momente beider Prozesse identisch.

\section{Konfidenzgebiete der Messwerte}\label{Kon}

Um die Qualit"at der Messergebnisse besser beurteilen zu k"onnen,
empfiehlt es sich, vor allem f"ur deren graphische Darstellung,
bei reellen Messwerten Intervalle und bei komplexen Messwerten
Gebiete in der komplexen Ebene anzugeben, von denen man sagen kann,
dass sich die wahren Gr"o"sen mit einer Wahrscheinlichkeit von
\mbox{$1\!-\!\alpha$} innerhalb dieser Intervalle bzw. Gebiete
befinden. Diese bezeichnet man als Konfidenzintervalle bzw. Konfidenzgebiete
zum Konfidenzniveau \mbox{$1\!-\!\alpha$}. Wir wollen nun f"ur den Fall,
dass die Mittelunganzahl $L$ gro"s ist, Sch"atzwerte f"ur die
Konfidenzintervalle bzw. Konfidenzgebiete der Messwerte herleiten.

Da die wahre Verteilung der Messwerte vom zu messenden System und den 
darin auftretenden St"orungen und deren Verteilung abh"angt, und somit 
unbekannt ist, nehmen wir an, dass alle Messwerte n"aherungsweise 
normalverteilt sind. Eine Normalverteilung ist durch ihre ersten und 
zweiten Momente vollst"andig beschrieben. Die Messwerte, die man bei 
einer konkreten Messung erh"alt, sind konkrete Sch"atzwerte f"ur die 
zu messenden theoretischen Gr"o"sen, die selbst wiederum die ersten 
Momente der Messwerte --- als Ergebnisse des Zufallsexperiments der 
Messung betrachtet --- sind. Desweiteren haben wir Sch"atzwerte f"ur 
die Messwertvarianzen und Kovarianzen berechnet, so dass wir auch die 
zweiten Momente der angenommenen Normalverteilung absch"atzen k"onnen. 
Wenn wir diese Sch"atzwerte f"ur die ersten und zweiten Momente in die 
Normalverteilung einsetzen, k"onnen wir Intervalle bzw. Gebiete angeben, 
innerhalb derer die Normalverteilungsdichte oberhalb einer Konstanten 
liegt, die so gew"ahlt wird, dass das Integral "uber die Verteilungsdichte 
innerhalb der Intervalle bzw. Gebiete gerade \mbox{$1\!-\!\alpha$} ist. 
Die so bestimmten Intervalle bzw. Gebiete sind also in zweierlei 
Hinsicht nur Sch"atzwerte f"ur die Kon"-fi"-denz"-inter"-valle bzw. Konfidenzgebiete 
der Messwerte. Zum einen wird eine Normalverteilung angenommen, und zum 
anderen werden die Momente dieser Normalverteilung lediglich abgesch"atzt.

Die Annahme, dass die Messwerte normalverteilt sind, erweist sich 
in der Praxis bei den Messwerten \mbox{$\Hat{H}(\mu)$}, die den 
Stichprobenvektor \mbox{$\Vec{N}_{\!f}(\mu)$} linear 
--- also als Term erster Ordnung --- enthalten, bei einer hinreichend 
gro"sen Anzahl $L$ von Einzelmessungen als in guter N"aherung zutreffend. 
Bei den Messwerten \mbox{$\Hat{\Phi}_{\boldsymbol{n}}(\mu)$} und
\mbox{$\Hat{\Psi}_{\boldsymbol{n}}(\mu)$} tritt der Stichprobenvektor
\mbox{$\Vec{N}_{\!f}(\mu)$} als Term zweiter Ordnung auf, und die 
Normalverteilungsannahme erweist sich in der Praxis manchmal als nicht 
zutreffend. Zur besseren Beurteilung wollen wir hier nun zwei Grenzf"alle 
unterscheiden.

Sind bei den Messwerten \mbox{$\Hat{\Phi}_{\boldsymbol{n}}(\mu)$} und
\mbox{$\Hat{\Psi}_{\boldsymbol{n}}(\mu)$} die Messwertstreuungen deutlich 
kleiner als die Betr"age der Messwerte, stellt man fest, dass sich auch 
bei diesen Messwerten in der Regel n"aherungsweise eine Normalverteilung 
ergibt. "`N"aherungsweise"' soll in diesem Zusammenhang hei"sen, dass
die Konfidenzgebiete, die man mit der exakten, aber in aller Regel
unbekannten Verteilung erhalten w"urde, sich von den mit der Normalverteilung
berechneten Konfidenzintervallen bei einer sinnvollen Wahl von
\mbox{$10^{-4}<\alpha<0.9$} nur um wenige Prozent unterscheiden. 
Gr"o"sere Werte von alpha wird man in der Praxis kaum verwenden, da 
es keinen Sinn macht, ein Konfidenzintervall bzw. -gebiet f"ur einen 
Messwert anzugeben, wenn die Wahrscheinlichkeit, dass sich der Messwert 
{\em nicht}\/ innerhalb des Konfidenzintervalls bzw. -gebietes liegt, 
mehr als 90 \% betr"agt. Kleinere Werte von \mbox{$\alpha$} k"onnten 
durchaus sinnvoll sein, wenn Wert auf besonders zuverl"assige Messwerte gelegt wird. 
Da dann jedoch die Konfidenzintervalle bzw. -gebiete erst mit sehr hohen 
Mittelungsanzahlen auf sinnvolle Gr"o"se schrumpfen, wird hier die Dauer 
der Messung eine Grenze setzen. 

Liegt bei den Messwerten \mbox{$\Hat{\Phi}_{\boldsymbol{n}}(\mu)$} und
\mbox{$\Hat{\Psi}_{\boldsymbol{n}}(\mu)$} die Messwertstreuung in derselben 
Gr"o"senordnung wie der Betrag der Messwerte, so ist die Annahme der 
n"aherungsweisen normalen Verteilung der Messwerte nicht mehr zutreffend.
Dennoch sollen aus zwei Gr"unden auch bei diesen Messwerten dieselben 
Sch"atzwerte f"ur die Konfidenzintervalle bzw. Konfidenzgebiete verwendet
werden. Der erste Grund ist, dass man die wahre Verteilung der 
Stichprobenelemente \mbox{$\boldsymbol{N}_{\!\!f,\lambda}(\mu)$}
nicht kennt, und somit auch die Verteilung der Messwerte
\mbox{$\Hat{\boldsymbol{\Phi}}_{\!\boldsymbol{n}}(\mu)$} und
\mbox{$\Hat{\boldsymbol{\Psi}}_{\!\boldsymbol{n}}(\mu)$}
nicht angeben kann. Wenn man f"ur die Stichprobenelemente 
\mbox{$\boldsymbol{N}_{\!\!f,\lambda}(\mu)$} eine Normalverteilung
annimmt, k"onnte man geeignetere Sch"atzwerte f"ur die Konfidenzgebiete
erhalten, die dann aber numerisch aufwendiger zu berechnen w"aren. Dass 
der Mehraufwand nicht vertretbar erscheint, da die Konfidenzintervalle 
bzw. Konfidenzgebiete, die mit der Normalverteilungsannahme f"ur die 
Messwerte gewonnen werden, genug Aussagekraft besitzen,
zeigt der zweite Grund, die einfacher zu berechnenden Werte zu benutzen.
Die gen"ahert gesch"atzten Konfidenzintervalle bzw. Konfidenzgebiete 
unterscheiden sich in aller Regel n"amlich nur dann von den 
theoretisch exakten Konfidenzintervallen bzw. -gebieten, wenn die 
Messwerte selbst sehr unzuverl"assig und somit unbrauchbar sind.
Liegt die Messwertstreuung in derselben Gr"o"senordnung wie der Betrag 
des Messwertes, so erh"alt man bei Verwendung der Annahme der 
Normalverteilung der Messwerte, wenn man $\alpha$ nicht zu gro"s w"ahlt, 
ein Konfidenzintervall bzw. Konfidenzgebiet das den Nullpunkt mit hoher 
Wahrscheinlichkeit enth"alt, oder bei dem der Nullpunkt in Vergleich zu 
der Ausdehnung des Konfidenzintervalls bzw. des Konfidenzgebietes
sehr nahe liegt. Will man sich nicht mit der pauschalen Aussage begn"ugen, 
dass es sich um einen unzuverl"assigen Messwert handelt, so wird man die 
Mittelungsanzahl $L$ erh"ohen, um die Messwertstreuung zu verringern. 
Bei dem dann erhaltenen Messwert und dessen Konfidenzintervall bzw. 
Konfidenzgebiet kann man nun wiederum zwei F"alle unterscheiden. 

Im ersten Fall erbrachte die Erh"ohung der 
Mittelungsanzahl nun einen Messwert, dessen Betrag nun 
aufgrund der verringerten Messwertstreuung deutlich gr"o"ser 
ist als die Messwertstreuung selbst. Das Konfidenzintervall bzw. 
Konfidenzgebiet liegt nun deutlich abseits des Nullpunktes,
so dass wieder von einer n"aherungsweise normalen Verteilung des 
Messwertes ausgegangen werden kann, und das Konfidenzintervall bzw. 
Konfidenzgebiet als zutreffend und aussagekr"aftig angesehen werden kann.

Im zweiten Fall, der z.B. dann gegeben ist, wenn die zu messende 
theoretische Gr"o"se Null ist, wird man auch nach einer Erh"ohung der 
Mittelungsanzahl $L$ mit hoher Wahrscheinlichkeit einen Messwert
erhalten, dessen Konfidenzintervall bzw. Konfidenzgebiet den Ursprung 
enth"alt, bzw. das in unmittelbarer Umgebung des Nullpunktes liegt.
Auch mit der gr"o"seren Mittelungsanzahl $L$ ist nun lediglich die 
Aussage m"oglich, dass der relative Fehler des Messwertes mit hoher 
Wahrscheinlichkeit sehr gro"s ist. Der absolute Fehler des Messwertes
ist nun jedoch mit der Erh"ohung der Mittelungsanzahl und der damit 
verbundenen Verringerung der Messwertstreuung, die sich weiterhin aus der
Gr"o"se des Konfidenzintervalls bzw. des Konfidenzgebietes leicht 
ablesen l"asst, deutlich verringert worden. Selbst wenn man die wahre
Lage des Konfidenzintervalls bzw. -gebietes, das sich aus der tats"achlichen 
Messwertverteilung ergeben w"urde, nicht kennt, kann man nun 
bei der konkreten Anwendung des RKM mit hoher Zuverl"assigkeit sagen, 
dass der zu messende Wert betraglich als hinreichend klein angesehen 
werden kann. 

Wenn es beispielsweise darum geht, bei einem System 
nachzuweisen, dass das Leistungsdichtespektrum der vom System
verursachten St"orungen in einem bestimmten Frequenzband unterhalb 
einer bestimmten Schwelle liegt, gen"ugt es zu wissen, dass der
Messwert des LDS mit hoher Wahrscheinlichkeit deutlich unterhalb dieser
Schwelle liegt. Die Kenntnis der genauen Lage des theoretisch exakten 
Konfidenzintervalls ist hierbei nicht von N"oten. In Kapitel \ref{Mess8} 
wird an einer Beispielmessung verdeutlicht, wie unterschiedlich sich
als unzuverl"assig erkannte Messwerte verhalten k"onnen, wenn man
die Mittelungsanzahl erh"oht.

Da die $M$ Messwerte \mbox{$\Hat{\Phi}_{\boldsymbol{n}}(\mu)$}, die einzigen
Messwerte sind, die immer reell sind, m"ussen wir diese getrennt behandeln.
Damit m"ochte ich beginnen. Wir gehen auch bei diesen stets positiven
Messwerten aus den oben genannten Gr"unden davon aus, dass die zuf"alligen
Messwerte \mbox{$\Hat{\boldsymbol{\Phi}}_{\!\boldsymbol{n}}(\mu)$}
normalverteilt seien, obwohl eine Normalverteilung auch
f"ur negative Werte immer echt positiv ist, und somit
niemals exakt vorliegen kann. 

Die Messwertabweichung
\mbox{$\Hat{\boldsymbol{\Phi}}_{\!\boldsymbol{n}}(\mu)\!-\!
\Tilde{\Phi}_{\boldsymbol{n}}(\mu)$} ist wegen der Erwartungstreue 
der Messwerte mittelwertfrei. Daher wird f"ur die Verteilung der 
auf die wahre Standardabweichung normierten Messwertabweichung 
eine zentrierte und normierte Normalverteilung angesetzt. "Uber die 
komplement"are Fehlerfunktion kann daher ein Konfidenzintervall f"ur 
ein vorgegebenes Konfidenzniveau f"ur die Messwerte angegeben werden. 
Da wir die wahre Varianz der Messwertabweichung nicht kennen, k"onnen 
wir das Konfidenzintervall nur absch"atzen, indem wir bei der Berechnung 
des Konfidenzintervalls nicht die wahre Varianz, sondern den Sch"atzwert
\mbox{$\Hat{C}_{\Hat{\boldsymbol{\Phi}}_{\!\boldsymbol{n}}(\mu),\Hat{\boldsymbol{\Phi}}_{\!\boldsymbol{n}}(\mu)}$}
nach Gleichung (\ref{3.70}) verwenden. Mit der komplement"aren Fehlerfunktion\vspace{-8pt}
\begin{equation}
\text{erfc}(x)\;=\;\frac{2}{\sqrt{\pi}}\cdoT\Int{x}{\infty}e^{\!-\xi^2}\cdot d\xi
\label{3.71}
\end{equation}
erh"alt man f"ur das Konfidenzniveau \mbox{$1\!-\!\alpha$} die
Sch"atzwerte
\begin{equation}
\Hat{A}_{\Phi}(\mu)\;=\;
\sqrt{2\;}\cdot
\sqrt{\:\Hat{C}_{\Hat{\boldsymbol{\Phi}}_{\!\boldsymbol{n}}(\mu),\Hat{\boldsymbol{\Phi}}_{\!\boldsymbol{n}}(\mu)}\:}
\;\cdot\;\text{erfc}^{-1}(\alpha)
\qquad\qquad\forall\qquad\mu=0\;(1)\;M\!-\!1
\label{3.72}
\end{equation}
f"ur die Konfidenzintervalle
\begin{equation}
\Big[\:\Hat{\Phi}_{\boldsymbol{n}}(\mu)-\Hat{A}_{\Phi}(\mu)\,;\,
\Hat{\Phi}_{\boldsymbol{n}}(\mu)+\Hat{A}_{\Phi}(\mu)\,\Big].
\label{3.73}
\end{equation}
Dabei ist \mbox{$\text{erfc}^{-1}(x)$}
die zur komplement"aren Fehlerfunktion inverse Funktion, die man
entweder numerisch oder mit Hilfe einer Tabelle auswertet.
Es l"asst sich somit absch"atzen, dass der wahre Wert
\mbox{$\Tilde{\Phi}_{\boldsymbol{n}}(\mu)$} mit einer Wahrscheinlichkeit
von $\alpha$ au"serhalb des angegebenen, zuf"alligen,
durch die Messung gewonnenen Intervalls liegt.

{\small Anmerkung: W"are der Messwert
\mbox{$\Hat{\boldsymbol{\Phi}}_{\!\boldsymbol{n}}(\mu)$} tats"achlich 
normalverteilt und die empirische Varianz von
\mbox{$\Hat{\boldsymbol{\Phi}}_{\!\boldsymbol{n}}(\mu)$} durch eine Messung
mit Hilfe einer Stichprobe vom Umfang $L$ dieser Zufallsgr"o"se gewonnen
worden, so w"are die auf den Wert \mbox{$\Tilde{\Phi}_{\boldsymbol{n}}(\mu)$}
zentrierte und auf die empirische Standardabweichung normierte Zufallsgr"o"se
STUDENT-verteilt mit \mbox{$L\!-\!1$} Freiheitsgraden.
Es erg"abe sich somit ein besonders f"ur kleine Mittelungsanzahlen $L$ und
kleine Werte \mbox{$\Tilde{\Phi}_{\boldsymbol{n}}(\mu)$} abweichendes
Konfidenzintervall, das auch f"ur kleines $L$ g"ultig ist. Diese Art der
Berechnung des Konfidenzintervalls ist in unserem Falle jedoch nicht
m"oglich, da \mbox{$\Hat{\boldsymbol{\Phi}}_{\!\boldsymbol{n}}(\mu)$}
nicht normalverteilt sein kann. Wenn ein perfekt lineares und zeitinvariantes
System vorliegen w"urde, das von einem Zufallsprozess gest"ort wird,
dessen Spektralwerte \mbox{$\boldsymbol{N}_{\!\!f}(\mu)$} normalverteilt
und unkorreliert \mbox{(\,$\Tilde{\Psi}_{\boldsymbol{n}}(\mu)=0$\,)} sind,
und wenn man eine Messung mit nicht zuf"alliger Erregung (\,siehe Kapitel \ref{linSys}\,)
durchf"uhren w"urde, so w"aren die mit \mbox{$2\CdoT(L\!-\!2)$} multiplizierten 
und auf die halbe Varianz von \mbox{$\boldsymbol{N}_{\!\!f}(\mu)$} normierten 
Messwerte \mbox{$\Hat{\boldsymbol{\Phi}}_{\!\boldsymbol{n}}(\mu)$}
mit \mbox{$2\CdoT(L\!-\!2)$} reellen Freiheitsgraden $\chi^2$-verteilt. 
Da der Mittelwert der $\chi^2$-Verteilung mit der Anzahl ihrer Freiheitsgrade 
linear steigt, w"ahrend die Streuung nur mit der Wurzel der Freiheitsgrade zunimmt, 
wird mit steigender Mittelungsanzahl $L$ das Konfidenzintervall bei
gleichbleibendem $\alpha$ den Messwert immer enger einschlie"sen,
und somit immer weiter vom Nullpunkt entfernt liegen. 
Dabei ist zu beachten, dass der Erwartungswert des Messwertes 
\mbox{$\Hat{\boldsymbol{\Phi}}_{\!\boldsymbol{n}}(\mu)$} immer von Null
verschieden ist, wenn die Varianz von \mbox{$\boldsymbol{N}_{\!\!f}(\mu)$}
nicht Null ist. Sollte die Varianz von \mbox{$\boldsymbol{N}_{\!\!f}(\mu)$}
Null sein, so erg"abe sich schon nach der Tschebyscheffschen
Ungleichung ein Konfidenzintervall der Breite Null. Nach \cite{Fisz}
wurde von \mbox{R.~A.~Fisher} gezeigt, dass die $\chi^2$-Verteilung
sich mit zunehmender Anzahl der Freiheisgrade der Normalverteilung
asymptotisch ann"ahert. Dabei wird die Anzahl von $30$ Freiheitsgraden
schon als ausreichend f"ur eine gute N"aherung angegeben. F"ur
hinreichend gro"se Werte von $L$ und sinnvolle Werte $\alpha$
werden sich die mit der $\chi^2$-Verteilung berechneten Werte
der Kon\-fi\-denz\-in\-ter\-valle nur unwesentlich von den mit der
Normalverteilung berechneten Konfidenzintervallen unterscheiden.
Wenn man nicht nur Sch"atzwerte f"ur die Konfidenzintervalle angeben
wollte, m"usste man diese "uber die inverse Verteilung des Quotienten
\mbox{$\big(\Hat{\boldsymbol{\Phi}}_{\!\boldsymbol{n}}(\mu)\!-\!
\Tilde{\Phi}_{\boldsymbol{n}}(\mu)\big)\;/\;
\Hat{\boldsymbol{C}}_{\Hat{\boldsymbol{\Phi}}_{\!\boldsymbol{n}}(\mu),\Hat{\boldsymbol{\Phi}}_{\!\boldsymbol{n}}(\mu)}$}
aus der Messwertabweichung und der empirischen
Messwertvarianz berechnen. Die empirische Messwertvarianz ist in
dem eben beschriebenen Spezialfall proportional zum Quadrat des
Messwertes. Damit l"asst sich die Verteilung dieses Quotienten
prinzipiell aus der $\chi^2$-Verteilung berechnen, und wenigstens
numerisch integrieren und invertieren. Somit k"onnte man in diesem
Spezialfall auch exakte Konfidenzgebiete angeben. Ob die
so berechneten Konfidenzintervalle besser geeignet w"aren, die
Zuverl"assigkeit der Messwerte abzusch"atzen, als die "uber
die Normalverteilung gesch"atzten Konfidenzintervalle, wenn man
bei einem realen System auf die Messung mit zuf"alliger und u.~U.
nicht normalverteilter Erregung angewiesen ist, darf bezweifelt
werden. Daher ist der Mehraufwand einer solchen Berechnung
nicht vertretbar.}

Alle anderen Messwerte sind komplex. f"ur diese Gr"o"sen kann man 
keine Konfidenzintervalle angeben. Man gibt daher ein Gebiet in 
der komplexen Ebene an, das dadurch gekennzeichnet ist, dass die
Wahrscheinlichkeit, dass die zu messende, theoretische Gr"o"se
au"serhalb des durch die Messung gewonnenen, und daher zuf"alligen
Gebietes liegt, kleiner als $\alpha$ ist. 

Da angenommen wird, dass
die Messwertabweichung, die aufgrund der Erwartungstreue der Messwerte
mittelwertfrei ist, f"ur eine hinreichend gro"se Mittelungsanzahl $L$
in guter N"aherung normalverteilt ist, und da die Linien konstanten
Wertes der Verbundverteilungsdichte bei einer Normalverteilung 
Ellipsen sind, w"ahlt man als Konfidenzgebiet jedes Messwertes ebenfalls
eine Ellipse, die durch die komplexen Zeiger der Hauptachsen vollst"andig
beschrieben ist. Am Beispiel des Messwertes \mbox{$\Hat{H}(\mu)$} f"ur die
"Ubertragungsfunktion soll nun hergeleitet werden, wie man die komplexen
Zeiger der Hauptachsen bestimmt.

Wir bilden zun"achst aus der zuf"alligen Messwertabweichung
\begin{equation}
\boldsymbol{\Delta}\Hat{\boldsymbol{H}}(\mu)\;=\;
\Hat{\boldsymbol{H}}(\mu)\!-\!
H\!\big({\T\mu\CdoT\frac{2\pi}{M}}\big) 
\label{3.74}
\end{equation}
durch eine Abbildung die neue Zufallsgr"o"se
\begin{equation}
\boldsymbol{\Delta}\Tilde{\boldsymbol{H}}(\mu)\;=\;
c_1(\mu)\CdoT\boldsymbol{\Delta}\Hat{\boldsymbol{H}}(\mu)-
c_2(\mu)\CdoT\boldsymbol{\Delta}\Hat{\boldsymbol{H}}(\mu)^{\Kk}
\qquad\qquad\forall\qquad\mu=0\;(1)\;M\!-\!1,\quad
\label{3.75}
\end{equation}
deren Varianz $1$ ist, deren Real- und Imagin"arteil unkorreliert sind 
und deren zweidimensionale Verbundverteilung eine mittelwertfreie 
Normalverteilung ist, wenn man die Konstanten
\begin{subequations}\label{3.76}
\begin{align}
c_1(\mu)=&\sqrt{\frac{\,
C_{\Hat{\boldsymbol{H}}(\mu),\Hat{\boldsymbol{H}}(\mu)}+
\sqrt{C_{\Hat{\boldsymbol{H}}(\mu),\Hat{\boldsymbol{H}}(\mu)}^2
\!-\big|C_{\Hat{\boldsymbol{H}}(\mu),\Hat{\boldsymbol{H}}(\mu)^{\Kk}}\big|^2\:}\:}
{C_{\Hat{\boldsymbol{H}}(\mu),\Hat{\boldsymbol{H}}(\mu)}^2
-\,\big|C_{\Hat{\boldsymbol{H}}(\mu),\Hat{\boldsymbol{H}}(\mu)^{\Kk}}\big|^2}\;}\Cdot
e^{\!-\frac{j}{2}\cdot\winkel{\T\{}
C_{\Hat{\boldsymbol{H}}(\mu),\Hat{\boldsymbol{H}}(\mu)^{\Kk}}{\T\}}}
\label{3.76.a}\\
\intertext{und}
c_2(\mu)=&\sqrt{\frac{\,
C_{\Hat{\boldsymbol{H}}(\mu),\Hat{\boldsymbol{H}}(\mu)}-
\sqrt{C_{\Hat{\boldsymbol{H}}(\mu),\Hat{\boldsymbol{H}}(\mu)}^2
\!-\big|C_{\Hat{\boldsymbol{H}}(\mu),\Hat{\boldsymbol{H}}(\mu)^{\Kk}}\big|^2\:}\;}
{C_{\Hat{\boldsymbol{H}}(\mu),\Hat{\boldsymbol{H}}(\mu)}^2
-\,\big|C_{\Hat{\boldsymbol{H}}(\mu),\Hat{\boldsymbol{H}}(\mu)^{\Kk}}\big|^2}\;}\Cdot
e^{\frac{j}{2}\cdot\winkel{\T\{}
C_{\Hat{\boldsymbol{H}}(\mu),\Hat{\boldsymbol{H}}(\mu)^{\Kk}}{\T\}}}
\label{3.76.b}
\end{align}
\end{subequations}
verwendet. Man beachte dabei, dass die unter den Wurzeln stehenden
Terme stets positiv sind (\,siehe Anhang \ref{Cauchy}\,). Die Linien
konstanter Werte der Verteilungsdichtefunktion der Zufallsgr"o"se
\mbox{$\boldsymbol{\Delta}\Tilde{\boldsymbol{H}}(\mu)$} sind 
konzentrische Kreise um den Nullpunkt. Wir bestimmen nun den 
Radius des Kreises, der ein Gebiet umschlie"st, in dem das Integral 
"uber die Verteilungsdichtefunktion den Wert \mbox{$1\!-\!\alpha$} 
liefert. Da das Betragsquadrat der Zufallsgr"o"se
\mbox{$\boldsymbol{\Delta}\Tilde{\boldsymbol{H}}(\mu)$}
mit zwei Freiheitsgraden $\chi^2$-verteilt ist, kann der gesuchte 
Radius aus der inversen Verteilungsfunktion der $\chi^2$-Verteilung
berechnet werden.
\begin{equation}
R = \sqrt{-2\cdot\ln(\alpha)\;}
\label{3.77}
\end{equation}
Die Winkel der komplexen Konstanten $c_1(\mu)$ und $c_2(\mu)$
wurden gegengleich gew"ahlt, so dass die Werte $R$ und $j\CdoT R$
bei der Umkehrabbildung
\begin{equation}
\boldsymbol{\Delta}\Hat{\boldsymbol{H}}(\mu)\;=\;
\frac{\;c_1(\mu)^{\Kk}\!\CdoT
\boldsymbol{\Delta}\Tilde{\boldsymbol{H}}(\mu)\,+\,
c_2(\mu)\CdoT\boldsymbol{\Delta}\Tilde{\boldsymbol{H}}(\mu)^{\Kk}\;}
{|c_1(\mu)|^2-\,|c_2(\mu)|^2}
\label{3.78}
\end{equation}
gerade die gro"se und die kleine Halbachse der Ellipse des Urbildes
des Kreises mit dem Radius $R$ ergeben. Diese Ellipse ist eine
Linie konstanter Werte der Verteilungsdichtefunktion der Zufallsgr"o"se
\mbox{$\boldsymbol{\Delta}\Hat{\boldsymbol{H}}(\mu)$}, weil der Betrag der
Funktionaldeterminante der Abbildung mit \mbox{$|c_1(\mu)|^2-|c_2(\mu)|^2$}
einen konstanten Wert annimmt. F"ur die Halbachsen erh"alt man

\begin{subequations}\label{3.79}
\begin{align}
A_{1,H}(\mu)&{}\,=\,\phantom{j\CdoT{}}\sqrt{-\ln(\!\alpha)\CdoT
\Big(C_{\Hat{\boldsymbol{H}}(\mu),\Hat{\boldsymbol{H}}(\mu)}+
\big|C_{\Hat{\boldsymbol{H}}(\mu),\Hat{\boldsymbol{H}}(\mu)^{\Kk}}\big|\Big)\,}\Cdot
e^{\frac{j}{2}\cdot\winkel{\T\{}
C_{\Hat{\boldsymbol{H}}(\mu),\Hat{\boldsymbol{H}}(\mu)^{\Kk}}{\T\}}}
\label{3.79.a}\\*
\intertext{und}
A_{2,H}(\mu)&{}\,=\,j\CdoT\sqrt{-\ln(\!\alpha)\CdoT
\Big(C_{\Hat{\boldsymbol{H}}(\mu),\Hat{\boldsymbol{H}}(\mu)}-
\big|C_{\Hat{\boldsymbol{H}}(\mu),\Hat{\boldsymbol{H}}(\mu)^{\Kk}}\big|\Big)\,}\Cdot
e^{\frac{j}{2}\cdot\winkel{\T\{}
C_{\Hat{\boldsymbol{H}}(\mu),\Hat{\boldsymbol{H}}(\mu)^{\Kk}}{\T\}}}\!.
\label{3.79.b}
\end{align}
\end{subequations}

F"ur die folgende Beschreibung der aus der Konfidenzellipse ablesbaren 
Werte kann Bild \ref{b5e1} herangezogen werden. Die Neigung der l"angeren 
Halbachse kann unmittelbar als der halbe Winkel der Kovarianz des Messwertes 
angegeben werden. Die k"urzere Halbachse steht wegen des Vorfaktors $j$ 
senkrecht auf der l"angeren Halbachse. Der Summe der Quadrate der L"angen 
der Halbachsen ist proportional zur Varianz des Messwertes.
Die Differenz der Quadrate der L"angen der Halbachsen
ist proportional zum Betrag der Kovarianz des Messwertes. W"ahlt
man \mbox{$\alpha=e^{\!-0,5}\approx 0,60653$} so erh"alt man \mbox{$R=1$}.
An die sich damit ergebende Ellipse kann man ein Rechteck in der Art
anpassen, dass jede Seite des Rechtecks die Ellipse ber"uhrt.
Unabh"angig davon, welcher Winkel zwischen den Halbachsen
und den Seiten des Rechtecks dabei gew"ahlt wird, ist die halbe
L"ange der Diagonale immer die Streuung des Messwertes.
F"ur Messwerte, deren Real- und Imagin"arteil linear abh"angig sind,
wird der Betrag des Korrelationskoeffizienten Eins. Dann wird auch
der Betrag der komplexen Kovarianz gleich der Varianz des Messwertes,
und die Ellipse entartet zu einem Geradenst"uck, deren Neigung die lineare
Abh"angigkeit des Real- und Imagin"arteils zeigt. 

Die Wahrscheinlichkeit
\mbox{$1\!-\!\alpha$} (\,Konfidenzniveau\,), dass die Zufallsgr"o"se
\mbox{$\boldsymbol{\Delta}\Tilde{\boldsymbol{H}}(\mu)$} innerhalb
des Kreises mit dem Radius $R$ liegt, ist gleich der Wahrscheinlichkeit,
dass die Zufallsgr"o"se \mbox{$\boldsymbol{\Delta}\Hat{\boldsymbol{H}}(\mu)$}
innerhalb der Ellipse mit den eben berechneten Halbachsen um den
Ursprung liegt, und ist somit gleich der Wahrscheinlichkeit,
dass der Messwert \mbox{$\Hat{\boldsymbol{H}}(\mu)$} innerhalb
der Ellipse mit denselben Halbachsen um den wahren Wert
\mbox{$H(\mu\CdoT2\pi/M)$} der "Ubertragungsfunktion
liegt. Diese Wahrscheinlichkeit ist wiederum gleich der Wahrscheinlichkeit,
dass der wahren Wert der "Ubertragungsfunktion innerhalb der zuf"allig
parallelverschobenen Ellipse mit diesen Halbachsen um den zuf"alligen Messwert
\mbox{$\Hat{\boldsymbol{H}}(\mu)$} liegt. Somit bildet die Ellipse mit diesen
Halbachsen und dem Messwert als Mittelpunkt ein Konfidenzgebiet f"ur den wahren
Wert der "Ubertragungsfunktion. 

Exakt k"onnen die Halbachsen dieser Ellipse
nicht angegeben werden, da weder die Varianz
\mbox{$C_{\Hat{\boldsymbol{H}}(\mu),\Hat{\boldsymbol{H}}(\mu)}$}
noch die Kovarianz
\mbox{$C_{\Hat{\boldsymbol{H}}(\mu),\Hat{\boldsymbol{H}}(\mu)^{\Kk}}$}
bekannt sind. Man kann diese Gr"o"sen jedoch anhand der
gemessenen Werte nach Gleichung (\ref{3.59}) und (\ref{3.64}) absch"atzen,
und erh"alt so eine Absch"atzung f"ur die Halbachsen der
Konfidenzellipse des Wertes der "Ubertragungsfunktion f"ur
das Konfidenzniveau \mbox{$1\!-\!\alpha$}.

\begin{subequations}\label{3.80}
\begin{align}
\Hat{A}_{1,H}(\mu)&{}\;=\;\phantom{j\cdot{}}\sqrt{\:-\ln(\alpha)\cdot
\Big(\:\Hat{C}_{\Hat{\boldsymbol{H}}(\mu),\Hat{\boldsymbol{H}}(\mu)}+
\big|\Hat{C}_{\Hat{\boldsymbol{H}}(\mu),\Hat{\boldsymbol{H}}(\mu)^{\Kk}}\big|\:\Big)\:}
\,\cdot\, e^{\frac{j}{2}\cdot\winkel{\T\{}
\Hat{C}_{\Hat{\boldsymbol{H}}(\mu),\Hat{\boldsymbol{H}}(\mu)^{\Kk}}{\T\}}}
\label{3.80.a}\\*
\Hat{A}_{2,H}(\mu)&{}\;=\;j\cdot\sqrt{\:-\ln(\alpha)\cdot
\Big(\:\Hat{C}_{\Hat{\boldsymbol{H}}(\mu),\Hat{\boldsymbol{H}}(\mu)}-
\big|\Hat{C}_{\Hat{\boldsymbol{H}}(\mu),\Hat{\boldsymbol{H}}(\mu)^{\Kk}}\big|\:\Big)\:}
\,\cdot\, e^{\frac{j}{2}\cdot\winkel{\T\{}
\Hat{C}_{\Hat{\boldsymbol{H}}(\mu),\Hat{\boldsymbol{H}}(\mu)^{\Kk}}{\T\}}}\label{3.80.b}\\*
&{}\;\;\,{\T\forall\qquad\qquad\mu=0\quad\vee\quad\mu=\frac{M}{2}}
\notag\\[-16pt]
\intertext{und}
\Hat{A}_{1,H}(\mu)&{}\;=\;-j\cdot\Hat{A}_{2,H}(\mu)\;=\;
\sqrt{\:-\ln(\alpha)\cdot
\Hat{C}_{\Hat{\boldsymbol{H}}(\mu),\Hat{\boldsymbol{H}}(\mu)}\;}
\qquad\qquad\text{ sonst }
\label{3.80.c}
\end{align}
\end{subequations}
Da wir bei der Berechnung der Messwertvarianzen und Kovarianzen 
die Messwerte \mbox{$\Hat{\Phi}_{\boldsymbol{n}}(\mu)$} und
\mbox{$\Hat{\Psi}_{\boldsymbol{n}}(\mu)$} nach Gleichung (\ref{3.56}) 
und (\ref{3.57}) verwenden, die der Gleichung~(\ref{3.38}) gen"ugen, 
sind die Sch"atzwerte f"ur die Varianzen gr"o"ser
als die Betr"age der Sch"atzwerte f"ur die Kovarianzen der Messwerte
der "Ubertragungsfunktion und es ist theoretisch sichergestellt,
dass unter den Wurzeln niemals eine negative Gr"o"se auftritt.
Dennoch sollte man die Radikanden auf nichtnegative Werte begrenzen,
da es bei der gesamten Berechnung der Radikanden zu Fehlern kommt,
die durch die begrenzte Wortl"ange des verwendeten Rechners verursacht
werden, und die --- abweichend von der Theorie --- zu schwach
negativen Radikanden f"uhren und numerische Probleme verursachen k"onnen.

Bei der Berechnung der Sch"atzwerte f"ur die Halbachsen der
Konfidenzellipsen tritt neben dem Betrag der Kovarianz
auch noch deren Winkel auf, der bei kleinen Werten der Kovarianz
aufgrund der Zuf"alligkeit der Messergebnisse nahezu gleichverteilt sein
wird. Da in diesem Fall die Konfidenzellipsen zu Kreisen werden,
ist der Winkel der Halbachsen bei verschwindender Kovarianz jedoch
bedeutungslos, sofern die Halbachsen nur senkrecht aufeinander stehen.
Dies ist durch den Faktor $j$ vor der Wurzel bei der zweiten Halbachse
der Ellipse immer sichergestellt. 

Die Konfidenzellipse kann in Parameterdarstellung mit dem 
Parameter $\varphi$ und den Sch"atzwerten
der Halbachsen folgenderma"sen angegeben werden:
\begin{equation}
\Hat{H}(\mu)\;+\;\Hat{A}_{1,H}(\mu)\cdot\cos(\varphi)\;+\;
\Hat{A}_{2,H}(\mu)\cdot\sin(\varphi)
\qquad\qquad\text{ mit }\qquad 0 \le \varphi < 2\pi
\label{3.81}
\end{equation}

Entsprechend erh"alt man die Sch"atzwerte f"ur die Konfidenzellipsen 
der Messwerte \mbox{$\Hat{\Psi}_{\boldsymbol{n}}(\mu)$}, indem man in 
den Gleichungen (\ref{3.79}) die wahre Varianz und Kovarianz der
Messwerte durch deren Sch"atzwerte gem"a"s der Gleichungen
(\ref{3.70}) ersetzt. Dabei verwenden wir wieder die Messwerte 
\mbox{$\Hat{\Phi}_{\boldsymbol{n}}(\mu)$} und
\mbox{$\Hat{\Psi}_{\boldsymbol{n}}(\mu)$} nach Gleichung (\ref{3.56}) 
und (\ref{3.57}), die der Gleichung~(\ref{3.38}) gen"ugen, und bei denen
keine Gefahr besteht, dass bei der Berechnung der L"ange der k"urzeren
Halbachse der Konfidenzellipse ein negativer Radikand entsteht.

In den Kapiteln \ref{Mess11} und \ref{Mess8} werden an zwei
Beispielen die Konfidenzgebiete veranschaulicht, und f"ur
das Konfidenzniveau \mbox{$\alpha\!=\!10\%$} dieses mit der
bei $1000$ Messungen gewonnenen relativen H"aufigkeit, dass
der theoretische Wert au"serhalb des Konfidenzgebiets liegt,
verglichen.

\section{Spezielle Testsignale}\label{Spesig}

Eine Reduktion des Aufwands des RKM kann dadurch erreicht
werden, dass man zur Messung spezielle Signale verwendet.
Diese Signale sind zum einen das zuf"allige Mehrtonsignal
und zum anderen das zuf"allige Chirpsignal. In Kapitel \ref{theo}
hatten wir festgestellt, dass die optimale L"osung der Regression, also
die optimale Systemapproximation, u.~U. von der Wahl des erregenden Prozesses
abh"angen kann. Die im folgenden vorgestellten Testsignale k"onnen daher
nur dann verwendet werden, wenn man sicher sein kann, dass
die Substitution des f"ur den Betriebszustand typischen, erregenden 
Zufallsprozesses durch einen der beiden bereichsweise periodischen
Zufallsprozesse des Mehrton- bzw. des Chirpsignals in der Art vorgenommen
werden kann, dass es f"ur diese Arten der Erregung nur L"osungen f"ur die
"Ubertragungsfunktion \mbox{$H(\mu\CdoT 2\pi/M)$} gibt, die auch
L"osungen im Fall der Erregung durch den f"ur den Betriebszustand
typischen Zufallsprozess sind.

\subsection{Das Mehrtonsignal}

In \cite{Dong} wurde vorgeschlagen, beim RKM sogenannte Mehrtonsignale
zu verwenden. Bei diesen Testsignalen ist das Betragsspektrum
\mbox{$|V_{\lambda}(\mu)|$} bei allen
Einzelmessungen konstant \mbox{$|V(\mu)|$} und von $\lambda$ unabh"angig.
Die Phase stammt aus einem reellen Zufallsvektor mit den $M$ Elementen
\mbox{$\boldsymbol{\varphi}(\mu)$}.
Meist wird man diese zwischen $-\pi$ und $\pi$ gleichverteilt und f"ur die $M$
Frequenzen unabh"angig w"ahlen. Die Testsignale der Einzelmessungen erh"alt
man mit den $L$ Elementen \mbox{$\varphi_{\lambda}(\mu)$} einer
konkreten Stichprobe der Zufallsgr"o"sen \mbox{$\boldsymbol{\varphi}(\mu)$}.
\begin{equation}
V_{\lambda}(\mu)\;=\;V(\mu)\cdot
e^{j\cdot \varphi_{\lambda}(\mu)}
\label{3.82}
\end{equation}
Die Wahl des bei allen Einzelmessungen konstanten Betragsspektrums
des Testsignals bietet im wesentlichen drei Vorteile. Zum einen
wird eine Reduzierung des Rechenaufwands erreicht, da dann die Summen
"uber alle Betragsquadrate der Spektralwerte des Eingangssignals, die
sonst mit Hilfe von Akkumulatoren berechnet werden m"ussten, konstant sind:
\begin{equation}
\Vec{V}(\mu)\CdoT\Vec{V}(\mu)^{\Hh}\;=\;
\Sum{\lambda=1}{L}V_{\lambda}(\mu)\CdoT V_{\lambda}(\mu)^{\Kk}\;=\;
L\CdoT |V(\mu)|^2\!.
\label{3.83}
\end{equation}
Der zweite Vorteil besteht darin, dass man vor Beginn der Messung die
Gesamtleistung der Erregung auf unterschiedliche Frequenzbereiche
unterschiedlich verteilen kann. Somit kann auch der
Erwartungswert des inversen zweiten empirischen Moments des Spektrums
des Mehrtonsignals in den unterschiedlichen Frequenzbereichen
variiert werden, wodurch erreicht werden kann, dass man in
einigen Frequenzbereichen eine geringere Varianz der Messwerte
der "Ubertragungsfunktion erzwingen kann. Bei konstanter
Gesamtleistung erkauft man sich diese Erh"ohung der Messgenauigkeit
auf Kosten einer h"oheren Messwertvarianz in den anderen
Frequenzbereichen. Schlie"slich ergibt sich noch der dritte Vorteil,
dass die Erwartungstreue der Messwerte der "Ubertragungsfunktion auch dann
gegeben ist, wenn die zuf"alligen Spektralwerte
\mbox{$\boldsymbol{N}_{\!\!f}(\mu)$} der gefensterten St"orung des
realen Systems {\em nicht}\/ unabh"angig von den Spektralwerten
\mbox{$\boldsymbol{V}\!(\mu)$} der Erregung bei derselben Frequenz sind.
Dies wurde in Gleichung (\ref{3.20}) gezeigt.

\subsection{Das Chirpsignal}

Man kann als Testsignal auch eine nicht zuf"allige bereichsweise
mit $M$ periodische Signalsequenz \mbox{$v_C(k)$} mit der
Fouriertransformierten \mbox{$V_C(\mu)$} verwenden, die man bei
jeder Einzelmessung $\lambda$ mit dem f"ur alle Frequenzen gleichen
Drehfaktor \mbox{$e^{j\cdot\varphi_{\lambda}}$} multipliziert.
\begin{equation}
V_{\lambda}(\mu)\;=\;V_C(\mu)\cdot e^{j\cdot\varphi_{\lambda}}
\label{3.84}
\end{equation}
$\varphi_{\lambda}$ ist dabei jeweils ein Element einer
konkreten Stichprobe vom Umfang $L$, die einer reellen Zufallsgr"o"se
\mbox{$\boldsymbol{\varphi}$} entnommen wird. Solch ein Testsignal weist
bei allen Einzelmessungen ebenfalls ein konstantes Betragsspektrum auf,
weil das Spektrum der Erregung bei der Einzelmessung $\lambda$ lediglich
gegen"uber dem konstanten Spektrum \mbox{$V_C(\mu)$} mit dem Faktor
\mbox{$e^{j\cdot\varphi_{\lambda}}$} verdreht wird. Die Vorteile,
die die Verwendung des Mehrtonsignals bietet, sind auch bei dieser
Art von Testsignal vorhanden. Ein weiterer Vorteil ist, dass sowohl
das Zeitsignal, als auch das Spektrum der Erregung durch einfache
Multiplikation mit dem Drehfaktor, und somit ohne eine DFT bei jeder
Einzelmessung durchf"uhren zu m"ussen, berechnet werden k"onnen.

In \cite{Sch/H} wurde vorgeschlagen f"ur das RKM als nicht zuf"allige
bereichsweise mit $M$ periodische Signalsequenz \mbox{$v_C(k)$} ein
Chirpsignal
\begin{equation}
\qquad v_C(k)\;=\;\frac{V_C}{\sqrt{M}}\cdot e^{j\cdot\frac{\pi}{M}\cdot k^2}
\label{3.85}
\end{equation}
mit der Fouriertransformierten
\begin{equation}
\qquad V_C(\mu)\;=\;
V_C\cdot e^{j\cdot\frac{\pi}{4}}\cdot e^{\!-j\cdot\frac{\pi}{M}\cdot\mu^2}
\label{3.86}
\end{equation}
zu verwenden, die dann wieder bei jeder Einzelmessung gem"a"s Gleichung
(\ref{3.84}) mit dem zuf"alligen Drehfaktor multipliziert wird.
Zu beachten ist, dass $M$ eine gerade Zahl sein muss, damit das
Chirpsignal die gew"unschte Periode $M$ aufweist.

Der so erzeugte erregende periodische Zufallsprozess ist nicht station"ar.
Verschiebt man jedoch das periodische Chirpsignal um eine zuf"allige Zeit 
\mbox{$\boldsymbol{\Delta}\boldsymbol{k}$} mit einem im Intervall 
\mbox{$[0; M\!-\!1]$} gleichverteiltem \mbox{$\boldsymbol{\Delta}\boldsymbol{k}$} 
so erh"alt man einen station"aren erregenden Prozess und die in dieser 
Abhandlung hergeleitete Theorie des Messverfahrens beh"alt ihrer G"ultigkeit.

Bei einem Chirpsignal weisen sowohl der Real- als auch der Imagin"arteil den 
guten Crest-Faktor\footnote{Verh"altnis aus dem Betrag des Signalspitzenwertes 
und dem Effektivwert} von ca. $\sqrt{2\,}$ auf. Da bei realen Systemen die 
Teilsysteme, die die Verarbeitung der Real- und Imagin"arteile bewerkstelligen,
in aller Regel einen begrenzten Bereich f"ur die eingangsseitige
Maximalaussteuerung besitzen, kann mit diesem Signal bei einem festen
Betrag des Spitzenwertes eine im Vergleich zu anderen Signalen hohe
Signalleistung am Eingang des zu messenden Systems erreicht werden. Da bei
der Varianz des Messwertes der "Ubertragungsfunktion nach Gleichung (\ref{3.58})
der Erwartungswert der inversen empirischen Varianz des Spektrums der Erregung
als Vorfaktor auftritt, kann die Mittelungsanzahl $L$ bei Verwendung des
Chirpsignals aufgrund der relativ gro"sen Varianz reduziert werden,
ohne dadurch die Qualit"at der Messung zu verringern. Die Verbesserung
der Frequenzselektivit"at des RKM, die durch die Fensterung erreicht wird,
bleibt durch die Wahl des Chirpsignals unbeeinflusst.

{\small Anmerkung: Das zuf"allige komplexe Chirpsignal besitzt 
nur den einen reellen Freiheitsgrad \mbox{$\boldsymbol{\varphi}$}. 
Dennoch besitzt bei der in dieser Abhandlung betrachteten Variante 
des RKM die theoretische Kovarianzmatrix 
\mbox{$\underline{C}_{\Hat{\Vec{\boldsymbol{V}}}(\mu),\Hat{\Vec{\boldsymbol{V}}}(\mu)} $}
nach Gleichung (\ref{3.46}) des Zufallsvektors
\mbox{$\Hat{\Vec{\boldsymbol{V}}}\!(\mu)$} nach Gleichung (\ref{3.47})
den h"ochstm"oglichen Rang $2$ und ist somit regul"ar. Somit ist das
Chirpsignal hier prinzipiell einsetzbar. Will man das komplexe Chirpsignal 
zur Messung zeitvarianter Systeme oder zyklostation"arer Approximationsfehler 
verwenden, wie dies in \cite{Erg} gezeigt wird, so reicht der eine Freiheitsgrad 
des Chirpsignals nicht aus, und man muss es modifizieren. Eine M"oglichkeit, 
die in \cite{Heinle} vorgeschlagen wird, besteht darin, die unterschiedlichen 
Polyphasenkomponenten des erregenden Signals mit unabh"angigen zuf"alligen 
Chirpsignalen zu besetzen. Auf den guten Crest-Faktor hat dies keinen Einfluss, 
jedoch geht die Konstanz des Spektrums verloren.}

\chapter{Messung reellwertiger Systeme}\label{Resys}

In diesem Kapitel wird untersucht, wie das in Kapitel 
\ref{RKM} vorgestellte Messverfahren zur Messung reellwertiger
realer Systeme benutzt werden kann, und welche Vereinfachungen sich ergeben, 
wenn man das RKM zur Messung komplexwertiger Systeme unver"andert auf 
reellwertige Systeme anwendet, wobei man lediglich die komplexen Signale 
durch reelle ersetzt. Es folgt eine Untersuchung, unter welchen 
Voraussetzungen man die zuf"alligen komplexen Mehrton- oder
Chirpsignale unver"andert verwenden kann, und wie diese zu modifizieren
sind, wenn die Voraussetzungen nicht erf"ullt sind. Auch in diesem Kapitel 
beschr"anken wir uns auf die Behandlung solcher Systeme, die von 
mittelwertfreien, station"aren Prozessen erregt und gest"ort werden. 
Drei weitere Varianten des RKM zur Messung reellwertiger
realer Systeme werden in \cite{Erg} kurz behandelt.

\section{Eine Variante des RKM zur Messung reellwertiger Systeme}\label{V1}

Zun"achst soll kurz dargestellt werden, welchen Einfluss die Reellwertigkeit
des zu messenden Systems auf die theoretischen Gr"o"sen der optimalen
Approximation des realen Systems durch das in Bild \ref{b1h} dargestellte
Systemmodell hat. Als Eingangssignal nehmen wir weiterhin die mit $M$
periodischen Signalsequenzen der L"ange \mbox{$E\!+\!F$}, die nun aber
reell sind, so dass f"ur die Zufallsgr"o"sen der Spektralwerte des
Eingangssignals die Symmetrie
\begin{equation}
\boldsymbol{V}(\mu)=\boldsymbol{V}(\!-\mu)^{\Kk}
\label{4.1}
\end{equation}
gilt. Bei reeller Erregung messen wir am Ausgang des realen Systems
ein ebenfalls reelles Signal \mbox{$\boldsymbol{y}(k)$}. Wenn wir eine
reelle Fensterfolge verwenden (\,die im Kapitel \ref{Algo} vorgestellten
Fensterfolgen sind reell\,), ist auch das gefensterte Ausgangssignal
reell und die Spektralwerte des gefensterten Systemausgangssignals
weisen dieselbe Symmetrie auf.
\begin{equation}
\boldsymbol{Y}_{\!\!\!f}(\mu)=\boldsymbol{Y}_{\!\!\!f}(\!-\mu)^{\Kk}
\label{4.2}
\end{equation}
Auch beim reellwertigen System werden die Parameter der Systemapproximation
so gew"ahlt, dass das zweite Moment des weiterhin nach Gleichung
(\ref{2.9}) definierten Approximationsfehlers minimiert wird. Der zu
minimierende Term ist also weiterhin unver"andert durch Gleichung
(\ref{2.10}) beschrieben. Wenn man voraussetzt, dass die Varianzen aller $M$
Zufallsgr"o"sen \mbox{$\boldsymbol{V}(\mu)$} von Null verschieden
sind, erh"alt man mit Gleichung (\ref{2.28}) die L"osung f"ur die theoretischen
Werte der "Ubertragungsfunktion, die hier ebenfalls die Symmetrie
\begin{equation}
H\big({\T\mu\CdoT\frac{2\pi}{M}}\big)=
H\big({\T-\mu\CdoT\frac{2\pi}{M}}\big)^{\Kk}
\label{4.3}
\end{equation}
aufweist. Bei einem reellwertigen System liefert also die optimale 
Approximierung immer ein reellwertiges Modellsystem, dem sich ausgangsseitig 
eine reelle St"orung "uberlagert. Zur vollst"andigen Beschreibung der
zweiten Momente des Approximationsfehlerprozesses gen"ugt daher die Angabe 
der reellen AKF, aus der man im hier angenommenen
Fall eines station"aren Approximationsfehlerprozesses durch diskrete
Fouriertransformation das geradesymmetrische reelle LDS
\mbox{$\Phi_{\boldsymbol{n}}(\Omega)$} gewinnt. Das bei einem
komplexwertigen System noch anzugebende MLDS
\mbox{$\Psi_{\boldsymbol{n}}(\Omega)$} ist hier identisch mit dem LDS.
Auf die Messung der Stufenapproximation
\mbox{$\Bar{\Psi}_{\boldsymbol{n}}(\mu)$} oder deren N"aherung
\mbox{$\Tilde{\Psi}_{\boldsymbol{n}}(\mu)$}
kann daher bei einem reellwertigen System verzichtet werden. 
Auch bei einem reellwertigen System liefert die Minimierung des
zweiten Moments des Approximationsfehlers eine L"osung, bei der die
Orthogonalit"at des Spektrums der Erregung und des Spektrums des
gefensterten Approximationsfehlers gegeben ist, so dass auch hier Gleichung
(\ref{2.30}) erf"ullt ist.

Um Sch"atzwerte f"ur die optimalen Regressionskoeffizienten durch eine Messung
zu erhalten, erregt man das System mit dem in Kapitel \ref{RKM} beschriebenen
Verfahren, also mit $L$ reellen Testsignalsequenzen, die bereichsweise
mit $M$ periodisch fortgesetzt sind. Was man misst, ist nur der Realteil
der $L$ Stichprobenelemente des Ausgangssignals des realen Systems.
Der Imagin"arteil der $L$ Systemausgangssignale wird bei der Berechnung
der Messwerte und ihrer Varianzen und Kovarianzen zu Null gesetzt.
Da jedes Stichprobenelement (\,$=$~Signalabschnitt einer Einzelmessung\,)
sowohl am Systemein- als auch am -ausgang reell ist, weisen beide
Stichprobenvektoren \mbox{$\Vec{V}(\mu)$} und
\mbox{$\Vec{Y}_{\!f}(\mu)$} die Symmetrie
\begin{equation}
\Vec{V}(\mu)=\Vec{V}(\!-\mu)^{\Kk}
\quad\text{ bzw. }\quad
\Vec{Y}_{\!f}(\mu)=\Vec{Y}_{\!f}(\!-\mu)^{\Kk}
\label{4.4}
\end{equation}
auf. Die $M$ Ausgleichsl"osungen (\ref{3.14}) der $M$
Gleichungssysteme (\ref{3.11}) besitzen dann ebenfalls die Symmetrieeigenschaft
\begin{equation}
\Hat{H}(\mu)=\Hat{H}(\!-\mu)^{\Kk}.
\label{4.5}
\end{equation}
Daher gen"ugt es bei geradem $M$ --- wovon wir im weiteren ausgehen ---
die jeweils \mbox{$M/2+1$} Werte der L"osung f"ur \mbox{$\mu=0\;(1)\;M/2$}
zu berechnen. Da wir bisher
keine Modifikationen im Messverfahren vorgenommen haben, braucht
die Erwartungstreue der Messwerte \mbox{$\Hat{\boldsymbol{H}}(\mu)$}
nicht gesondert gezeigt zu werden. 

Auch bei einem reellwertigen System verwenden wir f"ur die Absch"atzung der 
Stufenapproximation des LDS die immer reellen und erwartungstreuen Messwerte
\mbox{$\Hat{\Phi}_{\boldsymbol{n}}(\mu)$} nach Gleichung (\ref{3.34}).
Dabei verwenden wir weiterhin die Matrix \mbox{$\underline{V}_{\bot}\!(\mu)$}, 
die nach Gleichung (\ref{3.24}) den Vektor \mbox{$\Vec{V}(\mu)$} 
als Eigenvektor zum Eigenwert Null aufweist. Bei der Konstruktion dieser 
Matrix nach Gleichung (\ref{3.45}) ergibt sich nun jedoch ein wesentlicher
Unterschied. F"ur komplexwertige Systeme wurde diese mit Hilfe der in 
Gleichung (\ref{3.40}) definierten Matrix \mbox{$\Hat{\underline{V}}(\mu)$}
generiert. Da bei einem reellwertigen System die beiden Zeilenvektoren
\mbox{$\Vec{V}(\mu)$} und \mbox{$\Vec{V}(\!-\mu)^{\Kk}$} der 
Matrix \mbox{$\Hat{\underline{V}}(\mu)$} immer gleich sind, w"are die mit 
Gleichung (\ref{3.42}) berechenbare empirische Kovarianzmatrix immer singul"ar, 
und somit eine Berechnung der Matrix \mbox{$\underline{V}_{\bot}\!(\mu)$} mit 
Gleichung (\ref{3.45}) nicht m"oglich. Daher entfernen wir nun den Vektor
\mbox{$\Vec{V}(\!-\mu)^{\Kk}$} aus der Matrix \mbox{$\Hat{\underline{V}}(\mu)$}, 
so dass 
\begin{equation}
\Hat{\underline{V}}(\mu)\;=\;
\Vec{V}(\mu)
\qquad\qquad\forall\qquad\mu=0\;(1)\;M\!-\!1
\label{4.6}
\end{equation}
gilt. Die weitere Berechnung der Matrix \mbox{$\underline{V}_{\bot}\!(\mu)$} kann
dann wieder mit Hilfe der Gleichungen (\ref{3.42}) und (\ref{3.45}) erfolgen und ergibt:
\begin{equation}
\underline{V}_{\bot}\!(\mu)\;=\;\underline{E}-
\frac{\Vec{V}(\mu)^{\Hh}\!\CdoT\Vec{V}(\mu)}
{\,\Vec{V}(\mu)\CdoT\Vec{V}(\mu)^{\Hh}}
{\T\qquad\qquad\forall\qquad \mu=0\;(1)\;\frac{M}{2}}.
\label{4.7}
\end{equation}
Dass auch diese Matrix hermitesch und idempotent ist, und den Vektor \mbox{$\Vec{V}(\mu)$} 
als Eigenvektor zum Eigenwert Null aufweist, zeigt man wie im Fall des komplexwertigen 
Systems. Die Spur dieser Matrix ist nun jedoch \mbox{$L\!-\!1$} und somit um eins gr"o"ser.
Indem wir die modifiziert konstruierte Matrix \mbox{$\underline{V}_{\bot}\!(\mu)$} 
in die Gleichung (\ref{3.34}) einsetzen, erhalten wir die gegen"uber der Gleichung (\ref{3.56}) 
leicht modifizierten Messwerte:
\begin{gather}
\Hat{\Phi}_{\boldsymbol{n}}(\mu)\;=\;
\frac{1}{M\CdoT(L\!-\!1)}\cdot\Vec{N}_{\!f}(\mu)\CdoT
\underline{V}_{\bot}\!(\mu)\CdoT\Vec{N}_{\!f}(\mu)^{\Hh}\,=
\label{4.8}\\[10pt]
=\;\frac{1}{M\CdoT(L\!-\!1)}\cdot\Vec{Y}_{\!f}(\mu)\CdoT
\underline{V}_{\bot}\!(\mu)\CdoT\Vec{Y}_{\!f}(\mu)^{\Hh}\,=
\notag\\[10pt]
=\;\frac{1}{M\CdoT(L\!-\!1)}\cdot\Vec{Y}_{\!f}(\mu)\cdot
\Big(\,\underline{E}-\frac{1}{L}\CdoT
\Vec{V}(\mu)^{\Hh}\!\Cdot
\Hat{C}_{\boldsymbol{V}(\mu),\boldsymbol{V}(\mu)}^{\uP{0.4}{-1}}\!\CdoT
\Vec{V}(\mu)\,\Big)\cdot\Vec{Y}_{\!f}(\mu)^{\Hh}\;=
\notag\\[10pt]
=\frac{L}{M\CdoT(L\!-\!1)}\CdoT\Bigg(\frac{
\Vec{Y}_{\!f}(\mu)\CdoT
\Vec{Y}_{\!f}(\mu)^{\Hh}\!}{L}-\frac{
\Vec{Y}_{\!f}(\mu)\CdoT
\Vec{V}(\mu)^{\Hh}\!}{L}\cdot
\Hat{C}_{\boldsymbol{V}(\mu),\boldsymbol{V}(\mu)}^{\uP{0.4}{-1}}\Cdot
\frac{\Vec{V}(\mu)\CdoT
\Vec{Y}_{\!f}(\mu)^{\Hh}}{L}\Bigg)\;=
\notag\\[10pt]
=\;\frac{L}{M\CdoT(L\!-\!1)}\cdot\Big(\!
\Hat{C}_{\boldsymbol{Y}_{\!\!\!f}(\mu),\boldsymbol{Y}_{\!\!\!f}(\mu)}-
\Hat{C}_{\boldsymbol{Y}_{\!\!\!f}(\mu),\boldsymbol{V}(\mu)}\Cdot
\Hat{C}_{\boldsymbol{V}(\mu),\boldsymbol{V}(\mu)}^{\uP{0.4}{-1}}\CdoT
\Hat{C}_{\boldsymbol{Y}_{\!\!\!f}(\mu),\boldsymbol{V}(\mu)}^{\,\Kk}\Big)\;=
\notag\\[10pt]
=\;\frac{\Vec{Y}_{\!f}(\mu)\CdoT\Vec{Y}_{\!f}(\mu)^{\Hh}-
\big|\Hat{H}(\mu)\big|^2
\!\CdoT\Vec{V}(\mu)\CdoT\Vec{V}(\mu)^{\Hh}}
{M\CdoT(L\!-\!1)}
\notag\\[10pt]
\forall\qquad\mu=0\;(1)\;\frac{M}{2}.\notag
\end{gather}

Die Messwerte der "Ubertragungsfunktion sind f"ur die beiden diskreten
Frequenzen \mbox{$\mu\!=\!0$} und \mbox{$\mu\!=\!M/2$} immer reell. F"ur diese
beiden Messwerte kann in der Gleichung (\ref{3.63}) f"ur die Messwertkovarianz
aufgrund der Identit"at (\ref{4.1}) 
\mbox{$\Hat{\boldsymbol{C}}_{\boldsymbol{V}(\mu),\boldsymbol{V}(-\mu)^{\Kk}}^*$} mit 
\mbox{$\Hat{\boldsymbol{C}}_{\boldsymbol{V}(\mu),\boldsymbol{V}(\mu)}$} gek"urzt 
werden. Au"serden kann \mbox{$\Tilde{\Psi}_{\boldsymbol{n}}(\mu)$} durch 
\mbox{$\Tilde{\Phi}_{\boldsymbol{n}}(\mu)$} ersetzt werden. Damit ergibt 
sich in den Gleichungen (\ref{3.58}) und (\ref{3.63}) jeweils eine 
Messwertkovarianz, die gleich der Messwertvarianz ist. Ebenso sind deren 
Sch"atzwerte gem"a"s der Gleichungen (\ref{3.59}) und (\ref{3.64}) gleich. 
Die L"ange der k"urzeren Halbachse der Konfidenzellipse wird hier zu Null. 
Daher ist es f"ur diese beiden diskreten Frequenzen sinnvoll --- statt der 
Konfidenzellipsen --- Konfidenzintervalle analog zu den Konfidenzintervallen 
der LDS-Messwerte nach Gleichung (\ref{3.73}) anzugeben. 
Die halbe Intervallbreite wird mit Gleichung (\ref{3.72}) abgesch"atzt, 
wobei hier die Sch"atzwerte der Varianzen von 
\mbox{$\Hat{\boldsymbol{\Phi}}_{\boldsymbol{n}}(\mu)$} durch die Sch"atzwerte
der Varianzen von \mbox{$\Hat{\boldsymbol{H}}(\mu)$} zu ersetzen sind. 
Bei den komplexen Messwerten der "Ubertragungsfunktion \mbox{aller} anderen 
Frequenzen wird die Messwertkovarianz bei Verwendung einer hoch frequenzselektiven
Fensterfolge gegen"uber der Messwertvarianz wieder vernachl"assigbar klein,
so dass man auch beim reellwertigen System Konfidenzkreise erh"alt, deren
Radius man mit Gleichung (\ref{3.80.c}) aus der Messwertvarianz und dem 
gew"unschten Konfidenz\-niveau absch"atzt. 

Die Varianzen der Messwerte
\mbox{$\Hat{\boldsymbol{\Phi}}_{\boldsymbol{n}}(\mu)$} lassen sich mit
den Gleichungen (\ref{3.66.a}) und (\ref{3.68}) berechnen:
\begin{equation}
C_{\Hat{\boldsymbol{\Phi}}_{\!\boldsymbol{n}}(\mu),\Hat{\boldsymbol{\Phi}}_{\!\boldsymbol{n}}(\mu)}\;=\;
\begin{cases}
{\D\;\frac{2}{L\!-\!1}\cdot\Tilde{\Phi}_{\boldsymbol{n}}(\mu)^{\uP{0.4}{\!2}}}&
\text{ f"ur }\qquad\mu\in\big\{0\,;\frac{M}{2}\big\}\\[8pt]
{\D\;\frac{1}{L\!-\!1}\cdot\Tilde{\Phi}_{\boldsymbol{n}}(\mu)^{\uP{0.4}{\!2}}}&
\text{ f"ur}\qquad\mu=1\;(1)\;\frac{M-1}{2}.
\end{cases}
\label{4.9}
\end{equation}
Wir verwenden
\begin{equation}
\Hat{C}_{\Hat{\boldsymbol{\Phi}}_{\!\boldsymbol{n}}(\mu),\Hat{\boldsymbol{\Phi}}_{\!\boldsymbol{n}}(\mu)}\;=\;
\begin{cases}
{\D\;\frac{2}{L\!+\!1}\cdot\Hat{\Phi}_{\boldsymbol{n}}(\mu)^{\uP{0.4}{\!2}}}&
\text{ f"ur}\qquad\mu\in\big\{0\,;\frac{M}{2}\big\}\\[8pt]
{\D\;\frac{1}{L}\cdot\Hat{\Phi}_{\boldsymbol{n}}(\mu)^{\uP{0.4}{\!2}}}&
\text{ f"ur}\qquad\mu=1\;(1)\;\frac{M-1}{2}
\end{cases}
\label{4.10}
\end{equation}
als erwartungstreue Sch"atzwerte der Messwertvarianzen, mit deren 
Hilfe man die halbe Breite der Konfidenzintervalle nach Gleichung 
(\ref{3.73}) mit Gleichung (\ref{3.72}) absch"atzt.

Diese Variante des RKM ist in \cite{Dong} vorgestellt und untersucht worden.
Dort wurde allerdings keine Fensterung (\,$\widehat{=}$ Rechteckfenster der 
L"ange $M$\,) des Ausgangssignals des realen Systems vorgenommen
und es werden dort Messwerte und Sch"atzwerte f"ur die Messwertvarianzen
angegeben, die in einigen Details von den hier angegebenen abweichen.
F"ur gen"ugend gro"se Mittelungszahlen $L$ sind die Abweichungen
vernachl"assigbar.\newpage

\section{Spezielle Testsignale bei reellwertigen Systemen}\label{myChirp}

Bisher haben wir das reellwertige System in den unterschiedlichen
Zeitintervallen aller Einzelmessungen immer mit konkreten Realisierungen 
erregt, die unabh"angig aus einem Zufallsvektor $\Vec{\boldsymbol{v}}$ bzw. 
$\Vec{\boldsymbol{V}}$ gewonnen wurden. Bei den Stichprobenvektoren 
\mbox{$\Vec{V}(\mu)$} handelte es sich also um die Zeilenvektoren einer 
konkreten Realisierung einer zuf"alligen Stichprobenmatrix also einer 
mathematischen Stichprobe von Umfang $L$ des Zufallsvektors $\Vec{\boldsymbol{V}}$.
Unter Umst"anden kann man jedoch auch Zufallssignalsequenzen zur Erregung 
des Systems verwenden, die bei den Einzelmessungen voneinander abh"angig sind.
Man muss sich dann jedoch sicher sein, dass die gemessenen Werte
der "Ubertragungsfunktion --- bei geeigneter Wahl des erregenden 
Zufallsvektors $\Vec{\boldsymbol{V}}$ --- nicht durch diese Abh"angigkeit 
der einzelnen Stichprobenelemente verf"alscht werden und auch dann mit 
den Optimall"osungen der theoretischen Regression "ubereinstimmen. 
Trotz der Abh"angigkeit der Zufallssignalsequenzen der Erregung kann 
es sein, dass die einzelnen Stichprobenelemente des Approximationsfehlers 
dennoch --- wenigstens in guter N"aherung --- unabh"angig sind. 
In diesem Fall werden bei der Messung des Leistungsdichtespekrums 
des Approximationsfehlers keine Probleme auftreten. Wenn die in den 
Einzelmessungen abh"angige Erregung zu Abh"angigkeiten in den einzelnen 
Stichprobenelementen des Approximationsfehlers f"uhrt, m"ussen heuristische 
"Uberlegungen zeigen, dass sich die Messwerte des Leistungsdichtespektrums 
durch die ver"anderte Art der Erregung nicht beeinflussen lassen.

Eine M"oglichkeit das System mit voneinander abh"angigen Zufallssignalsequenzen 
zu erregen besteht bei gerader Anzahl von Einzelmessungen darin, in zwei 
aufeinanderfolgenden Einzelmessungen am realen, reellwertigen System jeweils 
den Real- und den Imagin"arteil einer komplexen Signalfolge zu verwenden, 
wobei die komplexe Signalfolge jeweils ein Element einer konkreten 
Stichprobe halben Umfangs einer komplexen zuf"alligen Erregung mit geeigneten
stochastischen Eigenschaften ist.

Wenn wir mit \mbox{$\lambda_g=1\;(1)\;L/2$} den Lauf"|index der $L/2$
Elemente \mbox{$V_{K,\lambda_g}(\mu)$} der konkreten Stichprobe
vom Umfang $L/2$ der komplexen Zufallsgr"o"se \mbox{$\boldsymbol{V}_{\!\!K}(\mu)$}
bezeichnen, erhalten wir die $L$ Spektralwerte \mbox{$V_{\lambda}(\mu)$}
\begin{equation}
\begin{aligned}
V_{2\cdot\lambda_g-1}(\mu)&=\;
\frac{V_{K,\lambda_g}(\mu)+V_{K,\lambda_g}(\!-\mu)^{\Kk}}{2}
\\*[4pt]
V_{2\cdot\lambda_g}(\mu)\;&=\;
\frac{V_{K,\lambda_g}(\mu)-V_{K,\lambda_g}(\!-\mu)^{\Kk}}{2\cdot j}
\end{aligned}\qquad\text{ mit}\quad{\T\lambda_g=1\;(1)\;\frac{L}{2}}
\label{4.11}
\end{equation}
der konkreten Erregung im Zeitintervall der $\lambda$-ten Einzelmessung,
die immer die Symmetrie aufweisen, die f"ur Spektren reeller Signale
typisch ist. 

Wenn bei dem zu vermessenden System bei Einhaltung gewisser
Dynamikgrenzen nicht zu erwarten ist, dass die Wahl des erregenden
Zufallsvektors einen Einfluss auf die theoretisch
optimalen Regressionskoeffizienten und die stochastischen Eigenschaften
des Approximationsfehlers hat, kann man als Zufallsvektor der Spektralwerte
der komplexen Erregung entweder das komplexe Mehrtonsignal oder das
komplexe Chirpsignal verwenden. Beide wurden in Kapitel \ref{Spesig}
vorgestellt, und weisen ein Spektrum auf, dessen Betrag \mbox{$|V_K(\mu)|$}
nicht zuf"allig ist. Mit diesen Signalen erh"alt man f"ur das empirische
zweite Moment
\begin{gather}
\Vec{V}(\mu)\CdoT\Vec{V}(\mu)^{\Hh}\;=\;
\Sum{\lambda=1}{L}\,\big|V_{\lambda}(\mu)\big|^2\;=\;
\Sum{\lambda_g=1}{L/2}\bigg(\big|V_{2\cdot\lambda_g-1}(\mu)\big|^2\!+
\big|V_{2\cdot\lambda_g}(\mu)\big|^2\bigg)\;=
\notag\\[8pt]
=\;\Sum{\lambda_g=1}{L/2}\Bigg(\bigg|
\frac{\,V_{K,\lambda_g}(\mu)\!+\!
V_{K,\lambda_g}(\!-\mu)^{\Kk}}{2}
\bigg|^2\!\!+\bigg|
\frac{\,V_{K,\lambda_g}(\mu)\!-\!
V_{K,\lambda_g}(\!-\mu)^{\Kk}}{2\cdot j}
\bigg|^2\Bigg)\;=
\notag\\[8pt]
=\;\Sum{\lambda_g=1}{L/2}\frac{\D\,|V_{K,\lambda_g}(\mu)|^2\!+
|V_{K,\lambda_g}(\!-\mu)|^2}{2}\;=\;
\Sum{\lambda_g=1}{L/2}\frac{\D\,|V_K(\mu)|^2\!+|V_K(\!-\mu)|^2}{2}\;=
\notag\\[8pt]
=\;\frac{\D\,|V_K(\mu)|^2\!+|V_K(\!-\mu)|^2}{4}\cdot L\notag\\*[2pt]
\label{4.12}
{\T\forall\qquad\qquad\mu=0\;(1)\;\frac{M}{2}}
\end{gather}
der Spektralwerte der Erregung einen nicht zuf"alligen Wert, der nur
mehr von der Anzahl der Einzelmessungen und dem nicht zuf"alligen
Betrag \mbox{$|V_K(\mu)|$} des Spektrums des Zufallsvektors
$\Vec{\boldsymbol{V}}_{\!\!K}$ abh"angt. Er muss daher nicht durch
Akkumulation berechnet werden. 

Wenn aber nun ein reales System zu vermessen ist, bei dem man sich nicht
sicher sein kann, dass die Abh"angigkeit der Erregung zweier
aufeinanderfolgender Einzelmessungen keinen Einfluss auf die
Messergebnisse haben wird, wenn man aber dennoch annehmen kann, dass eine
Messung mit geeignet ausgesteuerten Signalen mit nichtzuf"alligem
Betragsspektrum m"oglich sein m"usste, so kann man evtl. mit den 
modifizierten reellen Mehrton- oder Chirpsignalen arbeiten,
die nun vorgestellt werden.

Bei dem Mehrtonsignal nach Gleichung (\ref{3.82}) ergibt sich die
Modifikation durch die Symmetrieeigenschaft, die ein reelles Signal
erf"ullen muss. Man w"ahlt bei geradem $L$ bei jeder Einzelmessung 
die nichtzuf"alligen Teile der Spektralwerte der Mehrtonsignale in 
Gleichung (\ref{3.82}) gem"a"s
\begin{equation}
V(\!-\mu)=V(\mu)^{\Kk}
{\T\qquad\qquad\forall\qquad\mu=1\;(1)\;\frac{M}{2}\!-\!1,}
\label{4.13}
\end{equation}
und die Phasenwerte \mbox{$\varphi_{\lambda}(\mu)$} nur f"ur die
diskreten Frequenzen \mbox{$\mu=1\;(1)\;M/2\!-\!1$} als konkrete Stichprobe
aus einem reellen Zufallsvektor mit entsprechend vielen Elementen. Meist
wird man diese zwischen $-\pi$ und $\pi$ gleichverteilt und f"ur diese
Frequenzen unabh"angig w"ahlen. Da eine reelle Folge immer eine
schiefsymmetrische Phase aufweisen muss, ergeben sich die Phasenwerte
negativer diskreter Frequenzen zu
\begin{equation}
\varphi_{\lambda}(\!-\mu)=-\varphi_{\lambda}(\mu)
{\T\qquad\qquad\forall\qquad\mu=1\;(1)\;\frac{M}{2}\!-\!1.}
\label{4.14}
\end{equation}
Die Phasenwerte der beiden Frequenzen \mbox{$\mu\!=\!0$} und
\mbox{$\mu\!=\!M/2$} kann man gleichwahrscheinlich aus den zwei Werten $0$
und $\pi$ w"ahlen. Mit dem so konstruierten reellen
Mehrtonsignal kann man unabh"angige Signalsequenzen zur Erregung des Systems
in den Zeitintervallen der Einzelmessungen gewinnen, und man hat trotzdem
ein nicht zuf"alliges empirisches zweites Moment der Spektralwerte der Erregung, das
sich wieder nach Gleichung (\ref{3.83}) berechnet. Wenn die Freiheit der
Wahl des Vorzeichen der Spektralwerte \mbox{$V_{\lambda}(0)$} und
\mbox{$V_{\lambda}(M/2)$} nicht ausreichend erscheint, kann man
f"ur diese beiden Frequenzen auch den Kosinus einer Phase verwenden, die man
wie die anderen Phasenwerte aus einer Zufallsgr"o"se unabh"angig gewinnt.
Dann ist allerdings f"ur die Spektralwerte dieser beiden diskreten
Frequenzen das empirische zweite Moment durch Akkumulation zu ermitteln.

Das Chirpsignal nach Gleichung (\ref{3.85}) mit dem Spektrum nach Gleichung
(\ref{3.86}), das gem"a"s Gleichung (\ref{3.84}) mit einem bei allen
Frequenzen gleichen Drehfaktor mit einer Zufallsphase multipliziert
wird, weist immer ein Spektrum auf, bei dem sowohl der Real- als auch der
Imagin"arteil geradesymmetrisch ist. Daher gelingt es nicht, durch geeignete
Wahl einiger Nebenbedingungen daraus ein reelles zuf"alliges Testsignal
zu konstruieren, dessen Spektrum die Symmetrie aufweist, die bei reellen
Signalen immer vorhanden ist. Stattdessen kann man aber ein
modifiziertes Chirpsignal verwenden, dessen Spektrum
\begin{equation}
V_{\lambda}(\mu)\;=\;V_C\cdot 
e^{\mbox{\small$\!-j\CdoT\big(\frac{2\pi}{M}\CdoT\mu\CdoT|\mu|\!+\!
\varphi_{\lambda}\CdoT\frac{M}{\pi}\CdoT
\sin(\frac{2\pi}{M}\CdoT\mu)\big)$}}
{\T\qquad\qquad\forall\qquad\mu=1\!-\!\frac{M}{2}\;(1)\;\frac{M}{2}.}
\label{4.15}
\end{equation}
beim Betragsfrequenzgang einen konstanten reellen Wert $V_C$ aufweist
und daher geradesymmetrisch ist, und dessen Phasenverlauf schiefsymmetrisch
ist. Die mit Hilfe einer DFT daraus gewonnenen Testsignale der Einzelmessungen
sind daher immer reell. Der Phasenfrequenzgang setzt sich aus
zwei Anteilen zusammen. Der erste Anteil ist quadratisch, wie dies bei
Chirpsignalen "ublich ist, und ist mit dem Vorzeichnen von $\mu$ multipliziert,
so dass dieser Phasenanteil schiefsymmetrisch ist. Im Gegensatz zum
Phasengang des komplexen Chirpsignals nach Gleichung (\ref{3.86}) ist
der Koeffizient der quadratischen Phasenkomponente nun doppelt so gro"s.
Der zweite Anteil des Phasengangs ist sinusf"ormig und daher schiefsymmetrisch
und wird zum quadratischen Anteil mit einer Amplitude addiert, die einen
Faktor $\varphi_{\lambda}$ enth"alt, der bei dem Testsignal jeder
Einzelmessung ein Element einer konkreten Stichprobe einer reellen
Zufallsgr"o"se \mbox{$\boldsymbol{\varphi}$} ist.

Der gro"se Vorteil bei der Verwendung eines komplexen Chirpsignals
als Testsignal bestand darin, dass sowohl dessen Real- als auch
dessen Imagin"arteil den guten Crest-Faktor von ca. $\sqrt{2\,}$
aufweist, und man somit bei Systemen, die eingangsseitig einen
begrenzten Bereich f"ur die Maximalaussteuerung besitzen, 
bei Vollaussteuerung eine im Vergleich zu anderen Signalen hohe
Signalleistung am Eingang des zu messenden Systems erreicht.
Dadurch erreicht man, dass man bei gleicher Messwertvarianz mit
einer deutlich geringeren Mittelungsanzahl $L$ bei der Messung der
"Ubertragungsfunktion auskommt. Es zeigte sich, dass der
Crestfaktor des hier vorgestellten reellen Chirpsignals f"ur alle
konkreten Realisierungen des Parameters $\varphi_{\lambda}$
mit der Formel
\begin{equation}
\text{Cr}(\varphi_{\lambda})\,=\,
\frac{\max|v_{\lambda}(k)|}{\!
\D\sqrt{\frac{1}{M}\CdoT\!\Sum{k=1}{M-1}v_{\lambda}(k)^2}\,}\,=\,
\frac{\max|v_{\lambda}(k)|}{\!\D\sqrt{
\frac{1}{M^2}\CdoT\!\Sum{\mu=1}{M-1}\big|V_{\lambda}(\mu)\big|^2}\,}\,=\,
\frac{\D \sqrt{\!M\,}\!\CdoT\max|v_{\lambda}(k)|\,}{\D V_C}
\,\stackrel{\T<}{\sim}\,\sqrt{\frac{2}{1-\varphi_{\lambda}\!}\,}
\label{4.16}
\end{equation}
--- vor allen f"ur gro"se Werte von $M$ --- gut nach oben abgesch"atzt
werden kann. Wenn man eine Zufallsgr"o"se \mbox{$\boldsymbol{\varphi}$}
zur Erzeugung der konkreten Werte $\varphi_{\lambda}$ benutzt,
deren Wertebereich auf das Intervall
\begin{equation}
0 \le \varphi_{\lambda} \le \varphi_{\text{max}}
\label{4.17}
\end{equation}
mit \mbox{$\varphi_{\text{max}}<1$} beschr"ankt ist, erh"alt man
nur solche Testsignalsequenzen, deren Crestfaktor in guter N"aherung
auf den Maximalwert beschr"ankt ist, der sich mit
\mbox{$\varphi_{\text{max}}$} in der oberen Grenze der Ungleichung
(\ref{4.16}) ergibt. Andererseits kann man sich einen gew"unschten
Maximalwert \mbox{$\text{Cr}_{\text{max}}$} f"ur den Crestfaktor
gr"o"ser $\sqrt{2}$ vorgeben, und erh"alt mit der Umkehrung dieser
empirisch gewonnenen Formel eine Aussteuerungsgrenze f"ur die
Werte des zuf"alligen Phasenhubs des sinusf"ormigen Anteils an der
Gesamtphase. Gibt man sich beispielsweise den maximalen Crestfaktor
$1,\!5$ vor, so erh"alt man mit
\begin{figure}[btp]
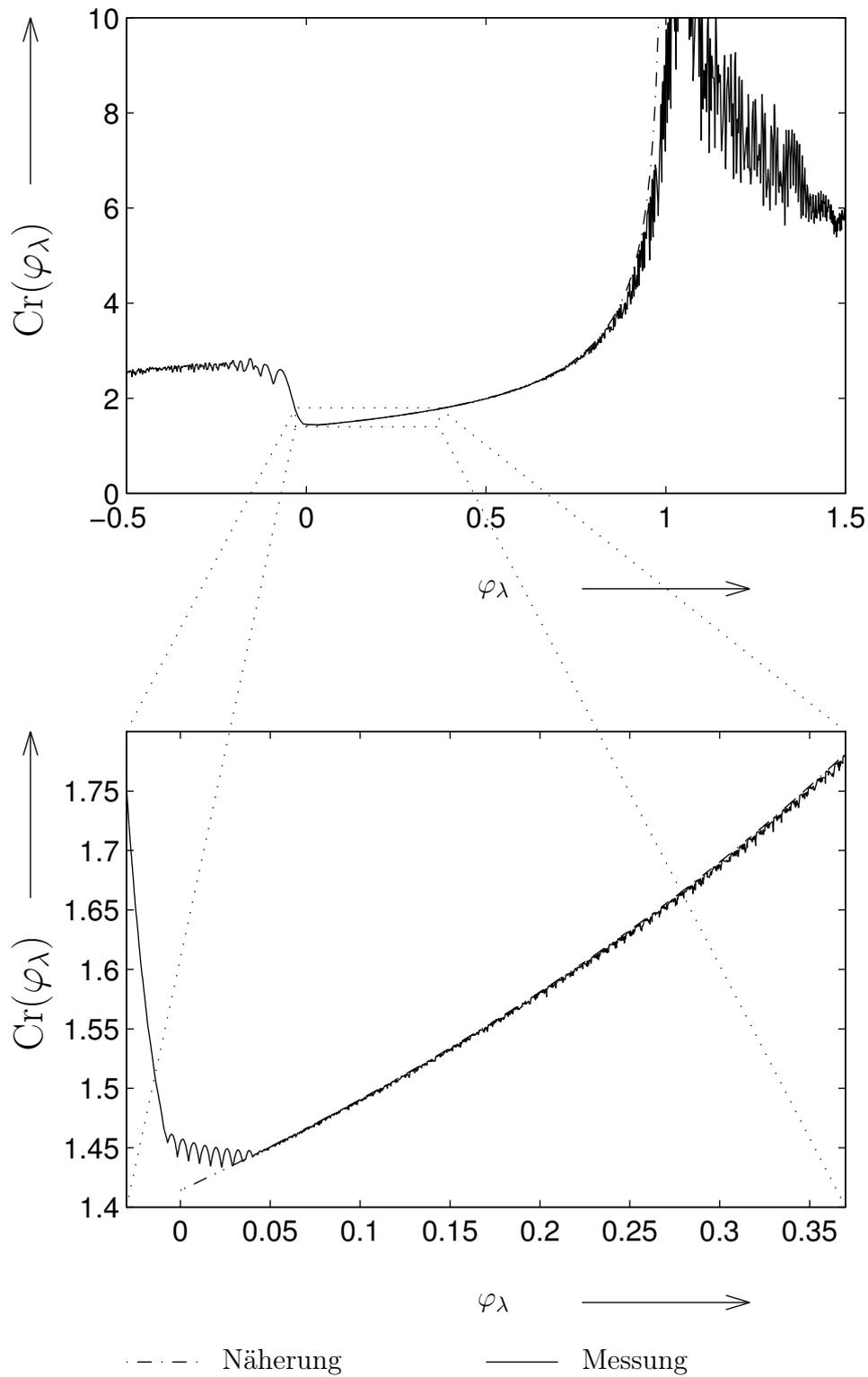

{ 
\begin{picture}(420,590)
\input{mbild3c1}
\input{mbild3c2}
\input{mbild3c3}
\put(250,30){\makebox(0,0)[r]{$\varphi_{\lambda}$}}
\put(250,330){\makebox(0,0)[r]{$\varphi_{\lambda}$}}
\put(42,180){\rotatebox[origin = cr]{90}{\Large Cr$(\varphi_{\lambda})$}}
\put(42,480){\rotatebox[origin = cr]{90}{\Large Cr$(\varphi_{\lambda})$}}
\put(130,5){\makebox(0,0)[l]{N"aherung}}
\put(280,5){\makebox(0,0)[l]{Messung}}
\end{picture}}
\caption{Crestfaktor des Chirpsignals f"ur M = 1024}
\label{b3c}
\end{figure}
\begin{figure}[btp]
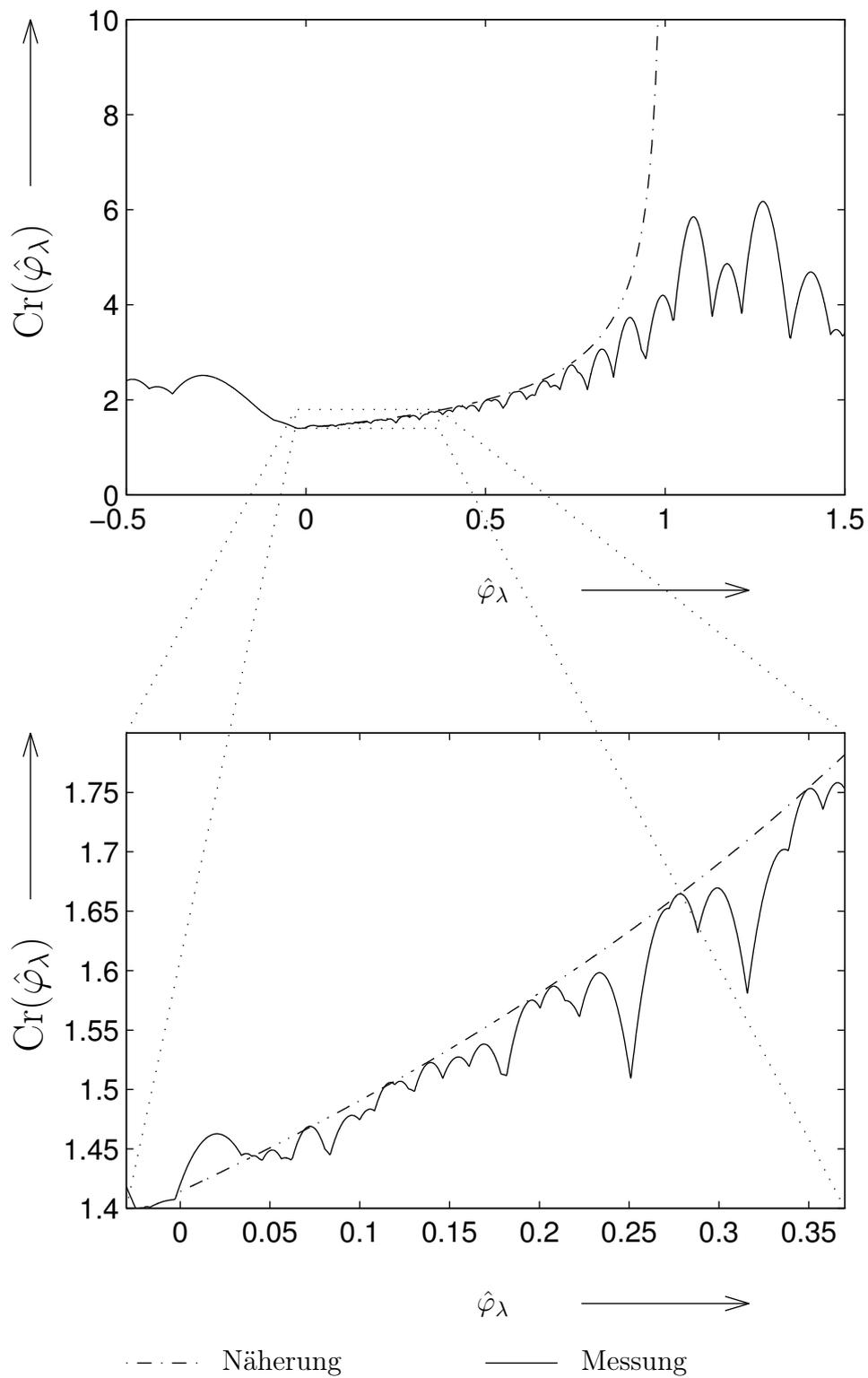

{ 
\begin{picture}(420,590)
\input{mbild3d1}
\input{mbild3d2}
\input{mbild3d3}
\put(250,30){\makebox(0,0)[r]{$\hat{\varphi}_{\lambda}$}}
\put(250,330){\makebox(0,0)[r]{$\hat{\varphi}_{\lambda}$}}
\put(42,180){\rotatebox[origin = cr]{90}{\Large Cr$(\hat{\varphi}_{\lambda})$}}
\put(42,480){\rotatebox[origin = cr]{90}{\Large Cr$(\hat{\varphi}_{\lambda})$}}
\put(130,5){\makebox(0,0)[l]{N"aherung}}
\put(280,5){\makebox(0,0)[l]{Messung}}
\end{picture}}
\caption{Crestfaktor des Chirpsignals f"ur M = 64}
\label{b3d}
\end{figure}
\begin{equation}
\varphi_{\text{max}}\;=\;1-\frac{2}{\D\;\text{Cr}_{\text{max}}^2\;}
\label{4.18}
\end{equation}
f"ur die obere Grenze \mbox{$\varphi_{\text{max}}$} der Verteilungsdichte
der Zufallsgr"o"se \mbox{$\boldsymbol{\varphi}$} den Wert $1/9$.
Bei einer DFT-L"ange von \mbox{$M\!=\!1024$} ergibt sich damit
f"ur die Aussteuerungsgrenze der Sinuskomponente im Phasenfrequenzgang
\mbox{$\varphi_{\text{max}}\CdoT M/\pi\approx 5,\!76\cdot2\pi$}.
Bei einem Crestfaktor, der nur unwesentlich schlechter als $\sqrt{2}$
ist, ergibt sich somit eine deutliche Phasendrehung im Spektrum, so dass
man f"ur unterschiedliche Realisierungen von $\varphi_{\lambda}$
durchaus deutlich abweichende Signalsequenzen erh"alt. So werden z.~B. die
in einem realen digitalen System vorhandenen Quantisierungsstellen bei den
Einzelmessungen wohl hinreichend zuf"allig und bei den Einzelmessungen
unterschiedlich ausgesteuert. In Bild \ref{b3c} ist der Verlauf des
Crestfaktors f"ur unterschiedliche Werte $\varphi_{\lambda}$, sowie
die Absch"atzung mit Gleichung (\ref{4.16}) am Beispiel \mbox{$M=1024$}
dargestellt. Man erkennt deutlich die Unsymmetrie des Crestfaktors
f"ur positive und negative Werte von $\varphi_{\lambda}$, weshalb
das nach Ungleichung (\ref{4.17}) empfohlene Intervall bei Null beginnt. 
F"ur kleine Werte von $M$ ergeben sich deutlichere Abweichungen von
der N"aherungsformel. Die entsprechenden Kurven f"ur \mbox{$M=64$}
sind in Bild \ref{b3d} dargestellt. Bei der Berechnung dieser Graphiken 
wurden die Phasenwerte $\varphi_{\lambda}$ zuf"allig innerhalb des
dargestellten Bereichs ausgew"ahlt, um so systematische Fehler zu 
vermeiden. Bisher ist noch nicht theoretisch gekl"art worden,
warum sich gerade ein sinusf"ormiger additiver Anteil im Phasengang 
des in Gleichung (\ref{4.15}) angegebenen Chirpsignalspektrums
so gut eignet, den Crestfaktor klein zu halten. Die Gr"unde f"ur
die Unsymmetrie des Crestfaktors bei positiven und negativen Werten
von $\varphi_{\lambda}$ konnten ebenso nicht gefunden werden,
wie die Mechanismen, die zu dem empirisch gewonnenen Zusammenhang
(\ref{4.16}) f"uhren. Die Tatsache, dass die Spektralwerte des reellen
Chirpsignals nach Gleichung (\ref{4.15}) besonders f"ur niedrige Frequenzen
nicht mittelwertfrei sind, da die zuf"allige Phasendrehung der Spektrallinien
sich nicht "uber mehrere Vielfache von \mbox{$2\pi$} erstreckt, stellt kein
Problem dar, da man das reelle Chirpsignal mit einem zuf"alligen und
gleichwahrscheinlichen Vorzeichen versehen kann.

\chapter{Spektralsch"atzung mittelwertfreier station"arer Prozesse}\label{LDS}

Wir beginnen dieses Kapitel mit einigen Vor"uberlegungen. 
Nach der folgenden Darstellung des Ablaufs des Spektralsch"atzverfahrens wird die 
Erwartungstreue der Sch"atzwerte gezeigt. Es folgt die Berechnung der Messwertvarianzen 
und es wird gezeigt wie auch diese sich erwartungstreu absch"atzen lassen. 
Abschlie"send wird der Spezialfall der Spektralsch"atzung reeller 
Prozesse behandelt.

\section{Vor"uberlegungen}

Bei der Spektralsch"atzung mittelwertfreier, station"arer 
Prozesse kann auf Ergebnisse zur"uckgegriffen werden, die bei der 
Herleitung des RKM gewonnen wurden. Dies ist deshalb m"oglich, weil 
das Spektralsch"atzverfahren den Sonderfall des RKM darstellt, bei dem das 
reale System, wie es in Bild \ref{b1h} dargestellt ist, nicht erregt wird. 
In diesem Fall ist das Signal \mbox{$\boldsymbol{x}(k)$} am Ausgang des 
linearen zeitinvarianten Modellsystems ebenfalls Null, und es gilt 
\mbox{$\boldsymbol{n}(k)=\boldsymbol{y}(k)$}. Daher sind die stochastischen 
Eigenschaften wie z.~B. das LDS oder das MLDS des Prozesses 
\mbox{$\boldsymbol{y}(k)$} zugleich die messtechnisch zu bestimmenden
Merkmale des Prozesses \mbox{$\boldsymbol{n}(k)$}. Eine optimale Approximation 
mit den Abtastwerten der "Ubertragungsfunktion als Freiheitsgrade der Approximation 
ist hier weder theoretisch noch praktisch beim Messverfahren durchzuf"uhren.

Wie wir in Gleichung (\ref{2.17}) gesehen haben, geht in die Erwartungswerte 
\mbox{$\Tilde{\Phi}_{\boldsymbol{n}}(\mu)$} der gefensterten Periodogramme, 
die wir zur Beschreibung des LDS \mbox{$\Phi_{\boldsymbol{n}}(\Omega)$} durch 
endlich viele Werte verwenden, der Phasengang des verwendeten Fensters nicht ein. 
Daher kann auch im Fall der Spektralsch"atzung mittelwertfreier station"arer 
Prozesse das in Kapitel \ref{Algo} vorgestellte Fenster, 
das einen nichtlinearen Phasengang aufweist, verwendet werden. 
Bei der Spektralsch"atzung --- also der empirischen Absch"atzung dieser 
Werte \mbox{$\Tilde{\Phi}_{\boldsymbol{n}}(\mu)$} --- muss die 
Nullstellenbedingung (\ref{2.27}) f"ur das Spektrum der Fensterfolge 
nicht erf"ullt werden. Es empfiehlt sich jedoch auch hier eine 
Fensterfolge zu verwenden, die der Bedingung (\ref{2.20}) gen"ugt, 
um auch im Spektralbereich die richtige Varianz als Mittelwert 
aller $M$ Werte \mbox{$\Tilde{\Phi}_{\boldsymbol{n}}(\mu)$} zu erhalten. 
Da diese Forderung von der im Kapitel \ref{Algo} vorgestellten Fensterfolge 
ebenso erf"ullt wird, wie auch die bei der Spektralsch"atzung gew"unschte 
hohe Frequenzselektivit"at, ist die Verwendung dieser Fensterfolge immer 
dann zu empfehlen, wenn das zu messende LDS "uber der Frequenz stark schwankt.

Wie wir in Kapitel \ref{W} gesehen haben, ist bei komplexwertigen 
Zufallsprozessen die Beschreibung der zweiten Momente erst vollst"andig, 
wenn man auch das modifizierte Leis"|tungs"|dich"|te"|spek"|trum (\,MLDS\,) 
\mbox{$\Psi_{\boldsymbol{n}}(\Omega)$} des Prozesses angibt. Dort wurden 
auch die $M$ Erwartungswerte \mbox{$\Tilde{\Psi}_{\boldsymbol{n}}(\mu)$}
vorgestellt, die wir zur Beschreibung des MLDS verwenden wollen.
Es wird daher hergeleitet, wie man neben den $M$ Sch"atzwerten 
\mbox{$\Hat{\Phi}_{\boldsymbol{n}}(\mu)$} f"ur die Erwartungswerte 
\mbox{$\Tilde{\Phi}_{\boldsymbol{n}}(\mu)$} auch $M$ Sch"atzwerte 
\mbox{$\Hat{\Psi}_{\boldsymbol{n}}(\mu)$} f"ur die Erwartungswerte 
\mbox{$\Tilde{\Psi}_{\boldsymbol{n}}(\mu)$}, mit Hilfe einer Messung 
gewinnt, die um die Fensterung erweitert ist. Wir beschr"anken uns 
nun dabei wieder auf den Fall eines mittelwertfreien station"aren 
Prozesses. Wie die Spektralsch"atzung f"ur mittelwertbehaftete 
und zyklostation"are Prozesse abl"auft, wird in \cite{Erg} erl"autert. 

\section{LDS- und MLDS-Messwerte}

Die Messung beginnt damit, dass wir eine Stichprobe vom Umfang $L$ erheben, 
indem wir $L$ Musterfolgen \mbox{$y_{\lambda}(k)$} mit \mbox{$\lambda=1\;(1)\;L$}
der L"ange $F$ (d.~h. \mbox{$k=0\;(1)\;F\!-\!1$}) unabh"angig aus dem Prozess
\mbox{$\boldsymbol{y}(k)$} entnehmen. Diese Musterfolgen werden gefenstert und einer 
DFT unterworfen, wie dies in der Aufz"ahlung auf Seite \pageref{yFen} beschrieben ist. 
Wir erhalten die $M$ nach Gleichung (\ref{3.12}) definierten $M$ Vektoren 
\mbox{$\Vec{Y}_{\!f}(\mu)$}, die hier gem"a"s Gleichung (\ref{3.18}) mit 
\mbox{$\Vec{V}(\mu)=\Vec{0}$} mit den $M$ Vektoren \mbox{$\Vec{N}_{\!f}(\mu)$} "ubereinstimmen. 
Mit Hilfe dieser Vektoren lassen sich die Messwerte wie folgt berechnen: 
\begin{equation}
\Hat{\Phi}_{\boldsymbol{n}}(\mu)\;=\;
\frac{1}{M\CdoT L}\cdot\Vec{Y}_{\!f}(\mu)\CdoT\Vec{Y}_{\!f}(\mu)^{\Hh}\;=\;
\frac{1}{M}\cdot\Hat{C}_{\boldsymbol{Y}(\mu),\boldsymbol{Y}(\mu)}
\qquad\forall\quad\mu=0\;(1)\;M\!-\!1
\label{5.1}
\end{equation}
und
\begin{equation}
\Hat{\Psi}_{\boldsymbol{n}}(\mu)\;=\;
\frac{1}{M\CdoT L}\cdot\Vec{Y}_{\!f}(\mu)\CdoT\Vec{Y}_{\!f}(\!-\mu)^{\Tt}\;=\;
\frac{1}{M}\cdot\Hat{C}_{\boldsymbol{Y}(\mu),\boldsymbol{Y}(-\mu)^{\Kk}}
\qquad\forall\quad\mu=0\;(1)\;M\!-\!1
\label{5.2}
\end{equation}
Aufgrund der Cauchy-Schwarzschen Ungleichung (\ref{3.36}) erf"ullen diese
Messwerte immer die Bedingung (\ref{3.38}).

Die Erwartungswerte dieser Messwerte berechnet man wieder mit den 
Gleichungen (\ref{3.30}). Dabei werden nun jedoch die Vektoren 
\mbox{$\Hat{\Vec{N}}_{\!f}(\mu)$} nicht nach Gleichung (\ref{3.25})
definiert, sondern sind identisch mit den Vektoren \mbox{$\Vec{N}_{\!f}(\mu)$},
was dort einer Abbildung mit der \mbox{$L\!\times\!L$} Einheitsmatrix 
\mbox{$\underline{E}$} entspricht. In den Gleichungen (\ref{3.30}) 
ist daher die Matrix \mbox{$\underline{V}_{\Phi}(\mu)$} durch die Matrix 
\mbox{$\underline{E}/(M\CdoT L)$} mit der Spur \mbox{$1/M$} zu ersetzen. 
Ein Vergleich der sich so ergebenden Erwartungswerte der Messwerte in 
Gleichung (\ref{3.30.h}) mit der Definition der zu messenden Werte nach 
Gleichung (\ref{2.17}) zeigt dann die Erwartungstreue der Messwerte 
\mbox{$\Hat{\Phi}_{\boldsymbol{n}}(\mu)$}. Analog l"asst sich auch 
die Erwartungstreue der Messwerte \mbox{$\Hat{\Psi}_{\boldsymbol{n}}(\mu)$}
zeigen.

\section{Messwertvarianzen und -kovarianzen}

Wir nehmen wieder an, dass die Zufallsgr"o"sentupel der Spektralwerte
\mbox{$\big[\boldsymbol{N}_{\!\!f}(\mu),\boldsymbol{N}_{\!\!f}(\!-\mu)\big]^{\TT}$}
verbundnormalverteilt sind. In diesem Fall erhalten wir die Varianzen und Kovarianzen der Messwerte 
\mbox{$\Hat{\Phi}_{\boldsymbol{n}}(\mu)$} und \mbox{$\Hat{\Psi}_{\boldsymbol{n}}(\mu)$} 
wie beim RKM mit Hilfe der Gleichungen (\ref{3.65}) bis (\ref{3.68}), wobei wir nun jedoch 
die Matrix \mbox{$\underline{\boldsymbol{V}}_{\bot}\!(\pm\mu)$} durch die 
Einheitsmatrix \mbox{$\underline{E}$}, deren Spur $L$ ist, zu ersetzen haben.
Wir erhalten f"ur die theoretischen Messwert(ko)varianzen:
\begin{subequations}\label{5.3}
\begin{align}
C_{\Hat{\boldsymbol{\Phi}}_{\!\boldsymbol{n}}(\mu),\Hat{\boldsymbol{\Phi}}_{\!\boldsymbol{n}}(\mu)}&{}\,=\,
\frac{1}{L}\cdot
\Big(\big|\Tilde{\Psi}_{\boldsymbol{n}}(\mu)\big|^2\!+
\Tilde{\Phi}_{\boldsymbol{n}}(\mu)^{\uP{0.4}{\!2}}\Big),
\label{5.3.a}\\[6pt]
C_{\Hat{\boldsymbol{\Psi}}_{\!\boldsymbol{n}}(\mu),\Hat{\boldsymbol{\Psi}}_{\!\boldsymbol{n}}(\mu)}&{}\,=\,
\frac{2}{L}\cdot
\Tilde{\Phi}_{\boldsymbol{n}}(\mu)^{\uP{0.4}{\!2}}\qquad\text{ und}
\label{5.3.b}\\[6pt]
C_{\Hat{\boldsymbol{\Psi}}_{\!\boldsymbol{n}}(\mu),\Hat{\boldsymbol{\Psi}}_{\!\boldsymbol{n}}(\mu)^{\Kk}}\!&{}\,=\,
\frac{2}{L}\cdot
\Tilde{\Psi}_{\boldsymbol{n}}(\mu)^{\uP{0.4}{\!2}}
\label{5.3.c}\\*[3pt]
&{}\;\;\,\forall{\T\qquad\qquad\mu=0\quad\vee\quad\mu=\frac{M}{2}}
\notag\\
\intertext{bzw.\vspace{-12pt}}
C_{\Hat{\boldsymbol{\Phi}}_{\!\boldsymbol{n}}(\mu),\Hat{\boldsymbol{\Phi}}_{\!\boldsymbol{n}}(\mu)}&{}\,=\,
\frac{1}{L}\cdot
\Tilde{\Phi}_{\boldsymbol{n}}(\mu)^{\uP{0.4}{\!2}}\!,
\label{5.3.d}\\[6pt]
C_{\Hat{\boldsymbol{\Psi}}_{\!\boldsymbol{n}}(\mu),\Hat{\boldsymbol{\Psi}}_{\!\boldsymbol{n}}(\mu)}&{}\,=\,
\frac{1}{L}\cdot
\Tilde{\Phi}_{\boldsymbol{n}}(\mu)\cdot
\Tilde{\Phi}_{\boldsymbol{n}}(\!-\mu)\qquad\text{ und}
\label{5.3.e}\\[6pt]
C_{\Hat{\boldsymbol{\Psi}}_{\!\boldsymbol{n}}(\mu),\Hat{\boldsymbol{\Psi}}_{\!\boldsymbol{n}}(\mu)^{\Kk}}\!&{}\,=\,
\frac{1}{L}\cdot
\Tilde{\Psi}_{\boldsymbol{n}}(\mu)^{\uP{0.4}{\!2}}
\label{5.3.f}\\*[3pt]
&{}\;\;\,\forall{\T\qquad\qquad\mu=1\;(1)\;M\!-\!1
\quad\wedge\quad\mu\neq\frac{M}{2}}.
\notag
\end{align}
\end{subequations}
Als Sch"atzwerte f"ur die Messwert(ko)varianzen verwenden wir:
\begin{subequations}\label{5.4}
\begin{align}
\Hat{C}_{\Hat{\boldsymbol{\Phi}}_{\!\boldsymbol{n}}(\mu),\Hat{\boldsymbol{\Phi}}_{\!\boldsymbol{n}}(\mu)}&{}\;=\;
\frac{L\!-\!2}{(L\!+\!2)\CdoT(L\!-\!1)}\cdot
\Hat{\Phi}_{\boldsymbol{n}}(\mu)^{\uP{0.4}{\!2}}+
\frac{L}{(L\!+\!2)\CdoT(L\!-\!1)}\cdot
\big|\Hat{\Psi}_{\boldsymbol{n}}(\mu)\big|^2
\label{5.4.a}\\*[8pt]
\Hat{C}_{\Hat{\boldsymbol{\Psi}}_{\!\boldsymbol{n}}(\mu),\Hat{\boldsymbol{\Psi}}_{\!\boldsymbol{n}}(\mu)}&{}\;=\;
\frac{2\CdoT L}{(L\!+\!2)\CdoT(L\!-\!1)}\cdot
\Hat{\Phi}_{\boldsymbol{n}}(\mu)^{\uP{0.4}{\!2}}-
\frac{2}{(L\!+\!2)\CdoT(L\!-\!1)}\cdot
\big|\Hat{\Psi}_{\boldsymbol{n}}(\mu)\big|^2
\label{5.4.b}\\[8pt]
\Hat{C}_{\Hat{\boldsymbol{\Psi}}_{\!\boldsymbol{n}}(\mu),\Hat{\boldsymbol{\Psi}}_{\!\boldsymbol{n}}(\mu)^{\Kk}}\!&{}\;=\;
\frac{2}{(L\!+\!2)}\cdot\Hat{\Psi}_{\boldsymbol{n}}(\mu)^{\uP{0.4}{\!2}}
\label{5.4.c}\\
&\qquad\qquad\qquad\qquad\qquad
\forall\qquad{\T\mu=0\quad\vee\quad\mu=\frac{M}{2}}
\notag\\[-6pt]
\intertext{bzw.\vspace{-3pt}}
\Hat{C}_{\Hat{\boldsymbol{\Phi}}_{\!\boldsymbol{n}}(\mu),\Hat{\boldsymbol{\Phi}}_{\!\boldsymbol{n}}(\mu)}&{}\;=\;
\frac{1}{L\!+\!1}\cdot
\Hat{\Phi}_{\boldsymbol{n}}(\mu)^{\uP{0.4}{\!2}}
\label{5.4.d}\\*[8pt]
\Hat{C}_{\Hat{\boldsymbol{\Psi}}_{\!\boldsymbol{n}}(\mu),\Hat{\boldsymbol{\Psi}}_{\!\boldsymbol{n}}(\mu)}&{}\;=\;
\frac{L}{(L\!+\!1)\CdoT(L\!-\!1)}\CdoT
\Hat{\Phi}_{\boldsymbol{n}}(\!-\mu)\CdoT\Hat{\Phi}_{\boldsymbol{n}}(\mu)-
\frac{1}{(L\!+\!1)\CdoT(L\!-\!1)}\CdoT
\big|\Hat{\Psi}_{\boldsymbol{n}}(\mu)\big|^2\!\!
\label{5.4.e}\\*[8pt]
\Hat{C}_{\Hat{\boldsymbol{\Psi}}_{\!\boldsymbol{n}}(\mu),\Hat{\boldsymbol{\Psi}}_{\!\boldsymbol{n}}(\mu)^{\Kk}}\!&{}\;=\;
\frac{1}{L\!+\!1}\cdot\Hat{\Psi}_{\boldsymbol{n}}(\mu)^{\uP{0.4}{\!2}}
\label{5.4.f}\\
&\qquad\qquad\qquad\qquad
\forall\qquad{\T\mu=1\;(1)\;M\!-\!1\quad\wedge\quad\mu\neq\frac{M}{2}}
\notag
\end{align}
\end{subequations}
Dass diese Sch"atzwerte erwartungstreu sind, zeigt man, indem man
zun"achst die Messwerte nach Gleichung (\ref{5.1}) und (\ref{5.2})
einsetzt, und dann mit Hilfe der Gleichung (\ref{A.5.9}) des Anhangs 
die Erwartungswerte berechnet. Dabei ist f"ur die Matrizen 
\mbox{$\underline{\boldsymbol{A}}=\underline{\boldsymbol{B}}=\underline{E}$}, 
f"ur alle Matrixspuren \mbox{$L$} und f"ur den skalaren Faktor jeweils 
\mbox{$\boldsymbol{c}=M^2\CdoT L^2$} einzusetzen.
Auch hier l"asst sich zeigen, dass die Sch"atzwerte der Messwertkovarianz
\mbox{$\Hat{C}_{\Hat{\boldsymbol{\Psi}}_{\!\boldsymbol{n}}(\mu),\Hat{\boldsymbol{\Psi}}_{\!\boldsymbol{n}}(\mu)^{\Kk}}$}
nie betraglich gr"o"ser sind als die Sch"atzwerte der Messwertvarianz
\mbox{$\Hat{C}_{\Hat{\boldsymbol{\Psi}}_{\!\boldsymbol{n}}(\mu),\Hat{\boldsymbol{\Psi}}_{\!\boldsymbol{n}}(\mu)}$},
da die Bedingung (\ref{3.38}) bei diesen Messwerten immer erf"ullt ist.

Zur Berechnung der Konfidenzintervalle (\ref{3.73}) der reellen Messwerte
\mbox{$\Hat{\boldsymbol{\Phi}}_{\boldsymbol{n}}(\mu)$} zu dem gew"unschten 
Konfidenzniveau \mbox{$1\!-\!\alpha$} setzt man die Sch"atzwerte der halben 
Intervallbreite nach Gleichung (\ref{3.72}) mit den eben berechneten 
Messwertvarianzsch"atzwerten ein. 
Die Sch"atzwerte der Halbachsen der Konfidenzellipsen der Messwerte
\mbox{$\Hat{\boldsymbol{\Psi}}_{\boldsymbol{n}}(\mu)$}
erh"alt man mit den eben berechneten Messwert(ko)varianzsch"atzwerten,
indem man diese statt der Messwert(ko)varianzsch"atzwerte
\mbox{$\Hat{C}_{\Hat{\boldsymbol{H}}(\mu),\Hat{\boldsymbol{H}}(\mu)}$} und
\mbox{$\Hat{C}_{\Hat{\boldsymbol{H}}(\mu),\Hat{\boldsymbol{H}}(\mu)^{\Kk}}$}
in die Gleichungen (\ref{3.80}) einsetzt.

\section{Spektralsch"atzung reellwertiger Zufallsprozesse}

Die Spektralsch"atzung reellwertiger, station"arer und mittelwertfreier 
Zufallsprozesse l"auft prinzipiell genau wie eben geschildert ab. 
Da bei den reellen Musterfolgen des Prozesses immer die bekannte 
Symmetrie der Spektren reeller Folgen auftritt, gilt
\begin{equation}
\Hat{\Psi}_{\boldsymbol{n}}(\mu)=
\Hat{\Phi}_{\boldsymbol{n}}(\mu)=
\Hat{\Phi}_{\boldsymbol{n}}(\!-\mu),
\label{5.5}
\end{equation}
und daher m"ussen die Messwerte \mbox{$\Hat{\Phi}_{\boldsymbol{n}}(\mu)$} 
nach Gleichung (\ref{5.1}) nur f"ur \mbox{$\mu=1\;(1)\;M/2$} und die 
Messwerte \mbox{$\Hat{\Psi}_{\boldsymbol{n}}(\mu)$} "uberhaupt nicht 
berechnet werden. Mit der letzten Gleichung erhalten wir die Messwertvarianzen 
\begin{equation}
C_{\Hat{\boldsymbol{\Phi}}_{\!\boldsymbol{n}}(\mu),\Hat{\boldsymbol{\Phi}}_{\!\boldsymbol{n}}(\mu)}\;=\;
\begin{cases}
{\D\;\frac{2}{L}\cdot\Tilde{\Phi}_{\boldsymbol{n}}(\mu)^{\uP{0.4}{\!2}}}&
\text{ f"ur }\qquad\mu\in\big\{0\,;\frac{M}{2}\big\}\\[8pt]
{\D\;\frac{1}{L}\cdot\Tilde{\Phi}_{\boldsymbol{n}}(\mu)^{\uP{0.4}{\!2}}}&
\text{ f"ur}\qquad\mu=1\;(1)\;\frac{M-1}{2}.
\end{cases}
\label{5.6}
\end{equation}
Wir verwenden
\begin{equation}
\Hat{C}_{\Hat{\boldsymbol{\Phi}}_{\!\boldsymbol{n}}(\mu),\Hat{\boldsymbol{\Phi}}_{\!\boldsymbol{n}}(\mu)}\;=\;
\begin{cases}
{\D\;\frac{2}{L\!+\!2}\cdot\Hat{\Phi}_{\boldsymbol{n}}(\mu)^{\uP{0.4}{\!2}}}&
\text{ f"ur}\qquad\mu\in\big\{0\,;\frac{M}{2}\big\}\\[8pt]
{\D\;\frac{1}{L\!+\!1}\cdot\Hat{\Phi}_{\boldsymbol{n}}(\mu)^{\uP{0.4}{\!2}}}&
\text{ f"ur}\qquad\mu=1\;(1)\;\frac{M-1}{2}
\end{cases}
\label{5.7}
\end{equation}
als erwartungstreue Sch"atzwerte der Messwertvarianzen, mit deren 
Hilfe man die halbe Breite der Konfidenzintervalle nach Gleichung 
(\ref{3.73}) mit Gleichung (\ref{3.72}) absch"atzt.

Insgesamt "ahnelt dieses Verfahren sehr stark dem Spektralsch"atzverfahren
nach Welch-Bartlett, das z.~B. in \cite{Dittrich} beschrieben ist. Der Unterschied
besteht darin, dass beim Verfahren nach Welch-Bartlett eine "Uberlappung der
Messintervalle der einzelnen Stichproben ausdr"ucklich zugelassen wird.
Die hier dargestellte Theorie geht immer von mathematischen Stichproben
der Signale am zu messenden System aus, so dass die einzelnen Elemente
der Stichprobe unabh"angig sein m"ussen, was bei einer "uberlappenden
Fensterung in der Regel nicht mehr erf"ullt sein d"urfte. Wie sich
das Welch-Bartlett Spektralsch"atzverfahren mit "uberlappender
Fensterung bei Verwendung der im Kapitel \ref{Algo} beschriebenen
Fenster\-folge verh"alt, und ob dabei die hohe Frequenzselektivit"at erhalten
bleibt, wurde im Rahmen dieser Arbeit nicht untersucht.

\chapter{Die Fensterfolge}\label{Algo}

In diesem Abschnitt wird nun ein Algorithmus vorgestellt,
mit dessen Hilfe Fensterfolgen berechnet werden k"onnen, 
deren Spektren die geforderte Nullstellenlage nach Gleichung 
(\ref{2.27}) aufweisen, und deren Betragsquadratspektrum nach 
der "Uberlagerung gem"a"s Gleichung (\ref{2.20}) konstant ist. 
Die im weiteren berechneten Fensterfolgen sind {\em reell}, 
so dass an einigen Stellen der weiteren Herleitung des Algorithmus 
auf das Konjugieren der Folgen im Zeitbereich verzichtet werden kann.
Auch wird die im Frequenzbereich vorhandene Symmetrie der Spektren 
reeller Folgen ausgenutzt, ohne dass darauf extra hingewiesen wird. 
Desweiteren sind alle --- bis auf den Trivialfall des Rechteckfensters ---
hier vorgestellten Fensterfolgen {\em nicht}\/ linearphasig, und
enthalten meist auch Werte kleiner Null. Dies steht in keinem
Widerspruch zur Anwendbarkeit der Fensterfolgen f"ur das RKM oder
f"ur eine reine Sch"atzung der spektralen Leistungsdichte 
stochastischer Prozesse. 

Zun"achst stelle ich die prinzipielle 
Vorgehensweise bei der Berechnung der Fensterfolgen vor. 
Anschlie"send wird der eigentliche Berechnungsalgorithmus im Detail 
dargestellt. Danach wird die Genauigkeit bei der Berechnung 
der Einzelschritte des Algorithmus untersucht, und es wird gezeigt, 
dass es mit diesem Algorithmus m"oglich ist, Fensterfolgen 
zu erhalten, die die geforderten Eigenschaften (\ref{2.27}) und 
(\ref{2.20}) mit einer Pr"azision erf"ullen, die nahe an die 
Genauigkeit heranreicht, die bei einer endlichen Zahlendarstellung 
an einem Computer mit Flie"skommaarithmetik "uberhaupt m"oglich ist. 
Im folgenden werden einige mit diesem Algorithmus berechnete 
Fensterfolgen beispielhaft dargestellt. Desweiteren werden die 
Eigenschaften so konstruierter Fensterfolgen n"aher untersucht. 
Abschlie"send wird bei einigen anderen aus der Literatur bekannten 
Fensterfolgen kurz ermittelt, inwieweit sie beim RKM eingesetzt werden k"onnen. 
Im Unterkapitel \ref{MatFen1} des Anhangs ist der ausf"uhrlich kommentierte,
wesentliche Rumpf eines Programms in der Interpretersprache {\tt MATLAB}
angegeben, mit dessen Hilfe sich die Fensterfolgen mit dem in diesem Kapitel
beschriebenen Verfahren berechnen lassen.

\section{Konstruktion der Fensterfolge}\label{Fen}

Bei einer reellwertigen Fensterfolge, die wie in Gleichung (\ref{2.15}) 
angegeben auf ein endlich langes Zeitintervall der L"ange $F$ begrenzt ist, 
hat man bei deren Entwurf genau $F$ reelle Freiheitsgrade um ihr gewisse 
gew"unschte Eigenschaften zu verleihen. In unserem Fall der Anwendung der 
Fensterfolge beim RKM sind diese gew"unschten Eigenschaften zum einen die 
geforderte Nullstellenlage nach Gleichung (\ref{2.27}) und zum anderen 
die Konstanz der "uberlagerten Betragsquadratspektren gem"a"s Gleichung 
(\ref{2.20}). Bei der weiter unten genannten Wahl der Fensterl"ange $F$ 
ben"otigen wir jedoch nicht alle Freiheitsgrade, um die geforderten 
Eigenschaften zu erf"ullen. 

Bei den im Rahmen dieser Abhandlung 
vorgestellten Fensterfolgen werden einige Freiheitsgrade nicht 
genutzt, und es werden nur solche Fensterfolgen konstruiert, bei denen 
die letzten Werte der Fensterfolge willk"urlich auf Null festgelegt werden. 
Somit ergeben sich Fensterfolgen deren "`wahre"' L"ange kleiner als $F$ 
ist\footnote{ Man beachte, dass die nicht im Widerspruch zu Gleichung 
(\ref{2.15}) steht.}. Die Anzahl der am Ende der Fensterfolge willk"urlich 
auf Null festgelegen Werte sei \mbox{$N\!-\!1$}. Dabei ist 
\mbox{$N\in\mathbb{N}$} der Parameter einer Schar von 
Fensterfolgen, die mit dem im weiteren vorgestellten Algorithmus erzeugt 
werden k"onnen. Der Parameter $N$ hat wesentlichen Einfluss auf die spektralen 
Eigenschaften der Fensterfolgen. In Kapitel \ref{FenBeisp} wird dies n"aher untersucht. 
 
{\small Anmerkung: Mit geringen Modifikationen kann man mit dem hier 
vorgestellten Algorithmus auch Fensterfolgen berechnen, bei denen auch 
die letzten \mbox{$N\!-\!1$} Werte von Null verschieden sein d"urfen, 
und die dennoch die Bedingungen (\ref{2.20}) und (\ref{2.27}) erf"ullen. 
Man hat dann mehr Freiheitsgrade, um die Eigenschaften der zu konstruierenden 
Fensterfolge je nach Applikation geeignet zu beeinflussen. Diese erweiterte 
Algorithmusvariante wird in \cite{Erg} vorgestellt. Dort wird auch ein 
Algorithmus zur Konstruktion einer kontinuierlichen Fensterfunktion 
endlicher L"ange mit zu (\ref{2.20}) und (\ref{2.27}) analogen Eigenschaften 
angegeben.}
 
Die Konstruktion der Fensterfolge \mbox{$f(k)$} erfolgt im wesentlichen 
in drei Schritten:\vspace{-4pt}
{\addtolength{\leftmargini}{27pt}%
\begin{itemize}
\item[1. Schritt:] Abh"angig von $N$ wird eine Basisfensterfolge \mbox{$g(k)$} vorgegeben.\vspace{-3pt}
\item[2. Schritt:] "Uber die Basisfenster-AKF \mbox{$\big(g(k)\!\ast\!g(-k)\big)/M$} wird die
      nach Gleichung (\ref{2.21}) definierte Fenster-AKF \mbox{$d(k)$} konstruiert. 
      Damit ist das Betragsquadrat des Spektrums der Fensterfolge \mbox{$f(k)$} bekannt.\vspace{-3pt}
\item[3. Schritt:] Mit Hilfe des in \cite{Boite/Leich} dargestellten Algorithmus --- der f"ur 
      diese Anwendung modifiziert und um eine geeignete Bilineartransformation 
      erweitert wurde --- wird zu dem Betragsfrequenzgang der Fensterfolge der 
      passende Phasenfrequenzgang in der Art bestimmt, dass sich daraus 
      die minimalphasige Fensterfolge \mbox{$f(k)$} berechnen l"asst.
\end{itemize}}
Zun"achst werden die Vorg"ange bei der schrittweisen Berechnung der
Fensterfolge theoretisch behandelt. Auch bei der detaillierteren Behandlung des 
Algorithmus in Unterkapitel \ref{Fenster1} werden wir soweit m"oglich alle 
Berechnungen theoretisch durchf"uhren, so dass die numerische Berechnung der 
Fensterfolge teils aus der numerischen Auswertung der Ergebnisse der theoretischen
Herleitung besteht. 

\subsection{Schritt 1: Konstruktion der Basisfensterfolge}

Wir beginnen damit, eine zeitdiskrete, reelle Basisfensterfolge \mbox{$g(k)$} 
der L"ange $F$ vorzugeben, deren letzte \mbox{$N\!-\!1$} Werte Null sind. 
\begin{equation}
g(k)\,=\,0\qquad\qquad\forall\qquad k<0\quad\vee\quad k>F\!-\!N.
\label{6.1}
\end{equation}
Die Basisfensterfolge \mbox{$g(k)$} darf also nur in genau demselben Bereich 
von Null verschieden sein, wie die zu konstruierende Fensterfolge \mbox{$f(k)$}.
Die L"ange $F$ soll dabei ein ganzzahliges Vielfaches der DFT-L"ange $M$ 
beim RKM sein. 
\begin{equation}
F=N\CdoT M\qquad\qquad\text{ mit }\qquad N\in\mathbb{N}
\label{6.2}
\end{equation}
Auch wenn die Fensterfolge \mbox{$f(k)$}, nicht im RKM eingesetzt werden soll, 
ist der Parameter $M$ von entscheidender Bedeutung, da er die gew"unschten 
Eigenschaften der Fensterfolge \mbox{$f(k)$} festlegt. In Gleichung (\ref{2.20}) 
legt er den Frequenzverschiebungsparameter \mbox{$2\pi/M$} fest, bei dem sich bei 
der "Uberlagerung der verschobenen Betragsquadratspektren eine Konstante ergibt. 
Dies f"uhrt in Gleichung (\ref{2.23}) dazu, dass die Nullstellen der 
Fenster-AKF \mbox{$d(\kappa)$} nach Gleichung (\ref{2.21}) 
in einem Abstand von $M$ liegen werden. Auch die Lage der Nullstellen des 
Frequenzgangs \mbox{$F(\Omega)$} der Fensterfolge wird --- wie noch gezeigt wird ---
mit $M$ in der Art festgelegt, dass Bedingung (\ref{2.27}) erf"ullt wird.

\begin{figure}[btp]
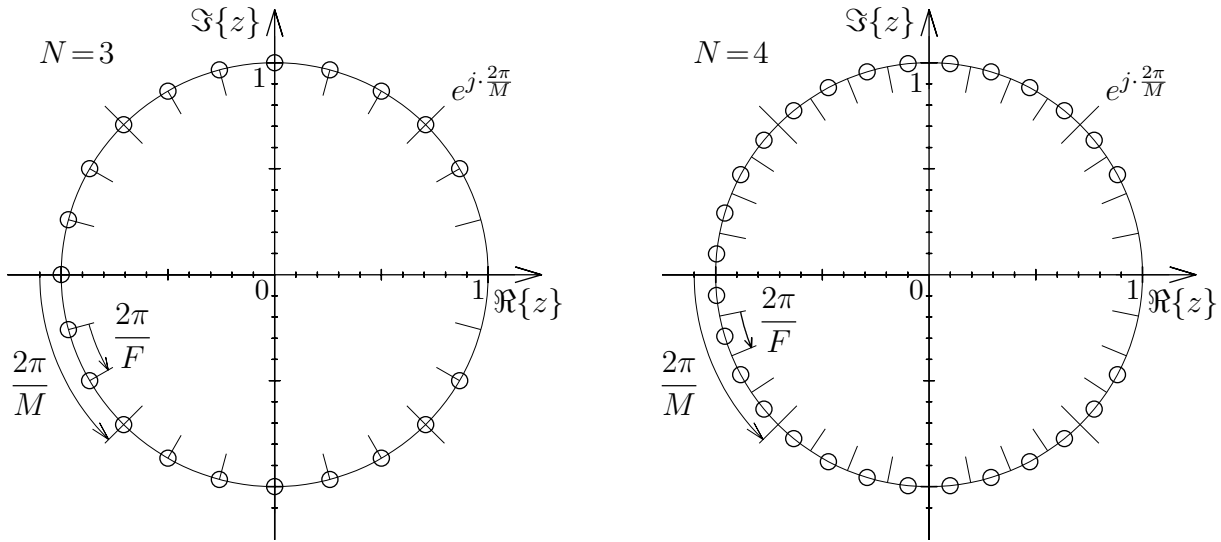

\begin{center}
{ 
\begin{picture}(454,200)

\input{mbild4a1.tex}
\put(196,95){\makebox(0,0)[t]{$\Re\{z\}$}}
\put(96,195){\makebox(0,0)[r]{$\Im\{z\}$}}
\put(99,98){\makebox(0,0)[rt]{\small$0$}}
\put(180,98){\makebox(0,0)[rt]{\small$1$}}
\put(98,178){\makebox(0,0)[rt]{\small$1$}}
\put(167,166){\makebox(0,0)[lb]{$e^{j\cdot\frac{2\pi}{M}}$}}
\put(40,65){\makebox(0,0)[lb]{${\D\frac{2\pi}{F}}$}}
\put(17,71){\makebox(0,0)[rt]{${\D\frac{2\pi}{M}}$}}
\put(41,180){\makebox(0,0)[rb]{$N\!=\!3$}}

\input{mbild4a2.tex}
\put(441,95){\makebox(0,0)[t]{$\Re\{z\}$}}
\put(341,195){\makebox(0,0)[r]{$\Im\{z\}$}}
\put(344,98){\makebox(0,0)[rt]{\small$0$}}
\put(425,98){\makebox(0,0)[rt]{\small$1$}}
\put(344,176){\makebox(0,0)[rt]{\small$1$}}
\put(412,166){\makebox(0,0)[lb]{$e^{j\cdot\frac{2\pi}{M}}$}}
\put(282,71){\makebox(0,0)[lb]{${\D\frac{2\pi}{F}}$}}
\put(260,71){\makebox(0,0)[rt]{${\D\frac{2\pi}{M}}$}}
\put(286,180){\makebox(0,0)[rb]{$N\!=\!4$}}

\end{picture}}
\end{center}\vspace{-15pt}
\setlength{\belowcaptionskip}{-3pt}
\caption[Nullstellen der Z-Transformierten der
Basisfensterfolge]{Festgelegte Nullstellen der Z-Transformierten
\mbox{$G(z)$} der Basisfensterfolge \mbox{$g(k)$} f"ur \mbox{$M\!=\!8$}
und die zwei Werte \mbox{$N\!=\!3$} und \mbox{$N\!=\!4$}.}
\label{b4a}
\rule{\textwidth}{0.5pt}\vspace{-4pt}
\end{figure}

Bis auf einen konstanten Faktor --- den man nicht fest vorgibt, da
er nie explizit bestimmt werden muss --- wird die Basisfensterfolge \mbox{$g(k)$}
durch die \mbox{$F\!-\!N$} Nullstellen des Polynoms 
\begin{equation}
z^{F-N}\Cdot G(z)\;=\;z^{F-N}\cdoT\Sum{k=0}{F-N}\,g(k)\CdoT z^{\!-k}
\label{6.3}
\end{equation}
der Z-Transformierten der Basisfensterfolge festgelegt. 
Die \mbox{$F\!-\!N$} Nullstellen des Polynoms werden nun fest 
vorgegeben. Sie sollen alle "aquidistant am Einheitskreis liegen:
\begin{equation}
G\big(e^{j\cdot\frac{\pi}{F}\cdot\nu}\big)\,=\,0
\qquad\qquad\forall\qquad\nu\,=\,N\!+\!1\;(2)\;2\CdoT F\!-\!N\!-\!1.
\label{6.4}
\end{equation}
F"ur ungerades $N$ liegen diese Nullstellen im Frequenzraster \mbox{$2\pi/F$},
w"ahrend sie f"ur gerades $N$ um \mbox{$\pi/F$} gegen"uber diesem Raster
versetzt (\,auf L"ucke\,) liegen. Bild \ref{b4a}
zeigt
die Lage der Nullstellen der Z-Transformierten der Basisfensterfolge 
\mbox{$g(k)$} f"ur \mbox{$M\!=\!8$} und die beiden Werte \mbox{$N\!=\!3$} 
und \mbox{$N\!=\!4$}. In der Umgebung des Punktes \mbox{$z\!=\!1$} fehlen 
also dem vollst"andigen Kreis aus Nullstellen (\,$\widehat{=}$ dem Polynom 
\mbox{$z^F\!+(-1)^N$}\,) gerade $N$ Nullstellen.
In beiden F"allen kann man f"ur das Polynom mit den festgelegten Nullstellen
\begin{equation}
z^{F-N}\Cdot G(z)\;=\;\frac{z^F\!+(-1)^N}
{\T\;\;\Prod{\nu_2=\frac{1-N}{2}}{\frac{N-1}{2}}\!\!\D
\big(z-e^{j\cdot\frac{2\pi}{F}\cdot\nu_2}\big)\;}
\label{6.5}
\end{equation}
schreiben, wobei f"ur das Produkt im Nenner die in der Liste der Formelzeichen
angegebene Definition verwendet wird, die es zul"asst, dass der Lauf"|index
auch Werte annehmen kann, die nicht ganzzahlig sind.

F"ur \mbox{${\D z\!=\!e^{j\Omega}}$} ist \mbox{${\D z^*\!=\!z^{\!-1}\!,}$} 
und da die Folge \mbox{$g(k)$} reell ist, gilt: 
\begin{equation}
|G(z)|^2\,=\;G(z)\CdoT G(z^{\!-1})
\qquad\qquad\text{ f"ur }\qquad z=e^{j\Omega}.
\label{6.6}
\end{equation}
Die Nullstellen von \mbox{${\D z^{F-N}\Cdot G(z)}$} liegen alle 
auf dem Einheitskreis, und sind daher als doppelte Nullstellen 
in \mbox{${\D z^{F-N}\Cdot G(z)\CdoT G(z^{\!-1})}$} vorhanden. 

\subsection{Schritt 2: Konstruktion der Fensterautokorrelationsfolge}

Durch "Uberlagerung der verschobenen Betragsquadrate des Spektrums
\mbox{${\D G\big(e^{j\Omega}\big)}$} wird nun das Spektrum\vspace{-12pt}
\begin{equation}
D\big(e^{j\Omega}\big)\;=\!
\Sum{\nu_1=\frac{1-N}{2}}{\frac{N-1}{2}}
\big|G\big(e^{j\cdot(\Omega-\frac{2\pi}{F}\cdot\nu_1)}\big)\big|^2
\label{6.7}
\end{equation}
der Fenster-AKF \mbox{$d(k)$} gewonnen. 
Diese "Uberlagerung im Spektralbereich ist die Faltung des
Betragsquadrates des Spektrums \mbox{$G\big(\!e^{j\Omega}\big)$}
mit einem endlich langen Impulskamm:\vspace{-2pt}
\begin{equation}
D\big(e^{j\Omega}\big)\;=\!
\big|G\big(\!e^{j\Omega}\big)\big|^2\ast
\Sum{\nu_1=\frac{1-N}{2}}{\frac{N-1}{2}}
\delta_0\big({\T\Omega\!-\!\nu_1\CdoT\frac{2\pi}{F}}\big).
\label{6.8}
\end{equation}
Im Zeitbereich entspricht das einer Multiplikation der $M$-fachen
zeitdiskreten endlichen Basisfenster-AKF\vspace{-5pt}
\begin{equation}
g_Q(k)\;=\;g(k)\!\ast\!g(-k)\;=\!
\Sum{\kappa=-\infty}{\infty}\!g(\kappa)\CdoT g(\kappa+k)\;=\;
\frac{1}{2\pi}\CdoT\!\Int{-\pi}{\pi}\big|G\big(e^{j\cdot\Omega}\big)\big|^2
\Cdot e^{j\cdot\Omega\cdot k}\cdot d\Omega
\label{6.9}
\end{equation}
mit der in $k$ kontinuierlichen\footnote{Die Summe in Gleichung (\ref{6.7}) 
ist endlich lang. Daher sind bei der Faltung in Gleichung (\ref{6.8}) 
nur endlich viele Impulse beteiligt. Daher ist die periodisch fortgesetzte 
si-Funktion in $k$ kontinuierlich und nicht zeitdiskret.}\!, periodisch 
fortgesetzten si-Funktion:
\begin{equation}
d(k)\;=\;\frac{\,\sin\!\big({\frac{\pi}{M}}\CdoT k\big)\,}
{\,\sin\!\big({\frac{\pi}{F}}\CdoT k\big)\,}\cdot g_Q(k).
\label{6.10}
\end{equation}
Die so entstandene Funktion \mbox{$d(k)$} ist zeitdiskret
(\,\mbox{$g_Q(k)$} ist zeitdiskret\,), von gleicher L"ange
wie die Basisfenster-AKF (\,\mbox{$g_Q(k)\!=\!0\,$}
f"ur \mbox{$\,k\!<\!N\!-\!F\;\vee\; k\!>\!F\!-\!N$}\,)
und weist au"ser bei \mbox{$k\!=\!0$} bei Vielfachen von $M$ 
die Nullstellen der periodisch fortgesetzten si-Funktion auf. 
F"uhrt man mit der Folge \mbox{$d(k)$} eine Unterabtastung mit dem 
Unterabtastfaktor $M$ durch, so erh"alt man --- wenn der konstante, 
nicht festgelegte Faktor der Basisfensterfunktion stimmt --- den 
diskreten Impuls 
\begin{equation}
d(\Tilde{k}\CdoT M)\;=\;\gamma_0(\Tilde{k})
\;=\;\begin{cases}
\quad 1&\quad\text{ f"ur }\quad\Tilde{k}=0\\
\quad 0&\quad\text{ sonst.}
\end{cases}
\label{6.11}
\end{equation}
Somit erf"ullt die Folge \mbox{$d(k)$} die in Gleichung (\ref{2.23}) 
an die Fensterkorrelationsfolge gestellte Forderung, die eine 
leistungsrichtige Messung beim RKM erm"oglicht. 
Nach \cite{Fliege} handelt es sich daher bei der Folge \mbox{$d(k)$}
um die Impulsantwort eines $M$-tel-Band-Filters. Nach Gleichung (\ref{2.22}) 
entspricht die Unterabtastung im Frequenzbereich einer "Uberlagerung der 
verschobenen Spektren. Man gewinnt daher die Aussage:
\begin{equation}
\Sum{\mu=0}{M-1}
D\big(e^{j\cdot(\Omega-\frac{2\pi}{M}\cdot\mu)}\big)
\;=\;\text{const.}
\label{6.12}
\end{equation}
Da \mbox{${\D D\big(e^{j\Omega}\big)}$} durch die "Uberlagerung der
verschobenen Spektren \mbox{$\big|G\big(e^{j\Omega}\big)\big|^2$} entstanden
ist, "uberlagern sich bei den Frequenzen \mbox{$\Omega=\nu\CdoT2\pi/F$}
mit \mbox{$\nu=N\;(1)\;F\!-\!N$} jeweils die doppelten Nullstellen,
die in \mbox{${\D z^{F-N}\Cdot G(z)\CdoT G(z^{\!-1})}$} vorhanden sind.
\mbox{${\D z^{F-1}\CdoT D(z)}$} hat daher bei diesen Frequenzen doppelte
Nullstellen. 

Falls es nun trotz der Multiplikation mit der
periodisch fortgesetzten si-Funktion gelingt, \mbox{$d(k)$} als
die auf $M$ normierte Selbstfaltung \mbox{$\big(f(k)\!\ast\!f(-k)\big)/M$}
einer Fensterfolge mit der L"ange $F$ (\,Forderung (\ref{2.15})\,)
darzustellen, ist die so gewonnene Fensterfolge prinzipiell
f"ur den Einsatz beim RKM geeignet, da dann die Bedingungen
(\ref{2.20}) und (\ref{2.27}) erf"ullt sind. 

\begin{figure}[btp]
\begin{center}
{ 
\begin{picture}(455,626)

\input{mbild4b}

\put(0,623){\makebox(0,0)[lb]{\footnotesize Basisfensterfolge $g(k)$:}}
\put(65,616){\makebox(0,0)[r]{\footnotesize $g(k)$}}
\put(135,559){\makebox(0,0)[t]{\footnotesize $k$}}
\put(90,559){\makebox(0,0)[t]{\footnotesize $M$}}
\put(68,559){\makebox(0,0)[tr]{\footnotesize $0$}}
\put(68,587){\makebox(0,0)[r]{\footnotesize $0.1$}}
\put(245,618){\makebox(0,0)[r]{\footnotesize $20*\log_{10}(|G(e^{j\Omega})|)$}}
\put(354,598){\makebox(0,0)[t]{\footnotesize $\Omega$}}
\put(252,600){\makebox(0,0)[tl]{\footnotesize $0$}}
\put(147,598){\makebox(0,0)[tl]{\footnotesize $-\pi$}}
\put(252,616){\makebox(0,0)[l]{\footnotesize $20$}}
\put(252,569){\makebox(0,0)[l]{\footnotesize $-50$}}
\put(440,612){\framebox(10,10){\footnotesize $z$}}
\put(398,585){\makebox(0,0)[tr]{\footnotesize $0$}}
\put(427,585){\makebox(0,0)[tl]{\footnotesize $1$}}
\put(398,610){\makebox(0,0)[tr]{\footnotesize $j$}}
\put(435,590){\makebox(0,0)[bl]{\footnotesize $\frac{2\pi}{F}$}}

\put(0,534){\makebox(0,0)[lb]{\footnotesize $M$-fache AKF des Basisfensters:}}
\put(62,526){\makebox(0,0)[r]{\footnotesize $g_Q(k)$}}
\put(135,467){\makebox(0,0)[t]{\footnotesize $k$}}
\put(90,467){\makebox(0,0)[t]{\footnotesize $M$}}
\put(68,467){\makebox(0,0)[tr]{\footnotesize $0$}}
\put(67,520){\makebox(0,0)[r]{\footnotesize $1$}}
\put(67,495){\makebox(0,0)[r]{\footnotesize $0.5$}}
\put(245,526){\makebox(0,0)[r]{\footnotesize $10*\log_{10}(|G(e^{j\Omega})|^2)$}}
\put(354,506){\makebox(0,0)[t]{\footnotesize $\Omega$}}
\put(252,508){\makebox(0,0)[tl]{\footnotesize $0$}}
\put(147,506){\makebox(0,0)[tl]{\footnotesize $-\pi$}}
\put(252,524){\makebox(0,0)[l]{\footnotesize $20$}}
\put(252,477){\makebox(0,0)[l]{\footnotesize $-50$}}
\put(440,520){\framebox(10,10){\footnotesize $z$}}
\put(398,493){\makebox(0,0)[tr]{\footnotesize $0$}}
\put(427,493){\makebox(0,0)[tl]{\footnotesize $1$}}
\put(398,518){\makebox(0,0)[tr]{\footnotesize $j$}}

\put(0,442){\makebox(0,0)[lb]{\footnotesize Um $2\pi/F$ nach links verschobenes Betragsquadratspektrum:}}
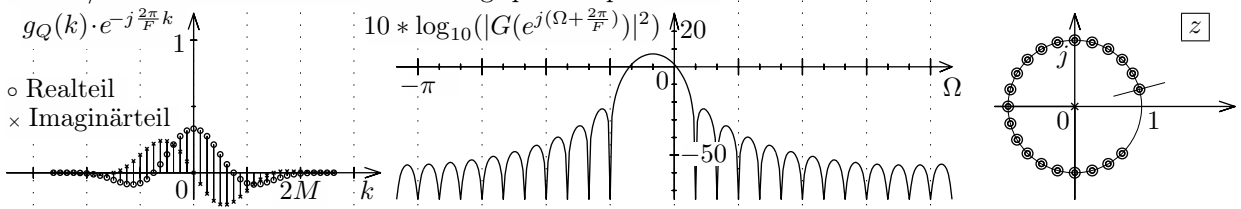
\put(63,434){\makebox(0,0)[r]{\footnotesize $g_Q(k)\CdoT e^{\!-j\frac{2\pi}{F}k}$}}
\put(0,410){\makebox(0,0)[l]{\scriptsize $\circ$ \footnotesize Realteil}}
\put(0,398){\makebox(0,0)[l]{\tiny $\times$ \footnotesize Imagin"arteil}}
\put(135,375){\makebox(0,0)[t]{\footnotesize $k$}}
\put(110,375){\makebox(0,0)[t]{\footnotesize $2M$}}
\put(68,375){\makebox(0,0)[tr]{\footnotesize $0$}}
\put(67,427){\makebox(0,0)[r]{\footnotesize $1$}}
\put(245,434){\makebox(0,0)[r]{\footnotesize $10*\log_{10}(|G(e^{j(\Omega+\frac{2\pi}{F})})|^2)$}}
\put(354,414){\makebox(0,0)[t]{\footnotesize $\Omega$}}
\put(248,416){\makebox(0,0)[tr]{\footnotesize $0$}}
\put(147,414){\makebox(0,0)[tl]{\footnotesize $-\pi$}}
\put(252,432){\makebox(0,0)[l]{\footnotesize $20$}}
\put(252,385){\makebox(0,0)[l]{\footnotesize $-50$}}
\put(440,428){\framebox(10,10){\footnotesize $z$}}
\put(398,401){\makebox(0,0)[tr]{\footnotesize $0$}}
\put(427,401){\makebox(0,0)[tl]{\footnotesize $1$}}
\put(398,426){\makebox(0,0)[tr]{\footnotesize $j$}}

\put(0,350){\makebox(0,0)[lb]{\footnotesize Um $2\pi/F$ nach rechts verschobenes Betragsquadratspektrum:}}
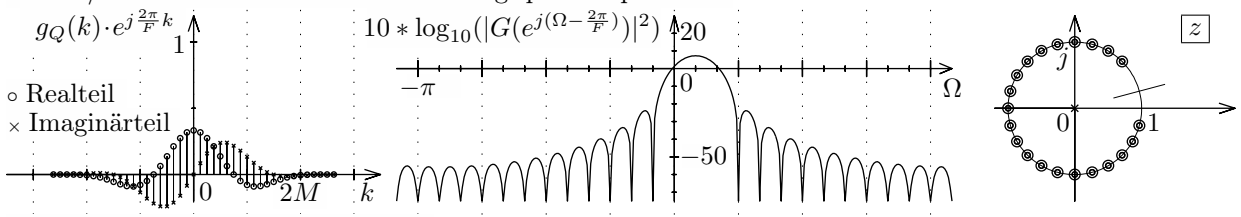
\put(63,342){\makebox(0,0)[r]{\footnotesize $g_Q(k)\CdoT e^{j\frac{2\pi}{F}k}$}}
\put(0,317){\makebox(0,0)[l]{\scriptsize $\circ$ \footnotesize Realteil}}
\put(0,305){\makebox(0,0)[l]{\tiny $\times$ \footnotesize Imagin"arteil}}
\put(135,283){\makebox(0,0)[t]{\footnotesize $k$}}
\put(110,283){\makebox(0,0)[t]{\footnotesize $2M$}}
\put(72,283){\makebox(0,0)[tl]{\footnotesize $0$}}
\put(67,335){\makebox(0,0)[r]{\footnotesize $1$}}
\put(245,342){\makebox(0,0)[r]{\footnotesize $10*\log_{10}(|G(e^{j(\Omega-\frac{2\pi}{F})})|^2)$}}
\put(354,322){\makebox(0,0)[t]{\footnotesize $\Omega$}}
\put(252,324){\makebox(0,0)[tl]{\footnotesize $0$}}
\put(147,322){\makebox(0,0)[tl]{\footnotesize $-\pi$}}
\put(252,340){\makebox(0,0)[l]{\footnotesize $20$}}
\put(252,293){\makebox(0,0)[l]{\footnotesize $-50$}}
\put(440,336){\framebox(10,10){\footnotesize $z$}}
\put(398,309){\makebox(0,0)[tr]{\footnotesize $0$}}
\put(427,309){\makebox(0,0)[tl]{\footnotesize $1$}}
\put(398,334){\makebox(0,0)[tr]{\footnotesize $j$}}

\put(0,258){\makebox(0,0)[lb]{\footnotesize periodisch fortgesetzte si-Funktion:}}
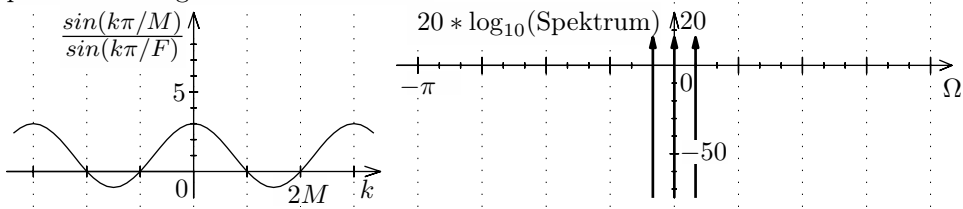
\put(67,245){\makebox(0,0)[r]{$\frac{sin(k\pi/M)}{sin(k\pi/F)}$}}
\put(135,191){\makebox(0,0)[t]{\footnotesize $k$}}
\put(113,189){\makebox(0,0)[t]{\footnotesize $2M$}}
\put(68,191){\makebox(0,0)[tr]{\footnotesize $0$}}
\put(67,224){\makebox(0,0)[r]{\footnotesize $5$}}
\put(245,250){\makebox(0,0)[r]{\footnotesize $20*\log_{10}($Spektrum$)$}}
\put(354,229){\makebox(0,0)[t]{\footnotesize $\Omega$}}
\put(252,231){\makebox(0,0)[tl]{\footnotesize $0$}}
\put(147,229){\makebox(0,0)[tl]{\footnotesize $-\pi$}}
\put(252,251){\makebox(0,0)[l]{\footnotesize $20$}}
\put(252,201){\makebox(0,0)[l]{\footnotesize $-50$}}

\put(0,166){\makebox(0,0)[lb]{\footnotesize AKF des Fensters:}}
\put(61,158){\makebox(0,0)[r]{\footnotesize $d(k)$}}
\put(135,99){\makebox(0,0)[t]{\footnotesize $k$}}
\put(110,99){\makebox(0,0)[t]{\footnotesize $2M$}}
\put(68,99){\makebox(0,0)[tr]{\footnotesize $0$}}
\put(66,153){\makebox(0,0)[r]{\footnotesize $1$}}
\put(245,158){\makebox(0,0)[r]{\footnotesize $10*\log_{10}(D(e^{j\Omega}))$}}
\put(354,138){\makebox(0,0)[t]{\footnotesize $\Omega$}}
\put(252,140){\makebox(0,0)[tl]{\footnotesize $0$}}
\put(147,138){\makebox(0,0)[tl]{\footnotesize $-\pi$}}
\put(252,156){\makebox(0,0)[l]{\footnotesize $20$}}
\put(252,109){\makebox(0,0)[l]{\footnotesize $-50$}}
\put(440,152){\framebox(10,10){\footnotesize $z$}}
\put(398,125){\makebox(0,0)[tr]{\footnotesize $0$}}
\put(427,125){\makebox(0,0)[tl]{\footnotesize $1$}}
\put(398,150){\makebox(0,0)[tr]{\footnotesize $j$}}

\put(0,74){\makebox(0,0)[lb]{\footnotesize Fensterfolge $f(k)$:}}
\put(63,66){\makebox(0,0)[r]{\footnotesize $f(k)$}}
\put(135,7){\makebox(0,0)[t]{\footnotesize $k$}}
\put(90,7){\makebox(0,0)[t]{\footnotesize $M$}}
\put(68,7){\makebox(0,0)[tr]{\footnotesize $0$}}
\put(68,35){\makebox(0,0)[r]{\footnotesize $0.5$}}
\put(68,60){\makebox(0,0)[r]{\footnotesize $1$}}
\put(245,66){\makebox(0,0)[r]{\footnotesize $20*\log_{10}(|F(\Omega)|)$}}
\put(354,46){\makebox(0,0)[t]{\footnotesize $\Omega$}}
\put(252,48){\makebox(0,0)[tl]{\footnotesize $0$}}
\put(147,46){\makebox(0,0)[tl]{\footnotesize $-\pi$}}
\put(252,66){\makebox(0,0)[l]{\footnotesize $20$}}
\put(252,17){\makebox(0,0)[l]{\footnotesize $-50$}}
\put(440,60){\framebox(10,10){\footnotesize $z$}}
\put(398,33){\makebox(0,0)[tr]{\footnotesize $0$}}
\put(427,33){\makebox(0,0)[tl]{\footnotesize $1$}}
\put(398,58){\makebox(0,0)[tr]{\footnotesize $j$}}

\end{picture}}
\end{center}\vspace{-14pt}
\caption{Konstruktion des Betragsquadrats des Spektrums der Fensterfolge}
\label{b4b}
\end{figure}
Bevor nun gezeigt werden soll, dass eine derartige Entfaltung von 
\mbox{$d(k)$} m"oglich ist, und wie man diese numerisch hochgenau 
realisieren kann, werden in Bild \ref{b4b} an dem Beispiel mit \mbox{$M\!=\!8$} 
und \mbox{$N\!=\!3$} die bei den einzelnen Schritten der Konstruktion 
von \mbox{${\D D\big(e^{j\Omega}\big)} $} ablaufenden Vorg"ange sowohl 
an den Zeitsignalen als auch an deren Spektren und Nullstellenlagen 
graphisch veranschaulicht. 

Im ersten Schritt wird die Basisfensterfolge 
im obersten linken Teilbild vorgegeben. Sie ergibt sich aus der im obersten 
rechten Teilbild dargestellten Lage der Nullstellen ihrer Z-Transformierten. 
Die "aquidistanten Nullstellen am Einheitskreis sind auch im Betragsspektrum 
der Basisfensterfolge im obersten mittleren Teilbild deutlich zu sehen. 

Die geradesymmetrische $M$-fache AKF \mbox{$g_Q(k)$} des Basisfensters, 
die durch Faltung des Basisfensters mit sich selbst gespiegelt entsteht,
ist in den Teilbildern der zweiten Reihe dargestellt. Die Nullstellen
ihrer Z-Transformierten liegen ebenfalls auf dem Einheitskreis, treten nun aber 
doppelt auf. Da die AKF des Basisfensters nun auch f"ur negatives $k$ 
von Null verschieden ist, erscheint nun bei \mbox{$z\!=\!0$} eine 
\mbox{$(F\!-\!N)$}-fache Polstelle. Durch die geeignete Normierung ist 
die halblogarithmische graphische Darstellung des Spektrums der 
$M$-fachen Basisfenster-AKF identisch mit der dar"uberliegenden 
Darstellung des Betragsspektrums der Basisfensterfolge. 

Die beiden folgenden Teilbildreihen zeigen die komplexwertigen Folgen, die 
man erh"alt, wenn man das Spektrum der $M$-fachen Basis"-fenster"=AKF 
nach links, bzw. nach rechts um \mbox{$2\pi/F$} verschiebt. 
Diese beiden Folgen weisen einen gegengleichen, schiefsymmetrischen 
und von Null verschiedenen Imagin"arteil auf. In der $z$-Ebene ist
die entsprechende Rotation der Nullstellenlage zu erkennen. 

Die beiden Folgen mit den verschobenen Spektren und die unver"anderte 
$M$-fache Basisfenster"=AKF "uberlagern sich additiv zu der Fenster-AKF, 
die in den Teilbildern der vorletzten Reihe dargestellt ist. 
Man erkennt, dass durch die Art der "Uberlagerung der Spektren nach 
Gleichung (\ref{6.7}) die gew"unschten Nullstellen bei Vielfachen der 
Frequenz \mbox{$2\pi/M$} (\,punktierte Linien in den mittleren 
Teilbildern, die "uber alle "ubereinander angeordneten Teilbilder 
verlaufen\,) erhalten bleiben. Durch die "Uberlagerung sind nun genau 
\mbox{$2\CdoT(N\!-\!1)=4$} Nullstellen bei der Z-Transformierten
der Fenster-AKF entstanden, die nicht mehr auf dem 
Einheitskreis, sondern spiegelsymmetrisch zu diesem liegen. 
Bei der "Uberlagerung der Folgen heben sich die 
Imagin"arteile der Folgen mit den verschobenen Spektren gerade auf,
so dass eine reelle Fenster-AKF entstanden ist.

Die "Uberlagerung der drei Anteile mit den gegeneinander verschobenen 
Spektren entspricht im Zeitbereich der Multiplikation der $M$-fachen 
Basisfenster-AKF mit der ebenfalls dargestellten periodisch 
fortgesetzten, kontinuierlichen si-Funktion, deren Spektrum \mbox{$N\!=\!3$} 
Impulse im Frequenzabstand \mbox{$2\pi/F$} aufweist. Da bei der periodisch 
fortgesetzten \mbox{si-Funktion} die Nullstellen des Zeitsignals im Abstand $M$ 
(\,punktierte Linien in den linken Teilbildern\,) liegen, hat auch die 
darunter dargestellte Fenster-AKF dieselben Nullstellen, 
wie dies in Gleichung (\ref{2.23}) gefordert war.

Indem man die innerhalb des Einheitskreises der $z$-Ebene 
liegenden \mbox{$N\!-\!1$} Nullstellen der Z-Transformierten 
der Fenster-AKF sowie von den doppelten 
Nullstellen am Einheitskreis nur jeweils eine nimmt, erh"alt 
man die Z-Transformierte der minimalphasigen Fensterfolge, 
die in den untersten Teilbildern dargestellt ist. 

\subsection{Schritt 3: Prinzip der Berechnung der Fensterfolge aus der Fenster-AKF}

Da \mbox{$\big|G\big(e^{j\Omega}\big)\big|^2$} eine reelle,
geradesymmetrische und nichtnegative Funktion in $\Omega$ ist, und da die
"Uberlagerung ebenfalls symmetrisch zu \mbox{$\Omega\!=\!0$} erfolgt, ist
auch \mbox{$D\big(e^{j\Omega}\big)$} eine reelle, geradesymmetrische und
nichtnegative Funktion in $\Omega$. Daher sind alle Nullstellen von
\mbox{$z^{F-N}\CdoT D(z)$} spiegelsymmetrisch zur reellen Achse, alle
Nullstellen auf dem Einheitskreis sind von gerader Vielfachheit
und alle Nullstellen, die nicht auf dem Einheitskreis liegen, sind
spiegelsymmetrisch zum Einheitskreis. Daher kann
\mbox{$M\CdoT D(z)$} immer als das Produkt
\begin{equation}
M\CdoT D(z)\;=\;
F\big(\!-j\Cdot\ln(z)\big)\CdoT F\big(j\Cdot\ln(z)\big)
\label{6.13}
\end{equation}
eines minimalphasigen Anteils \mbox{$F\big(\!-j\Cdot\ln(z)\big)$}
und eines Anteils \mbox{$F\big(j\Cdot\ln(z)\big)$}, dessen
Nullstellen gerade die am Einheitskreis gespiegelten Nullstellen des
minimalphasigen Anteils sind, geschrieben werden. 

Da am Einheitskreis \mbox{$\big(z\!=\!e^{j\Omega}\big)$}
\begin{equation}
M\CdoT D\big(e^{j\Omega}\big)\;=\;F(\Omega)\CdoT F(-\Omega)\;=\;
F(\Omega)\CdoT F(\Omega)^{\!\Kk}\;=\;\big|F(\Omega)\big|^2
\label{6.14}
\end{equation}
gilt, zeigt Gleichung (\ref{6.12}), dass die Bedingung (\ref{2.20}) 
vom Spektrum der Fenster-AKF erf"ullt wird.

Die oben genannten doppelten Nullstellen von \mbox{$z^{F-N}\CdoT D(z)$}
am Einheitskreis werden bei der Faktorisierung als einfache Nullstellen
dem Polynom \mbox{$z^{F-N}\CdoT F\big(\!-j\Cdot\ln(z)\big)$} zugeordnet.
Daher gilt:
\begin{equation}
F\big({\T\nu\CdoT\frac{2\pi}{F}}\big)=0
\qquad\qquad\forall\qquad\nu=N\;(1)\;F\!-\!N.
\label{6.15}
\end{equation}
Einerseits sind somit die nach Gleichung (\ref{2.27}) geforderten
Nullstellen im Spektrum vorhanden, und andererseits liegen zwischen
zwei geforderten Nullstellen am Einheitskreis jeweils \mbox{$N\!-\!1$}
weitere Nullstellen. 

Hat man die Faktorisierung von \mbox{$D(z)$}
durchgef"uhrt, so liefert die inverse Z-Transformation des
minimalphasigen Anteils \mbox{$F\big(\!-j\Cdot\ln(z)\big)$}
{\em eine} m"ogliche Fensterfolge \mbox{$f(k)$}, die die Forderungen
(\ref{2.20}) und (\ref{2.27}) zugleich erf"ullt, wenn man den konstanten,
nicht n"aher festgelegten Faktor der Basisfensterfolge geeignet w"ahlt.

Prinzipiell l"asst sich statt jeder Nullstelle innerhalb des Einheitskreises
auch die dazu spiegelsymmetrische Nullstelle der Z-Transformierten 
des Fensters zuordnen, ohne dass dadurch die geforderten spektralen 
Eigenschaften beeinflusst werden. Man w"urde so Fens"-ter"-fol"-gen erhalten,
deren Phasenverlauf zwischen dem Phasenverlauf des minimalphasigen 
Fensters und des dazu gespiegelten Fensters liegen w"urden.
Der im weiteren vorgestellte Algorithmus liefert immer die minimalphasige 
Fensterfolge, ohne dass die Nullstellen explizit berechnet und zugeordnet 
werden.

Da die Faktorisierung im Spektralbereich einer Entfaltung im Zeitbereich 
entspricht, l"asst sich die Folge \mbox{$d(k)$} als die auf $M$ normierte 
Faltung (\,\mbox{$f(k)\!\ast\!f(-k)$}\,) darstellen. Da \mbox{$d(k)$}
zeitlich auf das Intervall \mbox{$[N\!-\!F;F\!-\!N]$} begrenzt ist, muss 
die L"ange der Fensterfolge \mbox{$f(k)$} --- wie gew"unscht --- kleiner 
gleich $F\!-\!N\!+\!1$ sein, so dass \mbox{$f(k)$} au"serhalb
des in Gleichung (\ref{2.15}) angegebene Zeitintervalls Null ist.

Da \mbox{$f(k)$} zeitlich auf dieses Intervall begrenzt ist, und da 
\mbox{$F(\Omega)$} eine in $\Omega$ periodische Funktion ist, 
kann nach dem Abtasttheorem die inverse Z-Transformation bei der 
Berechnung von \mbox{$f(k)$} als inverse DFT von $F$ "aquidistanten 
Abtastwerten von \mbox{$F(\Omega)$} durchgef"uhrt werden. 
Sinnvollerweise w"ahlt man dabei die Abtastung im Raster
\mbox{$2\pi/F$}, da dann auf Grund der Nullstellen nach Gleichung (\ref{6.15})
nur die \mbox{$2\CdoT\!N\!-\!1$} Abtastwerte
bei \mbox{$\Omega=\nu\CdoT2\pi/F$} mit
\mbox{$\nu=0\;(1)\;N\!-\!1\,\;\vee\,\;\nu=F\!-\!N\!+\!1\;(1)\;F\!-\!1$}
von Null verschieden sind. Es gen"ugt daher einen Algorithmus anzugeben, 
der in der Lage ist die Faktorisierung von \mbox{$D(z)$} bei diesen 
\mbox{$2\CdoT\!N\!-\!1$} Frequenzen vorzunehmen, und \mbox{$F(\Omega)$} 
f"ur diese Werte von $\Omega$ mit hinreichender Genauigkeit zu berechnen. 

Die Fensterfolge \mbox{$f(k)$} l"asst sich dann mit Hilfe einer Fourierreihe 
gem"a"s
\begin{equation}
f(k)\;=\;\begin{cases}
{\D\;\Sum{\nu=1-N}{N-1}\frac{1}{F}\cdot
F\big({\T\nu\CdoT\frac{2\pi}{F}}\big)\cdot
e^{j\cdot\frac{2\pi}{F}\cdot\nu\cdot k}}\qquad
&\text{f"ur}\quad k=0\;(1)\;F\!-\!1\\
\quad0&\text{sonst}
\end{cases}
\label{6.16}
\end{equation}
aus diesen \mbox{$2\CdoT\!N\!-\!1$} Abtastwerten des Spektrums berechnen.
Da die Fensterfolge \mbox{$f(k)$} reell ist, braucht man wegen
der Symmetrie des Spektrums sogar nur die $N$ Abtastwerte des Spektrums
mit \mbox{$\nu=0\;(1)\;N\!-\!1$} zu berechnen.
Die Berechnung dieser Werte von \mbox{$F(\Omega)$} erfolgt dabei
getrennt nach Betrag und Phase.

\subsubsection{Betr"age der Fourierreihenkoeffizienten der Fensterfolge}

Den Betrag berechnet man durch Radizieren aus dem Betragsquadrat des 
Fensterspektrums, das wiederum nach Gleichung (\ref{6.7}) als die 
Summe der verschobenen Betragsquadrate der Basisfensterspektren 
berechnet wird. Diese Berechnung f"uhrt man f"ur die $N$ gesuchten 
Frequenzwerte \mbox{$\Omega=\nu\CdoT2\pi/F$} mit \mbox{$\nu=0\;(1)\;N\!-\!1$} 
durch. Betrachten wir nun die einzelnen Summanden f"ur eine 
dieser Frequenzen im Detail. F"ur diese diskrete Frequenz~$\nu$ 
--- also bei \mbox{$z\!=\!e^{j\cdot\frac{2\pi}{F}\cdot\nu}$} --- 
berechnen wir nun den in Gleichung (\ref{6.5}) angegebenen 
Pol-Nullstellenquotienten des Produktes 
\mbox{$G\big(z\CdoT e^{\!-j\cdot\frac{2\pi}{F}\cdot\nu_1}\big)\CdoT 
G\big(z^{\!-1}\!\CdoT e^{j\cdot\frac{2\pi}{F}\cdot\nu_1}\big)$}, wobei die 
komplexe Frequenz $z$ jeweils entsprechend dem Summenindex $\nu_1$ in 
Gleichung (\ref{6.7}) um Vielfache von \mbox{$2\pi/F$} rotiert ist. Bei 
einigen Summanden existiert f"ur die betrachtete diskrete Frequenz $\nu$ 
eine doppelte Nullstelle bei \mbox{$z\!=\!e^{j\cdot\frac{2\pi}{F}\cdot\nu}$}, 
die durch {\em keine}\/ identische doppelte Polstelle kompensiert wird. Diese 
Summanden sind daher Null, und tragen somit nichts zum Betragsquadrat 
des Fensterspektrums bei der Frequenz \mbox{$\Omega=\nu\CdoT2\pi/F$} bei. 
Bei den anderen Summanden kann jeweils eine doppelte Nullstelle 
mit einer identischen doppelten Polstelle bei
\mbox{$z\!=\!e^{j\cdot\frac{2\pi}{F}\cdot\nu}$} gek"urzt werden.
Das bei einem solchen Summanden verbleibende Produkt der Z"ahler der 
in Gleichung (\ref{6.5}) angegebenen Pol-Nullstellenquotienten ist 
immer \mbox{${\D F^2}$}. Dieses verbleibende Z"ahlerprodukt ist ja gerade 
das Betragsquadrat der periodisch fortgesetzten si-Funktion an der Stelle 
\mbox{$\Omega\!=\!0$}. Die Konstante \mbox{${\D F^2}$} ist dann 
noch durch das Produkt der doppelten Polstellen, die bei dem jeweiligen
Summanden verbleiben, zu dividieren, um den Summanden bei der Frequenz 
\mbox{$\Omega=\nu\CdoT2\pi/F$} vollst"andig zu berechnen. Die Summe aller 
dieser Summanden ergibt das gesuchte Betragsquadrat des Spektralwertes 
\mbox{$F(\nu\CdoT2\pi/F)$}.

{\small Beispiel: Betrachten wir in dem Beispiel des Bildes \ref{b4b} wie 
sich der der Betrag des Fensterspektrums bei der Frequenz \mbox{$\Omega=2\pi/F$} 
--- also mit \mbox{$\nu\!=\!1$} --- berechnet. In den rechten Teilbildern 
des Bildes \ref{b4b} habe ich diesen Frequenzpunkt am Einheitskreis jeweils 
durch eine Hilfslinie markiert. Der erste Summand in Gleichung (\ref{6.7})
mit \mbox{$\nu_1\!=\!-1$} ist in der dritten Reihe des Bildes \ref{b4b} 
dargestellt. Er hat bei der betrachteten komplexen Frequenz 
\mbox{$z\!=\!e^{j\cdot\frac{2\pi}{F}}$} eine doppelte Nullstelle und liefert 
daher keinen Beitrag. Der zweite Summand in Gleichung (\ref{6.7})
ist von Null verschieden, wie man in der zweiten Reihe des Bildes 
\ref{b4b} sehen kann. Es handelt sich dabei um das Betragsquadrat des
dar"uber dargestellten Spektrums, dessen dazugeh"orige Z-Transformierte in 
Gleichung (\ref{6.5}) angegebenen ist. Hier kann die Nullstelle des Z"ahlers 
bei \mbox{$e^{j\cdot\frac{2\pi}{F}}$} mit der Polstelle mit \mbox{$\nu_2\!=\!1$}
gek"urzt werden. Der Betrag des Z"ahlers ist dann $F$, so dass dieser Faktor
quadratisch in den Summanden mit \mbox{$\nu_1\!=\!0$} eingeht. Zur vollst"andigen 
Berechnung des Summanden ist noch durch das Betragsquadrat der Abst"ande 
des Punktes \mbox{$e^{j\cdot\frac{2\pi}{F}}$} zu den beiden noch verbleibenden 
Polstellen mit \mbox{$\nu_2\!=\!0$} und \mbox{$\nu_2\!=\!-1$} zu dividieren.
Bei dem in der vierten Reihe des Bildes \ref{b4b} dargestellten Summanden
ist aufgrund der Verschiebung in Gleichung (\ref{6.5}) $z$ durch 
\mbox{$z\CdoT e^{\!-j\cdot\frac{2\pi}{F}}$} zu ersetzen. Nun l"asst sich die 
Nullstelle bei \mbox{$e^{j\cdot\frac{2\pi}{F}}$} mit der Polstelle mit 
\mbox{$\nu_2\!=\!0$} k"urzen. Der sich auch hier ergebende Faktor 
\mbox{${\D F^2}$} ist wieder durch das Betragsquadrat der Abst"ande 
des Punktes \mbox{$e^{j\cdot\frac{2\pi}{F}}$} zu den verbleibenden 
Polstellen zu dividieren. Diese tragen nun die Produktindizes
\mbox{$\nu_2\!=\!1$} und \mbox{$\nu_2\!=\!-1$}. Abschlie"send wird
aus der Summe der beiden von Null verschiedenen Anteile (vorletzte 
Teilbildreihe) noch die Wurzel gezogen (unterste Teilbildreihe), 
und man erh"alt den gesuchten Betrag \mbox{$|F(2\pi/F)|$}} 

\subsubsection{Phasen der Fourierreihenkoeffizienten der Fensterfolge}

Bei der Berechnung der Phase des Spektrums der Fensterfolge
"uber die Faktorisierung von \mbox{$D(z)$} in einen minimal- 
und einen maximalphasigen Anteil kann der Anteil der doppelten 
Nullstellen von \mbox{$D(z)$} am Einheitskreis getrennt 
behandelt werden. Die Phase dieses Anteils, der im weiteren 
als \mbox{$D_E\big(e^{j\Omega}\big)$} bezeichnet wird, ist linear 
und kann je zur H"alfte dem minimal- und dem maximalphasigen Anteil 
zugeordnet werden. F"ur die $N$ Frequenzpunkte, f"ur die die
Abtastwerte des Spektrums berechnet werden m"ussen, kann man diesen 
linearen Anteil aufgrund der bekannten Nullstellenlage explizit angeben. 

Die Nullstellenlage des Restes von \mbox{$D(z)$}, ist unbekannt, da diese 
Nullstellen erst durch die additive "Uberlagerung der verschobenen Versionen 
\mbox{$G\big(z\CdoT e^{\!-j\cdot\frac{2\pi}{F}\cdot\nu_1}\big)\CdoT
G\big(z^{\!-1}\!\CdoT e^{j\cdot\frac{2\pi}{F}\cdot\nu_1}\big)$}
des Produkts der Z-Transformierten der Basisfensterfolge entstehen. 

Um den Anteil der Phase des Spektrums der Fensterfolge zu berechnen, 
der von dem Teil von \mbox{$D(z)$} verursacht wird, der die unbekannten 
Nullstellen enth"alt, und der im weiteren \mbox{$D_{\overline{E}}(z)$} 
genannt wird, w"ahlen wir einen Weg, bei dem wir die unbekannten 
Nullstellen nicht explizit berechnen m"ussen. Dabei nutzen wir 
die Tatsache, dass die invers Z-Transformierte
des Logarithmus der Z-Transformierten einer reellen minimalphasigen
Folge --- nach \cite{Opp} auch als komplexes Cepstrum bezeichnet ---
eine reelle kausale Folge ist. Wie jede reelle kausale Folge 
l"asst sich auch diese als die Summe einer geradesymmetrischen und 
einer schiefsymmetrischen Folge darstellen. F"ur positives $k$ sind 
der gerade- und der schiefsymmetrische Anteil identisch und halb 
so gro"s wie die reelle kausale Folge. Somit l"asst sich, wenn man 
den geradesymmetrischen Anteil kennt, der schiefsymmetrische Anteil bestimmen, 
indem man bei den Werten f"ur negatives $k$ das Vorzeichen invertiert 
und den Wert f"ur \mbox{$k\!=\!0$} zu Null setzt. 

Der geradesymmetrische Anteil --- nach \cite{Opp} auch als reelles 
Cepstrum bezeichnet --- ist die invers Fouriertransformierte der
D"ampfung mit invertierten Vorzeichen. Die D"ampfung ist n"amlich 
--- bis auf das Vorzeichen --- der Realteil des Logarithmus der 
Z-Transformierten der reellen minimalphasigen Folge f"ur 
\mbox{$z\!=\!e^{j\Omega}$}. Indem man nun daraus den 
schiefsymmetrischen Anteil berechnet und diesen fouriertransformiert, 
erh"alt man die mit $-j$ multiplizierte Phase. Diese ist n"amlich 
--- ebenfalls bis auf das Vorzeichen --- der Imagin"arteil des 
Logarithmus der Z-Transformierten der reellen minimalphasigen Folge 
f"ur \mbox{$z\!=\!e^{j\Omega}$}. 

Die D"ampfung des gesuchten minimalphasigen Anteils von 
\mbox{$D_{\overline{E}}(z)$} mit den unbekannten Nullstellen kann man 
f"ur \mbox{$z\!=\!e^{j\Omega}$} dadurch erhalten, dass man von einer 
Summe "uber die Betragsquadrate von Polynomen mit bekannten Nullstellen, 
die im Unterkapitel \ref{Fenster1} im Detail angegeben werden, den 
{\em halben}\footnote{Da hier die Quadrate der Betr"age der Polynome 
"uberlagert werden, und nicht die Betr"age selbst, ist der Logarithmus 
davon zu halbieren.} Logarithmus berechnet. Die Phase des 
minimalphasigen Anteils von \mbox{$D_{\overline{E}}\big(e^{j\Omega}\big)$} 
erh"alt man wie oben beschrieben, indem man die Hilbert-Transformierte 
der D"ampfung berechnet. Dazu wird die D"ampfung zun"achst 
fourierr"ucktransformiert, dann mit $j$ mal der Signumfunktion 
\mbox{sgn$(k)$} multipliziert und danach wieder fouriertransformiert. 

Damit sich diese beiden Fouriertransformationen mit einer FFT 
endlicher L"ange durchf"uhren lassen, m"usste die Fouriertransformierte 
der D"ampfung eine endliche L"ange aufweisen. Dies ist jedoch 
--- abgesehen von dem trivialen Fall der $\gamma_0$-Folge --- 
bei keiner Folge der Fall. Vielmehr setzt sich das komplexe Cepstrum einer
minimalphasigen Folge aus im wesentlichen exponentiell abklingenden und daher 
unbegrenzt lang andauernden Komponenten zusammen, die sich in der Form 
\mbox{$\zu^{\uP{-0.4}{\!k}}/\,k$} schreiben lassen. Dabei sind mit 
$\zu$ die in unserem Fall unbekannten Nullstellen bezeichnet. 
Das Fragezeichen im Formelzeichen soll hier andeuten, dass diese 
Nullstellen bei dem hier verwendeten Algorithmus an keiner Stelle 
explizit berechnet werden somit immer unbekannt bleiben. Wegen der 
Minimalphasigkeit liegen alle diese unbekannten Nullstellen innerhalb des 
Einheitskreises, und die exponentiellen Komponenten sind alle mit steigendem 
$k$ abklingend. Die Abklingkonstanten dieser Komponenten h"angen davon ab, 
wie nahe sich diese Nullstellen der Z-Transformierten der Folge am 
Einheitskreis befinden. Da an einem Computer die FFT sowieso nur mit 
einer endlichen Rechengenauigkeit berechnet werden kann, gen"ugt es die 
L"ange der FFT so gro"s zu w"ahlen, dass der Anteil der exponentiell
abklingenden Komponenten, der sich oberhalb des Rauschsockels der
Rechengenauigkeit befindet, innerhalb der halben FFT-L"ange liegt.
Bild \ref{b4c}
\begin{figure}[btp]
\begin{center}
{ 
\begin{picture}(454,190)

\input{mbild4c.tex}

\put(40,176){\makebox(0,0)[l]{$
20\cdot\log_{10}\Big(\,\Big|\,\text{ifft}_{4096}\big(\,\ln(\,|
X(\mu\CdoT2\pi/4096)|\,)\,\big)\,\Big|\,\Big)$}}
\put(25,160){\makebox(0,0)[r]{\small$50$}}
\put(25,100){\makebox(0,0)[r]{\small$-100$}}
\put(25,60){\makebox(0,0)[r]{\small$-200$}}
\put(25,20){\makebox(0,0)[r]{\small$-300$}}
\put(32,137){\makebox(0,0)[lt]{\small$0$}}
\put(230,136){\makebox(0,0)[t]{\small$1000$}}
\put(430,136){\makebox(0,0)[t]{\small$2000$}}
\put(450,144){\makebox(0,0)[br]{$k$}}

\end{picture}}
\end{center}\vspace{-11pt}
\setlength{\belowcaptionskip}{10pt}
\caption[Numerisch berechnetes Cepstrum einer minimalphasigen Folge]
{Numerisch berechnetes Cepstrum einer minimalphasigen Folge.\\[10pt]
\centering Beispiel mit der Folge
$x(k)=\begin{cases}
\;100\!-\!k&\quad\text{ f"ur } k=0\;(1)\;99\\
\;0&\quad\text{ sonst }
\end{cases}$}
\label{b4c}
\rule{\textwidth}{0.5pt}\vspace{-4pt}
\end{figure}
zeigt an einem Beispiel den Verlauf des
Betrags des reellen Cepstrums einer minimalphasigen Folge \mbox{$x(k)$},
das mit einem Computer mit einer FFT der L"ange $4096$ berechnet worden ist,
in halblogarithmischer Darstellung. Deutlich ist zu erkennen, dass
bei der gew"ahlten FFT-L"ange der Betrag des Cepstrums f"ur hinreichend
gro"se Werte von $k$ bereits in dem von der begrenzten Rechengenauigkeit
herr"uhrenden Rauschsockel verschwindet, so dass nicht zu erwarten ist,
dass die Phase des Spektrums dieser Folge mit einer l"angeren FFT
besser zu berechnen w"are.

Es zeigte sich, dass bei dem zu faktorisierenden Betragsquadratspektrum
\mbox{$D\big(e^{j\Omega}\big)$} die Fourierr"ucktransformierte der
D"ampfung des Anteils \mbox{$D_{\overline{E}}\big(e^{j\Omega}\big)$},
vor allem f"ur gro"se Werte von $M$ viel zu langsam abklingt, um damit
die Phase der Fensterfolge mit einer FFT vern"unftiger L"ange berechnen
zu k"onnen. Offensichtlich liegen also ein oder mehrere der unbekannten
Nullstellen viel zu nahe am Einheitskreis, so dass die im wesentlichen 
exponentiell abklingenden Komponenten \mbox{$\zu^{\uP{-0.4}{\!k}}/\,k$} 
selbst f"ur sehr gro"ses $k$ oberhalb des Rauschsockels der
Rechengenauigkeit liegen. Daher wird vor der Berechnung der Phase 
eine Bilineartransformation des Anteils \mbox{$D_{\overline{E}}(z)$} 
mit den unbekannten Nullstellen durchgef"uhrt. Bei dieser Bilineartransformation 
wird die komplexe Frequenzvariable $z$ durch die neue ebenfalls komplexe 
Frequenzvariable $\Tilde{z}$ ersetzt, und wir erhalten die bilinear 
Z-Transformierte \mbox{$\widetilde{D}_{\overline{E}}(\Tilde{z})$}. 

Durch die bilineare Transformation werden die Punkte 
\mbox{$e^{j\cdot\frac{2\pi}{F}\cdot\nu}$} am Einheitskreis der 
$z$-Ebene, die den Frequenzen entsprechen, f"ur die das Spektrum 
der Fensterfolge berechnet werden soll, auf Punkte am Einheitskreis 
der $\Tilde{z}$-Ebene abgebildet. F"ur diese Punkte ist nun die Phase 
von \mbox{$\widetilde{D}_{\overline{E}}(\Tilde{z})$} zu bestimmen. 

Die bilinear Z-Transformierte \mbox{$\widetilde{D}_{\overline{E}}(\Tilde{z})$} 
l"asst sich ebenfalls wieder in zwei Anteile zerlegen. Der eine Anteil 
\mbox{$\widetilde{D}_P(\Tilde{z})$} besteht dabei aus einem 
Polstellenpaar mehrfacher Vielfachheit mit {\em bekannter}\/ 
Polstellenlage. Der Einfluss dieser Polstellen auf die Phase des 
Spektrums der gesuchten Fensterfolge kann explizit angegeben werden. 

Der restlichen Anteil \mbox{$\widetilde{D}_N(\Tilde{z})$} enth"alt 
wieder nur unbekannte Nullstellen. Der Betrag dieses Anteils l"asst 
sich als eine Summe berechnen, wobei die einzelnen Summanden 
die Beitr"age bilineartransformierter, bekannter, verschobener 
Nullstellen sind. Die Phase des Anteils \mbox{$\widetilde{D}_N(\Tilde{z})$} kann 
man als Fouriersinusreihe darstellen. Die Koeffizienten der Fouriersinusreihe 
erh"alt man aus dem Cepstrum von \mbox{$\widetilde{D}_N(\Tilde{z})$}, das man 
in der oben beschriebenen Art "uber eine FFT des Logarithmus des Betrages von 
\mbox{$\widetilde{D}_N(\Tilde{z})$} berechnet. Da bei geeigneter Wahl des 
Bilineartransformationsparameters $c$ die unbekannten Nullstellen nach der 
Transformation nicht mehr so nahe am Einheitskreis liegen, klingt das Cepstrum 
rascher ab, und die L"ange $\widetilde{M}$ der zur Berechnung des Cepstrums 
eingesetzten FFT liegt in einer sinnvollen Gr"o"senordnung. Die Phase des Anteils 
\mbox{$\widetilde{D}_N(\Tilde{z})$} berechnet man nun "uber die Auswertung der 
Fouriersinusreihe f"ur die {\em nicht "aquidistanten}\footnote{Eine FFT 
kann hier nicht eingesetzt werden.} Frequenzen $\widetilde{\Omega}$, die durch 
die Bilineartransformation aus den Frequenzen \mbox{$\Omega=\nu\CdoT2\pi/F$} 
entstanden sind, f"ur die das Spektrum der Fensterfolge zu berechnen ist.

Die gesuchte Gesamtphase des Spektrums der Fensterfolge erh"alt man,
indem man diesen Phasenanteil zu dem Phasenanteil der Polstellen
von \mbox{$\widetilde{D}_P(\Tilde{z})$} und dem linearen
Phasenanteil der Nullstellen von \mbox{$D_E(z)$} auf dem
Einheitskreis addiert. Die Exponentialfunktion mit der so berechneten
Gesamtphase multipliziert man mit dem zuvor berechneten Betrag und man 
erh"alt die gesuchten Spektralwerte der Fensterfolge.
Nun normiert man diese Spektralwerte mit einem konstanten Faktor, so 
dass sich f"ur den Wert des Spektrums bei der Frequenz Null der nach 
Gleichung (\ref{2.27}) geforderte Wert $M$ ergibt.
Zuletzt wird die Fensterfolge aus den Spektralwerten berechnet.

{\small Anmerkung: Da die Zuordnung der Nullstellen nicht nach der
Minimalphasigkeit erfolgen muss, gibt es weitere m"ogliche L"osungen.
Die maximalphasige L"osung erh"alt man durch Spiegelung der minimalphasigen
L"osung. Die weiteren L"osungen mit dazwischenliegenden Phasenverl"aufen
k"onnen mit dem hier vorgestellten Algorithmus nicht berechnet werden.
Alle L"osungen w"aren f"ur die Anwendung beim RKM genauso gut geeignet.}

\subsection{Der dritte Schritt des Algorithmus im Detail}\label{Fenster1}

Wir beginnen damit, die gebrochen rationale Funktion \mbox{$D(z)$},
deren Betrag f"ur \mbox{$z\!=\!e^{j\Omega}$} das Betragsquadrat
des Spektrums der Fensterfolge ist, in eine geeignete 
Form zu bringen. Dazu ersetzen wir in der Definitionsgleichung (\ref{6.7})
zun"achst \mbox{$e^{j\Omega}$} durch~$z$, substituieren anschlie"send 
das Betragsquadrat mit Gleichung (\ref{6.6}) und setzen danach die 
gebrochen rationalen Funktionen aus Gleichung (\ref{6.5}) ein:
\begin{gather}
F\big(\!-j\Cdot\ln(z)\big)\CdoT F\big(j\Cdot\ln(z)\big)\;\sim\;
D(z)\;=\!
\Sum{\nu_1=\frac{1-N}{2}}{\frac{N-1}{2}}
G\big(z\CdoT e^{\!-j\cdot\frac{2\pi}{F}\cdot\nu_1}\big)\CdoT
G\big(z^{\!-1}\!\Cdot e^{j\cdot\frac{2\pi}{F}\cdot\nu_1}\big)\;=
\notag\\[6pt]
=\;\Sum{\nu_1=\frac{1-N}{2}}{\frac{N-1}{2}}
\frac{\D z^F\!\CdoT e^{\!-j\cdot2\pi\cdot\nu_1}+(-1)^N}
{\T\Prod{\nu_2=\frac{1-N}{2}}{\frac{N-1}{2}}\!\D
\Big(z\CdoT e^{\!-j\cdot\frac{2\pi}{F}\cdot\nu_1}-
e^{j\cdot\frac{2\pi}{F}\cdot\nu_2}\Big)}\cdot
\frac{\D z^{\!-F}\!\CdoT e^{j\cdot2\pi\cdot\nu_1}+(-1)^N}
{\T\Prod{\nu_2=\frac{1-N}{2}}{\frac{N-1}{2}}\!\D
\Big(z^{\!-1}\!\CdoT e^{j\cdot\frac{2\pi}{F}\cdot\nu_1}-
e^{j\cdot\frac{2\pi}{F}\cdot\nu_2}\Big)}\;=
\notag\\[10pt]
\qquad=\Sum{\nu_1=\frac{1-N}{2}}{\frac{N-1}{2}}
\frac{\D z^F\!\!\CdoT(-1)^{2\cdot\nu_1+N}\!+2+
z^{\!-F}\!\!\CdoT(-1)^{2\cdot\nu_1+N}}
{\T\Prod{\nu_2=\frac{1-N}{2}}{\frac{N-1}{2}}\!\D
\Big(\!-z\CdoT e^{\!-j\cdot\frac{2\pi}{F}\cdot(\nu_1+\nu_2)}+2-
z^{\!-1}\!\CdoT e^{ j\cdot\frac{2\pi}{F}\cdot(\nu_1+\nu_2)}\Big)}\;=
\notag\\[6pt]
=\Sum{\nu_1=\frac{1-N}{2}}{\frac{N-1}{2}}
\frac{\D-z^F\!+2-z^{\!-F}}
{\T\Prod{\nu_2=-\frac{N-1}{2}}{\frac{N-1}{2}}\!\D
\Big(\!-z\CdoT e^{\!-j\cdot\frac{2\pi}{F}\cdot(\nu_1+\nu_2)}+2-
z^{\!-1}\!\CdoT e^{j\cdot\frac{2\pi}{F}\cdot(\nu_1+\nu_2)}\Big)}
\label{6.17}
\end{gather}

\subsubsection{Betr"age der Fourierreihenkoeffizienten der Fensterfolge}

Setzt man \mbox{$z=e^{j\cdot\frac{2\pi}{F}\cdot\nu}$} mit
\mbox{$\nu=0\;(1)\;N\!-\!1$} ein, so erh"alt man f"ur das Betragsquadrat
der zu bestimmenden Spektralwerte, die als Koeffizienten der Fourierreihe
der Fensterfolge nach Gleichung (\ref{6.16}) auftreten,
bis auf einen konstanten Faktor den folgenden Ausdruck:
\begin{gather}
\big|F\big({\T\nu\CdoT\frac{2\pi}{F}}\big)\big|^2\,=\;
F\big({\T\nu\CdoT\frac{2\pi}{F}}\big)\CdoT
F\big({\T-\nu\CdoT\frac{2\pi}{F}}\big)\;\sim\;
D\big(e^{j\cdot\frac{2\pi}{F}\cdot\nu}\big)\;=
\label{6.18}\\*[10pt]
=\; F^2\cdot 4^{1-N}\cdoT\!\!\!
\Sum{\nu_1=\max(\nu,0)-\frac{N-1}{2}}{\min(\nu,0)+\frac{N-1}{2}}\;\;
\Prod{\substack{\nu_2=-\frac{N-1}{2}\\\nu_2\neq\nu-\nu_1}}
{\frac{N-1}{2}}\!\!
\sin\!\big({\T(\nu\!-\!\nu_2\!-\!\nu_1)\CdoT\frac{\pi}{F}}\big)^{\!-2}\!.
\notag
\end{gather}
Dabei wurde im Summationsindex ber"ucksichtigt, dass einige Summanden
Null werden, weil bei diesen die doppelte Nullstelle bei
\mbox{$z=e^{j\cdot\frac{2\pi}{F}\cdot\nu}$} {\em nicht}\/ durch eine 
entsprechende doppelte Polstelle kompensiert wird. Bei den verbleibenden 
Summanden wurde diese doppelte Nullstelle gek"urzt, was sich darin "au"sert, 
dass bei dem Produktindex der eine Wert \mbox{$\nu_2=\nu\!-\!\nu_1$} 
explizit ausgeschlossen ist. F"ur den verbleibenden Z"ahler wurde der Wert
\mbox{${\D F^2\!}$} eingesetzt, den man mit dem Satz von de l' Hospital
aus dem Quotienten des vollst"andigen Z"ahlers und der gek"urzten Polstelle
erh"alt. Es verbleibt im Nenner das Produkt der Quadrate der Abst"ande 
zu den verbleibenden doppelten Polstellen am Einheitskreis. Um die sich 
ergebende Formel "ubersichtlich zu halten, wurde nicht die Darstellung als 
Bruch mit dem Z"ahler $1$ gew"ahlt, sondern bei jedem Polstellenabstand der 
negative Exponent $-2$. Jeder Polstellenabstand wird als doppelter Sinus 
des halben Differenzwinkels zwischen den Polstellenwinkeln und dem 
Frequenzpunkt \mbox{$\frac{2\pi}{F}\cdot\nu$}, f"ur den das Produkt zu 
bestimmen ist, berechnet. Die Verdoppelung des Sinuswertes wurde f"ur alle 
Faktoren eines Produktes in dem Vorfaktor \mbox{$4^{1-N}$} zusammengefasst 
und vor die Summe gezogen. Die Halbierung des Differenzwinkels ist daran zu 
erkennen, dass im Argument der Sinusfunktion nicht \mbox{$2\pi/F$} sondern 
\mbox{$\pi/F$} steht. Da abschlie"send eine Normierung des Fensterspektrums in 
der Art erfolgt, dass \mbox{$F(0)\!=\!M$} ist, kann man auf eine explizite
Berechnung des Vorfaktors \mbox{$F^2\cdot 4^{1-N}$}, der f"ur alle zu berechnenden 
Frequenzpunkte gleich ist, verzichten.

Die positive Wurzel der mit der letzten Gleichung berechneten Werte ist der 
Betrag \mbox{$\big|F\big({\T\nu\CdoT\frac{2\pi}{F}}\big)\big|$} der gesuchten
Spektralwerte der Fensterfolge.

\subsubsection{Phasen der Fourierreihenkoeffizienten der Fensterfolge}

Etwas aufwendiger gestaltet sich die Berechnung der Phase des Spektrums der 
Fensterfolge mit der wir nun beginnen. Der $z^F\!$-fache Z"ahler der 
Summanden in Gleichung (\ref{6.17}) hat nur doppelte Nullstellen am 
Einheitskreis bei Vielfachen von \mbox{$2\pi/F$} und ist von $\nu_1$ 
unabh"angig. Er kann daher vor die Summe gezogen werden. Durch die 
anschlie"sende Erweiterung auf den Hauptnenner erh"alt man
\begin{gather}
\label{6.19}
F\big(\!-j\Cdot\ln(z)\big)\CdoT F\big(j\Cdot\ln(z)\big)\;\sim\;
D(z)\;=
\\*[7pt]
=\,\underbrace{\frac{\D-z^F\!+2-z^{\!-F}}{\!\T\Prod{\nu_3=1-N}{N-1}\!\!\D
\Big(\!\!-\!z\CdoT e^{\!-j\cdot\frac{2\pi}{F}\cdot\nu_3}+2-
z^{\!-1}\!\CdoT e^{j\cdot\frac{2\pi}{F}\cdot\nu_3}\Big)}}_{\D=D_E(z)}
\,\,\cdoT\!\underbrace{
\Sum{\nu_1=\frac{1-N}{2}}{\frac{N-1}{2}}\;\Prod{\nu_2}{}
\Big(\!-z\CdoT e^{\!-j\cdot\frac{2\pi}{F}\cdot\nu_2}+2-
z^{\!-1}\!\CdoT e^{j\cdot\frac{2\pi}{F}\cdot\nu_2}\Big)}
_{\D=D_{\overline{E}}(z)},
\notag
\end{gather}
wobei der Lauf"|index $\nu_2$ des in der Summe stehenden Produkts jeweils die
\mbox{$N\!-\!1$} Werte annimmt, die das Produkt entstehen lassen, das zur
Erweiterung auf den Hauptnenner beim jeweiligen Summanden ben"otigt wird. 
Da die Aufz"ahlung der Werte, die der Lauf"|index $\nu_2$ durchl"auft, zu 
umfangreich ist, um unterhalb des Produktsymbols Platz zu finden, wird diese 
nun explizit angegeben. Bei dem Summanden mit dem Lauf"|index $\nu_1$ kann 
$\nu_2$ dieselben Werte \mbox{$1\!-\!N\;(1)\;N\!-\!1$} annehmen, wie der Lauf"|index 
$\nu_3$ des Produkts im Nenner von \mbox{$D_E(z)$}. Die Werte
\mbox{$\nu_1\!+\!(1\!-\!N)/2\;\;(1)\;\;\nu_1\!+\!(N\!-\!1)/2$} sind jedoch 
ausgeschlossen, da diese den Polstellen in Gleichung (\ref{6.17}) entsprechen.

Wenn man die identischen Pol- und Nullstellen der gebrochen rationalen 
Funktion \mbox{$D_E(z)$} k"urzt, ist dieser Anteil mit \mbox{$z^{F+1-2\cdot N}$}
multipliziert, ein Polynom aus doppelten Nullstellen am Einheitskreis
bei \mbox{$\Omega=\nu\CdoT2\pi/F$} mit \mbox{$\nu=N\;(1)\;F\!-\!N$}.
Jede der doppelten Nullstellen tritt als einfache Nullstelle 
im minimalphasigen Anteil von \mbox{$D_E(z)$} auf. Die Phase
dieses minimalphasigen Anteils ist ---\,abgesehen von einem $\pi$-Sprung 
bei \mbox{jeder Nullstelle}\,--- in~$\Omega$ linear. Sie steigt mit 
\mbox{$(F\!-\!2\CdoT\!N\!+\!1)\CdoT\Omega/2$}, und wird sp"ater zu der 
Phase des minimalphasigen Anteils von \mbox{$D_{\overline{E}}(z)$} addiert. 
Im Bereich der interessierenden Frequenzen \mbox{$|\Omega|<2\pi/M$}
besitzt \mbox{$D_E(z)$} keine Nullstellen, so dass die Phase von
\mbox{$D_E(z)$} dort keine $\pi$-Spr"unge hat, die man andernfalls
ber"ucksichtigen m"usste.

Alle Summanden der Summe \mbox{$D_{\overline{E}}(z)$} der Produkte, die 
zur Erweiterung auf den Hauptnenner ben"otigt wurden, sind f"ur 
\mbox{$z\!=\!e^{j\Omega}$} nichtnegativ und bei keiner Frequenz $\Omega$ 
sind alle Summanden zugleich Null. Daher weist \mbox{$D_{\overline{E}}(z)$} 
am Einheitskreis keine Nullstellen auf.

Um die Phase dieses Anteils \mbox{$D_{\overline{E}}(z)$} "uber das Cepstrum 
mit einer FFT vern"unftiger L"ange berechnen zu k"onnen, empfiehlt sich vor 
der Berechnung des Cepstrums eine Bilineartransformation vorzunehmen, indem 
man die komplexe Frequenzvariable $z$ durch die gebrochen rationale Funktion\vspace{-6pt} 
\begin{equation}
z\;=\;\frac{\Tilde{z}+(1\!-\!c)}{{}\,1+(1\!-\!c)\CdoT \Tilde{z}\,{}}
\label{6.20}
\end{equation}
der neuen komplexen Frequenzvariable $\Tilde{z}$ substituiert.
Nach $z$ aufgel"ost erh"alt man f"ur die R"ucktransformation
die Substitution\vspace{-1pt}
\begin{equation}
\Tilde{z}\;=\;\frac{z-(1\!-\!c)}{{}\,1-(1\!-\!c)\CdoT z\,{}}.
\label{6.21}
\end{equation}
Diese Bilineartransformation bildet den Einheitskreis \mbox{$z=e^{j\Omega}$} 
der $z$-Ebene auf den Einheitskreis \mbox{$\Tilde{z}=e^{j\widetilde{\Omega}}$} 
der $\Tilde{z}$-Ebene ab. F"ur die Frequenzen $\Omega$ und $\widetilde{\Omega}$ 
ergibt sich folgender Zusammenhang:\vspace{-6pt} 
\begin{equation}
\widetilde{\Omega}\;=\;
\Omega+2\CdoT\arctan\bigg(\!\frac{(1\!-\!c)\cdot\sin(\Omega)}
{\;c+2\CdoT(1\!-\!c)\CdoT\sin\!\big(\frac{\Omega}{2}\big)^{\!2}}\!\bigg).
\label{6.22}
\end{equation}
Mit Hilfe des Parameters $c$ l"asst sich die Lage der unbekannten Nullstellen 
der bilineartransformierten Funktion in der $\Tilde{z}$-Ebene beeinflussen.
Wir werden diesen Parameter im Bereich \mbox{$0\!<\!c\!<\!2$} w"ahlen, so 
dass sowohl bei der Hin-, als auch bei der R"ucktransformation, die Polstelle
in den Gleichungen (\ref{6.20}) und (\ref{6.21}) immer au"serhalb des 
Einheitskreises liegt. Ein konkreter Vorschlag, wie der Parameter $c$ 
sinnvoll eingestellt werden kann, wird im Anschluss an Gleichung 
(\ref{6.29}) angegeben. 

F"ur \mbox{$D_{\overline{E}}(z)$} erh"alt man dann durch Substitution von~$z$\vspace{-4pt} 
\begin{gather}
D_{\overline{E}}(z)\;=
\Sum{\nu_1=\frac{1-N}{2}}{\frac{N-1}{2}}\;\Prod{\nu_2}{}
\Big(\!-z\CdoT e^{\!-j\cdot\frac{2\pi}{F}\cdot\nu_2}+2-
z^{\!-1}\!\CdoT e^{j\cdot\frac{2\pi}{F}\cdot\nu_2}\Big)\;=\;
\widetilde{D}_{\overline{E}}(\Tilde{z})\;=
\label{6.23}\\*
=\;\underbrace{\bigg(\frac{\Tilde{z}}
{\,\big(1\!+\!(1\!-\!c)\CdoT\Tilde{z}\big)\CdoT
\big(\Tilde{z}\!+\!(1\!-\!c)\big)\,}
\bigg)^{\!\!N-1}\!\!}_{\D=\widetilde{D}_P(\Tilde{z})}\,\cdoT\!
\underbrace{
\Sum{\nu_1=\frac{1-N}{2}}{\frac{N-1}{2}}\;\Prod{\;\nu_2}{}
\Big(K_{\nu_2}\!\CdoT(\Tilde{z}\!-\!\Tilde{z}_{\nu_2})\CdoT
(\Tilde{z}^{-1}\!\!-\!\Tilde{z}_{\nu_2}^{\,*})\Big)
}_{\D=\widetilde{D}_N(\Tilde{z})},\notag
\end{gather}
mit den Nullstellen $\Tilde{z}_{\nu_2}$, die durch die
Bilineartransformation aus den Nullstellen der einzelnen Summanden
am Einheitskreis bei \mbox{$z\!=\!e^{j\cdot\frac{2\pi}{F}\cdot\nu_2}$}
entstanden sind:
\begin{equation}
\Tilde{z}_{\nu_2}\;=\;\frac{\D e^{j\cdot\frac{2\pi}{F}\cdot\nu_2}-(1\!-\!c)}
{\D{}\;\;1-(1\!-\!c)\CdoT e^{j\cdot\frac{2\pi}{F}\cdot\nu_2}\;\;{}}\;=\;
e^{j\cdot2\cdot(\frac{\pi}{F}\cdot\nu_2+\Tilde{\psi}_{\nu_2})}
\label{6.24}
\end{equation}
F"ur die Konstante $K_{\nu_2}$ ergibt sich der Wert
\begin{equation}
K_{\nu_2}\;=\;
\big|1-(1\!-\!c)\CdoT e^{j\cdot\frac{2\pi}{F}\cdot\nu_2}\big|^2\;=\;
c^2+4\CdoT(1\!-\!c)\CdoT\sin\!\big({\T\nu_2\CdoT\frac{\pi}{F}}\big)^{\!2}\!,
\label{6.25}
\end{equation}
und f"ur den halben Winkel der Nullstellenrotation $\Tilde{\psi}_{\nu_2}$ 
mit Gleichung (\ref{6.22}) der Wert
\begin{equation}
\Tilde{\psi}_{\nu_2}\;=\;
\arctan\!\bigg(\!\frac{(1\!-\!c)\CdoT\sin\!\big(\frac{2\pi}{F}\CdoT\nu_2\big)}
{\;\;c+2\CdoT(1\!-\!c)\CdoT
\sin\!\big(\frac{\pi}{F}\CdoT\nu_2\big)^{\!2}\;}\!\bigg).
\label{6.26}
\end{equation} 
Weil der Einheitskreis bei der Bilineartransformation wieder auf den Einheitskreis 
abgebildet wird, ist \mbox{$\widetilde{D}_{\overline{E}}(\Tilde{z})$} ebenso 
wie \mbox{$D_{\overline{E}}(z)$} am Einheitskreis positiv und weist dort
{\em keine}\/ Nullstellen auf. 

In Gleichung (\ref{6.23}) steht der Term \mbox{$\widetilde{D}_P(\Tilde{z})$},
dessen Polstellen spiegelsymmetrisch zum Einheitskreis bei \mbox{$\Tilde{z}=c\!-\!1$} 
und \mbox{$\Tilde{z}=1/(c\!-\!1)$} liegen, vor der Summe.
Dieser Anteil wird bilinear zur"ucktransformiert. Man erh"alt
--- wieder bis auf einen konstanten Faktor --- den Term\vspace{-8pt} 
\begin{equation}
D_P(z)\;=\;\bigg(\frac{\,\big(1\!-\!(1\!-\!c)\CdoT z\big)\CdoT
\big(z\!-\!(1\!-\!c)\big)\,}{z}\bigg)^{\!N-1}\!\!\!.
\label{6.27}
\end{equation}
Dieser Term hat ein zum Einheitskreis symmetrisches Nullstellenpaar
der Vielfachheit \mbox{$N\!-\!1$} bei \mbox{$z=1\!-\!c$} und
\mbox{$z=1/(1\!-\!c)$} sowie eine Polstelle bei \mbox{$z\!=\!0$} mit
derselben Vielfachheit. Der minimalphasige Anteil dieses Terms
setzt sich aus den \mbox{$N\!-\!1$} Faktoren
\mbox{$\big(z\!-\!(1\!-\!c)\big)/z$} mit den Nullstellen innerhalb
des Einheitskreises zusammen und liefert f"ur jeden Faktor den
Phasenbeitrag\vspace{-6pt} 
\begin{equation}
\qquad-\arctan\!\bigg(\frac{(1\!-\!c)\CdoT\sin(\Omega)}
{\;c+2\CdoT(1\!-\!c)\CdoT\sin\!\big(\frac{\Omega}{2}\big)^{\!2}\,}\bigg),
\label{6.28}
\end{equation}
der f"ur \mbox{$\Omega = \nu\CdoT2\pi/F$} mit \mbox{$\nu=1\;(1)\;N\!-\!1$}
bei der Berechnung der Phase von \mbox{$F(\nu\CdoT2\pi/F)$} entsprechend
mit der Vielfachheit \mbox{$N\!-\!1$} zu addieren ist. 

Nun m"ussen wir noch die Phase des minimalphasigen Anteils des Summenanteils
\mbox{$\widetilde{D}_N(\Tilde{z})$} in Gleichung (\ref{6.23}) berechnen. Diese
Summe ist f"ur \mbox{$\Tilde{z}=e^{j\widetilde{\Omega}}$} positiv reell
und l"asst sich bis auf einen konstanten
Faktor in der Form
\begin{gather}
\widetilde{D}_N\big(e^{j\widetilde{\Omega}}\big)\;\sim\!
\Sum{\nu_1=\frac{1-N}{2}}{\frac{N-1}{2}}\;
\Prod{\nu_2}{}\Big(K_{\nu_2}\!\CdoT
\sin\!\big({\T\frac{\widetilde{\Omega}}{2}\!-\!\frac{\pi}{F}\CdoT\nu_2\!-\!
\Tilde{\psi}_{\nu_2}}\big)^{\!2}\Big)\;=
\label{6.29}\\*[4pt]
=\;\Sum{\nu_1=\frac{1-N}{2}}{\frac{N-1}{2}}\;
\Prod{\nu_2}{}\bigg(K_{\nu_2}\!\CdoT
\sin\!\bigg(\frac{\pi}{\widetilde{M}}\CdoT
\Big(\eta\!-\!\widetilde{M}\CdoT\big({\T\frac{\nu_2}{F}\!+\!
\frac{\Tilde{\psi}_{\nu_2}}{\pi}}\big)\Big)\bigg)^{\!\!2}\,\bigg)\notag
\end{gather}
berechnen. Hier wurden zuletzt die nach der Bilineartransformation 
f"ur ganzzahlige Werte~$\eta$ "aquidistanten Frequenzen 
\mbox{$\widetilde{\Omega}=\eta\CdoT2\pi/\widetilde{M}$} eingesetzt. 

Wenn wir nun den Logarithmus einer Periode der kontinuierlichen Funktion 
\mbox{$\widetilde{D}_N\big(e^{j\cdot\widetilde{\Omega}}\big)$} 
invers fouriertransformieren, erhalten wir das doppelte Cepstrum, das 
reell und geradesymmetrisch ist. Da das Cepstrum immer abklingt, kann 
man das doppelte Cepstrum im Rahmen der Rechengenauigkeit auch 
mit Hilfe einer inversen FFT des Logarithmus von 
\mbox{$\widetilde{D}_N\big(e^{j\cdot\eta\cdot2\pi/\widetilde{M}}\big)$} 
berechnen, wenn wir nur $\widetilde{M}$ gro"s genug w"ahlen. 
Wie gro"s $\widetilde{M}$ zu w"ahlen ist, h"angt neben der 
Zahlendarstellung des verwendeten Rechners vor allen von der Wahl 
des Bilineartransformationsparameters~$c$ ab. Damit das Cepstrum 
m"oglichst rasch abklingt, muss der Bilineartransformationsparameter~$c$ 
geeignet gew"ahlt werden. F"ur $c$ wird der empirisch ermittelte Wert 
\begin{equation}
c\;=\;\frac{\D2}
{\D{}\;\;1+\frac{{}\;\big(\frac{N}{2}\big)^{\frac{M}{3\cdot(1-M)}}\;{}}
{\tan\!\big(\frac{\pi}{2\cdot M}\big)}\;\;}
\label{6.30}
\end{equation}
eingestellt. Damit wird f"ur alle \mbox{$M\!\ge\!2$} und alle 
\mbox{$2\!\le\!N\!\le\!64$} nahezu das schnellste Abklingen des 
doppelten Cepstrums erreicht. Es ist anzunehmen, dass auch f"ur 
\mbox{$N\!>\!64$} diese Wahl des Bilineartransformationsparameters $c$ 
geeignet ist. Dies wurde jedoch nur stichprobenartig f"ur einige Werte von 
\mbox{$N\!\le\!256$} "uberpr"uft. Als n"otige L"ange $\widetilde{M}$ f"ur 
die inverse FFT konnte bei diesem $c$ der Mindestwert 
\begin{equation}
\widetilde{M}>\big(\text{ Mantissenwortl"ange}-1\,\big)\cdot
2^{{}^{\T\ln\!\big(\frac{N}{3}\big)}}\cdot3,\!6^{{}^{\T\frac{1}{M}}};
\label{6.31}
\end{equation}
empirisch bestimmt werden, der von der Mantissenwortl"ange des zur 
Berechnung der Fensterfolge verwendeten Rechners abh"angt. 
Die Mantissenwortl"ange des Rechners kann z.~B. in der 
Interpretersprache {\tt MATLAB} durch {\tt 1-log(eps)/log(2)} 
bestimmt werden. Nach dem IEEE-Standard 754 ist sie bei 8-Byte
Gleitkommaarithmetik 53 Bit. Von diesen 53 Bit werden nur 52 Bit
explizit in den acht Byte abgespeichert. Das h"ochstwertige
Mantissen-Bit ist bei der verwendeten normierten Zahlendarstellung
immer Eins, und wird daher nicht explizit abgespeichert. Damit eine
inverse FFT verwendet werden kann, wird die niedrigste Zweierpotenz
f"ur $\widetilde{M}$ gew"ahlt, die die Gleichung (\ref{6.31}) erf"ullt.

F"ur das doppelte Cepstrum erhalten wir:
\begin{equation}
2\CdoT C(k)\;=\; 
\frac{1}{\widetilde{M}}\cdot
\Sum{\eta=0}{\widetilde{M}-1}
\ln\Big(
\widetilde{D}_N\big(e^{j\cdot\eta\cdot\frac{2\pi}{\widetilde{M}}}\big)\Big)\cdot
e^{j\cdot\frac{2\pi}{\widetilde{M}}\cdot\eta\cdot k}.
\label{6.32}
\end{equation}
Die Werte des doppelten Cepstrums sind f"ur \mbox{$k\!>\!0$} die
Sinusreihenkoeffizienten der in der Frequenz $\widetilde{\Omega}$ periodischen
schiefsymmetrischen Phasenfunktion des minimalphasigen Anteils von 
\mbox{$\widetilde{D}_N(\Tilde{z})$}. Gesucht wird letztlich die 
Phase des Spektrums \mbox{$F(\Omega)$} bei den Frequenzen
\mbox{$\Omega=\nu\CdoT2\pi/F$} mit \mbox{$\nu=1\;(1)\;N\!-\!1$}.
Mit Gleichung (\ref{6.22}) erhalten wir die entsprechenden Frequenzen 
\mbox{$\widetilde{\Omega}_\nu$} nach der Bilineartransformation:
\begin{equation}
\widetilde{\Omega}_\nu\;=\;
\nu\CdoT{\T \frac{2\pi}{F}}+2\CdoT\arctan\bigg(\!
\frac{(1\!-\!c)\cdot\sin\big(\nu\CdoT\frac{2\pi}{F}\big)}
{\;c+2\CdoT(1\!-\!c)\CdoT\sin\!\big(\nu\CdoT\frac{\pi}{F}\big)^{\!2}}\!\bigg).
\label{6.33}
\end{equation}
Nun wird der Wert der in der Frequenz $\widetilde{\Omega}$ periodischen
schiefsymmetrischen Phasenfunktion des minimalphasigen Anteils von
\mbox{$\widetilde{D}_N\big(e^{j\widetilde{\Omega}}\big)$} durch
Auswertung der Sinusreihe\footnote{Eine FFT kommt hier nicht in Frage, 
da die Frequenzen \mbox{$\widetilde{\Omega}_\nu$} nicht "aquidistant sind.} 
bei diesen Frequenzen berechnet:\vspace{-6pt} 
\begin{equation}
\phi(\nu)\;=\; \Sum{k=1}{\frac{\widetilde{M}}{2}-1}
2\CdoT C(k)\cdot
\sin\big(\widetilde{\Omega}_\nu\CdoT k\big).
\label{6.34}
\end{equation}

Die anschlie"sende Addition dieses Phasenanteils, des Phasenanteils
nach Gleichung (\ref{6.28}) des in Gleichung (\ref{6.23}) vor die
Summe gezogenen Terms \mbox{$\widetilde{D}_P(\Tilde{z})$}
und des linearen Phasenanteils des in Gleichung (\ref{6.19})
vor die Summe gezogenen Terms \mbox{$D_E(z)$} ergibt die
gesuchte Phase von \mbox{$F(\nu\CdoT2\pi/F)$}. Der Betrag wurde
als die positive Wurzel des Ausdrucks in Gleichung (\ref{6.18})
berechnet. Der Betrag wird noch auf \mbox{$F(0)/M$} normiert
und mit \mbox{$e^{\!-j\cdot\text{Phase}}$} multipliziert.
Mit den so gewonnenen Fourierreihenkoeffizienten \mbox{$F(\nu\CdoT2\pi/F)/F$}
berechnet sich die gesuchte Fensterfolge nach Gleichung (\ref{6.16}).

{\small Anmerkung: Au"ser f"ur den trivialen Fall mit \mbox{$N\!=\!1$} 
(\,Rechteckfenster\,) gibt es auch f"ur \mbox{$N=2\;(1)\;5$}
und beliebiges $M$ eine geschlossene L"osung f"ur \mbox{$F(\nu\CdoT2\pi/F)$}
mit \mbox{$\nu=1\!-\!N\;(1)\;N\!-\!1$}. Man kann n"amlich das
Polynom \mbox{$D_{\overline{E}}(z)$}, das ein Teil des Polynoms
\mbox{$z^{F-N}\Cdot D(z)$} ist, und dessen Nullstellenlage unbekannt ist,
auch mit\vspace{0pt}
\begin{equation}
s=\frac{z-1}{z+1}
\label{6.35}
\end{equation}
bilineartransformieren. Diese Bilineartransformation bildet Punkte,
die in der $z$-Ebene zum Einheitskreis symmetrisch liegen, auf zur
imagin"aren Achse der $s$-Ebene symmetrische Punkte ab. Dies ist so,
weil eine Substitution von $z$ durch \mbox{${\D(z^*)^{-1}}$}
in Gleichung (\ref{6.35}) \mbox{$-s^*$} ergibt. Da die Nullstellen von
\mbox{$D_{\overline{E}}(z)$} zum Einheitskreis symmetrisch liegen,
ergeben sich durch die Bilineartransformation zur imagin"aren Achse
der $s$-Ebene spiegelsymmetrische Nullstellen. Die Spiegelsymmetrie
der Nullstellen zur reellen Achse bleibt erhalten. Abgesehen von
einer \mbox{$N\!-\!1$}-fachen Nullstelle bei \mbox{$s\!=\!0$} und
einem \mbox{$N\!-\!1$}-fachen Polstellenpaar bei \mbox{$s\!=\!\pm1$},
deren Phasenbeitrag geschlossen angegeben werden kann, erh"alt man
so ein Polynom mit zum Ursprung der $s$-Ebene punktsymmetrischen
Nullstellen. Jeweils zwei zueinander punktsymmetrische Nullstellen
liefern einen Polynomfaktor \mbox{$s^2-s_0^2$}, so dass sich insgesamt
ein Polynom in $s^2$ vom Grad \mbox{$N\!-\!1$} ergibt. Da sich die
Nullstellen von Polynomen bis zum Grad $4$ geschlossen berechnen
lassen, l"asst sich auch die Lage der Nullstellen vor der
Bilineartransformation, und somit auch die Phase des minimalphasigen
Anteils sowie die Fensterfolge selbst geschlossen berechnen. Diese
L"osungen sind aber au"ser f"ur \mbox{$N\!=\!2$} so umfangreich,
dass es auch f"ur \mbox{$N=3\;(1)\;5$} sinnvoller ist, diese
Fensterfolgen mit dem dargestellten Algorithmus zu berechnen,
zumal die Genauigkeit der Berechnung dadurch nicht schlechter
ist, als bei der Auswertung der geschlossenen L"osung. Die Berechnung
der Fensterfolge durch eine numerische Bestimmung der Lage der
unbekannten Nullstellen f"uhrt vor allem f"ur gro"se Werte von $N$
zu unbefriedigenden Ergebnissen. 

F"ur \mbox{$N\!=\!2$} ergibt
sich f"ur die gesuchten Spektralwerte der Fensterfolge die folgende
geschlossene L"osung:
\begin{align}
\qquad F\big({\T-\frac{2\pi}{F}}\big)&\;=\;
-M\cdot\frac{1\!-\!j}{2}\cdot e^{\!-j\cdot\frac{\pi}{M}}
\notag \\*[4pt]
F(0)&\;=\;M
\label{6.36}\\*[4pt]
F\big({\T\frac{2\pi}{F}}\big)&\;=\;
-M\cdot\frac{1\!+\!j}{2}\cdot e^{j\cdot\frac{\pi}{M}}
\notag
\end{align}
}

\section{Aspekte der notwendigen Rechengenauigkeit}\label{Rech}

In diesem Unterkapitel wird der eben beschriebene Weg
der Berechnung der Fensterfolge mit Hinblick auf die 
Genauigkeit der Berechnung untersucht. Ich beginne damit, 
eine Forderung f"ur die Genauigkeit der zu berechnenden 
Fensterfolge aufzustellen, und leite daraus die Forderungen 
f"ur die Genauigkeit der Berechnung der davorliegenden 
Zwischenergebnisse ab. 

Die Fensterfolge wird beim RKM mit der Folge am Ausgang des zu messenden 
Systems multipliziert und anschlie"send diskret fouriertransformiert. 
Da wir a priori keine Aussage "uber die zu fensternde Folge kennen 
und auch keine Einschr"ankungen hinsichtlich der Art der zu fensternden 
Folgen einf"uhren wollen, nehmen wir nun von der zu fensternden Folge an, 
dass deren Werte alle in derselben Gr"o"senordnung liegen k"onnen. 
Somit gehen alle absoluten Fehler der Werte der Fensterfolge im Mittel 
gleich stark in die aus der gefensterten Folge berechneten Spektralwerte 
ein. Daher macht es keinen Sinn, die Werte der Fensterfolge mit einer 
h"oheren Genauigkeit zu berechnen, als dies bei dem betraglichen 
Maximalwert der Fensterfolge auf Grund der endlichen Wortl"ange der 
Zahlendarstellung im Rechner m"oglich ist. Die Fensterfolge soll daher 
mit einem absoluten Fehler berechnet werden, der in der Gr"o"senordnung 
von \mbox{$\varepsilon\cdot\max(\,|f(k)|\,)$} liegt. Dabei ist 
$\varepsilon$ die relative Rechnergenauigkeit, also die Differenz
zwischen der Zahl Eins und der n"achstgr"o"seren Zahl, die am Rechner bei
Gleitkommaarithmetik darstellbar ist. Wenn hier von der Gr"o"senordnung des
Fehlers die Rede ist, so ist damit gemeint, dass der absolute Fehler
h"ochstens um einen konstanten Faktor wie etwa die Wurzel aus einer
DFT-L"ange von dem angegebenen Wert der Gr"o"senordnung abweicht, nicht
aber um einen Faktor, der in die Gr"o"senordnung von $\varepsilon^{-1}$ kommt,
was beispielsweise passieren kann, wenn man mit schlecht konditionierten
Gleichungssystemen rechnet.

Die Fensterfolge l"asst sich mit Hilfe der Fourierreihe gem"a"s 
Gleichung (\ref{6.16}) aus den $N$ Werten des Spektrums der 
Fensterfolge bei den "aquidistanten Frequenzen \mbox{$\Omega=\nu\CdoT2\pi/F$} 
mit \mbox{$\nu=0\;(1)\;N\!-\!1$} berechnen. Die Pr"azision, mit der die 
Nullstellenbedingung (\ref{2.27}) erf"ullt wird, ist von der Genauigkeit der 
Berechnung der $N$ Spektralwerte unabh"angig, da durch die Art der Berechnung
der Fensterfolge als Fourierreihe ausgeschlossen wird, dass Spektralanteile bei 
den Frequenzen \mbox{$\Omega=\mu\CdoT2\pi/M$} mit \mbox{$\mu=1\;(1)\;M\!-\!1$} 
in der Fensterfolge "uberhaupt vorhanden sein k"onnen. 

Die Fourierreihe in Gleichung (\ref{6.16}) berechnen wir nicht mit Hilfe einer DFT 
sondern als die Summe einer Kosinus- und einer Sinusreihe. 
\begin{equation}
f(k)\;=\;
\frac{1}{N}+\frac{2}{F}\cdoT\Sum{\nu=1}{N-1}
\Big(\Re\big\{F\big({\T\nu\CdoT\frac{2\pi}{F}}\big)\!\big\}\Cdot
\cos\big({\T\frac{2\pi}{F}\CdoT\nu\CdoT k}\big)-
\Im\big\{F\big({\T\nu\CdoT\frac{2\pi}{F}}\big)\!\big\}\Cdot
\sin\big({\T\frac{2\pi}{F}\CdoT\nu\CdoT k}\big)\!\Big)
\label{6.37}
\end{equation}
Die auf \mbox{$F/2$} normierten Realteile \mbox{$\Re\big\{F(\nu\CdoT2\pi/F)\big\}$} 
der $N$ Spektralwerte sind die Koeffizienten der Kosinusreihe, die genauso 
normierten Imagin"arteile \mbox{$\Im\big\{F(\nu\CdoT2\pi/F)\big\}$} sind die Koeffizienten 
der Sinusreihe. 

Da bei den einzelnen Summanden der Sinus- und Kosinusreihe 
h"aufig gleiche Werte der Sinus- und Kosinusfunktion auftreten, wird man die 
ben"otigten Funktionswerte vor Beginn der Reihenauswertung berechnen und 
abspeichern. Dabei werden die ben"otigten Funk\-tions\-werte mit Hilfe
trigonometrischer Umformungen als Sinusfunktionswerte dargestellt,
deren Argumente betraglich kleiner gleich \mbox{$\pi/2$} sind. Die 
Funktionswerte der Sinusfunktion lassen sich in diesem Wertebereich 
mit dem kleinstm"oglichen relativen Fehler von $\varepsilon$ berechnen. 
Die bei den trigonometrischen Umformungen auftretenden $\pi$-Reduktionen 
lassen sich immer als Reduktion der ganzen Zahl $\nu$ um Vielfache der 
ebenfalls ganzen Zahl \mbox{$F/2$} fehlerfrei durchf"uhren. 

Beispielsweise berechnen wir \mbox{$\frac{2\pi}{F}\CdoT\nu\CdoT k\!-\!3\CdoT\pi$} nicht als 
Differenz, weil der Fehler dieser Differenz zweier irrationaler und somit {\em nicht}\/ 
exakt darstellbarer Zahlen in einer Gr"o"senordnung liegt, die um $\varepsilon$ 
kleiner ist als die betraglich gr"o"sere der beiden Zahlen. Stattdessen berechnen wir 
das Produkt \mbox{$\frac{2\pi}{F}\CdoT(\nu\CdoT k\!-\!3\CdoT\frac{F}{2})$}, bei dem 
der eine Faktor die Differenz ganzer, am Rechner exakt darstellbarer Zahlen ist. Das 
Produkt kann so mit dem kleinstm"oglichen relativen Fehler von $\varepsilon$ berechnet 
werden. 

Nun kann man die Sinus- und Kosinusreihe auswerten, indem man sich 
aus den abgespeicherten Funktionswerten jeweils die bei dem Summanden gerade 
ben"otigten Funktionswerte herausgreift. Bei der Summation vertauschen wir nun 
die Reihenfolge der einzelnen Reihenglieder dadurch, dass wir 
\mbox{$\nu=N\!-\!1\;(-1)\;0$} nach und nach zur Berechnung der Teilsummen 
einsetzen. Die Erfahrung zeigte n"amlich, dass die Betr"age der 
Fourierreihenkoeffizienten eine mit $\nu$ streng monoton fallende Folge 
bilden. Durch die Vertauschung der Reihenfolge erreichen wir, dass zun"achst 
die betraglich kleinsten Summanden addiert werden und bei der Addition der 
weiteren, gr"o"seren Summanden zur bisher berechneten Teilsumme keine unn"otig 
grobe Rundung erfolgt. Somit bleibt der relative Fehler m"oglichst klein. 

Aufgrund der Quantisierungsfehler sind die letzten \mbox{$N\!-\!1$} Werte von 
\mbox{$f(k)$} nicht exakt Null, wie dies nach der Theorie der Fall sein m"usste.
Trotzdem empfiehlt es sich {\em nicht}, diese Werte abschlie"send auf Null
zu setzen. Durch die Berechnung als endliche Sinus- und Kosinusreihe
wird n"amlich sichergestellt, dass die Nullstellenbedingung (\ref{2.27}) 
vom Spektrum der Fensterfolge praktisch perfekt erf"ullt wird, auch wenn 
die Reihenkoeffizienten fehlerhaft berechnet worden sind. 
Setzt man die letzten \mbox{$N\!-\!1$} Werte der Fensterfolge jedoch
zu Null, so entspricht dies der additiven "Uberlagerung eines
zwar recht kleinen, aber doch stark zeitbegrenzten Signals, das
wegen der abrupten Begrenzung auf \mbox{$N\!-\!1$} Werte ein
nur mit \mbox{${\D \sin(\Omega/2)^{\!-1}}$} abfallendes Spektrum hat.
Daher w"urden bei den in der Nullstellenbedingung (\ref{2.27}) genannten
Frequenzen zus"atzliche, unn"otige Fehler entstehen, wenn man die letzten
\mbox{$N\!-\!1$} Werte der Fensterfolge zu Null setzen w"urde. 

Die relativen Fehler der Sinus- und Kosinusfunktionswerte gehen ebenso 
in die Fehler der Werte der Fensterfolge ein, wie die relativen Fehler 
der Fourierreihenkoeffizienten, da beide fehlerbehafteten Gr"o"sen bei 
der Berechnung der Sinus- und Kosinusreihe miteinander multipliziert werden. 
Da sich die Sinus- und Kosinusfunktionswerte mit einer relativen Genauigkeit 
berechnen lassen, die praktisch nur von der Mantissenwortl"ange des 
verwendeten Rechners begrenzt wird, ist anzunehmen, dass die relativen Fehler 
bei der Berechnung der Fourierreihenkoeffizienten den "uberwiegenden 
Beitrag zu den Fehlern der Werte der Fensterfolge liefern. Daher wollen wir 
ab nun die Fehler der Sinus- und Kosinusfunktionswerte vernachl"assigen.

Den $N$ Abtastwerten des Spektrums der Fensterfolge ist aufgrund der 
begrenzen Genauigkeit des Berechnungsalgorithmus eine St"orung 
additiv "uberlagert. Diese additive St"orung geht auf $F$ normiert 
"uber die Fourierreihenentwicklung nach Gleichung (\ref{6.16}) in die 
Werte der Fensterfolge ein, wie dies die Spektralwerte auch tun. Es wird 
nun gezeigt, dass der Fehler der Fensterfolge dann in der gew"unschten 
Gr"o"senordnung von \mbox{$\varepsilon\cdot\max(\,|f(k)|\,)$} liegt, 
wenn die $N$ Abtastwerte des Spektrums f"ur \mbox{$\nu=0\;(1)\;N\!-\!1$} 
mit einem absoluten Fehler in der Gr"o"senordnung von 
\mbox{$\max\big(|F(\nu\CdoT2\pi/F)|\big)\CdoT\varepsilon\,=\,
M\CdoT\varepsilon$} behaftet sind. Diese Fehler modellieren wir
als die additive "Uberlagerung von \mbox{$2\CdoT\!N\!-\!1$} 
reellen St"orungen von jeweils der Streuung \mbox{$M\CdoT\varepsilon$} 
zu den Real- und Imagin"arteilen der Spektralwerte. Nehmen wir noch die 
Unabh"angigkeit dieser St"orungen an, so erhalten wir 
bei allen Werten der Fensterfolge eine additive St"orung mit 
der Streuung \mbox{$\varepsilon/N\CdoT\sqrt{2\CdoT\!N\!-\!1\;}\approx
\varepsilon\CdoT\sqrt{2/N\;}$}. 

Der Effektivwert der Fensterfolge \mbox{$\sqrt{1/F\cdoT\sum_{}^{}f(k)^2\,}$}
ist nach Gleichung (\ref{2.21}) und (\ref{2.23}) \mbox{$\sqrt{M\CdoT d(0)/F\;}=\sqrt{1/N\;}$}. 
Die Streuung der Fehler in der Fensterfolge ist also um den Faktor 
\mbox{$\varepsilon\CdoT\sqrt{2\;}$} kleiner als der Effektivwert der 
Fensterfolge. Da erfahrungsgem"a"s der Maximalwert des Betrags der 
Fensterfolge immer gr"o"ser als der \mbox{$\sqrt{2\;}\!$}-fache 
Effektivwert ist, liegt der Fehler der Fensterfolge dann in der 
gew"unschten Gr"o"senordnung, wenn es gelingt, die $N$ Spektralwerte mit einem 
absoluten Fehler in der Gr"o"senordnung von \mbox{$M\CdoT\varepsilon$} 
zu berechnen. 

Betrag und Phase der gesuchten Spektralwerte der
Fensterfolge werden getrennt berechnet. F"ur den absoluten Fehler
des Betrags bedeutet das, dass dieser f"ur alle Frequenzen in
derselben Gr"o"senordnung \mbox{$M\CdoT\varepsilon$} liegen sollte,
w"ahrend der absolute Fehler der Phase in der Gr"o"senordnung
\mbox{$\varepsilon\CdoT M/|F(\nu\CdoT2\pi/F)|$} liegen darf.
Im Durchlassbereich des Spektrums der Fensterfolge darf der absolute
Fehler der Phase also nicht gr"o"ser als $\varepsilon$ sein, w"ahrend
er bei betraglich kleinen Spektralwerten auch gr"o"ser werden darf.

Um den Betrag eines gesuchten Spektralwertes zu berechnen, wird 
die Wurzel aus ihrem mit Gleichung (\ref{6.18}) berechneten 
Betragsquadrat gezogen. Wenn man zur Berechnung der Wurzel einen 
Algorithmus verwendet, der immer diejenige am Rechner darstellbare Zahl 
liefert, deren Quadrat am n"achsten bei der zu radizierenden Zahl liegt, 
ergibt sich im Idealfall (\,die zu radizierenden Betragsquadrate seien 
fehlerfrei\,) ein Fehler in den Betr"agen der Spektralwerte dessen Streuung 
in der Gr"o"senordnung von \mbox{$\varepsilon\CdoT|F(\nu\CdoT2\pi/F)|/2$} 
liegt. Die Fehler bei der Berechnung der Wurzel liegen somit nur f"ur 
die betraglich gr"o"sten Spektralwerte ann"ahernd in der Gr"o"senordnung 
\mbox{$M\CdoT\varepsilon$} des zul"assigen absoluten Fehlers und sind 
anderenfalls deutlich kleiner. 

Nun wird anhand zweier Grenzf"alle hergeleitet, wie gro"s der absolute 
Fehler des Betragsquadrates des Spektralwertes sein darf, wenn der Fehler 
des Betrags des gesuchten Spektralwertes der Fensterfolge in der gew"unschten 
Gr"o"senordnung von \mbox{$M\CdoT\varepsilon$} liegen soll. Wir 
vernachl"assigen dabei die im letzten Absatz untersuchten Fehler und 
nehmen an, dass die Wurzel fehlerfrei berechnet werden kann, und die 
Fehler des Betrages lediglich durch Fehler verursacht werden, die bereits 
im Betragsquadrat vorhanden sind. Der erste Grenzfall ist gegeben, 
wenn der absolute Fehler des Betragsquadrates deutlich kleiner als 
der Wert des Betragsquadrates ist. Dann kann man die Wurzelfunktion in guter N"aherung 
linearisieren und man erh"alt die Aussage, dass das zu radizierende 
Betragsquadrat einen absoluten Fehler in der Gr"o"senordnung von 
\mbox{$2\Cdot M\CdoT\varepsilon\CdoT|F(\nu\CdoT2\pi/F)|$} aufweisen darf, 
wenn der Fehler des Betrags in der gew"unschten Gr"o"senordnung liegen 
soll. Im zweiten Grenzfall dominiert der absolute Fehler des 
Betragsquadrates den zu radizierenden Wert. In diesem Fall 
vernachl"assigt man den theoretisch exakten Wert des Betragsquadrates 
und kommt zu dem Ergebnis, dass der absolute Fehler des Betragsquadrates 
in einer Gr"o"senordnung von \mbox{${\D M^2\CdoT\varepsilon^2}$} liegen 
darf. 

Die Betragsquadrate der gesuchten Spektralwerte der Fensterfunktion berechnen 
wir mit Gleichung (\ref{6.18}). Dabei wird eine Summe gebildet, bei der die 
Summanden jeweils Produkte von inversen Quadraten der Sinusfunktion sind. 
F"ur sinnvolle Werte von \mbox{$M\ge4$} bleibt dabei das Argument der 
Sinusfunktion immer betraglich kleiner als $\pi/2$. In diesem Bereich l"asst 
sich die Sinusfunktion mit einem relativen Fehler in der Gr"o"senordnung von 
$\varepsilon$ berechnen. Da jeder Summand aus dem Inversen eines Produktes von 
\mbox{$N\!-\!1$} Quadraten der Sinusfunktion besteht, liegt der relative Fehler 
jedes Summanden in der Gr"o"senordnung von \mbox{$\sqrt{2\CdoT N\;}\Cdot\varepsilon$}.
Je nach diskreter Frequenz $\nu$ besteht die Summe aus insgesamt maximal
\mbox{$N\!-\!1$} stets positiven Summanden. Als Absch"atzung f"ur den 
schlechtesten Fall k"onnen wir annehmen, dass ein Summand sehr viel gr"o"ser ist 
als alle anderen Summanden\footnote{Ansonsten w"urde die Summation zu einer 
Mittelung der als unabh"angig anzusehenden Fehler f"uhren.}. Dann ergibt 
sich f"ur den absoluten Fehler der Summe die obere Absch"atzung
\mbox{$\sqrt{2\CdoT N\;}\CdoT\varepsilon\CdoT|F(\nu\CdoT2\pi/F)|^2$}. 
Da hier der Fall vorliegt, dass der relative Fehler mit
\mbox{$\sqrt{2\CdoT N\;}\CdoT\varepsilon$} vor dem Radizieren deutlich 
kleiner als Eins ist, ist der im letzten Absatz hergeleitete maximal zul"assige 
absolute Fehler lediglich maximal um den Faktor $\sqrt{N/2\;}$ "uberschritten. 
Es sei hier darauf hingewiesen, dass $N$ "ublicherweise kleiner als $10$ sein d"urfte, 
und dass wir hier eine "`worst case"' Absch"atzung gemacht haben. Somit ist die 
"Uberschreitung der maximal zul"assigen Fehlers als akzeptabel einzustufen. 

Die Phase der $N$ Koeffizienten der Fourierreihendarstellung der Fensterfolge 
berechnet sich aus drei Anteilen. Da ist zun"achst der lineare Phasenanteil 
\mbox{$(F\!-\!2\CdoT\!N\!+\!1)\CdoT\Omega/2$}, der durch die bekannten 
"aquidistanten Nullstellen am Einheitskreis verursacht wird. Dieser l"asst 
sich mit einem relativen Fehler in der Gr"o"senordnung von $\varepsilon$
berechnen. Die Gr"o"senordnung des absoluten Fehlers dieses Phasenanteils 
steigt linear mit der Phase und somit mit der Frequenz \mbox{$\Omega$} an. 
Da wir die Phase f"ur die Frequenzen \mbox{$\Omega=\nu\CdoT2\pi/F$} mit 
\mbox{$\nu=1\!-\!N\;(1)\;N\!-\!1$} zu berechnen haben, steigt die Gr"o"senordnung 
des absoluten Fehlers mit $\nu$ bis zu einem Maximalwert von circa 
\mbox{$N\CdoT\pi\CdoT\varepsilon$} linear an. Da erfahrungsgem"a"s 
mit steigendem $\nu$ der Betrag der Fourierreihenkoeffizienten sinkt, 
und da f"ur kleine Betr"age der absolute Fehler der Phase auch gr"o"ser 
als $\varepsilon$ werden darf, bleibt der Fehler dieses Anteils der Phase 
der gesuchten Spektralwerte in der gew"unschten Gr"o"senordnung. 

Will man diesen Anteil der Phase mit einer noch h"oheren Genauigkeit 
berechnen, so kann man zun"achst die Phasenanteile als ein ganzzahliges 
Vielfaches von \mbox{$\pi/F$} schreiben. Von den am Rechner exakt darstellbaren, 
ganzen Zahlen zieht man dann vor der Multiplikation mit \mbox{$\pi/F$} ein 
ganzzahliges Vielfaches von \mbox{$2\CdoT F$} in der Art ab, dass der lineare
Phasenanteil immer im Bereich von \mbox{$(-\pi;\pi]$} liegt, und
somit immer mit einem absoluten Fehler in der Gr"o"senordnung von
$\varepsilon$ berechnet werden kann. Diese Art der $2\pi$-Reduktion, 
die wir auch bei der Auswertung der Fourierreihe in Gleichung (\ref{6.37})
verwenden, kommt jedoch bei dem im Anhang aufgelisteten Programm nicht 
zum Einsatz, da sich bei allen "uberpr"uften Beispielen
herausgestellt hat, dass auch die ohne die exakte $2\pi$-Reduktion
berechneten Spektralwerte der Fensterfolge ausreichend genau sind.

Der zweite Anteil der Phase der Fourierreihenkoeffizienten der
Fensterfolge ist die Phase des minimalphasigen Anteils 
\mbox{$\big(\big(z\!-\!(1\!-\!c)\big)/z\big)^{\uP{-0.3}{N-1}}$}
des Polynomquotienten \mbox{$D_P(z)$} nach Gleichung 
(\ref{6.27}). Dieser ist durch eine bilineare R"ucktransformation 
entstanden. Die Nullstellenlage von \mbox{$D_P(z)$} ist von $c$ abh"angig  
und somit bekannt. F"ur jedes der \mbox{$N\!-\!1$} identischen 
Pol- Nullstellenpaare (\,Pol jeweils bei \mbox{$z\!=\!0$}\,) berechnet 
sich der Phasenbeitrag nach Gleichung (\ref{6.28}). Vor allem f"ur 
die praktisch relevanten kleinen Werte von \mbox{$c \ll 1$} 
l"asst sich dieser Phasenanteil eines Pol- Nullstellenpaares
in der dort angegebenen Form nahezu mit dem kleinstm"oglichen 
relativen Fehler in der Gr"o"senordnung von $\varepsilon$ berechnen.
Dieser Phasenanteil ist betraglich immer kleiner als \mbox{$\pi/2$}.
Wegen der Vielfachheit der Pol- Nullstellenpaare ist der 
gesamte Phasenbeitrag des Polynomquotienten \mbox{$D_P(z)$} nach 
Gleichung (\ref{6.27}) immer kleiner als \mbox{$(N\!-\!1)\CdoT\pi/2$}. 
Er l"asst sich daher mit einem absoluten Fehler in der Gr"o"senordnung 
von \mbox{$\approx N\CdoT\varepsilon$} berechnen. Die Genauigkeit der 
Berechnung dieses Phasenanteils kann also f"ur die betraglich gr"o"sten 
Fourierreihenkoeffizienten der Fensterfolge lediglich um den Faktor $N$ 
(\,"ublicherweise kleiner $10$\,) zu gro"s sein. Die Praxis zeigt, 
dass der gesamte zweite Phasenbeitrag bei den betraglich gr"o"sten 
Fourierreihenkoeffizienten immer deutlich kleiner als der Maximalwert 
\mbox{$(N\!-\!1)\CdoT\pi/2$} ist, so dass man die Genauigkeit der 
Berechnung dieses Phasenanteils als ausreichend bewerten kann.

Der dritte Anteil der Phase wird nach \cite{Boite/Leich} bzw. \cite{Opp}
"uber das Cepstrum von \mbox{$\widetilde{D}_N(\Tilde{z})$} gem"a"s der 
Gleichungen (\ref{6.29}) und (\ref{6.32}) berechnet. Dazu werden zun"achst die 
konstanten Faktoren $K_{\nu_2}$ mit Gleichung (\ref{6.25}) und die halben 
Winkel der Nullstellenrotation $\Tilde{\psi}_{\nu_2}$ mit Gleichung (\ref{6.26}) 
berechnet. In der dort angegebenen Form lassen sich diese f"ur kleine Werte von 
\mbox{$c \ll 1$} nahezu mit einer relativen Genauigkeit von $\varepsilon$ 
berechnen. Da die hal\-ben Winkel der Nullstellenrotation im Argument der 
Sinusfunktion in Gleichung (\ref{6.29}) in einer Differenz auftreten, 
ist es im Allgemeinen nicht m"oglich, die Quadrate der Sinusfunktion 
dort mit einer {\em relativen} Genauigkeit von $\varepsilon$, sondern 
lediglich mit einer {\em absoluten} Genauigkeit in dieser Gr"o"senordnung 
zu berechnen. Der relative Fehler eines Produktes in der Summe in Gleichung 
(\ref{6.29}) wird also immer dann besonders gro"s, wenn sich eine Nullstelle 
in unmittelbarer N"ahe einer Frequenz \mbox{$\eta\CdoT2\pi/\widetilde{M}$} 
befindet, f"ur die das Produkt zu berechnen ist. In diesem Fall wird aber auch 
das jeweilige Produkt selbst sehr klein und f"ur die relative Genauigkeit der 
Berechnung von \mbox{$\widetilde{D}_N\big(e^{j\eta\cdot2\pi/\widetilde{M}}\big)$}
ist dann der relative Fehler der Berechnung des gr"o"sten Summanden ausschlaggebend. 

Es zeigte sich, dass durch die Wahl des Parameters $c$ gem"a"s Gleichung (\ref{6.30}) 
nicht nur das Cepstrum besonders rasch abklingt, sondern auch, dass die Nullstellen 
$\Tilde{z}_{\nu_2}$ in den Produkten der Gleichung (\ref{6.29}) durch die 
Bilineartransformation einigerma"sen gleichm"a"sig am Einheitskreis im 
Bereich von \mbox{$|\widetilde{\Omega}|<2\pi/3$} verteilt werden. 
Bei dem in einem bestimmten Frequenzbereich jeweils dominierenden Summanden 
ergibt sich erfahrungsgem"a"s ein minimaler Abstand zur n"achstgelegenen 
Nullstelle von wenigstens \mbox{$\pi/4$}. Damit l"asst sich der jeweils 
dominierende Summand mit einem relativen Fehler in der Gr"o"senordnung von 
einigen $\varepsilon$ berechnen. Pauschal kann man sagen, dass der relative 
Fehler eines Summanden zunimmt, je geringer dessen Beitrag zur Gesamtsumme 
wird. Es ist daher anzunehmen, dass der relative Fehler von 
\mbox{$\widetilde{D}_N\big(e^{j\eta\cdot2\pi/\widetilde{M}}\big)$} 
im gesamten Frequenzbereich $\varepsilon$ nicht wesentlich "ubersteigt.

Bevor wir nun mit Hilfe einer FFT gem"a"s Gleichung (\ref{6.32}) 
aus den Logarithmen dieser $\widetilde{M}$ Werte der Funktion 
\mbox{$\widetilde{D}_N(\Tilde{z})$} das doppelte reelle Cepstrum 
berechnen, empfiehlt es sich, diese Werte mit einem konstanten 
multiplikativen Faktor in der Art zu normieren, dass der Maximalwert 
dieser Werte gleich dem Reziproken des Minimalwertes ist. Dies h"alt, 
wie in Anhang \ref{log} gezeigt wird, die Fehler klein, die selbst bei 
idealer Berechnung des Logarithmus durch die Quantisierung der berechneten 
Logarithmuswerte bei Flie"skommaarithmetik zwangsweise entstehen.
Dieser konstante Normierungsfaktor bewirkt lediglich eine Modifikation 
des Cepstralwertes \mbox{$C(0)$}, der bei der Berechnung der Phase 
nach Gleichung (\ref{6.34}) sowieso nicht auftritt. Der relative Fehler 
von \mbox{$\widetilde{D}_N\big(e^{j\eta\cdot2\pi/\widetilde{M}}\big)$}, 
der einige $\varepsilon$ nicht "ubersteigt, wird im Idealfall durch die 
Bildung des Logarithmus zu einem absoluten Fehler, der im gesamten 
Frequenzbereich in der Gr"o"senordnung einiger $\varepsilon$ liegt. 

Es zeigte sich, dass bei der oben angegeben Wahl des 
Bilineartransformationsparameters $c$ nach der Normierung der Logarithmus 
f"ur gebr"auchliche Werte von \mbox{$N\!<\!10$} betraglich immer maximal 
im einstelligen Zahlenbereich liegt. Daher "ubersteigen die absoluten 
Fehler, die durch die begrenzte Zahlendarstellung des Ergebnisses des 
Logarithmierens entstehen, die absoluten Fehler, die von der Berechnung 
der Werte \mbox{$\widetilde{D}_N\big(e^{j\eta\cdot2\pi/\widetilde{M}}\big)$} 
verursacht werden, allenfalls unwesentlich. 

Eine FFT der Logarithmuswerte 
liefert uns das doppelte reelle Cepstrum. Es kann angenommen werden, dass
bei der Berechnung der FFT alle Fehler, die bereits vor der FFT vorhanden 
sind, auf alle Werte des Cepstrums einen einigerma"sen gleich gro"sen 
Einfluss haben, so dass alle Phasenwerte, die man mit Gleichung (\ref{6.34})
berechnet, einen absoluten Fehler in derselben Gr"o"senordnung haben werden. 

Weitere Fehler entstehen bei der Berechnung der FFT selbst\footnote{Eine FFT 
verursacht --- wie auch in \cite{Erg} an einem Beispiel gezeigt wird --- 
in der Regel wesentlich gr"o"sere Fehler, als man dies bei einer bis auf 
den Faktor \mbox{$\sqrt{\text{FFT-L"ange}\,}$} orthogonalen Transformation 
zun"achst erwarten w"urde \cite{Heute}. Dennoch stellte sich heraus, dass 
eine Berechnung durch direkte Auswertung der DFT-Summenformel bei optimaler 
Genauigkeit der Drehfaktoren und bestm"oglicher Reihenfolge der Summanden, 
die Genauigkeit bei der Berechnung der Werte der Fensterfolge nur so 
unwesentlich verbessert, dass der damit gegen"uber der FFT verbundene 
Mehraufwand nicht zu rechtfertigen ist, und somit wird hier eine FFT 
verwendet.}, von denen man annehmen kann, dass sie einerseits 
gegen"uber den bisher gemachten Fehlern vernachl"assigbar sind, und dass 
sie andererseits ebenfalls einen gleichm"a"sigen Einfluss auf die absoluten 
Fehler aller nach Gleichung (\ref{6.34}) zu berechnenden Phasenwerte haben. 
Weitere Fehler entstehen bei der Berechnung der Sinusreihe in letztgenannter 
Gleichung. Wie sich diese klein halten lassen, wird nun gezeigt.

Da sich nach \cite{Boite/Leich} bzw. \cite{Opp} die Werte des Cepstrums f"ur 
\mbox{$k\!>\!0$} als \mbox{$1/k\cdot\sum\zu^{\uP{-0.4}{\!k}}\,$}
schreiben lassen, wobei die Summe "uber alle unbekannten Nullstellen $\zu$ 
des Mindestphasenanteils zu erstrecken ist, und da bei einem
Mindestphasensystem, das keine Nullstellen am Einheitskreis hat, die 
Nullstellen innerhalb des Einheitskreises liegen, gilt f"ur \mbox{$k\!>\!1$}:
\begin{equation}
\Big|\,\frac{1}{k}\cdoT\sum\zu^{\uP{-0.4}{\!k}}\,\Big|\;\le\;
\frac{1}{k}\cdoT\sum|\zu|^k\;<\; 
\frac{1}{k}\cdoT\sum|\zu|.
\label{6.38}
\end{equation}
Die Koeffizienten der Sinusreihe fallen also betraglich mindestens mit
\mbox{$1/k$} ab. In der Sinusreihe der Phase sind die Argumente der
Sinusfunktionen \mbox{$\widetilde{\Omega}_\nu\CdoT k$} proportional zu $k$.
Der maximale absolute Fehler der einzelnen Sinusfunktionen ist daher
\mbox{$\widetilde{\Omega}_\nu\CdoT k\CdoT\varepsilon$}. Diese absoluten Fehler
werden mit den wenigstens mit \mbox{$1/k$} abfallenden Koeffizienten
"uberlagert. Daher ist der Einfluss des Fehler auf die Phase f"ur Terme
mit steigendem $k$ insgesamt abnehmend. Da bei Flie"skommaarithmetik
kleine Werte mit kleinen absoluten Fehlern berechnet werden k"onnen,
wird die Auswertung der Sinusreihe in Gleichung (\ref{6.34}) in 
umgekehrter Reihenfolge vorgenommen, so dass mit den kleinsten 
Summanden begonnen wird. Dadurch wird erreicht, dass bei den kleinen 
Summanden keine unn"otig gro"sen absoluten Quantisierungsfehler 
akkumuliert werden. Bei dieser Art der Berechnung der Phase bei den 
Frequenzen \mbox{$\Omega=\nu\CdoT2\pi/F$} mit \mbox{$\nu=1\;(1)\;N\!-\!1$} 
ist $\widetilde{\Omega}$ immer kleiner $\pi$ und die Sinusreihe 
kann mit einem absoluten Fehler bis fast in der Gr"o"senordnung von 
$\varepsilon$ berechnet werden. 

Zusammenfassend kann man sagen, dass der dritte Anteil der Phase der 
Fourierreihenkoeffizienten der Fensterfolge am st"arksten fehlerbehaftet 
ist. Jedoch bleibt auch hier der absolute Fehler in einer Gr"o"senordnung 
von einigen Vielfachen von $\varepsilon$. Er ist damit nur bei den 
betraglich gr"o"sten Fourierreihenkoeffizienten unwesentlich gr"o"ser 
als dies am Anfang dieses Unterkapitels gefordert worden ist. Der 
Fehler wird sicherlich mit $N$ ansteigen\footnote{Es wird vermutet, 
dass der Anstieg der Fehlerstreuung maximal proportional zu $N$ erfolgt. 
Dies wurde jedoch nicht n"aher "uberpr"uft.}, jedoch nicht so stark, 
dass bereits f"ur kleine Werte von $N$ (\,in der Gr"o"senordnung von 
circa \mbox{$N\approx 10$}\,) eine Berechnung der Fensterfunktion unm"oglich 
wird. Es wurden vorher auch einige andere Algorithmen entwickelt, 
eine Fensterfolge zu berechnen, die die Forderungen (\ref{2.20}) und 
(\ref{2.27}) erf"ullen und somit beim RKM eingesetzt werden k"onnen.
Diese Algorithmen minimierten die Fehler der Gleichung (\ref{2.20}) 
iterativ und lieferten nur f"ur sehr kleine Werte von $N$ brauchbare 
Ergebnisse. Sp"atestens bei \mbox{$N\!=\!16$} versagten diese Algorithmen, 
da die Funktionalmatrix (Jacobi-Matrix) trotz geeigneter Skalierungsschritte 
immer singul"ar wurde. Mit dem hier vorgestellten nicht iterativen 
Algorithmus k"onnen sogar Fensterfolgen mit extrem hohen Werten von $N$ 
mit erstaunlich guter Genauigkeit berechnet werden. Selbst f"ur so unsinnig 
gro"se Werte wie z.~B. \mbox{$N\!=\!256$} traten keinerlei numerische 
Schwierigkeiten auf. Da die Berechnungszeit mit $N$ "uberproportional 
ansteigt, wurden noch gr"o"sere Werte von $N$ nicht ausprobiert. 

Viele Gleichungen im letzten Unterkapitel wurden so optimiert, dass 
man f"ur kleine Werte des Bilineartransformationsparameters $c$ die 
h"ochstm"ogliche Genauigkeit erh"alt. F"ur extrem kleine Werte von $M$ 
ist die Annahme \mbox{$c \ll 1$} jedoch nicht gegeben. Dann empfiehlt 
es sich eventuell statt des Parameters $c$ den neuen Parameter 
\mbox{$(1\!-\!c)$} zu verwenden und einige Gleichungen in eine 
andere Form zu "uberf"uhren, die mit dem neuen Parameter wieder 
eine optimale numerische Genauigkeit gew"ahrleistet. Diese 
Algorithmusvariante wurde jedoch nicht n"aher untersucht, 
da beim RKM eine Anwendung einer Fensterfolge mit solch kleinen 
Werten von $M$ keinen Sinn macht, und da bei der Erstellung dieser 
Arbeit auch f"ur kleine Werte von $M$ keinerlei numerische Probleme 
feststellbar waren.

\section{Anmerkungen und Beispiele zu den Fensterfolgen}\label{FenBeisp}

In diesem Abschnitt werden zuerst beispielhaft einige Fensterfolgen sowie 
deren Spektren im Durchlassbereich graphisch dargestellt. F"ur eine Schar
weiterer Beispielfensterfolgen mit gro"sem $M$ wird dann nachgewiesen, 
dass f"ur die Sperrd"ampfung des Fensterspektrums der in Kapitel \ref{W} 
gew"unschte potenzm"a"sige Zusammenhang mit der Frequenz existiert, 
und dass dieser mit Hilfe des Parameters $N$ einstellbar ist. Dann wird 
an einem weiteren Beispiel gezeigt, wie gut die Nullstellenbedingung 
(\ref{2.27}) f"ur das Spektrum der Fensterfolge erf"ullt wird. 
Die Lage der Nullstellen der Z-Transformierten der Fensterfolge wird 
anhand einiger weiterer Beispiele gezeigt. 

Im Folgenden werden die AKFs 
einiger Beispielfensterfolgen graphisch dargestellt. Es folgt eine 
Untersuchung zur Pr"azision der Erf"ullung der Nullstellenforderung 
(\ref{2.23}) f"ur die AKF sowie der "aquivalenten Forderung (\ref{2.20}), 
nach der das Betragsquadrat des Spektrums der Fensterfolge sich zu einer 
Konstanten "uberlagern l"asst. Abschlie"send wird in diesem Unterkapitel 
gezeigt, dass sich mit dem hier vorgestellten Algorithmus auch 
Halbbandfilter berechnen lassen. 

\subsection{Beispiele einiger Fensterfolgen}

Bild \ref{b5h}
\begin{figure}[btp]
{ 
\begin{picture}(454,600)

\input{mbild5h}
\put(0,600){\makebox(0,0)[lt]{Fensterfolgen \mbox{$f(k)$} mit \mbox{$M\!=\!4$}:}}
\put(315,595){\makebox(0,0)[r]{\small $f(k)$}}
\put(317,580){\makebox(0,0)[r]{\footnotesize $1$}}
\put(318,558){\makebox(0,0)[rt]{\footnotesize $0$}}
\put(420,557){\makebox(0,0)[t]{\footnotesize $4$}}
\put(445,556){\makebox(0,0)[tr]{\small $k$}}
\put(445,570){\makebox(0,0)[rb]{$N\!=\!1$}}

\put(215,547){\makebox(0,0)[r]{\small $f(k)$}}
\put(217,532){\makebox(0,0)[r]{\footnotesize $1$}}
\put(218,510){\makebox(0,0)[rt]{\footnotesize $0$}}
\put(320,509){\makebox(0,0)[t]{\footnotesize $4$}}
\put(420,509){\makebox(0,0)[t]{\footnotesize $8$}}
\put(445,508){\makebox(0,0)[tr]{\small $k$}}
\put(445,522){\makebox(0,0)[rb]{$N\!=\!2$}}

\put(115,499){\makebox(0,0)[r]{\small $f(k)$}}
\put(117,484){\makebox(0,0)[r]{\footnotesize $1$}}
\put(118,462){\makebox(0,0)[rt]{\footnotesize $0$}}
\put(220,461){\makebox(0,0)[t]{\footnotesize $4$}}
\put(320,459){\makebox(0,0)[t]{\footnotesize $8$}}
\put(420,461){\makebox(0,0)[t]{\footnotesize $12$}}
\put(445,460){\makebox(0,0)[tr]{\small $k$}}
\put(445,474){\makebox(0,0)[rb]{$N\!=\!3$}}

\put(15,451){\makebox(0,0)[r]{\small $f(k)$}}
\put(17,436){\makebox(0,0)[r]{\footnotesize $1$}}
\put(18,414){\makebox(0,0)[rt]{\footnotesize $0$}}
\put(120,413){\makebox(0,0)[t]{\footnotesize $4$}}
\put(220,408){\makebox(0,0)[t]{\footnotesize $8$}}
\put(320,413){\makebox(0,0)[t]{\footnotesize $12$}}
\put(420,413){\makebox(0,0)[t]{\footnotesize $16$}}
\put(445,412){\makebox(0,0)[tr]{\small $k$}}
\put(445,426){\makebox(0,0)[rb]{$N\!=\!4$}}

\put(0,393){\makebox(0,0)[lt]{Fensterfolgen \mbox{$f(k)$} mit \mbox{$M\!=\!64$}:}}
\put(315,393){\makebox(0,0)[r]{\small $f(k)$}}
\put(317,378){\makebox(0,0)[r]{\footnotesize $1$}}
\put(318,356){\makebox(0,0)[rt]{\footnotesize $0$}}
\put(420,355){\makebox(0,0)[t]{\footnotesize $64$}}
\put(445,354){\makebox(0,0)[tr]{\small $k$}}
\put(445,368){\makebox(0,0)[rb]{$N\!=\!1$}}

\put(215,345){\makebox(0,0)[r]{\small $f(k)$}}
\put(217,330){\makebox(0,0)[r]{\footnotesize $1$}}
\put(218,308){\makebox(0,0)[rt]{\footnotesize $0$}}
\put(320,307){\makebox(0,0)[t]{\footnotesize $64$}}
\put(420,307){\makebox(0,0)[t]{\footnotesize $128$}}
\put(445,307){\makebox(0,0)[tr]{\small $k$}}
\put(445,320){\makebox(0,0)[rb]{$N\!=\!2$}}

\put(115,297){\makebox(0,0)[r]{\small $f(k)$}}
\put(117,282){\makebox(0,0)[r]{\footnotesize $1$}}
\put(118,260){\makebox(0,0)[rt]{\footnotesize $0$}}
\put(220,259){\makebox(0,0)[t]{\footnotesize $64$}}
\put(320,255){\makebox(0,0)[t]{\footnotesize $128$}}
\put(420,259){\makebox(0,0)[t]{\footnotesize $192$}}
\put(445,258){\makebox(0,0)[tr]{\small $k$}}
\put(445,272){\makebox(0,0)[rb]{$N\!=\!3$}}

\put(15,249){\makebox(0,0)[r]{\small $f(k)$}}
\put(17,234){\makebox(0,0)[r]{\footnotesize $1$}}
\put(18,212){\makebox(0,0)[rt]{\footnotesize $0$}}
\put(120,211){\makebox(0,0)[t]{\footnotesize $64$}}
\put(220,211){\makebox(0,0)[t]{\footnotesize $128$}}
\put(320,211){\makebox(0,0)[t]{\footnotesize $192$}}
\put(420,211){\makebox(0,0)[t]{\footnotesize $256$}}
\put(445,210){\makebox(0,0)[tr]{\small $k$}}
\put(445,224){\makebox(0,0)[rb]{$N\!=\!4$}}

\put(0,191){\makebox(0,0)[lt]{Fensterfolgen \mbox{$f(k)$} mit \mbox{$M\!=\!1024$}:}}
\put(315,191){\makebox(0,0)[r]{\small $f(k)$}}
\put(317,176){\makebox(0,0)[r]{\footnotesize $1$}}
\put(318,154){\makebox(0,0)[rt]{\footnotesize $0$}}
\put(420,153){\makebox(0,0)[t]{\footnotesize $1024$}}
\put(445,152){\makebox(0,0)[tr]{\small $k$}}
\put(445,166){\makebox(0,0)[rb]{$N\!=\!1$}}

\put(215,143){\makebox(0,0)[r]{\small $f(k)$}}
\put(217,128){\makebox(0,0)[r]{\footnotesize $1$}}
\put(218,106){\makebox(0,0)[rt]{\footnotesize $0$}}
\put(320,105){\makebox(0,0)[t]{\footnotesize $1024$}}
\put(420,104){\makebox(0,0)[t]{\footnotesize $2048$}}
\put(445,104){\makebox(0,0)[tr]{\small $k$}}
\put(445,118){\makebox(0,0)[rb]{$N\!=\!2$}}

\put(115,95){\makebox(0,0)[r]{\small $f(k)$}}
\put(117,80){\makebox(0,0)[r]{\footnotesize $1$}}
\put(118,58){\makebox(0,0)[rt]{\footnotesize $0$}}
\put(220,57){\makebox(0,0)[t]{\footnotesize $1024$}}
\put(320,53){\makebox(0,0)[t]{\footnotesize $2048$}}
\put(420,57){\makebox(0,0)[t]{\footnotesize $3072$}}
\put(445,56){\makebox(0,0)[tr]{\small $k$}}
\put(445,70){\makebox(0,0)[rb]{$N\!=\!3$}}

\put(15,47){\makebox(0,0)[r]{\small $f(k)$}}
\put(17,32){\makebox(0,0)[r]{\footnotesize $1$}}
\put(18,10){\makebox(0,0)[rt]{\footnotesize $0$}}
\put(120,9){\makebox(0,0)[t]{\footnotesize $1024$}}
\put(220,9){\makebox(0,0)[t]{\footnotesize $2048$}}
\put(320,9){\makebox(0,0)[t]{\footnotesize $3072$}}
\put(420,9){\makebox(0,0)[t]{\footnotesize $4096$}}
\put(445,8){\makebox(0,0)[tr]{\small $k$}}
\put(445,22){\makebox(0,0)[rb]{$N\!=\!4$}}

\end{picture}}
\caption{Beispiele einiger Fensterfolgen.}
\label{b5h}
\end{figure}
zeigt an den Beispielen
mit den DFT-L"angen \mbox{$M\in\{4;64;1024\}$} und den
Fensterl"angenfaktoren \mbox{$N=1\;(1)\;4$}, welche Fensterfolgen sich bei
der von mir vorgeschlagenen Berechnungsmethode ergeben. Dabei wurde
f"ur \mbox{$M\in\{64;1024\}$} nicht die graphische Darstellung der
zeitdiskreten Wertefolge wie f"ur \mbox{$M\!=\!4$} gew"ahlt, sondern
es wurde eine quasikontinuierliche Darstellung angewendet. Dabei werden die 
Werte durch gerade Linien verbunden. Dadurch wird vermieden, dass schwarze
Fl"achen zwischen der Abszisse und der Kurve der Werte der Fensterfolge
entstehen.

\begin{figure}[t]
\begin{center}
{ 
\begin{picture}(454,260)

\input{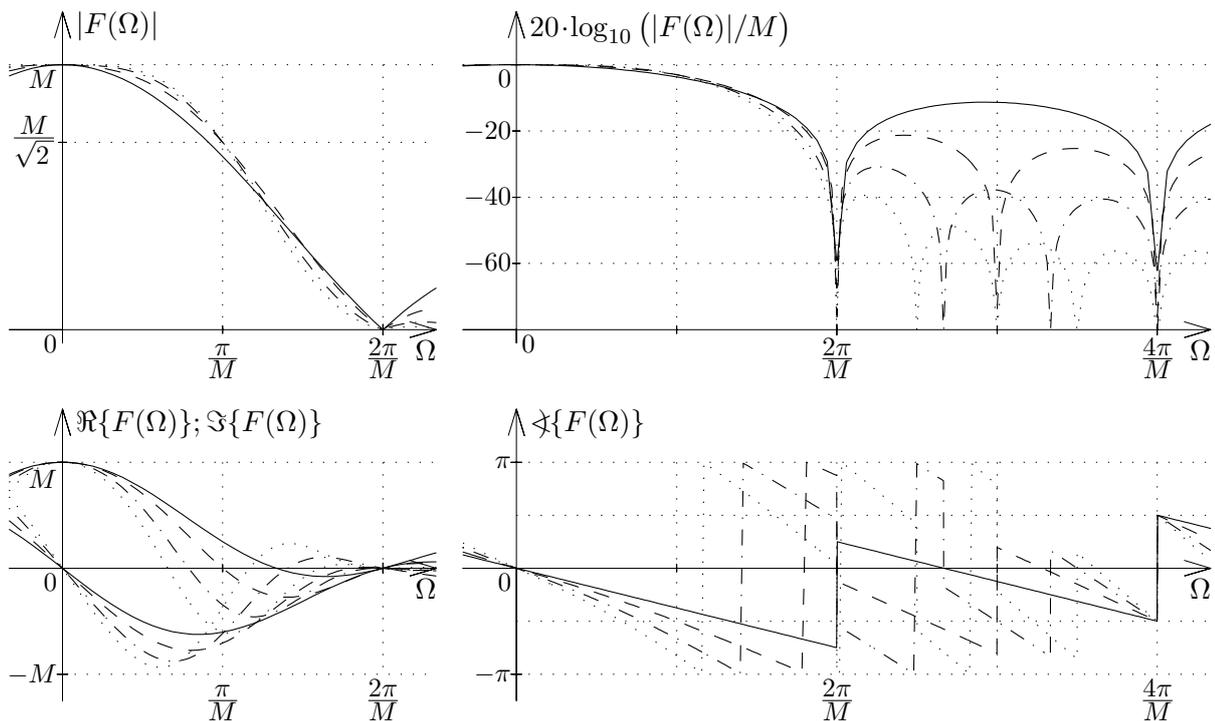}
\put(25,260){\makebox(0,0)[lt]{\small $|F(\Omega)|$}}
\put(18,238){\makebox(0,0)[rt]{\footnotesize $M$}}
\put(18,211){\makebox(0,0)[r]{\footnotesize $\frac{\T M}{\T\sqrt{2\,}}$}}
\put(18,138){\makebox(0,0)[rt]{\footnotesize $0$}}
\put(80,137){\makebox(0,0)[t]{\footnotesize $\frac{\T\vphantom{2}\pi}{\T M}$}}
\put(140,137){\makebox(0,0)[t]{\footnotesize $\frac{\T2\pi}{\T M}$}}
\put(160,136){\makebox(0,0)[tr]{\small $\Omega$}}

\put(195,260){\makebox(0,0)[lt]{\small $20\CdoT\log_{10}\big(|F(\Omega)|/M\big)$}}
\put(188,238){\makebox(0,0)[rt]{\footnotesize $0$}}
\put(188,215){\makebox(0,0)[r]{\footnotesize $-20$}}
\put(188,190){\makebox(0,0)[r]{\footnotesize $-40$}}
\put(188,165){\makebox(0,0)[r]{\footnotesize $-60$}}
\put(192,138){\makebox(0,0)[lt]{\footnotesize $0$}}
\put(310,137){\makebox(0,0)[t]{\footnotesize $\frac{\T2\pi}{\T M}$}}
\put(430,137){\makebox(0,0)[t]{\footnotesize $\frac{\T4\pi}{\T M}$}}
\put(450,136){\makebox(0,0)[tr]{\small $\Omega$}}

\put(25,110){\makebox(0,0)[lt]{\small $\Re\{F(\Omega)\};\Im\{F(\Omega)\}$}}
\put(18,87){\makebox(0,0)[rt]{\footnotesize $M$}}
\put(18,10){\makebox(0,0)[r]{\footnotesize $-M$}}
\put(18,48){\makebox(0,0)[rt]{\footnotesize $0$}}
\put(80,8){\makebox(0,0)[t]{\footnotesize $\frac{\T\vphantom{2}\pi}{\T M}$}}
\put(140,8){\makebox(0,0)[t]{\footnotesize $\frac{\T2\pi}{\T M}$}}
\put(160,46){\makebox(0,0)[tr]{\small $\Omega$}}

\put(195,110){\makebox(0,0)[lt]{\small $\winkel\{F(\Omega)\}$}}
\put(188,90){\makebox(0,0)[r]{\footnotesize $\pi$}}
\put(188,10){\makebox(0,0)[r]{\footnotesize $-\pi$}}
\put(188,48){\makebox(0,0)[rt]{\footnotesize $0$}}
\put(310,8){\makebox(0,0)[t]{\footnotesize $\frac{\T2\pi}{\T M}$}}
\put(430,8){\makebox(0,0)[t]{\footnotesize $\frac{\T4\pi}{\T M}$}}
\put(450,46){\makebox(0,0)[tr]{\small $\Omega$}}

\end{picture}}
\end{center}\vspace{-3pt}
\setlength{\belowcaptionskip}{-6pt}
\caption{Spektren der Fensterfolgen in Bild \protect{\ref{b5h}}
mit \mbox{$M\!=\!4$}.\protect\\
\mbox{$-\!\!\!-\!\!\!-\!\!\!-\!\!\!-\;N\!=\!1;\;
----\;N\!=\!2;\;
-\cdot-\cdot\;N\!=\!3;\;
\cdot\cdot\cdot\cdot\cdot\;N\!=\!4$.}}
\label{b5i1}
\rule{\textwidth}{0.5pt}\vspace{-10pt}
\end{figure}

Von den in Bild \ref{b5h} dargestellten Fenstern werden in 
Bild \ref{b5i1}
\begin{figure}[tp]
\begin{center}
{ 
\begin{picture}(454,260)

\input{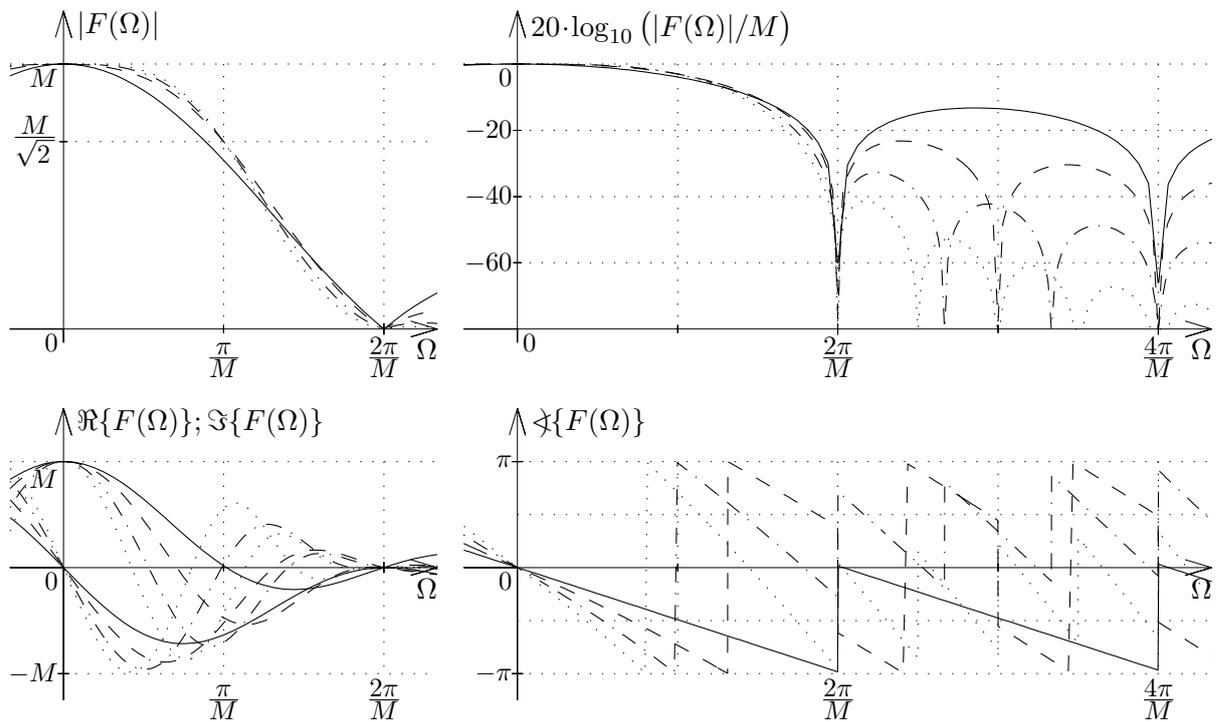}
\put(25,260){\makebox(0,0)[lt]{\small $|F(\Omega)|$}}
\put(18,238){\makebox(0,0)[rt]{\footnotesize $M$}}
\put(18,211){\makebox(0,0)[r]{\footnotesize $\frac{\T M}{\T\sqrt{2\,}}$}}
\put(18,138){\makebox(0,0)[rt]{\footnotesize $0$}}
\put(80,137){\makebox(0,0)[t]{\footnotesize $\frac{\T\vphantom{2}\pi}{\T M}$}}
\put(140,137){\makebox(0,0)[t]{\footnotesize $\frac{\T2\pi}{\T M}$}}
\put(160,136){\makebox(0,0)[tr]{\small $\Omega$}}

\put(195,260){\makebox(0,0)[lt]{\small $20\CdoT\log_{10}\big(|F(\Omega)|/M\big)$}}
\put(188,238){\makebox(0,0)[rt]{\footnotesize $0$}}
\put(188,215){\makebox(0,0)[r]{\footnotesize $-20$}}
\put(188,190){\makebox(0,0)[r]{\footnotesize $-40$}}
\put(188,165){\makebox(0,0)[r]{\footnotesize $-60$}}
\put(192,138){\makebox(0,0)[lt]{\footnotesize $0$}}
\put(310,137){\makebox(0,0)[t]{\footnotesize $\frac{\T2\pi}{\T M}$}}
\put(430,137){\makebox(0,0)[t]{\footnotesize $\frac{\T4\pi}{\T M}$}}
\put(450,136){\makebox(0,0)[tr]{\small $\Omega$}}

\put(25,110){\makebox(0,0)[lt]{\small $\Re\{F(\Omega)\};\Im\{F(\Omega)\}$}}
\put(18,87){\makebox(0,0)[rt]{\footnotesize $M$}}
\put(18,10){\makebox(0,0)[r]{\footnotesize $-M$}}
\put(18,48){\makebox(0,0)[rt]{\footnotesize $0$}}
\put(80,8){\makebox(0,0)[t]{\footnotesize $\frac{\T\vphantom{2}\pi}{\T M}$}}
\put(140,8){\makebox(0,0)[t]{\footnotesize $\frac{\T2\pi}{\T M}$}}
\put(160,46){\makebox(0,0)[tr]{\small $\Omega$}}

\put(195,110){\makebox(0,0)[lt]{\small $\winkel\{F(\Omega)\}$}}
\put(188,90){\makebox(0,0)[r]{\footnotesize $\pi$}}
\put(188,10){\makebox(0,0)[r]{\footnotesize $-\pi$}}
\put(188,48){\makebox(0,0)[rt]{\footnotesize $0$}}
\put(310,8){\makebox(0,0)[t]{\footnotesize $\frac{\T2\pi}{\T M}$}}
\put(430,8){\makebox(0,0)[t]{\footnotesize $\frac{\T4\pi}{\T M}$}}
\put(450,46){\makebox(0,0)[tr]{\small $\Omega$}}

\end{picture}}
\end{center}\vspace{-8pt}
\setlength{\belowcaptionskip}{-6pt}
\caption{Spektren der Fensterfolgen in Bild \protect{\ref{b5h}}
mit \mbox{$M\!=\!64$}.\protect\\
\mbox{$-\!\!\!-\!\!\!-\!\!\!-\!\!\!-\;N\!=\!1;\;
----\;N\!=\!2;\;
-\cdot-\cdot\;N\!=\!3;\;
\cdot\cdot\cdot\cdot\cdot\;N\!=\!4$.}}
\label{b5i2}
\rule{\textwidth}{0.5pt}\vspace{-10pt}
\end{figure}%
\begin{figure}[bp]
\begin{center}
{ 
\begin{picture}(454,260)

\input{mbild5i3}
\put(25,260){\makebox(0,0)[lt]{\small $|F(\Omega)|$}}
\put(18,238){\makebox(0,0)[rt]{\footnotesize $M$}}
\put(18,211){\makebox(0,0)[r]{\footnotesize $\frac{\T M}{\T\sqrt{2\,}}$}}
\put(18,138){\makebox(0,0)[rt]{\footnotesize $0$}}
\put(80,137){\makebox(0,0)[t]{\footnotesize $\frac{\T\vphantom{2}\pi}{\T M}$}}
\put(140,137){\makebox(0,0)[t]{\footnotesize $\frac{\T2\pi}{\T M}$}}
\put(160,136){\makebox(0,0)[tr]{\small $\Omega$}}

\put(195,260){\makebox(0,0)[lt]{\small $20\CdoT\log_{10}\big(|F(\Omega)|/M\big)$}}
\put(188,238){\makebox(0,0)[rt]{\footnotesize $0$}}
\put(188,215){\makebox(0,0)[r]{\footnotesize $-20$}}
\put(188,190){\makebox(0,0)[r]{\footnotesize $-40$}}
\put(188,165){\makebox(0,0)[r]{\footnotesize $-60$}}
\put(192,138){\makebox(0,0)[lt]{\footnotesize $0$}}
\put(310,137){\makebox(0,0)[t]{\footnotesize $\frac{\T2\pi}{\T M}$}}
\put(430,137){\makebox(0,0)[t]{\footnotesize $\frac{\T4\pi}{\T M}$}}
\put(450,136){\makebox(0,0)[tr]{\small $\Omega$}}

\put(25,110){\makebox(0,0)[lt]{\small $\Re\{F(\Omega)\};\Im\{F(\Omega)\}$}}
\put(18,87){\makebox(0,0)[rt]{\footnotesize $M$}}
\put(18,10){\makebox(0,0)[r]{\footnotesize $-M$}}
\put(18,48){\makebox(0,0)[rt]{\footnotesize $0$}}
\put(80,8){\makebox(0,0)[t]{\footnotesize $\frac{\T\vphantom{2}\pi}{\T M}$}}
\put(140,8){\makebox(0,0)[t]{\footnotesize $\frac{\T2\pi}{\T M}$}}
\put(160,46){\makebox(0,0)[tr]{\small $\Omega$}}

\put(195,110){\makebox(0,0)[lt]{\small $\winkel\{F(\Omega)\}$}}
\put(188,90){\makebox(0,0)[r]{\footnotesize $\pi$}}
\put(188,10){\makebox(0,0)[r]{\footnotesize $-\pi$}}
\put(188,48){\makebox(0,0)[rt]{\footnotesize $0$}}
\put(310,8){\makebox(0,0)[t]{\footnotesize $\frac{\T2\pi}{\T M}$}}
\put(430,8){\makebox(0,0)[t]{\footnotesize $\frac{\T4\pi}{\T M}$}}
\put(450,46){\makebox(0,0)[tr]{\small $\Omega$}}

\end{picture}}
\end{center}\vspace{-8pt}
\setlength{\belowcaptionskip}{-8pt}
\caption{Spektren der Fensterfolgen in Bild \protect{\ref{b5h}}
mit \mbox{$M\!=\!1024$}.\protect\\
\mbox{$-\!\!\!-\!\!\!-\!\!\!-\!\!\!-\;N\!=\!1;\;
----\;N\!=\!2;\;
-\cdot-\cdot\;N\!=\!3;\;
\cdot\cdot\cdot\cdot\cdot\;N\!=\!4$.}}
\label{b5i3}
\end{figure}
bis Bild \ref{b5i3} jeweils die
Spektren f"ur niedrige Frequenzen gezeigt. Rechts oben ist jeweils
der Betrag des Spektrums im Frequenzbereich der Hauptkeule in einem
linearen Ma"sstab dargestellt. Das darunterliegende Teilbild zeigt
jeweils den Real- und den Imagin"arteil innerhalb desselben Frequenzbereichs.
Im linken oberen Teilbild ist der Betrag des Spektrums in einem
logarithmischen Ma"sstab zu sehen, wobei der doppelte Frequenzbereich
gew"ahlt wurde, um auch die D"ampfung der ersten und zugleich gr"o"sten
Nebenkeulen darzustellen. Das darunterliegende Teilbild zeigt den
negativen Phasenfrequenzgang in demselben Frequenzbereich.

\subsection{Beispiele zum Anstieg der Sperrd"ampfung}

In Kapitel \ref{W} wurde bei der Festlegung des Wunschverlaufs des 
Betragsquadrats des Spektrums der Fensterfolge gezeigt, dass 
\mbox{$|F(\Omega)|^2$} n"aherungsweise wenigstens mit 
\mbox{$\sin(\Omega/2)^{-2\cdot B}$} abfallen sollte, wenn man eine
\mbox{$2\CdoT B$}-fache Nullstelle im LDS der St"orung messen will. 
Es wird nun gezeigt, dass die hier entworfenen Fensterfolgen diese 
Forderung n"aherungsweise erf"ullen, wenn $M$ gro"s ist --- beim RKM 
ist das "ublicherweise der Fall --- und der Fensterl"angenfaktor 
\mbox{$N\!\ge\!B$} ist. 

F"ur das Polynom \mbox{$z^{F-N}\Cdot G(z)$} 
der Z-Transformierten der Basisfensterfolge erhalten wir nach
Gleichung (\ref{6.5}) nach einer Multiplikation mit dem Term
\mbox{$z^{\frac{N-F}{2}}$} am Einheitskreis (\,\mbox{$z\!=\!e^{j\Omega}$}\,)
folgenden reellen Ausdruck:
\begin{equation}
e^{j\cdot\frac{F-N}{2}\cdot\Omega}\cdot G\big(e^{j\Omega}\big)\;=\;
2^{1-N}\cdot\frac{
\sin\!\big(\frac{\Omega}{2}\CdoT F\!-\!\pi\CdoT\frac{N-1}{2}\big)}
{\;\Prod{\nu_2=\frac{1-N}{2}}{\frac{N-1}{2}}\!\!
\sin\!\big(\frac{\Omega}{2}\!-\!\frac{\pi}{F}\CdoT\nu_2\big)\;}
\label{6.39}
\end{equation}
Das Betragsquadrat des Z"ahlers ist eine schnelle Kosinusschwingung in 
$\Omega$ mit der \glqq{}Kreisfrequenz\grqq{} $F$, deren Wert immer 
zwischen $0$ und $1$ liegt. Der Reziprokwert des Quadrats des Nenners ist 
daher eine obere Grenze f"ur \mbox{$\big|G\big(e^{j\Omega}\big)\big|^2\!$}, 
die jeweils im Abstand \mbox{$2\pi/F$} erreicht wird. 

F"ur jeden einzelnen Faktor des Quadrats des Nenners in Gleichung (\ref{6.39}) 
kann man auch \mbox{$\cos(\pi/F\CdoT\nu_2)\CdoT\sin(\Omega/2)-
\sin(\pi/F\CdoT\nu_2)\CdoT\cos(\Omega/2)$} schreiben. Bei jedem dieser
Faktoren "uberwiegt f"ur \mbox{$2\pi/M\ll|\Omega|\le\pi$} der Anteil der
mit \mbox{$\sin(\Omega/2)$} ansteigt. Das Produkt dieser einzelnen 
Faktoren ist ein Polynom in \mbox{$\sin(\Omega/2)$} mit der h"ochsten
Potenz \mbox{$2\CdoT\!N$}. F"ur den angegebenen Bereich von $\Omega$ wird
der Nenner im wesentlichen durch die h"ochste Potenz in \mbox{$\sin(\Omega/2)$}
bestimmt. \mbox{$\big|G\big(e^{j\Omega}\big)\big|^2$} f"allt daher
n"aherungsweise mit \mbox{$\sin(\Omega/2)^{-2\cdot N}$} ab. 

Das Betragsquadrat
\mbox{$|F(\Omega)|^2$} der Fensterfolge ergibt sich nach Gleichung~(\ref{6.7})
und Gleichung (\ref{6.14}) als "Uberlagerung der verschobenen Betrags\-quadrate 
des Spektrums \mbox{$G\big(e^{j\Omega}\big)$}. Wenn die dabei maximal 
auftretende Verschiebung \mbox{$(N\!-\!1)\CdoT\pi/F$} wieder klein gegen 
$|\Omega|$ ist, f"allt auch \mbox{$|F(\Omega)|^2$} n"aherungsweise mit 
\mbox{$\sin(\Omega/2)^{-2\cdot N}$} ab. 

\begin{figure}[btp]
\begin{center}
{ 
\begin{picture}(450,450)

\input{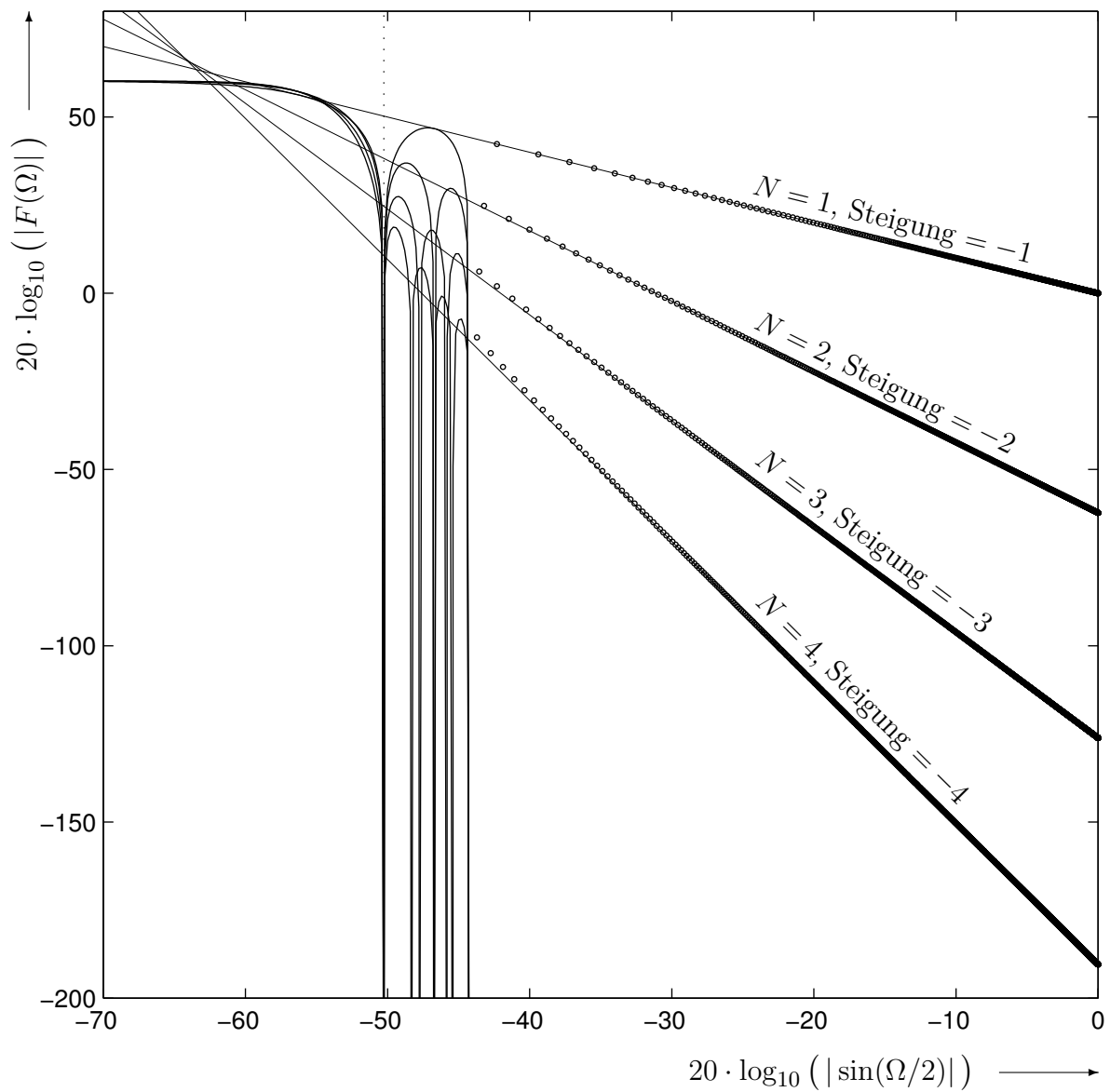}
\put(10,390){\makebox(0,0)[t]{\rotatebox{90}
{$20\cdot\log_{10}\big(\,|F(\Omega)|\,\big)$}}}
\put(10,400){\vector(0,1){40}}
\put(390,10){\makebox(0,0)[r]
{$20\cdot\log_{10}\big(\,|\sin(\Omega/2)|\,\big)$}}
\put(400,10){\vector(1,0){40}}
\put(300,366){\rotatebox{-14}{$N=1$, Steigung $= -1$}}
\put(300,312){\rotatebox{-26.6}{$N=2$, Steigung $= -2$}}
\put(300,257){\rotatebox{-36.9}{$N=3$, Steigung $= -3$}}
\put(300,200){\rotatebox{-45}{$N=4$, Steigung $= -4$}}

\end{picture}}
\end{center}\vspace{-18pt}
\setlength{\belowcaptionskip}{-3pt}
\caption{Sperrd"ampfung der Nebenmaxima des Spektrums der Fensterfolge}
\label{b4d}
\rule{\textwidth}{0.5pt}\vspace{-7pt}
\end{figure}

In Bild~\ref{b4d}
ist 
der Betrag des Spektrums der Fensterfolge im doppelt logarithmischen Ma"sstab 
f"ur \mbox{$M\!=\!1024$} und f"ur \mbox{$N=1\;(1)\;4$} dargestellt. Um in dem 
Bereich, in dem die Nullstellen des Fensterspektrums zu dicht liegen, eine 
graphische Darstellung zu erm"oglichen, wurden f"ur \mbox{$\Omega>4\pi/M$} 
nur Abtastwerte des Spektrums dargestellt. Dabei wurde das Spektrum bei 
ungeradzahligen Vielfachen von \mbox{$\pi/F$} abgetastet, also immer dort, 
wo die Sinusfunktion im Z"ahler ihr betragliches Maximum erreicht und somit 
die "`Einh"ullende"' des Spektrums gerade ber"uhrt wird. Um den potenzm"a"sigen 
Zusammenhang besser "uberpr"ufen zu \mbox{k"onnen,} ist auf der Abszisse der 
Betrag des Sinus der halben Kreisfrequenz aufgetragen. Zur besseren Orientierung 
ist bei der Frequenz \mbox{$\Omega=2\pi/M$} --- also bei der Frequenz der ersten
Nullstelle im Spektrum --- eine senkrechte punktierte Hilfslinie eingetragen.
Die Achsen sind im Ma"sstab \mbox{$1\!:\!4$} gew"ahlt worden, um ein 
sinnvolles Verh"altnis von Breite zu H"ohe der Graphik zu erzielen. 
Durch die Abtastwerte bei den Frequenzen \mbox{$\Omega=\pi\CdoT(F\!-\!1)/F$} 
--- also jeweils durch den gr"o"sten Abszissenwert --- wurden zum Vergleich 
die Geraden mit den Steigungen $-N$ eingezeichnet. Wie man sieht, hat man 
durch die Wahl des Parameters $N$ die M"oglichkeit den potenzm"a"sigen 
Anstieg der Sperrd"ampfung wie gew"unscht einzustellen. 

Wenn \mbox{$N\!>\!B$} 
gew"ahlt wird, f"allt der Integrand in Gleichung (\ref{2.17}) bei einem LDS mit 
einer \mbox{$2\CdoT B$}-fachen Nullstelle mit steigender Abweichung von der zu 
messenden Frequenz ab, und es ist zu erwarten, dass die Messergebnisse des LDS 
einigerma"sen zuverl"assig sind. In einem Beispiel in Kapitel \ref{Mess7} weist 
das zu messende LDS eine sechsfache Nullstelle bei \mbox{$\Omega\!=\!0$} auf. 
Deutlich ist zu sehen, dass erst f"ur \mbox{$N\!=\!3$} die Werte des LDS in der 
Umgebung von \mbox{$\Omega\!=\!0$} gemessen werden k"onnen. 
\newpage

\subsection{Pr"azision der Nullstellen im Fensterspektrum}

In Kapitel \ref{W} hatten wir festgestellt, dass das Spektrum bei den in 
Gleichung (\ref{2.27}) angegebenen Frequenzen Nullstellen aufzuweisen hat. 
Um zu demonstrieren, das diese Bedingung von den Fensterfolgen, die mit 
dem in den vorherigen Unterkapiteln beschriebenen Algorithmus berechnet 
worden sind, mit Fehlern erf"ullt werden, die in der Gr"o"senordnung 
liegen, die aufgrund der endlichen Wortl"ange bei der Abspeicherung der 
Fensterfolge unvermeidbar sind, werden f"ur alle Vielfachen der Frequenz 
\mbox{$2\pi/M$} die Spektralwerte eines Beispielfensters berechnet, 
und deren Betrag in Bild \ref{b5j}
\begin{figure}[btp]
\begin{center}
{ 
\begin{picture}(454,580)

\input{mbild5j1}
\put(0,580){\makebox(0,0)[lt]{
Berechnung durch Auswertung der DFT-Summenformel:}}
\put(47,562){\makebox(0,0)[lt]{$20\CdoT\log_{10}\Big(\big|
F\big({\T\mu\CdoT\frac{2\pi}{M}}\big)\big|\Big)$}}
\put(35,530){\makebox(0,0)[r]{\small$50$}}
\put(35,498){\makebox(0,0)[rt]{\small$0$}}
\put(35,440){\makebox(0,0)[r]{\small$-100$}}
\put(35,380){\makebox(0,0)[r]{\small$-200$}}
\put(35,320){\makebox(0,0)[r]{\small$-300$}}
\put(130,497){\makebox(0,0)[t]{\small$\pi/4$}}
\put(220,497){\makebox(0,0)[t]{\small$\pi/2$}}
\put(310,497){\makebox(0,0)[t]{\small$3\pi/4$}}
\put(400,493){\makebox(0,0)[t]{\small$\pi$}}
\put(145,410){\makebox(0,0)[l]{${\D20\CdoT\log_{10}\Big(
\sqrt{F\,}\Cdot\varepsilon\cdot\max_{k}\big(\,|f(k)|\,\big)\Big)}$}}
\put(454,505){\makebox(0,0)[rb]{$\Omega=\frac{2\pi}{M}\CdoT\mu$}}

\input{mbild5j2}
\put(0,280){\makebox(0,0)[lt]{
Berechnung mit Hilfe einer FFT:}}
\put(47,262){\makebox(0,0)[lt]{$20\CdoT\log_{10}\Big(\big|
F\big({\T\mu\CdoT\frac{2\pi}{M}}\big)\big|\Big)$}}
\put(35,230){\makebox(0,0)[r]{\small$50$}}
\put(35,198){\makebox(0,0)[rt]{\small$0$}}
\put(35,140){\makebox(0,0)[r]{\small$-100$}}
\put(35,80){\makebox(0,0)[r]{\small$-200$}}
\put(35,20){\makebox(0,0)[r]{\small$-300$}}
\put(130,197){\makebox(0,0)[t]{\small$\pi/4$}}
\put(220,197){\makebox(0,0)[t]{\small$\pi/2$}}
\put(310,197){\makebox(0,0)[t]{\small$3\pi/4$}}
\put(400,193){\makebox(0,0)[t]{\small$\pi$}}
\put(145,110){\makebox(0,0)[l]{${\D20\CdoT\log_{10}\Big(
\sqrt{F\,}\Cdot\varepsilon\cdot\max_{k}\big(\,|f(k)|\,\big)\Big)}$}}
\put(454,205){\makebox(0,0)[rb]{$\Omega=\frac{2\pi}{M}\CdoT\mu$}}

\end{picture}}
\end{center}\vspace{-0pt}
\caption{Fehler bei den nach Gleichung (\protect\ref{2.27})
geforderten Nullstellen\protect\\
Beispiel mit: \mbox{$M\!=\!231$}; \mbox{$N\!=\!17$}}
\label{b5j}
\end{figure}
halblogarithmisch 
"uber der Frequenz aufgetragen. 

Als Beispiel wurde das Fenster mit, 
\mbox{$M\!=\!231$} und \mbox{$N\!=\!17$} ausgew"ahlt, um so auch zu 
zeigen, dass weder $M$ noch $N$ gerade oder gar eine Zweierpotenz 
sein m"ussen. Zum Vergleich wird auch der Wert \mbox{${\D20\CdoT\log_{10}\big(
\sqrt{F\,}\Cdot\varepsilon\cdot\max_{k}\big(\,|f(k)|\,\big)\big)}$} als 
waagrechte Linie eingetragen. Ein Rauschsockel dieser Gr"o"senordnung w"urde 
sich theoretisch ergeben, wenn den Werten der Fensterfolge ein additiver 
St"orer der Streuung \mbox{${\D\varepsilon\cdot\max_{k}\big(\,|f(k)|\,\big)}$} 
"uberlagert w"are, und eine fehlerfreie Berechnung der Spektralwerte 
m"oglich w"are. 

Die Spektral\-werte wurden f"ur Vielfache der Frequenz 
\mbox{$2\pi/M$} auf zwei unterschiedliche Weisen berechnet. 
Die Spektralwerte des unteren Teilbildes wurden gewonnen, indem 
\mbox{$f(k)$} zun"achst einer FFT der L"ange $F$ 
(\,{\tt MATLAB: F\_Omega = fft(f\_k)}\,) unterworfen wurde, und 
daraus die $M$ gew"unschten Spektralwerte entnommen wurden 
(\,{\tt MATLAB: F\_Omega(1:N:F)}\,). Aus dem Wert bei der Frequenz 
\mbox{$\mu\CdoT2\pi/M$} und dem Konjugierten des Wertes bei der 
Frequenz \mbox{$(M\!-\!\mu)\cdoT2\pi/M$} wurde abschlie"send jeweils der 
Mittelwert gebildet, um eine Reduktion der Fehler bei der Berechnung 
der DFT zu erreichen. Man erkennt, dass die Betr"age der Fehler der
Spektralwerte doch deutlich "uber der Referenzlinie liegen. Um zu
zeigen, dass diese Fehler nicht durch Fehler in der Fensterfolge, sondern
durch die relativ stark fehlerbehaftete Berechnung der Spektralwerte
verursacht werden, wurden die Spektralwerte des oberen Teilbildes
durch Auswertung der DFT-Summenformel nach Gleichung (\ref{2.18}) mit
\mbox{$\Omega=\mu\CdoT2\pi/M$} berechnet. Dabei wurden die ganzzahligen
Produkte \mbox{$\mu\CdoT k$} in den Exponenten der Drehfaktoren
\begin{multline*}
e^{\!-j\cdot\frac{2\pi}{M}\cdot\mu\cdot k}\,=\;
\cos\big(\frac{2\pi}{M}\CdoT\mu\CdoT k\big)-
j\CdoT\sin\big(\frac{2\pi}{M}\CdoT\mu\CdoT k\big)\;=\\
=\;\sin\big(\frac{\pi}{2\CdoT M}\CdoT(4\CdoT\mu\CdoT k+M)\big)-
j\CdoT\sin\big(\frac{2\pi}{M}\CdoT\mu\CdoT k\big)\;=\\
=\;\sin\big(\frac{\pi}{2\CdoT M}\CdoT(M - 4\CdoT\mu\CdoT k)\big)-
j\CdoT\sin\big(\frac{\pi}{2\CdoT M}\CdoT(2\CdoT M - 4\CdoT\mu\CdoT k)\big)
\end{multline*}
um ganzzahlige Vielfache von $M$ jeweils in der Art reduziert,
dass sich die Real- und Imagin"arteile der Drehfaktoren nach
der jeweils geeignetsten\footnote{Das Argument der Sinusfunktion 
ist dann betraglich immer kleiner als \mbox{$\pi/2$}.} Darstellung 
in der letzten Gleichung mit h"ochstm"oglicher Genauigkeit berechnen 
lassen. Die Reihenfolge der Summanden in Gleichung (\ref{2.18}) wurde 
--- beim Real- und Imagin"arteil getrennt --- jeweils so gew"ahlt, 
dass zu der bisher berechneten Zwischensumme jeweils der Summand als 
n"achstes addiert wird, der zu dem minimalen Betrag der folgenden 
Zwischensumme f"uhrt. Auf diese Art wird vermieden, dass bei Spektralwerten, die selbst
einen kleinen Betrag haben (\,in unserem Fall sind das alle, bis
auf den Spektralwert bei der Frequenz Null\,), gro"se und damit
ungenaue Zwischensummen entstehen. 

Man erkennt bei den so berechneten
Spektralwerten in dem oberen Teilbild, dass die Fehler in der
Fensterfolge so gering sind, dass die Gleichung (\ref{2.27})
mit maximal m"oglicher Genauigkeit erf"ullt wird.

\subsection{Nullstellenlage der Z-Transformierten der Fensterfolgen}

In Bild \ref{b5k}
\begin{figure}[btp]
\begin{center}
{ 
\begin{picture}(454,440)

\input{mbild5k1}
\put(196,335){\makebox(0,0)[t]{$\Re\{z\}$}}
\put(96,435){\makebox(0,0)[r]{$\Im\{z\}$}}
\put(99,338){\makebox(0,0)[rt]{\small$0$}}
\put(180,338){\makebox(0,0)[rt]{\small$1$}}
\put(98,418){\makebox(0,0)[rt]{\small$1$}}
\put(167,406){\makebox(0,0)[lb]{$e^{j\cdot\frac{2\pi}{M}}$}}
\put(40,305){\makebox(0,0)[lb]{${\D\frac{2\pi}{F}}$}}
\put(17,311){\makebox(0,0)[rt]{${\D\frac{2\pi}{M}}$}}
\put(41,420){\makebox(0,0)[rb]{$N\!=\!1$}}

\input{mbild5k2.tex}
\put(441,335){\makebox(0,0)[t]{$\Re\{z\}$}}
\put(341,435){\makebox(0,0)[r]{$\Im\{z\}$}}
\put(344,338){\makebox(0,0)[rt]{\small$0$}}
\put(425,338){\makebox(0,0)[rt]{\small$1$}}
\put(344,416){\makebox(0,0)[rt]{\small$1$}}
\put(412,406){\makebox(0,0)[lb]{$e^{j\cdot\frac{2\pi}{M}}$}}
\put(282,311){\makebox(0,0)[lb]{${\D\frac{2\pi}{F}}$}}
\put(260,311){\makebox(0,0)[rt]{${\D\frac{2\pi}{M}}$}}
\put(286,420){\makebox(0,0)[rb]{$N\!=\!2$}}

\input{mbild5k3.tex}
\put(196,95){\makebox(0,0)[t]{$\Re\{z\}$}}
\put(96,195){\makebox(0,0)[r]{$\Im\{z\}$}}
\put(99,98){\makebox(0,0)[rt]{\small$0$}}
\put(180,98){\makebox(0,0)[rt]{\small$1$}}
\put(98,178){\makebox(0,0)[rt]{\small$1$}}
\put(167,166){\makebox(0,0)[lb]{$e^{j\cdot\frac{2\pi}{M}}$}}
\put(40,65){\makebox(0,0)[lb]{${\D\frac{2\pi}{F}}$}}
\put(17,71){\makebox(0,0)[rt]{${\D\frac{2\pi}{M}}$}}
\put(41,180){\makebox(0,0)[rb]{$N\!=\!3$}}

\input{mbild5k4.tex}
\put(441,95){\makebox(0,0)[t]{$\Re\{z\}$}}
\put(341,195){\makebox(0,0)[r]{$\Im\{z\}$}}
\put(344,98){\makebox(0,0)[rt]{\small$0$}}
\put(425,98){\makebox(0,0)[rt]{\small$1$}}
\put(344,176){\makebox(0,0)[rt]{\small$1$}}
\put(412,166){\makebox(0,0)[lb]{$e^{j\cdot\frac{2\pi}{M}}$}}
\put(282,71){\makebox(0,0)[lb]{${\D\frac{2\pi}{F}}$}}
\put(260,71){\makebox(0,0)[rt]{${\D\frac{2\pi}{M}}$}}
\put(286,180){\makebox(0,0)[rb]{$N\!=\!4$}}

\end{picture}}
\end{center}\vspace{-12pt}
\setlength{\belowcaptionskip}{-7pt}
\caption{Nullstellenlage der Z-Transformierten der Fensterfolge mit
\mbox{$M\!=\!8$}.\protect\\
Die Nullstelle bei \mbox{$z\!=\!0$} hat die Vielfachheit \mbox{$N\!-\!1$}.}
\label{b5k}
\rule{\textwidth}{0.5pt}\vspace{-11pt}
\end{figure}
sind f"ur \mbox{$M\!=\!8$} und 
\mbox{$N=1\;(1)\;4$} jeweils die Nullstellen der 
Z-Transformierten der Fensterfolge in der $z$-Ebene 
dargestellt. Diese wurden mit {\tt MATLAB} mit dem Befehl 
{\tt roots(f\_k)} berechnet. 

Die Nullstellen im 
inneren des Einheitskreises (\,ohne die Nullstellen bei 
\mbox{$z\!=\!0$}\,) bewirken, dass sich f"ur das Spektrum 
der Fensterfolge ein Betrags- und Phasenfrequenzgang in der Art 
ergibt, dass die Bedingung (\ref{2.20}) erf"ullt wird. Da der 
Betrag des Spektrums der Fensterfolge bei deren Berechnung 
durch die "Uberlagerung der Betragsquadrate verschobener Spektren 
vorgegeben wird, und die Phase des Spektrums der Fensterfolge 
mit Hilfe des Cepstrums berechnet wird, braucht die Lage dieser 
Nullstellen bei der Berechnung der Fensterfolge nicht explizit 
bestimmt zu werden. Es wurde auch ein Algorithmus ausprobiert, 
bei dem zun"achst die Lage dieser Nullstellen numerisch bestimmt 
wurde, und daraus die Fensterfolge "uber deren Spektralwerte 
berechnet wird. Es zeigte sich dabei, dass --- vor allem f"ur gro"ses 
$N$ --- diese Art der Berechnung zu unbefriedigenden Ergebnissen f"uhrte. 

\subsection{Beispiele der Fenster-AKF}

Bild \ref{b5l}
\begin{figure}[btp]
\begin{center}
{ 
\begin{picture}(454,600)

\input{mbild5l}
\put(0,600){\makebox(0,0)[lt]{Fenster-AKF \mbox{$d(k)$} mit \mbox{$M\!=\!4$}:}}
\put(365,591){\makebox(0,0)[r]{\small $d(k)$}}
\put(367,580){\makebox(0,0)[r]{\footnotesize $1$}}
\put(320,557){\makebox(0,0)[t]{\footnotesize $\!\!-4$}}
\put(368,558){\makebox(0,0)[rt]{\footnotesize $0$}}
\put(420,557){\makebox(0,0)[t]{\footnotesize $4$}}
\put(445,556){\makebox(0,0)[tr]{\small $k$}}
\put(445,570){\makebox(0,0)[rb]{$N\!=\!1$}}

\put(315,541){\makebox(0,0)[r]{\small $d(k)$}}
\put(317,532){\makebox(0,0)[r]{\footnotesize $1$}}
\put(220,509){\makebox(0,0)[t]{\footnotesize $\!\!-8$}}
\put(270,509){\makebox(0,0)[t]{\footnotesize $\!\!-4$}}
\put(318,510){\makebox(0,0)[rt]{\footnotesize $0$}}
\put(370,509){\makebox(0,0)[t]{\footnotesize $4$}}
\put(420,509){\makebox(0,0)[t]{\footnotesize $8$}}
\put(445,508){\makebox(0,0)[tr]{\small $k$}}
\put(445,522){\makebox(0,0)[rb]{$N\!=\!2$}}

\put(265,495){\makebox(0,0)[r]{\small $d(k)$}}
\put(267,484){\makebox(0,0)[r]{\footnotesize $1$}}
\put(120,461){\makebox(0,0)[t]{\footnotesize $\!\!-12$}}
\put(170,461){\makebox(0,0)[t]{\footnotesize $\!\!-8$}}
\put(220,461){\makebox(0,0)[t]{\footnotesize $\!\!-4$}}
\put(268,462){\makebox(0,0)[rt]{\footnotesize $0$}}
\put(320,461){\makebox(0,0)[t]{\footnotesize $4$}}
\put(370,461){\makebox(0,0)[t]{\footnotesize $8$}}
\put(420,461){\makebox(0,0)[t]{\footnotesize $12$}}
\put(445,460){\makebox(0,0)[tr]{\small $k$}}
\put(445,474){\makebox(0,0)[rb]{$N\!=\!3$}}

\put(215,447){\makebox(0,0)[r]{\small $d(k)$}}
\put(217,436){\makebox(0,0)[r]{\footnotesize $1$}}
\put(20,413){\makebox(0,0)[t]{\footnotesize $\!\!-16$}}
\put(70,413){\makebox(0,0)[t]{\footnotesize $\!\!-12$}}
\put(120,413){\makebox(0,0)[t]{\footnotesize $\!\!-8$}}
\put(170,413){\makebox(0,0)[t]{\footnotesize $\!\!-4$}}
\put(218,414){\makebox(0,0)[rt]{\footnotesize $0$}}
\put(270,413){\makebox(0,0)[t]{\footnotesize $4$}}
\put(320,413){\makebox(0,0)[t]{\footnotesize $8$}}
\put(370,413){\makebox(0,0)[t]{\footnotesize $12$}}
\put(420,413){\makebox(0,0)[t]{\footnotesize $16$}}
\put(445,412){\makebox(0,0)[tr]{\small $k$}}
\put(445,426){\makebox(0,0)[rb]{$N\!=\!4$}}

\put(0,393){\makebox(0,0)[lt]{Fenster-AKF \mbox{$d(k)$} mit \mbox{$M\!=\!64$}:}}
\put(365,389){\makebox(0,0)[r]{\small $d(k)$}}
\put(367,378){\makebox(0,0)[r]{\footnotesize $1$}}
\put(320,355){\makebox(0,0)[t]{\footnotesize $\!\!-64$}}
\put(368,356){\makebox(0,0)[rt]{\footnotesize $0$}}
\put(420,355){\makebox(0,0)[t]{\footnotesize $64$}}
\put(445,354){\makebox(0,0)[tr]{\small $k$}}
\put(445,368){\makebox(0,0)[rb]{$N\!=\!1$}}

\put(315,341){\makebox(0,0)[r]{\small $d(k)$}}
\put(317,330){\makebox(0,0)[r]{\footnotesize $1$}}
\put(220,307){\makebox(0,0)[t]{\footnotesize $\!\!-128$}}
\put(270,307){\makebox(0,0)[t]{\footnotesize $\!\!-64$}}
\put(318,308){\makebox(0,0)[rt]{\footnotesize $0$}}
\put(370,307){\makebox(0,0)[t]{\footnotesize $64$}}
\put(420,307){\makebox(0,0)[t]{\footnotesize $128$}}
\put(445,307){\makebox(0,0)[tr]{\small $k$}}
\put(445,320){\makebox(0,0)[rb]{$N\!=\!2$}}

\put(265,293){\makebox(0,0)[r]{\small $d(k)$}}
\put(267,282){\makebox(0,0)[r]{\footnotesize $1$}}
\put(120,259){\makebox(0,0)[t]{\footnotesize $\!\!-192$}}
\put(170,259){\makebox(0,0)[t]{\footnotesize $\!\!-128$}}
\put(220,259){\makebox(0,0)[t]{\footnotesize $\!\!-64$}}
\put(268,260){\makebox(0,0)[rt]{\footnotesize $0$}}
\put(320,259){\makebox(0,0)[t]{\footnotesize $64$}}
\put(370,259){\makebox(0,0)[t]{\footnotesize $128$}}
\put(420,259){\makebox(0,0)[t]{\footnotesize $192$}}
\put(445,258){\makebox(0,0)[tr]{\small $k$}}
\put(445,272){\makebox(0,0)[rb]{$N\!=\!3$}}

\put(215,245){\makebox(0,0)[r]{\small $d(k)$}}
\put(217,234){\makebox(0,0)[r]{\footnotesize $1$}}
\put(20,211){\makebox(0,0)[t]{\footnotesize $\!\!-256$}}
\put(70,211){\makebox(0,0)[t]{\footnotesize $\!\!-192$}}
\put(120,211){\makebox(0,0)[t]{\footnotesize $\!\!-128$}}
\put(170,211){\makebox(0,0)[t]{\footnotesize $\!\!-64$}}
\put(218,212){\makebox(0,0)[rt]{\footnotesize $0$}}
\put(270,211){\makebox(0,0)[t]{\footnotesize $64$}}
\put(320,211){\makebox(0,0)[t]{\footnotesize $128$}}
\put(370,211){\makebox(0,0)[t]{\footnotesize $192$}}
\put(420,211){\makebox(0,0)[t]{\footnotesize $256$}}
\put(445,210){\makebox(0,0)[tr]{\small $k$}}
\put(445,224){\makebox(0,0)[rb]{$N\!=\!4$}}

\put(0,191){\makebox(0,0)[lt]{Fenster-AKF \mbox{$d(k)$} mit \mbox{$M\!=\!1024$}:}}
\put(365,187){\makebox(0,0)[r]{\small $d(k)$}}
\put(367,176){\makebox(0,0)[r]{\footnotesize $1$}}
\put(320,153){\makebox(0,0)[t]{\footnotesize $\!\!-1024$}}
\put(368,154){\makebox(0,0)[rt]{\footnotesize $0$}}
\put(420,153){\makebox(0,0)[t]{\footnotesize $1024$}}
\put(445,152){\makebox(0,0)[tr]{\small $k$}}
\put(445,166){\makebox(0,0)[rb]{$N\!=\!1$}}

\put(315,139){\makebox(0,0)[r]{\small $d(k)$}}
\put(317,128){\makebox(0,0)[r]{\footnotesize $1$}}
\put(220,105){\makebox(0,0)[t]{\footnotesize $\!\!-2048$}}
\put(270,105){\makebox(0,0)[t]{\footnotesize $\!\!-1024$}}
\put(318,106){\makebox(0,0)[rt]{\footnotesize $0$}}
\put(370,105){\makebox(0,0)[t]{\footnotesize $1024$}}
\put(420,105){\makebox(0,0)[t]{\footnotesize $2048$}}
\put(445,104){\makebox(0,0)[tr]{\small $k$}}
\put(445,118){\makebox(0,0)[rb]{$N\!=\!2$}}

\put(265,91){\makebox(0,0)[r]{\small $d(k)$}}
\put(267,80){\makebox(0,0)[r]{\footnotesize $1$}}
\put(120,57){\makebox(0,0)[t]{\footnotesize $\!\!-3072$}}
\put(170,57){\makebox(0,0)[t]{\footnotesize $\!\!-2048$}}
\put(220,57){\makebox(0,0)[t]{\footnotesize $\!\!-1024$}}
\put(268,58){\makebox(0,0)[rt]{\footnotesize $0$}}
\put(320,57){\makebox(0,0)[t]{\footnotesize $1024$}}
\put(370,57){\makebox(0,0)[t]{\footnotesize $2048$}}
\put(420,57){\makebox(0,0)[t]{\footnotesize $3072$}}
\put(445,56){\makebox(0,0)[tr]{\small $k$}}
\put(445,70){\makebox(0,0)[rb]{$N\!=\!3$}}

\put(215,43){\makebox(0,0)[r]{\small $d(k)$}}
\put(217,32){\makebox(0,0)[r]{\footnotesize $1$}}
\put(20,9){\makebox(0,0)[t]{\footnotesize $\!\!-4096$}}
\put(70,9){\makebox(0,0)[t]{\footnotesize $\!\!-3072$}}
\put(120,9){\makebox(0,0)[t]{\footnotesize $\!\!-2048$}}
\put(170,9){\makebox(0,0)[t]{\footnotesize $\!\!-1024$}}
\put(218,10){\makebox(0,0)[rt]{\footnotesize $0$}}
\put(270,9){\makebox(0,0)[t]{\footnotesize $1024$}}
\put(320,9){\makebox(0,0)[t]{\footnotesize $2048$}}
\put(370,9){\makebox(0,0)[t]{\footnotesize $3072$}}
\put(420,9){\makebox(0,0)[t]{\footnotesize $4096$}}
\put(445,8){\makebox(0,0)[tr]{\small $k$}}
\put(445,22){\makebox(0,0)[rb]{$N\!=\!4$}}

\end{picture}}
\end{center}\vspace{0pt}
\caption{Autokorrelationsfolgen einiger Fenster.}
\label{b5l}
\end{figure}
zeigt beispielhaft die 
Fenster-AKF \mbox{$d(k)$} nach Gleichung (\ref{2.21}), 
wobei die DFT-L"angen \mbox{$M\in\{4;64;1024\}$} und die 
Fensterl"angenfaktoren \mbox{$N=1\;(1)\;4$} ausgew"ahlt wurden. 
Wieder wurden aus dem obengenannten Grund (\,vgl. Bild \ref{b5h}\,) 
nur f"ur \mbox{$M\!=\!4$} die zeitdiskreten Wertefolgen dargestellt, 
w"ahrend f"ur \mbox{$M\in\{64;1024\}$} eine quasikontinuierliche 
Darstellung der Folgen gew"ahlt wurde. 

\subsection{Pr"azision der Nullstellen der Fenster-AKF}

Um beurteilen zu k"onnen, wie gut die Bedingung (\ref{2.23}) erf"ullt wird, 
nach der die Fenster-AKF \mbox{$d(k)$} Nullstellen haben 
soll, wenn $k$ ein Vielfaches von $M$ ist, und nach der \mbox{$d(0)$} 
Eins sein soll, werden f"ur \mbox{$M\in\{4;64;1024\}$} und 
\mbox{$N=2\;(1)\;4$} diese Werte in der Tabelle~\ref{T6.1} aufgelistet. 
\begin{table}[t]
\begin{center}$
{\renewcommand{\arraystretch}{1.2}
\setlength{\arraycolsep}{0pt}
\begin{array}{||c|c||r@{,}l@{{}\CdoT{}}l||r@{,}l@{{}\CdoT{}}l|
r@{,}l@{{}\CdoT{}}l|r@{,}l@{{}\CdoT{}}l|r@{,}l@{{}\CdoT{}}l||}
\hline
\hline
M&\;N\;&
\multicolumn{3}{c||}{\sigma\big(\varepsilon_{(\ref{2.20})}(\Omega)\big)}&
\multicolumn{3}{c|}{d(0)-1}&
\multicolumn{3}{c|}{d(M)}&
\multicolumn{3}{c|}{d(2\CdoT\!M)}&
\multicolumn{3}{c||}{d(3\CdoT\!M)}\\
\hline
\hline
&2&
9&8131&10^{-18}&
\multicolumn{3}{c|}{0}&
6&9389&10^{-18}&
\multicolumn{3}{c|}{\frac{\hphantom{MM}}{\hphantom{MM}}}&
\multicolumn{3}{c||}{\frac{\hphantom{MM}}{\hphantom{MM}}}\\
\cline{2-17}
4&3&
\;4&6278&10^{-16}\;&
\;-4&4409&10^{-16}\;&
6&2450&10^{-17}&
6&7654&10^{-17}&
\multicolumn{3}{c||}{\frac{\hphantom{MM}}{\hphantom{MM}}}\\
\cline{2-17}
&4&
2&5257&10^{-16}&
\multicolumn{3}{c|}{0}&
1&2143&10^{-16}&
1&7347&10^{-18}&
-1&3095&10^{-16}\\
\hline
&2&
2&2684&10^{-16}&
2&2204&10^{-16}&
3&2797&10^{-17}&
\multicolumn{3}{c|}{\frac{\hphantom{MM}}{\hphantom{MM}}}&
\multicolumn{3}{c||}{\frac{\hphantom{MM}}{\hphantom{MM}}}\\
\cline{2-17}
64&3&
1&2969&10^{-16}&
-1&1102&10^{-16}&
3&2277&10^{-17}&
\;-3&4704&10^{-17}\;&
\multicolumn{3}{c||}{\frac{\hphantom{MM}}{\hphantom{MM}}}\\
\cline{2-17}
&4&
1&9109&10^{-16}&
-1&1102&10^{-16}&
\;-7\!&4029&10^{-17}\;&
8&0934&10^{-17}&
\;-7&9519&10^{-18}\;\\
\hline
&2&
2&2566&10^{-16}&
2&2204&10^{-16}&
2&8435&10^{-17}&
\multicolumn{3}{c|}{\frac{\hphantom{MM}}{\hphantom{MM}}}&
\multicolumn{3}{c||}{\frac{\hphantom{MM}}{\hphantom{MM}}}\\
\cline{2-17}
\;1024\;&3&
1&1946&10^{-15}&
1&1102&10^{-15}&
-3&1168&10^{-16}&
1&0922&10^{-17}&
\multicolumn{3}{c||}{\frac{\hphantom{MM}}{\hphantom{MM}}}\\
\cline{2-17}
&4&
6&7237&10^{-16}&
-6&6613&10^{-16}&
-6&0996&10^{-17}&
1&2500&10^{-17}&
1&7279&10^{-17}\\
\hline
\hline
\end{array}}$
\end{center}\vspace{5pt}
\setlength{\abovecaptionskip}{0pt}
\caption{Werte der Fenster-AKF \mbox{$d(k)$}
f"ur \mbox{$k=\protect\Tilde{k}\CdoT M$} mit
\mbox{$\protect\Tilde{k}=0\;(1)\;N\!-\!1$}.}
\label{T6.1}
\rule{\textwidth}{0.5pt}\vspace{0pt}
\end{table}
F"ur \mbox{$N\!=\!1$} ergibt sich das Rechteckfenster, das die 
Bedingung (\ref{2.23}) f"ur \mbox{$k\!=\!0$} fehlerfrei erf"ullt, und 
das wegen seiner kurzen L"ange \mbox{$F\!=\!M$} keine weiteren Nullstellen 
zu haben braucht. Der Fall \mbox{$N\!=\!1$} ist daher in der Tabelle~\ref{T6.1} 
nicht aufgef"uhrt. Zum Vergleich sei noch die relative Genauigkeit des verwendeten 
Rechners mit 8-Byte-Gleitkomma-Zahlendarstellung angegeben: 
\mbox{$\varepsilon=2,\!22045\CdoT10^{-16}$}. Wie man sieht liegen alle Werte 
\mbox{$d(\Tilde{k}\CdoT M)$} in der gew"unschten Gr"o"senordnung.

\begin{figure}[btp]
\begin{center}
{ 
\begin{picture}(440,240)

\input{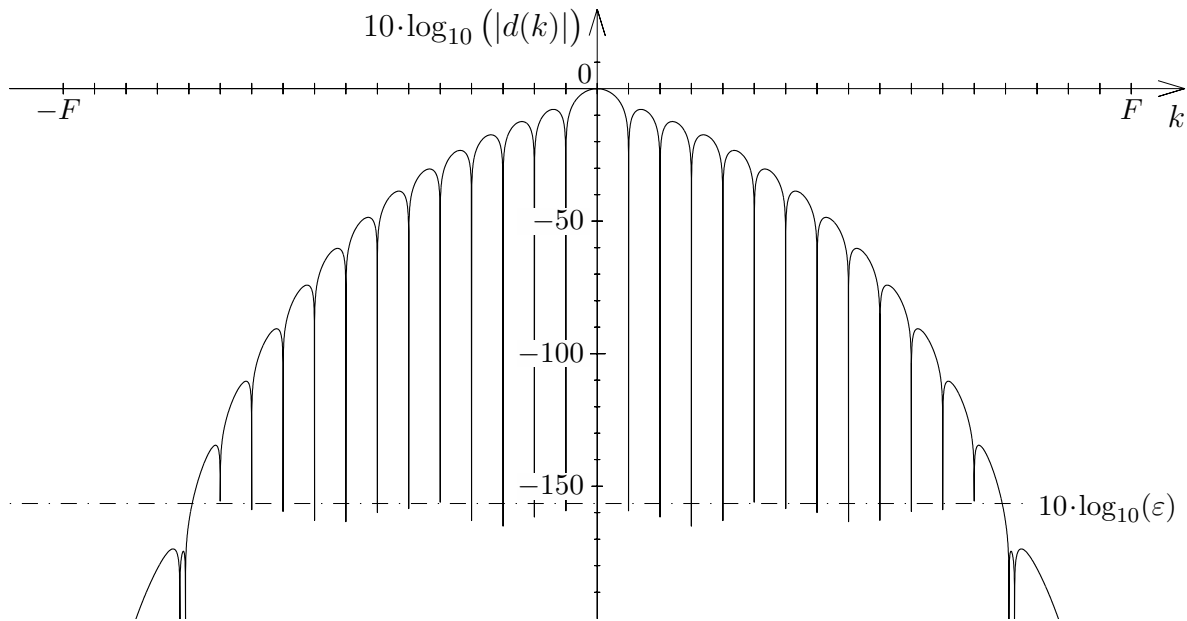}
\put(215,230){\makebox(0,0)[rt]{$10\CdoT\log_{10}\big(|d(k)|\big)$}}
\put(218,202){\makebox(0,0)[rb]{\small$0$}}
\put(20,196){\makebox(0,0)[t]{\small$\!\!-F$}}
\put(420,196){\makebox(0,0)[t]{\small$F$}}
\put(215,150){\makebox(0,0)[r]{\small$-50$}}
\put(215,100){\makebox(0,0)[r]{\small$-100$}}
\put(215,50){\makebox(0,0)[r]{\small$-150$}}
\put(440,194){\makebox(0,0)[tr]{$k$}}
\put(385,42){\makebox(0,0)[l]{\small$10\CdoT\log_{10}(\varepsilon)$}}

\end{picture}}
\end{center}\vspace{-10pt}
\setlength{\belowcaptionskip}{-6pt}
\caption{Betrag der Fenster-AKF \mbox{$d(k)$} am Beispiel
der Fensterfolge mit \mbox{$M\!=\!231$}, \mbox{$N\!=\!17$}.}
\label{b5m}
\rule{\textwidth}{0.5pt}\vspace{-6pt}
\end{figure}

F"ur das etwas ungew"ohnliche Beispiel mit \mbox{$M\!=\!231$} und 
\mbox{$N\!=\!17$} ist in Bild \ref{b5m}
der 
Betrag der Fenster-AKF halblogarithmisch dargestellt. 
Zum Vergleich dient die waagrecht eingetragene Linie bei dem Wert 
\mbox{$10\CdoT\log_{10}(\varepsilon)$}. Man erkennt deutlich die 
"aquidistante Lage der Nullstellen, und dass die Werte zu diesen 
Zeitpunkten selbst bei dem relativ gro"sen Wert \mbox{$N\!=\!17$} 
noch in der gew"unschten Gr"o"senordnung liegen.
 
Die Berechnung der Fenster-AKF erfolgte \mbox{dabei} durch 
Auswertung der die Faltung diskreter Folgen definierenden Summe 
(\,{\tt MATLAB: conv(f\_k,f\_k(F:-1:1))}\,). Eine schnelle Faltung 
mit Hilfe der FFT wurde {\em nicht}\/ angewandt, da hier die Qualit"at 
der Fensterfolge, und nicht die Qualit"at der FFT beurteilt werden soll. 
Es zeigte sich, dass es hier {\em nicht}\/ notwendig ist die Reihenfolge 
der Summanden in der Faltungssumme geschickt zu w"ahlen, um die Fehler 
bei der Berechnung der Summe klein zu halten. W"ahlt man die Reihenfolge 
der Summanden in der Art, dass sich jeweils die betraglich kleinste 
Zwischensumme ergibt, so erh"alt man Werte f"ur die Faltungssumme, die 
sich f"ur gro"se Werte der Fenster-AKF nur unwesentlich 
von den Werten unterscheiden, die man erh"alt, wenn man die Reihenfolge der 
Summanden nicht ver"andert. F"ur kleine Werte der Fenster-AKF 
liegen die Ergebnisse auch dann in derselben Gr"o"senordnung, wenn man die 
Reihenfolge der Summanden optimal w"ahlt. 
\newpage

\subsection{Pr"azision der "Uberlagerung der Betragsquadrate der Spektren der Fensterfolgen}

In Kapitel \ref{W} hatten wir festgestellt, dass die Bedingung 
(\ref{2.23}) "aquivalent zu der Bedingung (\ref{2.20}) ist, nach 
der die Summe der Betragsquadrate der verschobenen Spektren der 
Fensterfolge konstant ${\D M^2}$ sein soll. Als relativen Fehler 
dieser Summe erhalten wir mit 
\begin{equation}
\varepsilon_{(\ref{2.20})}(\Omega)\;=\;
\frac{\D\;\Sum{\mu=0}{M-1}\:
\big|\,F\big({\T\Omega\!-\!\mu\CdoT\frac{2\pi}{M}}\big)\,\big|^2-M^2\;}
{\D M^2}\;=
\Sum{\Tilde{k}=1-N}{N-1}d(\Tilde{k}\CdoT M)
\cdot e^{\!-j\cdot\Omega\cdot M\cdot\Tilde{k}}-1
\label{6.40}
\end{equation}
einen von $\Omega$ abh"angigen Term, der sich als die Fouriertransformierte 
der Abtastwerte der AKF des tats"achlich berechneten 
Fensters angeben l"asst. 

F"ur die Fensterfolge mit \mbox{$M\!=\!231$} und 
\mbox{$N\!=\!17$} ist in Bild \ref{b5n}
\begin{figure}[btp]
\begin{center}
{ 
\begin{picture}(454,160)(-4,0)

\input{mbild5n}
\put(43,162){\makebox(0,0)[rt]{$\varepsilon_{(\ref{2.20})}(\Omega)$}}
\put(47,141){\makebox(0,0)[r]{$10^{-15}$}}
\put(48,38){\makebox(0,0)[rt]{$0$}}
\put(140,36){\makebox(0,0)[t]{$\frac{\vphantom{M}\pi}{2\cdot M}$}}
\put(230,36){\makebox(0,0)[t]{$\frac{\vphantom{M}\pi}{M}$}}
\put(320,36){\makebox(0,0)[t]{$\frac{\vphantom{M}3\pi}{2\cdot M}$}}
\put(410,36){\makebox(0,0)[t]{$\frac{\vphantom{M}2\pi}{M}$}}
\put(450,35){\makebox(0,0)[rt]{$\Omega$}}

\end{picture}}
\end{center}\vspace{-26pt}
\setlength{\belowcaptionskip}{-8pt}
\caption{Relativer Fehler in Gleichung (\protect\ref{2.20}). Beispiel
mit \mbox{$M\!=\!231$} und \mbox{$N\!=\!17$}.}
\label{b5n}
\rule{\textwidth}{0.5pt}\vspace{-12pt}
\end{figure}
der Verlauf des 
relativen Fehlers exemplarisch "uber der Frequenz $\Omega$ dargestellt. 
Dabei ist jedoch zu ber"ucksichtigen, dass dieser Verlauf selbst 
nur sehr ungenau berechnet werden kann. Die Werte \mbox{$d(\Tilde{k}\CdoT M)$}, 
die in der letzten Gleichung als Koeffizienten der Fourierreihe auftreten, 
wurden dabei wieder mit Hilfe der Selbstfaltung aus der Fensterfolge berechnet, 
und sind daher mit zus"atzlichen Fehlern behaftet, die in der Gr"o"senordnung 
der Werte liegen, die sich bei idealer Berechnung der Faltungs\-operation aus 
den fehlerbehafteten Werten der Fensterfolge ergeben w"urden. 

Dennoch wurde aus zwei Gr"unden auf eine Darstellung dieses Bildes nicht 
verzichtet. Einerseits erkennt man, dass der relative Fehler 
prinzipiell mit \mbox{$2\pi/M$} periodisch und zu allen Vielfachen 
der Frequenz \mbox{$\pi/M$} geradesymmetrisch ist. Das liegt daran, dass $d(k)$ eine 
gerade reelle Folge ist, und somit beschreibt Gleichung (\ref{6.40}) eine Kosinusreihe 
in $\Omega$, die mit \mbox{$2\pi/M$} periodisch ist. Andererseits kann man feststellen, 
dass der relative Fehler in einer Gr"o"senordnung liegt, die die relative 
Rechnergenauigkeit \mbox{$\varepsilon=2,\!22045\cdoT10^{-16}$} nur unwesentlich 
"ubersteigt. 

Als mittleren quadratischen relativen Fehler erh"alt man 
\begin{equation}
\sigma\big(\varepsilon_{(\ref{2.20})}(\Omega)\big)^2\;=\;
\frac{1}{2\pi}\cdoT\Int{-\pi}{\pi}
\big|\varepsilon_{(\ref{2.20})}(\Omega)\big|^2\Cdot d\Omega\;=\;
\Sum{\Tilde{k}=1-N}{N-1}\big|d(\Tilde{k}\CdoT M)-\gamma_0(\Tilde{k})\big|^2,
\label{6.41}
\end{equation}
deren Wurzel in Tabelle \ref{T6.1} f"ur \mbox{$M\in\{4;64;1024\}$} und
\mbox{$N=2\;(1)\;4$} ebenfalls eingetragen ist. Auch diese Werte
k"onnen nur zur Absch"atzung der Gr"o"senordnung dienen, da sie
ebenfalls aus den fehlerbehafteten Werten
\mbox{$d(\Tilde{k}\CdoT M)$} berechnet werden.

Somit k"onnen wir feststellen, dass sich mit dem in 
diesem Kapitel angegeben Algorithmus Fensterfolgen berechnen lassen, 
die selbst f"ur unsinnig hohe Werte von $N$ die Bedingungen (\ref{2.20}) 
und (\ref{2.27}) mit nahezu der maximal m"oglichen Pr"azision
erf"ullen.

\subsection{Halbbandfilter}

Bereits in Kapitel \ref{W} hatten wir festgestellt, dass die FIR-Filter, 
die man erh"alt, wenn man die Werte der Fenster-AKF 
\mbox{$d(k)$} als Filterkoeffizienten verwendet, \mbox{$M$-tel}-Band-Filter 
sind. Im Allgemeinen wird man zur Konstruktion eines $M$-tel-Band-Filters 
nat"urlich nicht erst eine Fensterfolge berechnen, bei der die 
Phase der Spektralwerte "uber das Cepstrum berechnet werden muss, 
um dann durch die Selbstfaltung der Fensterfolge bei der Berechnung 
der Fenster-AKF die Phaseninformation wieder zu verlieren. 
Man wird sich normalerweise lediglich analog zu Gleichung (\ref{6.17}) und 
(\ref{6.18}) die $N$ Werte \mbox{$\big|F\big({\T\nu\CdoT\frac{\pi}{F}}\big)\big|^2$} 
mit \mbox{$\nu=0\;(1)\;F$} des Betragsquadrats des Spektrums der 
Fensterfolge und damit mittels einer diskreten Kosinustransformation 
die Fenster-AKF \mbox{$d(k)$} berechnen. Es gibt jedoch 
Anwendungen vor allem bei Nachrichten"ubertragungssystemen, die in der 
"Ubertragungskette zwei Filter enthalten, von denen man w"unscht, dass 
sie zueinander spiegelsymmetrische Koeffizienten aufweisen sollen, die 
derart gew"ahlt sein sollen, dass das Gesamtsystem die $M$-tel-Band-Eigenschaft 
erf"ullt. Dann kann es sinnvoll erscheinen, die Werte der hier vorgestellten 
Fensterfolge als Koeffizienten beider Filter zu verwenden. 

\begin{figure}[btp]
\begin{center}
{ 
\begin{picture}(450,295)

\input{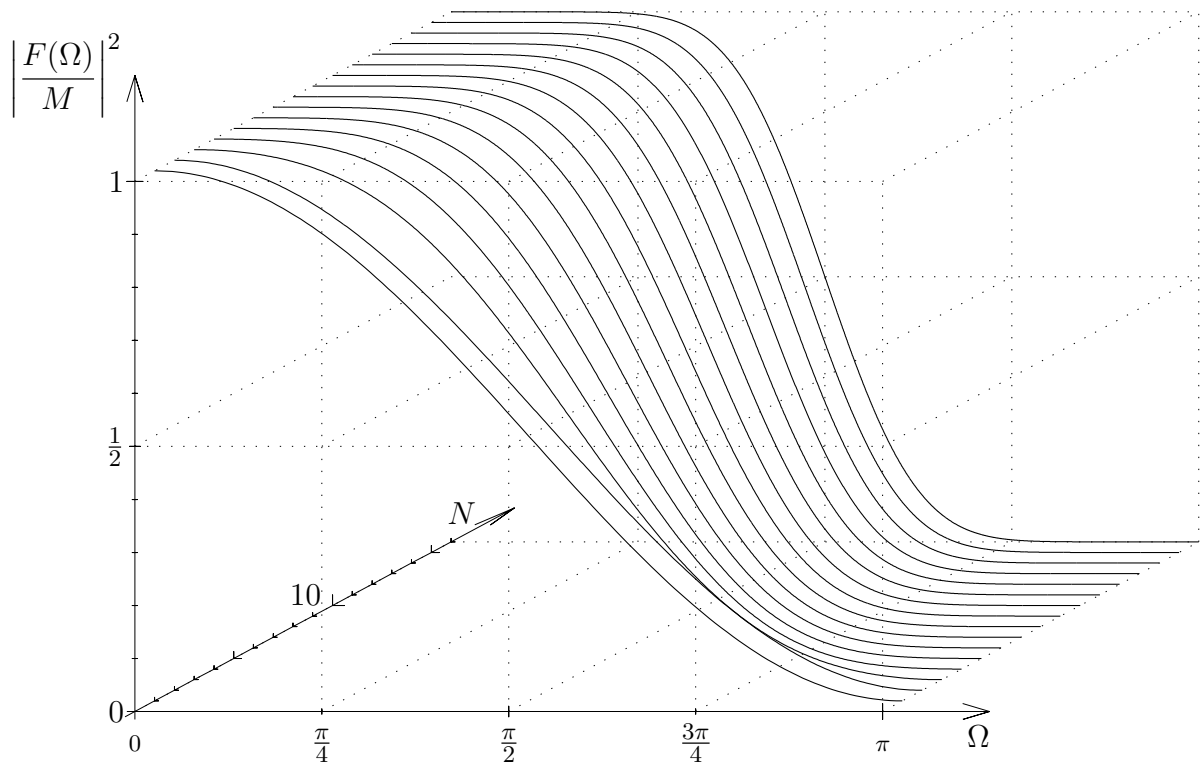}

\put(45,260){\makebox(0,0)[r]{${\D\bigg|\frac{F(\Omega)}{M}\bigg|^2}$}}
\put(46,220){\makebox(0,0)[r]{$1$}}
\put(46,120){\makebox(0,0)[r]{$\frac{1}{2}$}}
\put(46,20){\makebox(0,0)[r]{$0$}}
\put(50,16){\makebox(0,0)[t]{${\scriptstyle0}\vphantom{\frac{M}{M}}$}}
\put(120,16){\makebox(0,0)[t]{$\frac{\vphantom{M}\pi}{4\vphantom{M}}$}}
\put(190,16){\makebox(0,0)[t]{$\frac{\vphantom{M}\pi}{2\vphantom{M}}$}}
\put(260,16){\makebox(0,0)[t]{$\frac{\vphantom{M}3\pi}{4\vphantom{M}}$}}
\put(330,16){\makebox(0,0)[t]{${\scriptstyle\pi}\vphantom{\frac{M}{M}}$}}
\put(370,15){\makebox(0,0)[rt]{$\Omega$}}
\put(167,95){\makebox(0,0)[l]{$N$}}
\put(108,64){\makebox(0,0)[l]{$10$}}

\end{picture}}
\end{center}\vspace{-13pt}
\setlength{\belowcaptionskip}{-4pt}
\caption{Spektren der Autokorrelationsfolgen der Fenster mit \mbox{$M\!=\!2$} (\,Halbbandfilter\,).
\mbox{$N=1\;(1)\;16$} als Parameter.}
\label{b5q}
\rule{\textwidth}{0.5pt}\vspace{-8pt}
\end{figure}

Im Fall \mbox{$M\!=\!2$} handelt es sich bei \mbox{$d(k)$} um ein Halbbandfilter, 
da sich das Spektrum der Fenster-AKF mit dem um $\pi$ verschobenen 
Spektrum gerade zu einer Konstanten erg"anzt. Bild \ref{b5q}
zeigt 
das Spektrum des Halbbandfilters f"ur unterschiedliche Werte des Parameters $N$. 
Das Spektrum wurde dabei jeweils so normiert, dass sich bei \mbox{$\Omega\!=\!0$} 
der Wert Eins ergibt, wie dies bei Halbbandfiltern "ublich ist. 
Man erkennt, dass sich mit zunehmendem $N$ eine bessere Trennung 
der beiden Frequenzb"ander ergibt. 

Wenn man bedenkt, dass die letzten
\mbox{$N\!-\!1$} Werte der zugrunde liegenden Fensterfolge wenigstens 
theoretisch gleich Null sind, ergibt sich im Allgemeinen ein Halbbandfilter 
mit \mbox{$2\CdoT N\!+\!1$} von Null verschiedenen Koeffizienten. 
Eine Ausnahme ergibt sich bei \mbox{$N\!=\!2$}. Hier besteht die
zugrunde liegende Fensterfolge aus den Abtastwerten der Funk\-tion 
\mbox{$0,\!5\CdoT\big(1\!+\!\cos(x)\!+\!\sin(x)\big)$} mit \mbox{$x=k\CdoT\pi/2$}. 
Dies ist die Fensterfolge (\,1, 1, 0, 0\,), die gleich dem Rechteckfenster 
(\,1, 1\,) ist, das man mit \mbox{$N\!=\!1$} erh"alt. Daher sind auch die 
beiden Halbbandfilter f"ur \mbox{$N\!=\!1$} und \mbox{$N\!=\!2$} und somit 
auch deren Spektren identisch, wie man auch in Bild \ref{b5q} sehen kann.

\section{Andere Fensterfolgen}\label{Andere}

Au"ser dem in Kapitel \ref{Algo} beschriebenen Fenster sind noch 
einige andere Fensterfolgen auf Anwendbarkeit beim RKM untersucht 
worden. In folgenden werden diese Fensterfolgen kurz beschrieben, 
und festgestellt, ob oder unter welchen Umst"anden
\begin{itemize}
\item ihr Spektrum die nach Gleichung (\ref{2.27}) geforderte Nullstellenlage 
      aufweist,
\item ihre Fenster-AKF die gew"unschte M-tel-Band-Eigenschaft 
      nach Gleichung (\ref{2.23}) besitzt, so dass deren Spektrum die Gleichung 
      (\ref{2.20}) erf"ullt, und
\item mit welcher Potenz die Sperrd"ampfung f"ur Frequenzen deutlich 
      au"serhalb des\\ Durch\-lassbereichs des Spektrums asymptotisch 
      ansteigt, wenn die Fensterl"ange\\ gro"s genug gew"ahlt wird.
\end{itemize}
Dabei wird bei allen Fenstern ein nicht n"aher festgelegter 
konstanter Faktor $c_f$ verwendet, von dem angenommen wird, dass er 
so gew"ahlt sei, dass das Spektrum der Fensterfolge bei der 
Frequenz \mbox{$\Omega\!=\!0$} den geforderten Wert $M$ annimmt. 
Die meisten Fensterfolgen entstehen durch Abtastung einer zeitbegrenzten 
kontinuierlichen Fensterfunktion. Dabei werden zur Festlegung der 
kontinuierlichen Fensterfunktion und zur Beschreibung der Abtastung 
oft Parameter (\,meist sind dies $\alpha$ und $\Tilde{N}$\,) eingef"uhrt, 
deren Bedeutung bei den unterschiedlichen Fenstern verschieden sein 
kann.

{\bf Rechteck- oder Dirichlet-Fenster} (\,z.~B.~\cite{Kam}\,)\\
Dieses diskrete Fenster \mbox{$f(k)$} ist innerhalb des Intervalls 
\mbox{$[0;F)$} konstant $c_f$, und sonst Null. Wenn die Fensterl"ange $F$ ein 
ganzzahliges Vielfaches $N$ der RKM-DFT-L"ange $M$ ist und der konstante 
Faktor \mbox{$c_f\!=\!1/N$} ist, erf"ullt das Rechteckfenster die 
Nullstellenbedingung (\ref{2.27}). F"ur \mbox{$F\!\le\!M$} und 
\mbox{$c_f\!=\!\sqrt{M/F\,}$} ist die M-tel-Band-Eigenschaft (\ref{2.23}) 
der dreieckf"ormigen Fenster-AKF gegeben, weil diese 
auf das Intervall \mbox{$(-F;F)$} begrenzt ist, so dass Gleichung 
(\ref{2.23}) immer erf"ullt ist. F"ur \mbox{$F\!=\!M$} und 
\mbox{$c_f\!=\!1$} ist das Rechteckfenster also beim RKM einsetzbar. 
Dieses Fenster entspricht genau der inh"arenten Fensterung bei dem 
bisher "ublichen RKM "`ohne Fensterung"' und stellt den Spezialfall 
des Fensters nach Kapitel \ref{Algo} f"ur \mbox{$N\!=\!1$} dar. Die 
Sperrd"ampfung steigt linear mit \mbox{$\sin(\Omega/2)$} an.

{\bf Dreieck-, Frej\'{e}r- oder Bartlett-Fenster} (\,z.~B.~\cite{Kam}\,)\\
Dieses diskrete Fenster entsteht durch Abtastung einer kontinuierlichen 
Dreieckfunktion, die z.~B. durch die Faltung einer Rechteckfunktion 
mit sich selbst entsteht.
\[
f(t)\;=\;\begin{cases}
c_f\CdoT\big(1\!-\!|t|\,\big)\qquad&\text{ f"ur }\quad |t|\!<\!1\\
0&\text{ sonst}
\end{cases}
\]
Wenn diese Dreiecksfunktion mit einer Abtastfrequenz abgetastet wird, 
die ein ganzzahliges Vielfaches (\,$=N/2$\,) von $M$ ist, und wenn ein 
passender Normierungsfaktor $c_f$ eingestellt wird, ergibt sich eine 
Fensterl"ange von \mbox{$F\!=\!N\CdoT M$}, und die Fensterfolge erf"ullt die 
Nullstellenbedingung (\ref{2.27}). Die M-tel-Band-Eigenschaft (\ref{2.23}) 
der Fenster-AKF ist bei dieser Abtastung nicht gegeben. 
Die Sperrd"ampfung steigt quadratisch mit \mbox{$\sin(\Omega/2)$} an.

{\bf Parzen-, de la Vall\'{e}-Poussin- oder Jackson-Fenster}
(\,z.~B.~\cite{Thomae}\,)\\
Dieses diskrete Fenster entsteht durch Abtastung einer kontinuierlichen 
Funktion, die durch die Faltung des kontinuierlichen Dreieckfensters mit 
sich selbst, bzw. durch dreimalige Faltung eines kontinuierlichen 
Rechteckfensters mit sich selbst entsteht. 
\[
f(t)\;=\;c_f\cdot\begin{cases}
1-\frac{3}{4}\CdoT{\D t^2}\!\CdoT\big(2\!-\!|t|\,\big)\qquad&
\text{ f"ur }\quad |t|\!<\!1\\
\frac{1}{4}\CdoT{\D\big(2\!-\!|t|\,\big)^3}\qquad&
\text{ f"ur }\quad 1\le|t|<2\\
0&\text{ sonst}
\end{cases}
\]
Wenn diese Funktion mit einer Abtastfrequenz abgetastet wird, die ein 
ganzzahliges Vielfaches (\,$=N/4$\,) von $M$ ist, und wenn ein passender 
Normierungsfaktor $c_f$ eingestellt wird, ergibt sich eine 
Fensterl"ange von \mbox{$F\!=\!N\CdoT M$}, und die Fensterfolge erf"ullt die 
Nullstellenbedingung (\ref{2.27}). Die M-tel-Band-Eigenschaft (\ref{2.23}) 
der Fenster-AKF ist bei dieser Abtastung nicht gegeben. Die 
Sperrd"ampfung steigt mit der vierten Potenz in \mbox{$\sin(\Omega/2)$} an.

{\bf Hamming-Fenster} (\,z.~B.~\cite{Kam}\,)\\
Dieses diskrete Fenster entsteht durch Abtastung des kontinuierlichen 
Hamming-Fensters:
\[
f(t)\;=\;\begin{cases}
c_f\cdot\big(\,0,\!54 + 0,\!46 \CdoT\cos(2\pi\CdoT t)\,\big)\qquad&
\text{ f"ur }\quad |t|\!<\!\frac{1}{2}\\
0&\text{ sonst.}
\end{cases}
\]
Wenn diese Funktion mit einer Abtastfrequenz abgetastet wird, die 
ein ganzzahliges Vielfaches (\,$=\!N$\,) von $M$ mit \mbox{$N\!\ge\!2$} ist, 
und wenn ein passender Normierungsfaktor $c_f$ eingestellt wird, 
ergibt sich eine Fensterl"ange von \mbox{$F\!=\!N\CdoT M$}, und 
die Fensterfolge erf"ullt die Nullstellenbedingung (\ref{2.27}). 
Die M-tel-Band-Eigenschaft (\ref{2.23}) der Fenster-AKF 
ist bei dieser Abtastung nicht gegeben. Die Sperrd"ampfung steigt 
asymptotisch linear mit \mbox{$\sin(\Omega/2)$} an.

{\bf Hann-Fenster} (\,z.~B.~\cite{Kam}\,)\\
Dieses diskrete Fenster entsteht durch Abtastung des kontinuierlichen 
Hann-Fensters:
\[
f(t)\;=\;\begin{cases}
c_f\cdot\big(\,1 + \cos(2\pi\CdoT t)\,\big)\qquad&
\text{ f"ur }\quad |t|\!<\!\frac{1}{2}\\
0&\text{ sonst.}
\end{cases}
\]
Wenn diese Funktion mit einer Abtastfrequenz abgetastet wird, die ein 
ganzzahliges Vielfaches (\,$=\!N$\,) von $M$ mit \mbox{$N\!\ge\!2$} 
ist, und wenn ein passender Normierungsfaktor $c_f$ eingestellt wird, 
ergibt sich eine Fensterl"ange von \mbox{$F\!=\!N\CdoT M$}, und 
die Fensterfolge erf"ullt die Nullstellenbedingung (\ref{2.27}). 
Die M-tel-Band-Eigenschaft (\ref{2.23}) der Fenster-AKF 
ist bei dieser Abtastung nicht gegeben. Die Sperrd"ampfung steigt 
asymptotisch mit der dritten Potenz in \mbox{$\sin(\Omega/2)$} an.

{\bf Blackman-Fenster} (\,z.~B.~\cite{Harris}\,)\\
Dieses diskrete Fenster entsteht durch Abtastung des kontinuierlichen 
Blackman-Fensters. Dabei handelt es sich um ein Fenster, das sich als 
Kosinusreihe
\begin{equation}
f(t)\;=\;\begin{cases}
c_f\cdot\bigg(c_0+\!
\Sum{\nu=1}{\Tilde{N}-1}c_{\nu}\CdoT2\CdoT\cos(\nu\CdoT2\pi\CdoT t)\bigg)
\qquad&\text{ f"ur }\quad |t|\!<\!\frac{1}{2}\\
0&\text{ sonst}
\end{cases}
\label{6.42}
\end{equation}
mit \mbox{$\Tilde{N}\!=\!3$} Gliedern darstellen l"asst. 
Es gibt zwei verschiedene Versionen des Blackman-Fensters, 
die sich in den Reihenkoeffizienten unterscheiden. 

In \cite{Harris} wird das sog. "`exakte"' Blackman-Fenster 
mit den Koeffizienten \mbox{$c_0\!=\!7938$}, \mbox{$c_1\!=\!4620$} 
und \mbox{$c_2\!=\!715$} angegeben. Zur besseren D"ampfung der 
ersten Nebenmaxima werden bei diesem Fenster die Koeffizienten 
so gew"ahlt, dass sich im Spektrum der kontinuierlichen Fensterfunktion 
bei den Frequenzen \mbox{$\omega\!=\!7\pi$} und \mbox{$\omega\!=\!9\pi$} 
Nullstellen ergeben. Damit steigt die Sperrd"ampfung asymptotisch 
linear mit \mbox{$\omega$} an. 

Werden die Koeffizienten unter 
der Nebenbedingung \mbox{$c_0\!+\!2\CdoT c_2\!=\!2\CdoT c_1$} gerundet, 
so erh"alt man das "`"ubliche"' Blackman-Fenster, dessen Koeffizienten 
\mbox{$c_0\!=\!42$}, \mbox{$c_1\!=\!25$} und \mbox{$c_2\!=\!4$} 
sind, und dessen Sperrd"ampfung asymptotisch mit der dritten Potenz 
in \mbox{$\omega$} ansteigt. 

Wenn diese Funktion mit einer Abtastfrequenz 
abgetastet wird, die ein ganzzahliges Vielfaches (\,$=\!N$\,) von $M$ mit 
\mbox{$N\!\ge\!3$} ist, und wenn ein passender Normierungsfaktor $c_f$ 
eingestellt wird, ergibt sich eine Fensterl"ange von \mbox{$F\!=\!N\CdoT M$}, 
und die Fensterfolge erf"ullt unabh"angig von der Wahl der Koeffizienten die 
Nullstellenbedingung (\ref{2.27}). Die M-tel-Band-Eigenschaft (\ref{2.23}) der 
Fenster-AKF ist bei dieser Abtastung bei keiner der beiden 
Versionen des Blackman-Fensters gegeben. F"ur hinreichend gro"se Werte von 
$M$, wie sie f"ur das RKM typisch sind, ergibt sich ein asymptotischer Anstieg 
der Sperrd"ampfung mit derselben Potenz in \mbox{$\sin(\Omega/2)$}, wie er 
sich bei dem Spektrum der kontinuierlichen Fensterfunktion in $\omega$ ergibt.

{\bf Harris-Fenster} (\,\cite{Harris}\,)\\
Dieses diskrete Fenster entsteht durch Abtastung des kontinuierlichen 
Harris-Fensters. Es handelt sich dabei um ein Kosinusreihenfenster 
nach Gleichung (\ref{6.42}), das das Kaiser-$I_0$-Fenster approximiert. 
Die Anzahl $\Tilde{N}$ der Glieder der Kosinusreihe ist drei oder vier. 
Die Koeffizienten berechnen sich mit einem w"ahlbaren Parameter $\alpha$, 
der in die Bandbreite des Fensterspektrums und die Sperrd"ampfung 
eingeht, und f"ur den ein Bereich von \mbox{$\Tilde{N}\!-\!1$} 
bis vier angegeben wird. Die Koeffizienten sind dann:
\[
c_{\nu}\;=\;\text{si}\big(j\CdoT\pi\CdoT\sqrt{\alpha^2-\nu^2\,}\,\big)
\qquad\qquad\text{ f"ur }\qquad \nu\!=\!0\;(1)\;\Tilde{N}\!-\!1.
\]
Wenn die endliche Kosinusreihe mit einer Abtastfrequenz abgetastet wird, die 
ein ganzzahliges Vielfaches (\,$=\!N$\,) von $M$ mit \mbox{$N\!\ge\!\Tilde{N}$} 
ist, und wenn ein passender Normierungsfaktor $c_f$ eingestellt wird, 
ergibt sich eine Fensterl"ange von \mbox{$F\!=\!N\CdoT M$}, und die 
Fensterfolge erf"ullt die Nullstellenbedingung (\ref{2.27}). 
Bei dieser Abtastung wird die Konstanz der nach Gleichung (\ref{2.20}) 
"uberlagerten, verschobenen Betragsquadratspektren nicht erzielt. 
Dies ist auch nicht anders zu erwarten, da beim Entwurf dieser Fensterfolge 
die Erf"ullung der Gleichung (\ref{2.20}) keine Randbedingung ist. 
Die Sperrd"ampfung steigt asymptotisch linear mit \mbox{$\sin(\Omega/2)$} an. 

Nach \cite{Thomae} soll es auch eine Kosinusreihenapproximation 
eines modifizierten Kaiser-$I_0$-Fensters geben, dessen Sperrd"ampfung 
mit der dritten Potenz in \mbox{$\sin(\Omega/2)$} ansteigt. Dieses Fenster 
wurde nicht auf seine Anwendbarkeit beim RKM "uberpr"uft.

{\bf Prabhu-Fenster} (\,\cite{Thomae}\,)\\
Dieses diskrete Fenster entsteht durch Abtastung des kontinuierlichen 
Prabhu-Fensters. Es handelt sich dabei um ein Kosinusreihenfenster 
nach Gleichung (\ref{6.42}) mit \mbox{$\Tilde{N}\!=\!3$}. 
Die ersten beiden Koeffizienten sind mit \mbox{$c_0\!=\!2$} und 
\mbox{$c_1\!=\!1$} fest, w"ahrend man den Koeffizienten $c_2$ 
frei w"ahlt. Nach \cite{Thomae} ergibt sich eine Frequenzcharakteristik, 
die f"ur den typischen Bereich von $0,\!0002$ bis $0,\!03$ 
der Familie der Kaiser-Fenster sehr "ahnlich sein soll. 

Wenn die 
endliche Kosinusreihe mit einer Abtastfrequenz abgetastet wird, die ein 
ganzzahliges Vielfaches (\,$=\!N$\,) von $M$ mit \mbox{$N\!\ge\!3$} ist,
und wenn ein passender Normierungsfaktor $c_f$ eingestellt wird, 
ergibt sich eine Fensterl"ange von \mbox{$F\!=\!N\CdoT M$}, und die 
Fensterfolge erf"ullt die Nullstellenbedingung (\ref{2.27}). 

Im Fall, dass die Abtastfrequenz \mbox{$3\CdoT M$} ist, und 
dass f"ur den frei w"ahlbaren Koeffizienten \mbox{$c_2\!=\!\pm\sqrt{3\,}$} eingestellt 
wird, wird die Bedingung (\ref{2.20}) f"ur ganzzahlige Vielfache der Frequenz
\mbox{$\Omega=2\pi/(3\CdoT M)$} erf"ullt. Einerseits gilt dann
\mbox{${\D c_0^2=c_1^2+c_2^2}$}, und andererseits "uberlagert sich
in Gleichung (\ref{2.20}) bei diesen Frequenzen $\Omega$ das Quadrat
des Spektralwertes, der dem Kosinusreihenkoeffizienten $c_1$ entspricht,
mit dem Quadrat des Spektralwertes, der dem Kosinusreihenkoeffizienten
$c_2$ entspricht, zu dem dann gleichen Quadrat des Spektralwertes bei
\mbox{$\Omega\!=\!0$}, der dem Kosinusreihenkoeffizienten $c_0$ entspricht. 

F"ur Frequenzen die nicht im Raster 
\mbox{$2\pi/(3\CdoT M)$} liegen, ergeben sich in Gleichung (\ref{2.20})
jedoch erhebliche Abweichungen. So liegt der relative Fehler bei der
Frequenz \mbox{$\Omega\!=\!\pi/M$} etwa bei $60\%$. W"ahlt man die
Abtastfrequenz als ein h"oheres ganzzahliges Vielfaches von $M$, oder
einen anderen Kosinusreihenkoeffizienten $c_2$, so werden die
Abweichungen in Gleichung (\ref{2.20}) in der Regel eher noch gr"o"ser.

W"ahlt man \mbox{$c_2\!\neq\!0$} so steigt die Sperrd"ampfung asymptotisch
linear mit \mbox{$\sin(\Omega/2)$} an. F"ur \mbox{$c_2\!=\!0$} ergibt
sich das Hann-Fenster, dessen Sperrd"ampfung asymptotisch mit der dritten
Potenz in \mbox{$\sin(\Omega/2)$} ansteigt.

{\bf Taylor-Fenster} (\,\cite{Thomae}\,)\\
Dieses diskrete Fenster entsteht durch Abtastung des kontinuierlichen 
Taylor-Fensters. Es handelt sich dabei um ein Kosinusreihenfenster 
nach Gleichung (\ref{6.42}), das ein Tsche"-by"-scheff-Fenster approximiert. 
In \cite{Thomae} wird angegeben, wie sich die Koeffizienten \mbox{$c_{\nu}$} 
berechnen lassen. Wieviele Koeffizienten $\Tilde{N}$ f"ur eine brauchbare 
Approximation ben"otigt werden, h"angt davon ab, welche Sperrd"ampfung 
man zu erzielen w"unscht. 

Wenn die abgebrochene Kosinusreihe mit einer 
Abtastfrequenz abgetastet wird, die ein ganzzahliges Vielfaches (\,$=\!N$\,) 
von $M$ mit \mbox{$N\!\ge\!\Tilde{N}$} ist, und wenn ein passender 
Normierungsfaktor $c_f$ eingestellt wird, ergibt sich eine Fensterl"ange 
von \mbox{$F\!=\!N\CdoT M$}, und die Fensterfolge erf"ullt die 
Nullstellenbedingung (\ref{2.27}). Bei solch einer Abtastung sinkt aber 
die "Uberlagerung der verschobenen Betragsquadratspektren in Gleichung 
(\ref{2.20}) im Bereich der Frequenzen um \mbox{$\Omega\!=\!\pi/M$} 
bereits in die Gr"o"senordnung des Betragsquadrats des Spektrums 
im Sperrbereich ab, so dass die Bedingung (\ref{2.20}) "uberhaupt nicht 
erf"ullt wird. 

Bei einer Abtastung mit einer Abtastfrequenz 
\mbox{$\le\!M$} wird die Bedingung (\ref{2.20}) erf"ullt. Da dann aber 
die Nullstellenbedingung (\ref{2.27}) auf gr"oblichste verletzt wird, 
ist ein Einsatz des sich so ergebenden Fensters beim RKM unm"oglich. 

F"ur relativ kleine Werte von \mbox{$F/\Tilde{N}$} unterhalb von etwa $10$ 
ergibt sich eine ganz gute N"aherung f"ur das typische Tschebyscheff 
Verhalten im Sperrbereich mit einer konstanten Sperrd"ampfung. 
Wird die Abtastung jedoch so gew"ahlt, dass sich eine Fensterl"ange $F$ 
ergibt, die um deutlich mehr als das zehnfache gr"o"ser ist als 
die Anzahl $\Tilde{N}$ der Kosinusreihenkoeffizienten --- was 
beim RKM typischerweise der Fall ist ---, so hat die Erfahrung gezeigt, 
dass die Sperrd"ampfung dann asymptotisch linear mit \mbox{$\sin(\Omega/2)$} 
ansteigt.

{\bf Ge\c{c}kinli-Yavuz-Fenster} (\,\cite{Geck}\,)\\
Dieses diskrete Fenster entsteht durch Abtastung des kontinuierlichen, zu 
\mbox{$t\!=\!0$} symmetrischen Ge\c{c}kinli-Yavuz-Fensters. Es handelt sich 
dabei um ein Kosinusreihenfenster nach Gleichung (\ref{6.42}), das alle Momente 
\mbox{${\T\int_{-\infty}^{\infty}|\omega^n\Cdot F(\omega)|\Cdot d\omega}$} 
des Spektrums unter einigen Randbedingungen minimiert. 

Die erste Randbedingung ist, dass die Fensterfunktion f"ur \mbox{$t\!=\!0$} 
Eins ist, und dass der Betrag des Fensters nur innerhalb eines 
symmetrischen Bereiches $\pm\Delta T$ um den Nullpunkt gr"o"ser als 
eine vorgebbare Schranke $\epsilon$ sein darf. Die zweite Randbedingung ist, 
dass der Wert der kontinuierlichen Fensterfunktion am Anfang und Ende Null ist. 
Drittens sollen alle Kosinusreihenkoeffizienten gr"o"ser oder gleich Null 
sein. 

Als L"osung ergibt sich ein Tschebyscheff-Polynom 
\mbox{$\epsilon\CdoT T_{\Tilde{N}-1}(a\CdoT\cos(2\pi\CdoT t)+b)$} 
vom Grad \mbox{$\Tilde{N}\!-\!1$}, dessen Argument eine linear gestreckte 
und verschobene Kosinusfunktion ist. Durch die lineare Skalierung des 
Arguments wird einerseits erreicht, dass das Minimum des Arguments (\,$b\!-\!a$\,)
genau in der kleinsten Nullstelle des Tschebyscheff-Polynoms liegt. 
Andererseits wird das Maximum des Arguments (\,$b\!+\!a$\,) so eingestellt, 
dass sich dort f"ur den Wert des Tschebyscheff-Polynoms $1/\epsilon$ ergibt. 
Man erh"alt so eine Fensterfunktion, die au"serhalb der Nullpunktsumgebung 
$\pm\Delta T$ zwischen $\pm\epsilon$ oszilliert, und die am Anfang und Ende 
doppelte Nullstellen aufweist. 

Wenn man $\epsilon$ vorgibt, ergibt sich 
abh"angig von der Anzahl $\Tilde{N}$ der Glieder der Kosinusreihe ein Wert 
f"ur die Breite \mbox{$2\CdoT\Delta T$} der Nullpunktsumgebung innerhalb derer 
die Fensterfunktion gr"o"ser $\epsilon$ ist. F"ur \mbox{$\epsilon\!=\!0.001$} 
sind die Reihenkoeffizienten in \cite{Geck} f"ur die Werte 
\mbox{$\Tilde{N}\!=\!2\;(1)\;14$} tabelliert, und es wird dort der 
sich jeweils ergebende Wert $\Delta T$ relativ zur Fensterl"ange angegeben. 

Wenn die endliche Kosinusreihe mit einer Abtastfrequenz abgetastet wird, 
die ein ganzzahliges Vielfaches (\,$=\!N$\,) von $M$ mit \mbox{$N\!\ge\!\Tilde{N}$} 
ist, und wenn ein passender Normierungsfaktor $c_f$ eingestellt wird, 
ergibt sich eine Fensterl"ange von \mbox{$F\!=\!N\CdoT M$}, und die
Fensterfolge erf"ullt die Nullstellenbedingung (\ref{2.27}). 
Bei solch einer Abtastung wird die Konstanz der nach Gleichung (\ref{2.20}) 
"uberlagerten, verschobenen Betragsquadratspektren bei keinem 
Ge\c{c}kinli-Yavuz-Fenster mit \mbox{$\epsilon\!=\!0.001$} erreicht. 
Je kleiner der Wert von $\epsilon$ wird, desto gr"o"ser wird der Zeitbereich 
\mbox{$|t|\!<\!\Delta T$}, in dem die kontinuierliche Fensterfunktion stets 
positiv ist. Wenn $\epsilon$ hinreichend klein ist (\,f"ur 
\mbox{$\epsilon\!<\!0.001$} trifft dies zu\,), setzt sich 
der Wert der Fenster-AKF $d(k)$ an der Stelle 
\mbox{$k\!=\!M$} bei einer mit der Nullstellenbedingung (\ref{2.27}) 
kompatiblen Abtastung aus zwei Anteilen zusammen. Der eine Anteil 
besteht aus den Produkten, die wenigstens jeweils einen Abtastwert 
der Fensterfunktion enthalten, der im Intervall \mbox{$|t|\!\ge\!\Delta T$} 
liegt. Diese Produkte sind sehr klein, wenn $\epsilon$ klein ist. 
Der andere Anteil enth"alt nur die Produkte zweier Abtastwerte, 
die beide dem Zeitintervall \mbox{$|t|\!<\!\Delta T$} entnommen sind. 
Da die Fensterfunktion sehr rasch anw"achst, wenn man sich von 
\mbox{$|t|\!=\!\Delta T$} zu \mbox{$t\!=\!0$} bewegt, wird der zweite 
Anteil der Fenster-AKF $d(k)$ an der Stelle 
\mbox{$k\!=\!M$} f"ur kleine Werte von $\epsilon$ deutlich 
"uberwiegen. Da dieser Anteil nur positive Produkte enth"alt, 
wird $d(M)$ stets positiv sein, so dass die Forderung 
(\ref{2.23}), und somit auch die Konstanz der nach Gleichung (\ref{2.20})
"uberlagerten, verschobenen Betragsquadratspektren, nicht erf"ullt 
sein kann. F"ur Werte von $\epsilon$ in der Gr"o"senordnung von Eins, konnte
weder ein Beweis daf"ur gefunden werden, dass die Bedingung (\ref{2.20}) 
nicht erf"ullt sein kann, noch ein Fall bei dem diese Bedingung erf"ullt 
ist. 

Durch die doppelten Nullstellen im Zeitbereich am Anfang und Ende 
der kontinuierlichen Fensterfunktion ergibt sich nach der Abtastung 
eine Fensterfolge, deren Sperrd"ampfung bei gro"ser Fensterl"ange 
asymptotisch mit der dritten Potenz in \mbox{$\sin(\Omega/2)$} ansteigt.

{\bf Babi\'{c}-Temes-Fenster} (\,\cite{Thomae}\,)\\
Hierbei handelt es sich um ein diskretes 
Kosinusreihen-Fenster nach Gleichung (\ref{6.42}). Die Koeffizienten 
\mbox{$c_{\nu}$} gewinnt man abh"angig von der gew"unschten Fensterl"ange 
$F$, die bei diesem Fenster als gerade angenommen wird, und der gew"unschten 
Anzahl $\Tilde{N}$ der Reihenkoeffizienten durch Interpolation aus einer 
Tabelle, in der f"ur einige Werte der Fensterl"ange die Koeffizienten als 
St"utzstellen angegeben sind. F"ur die Anzahl \mbox{$\Tilde{N}=2\;(1)\;5$} 
der Reihenkoeffizienten und die Fensterl"angen\footnote{F"ur \mbox{$F\!\ge\!64$} 
wird in \cite{Thomae} empfohlen, die Werte der St"utzstellen f"ur \mbox{$F\!=\!64$} 
zu verwenden.} \mbox{$F\in\{16;32;64\}$} ist eine derartige St"utzstellentabelle 
in \cite{Thomae} angegeben. 

Die damit gewonnenen Fensterfolgen maximieren --- wie 
das Kaiser-Fenster bei einer kontinuierlichen Fensterfunktion --- die auf die 
Gesamtenergie bezogene Energie des Spektrums innerhalb des Durchlassbereichs 
\mbox{$|\Omega|\!<\!\Omega_g$}, wobei je nach gew"unschter Grenzfrequenz 
$\Omega_g$ eine andere St"utzstellentabelle zu verwenden ist. In 
\cite{Thomae} sind die zwei Tabellen f"ur die beiden Grenzfrequenzen
\mbox{$\Omega_g\!=\!2\pi/F$} und \mbox{$\Omega_g\!=\!4\pi/F$}
enthalten. 

Wenn eine Fensterl"ange $F$ verwendet wird, die ein
ganzzahliges Vielfaches (\,$=\!N$\,) von $M$ mit \mbox{$N\!\ge\!\Tilde{N}$} 
ist, und wenn ein passender Normierungsfaktor $c_f$ eingestellt wird, 
erf"ullt die Fensterfolge die Nullstellenbedingung (\ref{2.27}). 
Bei solch einer Abtastung wird die Konstanz der nach Gleichung (\ref{2.20}) 
"uberlagerten, verschobenen Betragsquadratspektren bei dieser Fensterfolge 
nicht erreicht. Je kleiner n"amlich die Grenzfrequenz $\Omega_g$ im Vergleich 
zu \mbox{$2\pi/M$} wird, desto kleiner wird die Summe in Gleichung (\ref{2.20}) 
bei der Frequenz \mbox{$\Omega\!=\!\pi/M$}, w"ahrend sie bei geeigneter
Fensterl"ange \mbox{$F\!=\!N\CdoT M$} bei der Frequenz \mbox{$\Omega\!=\!0$}
unabh"angig von der Grenzfrequenz $\Omega_g$ ist. Die Sperrd"ampfung 
steigt asymptotisch linear mit \mbox{$\sin(\Omega/2)$} an.

{\bf Nuttall-Fenster} (\,\cite{Thomae}\,)\\
Dieses diskrete Fenster entsteht durch Abtastung des kontinuierlichen 
Nuttall-Fensters. Es handelt sich dabei um ein Kosinusreihenfenster 
nach Gleichung (\ref{6.42}), bei dem die Koeffizienten der Kosinusreihe 
von der gew"unschten Optimierung abh"angen. Entweder kann dabei ein 
Anstieg der Sperrd"ampfung mit einer m"oglichst hohen Potenz in 
$\omega$ erreicht werden, oder andererseits eine m"oglichst hohe 
D"ampfung der Nebenmaxima des Spektrums der Fensterfunktion. F"ur 
\mbox{$\Tilde{N}\!\ge\!3$} kann auch ein Kompromiss aus beiden 
Optimierungszielen verwirklicht werden. F"ur \mbox{$\Tilde{N}=2\;(1)\;4$} 
sind die Kosinusreihenkoeffizienten f"ur die unterschiedlichen 
Arten der Optimierung in \cite{Thomae} tabelliert. 

Wenn die endliche 
Kosinusreihe mit einer Abtastfrequenz abgetastet wird, die ein 
ganzzahliges Vielfaches (\,$=\!N$\,) von $M$ mit \mbox{$N\!\ge\!\Tilde{N}$} 
ist, und wenn ein passender Normierungsfaktor $c_f$ eingestellt wird, 
ergibt sich eine Fensterl"ange von \mbox{$F\!=\!N\CdoT M$}, und die 
Fensterfolge erf"ullt die Nullstellenbedingung (\ref{2.27}). 
Unabh"angig von der Art der Optimierung, ist bei solch einer Abtastung 
die Konstanz der nach Gleichung (\ref{2.20}) "uberlagerten, verschobenen 
Betragsquadratspektren bei \mbox{keinem} Nuttall-Fenster gegeben. 
F"ur hinreichend gro"se Werte von $M$, wie sie f"ur das RKM typisch 
sind, ergibt sich ein asymptotischer Anstieg der Sperrd"ampfung mit 
derselben Potenz in \mbox{$\sin(\Omega/2)$}, wie er bei dem 
Spektrum der kontinuierlichen Fensterfunktion in $\omega$ 
festzustellen ist.

{\bf Flat-Top-Fenster} (\,\cite{Thomae}\,)\\
Dieses diskrete Fenster entsteht durch Abtastung eines kontinuierlichen 
Flat-Top-Fensters. Es handelt sich dabei um ein Kosinusreihenfenster 
nach Gleichung (\ref{6.42}), bei dem das Spektrum der Fensterfunktion im 
Durchlassbereich \mbox{$|\omega|\!<\!\pi/2$} genau einen "Uberschwinger 
aufweist. Dieser "Uberschwinger wird dabei so gew"ahlt, dass der Wert 
des Maximums um genausoviel gr"o"ser ist als der Wert des Spektrums 
bei der Frequenz \mbox{$\omega\!=\!0$}, wie der Wert des Spektrums 
bei der Frequenz \mbox{$\omega\!=\!\pi/2$} kleiner ist. In Abh"angigkeit 
von der gew"ahlten Anzahl $\Tilde{N}$ der Glieder der Kosinusreihe 
bleiben einige Freiheitsgrade bei der Wahl der Kosinusreihenkoeffizienten 
unbestimmt, die man dazu nutzen kann, entweder einen Anstieg der 
Sperrd"ampfung mit einer m"oglichst hohen Potenz in $\omega$, oder 
eine m"oglichst hohe D"ampfung der Nebenmaxima des Spektrums der 
Fensterfunktion zu erreichen. Auch ein Kompromiss aus diesen beiden 
Zielen ist m"oglich. F"ur \mbox{$\Tilde{N}=3\;(1)\;5$} sind die 
Kosinusreihenkoeffizienten f"ur die unterschiedlichen Arten der 
Optimierung in \cite{Thomae} tabelliert. 

Wenn die endliche Kosinusreihe 
mit einer Abtastfrequenz abgetastet wird, die ein ganzzahliges Vielfaches 
(\,$=\!N$\,) von $M$ mit \mbox{$N\!\ge\!\Tilde{N}$} ist, und wenn ein 
passender Normierungsfaktor $c_f$ eingestellt wird, ergibt sich eine 
Fensterl"ange von \mbox{$F\!=\!N\CdoT M$}, und die Fensterfolge erf"ullt 
die Nullstellenbedingung (\ref{2.27}). Unabh"angig von der Art der Optimierung, 
ist bei solch einer Abtastung die Konstanz der nach Gleichung (\ref{2.20}) 
"uberlagerten, verschobenen Betragsquadratspektren bei keinem Flat-Top-Fenster 
gegeben. F"ur hinreichend gro"se Werte von $M$, wie sie f"ur das RKM typisch 
sind, erh"alt man einen asymptotischen Anstieg der Sperrd"ampfung mit 
derselben Potenz in \mbox{$\sin(\Omega/2)$}, wie er sich bei dem 
Spektrum der kontinuierlichen Fensterfunktion in $\omega$ ergibt. 
Es sei noch darauf hingewiesen, dass der Begriff Flat-Top-Fenster 
auch f"ur das Tukey-Fenster verwendet wird, bei dem die Fensterfunktion 
im mittleren Zeitbereich einen konstanten Maximalwert annimmt.

\vspace{10pt}

{\bf Kosinus-Halbwellen-Fenster} (\,z.~B.~\cite{Harris}\,)\\
Dieses diskrete Fenster entsteht durch Abtastung eines kontinuierlichen
Kosinus-Halb\-wel\-len-Fensters:\vspace{-5pt}
\[
f(t)\;=\;\begin{cases}
c_f\cdot\cos(\pi\CdoT t)\qquad&
\text{ f"ur }\quad |t|\!<\!\frac{1}{2}\\
0&\text{ sonst.}
\end{cases}
\]
Die Nullstellen des Spektrums des kontinuierlichen
Kosinus-Halbwellen-Fensters liegen bei ungeradzahligen Vielfachen
(\,gr"o"ser eins\,) der Frequenz \mbox{$\omega\!=\!\pi$}. Daher kann die
Abtastung dieses Fensters nicht in der Art erfolgen, dass sich bei allen
ganzzahligen Vielfachen der Frequenz \mbox{$\Omega\!=\!2\pi/M$}
immer nur Nullstellen "uberlagern, so dass die Nullstellenbedingung
(\ref{2.27}) erf"ullt w"are. Wegen der Unstetigkeit der ersten Ableitung
der kontinuierlichen Fensterfunktion bei \mbox{$t\!=\!\pm1/2$}
steigt die Sperrd"ampfung quadratisch in $\omega$ an. 

F"ur hinreichend
feine Abtastung des kontinuierlichen Kosinus-Halbwellen-Fensters
--- also f"ur gro"se Werte von $F$ --- ergibt sich somit auch bei der
zeitdiskreten Fensterfolge ein asymptotisch quadratischer
Anstieg der Sperrd"ampfung in \mbox{$\sin(\Omega/2)$}.
Nur bei einer Abtastung mit einer Abtastfrequenz kleiner gleich $M$
ist die Konstanz der nach Gleichung (\ref{2.20}) "uberlagerten,
verschobenen Betragsquadratspektren erf"ullt.

Man kann auch Potenzen des Kosinus-Halbwellen-Fensters verwenden.
Das Quadrat des Kosinus-Halbwellen-Fensters ergibt das Hann-Fenster.
Die weiteren geraden Potenzen ergeben die Nuttall-Fenster, die man
bei einer Optimierung der Potenz des asymptotischen Anstiegs der
Sperrd"ampfung erh"alt. Sie lassen sich problemlos in der Art abtasten,
dass die Nullstellenbedingung (\ref{2.27}) erf"ullt wird.
Bei den ungeraden Potenzen des Kosinus-Halbwellen-Fensters l"asst sich
eine derartige Abtastung nicht finden. Die Sperrd"ampfung steigt
f"ur gro"se Fensterl"angen $F$ asymptotisch mit einer Potenz in
\mbox{$\sin(\Omega/2)$} an, die um Eins h"oher ist, als die Potenz,
mit der das Kosinus-Halbwellen-Fensters potenziert wird.

Die Konstanz der nach Gleichung (\ref{2.20}) "uberlagerten,
verschobenen Betragsquadratspektren wird nur f"ur eine Abtastung mit
einer Abtastfrequenz kleiner gleich $M$ immer erf"ullt, weil dann
die Fensterl"ange $F$ so klein ist, dass die Bedingung (\ref{2.23}) immer
erf"ullt ist. F"uhrt die Abtastung zu einer Fensterl"ange \mbox{$F\!>\!M$},
so ist der Wert der Fenster-AKF $d(k)$ f"ur \mbox{$k\!=\!M$}
stets positiv, weil auch die Fensterfolge stets positiv ist. Daher kann
die Forderung (\ref{2.23}), und somit auch die Forderung (\ref{2.20}) nicht
erf"ullt werden.

{\bf Tukey-Fenster} (\,z.~B.~\cite{Kam}\,)\\
Dieses diskrete Fenster entsteht durch Abtastung des kontinuierlichen 
Tukey-Fensters
\[
f(t)\;=\;c_f\cdot\begin{cases}
2&\text{ f"ur }\quad |t|\le\frac{1}{2}\!-\!\alpha\\
1\!-\!\cos\big(\frac{\pi}{2\cdot\alpha}\CdoT(1\!-\!2\CdoT|t|\,)\big)\qquad&
\text{ f"ur }\quad \frac{1}{2}\!-\!\alpha<|t|<\frac{1}{2}\\
0&\text{ sonst,}
\end{cases}
\]
das im Intervall \mbox{$[\alpha\!-\!\frac{1}{2};\frac{1}{2}\!-\!\alpha]$} 
konstant ist, und anschlie"sende Kosinus-Flanken aufweist. F"ur 
\mbox{$\alpha\!=\!0$} erh"alt man das Rechteckfenster und f"ur 
\mbox{$\alpha\!=\!\frac{1}{2}$} das Hann-Fenster. 

F"ur 
\mbox{$0\!<\!\alpha\!<\!\frac{1}{2}$} kann man sich das kontinuierliche 
Tukey-Fenster durch eine Faltung zweier zeitbegrenzter Fensterfunktionen 
entstanden vorstellen. Die eine Funktion ist dabei ein symmetrisches Rechteck 
der gesamten Breite \mbox{$1\!-\!\alpha$}. Die andere Funktion ist das 
in der Zeitvariablen gestauchte Kosinus-Halbwellen-Fenster, das nach der 
Stauchung eine Gesamtbreite von $\alpha$ aufweist. Die Nullstellen 
des Spektrums des Rechteckfensters liegen bei ganzzahligen Vielfachen 
der Frequenz \mbox{$2\pi/(1\!-\!\alpha)$} mit Ausnahme von 
\mbox{$\omega\!=\!0$}. Die Nullstellen des Spektrums des gestauchten 
Kosinus-Halbwellen-Fensters liegen bei ungeradzahligen Vielfachen 
der Frequenz \mbox{$\pi/\alpha$} mit Ausnahme der beiden Frequenzen 
\mbox{$\omega\!=\!\pm\pi/\alpha$}. Das Spektrum des kontinuierlichen 
Tukey-Fensters besitzt die Nullstellen beider Spektralfaktoren. 

Bei der Abtastung des kontinuierlichen Tukey-Fensters mit der 
Abtastfrequenz $F$ erh"alt man eine Fensterfolge der L"ange $F$. 
Man wird nun versuchen $F$ so zu w"ahlen, dass sich die Nullstellen 
der verschobenen Spektren des kontinuierlichen Tukey-Fensters zu 
Nullstellen im Spektrum des zeitdiskreten Tukey-Fensters bei 
ganzzahligen Vielfachen der Frequenz \mbox{$\Omega\!=\!2\pi/M$} 
"uberlagern, um so sicherzustellen, dass die Nullstellenbedingung 
(\ref{2.27}) erf"ullt wird. Die Nullstellen des Spektrums des gestauchten 
Kosinus-Halbwellen-Fensters k"onnen dabei nicht genutzt werden, weil 
diese auch nach der "Uberlagerung der verschobenen Spektren des 
kontinuierlichen Fensters allenfalls in einem Raster liegen k"onnen, 
das ein ungeradzahliges Vielfaches einer Grundfrequenz ist. Somit 
k"onnen diese Nullstellen \mbox{niemals} in dem gew"unschten Raster liegen. 
Man wird daher die Abtastfrequenz $F$ so w"ahlen, dass $F$ ein ganzzahliges 
Vielfaches von \mbox{$1\!-\!\alpha$} ist, so dass sich die Nullstellen 
des kontinuierlichen Rechteckfensters zu Nullstellen im Spektrum des 
zeitdiskreten Tukey-Fensters "uberlagern, die bei ganzzahligen 
Vielfachen einer Grundfrequenz liegen. Desweiteren ist $F$ so 
zu w"ahlen, dass diese Grundfrequenz ein Teiler der Frequenz 
\mbox{$\Omega\!=\!2\pi/M$} ist. Mit den ganzzahligen Werten $M$, 
$\Tilde{N}$ und $F$ ist also $F$ und $\alpha$ so zu w"ahlen, dass 
\mbox{$F\CdoT(1\!-\!\alpha)\!=\!\Tilde{N}\CdoT M$} gilt. 
Zwischen zwei nach Gleichung (\ref{2.27}) notwendigen Nullstellen 
liegen dann wenigstens $\Tilde{N}$ weitere Nullstellen der 
Z-Transformierten des zeitdiskreten Tukey-Fensters am Einheitskreis. 
Bei dieser Art der Abtastung ist die Nullstellenbedingung 
(\ref{2.27}) erf"ullt, und $F$ ist immer gr"o"ser als $M$. 

Weil die stets positive Fensterfolge $f(k)$ eine stets positive 
Fenster-AKF $d(k)$ aufweist, kann die Forderung 
(\ref{2.23}), dass die Fenster-AKF f"ur Vielfache 
von $M$ Null ist, f"ur \mbox{$F\!>\!M$} nicht erf"ullt werden. 
Daher kann auch die "aquivalente Forderung (\ref{2.20}) nach der 
Konstanz der "uberlagerten, verschobenen Betragsquadratspektren 
der Fensterfolge nicht erf"ullt werden. F"ur eine hinreichend gro"se 
Fensterl"ange $F$ steigt die Sperrd"ampfung asymptotisch mit 
der dritten Potenz in \mbox{$\sin(\Omega/2)$} an.

\vspace{10pt}

{\bf Riesz-Fenster} (\,\cite{Harris}\,)\\
Dieses diskrete Fenster entsteht durch Abtastung eines kontinuierlichen 
Riesz-Fensters:\vspace{-5pt}
\[
f(t)\;=\;\begin{cases}
c_f\cdot\big(1\!-\!4\CdoT t^2\big)\qquad&\text{ f"ur }\quad |t|\!\le\!\frac{1}{2}\\
0&\text{ sonst.}
\end{cases}
\]
Die Fouriertransformierte des Riesz-Fensters ist proportional zu 
\mbox{$\big(\cos(\omega/2)-\text{si}(\omega/2)\big)/\omega^2$} 
und weist keine "aquidistanten Nullstellen auf. Es wird daher 
--- abgesehen von einigen uninteressanten F"allen f"ur extrem 
kleine Werte von \mbox{$M\!\le\!3$} --- nicht gelingen, die 
Abtastung in der Art zu w"ahlen, dass das zeitdiskrete Riesz-Fenster 
im Spektrum die in Gleichung (\ref{2.27}) geforderten "aquidistanten 
Nullstellen aufweist. Bei einer Abtastung mit einer Abtastfrequenz $F$ 
erh"alt man eine Fensterfolge der L"ange $F$. Weil die stets positive 
Fensterfolge $f(k)$ eine stets positive Fenster-AKF $d(k)$ 
aufweist, kann die Forderung (\ref{2.23}), dass die Fenster-AKF 
f"ur Vielfache von $M$ Null ist, f"ur \mbox{$F\!>\!M$} nicht erf"ullt werden, 
w"ahrend sie f"ur \mbox{$F\!\le\!M$} immer erf"ullt ist. Entsprechendes 
gilt f"ur die "aquivalente Forderung (\ref{2.20}) nach der Konstanz der 
"uberlagerten, verschobenen Betragsquadratspektren der Fensterfolge. 
F"ur eine hinreichend gro"se Fensterl"ange $F$ steigt die Sperrd"ampfung 
asymptotisch quadratisch mit \mbox{$\sin(\Omega/2)$} an.

{\bf Papoulis- oder Bohman-Fenster} (\,z.~B.~\cite{Thomae}\,)\\
Dieses diskrete Fenster entsteht durch Abtastung eines kontinuierlichen 
Papoulis-Fensters
\[
f(t)\;=\;\begin{cases}
c_f\cdot\big(\sin(2\pi\CdoT|t|\,)+
\pi\CdoT(1\!-\!2\CdoT|t|\,)\CdoT\cos(2\pi\CdoT t)\big)\qquad&
\text{ f"ur }\quad |t|\!\le\!\frac{1}{2}\\
0&\text{ sonst,}
\end{cases}
\]
das, abgesehen von einer Zeit- und Wertskalierung, die 
Fensterautokorrelationsfunktion des kontinuierlichen 
Kosinus-Halbwellen-Fensters ist. Daher kann die Abtastung 
dieses Fensters --- wie beim Kosinus-Halbwellen-Fenster --- 
nicht in der Art erfolgen, dass die Nullstellenbedingung 
(\ref{2.27}) erf"ullt wird. F"ur gro"se Werte von $F$ 
ist wegen der engen Abtastung des kontinuierlichen 
Papoulis-Fensters auch bei der zeitdiskreten Fensterfolge 
ein asymptotischer Anstieg der Sperrd"ampfung mit der vierten 
Potenz in \mbox{$\sin(\Omega/2)$} zu beobachten. Nur bei 
einer Abtastung mit einer Abtastfrequenz kleiner gleich $M$ 
ist die Konstanz der nach Gleichung (\ref{2.20}) "uberlagerten, 
verschobenen Betragsquadratspektren erf"ullt.

Varianten des Papoulis-Fensters erh"alt man, wenn man 
die geeignet skalierten Fensterautokorrelationsfunktionen von 
Potenzen des Kosinus-Halbwellen-Fensters verwendet. 
Bei geraden Potenzen ergeben sich Fenster, die sich problemlos 
in der Art abtasten lassen, dass die Nullstellenbedingung 
(\ref{2.27}) erf"ullt wird. Bei den ungeraden Potenzen des 
Kosinus-Halbwellen-Fensters l"asst sich eine derartige Abtastung 
der Fensterautokorrelationsfunktion nicht finden. Die Sperrd"ampfung 
steigt f"ur gro"se Fensterl"angen $F$ asymptotisch mit einer Potenz 
in \mbox{$\sin(\Omega/2)$} an, die um zwei h"oher ist als die doppelte 
Potenz, mit der das zugrunde liegende Kosinus-Halbwellen-Fensters 
potenziert wird. Die Konstanz der nach Gleichung~(\ref{2.20}) "uberlagerten, 
verschobenen Betragsquadratspektren wird nur f"ur eine Abtastung mit 
einer Abtastfrequenz kleiner gleich $M$ immer erf"ullt. Mit der 
zweiten Potenz des Kosinus-Halbwellen-Fensters erh"alt man bei 
entsprechender Skalierung die Fensterfunktion 
\[
\qquad f(t)\;=\;\begin{cases}
c_f\cdot\Big(2\pi\CdoT(1\!-\!2\CdoT|t|\,)\CdoT\big(2\!+\!\cos(4\pi\CdoT t)\big)+
3\CdoT\sin(4\pi\CdoT|t|\,)\Big)\qquad&
\text{ f"ur }\quad |t|\!\le\!\frac{1}{2}\\
0&\text{ sonst,}
\end{cases}
\]
die zugleich die Fensterautokorrelationsfunktion des Hann-Fensters ist.

{\bf Kosinus-Rolloff-Fenster} (\,\cite{Fliege}\,)\\
Zun"achst hat man wieder eine kontinuierliche Fensterfunktion, die 
nun zeitlich unbegrenzt ist:
\[
f(t)\;=\;c_f\cdot\text{si}(\pi\CdoT t)\cdot
\frac{\cos(\pi\CdoT\alpha\CdoT t)}{\D \;1-(2\CdoT\alpha\CdoT t)^2\,}.
\]
Das Spektrum des kontinuierlichen Kosinus-Rolloff-Fensters 
ist im Intervall \mbox{$|\omega|\!<\!\pi\CdoT(1\!-\!\alpha)$} 
konstant, und f"allt anschlie"send mit einer Kosinus-Flanke ab. 
F"ur \mbox{$\omega\!=\!\pi$} ist das Spektrum nur halb so gro"s wie f"ur 
\mbox{$\omega\!=\!0$}, und f"ur \mbox{$|\omega|\!>\!\pi\CdoT(1\!+\!\alpha)$} 
ist es Null. 

Aus dem kontinuierlichen Kosinus-Rolloff-Fenster kann man auf 
zweierlei Weisen ein zeitdiskretes Fenster der L"ange $F$ erhalten. Die erste 
M"oglichkeit besteht darin, einen Ausschnitt der endlichen L"ange $N$ des 
kontinuierlichen Kosinus-Rolloff-Fensters mit einer Abtastfrequenz $F/N$ 
abzutasten. Dies entspricht im Frequenzbereich einer Faltung des Spektrums 
des kontinuierlichen Kosinus-Rolloff-Fensters mit einer periodisch 
fortgesetzten si-Funktion. Zwei benachbarte absolute Betragsmaxima 
der periodisch fortgesetzten si-Funktion haben dabei einen \mbox{Abstand} 
von \mbox{$2\pi\CdoT F/N$}. Zwischen zwei benachbarten absoluten 
Betragsmaxima liegen jeweils \mbox{$F\!-\!1$} "aquidistante Nullstellen 
in einem Raster mit einem Frequenzabstand von \mbox{$2\pi/N$}. 
Wo das Spektrum der zeitdiskreten Fensterfolge Nullstellen aufweist, 
kann allgemein nicht gesagt werden, da die Nullstelleneigenschaft 
der periodisch fortgesetzten si-Funktion bei der Faltung mit dem 
Spektrum des kontinuierlichen Kosinus-Rolloff-Fensters nicht erhalten 
bleibt. 

Um dies zu vermeiden, kann man die zweite M"oglichkeit, 
ein zeitdiskretes Fenster aus dem kontinuierlichen Kosinus-Rolloff-Fenster 
zu erhalten, anwenden. Dabei wird die kontinuierliche Fensterfunktion 
zun"achst mit der Periode $\Tilde{N}$ fortgesetzt, indem man sie mit einem 
Impulskamm mit dem Impulsabstand $\Tilde{N}$ faltet. Man erh"alt so zun"achst 
ein Linienspektrum mit dem Impulsabstand \mbox{$2\pi/\Tilde{N}$}, bei dem 
die St"arken der Impulse des Spektrums proportional zu den entsprechenden 
Abtastwerten des Spektrums des kontinuierlichen Kosinus-Rolloff-Fensters 
sind. Da das Spektrum der kontinuierlichen Fensterfunktion begrenzt ist, 
wird das Spektrum der periodisch fortgesetzten Fensterfunktion f"ur 
\mbox{$0\!<\!\alpha\!\le\!1$} maximal \mbox{$2\CdoT\Tilde{N}\!-\!1$} Linien 
enthalten. Tastet man bei dieser periodisch fortgesetzten, kontinuierlichen 
Fensterfunktion einen Ausschnitt der L"ange $N$ mit der Abtastfrequenz 
$F/N$ ab, so erh"alt man eine zeitdiskrete Fensterfolge der L"ange $F$, 
deren Spektrum sich nun als die Faltung des Linienspektrums mit der oben 
angegebenen periodisch fortgesetzten si-Funktion berechnen l"asst. 
Wenn man nun $N$ als ein ganzzahliges Vielfaches von $\Tilde{N}$ 
w"ahlt, bleiben alle Nullstellen der periodisch fortgesetzten si-Funktion, 
die bei der Faltung nicht auf den Frequenzen der Linien zu liegen kommen, 
erhalten. Mit der auf die Abtastfrequenz normierten Frequenz $\Omega$ besitzt 
das Spektrum der zeitdiskreten Fenster\-folge Nullstellen im Abstand 
\mbox{$2\pi/F$}, au"ser bei den maximal \mbox{$2\CdoT N\!-\!1$} niedrigen 
Frequenzen im Abstand \mbox{$2\pi\CdoT N/(\Tilde{N}\CdoT F)$}, die den 
Frequenzen der Linien des Spektrums der periodisch fortgesetzten 
Fensterfunktion entsprechen. Werden nun $N$, $\Tilde{N}$ und $F$ geeignet 
gew"ahlt (\,z.~B. \mbox{$N\!=\!\Tilde{N}$} und \mbox{$F\!=\!N\CdoT M$}\,), 
so wird die Nullstellenbedingung (\ref{2.27}) erf"ullt. 

Wenn man f"ur die 
Periode $\Tilde{N}$ der periodisch fortgesetzten Fensterfunktion eine gerade 
Zahl w"ahlt, weist die si-Funktion, die als Faktor in der urspr"unglichen, 
kontinuierlichen Kosinus-Rolloff-Fensterfunktion vorhanden ist, bei 
ungeradzahligen Vielfachen von \mbox{$t\!=\!\pm\Tilde{N}/2$} Nullstellen auf. 
Aus diesen werden nach der periodischen Fortsetzung doppelte Nullstellen, 
da sich zu diesen Zeitpunkten immer einfache Nullstellen "uberlagern, 
wobei jeweils zwei "uberlagerte Nullstellen gegengleiche Ableitungen 
besitzen. Die Sperrd"ampfung des Spektrums der zeitdiskreten 
Kosinus-Rolloff-Fensterfolge wird daher f"ur gerades $\Tilde{N}$ und 
gro"se Fensterl"ange $F$ asymptotisch mit der dritten Potenz in 
\mbox{$\sin(\Omega/2)$} ansteigen, wenn der abgetastete Ausschnitt 
symmetrisch zu \mbox{$t\!=\!0$} gew"ahlt wird. 

F"ur ungerades $\Tilde{N}$ 
ergibt sich bei symmetrischer Wahl des Ausschnitts derselbe 
Anstieg der Sperrd"ampfung, wenn $\alpha$ ein ungerades, ganzzahliges 
Vielfaches gr"o"ser Eins von \mbox{$1/\Tilde{N}$} ist, so dass die 
Kosinusfunktion, die als Faktor in der urspr"unglichen, 
kontinuierlichen Kosinus-Rolloff-Fensterfunktion vorhanden ist, bei 
ungeradzahligen Vielfachen von \mbox{$t\!=\!\pm\Tilde{N}/2$} Nullstellen 
besitzt. Wird ein unsymmetrischer Ausschnitt zur Abtastung aus der 
periodisch fortgesetzten kontinuierlichen Fensterfunktion ausgew"ahlt, 
so erh"alt man eine mit \mbox{$\sin(\Omega/2)$} linear ansteigende 
Sperrd"ampfung, es sei denn, der abgetastete Ausschnitt ist so gew"ahlt 
worden, dass dieser bei einer einfachen Nullstelle der periodisch 
fortgesetzten Fensterfunktion beginnt und endet. Dann steigt die 
Sperrd"ampfung des Spektrums der zeitdiskreten Fensterfolge quadratisch 
mit \mbox{$\sin(\Omega/2)$} an. 

Die Konstanz der nach Gleichung (\ref{2.20})
"uberlagerten, verschobenen Betragsquadratspektren kann bei einer mit der 
Nullstellenbedingung (\ref{2.27}) kompatiblen Abtastung, nur f"ur 
den trivialen Fall des Rechteckfensters mit \mbox{$N\!=\!\Tilde{N}\!=\!1$} 
und \mbox{$F\!=\!M$} erreicht werden.

\vspace{10pt}

{\bf Wurzel-Kosinus-Rolloff-Fenster} (\,\cite{Fliege}\,)\\
Wieder hat man eine kontinuierliche, zeitlich unbegrenzte Fensterfunktion:
\[
f(t)\;=\;c_f\cdot\frac{
4\CdoT\alpha\CdoT t\CdoT\cos\big((1\!+\!\alpha)\CdoT\pi\CdoT t\big)+
\sin\big((1\!-\!\alpha)\CdoT\pi\CdoT t\big)}{\D
\big(1\!-\!(4\CdoT\alpha\CdoT t)^2\big)\CdoT\pi\CdoT t}.
\]
Die positive Wurzel des Spektrums des kontinuierlichen 
Kosinus-Rolloff-Fensters ist das Spektrum des kontinuierlichen 
Wurzel-Kosinus-Rolloff-Fensters. Bez"uglich der Art der Abtastung, die
gew"ahrleistet, dass die Nullstellenbedingung (\ref{2.27}) erf"ullt wird, 
gilt im wesentlichen das zum Kosinus-Rolloff-Fenster gesagte. Im Allgemeinen
wird es nur f"ur einige spezielle Werte von $\alpha$ gelingen, dass 
der symmetrisch zu \mbox{$t\!=\!0$} gew"ahlte, abzutastende Ausschnitt 
der periodifizierten, kontinuierlichen Fensterfunktion mit einer 
doppelten Nullstelle beginnt und endet, so dass die Sperrd"ampfung 
des Spektrums der zeitdiskreten Wurzel-Kosinus-Rolloff-Fensterfolge 
f"ur gro"se Fensterl"ange $F$ asymptotisch mit der dritten Potenz in 
\mbox{$\sin(\Omega/2)$} ansteigt. Einen Ausschnitt zu finden, der mit einer 
einfachen Nullstelle in der periodifizierten, kontinuierlichen Fensterfunktion 
beginnt und endet, ist relativ einfach, wenn man den Ausschnitt nicht 
symmetrisch zu \mbox{$t\!=\!0$} w"ahlt. Dann erreicht man einen 
quadratischen Anstieg der Sperrd"ampfung mit \mbox{$\sin(\Omega/2)$}. 
Im Allgemeinen beginnt und endet der Ausschnitt jedoch meist mit einem 
Sprung, so dass sich ein a\-symp\-to\-tisch linearer Sperrd"ampfungsanstieg 
ergibt. 

Der Sinn des bei der kontinuierlichen Fensterfunktion 
gew"ahlten Spektrums besteht darin, dass die "Uberlagerung aller 
um Vielfache von $2\pi$ verschobenen Betragsquadratspektren 
eine Konstante ergibt. Wenn man aber nun einen Ausschnitt 
der kontinuierlichen Fensterfunktion oder der periodifizierten 
Fensterfunktion betrachtet, geht diese der Gleichung (\ref{2.20}) 
entsprechende Eigenschaft bereits vor der Abtastung verloren, und 
kann auch durch die Abtastung nicht wiederhergestellt werden. 
Lediglich f"ur die diskreten Frequenzen im Raster \mbox{$2\pi/F$} 
wird die Konstanz der nach Gleichung (\ref{2.20}) "uberlagerten, 
verschobenen Betragsquadratspektren erreicht, wenn man die oben 
beschriebene, mit der Nullstellenbedingung (\ref{2.27}) vertr"agliche 
Art der Abtastung verwendet. Bei den anderen Frequenzen ergeben sich 
Abweichungen, die z.~B. f"ur \mbox{$N\!=\!\Tilde{N}\!=\!8$} und 
\mbox{$M\!=\!64$} bei maximal \mbox{$0,\!5\%$} liegen, und somit 
um viele Zehnerpotenzen "uber den Fehlern liegen, die bei dem nach 
Kapitel \ref{Algo} konstruierten Fenster verbleiben.

{\bf Daniell-Fenster} (\,\cite{Dittrich}\,)\\
In Kapitel \ref{W} hatten wir festgestellt, dass es w"unschenswert w"are, 
das in $\Omega$ kontinuierliche LDS mit Hilfe der $M$ Werte 
\mbox{$\Bar{\Phi}_{\boldsymbol{n}}(\mu)$} der fl"achengleichen 
Stufenapproximation nach Gleichung (\ref{2.13}) zu beschreiben. Die 
Erwartungswerte \mbox{$\Tilde{\Phi}_{\boldsymbol{n}}(\mu)$} der 
Zufallsgr"o"sen \mbox{$|\boldsymbol{N}_{\!\!f}(\mu)|^2/M$} entsprechen 
genau den gew"unschten Gr"o"sen \mbox{$\Bar{\Phi}_{\boldsymbol{n}}(\mu)$}, 
wenn man als Fensterfolge die auch als Daniell-Fenster bezeichnete abgetastete 
si-Funktion \mbox{$\text{si}(\pi\CdoT k/M)$}, verwendet. 

Da man wieder nur 
mit endlicher Fensterl"ange $F$ arbeiten kann, muss man auch hier 
wieder eine geeignete Art finden, wie man aus der kontinuierlichen 
si-Funktion eine entsprechende zeitbegrenzte und zeitdiskrete 
Fensterfolge gewinnen kann. Wenn man beim kontinuierlichen 
Kosinus-Rolloff-Fenster oder beim Wurzel-Kosinus-Rolloff-Fenster 
den Parameter $\alpha$ zu Null setzt, erh"alt man die si-Funktion. 
Daher k"onnen die beiden dort beschriebenen Verfahren, eine zeitdiskrete 
Fensterfolge zu gewinnen, auch hier angewendet werden, und die dort 
gewonnenen Erkenntnisse bez"uglich der Bedingungen (\ref{2.20}) und
(\ref{2.27}) sowie des Anstiegs der Sperrd"ampfung behalten auch hier 
ihre G"ultigkeit. F"ur den Fall, dass man die si-Funktion periodifiziert 
bevor man sie abtastet, indem man sie mit einem Impulskamm faltet, 
ergeben sich hier jedoch zwei Besonderheiten. 

Wenn man die L"ange des 
Ausschnittes, den man bei der periodifizierten si-Funktion abtastet, 
als ein Vielfaches der Periode des Impulskamms w"ahlt, so dass die 
Nullstellenbedingung (\ref{2.27}) erf"ullt wird, liegen die Linien 
des Spektrums der periodifizierten si-Funktion immer symmetrisch zu 
\mbox{$\omega\!=\!0$} und alle sind gleich hoch. Die periodifizierte 
si-Funktion hat daher innerhalb einer Periode immer eine gerade Anzahl 
von einfachen Nullstellen. Daher wird bei einer zum Hauptmaximum der 
periodifizierten si-Funktion symmetrischen Abtastung immer ein Sprung 
an den Enden des herausgeschnittenen Ausschnitts auftreten, weshalb 
die Sperrd"ampfung dann asymptotisch linear mit \mbox{$\sin(\Omega/2)$} 
ansteigt. 

Bei unsymmetrischer Abtastung kann man allenfalls einen 
quadratischen Sperrd"ampfungsanstieg erreichen. Bei dieser Art der 
Periodifizierung kann man leicht zeigen, dass die Konstanz der nach 
Gleichung (\ref{2.20}) "uberlagerten, verschobenen Betragsquadratspektren 
nie gegeben sein kann. Da sich das Spektrum des zeitdiskreten 
Daniell-Fensters als Faltung der Linien des Spektrums der 
periodifizierten si-Funktion mit einer periodisch fortgesetzten 
si-funktion berechnet, deren Nullstellenabstand mit dem Abstand der 
Spektrallinien "ubereinstimmt, tritt das Gibbssche Ph"anomen auf. Daher 
ist der Maximalwert des Betragsquadrats des Spektrums gr"o"ser als 
bei der Frequenz \mbox{$\Omega\!=\!0$}. Bei dieser Frequenz werden 
wegen der Nullstellenlage nach Gleichung (\ref{2.27}) bei der "Uberlagerung 
der verschobenen Betragsquadratspektren nach Gleichung (\ref{2.20}) 
Nullen addiert. Somit ergibt sich f"ur \mbox{$\Omega\!=\!0$} in 
Gleichung (\ref{2.20}) ein kleinerer Wert, als bei der Frequenz, bei 
der als Summand der Maximalwert des Betragsquadrats des Spektrums des 
zeitdiskreten Daniell-Fensters enthalten ist. Es sei noch darauf hingewiesen,
dass das Fenster, das man erh"alt, wenn man aus der si-Funktion nur den
Bereich zwischen $-\pi$ und $\pi$ --- also die Hauptkeule ---
ausschneidet, auch unter dem Namen Riemann-Fenster bekannt ist.

{\bf Gau"s-Jones- oder Weierstra"s-Fenster} (\,z.~B.~\cite{Dittrich}\,)\\
Der kontinuierliche Gau"simpuls\vspace{-6pt}
\[
f(t)\;=\;c_f\cdot e^{\!-\frac{\alpha}{2}\cdot t^2}
\]
besitzt das kleinstm"ogliche Produkt aus Zeitdauer und Bandbreite, wobei 
diese als die zweiten Momente des Zeitsignals und des Spektrums definiert 
sind, wie dies z.~B. in \cite{Franks} angegeben ist. Daher erscheint es 
auf den ersten Blick sinnvoll einen Gau"simpuls beim RKM zu verwenden, 
wenn man bei einer m"oglichst kurzen Messdauer eine m"oglichst 
gute Frequenzauf"|l"osung bei den Messergebnissen erzielen will. 

Um ein zeitdiskretes Gau"s-Jones-Fenster zu erhalten, tastet man 
einen Ausschnitt des Gau"s-Impulses ab, so dass man eine Fensterfolge 
der endlichen L"ange $F$ erh"alt. Da die Messdauer beim RKM 
im wesentlichen von der Anzahl der gemessenen Abtastwerte, also vom 
Produkt aus der Mittelungsanzahl $L$ und der L"ange $F$ der Fensters 
abh"angt, ist hier nicht die als zweites Moment definierte 
Zeitdauer entscheidend. Man wird eher daran interessiert sein, 
bei einer vorgegebenen Fensterl"ange $F$ die Bandbreite des 
Spektrums der Fensterfolge klein zu halten. Ob sich dabei die 
Energie der Fensterfolge um deren Schwerpunkt herum konzentriert, 
ist dabei von untergeordneter Bedeutung. Zudem ist es hier empfehlenswert, 
bei der Definition der Bandbreite das Betragsquadrat des Spektrums der 
Fensterfolge nicht mit ${\D \Omega^2}$ zu gewichten, sondern eine 
Gewichtung zu verwenden, die ber"ucksichtigt, dass man nicht an den 
Abtastwerten des LDS, sondern an den nach Gleichung (\ref{2.13}) definierten 
Werten der Stufenapproximation interessiert ist. Der Wunschverlauf 
des Spektrums der Fensterfolge ist ein Rechteck, und kein Dirac-Impuls, 
der die bei \cite{Franks} verwendete minimal m"ogliche Bandbreite Null 
besitzt. 

Beim Gau"s-Impuls ist es --- abgesehen von einigen F"allen 
mit extrem kleiner Fensterl"ange \mbox{$F\!\le\!4$} --- nicht m"oglich, 
die Abtastung so zu w"ahlen, dass die Nullstellenbedingung (\ref{2.27}) 
erf"ullt wird. Dabei kann auch eine Periodifizierung des Gau"s-Impulses 
vor der Abtastung nichts "andern. Der asymptotische Anstieg der 
Sperrd"ampfung f"ur gro"se Fensterl"angen $F$ erfolgt in jedem Fall 
linear mit \mbox{$\sin(\Omega/2)$}, weil am Anfang und Ende des Ausschnitts 
immer ein Sprung auftritt. Es kann auch keine Art der Abtastung 
angegeben werden, die die Konstanz der nach Gleichung (\ref{2.20}) 
"uberlagerten, verschobenen Betragsquadratspektren garantiert.

{\bf Poisson-Fenster} (\,z.~B.~\cite{Harris}\,)\\
Dieses diskrete Fenster entsteht durch Abtastung eines kontinuierlichen 
Poisson-Fensters
\[
f(t)\;=\;c_f\cdot e^{\!-\alpha\cdot |t|}
\qquad\qquad\text{mit}\qquad\alpha\!>\!0,
\]
dessen Spektrum \mbox{$2\CdoT f\CdoT\alpha/(\alpha^2\!+\!\omega^2)$} 
asymptotisch mit $\omega^2$ abf"allt, weil das kontinuierliche 
Fenster bei \mbox{$t\!=\!0$} eine Unstetigkeit in der Ableitung 
besitzt. Auch bei diesem Fenster ist es --- abgesehen von einigen F"allen
mit extrem kleiner Fensterl"ange --- nicht m"oglich, die Abtastung eines 
Ausschnitts in der Art vorzunehmen, dass die Nullstellenbedingung (\ref{2.27})
oder die Konstanz der nach Gleichung (\ref{2.20}) "uberlagerten, 
verschobenen Betragsquadratspektren erf"ullt wird. Eine eventuelle 
Periodifizierung der kontinuierlichen Poisson-Fensterfunktion vor der 
Abtastung "andert daran nichts. Da der abzutastende Ausschnitt der 
Poisson-Fensterfunktion immer mit einem Sprung beginnt und endet, 
steigt die Sperrd"ampfung der durch die Abtastung entstandenen 
zeitdiskreten Poisson-Fensterfolge f"ur gro"se Fensterl"angen $F$ 
asymptotisch nur linear mit \mbox{$\sin(\Omega/2)$} an. 

Es wird in \cite{Harris} auch die M"oglichkeit genannt, das 
kontinuierliche Poisson-Fenster vor der Abtastung nicht mit einem 
Rechteckfenster zu begrenzen, sondern dazu ein kontinuierliches 
Hann-Fenster zu verwenden, um so das sog. Hanning-Poisson-Fenster zu 
erhalten. Auf die Erf"ullung der Bedingungen (\ref{2.27}) und (\ref{2.20}) 
hat dies keinen Einfluss. Man erreicht jedoch dadurch den 
asymptotisch quadratischen Anstieg der Sperrd"ampfung, der auch 
beim kontinuierlichen Poisson-Fenster zu beobachten ist. 

{\bf Cauchy-, Poisson- oder Abel-Fenster}\\[1pt]
Das Spektrum der kontinuierlichen Cauchy-Fensterfunktion
\[
f(t)\;=\;\frac{c_f}{\alpha^2\!+\!t^2}
\]
ist ein kontinuierliches Poisson-Fenster in $\omega$. 
Auch bei diesem Fenster ist es --- abgesehen von einigen F"allen 
mit extrem kleiner Fensterl"ange --- nicht m"oglich, die Ab"-tas"-tung eines 
Ausschnitts in der Art vorzunehmen, dass die Nullstellenbedingung (\ref{2.27}) 
oder die Konstanz der nach Gleichung (\ref{2.20}) "uberlagerten, 
verschobenen Betragsquadratspektren garantiert wird. Eine eventuelle 
Periodifizierung der kontinuierlichen Cauchy-Fensterfunktion vor der 
Abtastung "andert daran nichts. Da der abzutastende Ausschnitt der 
Cauchy-Fensterfunktion immer mit einem Sprung beginnt und endet, 
steigt die Sperrd"ampfung der durch die Abtastung entstandenen 
zeitdiskreten Cauchy-Fensterfolge f"ur gro"se Fensterl"angen $F$ 
asymptotisch nur linear mit \mbox{$\sin(\Omega/2)$} an.

{\bf Kaiser-Fenster} (\,z.~B.~\cite{Thomae}\,)\\
Dieses diskrete Fenster entsteht durch Abtastung eines kontinuierlichen 
Kaiser-$I_0$-Fensters
\[
f(t)\;=\;\begin{cases}
c_f\cdot I_0\big(\alpha\CdoT\sqrt{1-4\CdoT t^2\,}\,\big)\qquad&
\text{ f"ur }\quad |t|\!\le\!\frac{1}{2}\\
0&\text{ sonst,}
\end{cases}
\]
das sich mit Hilfe der modifizierten Besselfunktion $I_0(x)$ erster 
Art, nullter Ordnung berechnen l"asst, und das an seinen Enden 
Sprungstellen aufweist. Das Kaiser-$I_0$-Fenster stellt eine N"aherung 
f"ur eine Optimierung eines zeitbegrenzten Fensters dar, das bei vorgegebener 
Gesamtenergie den Anteil der Energie maximiert, der innerhalb des 
Durchlassbereichs des Spektrums des Fensters liegt. Da das Spektrum des 
kontinuierlichen Kaiser-$I_0$-Fensters keine "aquidistanten Nullstellen 
aufweist, kann keine Abtastmethode angegeben werden, die die Erf"ullung 
der Nullstellenbedingung (\ref{2.27}) garantiert. Die Sprungstellen der 
kontinuierlichen Fensterfunktion bewirken, dass auch die Sperrd"ampfung 
der durch die Abtastung entstandenen zeitdiskreten Kaiser-$I_0$-Fensterfolge 
f"ur gro"se Fensterl"angen $F$ asymptotisch nur linear mit 
\mbox{$\sin(\Omega/2)$} ansteigt. 

In \cite{Thomae} wird als Variante 
daher vorgeschlagen, von dem kontinuierlichen Kaiser-$I_0$-Fenster ein 
Rechteckfenster zu subtrahieren, dessen Amplitude gerade die 
Sprungh"ohe ist. Die Sperrd"ampfung steigt bei dieser Variante dann 
asymptotisch quadratisch mit \mbox{$\sin(\Omega/2)$} an, weil an den 
Enden des kontinuierlichen Fensters die Ableitung von Null verschieden ist. 
Auch bei dieser Variante l"asst sich keine mit der Nullstellenbedingung 
(\ref{2.27}) kompatible Abtastmethode angeben. 

Bei einer weiteren Variante 
wird nicht die modifizierte Besselfunktion $I_0(x)$, sondern die 
modifizierte Besselfunktion $I_1(x)$ erster Art, erster Ordnung 
verwendet, die durch ihr Argument $x$ dividiert wird:
\[
f(t)\;=\;\begin{cases}
c_f\cdot\frac{\D I_1\big(\alpha\CdoT\sqrt{1-4\CdoT t^2\,}\,\big)}
{\D\alpha\CdoT\sqrt{1-4\CdoT t^2\,}}\qquad&
\text{ f"ur }\quad |t|\!\le\!\frac{1}{2}\\
0&\text{ sonst.}
\end{cases}
\]
Auch bei dieser Variante weist die kontinuierliche Fensterfunktion 
an ihren Enden Sprungstellen auf, was einen asymptotisch linearen 
Anstieg der Sperrd"ampfung auch bei der abgetasteten Fensterfolge bewirkt. 
Die Nullstellen des Spektrums des kontinuierlichen Fensters sind 
--- wie bei den anderen Varianten --- nicht "aquidistant, wodurch keine 
Abtastmethode angegeben werden kann, die die Erf"ullung der 
Nullstellenbedingung (\ref{2.27}) garantiert. Bez"uglich der 
Konstanz der nach Gleichung (\ref{2.20}) "uberlagerten, verschobenen 
Betragsquadratspektren gilt bei allen drei Varianten die folgende 
Feststellung: Wenn man den Parameter $\alpha$ auf \mbox{$\pi\CdoT F/M$} 
setzt, beginnt der Sperrbereich des Spektrums --- \mbox{also} der Bereich, in 
dem die Nullstellen des Spektrums der Fensterfolge liegen --- \mbox{etwa} bei 
der Frequenz \mbox{$\Omega\!=\!2\pi/M$}, bei der die erste nach Gleichung 
(\ref{2.27}) geforderte Nullstelle im Spektrum zu liegen hat. Wenn man sich 
damit begn"ugt, bei den in Gleichung (\ref{2.27}) angegebenen Frequenzen 
statt der exakten Nullstellen ein hohe Sperrd"ampfung zu haben, 
darf man den Parameter $\alpha$ nicht gr"o"ser w"ahlen. Bereits bei 
dieser Wahl ist aber die Hauptkeule des Spektrums des Kaiser-Fensters so 
schmal, dass die "Uberlagerung der verschobenen Betragsquadratspektren 
bei der Frequenz \mbox{$\Omega\!=\!\pi/M$} bei gro"ser Fensterl"ange um 
einige Zehnerpotenzen kleiner ist als bei \mbox{$\Omega\!=\!0$}.

{\bf Dolph-Tschebyscheff-Fenster}\label{Dolph}\\
Zun"achst sei kurz die Konstruktion des Dolph-Tschebyscheff-Fensters 
leicht abweichend von \cite{Harris} und \cite{Thomae} erl"autert. 
Das Tschebyscheff-Polynom 
\mbox{$T(x)=\cos\!\big((F\!-\!1)\CdoT\arccos(x)\big)$} 
hat \mbox{$F\!-\!1$} Nullstellen im Intervall \mbox{$[-1;1]$}, 
die so verteilt sind, dass die jeweils dazwischenliegenden 
Maxima alle den Betrag Eins haben und, dass bei \mbox{$|x|\!=\!1$} 
der Betrag des Tschebyscheff-Polynoms ebenfalls Eins ist. 
W"ahlt man als Argument des Tschebyscheff-Polynoms die Ko"-si"-nus-Funktion 
\mbox{$x=\cos(\Omega/2)/\cos(\Omega_1/2)$}, so erh"alt man das Spektrum des 
Dolph-Tschebyscheff-Fensters. Wenn \mbox{$0\!\le\!\Omega_1\!<\!\pi$} 
eingehalten wird, wird f"ur \mbox{$0\!\le\!\Omega\!<\!\Omega_1$} das 
Tschebyscheff-Po"-ly"-nom im Bereich \mbox{$|x|\!>\!1$} ausgesteuert, bei 
dem der Betrag des Tsche"-by"-scheff-Polynoms gr"o"ser als eins ist, was dem 
Durchlassbereich des Spektrums des Fensters entspricht. F"ur 
\mbox{$\Omega_1<\Omega\le2\pi\!-\!\Omega_1$} alterniert das 
Tschebyscheff-Polynom zwischen den Werten -1 und 1, was im Sperrbereich 
zum typischen oszillierenden Tschebyscheff-Verhalten mit \mbox{$F\!-\!1$} 
Nullstellen f"uhrt. Ein potenzm"a"siger Anstieg der Sperrd"ampfung kann hier 
also nicht eingestellt werden. 

Da das Tschebyscheff-Polynom von endlichem 
Grad \mbox{$F\!-\!1$} ist, l"asst sich das mit $4\pi$ periodische, reelle 
und geradesymmetrische Spektrum des Dolph-Tschebyscheff-Fensters in eine 
Kosinus-Reihe entwickeln, deren h"ochstfrequenter Anteil 
\mbox{$\cos\!\big((F\!-\!1)\CdoT\Omega/2\big)$} ist. Da das 
Spektrum des Tschebyscheff-Fensters f"ur gerades $F$ bez"uglich der 
Frequenz \mbox{$\Omega\!=\!2\pi$} schiefsymmetrisch und f"ur ungerades 
$F$ geradesymmetrisch ist, ist jeder zweite Koeffizient der Kosinus-Reihe 
Null. Daher l"asst sich die mit \mbox{$e^{\!-j\cdot(F-1)\cdot\Omega/2}$} 
multiplizierte Kosinus-Reihe in jedem Fall als eine Fourierreihe 
mit $F$ Gliedern schreiben, die bei Null beginnt, und deren 
Koeffizienten die Werte \mbox{$f(k)$} des Tschebyscheff-Fensters sind. 
Diese lassen sich somit durch eine inverse DFT der L"ange $F$ aus 
den $F$ Abtastwerten des Tschebyscheff Polynoms und des Drehterms 
\mbox{$e^{\!-j\cdot(F-1)\cdot\Omega/2}$} bei Vielfachen der 
Frequenz \mbox{$2\pi/F$} berechnen:\vspace{-6pt}
\begin{align*}
F\big({\T\nu\CdoT\frac{2\pi}{F}}\big)\;=\;&
\cos\!\Bigg((F\!-\!1)\cdot\arccos\!\bigg(
\frac{\cos(\nu\CdoT\frac{\pi}{F})}{\cos(\frac{\Omega_1}{2})}\bigg)\Bigg)\cdot
e^{\!-j\cdot(F-1)\cdot\frac{\pi}{F}\cdot\nu}\\[4pt]
f(k)\;=\;&\mbox{FFT}^{-1}\!\big\{F\big({\T\nu\CdoT\frac{2\pi}{F}}\big)\big\}
\qquad\qquad\text{ mit }\qquad\nu=0\;(1)\;F\!-\!1.
\end{align*}
Da die Nullstellen im Spektrum des Tschebyscheff-Fensters nicht 
"aquidistant liegen, kann die Nullstellenbedingung (\ref{2.27}) 
nicht erf"ullt werden. Wenn man sich damit begn"ugt, bei den in 
Gleichung (\ref{2.27}) angegebenen Frequenzen statt der exakten 
Nullstellen ein hohe Sperrd"ampfung zu haben, darf man die Grenzfrequenz 
$\Omega_1$ nicht gr"o"ser als \mbox{$2\pi/M$} w"ahlen. Bereits bei 
dieser Wahl ist aber die Hauptkeule des Spektrums des Tschebyscheff-Fensters 
so schmal, dass die "Uberlagerung der verschobenen Betragsquadratspektren 
bei der Frequenz \mbox{$\Omega\!=\!\pi/M$} bei gro"ser Fensterl"ange um 
ein paar Zehnerpotenzen kleiner ist als bei \mbox{$\Omega\!=\!0$}. 
Daher kann die Konstanz der nach Gleichung (\ref{2.20}) "uberlagerten,
verschobenen Betragsquadratspektren praktisch "uberhaupt nicht erf"ullt 
werden.

{\bf Zusammenfassung}\\
Man kann sagen, dass keines der hier untersuchten Fenster 
die beiden Bedingungen (\ref{2.20}) und (\ref{2.27}) 
bei beliebiger Wahl von $M$ zugleich erf"ullt, und dass der 
Sperrd"ampfungsanstieg bei fast keiner Fensterfolge beliebig gew"ahlt 
werden kann. Nat"urlich gibt es weitere hier nicht aufgef"uhrte 
Fensterfolgen, sowie Fenster, die mit \mbox{Hilfe} eines allgemeinen 
Tiefpassentwurfs f"ur nichtrekursive Filter (\,z.~B. Remez-Algorithmus 
nach Parks-McClellan\,) gewonnen werden. Dem Autor ist aber auch bei diesen 
Varianten kein Fenster bekannt, dass die geforderten Eigenschaften 
erf"ullt. Es besteht auch die M"oglichkeit, ausgehend von dem wohl 
noch am ehesten geeigneten Fenster --- dem Wurzel-Kosinus-Rolloff-Fenster --- 
eine iterative Verbesserung bei der Erf"ullung der Gleichung (\ref{2.20})
unter der Randbedingung (\ref{2.27}) zu erhalten. Dieses Verfahren wird 
gelegentlich bei Anwendungen im Bereich der Multiraten-Signalverarbeitung 
verwendet. Die damit gewonnenen Fenster sind dann prinzipiell beim 
RKM einsetzbar. Der Autor hat jedoch die Erfahrung gemacht, dass 
die nach der iterativen Verbesserung verbleibenden Restfehler 
deutlich gr"o"ser sind als bei den Fenstern, die mit dem in Kapitel 
\ref{Algo} beschriebenen Verfahren berechnet worden sind. Au"serdem 
stellte sich zumindest bei einem solchen Verfahren heraus, dass 
die Iteration zu keinem sinnvollen Ergebnis kam, wenn das Fenster 
um mehr als das $16$-fache l"anger als $M$ war.

Bei den meisten Fenstern und bei deren Anwendung werden drei 
Voraussetzungen getroffen, die von vornherein die Freiheit bei der 
Konstruktion der Fensterfolge einschr"anken, und die f"ur die Anwendung 
beim RKM nicht notwendig sind. 

Die erste Voraussetzung ist dabei, dass 
von einigen Fenstern gefordert wird, dass sie nie negativ werden. F"ur 
diese Einschr"ankung konnte keine plausible Motivation gefunden werden. 
Da es zudem durch diese Forderung f"ur Fenster, deren L"ange gr"o"ser $M$ 
ist, unm"oglich wird, die Gleichung (\ref{2.23}) zu erf"ullen, wurde diese 
Forderung auch nicht f"ur die von mir entworfene Fensterfolge "ubernommen. 

Die zweite Voraussetzung, die oft gemacht wird, ist gerade Symmetrie 
der Fensterfolge und die daraus resultierende Linearphasigkeit 
des Spektrums. Diese Voraussetzung ist wohl eher historisch 
bedingt, da man fr"uher die Fensterung "ublicherweise auf eine 
gesch"atzte AKF angewendet hat, um daraus mit Hilfe einer DFT ein 
gegl"attetes LDS zu berechnen (\,Methode nach Blackman-Tukey\,). 
Wenn die gesch"atzte AKF dabei symmetrisch ist, 
w"urde bei einer unsymmetrischen Fensterung ein Spektrum mit einem 
nicht verschwindenden Imagin"arteil, und somit ein unsinniges Messergebnis 
entstehen. Da jedoch beim RKM --- wie auch bei der Spektralsch"atzung 
nach Welch-Bartlett --- an allen entscheidenden Stellen das Spektrum 
immer als Betragsquadrat auftritt, braucht die Fensterfolge beim 
RKM diese Forderung auch nicht zu erf"ullen. 

Bei allen mir bekannten 
Spektralsch"atzverfahren wurde impliziert, dass die L"ange der 
Fens"-ter"-funk"-tion die Anzahl der berechneten Frequenzpunkte des LDS 
nicht "ubersteigen darf. Da jedoch bei der DFT, die sowohl bei der 
Spektralsch"atzung als auch beim RKM zum Einsatz kommt, bei Abtastwerten, 
die zeitlich um $M$ Takte verschoben liegen, immer derselbe Drehfaktor 
auftritt, spricht nichts dagegen, die mit diesem Drehfaktor multiplizierten 
Summanden auch unterschiedlich zu gewichten. Erst dadurch, dass man eine 
Fensterfolge verwendet, die l"anger als $M$ ist, wurde es m"oglich, bei 
hoher Sperrd"ampfung zugleich die Bedingung (\ref{2.20}) zu erf"ullen. 
Deshalb wurde die Einschr"ankung \mbox{$F\!\le\!M$} von mir auch nicht verwendet.

\chapter{Beispiele f"ur RKM-Messergebnisse}\label{RKMBeisp}
\vspace*{20pt}
In diesem Kapitel, werden nun anhand einiger Beispiele die 
in den vorhergehenden Kapiteln beschriebenen Eigenschaften 
des Rauschklirrmessverfahrens mit Fensterung n"aher beleuchtet. 
Neben einer Untersuchung, welche Fehler durch die Berechnungen 
des Messverfahrens selbst entstehen, und wie sich eine zu kurze 
Einschwingzeit auf die Messergebnisse auswirkt, werden vor allen 
die Vorteile der Verwendung einer Fensterfolge beim RKM demonstriert. 
Dies wird sowohl f"ur die Messwerte der "Ubertragungsfunktion als auch 
f"ur die Messwerte des LDS gezeigt. Zwei Unterkapitel dienen dazu, 
dem Leser eine Hilfestellung bei der Interpretation der mit dem RKM 
gewonnenen Konfidenzgebiete zu bieten.

\section{Einfluss der endlichen Wortl"ange bei der Messwertberechnung}\label{Mess0}

Um die Auswirkungen absch"atzen zu k"onnen, die durch die endliche
Wortl"ange bei der Berechnung des Spektrums der Erregung, und der 
Messergebnisse entstehen, wurde eine Verz"ogerung um einen Takt mit
\mbox{$H(\Omega)=e^{-j\Omega}$} simuliert und vermessen.
Es wurde keine St"orung "uberlagert, so dass das theoretische LDS
\mbox{$\Phi_{\boldsymbol{n}}(\Omega)$} konstant Null war.

Bei der Simulation dieses Systems entstehen
keinerlei Fehler, da die Werte der Erregung unver"andert und
lediglich um einen Takt verz"ogert an den Ausgang zur Berechnung
der Messwerte weitergeleitet werden. Alle Fehler in den Messwerten
werden daher ausschlie"slich durch die endliche Wortl"ange bei der 
Messwertberechnung verursacht. 
\begin{figure}[b]
\rule{\textwidth}{0.5pt}\vspace{+15pt}
\begin{center}
{ 
\begin{picture}(454,284)
\input{mbild5a}
\put(35,238){\makebox(0,0)[rt]{\small$0$}}
\put(35,180){\makebox(0,0)[r]{\small$-100$}}
\put(35,120){\makebox(0,0)[r]{\small$-200$}}
\put(35,60){\makebox(0,0)[r]{\small$-300$}}
\put(42,17){\makebox(0,0)[lb]{\small$0$}}
\put(130,17){\makebox(0,0)[b]{\small$\pi/2$}}
\put(220,17){\makebox(0,0)[b]{\small$\pi$}}
\put(310,17){\makebox(0,0)[b]{\small$3\pi/2$}}
\put(395,18){\makebox(0,0)[b]{\small$2\pi$}}
\put(0,270){\makebox(0,0)[l]{
$20\CdoT\log_{10}\bigl(\bigl|\Hat{H}(\mu)\bigr|\bigr),
20\CdoT\log_{10}\bigl(\bigl|\Hat{H}(\mu)\!-\!
H\Left({\T\mu\CdoT\frac{2\pi}{M}}\right)\bigr|\bigr)\text{ und }
10\CdoT\log_{10}\bigl(\Hat{\Phi}_{\boldsymbol{n}}(\mu)\bigr)$}}
\put(220,40){\makebox(0,0)[l]{$20\CdoT\log_{10}\bigl(\,\bigl|\Hat{H}(\mu)-
H\Left({\T\mu\CdoT\frac{2\pi}{M}}\right)\bigr|\,\bigr)$}}
\put(220,160){\makebox(0,0)[l]{
$10\CdoT\log_{10}\bigl(\,\Hat{\Phi}_{\boldsymbol{n}}(\mu)\,\bigr)$}}
\put(220,230){\makebox(0,0)[l]{
$20\CdoT\log_{10}\bigl(\,\bigl|\Hat{H}(\mu)\bigr|\,\bigr)$}}
\put(105,100){\makebox(0,0)[l]{$20\CdoT\log_{10}(\varepsilon)$}}
\put(105,180){\makebox(0,0)[l]{$10\CdoT\log_{10}(\varepsilon)$}}
\put(455,22){\makebox(0,0)[tr]{${\T\Omega=\mu\CdoT\frac{2\pi}{M}}$}}

\end{picture}}
\end{center}\vspace{-15pt}
\setlength{\belowcaptionskip}{8pt}
\caption[Einfluss der endlichen Wortl"ange bei der Messwertberechnung]{
Einfluss der endlichen Wortl"ange bei der Messwertberechnung.\protect\\
Beispielsystem: Verz"ogerungsglied.\protect\\
Messung mit: \mbox{$M\!=\!1024$}, \mbox{$E\!=\!1$}, \mbox{$L\!=\!50$}
und Fenster nach Kapitel \protect\ref{Algo} mit \mbox{$N\!=\!4$}.}
\label{b5a}
\end{figure}

Erregt wurde mit einem
reellen, mittelwertfreien, wei"sen und normalverteilten Zufallsvektor der
L"ange \mbox{$M\!=\!1024$}, dessen Elemente die Varianz Eins aufwiesen.
Dieser wurde im Zeitintervall \mbox{$[-E;F\!-\!1]$} mit \mbox{$E\!=\!1$} und
\mbox{$F\!=\!4096$} periodisch fortgesetzt, so dass keine St"orungen
aufgrund einer zu kurzen Einschwingzeit entstehen konnten. Als Fensterfolge
wurde das mit dem in Kapitel~\ref{Algo} beschriebenen Algorithmus
berechnete Fenster verwendet, wobei dort der Parameter \mbox{$N\!=\!4$}
eingestellt wurde, was zu einem Fenster der L"ange \mbox{$F\!=\!4096$}
f"uhrte. Die letzten drei Werte dieser Fensterfolge sind theoretisch
Null, wurden jedoch aus den in Kapitel \ref{Rech} beschriebenen Gr"unden 
nach der Berechnung der Fensterfolge nicht explizit auf Null gesetzt. 
Es wurde ein reellwertiges, lineares und zeitinvariantes Modellsystem
angesetzt, und dessen "Ubertragungsfunktion gemessen. Von den Fehlern
bei der Messwertberechnung wurde angenommen, dass sie station"ar sind,
so dass es sinnvoll erschien ein eindimensionales LDS zu messen. Da es 
sich hier um ein reellwertiges System handelt, wurde auf die Messung 
des MLDS \mbox{$\Hat{\Psi}_{\boldsymbol{n}}(\mu)$} verzichtet. Der
Stichprobenumfang (\,=~Anzahl der Einzelmessungen\,) betrug
\mbox{$L\!=\!50$}. Die Berechnung erfolgte mit der 8-Byte
Gleitkommaarithmetik nach IEEE-Standard 754, bei der sich
f"ur die relative Genauigkeit der Zahlendarstellung der Wert
\mbox{$\varepsilon=2^{-52}\approx2,22\CdoT10^{-16}$} ergibt.

Die Messergebnisse sind in Bild~\ref{b5a} dargestellt. Als Hilfslinien 
sind die Werte \mbox{$20\CdoT\log_{10}(\varepsilon)$} und 
\mbox{$10\CdoT\log_{10}(\varepsilon)$} eingetragen. Ein 
Rauschsockel dieser Gr"o"senordnung w"urde sich bei einer 
St"orung der Streuung $\varepsilon$ theoretisch ergeben, 
die sich dem Signal am untersuchten Verz"ogerungsglied 
additiv "uberlagert. Die Messwerte des LDS weisen bei vielen 
Frequenzen den Wert Null auf. Da diese Werte im logarithmischen 
Ma"sstab graphisch nicht dargestellt werden k"onnen, erscheint 
diese Messkurve l"uckenhaft.
\newpage

\section{Einfluss eines periodischen St"orers auf die Messung der
"Ubertragungs\protect\-funktion}\label{Mess05}

\begin{figure}[btp]
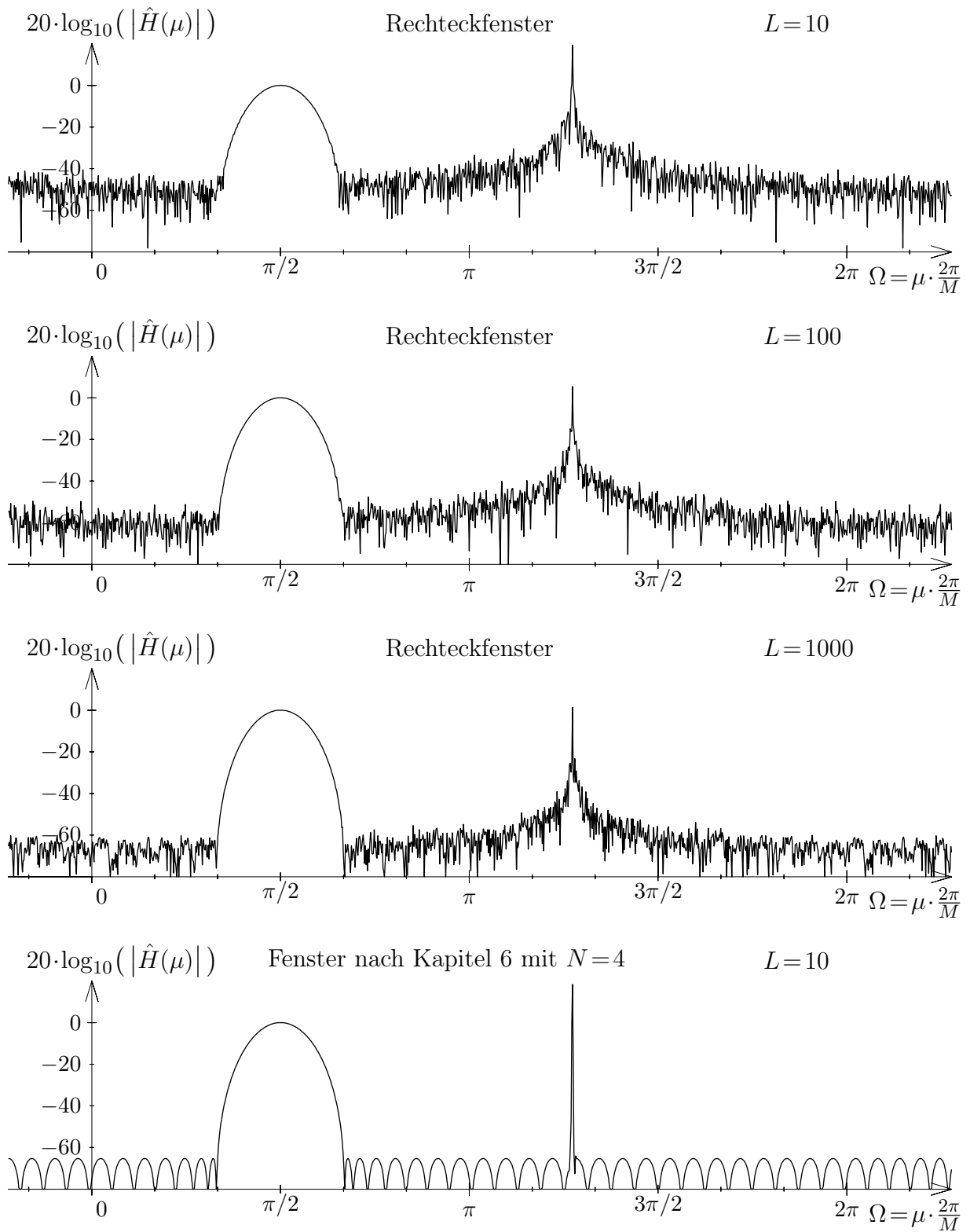

\begin{center}
{ 
\begin{picture}(454,599)

\input{mbild5b1}
\put(35,560){\makebox(0,0)[r]{\small$0$}}
\put(35,540){\makebox(0,0)[r]{\small$-20$}}
\put(35,520){\makebox(0,0)[r]{\small$-40$}}
\put(35,500){\makebox(0,0)[r]{\small$-60$}}
\put(42,467){\makebox(0,0)[lb]{\small$0$}}
\put(130,467){\makebox(0,0)[b]{\small$\pi/2$}}
\put(220,467){\makebox(0,0)[b]{\small$\pi$}}
\put(310,467){\makebox(0,0)[b]{\small$3\pi/2$}}
\put(400,467){\makebox(0,0)[b]{\small$2\pi$}}
\put(455,475){\makebox(0,0)[tr]{${\T\Omega\!=\!\mu\CdoT\frac{2\pi}{M}}$}}
\put(5,590){\makebox(0,0)[l]{
$20\CdoT\log_{10}\bigl(\,\bigl|\Hat{H}(\mu)\bigr|\,\bigr)$}}
\put(180,590){\makebox(0,0)[l]{Rechteckfenster}}
\put(360,590){\makebox(0,0)[l]{$L\!=\!10$}}

\input{mbild5b2}
\put(35,410){\makebox(0,0)[r]{\small$0$}}
\put(35,390){\makebox(0,0)[r]{\small$-20$}}
\put(35,370){\makebox(0,0)[r]{\small$-40$}}
\put(35,350){\makebox(0,0)[r]{\small$-60$}}
\put(42,317){\makebox(0,0)[lb]{\small$0$}}
\put(130,317){\makebox(0,0)[b]{\small$\pi/2$}}
\put(220,317){\makebox(0,0)[b]{\small$\pi$}}
\put(310,317){\makebox(0,0)[b]{\small$3\pi/2$}}
\put(400,317){\makebox(0,0)[b]{\small$2\pi$}}
\put(455,325){\makebox(0,0)[tr]{${\T\Omega\!=\!\mu\CdoT\frac{2\pi}{M}}$}}
\put(5,440){\makebox(0,0)[l]{
$20\CdoT\log_{10}\bigl(\,\bigl|\Hat{H}(\mu)\bigr|\,\bigr)$}}
\put(180,440){\makebox(0,0)[l]{Rechteckfenster}}
\put(360,440){\makebox(0,0)[l]{$L\!=\!100$}}

\input{mbild5b3}
\put(35,260){\makebox(0,0)[r]{\small$0$}}
\put(35,240){\makebox(0,0)[r]{\small$-20$}}
\put(35,220){\makebox(0,0)[r]{\small$-40$}}
\put(35,200){\makebox(0,0)[r]{\small$-60$}}
\put(42,167){\makebox(0,0)[lb]{\small$0$}}
\put(130,167){\makebox(0,0)[b]{\small$\pi/2$}}
\put(220,167){\makebox(0,0)[b]{\small$\pi$}}
\put(310,167){\makebox(0,0)[b]{\small$3\pi/2$}}
\put(400,167){\makebox(0,0)[b]{\small$2\pi$}}
\put(455,175){\makebox(0,0)[tr]{${\T\Omega\!=\!\mu\CdoT\frac{2\pi}{M}}$}}
\put(5,290){\makebox(0,0)[l]{
$20\CdoT\log_{10}\bigl(\,\bigl|\Hat{H}(\mu)\bigr|\,\bigr)$}}
\put(180,290){\makebox(0,0)[l]{Rechteckfenster}}
\put(360,290){\makebox(0,0)[l]{$L\!=\!1000$}}

\input{mbild5b4}
\put(35,110){\makebox(0,0)[r]{\small$0$}}
\put(35,90){\makebox(0,0)[r]{\small$-20$}}
\put(35,70){\makebox(0,0)[r]{\small$-40$}}
\put(35,50){\makebox(0,0)[r]{\small$-60$}}
\put(42,17){\makebox(0,0)[lb]{\small$0$}}
\put(130,17){\makebox(0,0)[b]{\small$\pi/2$}}
\put(220,17){\makebox(0,0)[b]{\small$\pi$}}
\put(310,17){\makebox(0,0)[b]{\small$3\pi/2$}}
\put(400,17){\makebox(0,0)[b]{\small$2\pi$}}
\put(455,25){\makebox(0,0)[tr]{${\T\Omega\!=\!\mu\CdoT\frac{2\pi}{M}}$}}
\put(5,140){\makebox(0,0)[l]{
$20\CdoT\log_{10}\bigl(\,\bigl|\Hat{H}(\mu)\bigr|\,\bigr)$}}
\put(120,140){\makebox(0,0)[l]{
Fenster nach Kapitel \protect\ref{Algo} mit \mbox{$N\!=\!4$}}}
\put(360,140){\makebox(0,0)[l]{$L\!=\!10$}}

\end{picture}}
\end{center}\vspace{-30pt}
\caption[Komplexwertiger Bandpass mit Tschebyscheff-Verhalten im Sperrbereich,
der von einem komplexen Eintonst"orer mit Zufallsphase gest"ort wird.]{
Komplexwertiger Bandpass mit Tschebyscheff-Verhalten im Sperrbereich, der
von einem komplexen Eintonst"orer mit Zufallsphase gest"ort wird.\protect\\
Messung mit: \mbox{$M\!=\!1024$} und \mbox{$E\!=\!31$}.}
\label{b5b}
\end{figure}
Um deutlich zu machen, wie sich ein schmalbandiges St"orsignal bei der
Messung der "Ubertragungsfunktion auswirkt, wurde ein komplexwertiger
FIR-Bandpass mit Tschebyscheff-Verhalten im Sperrbereich simuliert,
dem ausgangsseitig ein komplexer Eintonst"orer 
\[
n(k) = e^{\T j\CdoT4\CdoT k+\boldsymbol{\varphi}}
\]
mit der Kreisfrequenz $4$ und einer in \mbox{$[0;2\pi)$}
gleichverteilten zuf"alligen Phase "uberlagert wurde.
Das LDS dieser station"aren St"orung ist ein Dirac-Impuls der
St"arke $2\pi$ bei der Kreisfrequenz $4$. Die Filterkoeffizienten
des Bandpasses erh"alt man, indem man die Werte eines
Dolph-Tschebyscheff-Fensters der L"ange $32$ mit einer
Sperrkreisfrequenz von $\pi/6$, das z.~B. im Unterkapitel 
\ref{Andere} auf Seite \pageref{Dolph}
oder in \cite{Kam} oder \cite{Thomae} beschrieben ist, mit $j^k$ 
multipliziert, und so normiert, dass sich f"ur \mbox{$\Omega=\pi/2$} 
der Wert Eins f"ur die "Ubertragungsfunktion des Bandpasses ergibt. 

Erregt wurde mit einem komplexen mittelwertfreien wei"sen Gau"szufallsvektor
der L"ange \mbox{$M\!=\!1024$} mit unkorreliertem Real- und Imagin"arteil mit 
jeweils der Varianz $1/2$, so dass der komplexe Zufallsvektor die 
zeitunabh"angige Varianz Eins aufwies. Dieser Zufallsvektor wurde 
im Zeitintervall \mbox{$[-E;F\!-\!1]$} mit $M$ periodisch fortgesetzt.
Als Einschwingzeit \mbox{wurde} \mbox{$E\!=\!31$} gew"ahlt, so dass keine
St"orungen aufgrund zu kurzer Einschwingzeit entstehen konnten.
Es wurde ein komplexwertiges, lineares und zeitinvariantes Modellsystem
angesetzt, und dessen "Ubertragungsfunktion \mbox{$\Hat{H}(\mu)$} f"ur
die $M$ "aquidistanten Frequenzen \mbox{$\Omega=\mu\CdoT2\pi/M$} gemessen.
Die Messung erfolgte mit zwei verschiedenen Fensterfolgen. Zum einen 
wurde mit einem Rechteckfenster der L"ange \mbox{$F\!=\!M\!=\!1024$},
und zum anderen mit dem nach Kapitel \ref{Algo} mit \mbox{$N\!=\!4$}
berechneten Fenster der L"ange \mbox{$F=N\CdoT M=4096$} ge"-fens"-tert.

Die Messergebnisse sind in Bild \ref{b5b} f"ur die
Mittelungsanzahlen \mbox{$L\in\{10;100;1000\}$} f"ur das Rechteckfenster und
f"ur \mbox{$L\!=\!10$} f"ur das Fenster nach Kapitel \ref{Algo}
dargestellt. Bei Verwendung des Rechteckfensters ist die Varianz der Messwerte
\mbox{$\Hat{H}(\mu)$} aufgrund der schlechten Unterdr"uckung des
Impulses des LDS der St"orung im gesamten Frequenzbereich so gro"s,
dass selbst bei einer Mittelungsanzahl von \mbox{$L\!=\!1000$} die
Messwerte der "Ubertragungsfunktion lediglich im Durchlassbereich
einigerma"sen zuverl"assig bestimmt werden konnten. Bei Verwendung
des nach Kapitel \ref{Algo} berechneten Fensters sind einzig die
Messwerte in unmittelbarer Umgebung der Kreisfrequenz des St"orers
so stark gest"ort, dass diese nicht brauchbar sind. So ist selbst bei
einer Mittelungsanzahl von \mbox{$L\!=\!10$} bereits deutlich das typische
"`equiripple\,"' Verhalten der Sperrd"ampfung des Tschebyscheff-Bandpasses
zu erkennen.

\section{Messung mit zu kurzer Einschwingzeit}

Bei der im letzten Unterkapitel beschriebenen Messung des simulierten 
komplexwertigen Tschebyscheff-FIR-Bandpasses wurde nun der komplexe 
Eintonst"orer mit der Zufallsphase weggelassen, und somit das ungest"orte 
System gemessen. Die Einschwingzeit $E$ wurde dabei bewusst falsch gew"ahlt 
und auf Null gesetzt. Der gesamte Einschwingvorgang fiel also immer in 
den Zeitbereich \mbox{$[0;F\!-\!1]$} der Messung. Bei jeder Einzelmessung 
wurden die Zustandsgr"o"sen des FIR-Bandpasses mit Null initialisiert. 
Nach \mbox{$L\!=\!100$} Einzelmessungen wurde die "Ubertragungsfunktion 
bestimmt, wobei zum einen ein Rechteckfenster der L"ange 
\mbox{$F\!=\!M\!=\!1024$} und zum anderen das nach Kapitel \ref{Algo} 
mit \mbox{$N\!=\!4$} berechnete Fenster der L"ange \mbox{$F=N\CdoT M=4096$} 
verwendet wurde. 

Die beiden oberen Teilbilder in 
Bild~\ref{b5w}
\begin{figure}[btp]
\begin{center}
{ 
\begin{picture}(454,599)

\input{mbild5w1}
\put(35,560){\makebox(0,0)[r]{\small$0$}}
\put(35,540){\makebox(0,0)[r]{\small$-20$}}
\put(35,520){\makebox(0,0)[r]{\small$-40$}}
\put(35,500){\makebox(0,0)[r]{\small$-60$}}
\put(42,467){\makebox(0,0)[lb]{\small$0$}}
\put(130,467){\makebox(0,0)[b]{\small$\pi/2$}}
\put(220,467){\makebox(0,0)[b]{\small$\pi$}}
\put(310,467){\makebox(0,0)[b]{\small$3\pi/2$}}
\put(400,467){\makebox(0,0)[b]{\small$2\pi$}}
\put(455,475){\makebox(0,0)[tr]{${\T\Omega\!=\!\mu\CdoT\frac{2\pi}{M}}$}}
\put(5,590){\makebox(0,0)[l]{
$20\CdoT\log_{10}\bigl(\,\bigl|\Hat{H}(\mu)\bigr|\,\bigr)$}}
\put(180,590){\makebox(0,0)[l]{Rechteckfenster mit \mbox{$N\!=\!1$}}}
\put(360,590){\makebox(0,0)[l]{$L\!=\!100$}}

\input{mbild5w2}
\put(35,410){\makebox(0,0)[r]{\small$0$}}
\put(35,390){\makebox(0,0)[r]{\small$-20$}}
\put(35,370){\makebox(0,0)[r]{\small$-40$}}
\put(35,350){\makebox(0,0)[r]{\small$-60$}}
\put(42,317){\makebox(0,0)[lb]{\small$0$}}
\put(130,317){\makebox(0,0)[b]{\small$\pi/2$}}
\put(220,317){\makebox(0,0)[b]{\small$\pi$}}
\put(310,317){\makebox(0,0)[b]{\small$3\pi/2$}}
\put(400,317){\makebox(0,0)[b]{\small$2\pi$}}
\put(455,325){\makebox(0,0)[tr]{${\T\Omega\!=\!\mu\CdoT\frac{2\pi}{M}}$}}
\put(5,440){\makebox(0,0)[l]{
$20\CdoT\log_{10}\bigl(\,\bigl|\Hat{H}(\mu)\bigr|\,\bigr)$}}
\put(120,440){\makebox(0,0)[l]{Fenster mit \mbox{$N\!=\!4$}}}
\put(360,440){\makebox(0,0)[l]{$L\!=\!100$}}

\input{mbild5w3}
\put(35,260){\makebox(0,0)[r]{\small$0$}}
\put(35,240){\makebox(0,0)[r]{\small$-20$}}
\put(35,220){\makebox(0,0)[r]{\small$-40$}}
\put(35,200){\makebox(0,0)[r]{\small$-60$}}
\put(42,167){\makebox(0,0)[lb]{\small$0$}}
\put(130,167){\makebox(0,0)[b]{\small$\pi/2$}}
\put(220,167){\makebox(0,0)[b]{\small$\pi$}}
\put(310,167){\makebox(0,0)[b]{\small$3\pi/2$}}
\put(400,167){\makebox(0,0)[b]{\small$2\pi$}}
\put(455,175){\makebox(0,0)[tr]{${\T\Omega\!=\!\mu\CdoT\frac{2\pi}{M}}$}}
\put(5,290){\makebox(0,0)[l]{
$10\CdoT\log_{10}\bigl(\,\bigl|\Hat{\Phi}_{\boldsymbol{n}}(\mu)\bigr|\,\bigr)$}}
\put(180,290){\makebox(0,0)[l]{Rechteckfenster mit \mbox{$N\!=\!1$}}}
\put(360,290){\makebox(0,0)[l]{$L\!=\!100$}}

\input{mbild5w4}
\put(35,110){\makebox(0,0)[r]{\small$-120$}}
\put(35,90){\makebox(0,0)[r]{\small$-140$}}
\put(35,70){\makebox(0,0)[r]{\small$-160$}}
\put(35,50){\makebox(0,0)[r]{\small$-180$}}
\put(42,17){\makebox(0,0)[lb]{\small$0$}}
\put(130,17){\makebox(0,0)[b]{\small$\pi/2$}}
\put(220,17){\makebox(0,0)[b]{\small$\pi$}}
\put(310,17){\makebox(0,0)[b]{\small$3\pi/2$}}
\put(400,17){\makebox(0,0)[b]{\small$2\pi$}}
\put(455,25){\makebox(0,0)[tr]{${\T\Omega\!=\!\mu\CdoT\frac{2\pi}{M}}$}}
\put(5,140){\makebox(0,0)[l]{
$10\CdoT\log_{10}\bigl(\,\bigl|\Hat{\Phi}_{\boldsymbol{n}}(\mu)\bigr|\,\bigr)$}}
\put(120,140){\makebox(0,0)[l]{Fenster nach Kapitel \protect\ref{Algo} mit \mbox{$N\!=\!4$}}}
\put(360,140){\makebox(0,0)[l]{$L\!=\!100$}}

\end{picture}}
\end{center}\vspace{-30pt}
\caption[Komplexwertiger Bandpass mit Tschebyscheff-Verhalten im Sperrbereich.
Messung mit zu kurzer Einschwingzeit.]{
Komplexwertiger Bandpass mit Tschebyscheff-Verhalten im Sperrbereich.
\protect\\
Messung mit: \mbox{$M\!=\!1024$} und \mbox{$E\!=\!0$}.}
\label{b5w}
\end{figure}
zeigen die damit gemessenen 
"Ubertragungsfunktionen. Deutlich ist zu erkennen, dass das Messergebnis 
bei Verwendung des Rechteckfensters zumindest im Sperrbereich 
des Tschebyscheff-Bandpasses unbrauchbar ist, w"ahrend bei der 
Messung mit dem Fenster nach Kapitel \ref{Algo} die 
"Ubertragungsfunktion praktisch fehlerfrei messbar ist. 
Dieser deutliche Unterschied liegt darin begr"undet, dass das 
Fenster mit \mbox{$N\!=\!4$} am Anfang des Zeitintervalls 
\mbox{$[0;F\!-\!1]$}, also dort, wo sich die Einschwingvorg"ange auswirken, 
sehr kleine Werte aufweist, die mit zunehmendem $k$ nur sehr langsam 
ansteigen. Der Einfluss des Einschwingvorgangs, der sich in diesem Beispiel 
auf die ersten $31$ Werte des Signals am Systemausgang begrenzt, ist daher 
deutlich geringer als beim Rechteckfenster. Erfahrungsgem"a"s 
steigt die gespiegelte, maximalphasige Fensterfolge noch langsamer 
an, so dass man, wenn man nicht sicher ist, dass die Einschwingzeit 
gro"s genug gew"ahlt worden ist, die gespiegelte Fensterfolge bevorzugen 
sollte. 

Die beiden unteren Teilbilder in Bild~\ref{b5w} zeigen die 
bei der Messung der dar"uber dargestellten "Ubertragungsfunktionen 
zugleich gemessenen Leistungsdichtespektren. Die 
Einschwingvorg"ange bewirken offensichtlich eine Verf"alschung 
der Messergebnisse des LDS, das in unserem Fall eigentlich im 
gesamten Frequenzbereich Null sein m"usste. Bei der Messung wurde 
f"alschlicherweise angenommen, dass --- wenn "uberhaupt --- eine 
station"are St"orung vorliegt. Diese Annahme, auf der die gesamte 
Theorie der Kapitel \ref{SysMod} und \ref{RKM} aufbaut, wird hier auf das 
gr"oblichste verletzt. Wenn man die Einschwingvorg"ange als eine additiv 
"uberlagerte St"orung modelliert, so ist diese, abgesehen davon, dass sie 
von der Erregung sicher {\em nicht}\/ unabh"angig ist, auch noch instation"ar. 
Daher k"onnen die Messwerte \mbox{$\Hat{\Phi}_{\boldsymbol{n}}(\mu)$} 
den wahren Wert Null des LDS nicht erwartungstreu absch"atzen. 

Obwohl bei der Messung ein komplexer mittelwertfreier \mbox{wei"ser} 
Gau"szufallsvektor mit unkorreliertem Real- und Imagin"arteil zur Erregung 
verwendet wurde, ist das gemessene LDS {\em nicht} wei"s also "uber der Frequenz 
konstant. Es weist im Bereich des Maximums der "Ubertragungsfunktion ebenfalls 
ein Maximum auf. "Ahnliches Verhalten konnte auch bei anderen Messungen 
mit zu kurzer Einschwingzeit beobachtet werden. Als Faustregel, die theoretisch 
nicht weiter begr"undet ist, kann man, wenn man beim gemessenen LDS einen 
"ahnlichen Frequenzverlauf wie bei der "Ubertragungsfunktion erh"alt, sagen, dass 
dies ein Hinweis auf eine zu kurze Einschwingzeit sein kann. Es empfiehlt sich dann, 
die Messung mit jeweils zunehmender Einschwingzeit solange zu wiederholen, bis sich 
die Messergebnisse nicht mehr ver"andern.

\section{Messung des LDS mit Fenstern
unterschiedlicher L"ange}\label{Mess7}

Um einerseits zu demonstrieren, dass es ein Anstieg der Sperrd"ampfung 
der Fensterfolge mit der Potenz $N$ in \mbox{$\sin(\Omega/2)$} erm"oglicht, 
eine \mbox{$2\CdoT N$}-fache Nullstelle im LDS zu messen, und andererseits 
zu zeigen, wie man auch ohne die Kenntnis des wahren LDS erkennen kann, 
dass mit einer zu kurzen Fensterfolge gemessen wurde, wird bei 
diesem Beispiel das LDS eines gefilterten, wei"sen, station"aren, 
normierten und zentrierten Gau"sprozesses gemessen. Als Filter des 
Gau"sprozesses wird dabei ein Filter mit der Z-Transformierten 
\mbox{$H_{\boldsymbol{n}}(z)=(z^{-1}\!\!-\!1)^3$} verwendet, so 
dass sich im LDS des Prozesses am Ausgang des Filters eine sechsfache 
Nullstelle bei \mbox{$\Omega\!=\!0$} ergibt. 

F"ur diesen Prozess wird nun 
eine Spektralsch"atzung nach dem in Kapitel \ref{LDS} beschriebenen 
Verfahren mit \mbox{$M\!=\!1024$} Frequenzmesspunkten durchgef"uhrt. 
In Bild~\ref{b5f}
\begin{figure}[btp]
\begin{center}
{ 
\begin{picture}(450,600)

\input{mbild5f}
\put(65,579){\makebox(0,0)[lb]{Rechteckfenster = Fenster mit \mbox{$N\!=\!1$}}}
\put(15,550){\vector(0,1){24}}
\put(15,540){\makebox(0,0)[t]{\rotatebox{90}
{$10\cdot\log_{10}\big(\,\big|\Phi_{\boldsymbol{n}}(\Omega)\big|\,\big)$}}}
\put(425,394){\vector(1,0){18}}
\put(420,394){\makebox(0,0)[r]{$20\cdot\log_{10}\big(\sin(\Omega/2)\big)$}}
\put(131,562){\makebox(0,0)[l]{gemessen}}
\put(131,545){\makebox(0,0)[l]{theoretisch}}

\put(65,387){\makebox(0,0)[lb]{Fenster mit \mbox{$N\!=\!2$}}}
\put(15,358){\vector(0,1){24}}
\put(15,348){\makebox(0,0)[t]{\rotatebox{90}
{$10\cdot\log_{10}\big(\,\big|\Phi_{\boldsymbol{n}}(\Omega)\big|\,\big)$}}}
\put(425,202){\vector(1,0){18}}
\put(420,202){\makebox(0,0)[r]{$20\cdot\log_{10}\big(\sin(\Omega/2)\big)$}}
\put(131,370){\makebox(0,0)[l]{gemessen}}
\put(131,353){\makebox(0,0)[l]{theoretisch}}

\put(65,195){\makebox(0,0)[lb]{Fenster mit \mbox{$N\!=\!3$}}}
\put(15,166){\vector(0,1){24}}
\put(15,156){\makebox(0,0)[t]{\rotatebox{90}
{$10\cdot\log_{10}\big(\,\big|\Phi_{\boldsymbol{n}}(\Omega)\big|\,\big)$}}}
\put(425,10){\vector(1,0){18}}
\put(420,10){\makebox(0,0)[r]{$20\cdot\log_{10}\big(\sin(\Omega/2)\big)$}}
\put(131,178){\makebox(0,0)[l]{gemessen}}
\put(131,161){\makebox(0,0)[l]{theoretisch}}

\end{picture}}
\end{center}\vspace{-15pt}
\caption{LDS mit sechsfacher Nullstelle im LDS bei
\mbox{$\Omega\!=\!0$}.\protect\\
Messung mit: \mbox{$M\!=\!1024$} und \mbox{$L\!=\!10$}.}
\label{b5f}
\end{figure}
sind die Ergebnisse der Messung nach 
\mbox{$L\!=\!10$} Einzelmessungen f"ur die drei Werte \mbox{$N\in\{1;2;3\}$} 
der nach Kapitel \ref{Algo} berechneten Fensterfolgen in einem 
doppelt logarithmischen Ma"sstab dargestellt. Da es sich hier um ein 
zeitdiskretes System handelt, bei dem das LDS eine periodische 
Funktion der normierten Kreisfrequenz \mbox{$\Omega$} ist, wurde bei 
der Abszisse der Sinus der halben Kreisfrequenz logarithmiert. F"ur 
das theoretische LDS ergibt sich bei dieser Art der Darstellung 
eine Gerade mit der Steigung drei, die durch die dreifache Nullstelle 
der Z-Transformierten des Filters des Gau"sprozesses bestimmt ist. Man 
kann erkennen, dass erst mit der Fensterfolge mit \mbox{$N\!=\!3$} diese 
Gerade "uber dem gesamten Dynamikbereich der Rechengenauigkeit bei 8-Byte 
Gleitkommadarstellung (\,entspricht ca. 156~dB\,) gemessen werden kann. 

Desweiteren ist zu sehen, dass in dem Bereich, in dem die gemessene Kurve 
von der theoretischen Kurve abweicht, das gemessene LDS weniger stark 
verrauscht erscheint. Nach der in den Kapiteln \ref{Kova} und \ref{LDS} 
dargestellten Theorie besteht zwischen der Varianz der Messwerte und 
dem theoretischen LDS eine Proportionalit"at, deren Faktor nach Gleichung 
(\ref{5.6}) nur von der Mittelungsanzahl abh"angt, sofern eine hoch 
frequenzselektive Fensterfolge verwendet wird. Nach Gleichung (\ref{3.67})
sind in diesem Fall bei einem station"aren Prozess benachbarte Spektralwerte 
unkorreliert. Daher m"usste die Messkurve bei logarithmischer Darstellung der 
Ordinate im gesamten Frequenzbereich etwa gleich stark verrauscht erscheinen. 
Da jedoch auch die Abszisse logarithmisch aufgetragen wurde, liegen 
die linear "aquidistanten Frequenzpunkte bei niedrigen Frequenzen, 
also in dem Bereich, in dem die Messergebnisse bei zu geringer 
Fensterl"ange von der theoretischen Geraden abweichen, weiter 
auseinander als im Bereich guter "Ubereinstimmung. Dadurch ist zu 
erwarten, dass die Kurve bei niedrigen Frequenzen glatter erscheint. 

Dass dies jedoch nicht der Hauptgrund f"ur dieses Ph"anomen ist, wird an 
den in Bild \ref{b5g}
\begin{figure}[btp]
\begin{center}
{ 
\begin{picture}(450,600)

\input{mbild5g}
\put(65,579){\makebox(0,0)[lb]{Rechteckfenster = Fenster mit \mbox{$N\!=\!1$}}}
\put(15,550){\vector(0,1){24}}
\put(15,540){\makebox(0,0)[t]{\rotatebox{90}
{$10\cdot\log_{10}\big(\,\big|\Phi_{\boldsymbol{n}}(\Omega)\big|\,\big)$}}}
\put(400,398){\vector(1,0){48}}
\put(395,398){\makebox(0,0)[r]{$\Omega$}}
\put(131,482){\makebox(0,0)[l]{gemessen}}
\put(131,465){\makebox(0,0)[l]{theoretisch}}

\put(65,387){\makebox(0,0)[lb]{Fenster mit \mbox{$N\!=\!2$}}}
\put(15,358){\vector(0,1){24}}
\put(15,348){\makebox(0,0)[t]{\rotatebox{90}
{$10\cdot\log_{10}\big(\,\big|\Phi_{\boldsymbol{n}}(\Omega)\big|\,\big)$}}}
\put(400,206){\vector(1,0){48}}
\put(395,206){\makebox(0,0)[r]{$\Omega$}}
\put(131,290){\makebox(0,0)[l]{gemessen}}
\put(131,273){\makebox(0,0)[l]{theoretisch}}

\put(65,195){\makebox(0,0)[lb]{Fenster mit \mbox{$N\!=\!3$}}}
\put(15,166){\vector(0,1){24}}
\put(15,156){\makebox(0,0)[t]{\rotatebox{90}
{$10\cdot\log_{10}\big(\,\big|\Phi_{\boldsymbol{n}}(\Omega)\big|\,\big)$}}}
\put(400,14){\vector(1,0){48}}
\put(395,14){\makebox(0,0)[r]{$\Omega$}}
\put(131,98){\makebox(0,0)[l]{gemessen}}
\put(131,81){\makebox(0,0)[l]{theoretisch}}

\end{picture}}
\end{center}\vspace{-20pt}
\caption{LDS des mit einem Butterworth-Halbbandfilter gefilterten
Gau"sprozesses.\protect\\
Messung mit: \mbox{$M\!=\!1024$} und \mbox{$L\!=\!10$}.}
\label{b5g}
\end{figure}
dargestellten Ergebnissen deutlich. 
Hier wurde f"ur die Abszisse ein linearer Ma"sstab gew"ahlt. Um diesen
Effekt besonders deutlich darstellen zu k"onnen, wurde einerseits die
kleine Mittelungsanzahl \mbox{$L\!=\!10$} und andererseits ein anderes
Filter f"ur den Gau"sprozess verwendet. Als Filterimpulsantwort wurde
\begin{equation}
h_{\boldsymbol{n}}(k)\;=\;\begin{cases}
{\D \quad \binom{n}{\frac{n-1}{2}}\cdot\binom{n}{\frac{k}{2}}\cdot
\frac{(n\!+\!1)\cdot(-1)^{\frac{k}{2}}}{(n\!-\!k)\cdot 2^{2\cdot n+1}}}&
\quad\text{ f"ur }\quad k\text{ gerade }\;\wedge\;0\le k \le 2\CdoT n\\
\quad 0,\!500001&\quad\text{ f"ur }\quad k=n\\
\quad 0&\quad\text{ sonst, }
\end{cases}
\raisetag{20pt}\label{7.1}
\end{equation}
mit \mbox{$n\!=\!25$} gew"ahlt. W"are der mittlere Filterkoeffizient 
\mbox{$0,\!5$}, so w"urde es sich hier um ein linearphasiges 
Butterworth-Halbband-FIR-Filter handeln. Die Sperrd"ampfung dieses 
Filters w"are so hoch, dass diese schon wegen der endlichen 
Berechnungswortl"ange nicht mehr messbar w"are. Daher wurde 
der mittlere Filterkoeffizient um $10^{-6}$ erh"oht, um eine 
noch messbare Sperrd"ampfung von -120~dB zu erhalten. 

Die Messergebnisse zeigen auch bei diesem System, dass weder mit 
dem Rechteckfenster mit \mbox{$N\!=\!1$}, noch mit dem Fenster mit 
\mbox{$N\!=\!2$} eine Messung der Sperrd"ampfung m"oglich ist. 
Au"serdem sieht man, dass trotz der linearen Abszissenskalierung,
die Messkurve im Sperrbereich des Filters bei diesen beiden
Werten von $N$ glatt erscheint, w"ahrend sie f"ur \mbox{$N\!=\!3$}
das zu erwartende Rauschverhalten aufweist. Erkl"aren l"asst sich
dieser Effekt folgenderma"sen: Bei jeder Einzelmessung wird das 
Spektrum der Musterfolge \mbox{$y_{\lambda}(k)$} mit dem Spektrum 
der Fensterfolge gefaltet. Bei einem Messwert in dem Frequenzbereich,
in dem die Messung des LDS nicht m"oglich ist, "uberwiegt bei dem 
Faltungsintegral wegen der zu geringen Sperrd"ampfung des Spektrums 
der Fensterfolge der Beitrag des Integrationsbereichs, der dem 
Durchlassbereich des Filters \mbox{$H_{\boldsymbol{n}}(z)$} entspricht, 
und in dem das Spektrum der Musterfolge betraglich gro"s ist. 
Der Teil des Faltungsintegrals, der durch das Fensterspektrum 
eigentlich ausgeschnitten werden soll, und dessen Varianz eigentlich 
gemessen werden soll, liefert nur einen unwesentlichen Beitrag.
Der dominierende Anteil des Faltungsintegrals ist bei allen benachbarten 
Frequenzen in dem Spektralbereich, in dem die Messung des LDS nicht 
m"oglich ist, betraglich etwa gleich, so dass die Betragsquadrate dieser 
gefalteten Spektralwerte stark korreliert sind. Durch diese Korrelationen 
erscheint die Messkurve in diesem Frequenzbereich gegl"attet. 

Wenn man also 
bei einer Messung mit dem RKM im Bereich kleiner Messwerte des LDS 
auf"|f"allig glatte Messkurven erh"alt, ist zu vermuten, dass diese 
Messwerte nicht in Ordnung sind, und die Messung ist gegebenfalls mit einer
frequenzselektiveren Fensterfolge zu wiederholen. Um dies entscheiden
zu k"onnen, ben"otigt man nur die Messwerte und nicht das im Allgemeinen 
unbekannte LDS.\vspace{-3pt minus 6pt}

\section{Konfidenzgebiete der Messwerte}\label{Mess11}

Bei dieser Beispielmessung soll anhand des Messwertes
\mbox{$\Hat{\boldsymbol{H}}({\T\frac{M}{2}})$} die Bedeutung der
nach Kapitel \ref{Kon} berechneten Konfidenzellipsen demonstriert
werden. Die Messung erfolgte dabei nach der in Kapitel \ref{RKM}
hergeleiteten Methode. Bei dem dort vorliegenden Fall eines 
station"aren Approximationsfehlerprozesses kann es nach Gleichung 
(\ref{3.63}) bei den Messwerten der "Ubertragungsfunktion nur f"ur die
beiden Frequenzpunkte \mbox{$\Omega=0$} und \mbox{$\Omega=\pi$}
zu einer von Null verschiedenen Messwertkovarianz kommen. Nach Gleichung 
(\ref{3.79}) kann man somit nur f"ur diese beiden Frequenzen eine "`echte"' 
Ellipse mit einer von Null verschiedenen Exzentrizit"at als Konfidenzgebiet 
erhalten. 

Damit sich f"ur \mbox{$\Hat{\boldsymbol{H}}({\T\frac{M}{2}})$} eine 
von Null verschiedene Messwertkovarianz ergibt, muss in Gleichung (\ref{3.63}) 
sowohl der von der Erregung abh"angige Erwartungswert als auch der das MLDS 
der St"orung beschreibende Wert \mbox{$\Tilde{\Psi}_{\boldsymbol{n}}(\frac{M}{2})$} 
von Null verschieden sein. Daher wurde bei der Messung mit einem Zufallsvektor 
erregt, bei dem der Real- und der Imagin"arteil des zuf"alligen Spektralwertes 
\mbox{$\boldsymbol{V}(\frac{M}{2})$} korreliert waren. Auch wurde eine 
St"orung simuliert, bei der der zuf"allige Spektralwert 
\mbox{$\boldsymbol{N}_{\!\!f}(\frac{M}{2})$} einen korrelierten 
Real- und Imagin"arteil aufwies. 

Es seien nun wieder kurz das simulierte
System, die simulierte St"orung, der bei der Messung verwendete
erregende Zufallsvektor und die weiteren Parameter der Messung beschrieben.

Als zu messendes System wurde ein komplexes linearphasiges
Butterworth-Halbband-FIR-Filter mit der Impulsantwort
\begin{equation}
h(k)\;=\;\begin{cases}
{\D\quad j\cdot\binom{n}{\frac{n-1}{2}}\cdot\binom{n}{\frac{k}{2}}\cdot
\frac{n+1}{(k\!-\!n)\cdot 2^{2\cdot n+1}}}&
\quad\text{ f"ur }\quad k\text{ gerade }\;\wedge\;0\le k \le 2\CdoT n\\
\quad 0.5&\quad\text{ f"ur }\quad k=n\\
\quad 0&\quad\text{ sonst }
\end{cases}
\raisetag{20pt}\label{Mess11.1}
\end{equation}
gew"ahlt, das die Signalanteile der
Frequenzen \mbox{$0\!<\!\Omega\!<\!\pi$} durchl"asst, w"ahrend
es die Signalanteile bei den anderen Frequenzen sperrt.
Der Filtergrad wurde ebenso wie die Einschwingzeit auf
\mbox{$E\!=\!2\CdoT n\!=\!50$} eingestellt, so dass keine St"orungen
aufgrund zu kurzer Einschwingzeit entstehen konnten. Die theoretischen
und die bei einer kompletten RKM-Messung ermittelten Betr"age der 
"Ubertragungsfunktion sind in Bild \ref{b5e2}
\begin{figure}[btp]
\begin{center}
{ 
\begin{picture}(454,163)
\input{mbild5e2}
\put(35,130){\makebox(0,0)[r]{\small$1$}}
\put(35,32){\makebox(0,0)[b]{\small$0$}}
\put(42,17){\makebox(0,0)[lb]{\small$0$}}
\put(130,17){\makebox(0,0)[b]{\small$\pi/2$}}
\put(220,17){\makebox(0,0)[b]{\small$\pi$}}
\put(310,17){\makebox(0,0)[b]{\small$3\pi/2$}}
\put(400,17){\makebox(0,0)[b]{\small$2\pi$}}
\put(455,25){\makebox(0,0)[tr]{${\T\Omega=\mu\CdoT\frac{2\pi}{M}}$}}
\put(48,154){\makebox(0,0)[l]{$|H(\Omega)|,\,|\Hat{H}(\mu)|$}}
\end{picture}}
\end{center}\vspace{-30pt}
\setlength{\belowcaptionskip}{-5pt}
\caption["Ubertragungsfunktion des Beispielsystems zur Angabe einer Konfidenzellipse]{
"Ubertragungsfunktion des Beispielsystems zur Angabe einer Konfidenzellipse.\protect\\
Messung mit: $M\!=\!128$, $E\!=\!50$, $L\!=\!200$ 
und Fenster nach Kapitel \protect\ref{Algo} mit $N\!=\!4$.}
\label{b5e2}
\rule{\textwidth}{0.5pt}\vspace{-7pt}
\end{figure}
\begin{figure}[btp]
\begin{center}
{ 
\begin{picture}(454,301)
\input{mbild5e1}
\put(220,269){\makebox(0,0)[r]{\small$0,\!04$}}
\put(220,29){\makebox(0,0)[r]{\small$-0,\!04$}}
\put(72,137){\makebox(0,0)[b]{\small$-0,\!55$}}
\put(372,137){\makebox(0,0)[b]{\small$-0,\!45$}}
\put(420,143){\makebox(0,0)[t]{${\T
 \Re\big\{\Hat{H}\Left(\frac{M}{2}\right)\big\}}$}}
\put(217,292){\makebox(0,0)[r]{${\T
\Im\big\{\Hat{H}\Left(\frac{M}{2}\right)\big\}}$}}
\end{picture}}
\end{center}\vspace{-16pt}
\setlength{\belowcaptionskip}{-6pt}
\caption[Beispiel zur Angabe einer Konfidenzellipse]{
Beispiel zur Angabe einer Konfidenzellipse.\protect\\
Messung mit: $M\!=\!128$, $E\!=\!50$, $L\!=\!200$, $\alpha\!=\!0,\!1$
und Fenster nach Kapitel \protect\ref{Algo} mit $N\!=\!4$.\protect\\
$1000$ komplette RKM-Messungen.}
\label{b5e1}
\rule{\textwidth}{0.5pt}\vspace{-9pt}
\end{figure}
\begin{figure}[btp]
\begin{center}
{ 
\begin{picture}(454,359)

\input{mbild5e6}
\put(35,339){\makebox(0,0)[r]{\small$0,\!2$}}
\put(35,262){\makebox(0,0)[b]{\small$0$}}
\put( 42,247){\makebox(0,0)[lb]{\small$0$}}
\put(130,247){\makebox(0,0)[b]{\small$\pi/2$}}
\put(220,247){\makebox(0,0)[b]{\small$\pi$}}
\put(310,247){\makebox(0,0)[b]{\small$3\pi/2$}}
\put(400,247){\makebox(0,0)[b]{\small$2\pi$}}
\put(455,255){\makebox(0,0)[tr]{${\T\Omega=\mu\CdoT\frac{2\pi}{M}}$}}
\put(50,350){\makebox(0,0)[l]{Anzahl der Werte $H\Left({\T\mu\CdoT\frac{2\pi}{M}}\right)$ au"serhalb des Konfidenzgebietes / 1000}}

\input{mbild5e7}
\put(35,219){\makebox(0,0)[r]{\small$0,\!2$}}
\put(35,142){\makebox(0,0)[b]{\small$0$}}
\put( 42,127){\makebox(0,0)[lb]{\small$0$}}
\put(130,127){\makebox(0,0)[b]{\small$\pi/2$}}
\put(220,127){\makebox(0,0)[b]{\small$\pi$}}
\put(310,127){\makebox(0,0)[b]{\small$3\pi/2$}}
\put(400,127){\makebox(0,0)[b]{\small$2\pi$}}
\put(455,135){\makebox(0,0)[tr]{${\T\Omega=\mu\CdoT\frac{2\pi}{M}}$}}
\put(50,230){\makebox(0,0)[l]{Anzahl der Werte $\Tilde{\Phi}_{\boldsymbol{n}}(\mu)$ au"serhalb des Konfidenzgebietes / 1000}}

\input{mbild5e8}
\put(35,99){\makebox(0,0)[r]{\small$0,\!2$}}
\put(35,22){\makebox(0,0)[b]{\small$0$}}
\put( 42,7){\makebox(0,0)[lb]{\small$0$}}
\put(130,7){\makebox(0,0)[b]{\small$\pi/2$}}
\put(220,7){\makebox(0,0)[b]{\small$\pi$}}
\put(310,7){\makebox(0,0)[b]{\small$3\pi/2$}}
\put(400,7){\makebox(0,0)[b]{\small$2\pi$}}
\put(455,15){\makebox(0,0)[tr]{${\T\Omega=\mu\CdoT\frac{2\pi}{M}}$}}
\put(50,110){\makebox(0,0)[l]{Anzahl der Werte $\Tilde{\Psi}_{\boldsymbol{n}}(\mu)$ au"serhalb des Konfidenzgebietes / 1000}}

\end{picture}}
\end{center}\vspace{-21pt}
\setlength{\belowcaptionskip}{-6pt}
\caption[Zur Kontrolle des Konfidenzniveaus]{
Zur Kontrolle des Konfidenzniveaus.\protect\\
Messung mit: $M\!=\!128$, $E\!=\!50$, $L\!=\!200$, $\alpha\!=\!0,\!1$
und Fenster nach Kapitel \protect\ref{Algo} mit $N\!=\!4$.\protect\\
$1000$ komplette RKM-Messungen.}
\label{b5e6}
\rule{\textwidth}{0.5pt}\vspace{-9pt}
\end{figure}
dargestellt. 
F"ur \mbox{$\Omega\!=\!\pi$} ergibt sich der Wert \mbox{$H(\pi)=-\frac{1}{2}$}. 

Erregt wurde mit einem normalverteilten mittelwertfreien station"aren 
Zufallsvektor, wobei sowohl der Real- als auch der Imagin"arteil jeweils 
eine Varianz von $1/2$ aufwies. Die Elemente \mbox{$\boldsymbol{v}(k)$} 
waren f"ur unterschiedliche Werte von $k$ voneinander unabh"angig. Der Real-
und der Imagin"arteil desselben Wertes von $k$ wurden so kombiniert, dass
sich f"ur den Korrelationskoeffizienten der von $k$ unabh"angige Wert
\mbox{$\text{E}\big\{\boldsymbol{v}(k)^2\big\}/
\text{E}\big\{|\boldsymbol{v}(k)|^2\big\}=(-1\!+\!j)/2$} ergab. Unabh"angig von
$\mu$ betrug damit die Kovarianz \mbox{$\text{E}\big\{\boldsymbol{V}(\mu)^2\big\}$}
der zuf"alligen Spektralwerte \mbox{$1/2\CdoT M\CdoT(-1\!+\!j)$}. Es wurden 
\mbox{$M\!=\!128$} Frequenzpunkte gemessen. 

Dem simulierten System wurde eine 
St"orung "uberlagert, die in zwei Schritten erzeugt wurde. Zun"achst wurde 
ein komplexer wei"ser station"arer mittelwertfreier Prozess mit unkorrelierten 
Real- und Imagin"arteilprozessen erzeugt, wobei beide Prozessanteile eine Streuung von 
\mbox{$\sigma_{\Re\{\boldsymbol{n}\}}=\sigma_{\Im\{\boldsymbol{n}\}}=0,\!1$}
hatten. Danach wurde diesem komplexen Prozess der um einen Takt verz"ogerte
und konjugierte Prozess, sowie ein unabh"angiger station"arer Eintonst"orer
der Amplitude Eins mit der Kreisfrequenz Eins und einer in \mbox{$[0;2\pi)$}
gleichverteilten zuf"alligen Phase "uberlagert. Das LDS und das MLDS
dieser St"orung sind im Unterkapitel \ref{Mess8} diskutiert.

Es wurde das nach Kapitel \ref{Algo} mit \mbox{$N\!=\!4$} berechnete
Fenster verwendet und es erfolgte eine Mittelung "uber \mbox{$L\!=\!200$}
Einzelmessungen. Die Konfidenz\-ellipsen wurden f"ur das Konfidenzniveau
\mbox{$1\!-\!\alpha=90\%$} ermittelt. Es wurden insgesamt $1000$
komplette RKM-Messungen zu je $200$ Einzelmessungen durchgef"uhrt.
Die dabei gemessenen $1000$ Werte \mbox{$\Hat{H}({\T\frac{M}{2}})$}
sind als Punkte in Bild \ref{b5e1} zu sehen. Die Konfidenzellipse {\em einer} 
kompletten RKM-Messung ist ebenfalls eingetragen. 

Offensichtlich ist die
Kon\-fi\-denz\-el\-lip\-se nicht zur Absch"atzung der H"ohenlinien der Verteilung
der Messwerte geeignet, da sie deutlich gegen"uber der Punkteh"aufung
der gemessenen Werte verschoben liegt. Dies ist aber auch durch die Angabe
einer Konfidenzellipse nicht beabsichtigt. Die zuf"allig bei einer
Messung erhaltene Konfidenzellipse "uberdeckt lediglich mit
einer vorgegebenen Wahrscheinlichkeit den konstanten zu messenden Wert.

Es verh"alt sich dabei ganz "ahnlich wie bei dem Spiel, bei dem man versucht,
Ringe aus einer gewissen Entfernung auf einen senkrechten
Stab zu werfen. Je gr"o"ser der Ring ist, desto gr"o"ser die
Wahrscheinlichkeit den Stab zu treffen. Wenn man von einem Werfer wei"s,
dass der im Mittel $90$\% der Ringe auf den Stab wirft, so kann
man nach einem Wurf, nachdem der Stab entfernt wurde alleine anhand der
Lage des Ringes, und ohne die Kenntnis wo sich der Stab wirklich befand,
sagen, dass sich der Stab mit einer Wahrscheinlichkeit von $90$\% innerhalb
des Ringes befand. Man kann jedoch aus der Lage eines konkret geworfenen
Ringes nicht auf die Verteilungsdichte der Endlage des Mittelpunktes eines
beliebigen anderen noch zu werfenden Ringes schlie"sen. 

Um die Qualit"at der mit dem RKM berechneten Konfidenzellipsen absch"atzen zu 
k"onnen, wurde bei den $1000$ kompletten RKM-Messungen als Sch"atzwert f"ur die
Wahrscheinlichkeit $\alpha$, mit der die Konfidenzellipse den Messwert nicht 
beinhaltet, die entsprechende relative H"aufigkeit gemessen. Dies wurde f"ur 
alle Messwerte durchgef"uhrt. Die Ergebnisse sind in Bild~\ref{b5e6} zu sehen. 
Man erkennt, dass mit der im folgenden diskutierten Ausnahme die Gr"o"se der 
Konfidenzintervalle bzw. -gebiete mit dem in Kapitel \ref{Kon} beschriebenen 
Verfahren offensichtlich geeignet abgesch"atzt werden kann. 

Lediglich bei den Messwerten \mbox{$\Hat{\Phi}_{\boldsymbol{n}}(\mu)$} f"ur Frequenzen
in unmittelbarer Umgebung der Frequenz \mbox{$\Omega\!=\!1$} f"allt auf, dass 
das Konfidenzgebiet offenbar zu gro"s gew"ahlt worden ist, da bei
diesen Messpunkten die Konfidenzellipse bei allen $1000$ Messungen
den wahren Wert \mbox{$\Tilde{\Psi}_{\boldsymbol{n}}(\mu)$} der
LDS-N"aherung beinhaltet. Daf"ur gibt es eine plausible Erkl"arung.

Bei der Berechnung der Messwertvarianzen hatten wir angenommen,
dass die Spektralwerte der St"orung normalverteilt seien, und konnten
mit dieser Annahme die vierten Momente der Spektralwerte durch ihre
zweiten Momente ausdr"ucken. Bei deutlich abweichender Verteilung,
wird diese Absch"atzung nicht mehr g"ultig sein. Dies ist hier der Fall,
weil bei diesen Frequenzen der Eintonst"orer mit der Zufallsphase,
dessen Amplitude konstant gehalten wurde, "uberwiegt. Da der andere
St"oreinfluss hier fast vernachl"assigt werden kann, nimmt auch das
Betragsquadrat dieser Spektralwerte einen praktisch konstanten Wert
an. Die Verteilungsdichte dieser Spektralwerte ist daher auf einem 
schmalen Ring um den Ursprung der komplexen Ebene besonders hoch, 
w"ahrend sie sonst fast vernachl"assigbar klein ist. Das vierte Moment 
einer solchen Verteilung steht in einem ganz anderen Verh"altnis zum 
zweiten Moment als bei der Normalverteilung. Dadurch wird die wahre 
Messwertvarianz von der gemessenen Messwertvarianz, und damit auch die 
Breite des Konfidenz\-intervalls, deutlich abweichen. 

Um dies zu best"atigen
wurde aus allen $1000$ Messwerten aller RKM-Messungen die empirische
Varianz \mbox{$\overline{\big|\Hat{\Phi}_{\boldsymbol{n}}(\mu)\!-\!
\Tilde{\Phi}_{\boldsymbol{n}}(\mu)\big|^2}$} der Messwerte
bestimmt und diese mit dem gemittelten Messwertvarianzsch"atzwerten
\mbox{$\overline{\Hat{C}_{\Hat{\boldsymbol{\Phi}}_{\!\boldsymbol{n}}(\mu),
\Hat{\boldsymbol{\Phi}}_{\!\boldsymbol{n}}(\mu)}}$} verglichen. In der
Tabelle \ref{T7.1}
\begin{table}[b]
\rule{\textwidth}{0.5pt}\vspace{5pt}
\begin{center}$
{\setlength{\arraycolsep}{0pt}
\begin{array}{||c|r@{,}l||r@{,}l@{{}\CdoT{}}l|r@{,}l@{{}\CdoT{}}l|
r@{,}l@{{}\CdoT{}}l|r@{,}l@{{}\CdoT{}}l|r@{,}l@{{}\CdoT{}}l||}
\hline
\hline
\rule[-8pt]{0pt}{26pt}\mu&\multicolumn{2}{c||}{\;\Omega\!=\!
\mu\CdoT\frac{2\pi}{M}\;}&
\multicolumn{3}{c|}{\;\overline{\big|\Hat{\Phi}_{\boldsymbol{n}}(\mu)\!-\!
\Tilde{\Phi}_{\boldsymbol{n}}(\mu)\big|^2}\;}&
\multicolumn{3}{c|}{\;\overline{
\Hat{C}_{\Hat{\boldsymbol{\Phi}}_{\!\boldsymbol{n}}(\mu),\Hat{\boldsymbol{\Phi}}_{\!\boldsymbol{n}}(\mu)}}\;}&
\multicolumn{3}{c|}{A_{\overline{\Phi}}(\mu)}&
\multicolumn{3}{c|}{\overline{\Hat{A}_{\Phi}(\mu)}}&
\multicolumn{3}{c||}{\overline{\alpha}}\\
\hline
\vdots&\multicolumn{2}{c||}{\vdots}&
\multicolumn{3}{c|}{\vdots}&
\multicolumn{3}{c|}{\vdots}&
\multicolumn{3}{c|}{\vdots}&
\multicolumn{3}{c|}{\vdots}&
\multicolumn{3}{c||}{\vdots}\\
\;18\;&\;\;\;0&88357\;&\;\;\;\;\;8&2649&10^{-6}\;&\;\;8&0978&10^{-6}\;&
\;4&7288&10^{-3}\;&\;4&6807&10^{-3}\;&\;1&0350&10^{-1}\\
19&0&93266&8&3398&10^{-6}&8&3176&10^{-6}&
4&7501&10^{-3}&4&7438&10^{-3}&1&0045&10^{-1}\\
20&0&98175&5&3426&10^{-1}&5&1999&10^1&
1&2023&10^0&1&1861&10^1&3&2271&10^{-59}\;\\
21&1&0308&4&5913&10^{-2}&3&5012&10^0&
3&5245&10^{-1}&3&0778&10^0&8&7521&10^{-47}\\
22&1&0799&7&8135&10^{-6}&8&1309&10^{-6}&
4&5978&10^{-3}&4&6903&10^{-3}&9&3361&10^{-2}\\
23&1&1290&7&9741&10^{-6}&8&0798&10^{-6}&
4&6448&10^{-3}&4&6755&10^{-3}&9&7780&10^{-2}\\
\vdots&\multicolumn{2}{c||}{\vdots}&
\multicolumn{3}{c|}{\vdots}&
\multicolumn{3}{c|}{\vdots}&
\multicolumn{3}{c|}{\vdots}&
\multicolumn{3}{c|}{\vdots}&
\multicolumn{3}{c||}{\vdots}\\
\hline
\hline
\end{array}}$
\end{center}\vspace{0pt}
\setlength{\abovecaptionskip}{0pt}
\caption{Beispiel f"ur unzutreffende Sch"atzwerte der Messwertvarianzen des LDS.}
\label{T7.1}
\end{table}
sind auch noch die halben Konfidenzintervallbreiten
\mbox{$A_{\overline{\Phi}}(\mu)$} und \mbox{$\overline{\Hat{A}_{\Phi}(\mu)}$}
berechnet, die sich bei demselben \mbox{$\alpha=0,\!1$} einerseits mit der
mittleren tats"achlichen und andererseits mit der mittleren gesch"atzten
Messwertvarianz ergeben, sowie die sich aus der mittleren tats"achlichen
Messwertvarianz und der Konfidenzintervallbreite
\mbox{$\overline{\Hat{A}_{\Phi}(\mu)}$} ergebende
Wahrscheinlichkeit \mbox{$\overline{\alpha}$} f"ur einen
Messwert au"serhalb des gesch"atzten Intervalls. 
Man erkennt, dass die gesch"atzte Varianz der beiden Messwerte ober- 
und unterhalb der Frequenz des periodischen Eintonst"orers fast um 
zwei Zehnerpotenzen falsch gesch"atzt wurde, so dass das korrekte
Konfidenzintervall, in das die Messwertstreuung eingeht, ca. um den
Faktor zehn kleiner ist als das verwendete Konfidenzintervall,
bei dem es dadurch praktisch unm"oglich wird, dass das Konfidenzintervall
den wahren Wert nicht "uberdeckt. 

Was man an diesen Messwerten auch erkennen kann, ist, dass von diesem Effekt 
wirklich nur die beiden zur Frequenz \mbox{$\Omega\!=\!1$} unmittelbar 
benachbarten Frequenzen betroffen sind. Auch dies ist ein Vorteil der 
Verwendung einer hoch frequenzselektiven Fensterfolge.\vspace{-4pt minus 6pt}

\section{Beispiel zur Interpretation der Konfidenzgebiete}\label{Mess8}

\begin{figure}[btp]
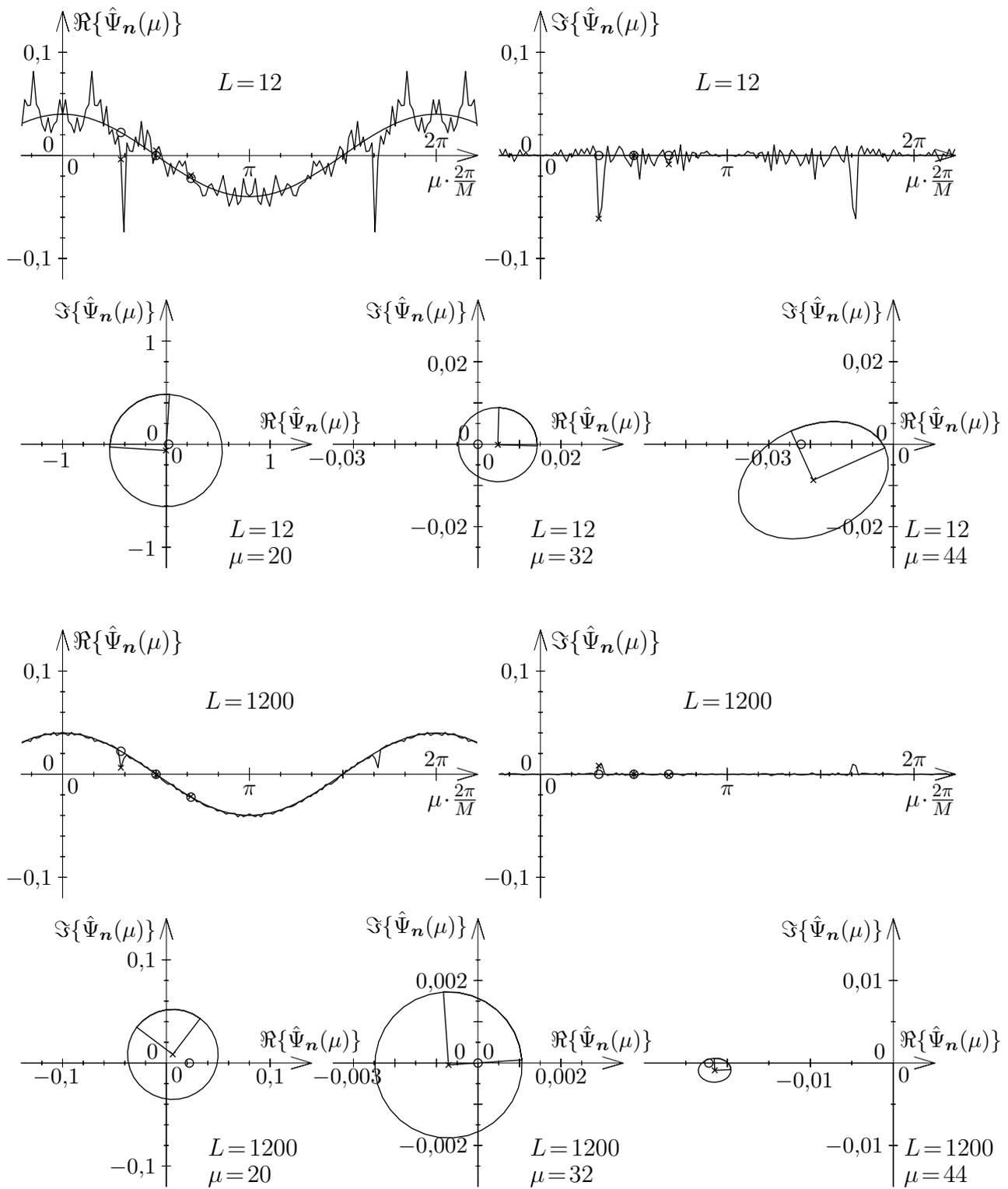

\begin{center}
{ 
\begin{picture}(454,599)

\input{mbild5d1}
\put(15,559){\makebox(0,0)[r]{\small$0,\!1$}}
\put(16,512){\makebox(0,0)[rb]{\small$0$}}
\put(15,459){\makebox(0,0)[r]{\small$-0,\!1$}}
\put(25,501){\makebox(0,0)[b]{\small$0$}}
\put(110,500){\makebox(0,0)[b]{\small$\pi$}}
\put(200,514){\makebox(0,0)[b]{\small$2\pi$}}
\put(220,505){\makebox(0,0)[tr]{${\T\mu\CdoT\frac{2\pi}{M}}$}}
\put(25,575){\makebox(0,0)[l]{$\Re\{\Hat{\Psi}_{\boldsymbol{n}}(\mu)\}$}}
\put(110,550){\makebox(0,0)[t]{$L\!=\!12$}}

\input{mbild5d2}
\put(245,559){\makebox(0,0)[r]{\small$0,\!1$}}
\put(246,512){\makebox(0,0)[rb]{\small$0$}}
\put(245,459){\makebox(0,0)[r]{\small$-0,\!1$}}
\put(255,501){\makebox(0,0)[b]{\small$0$}}
\put(340,500){\makebox(0,0)[b]{\small$\pi$}}
\put(430,514){\makebox(0,0)[b]{\small$2\pi$}}
\put(450,505){\makebox(0,0)[tr]{${\T\mu\CdoT\frac{2\pi}{M}}$}}
\put(255,575){\makebox(0,0)[l]{$\Im\{\Hat{\Psi}_{\boldsymbol{n}}(\mu)\}$}}
\put(340,550){\makebox(0,0)[t]{$L\!=\!12$}}

\input{mbild5d3}
\put(65,419){\makebox(0,0)[r]{\small$1$}}
\put(66,372){\makebox(0,0)[rb]{\small$0$}}
\put(65,319){\makebox(0,0)[r]{\small$-1$}}
\put(17,358){\makebox(0,0)[b]{\small$-1$}}
\put(75,361){\makebox(0,0)[b]{\small$0$}}
\put(120,358){\makebox(0,0)[b]{\small$1$}}
\put(115,375){\makebox(0,0)[lb]{\small$\Re\{\Hat{\Psi}_{\boldsymbol{n}}(\mu)\}$}}
\put(65,435){\makebox(0,0)[r]{\small$\Im\{\Hat{\Psi}_{\boldsymbol{n}}(\mu)\}$}}
\put(100,325){\makebox(0,0)[lb]{$L\!=\!12$}}
\put(100,310){\makebox(0,0)[lb]{$\mu\!=\!20$}}

\input{mbild5d4}
\put(215,408){\makebox(0,0)[r]{\small$0,\!02$}}
\put(216,372){\makebox(0,0)[rb]{\small$0$}}
\put(215,329){\makebox(0,0)[r]{\small$-0,\!02$}}
\put(152,358){\makebox(0,0)[b]{\small$-0,\!03$}}
\put(225,358){\makebox(0,0)[b]{\small$0$}}
\put(260,358){\makebox(0,0)[b]{\small$0,\!02$}}
\put(255,375){\makebox(0,0)[lb]{\small$\Re\{\Hat{\Psi}_{\boldsymbol{n}}(\mu)\}$}}
\put(215,435){\makebox(0,0)[r]{\small$\Im\{\Hat{\Psi}_{\boldsymbol{n}}(\mu)\}$}}
\put(245,325){\makebox(0,0)[lb]{$L\!=\!12$}}
\put(245,310){\makebox(0,0)[lb]{$\mu\!=\!32$}}

\input{mbild5d5}
\put(415,409){\makebox(0,0)[r]{\small$0,\!02$}}
\put(417,372){\makebox(0,0)[rb]{\small$0$}}
\put(415,329){\makebox(0,0)[r]{\small$-0,\!02$}}
\put(357,358){\makebox(0,0)[b]{\small$-0,\!03$}}
\put(425,361){\makebox(0,0)[b]{\small$0$}}
\put(423,375){\makebox(0,0)[lb]{\small$\Re\{\Hat{\Psi}_{\boldsymbol{n}}(\mu)\}$}}
\put(415,435){\makebox(0,0)[r]{\small$\Im\{\Hat{\Psi}_{\boldsymbol{n}}(\mu)\}$}}
\put(425,325){\makebox(0,0)[lb]{$L\!=\!12$}}
\put(425,310){\makebox(0,0)[lb]{$\mu\!=\!44$}}

\input{mbild5d6}
\put(15,259){\makebox(0,0)[r]{\small$0,\!1$}}
\put(16,212){\makebox(0,0)[rb]{\small$0$}}
\put(15,159){\makebox(0,0)[r]{\small$-0,\!1$}}
\put(25,201){\makebox(0,0)[b]{\small$0$}}
\put(110,200){\makebox(0,0)[b]{\small$\pi$}}
\put(200,214){\makebox(0,0)[b]{\small$2\pi$}}
\put(220,205){\makebox(0,0)[tr]{${\T\mu\CdoT\frac{2\pi}{M}}$}}
\put(25,275){\makebox(0,0)[l]{$\Re\{\Hat{\Psi}_{\boldsymbol{n}}(\mu)\}$}}
\put(110,250){\makebox(0,0)[t]{$L\!=\!1200$}}

\input{mbild5d7}
\put(245,259){\makebox(0,0)[r]{\small$0,\!1$}}
\put(246,212){\makebox(0,0)[rb]{\small$0$}}
\put(245,159){\makebox(0,0)[r]{\small$-0,\!1$}}
\put(255,201){\makebox(0,0)[b]{\small$0$}}
\put(340,200){\makebox(0,0)[b]{\small$\pi$}}
\put(430,214){\makebox(0,0)[b]{\small$2\pi$}}
\put(450,205){\makebox(0,0)[tr]{${\T\mu\CdoT\frac{2\pi}{M}}$}}
\put(255,275){\makebox(0,0)[l]{$\Im\{\Hat{\Psi}_{\boldsymbol{n}}(\mu)\}$}}
\put(340,250){\makebox(0,0)[t]{$L\!=\!1200$}}

\input{mbild5d8}
\put(65,119){\makebox(0,0)[r]{\small$0,\!1$}}
\put(66,72){\makebox(0,0)[rb]{\small$0$}}
\put(65,19){\makebox(0,0)[r]{\small$-0,\!1$}}
\put(17,58){\makebox(0,0)[b]{\small$-0,\!1$}}
\put(75,61){\makebox(0,0)[b]{\small$0$}}
\put(120,58){\makebox(0,0)[b]{\small$0,\!1$}}
\put(115,75){\makebox(0,0)[lb]{\small$\Re\{\Hat{\Psi}_{\boldsymbol{n}}(\mu)\}$}}
\put(65,135){\makebox(0,0)[r]{\small$\Im\{\Hat{\Psi}_{\boldsymbol{n}}(\mu)\}$}}
\put(90,25){\makebox(0,0)[lb]{$L\!=\!1200$}}
\put(90,10){\makebox(0,0)[lb]{$\mu\!=\!20$}}

\input{mbild5d9}
\put(215,109){\makebox(0,0)[r]{\small$0,\!002$}}
\put(214,72){\makebox(0,0)[rb]{\small$0$}}
\put(215,29){\makebox(0,0)[r]{\small$-0,\!002$}}
\put(157,58){\makebox(0,0)[b]{\small$-0,\!003$}}
\put(225,72){\makebox(0,0)[b]{\small$0$}}
\put(260,58){\makebox(0,0)[b]{\small$0,\!002$}}
\put(255,75){\makebox(0,0)[lb]{\small$\Re\{\Hat{\Psi}_{\boldsymbol{n}}(\mu)\}$}}
\put(215,137){\makebox(0,0)[r]{\small$\Im\{\Hat{\Psi}_{\boldsymbol{n}}(\mu)\}$}}
\put(245,25){\makebox(0,0)[lb]{$L\!=\!1200$}}
\put(245,10){\makebox(0,0)[lb]{$\mu\!=\!32$}}

\input{mbild5d10}
\put(415,109){\makebox(0,0)[r]{\small$0,\!01$}}
\put(416,72){\makebox(0,0)[rb]{\small$0$}}
\put(415,29){\makebox(0,0)[r]{\small$-0,\!01$}}
\put(377,56){\makebox(0,0)[b]{\small$-0,\!01$}}
\put(425,61){\makebox(0,0)[b]{\small$0$}}
\put(423,75){\makebox(0,0)[lb]{\small$\Re\{\Hat{\Psi}_{\boldsymbol{n}}(\mu)\}$}}
\put(415,135){\makebox(0,0)[r]{\small$\Im\{\Hat{\Psi}_{\boldsymbol{n}}(\mu)\}$}}
\put(425,25){\makebox(0,0)[lb]{$L\!=\!1200$}}
\put(425,10){\makebox(0,0)[lb]{$\mu\!=\!44$}}
\end{picture}}
\end{center}\vspace{-10pt}
\caption[Beispiel zur Interpretation der Konfidenzgebiete]{
Beispiel zur Interpretation der Konfidenzgebiete von 
\mbox{$\protect\Hat{\Psi}_{\boldsymbol{n}}(\mu)$}.\protect\\
Messung mit \mbox{$M\!=\!128$}, \mbox{$\alpha\!=\!0,\!1$} und
Fenster nach Kapitel \protect\ref{Algo} mit \mbox{$N\!=\!4$}.}
\label{b5d}
\end{figure}
In Kapitel \ref{Kon} wurden die Konfidenzgebiete mit Hilfe der gemessenen
Mess"-wert(ko)va"-ri"-an"-zen abgesch"atzt, wobei zugrunde gelegt wurde,
dass die Messwerte n"aherungsweise normalverteilt seien. Es ist zu vermuten,
dass diese Annahme wenigstens f"ur die Messwerte
\mbox{$\Hat{\Phi}_{\boldsymbol{n}}(\mu)$} und
\mbox{$\Hat{\Psi}_{\boldsymbol{n}}(\mu)$} nur schlecht zutrifft,
wenn der Betrag der Messwerte in derselben Gr"o"senordnung liegt 
wie die gemessene Messwertstreuung. Dass die damit ermittelten
Konfidenzellipsen auch dann noch einen Sinn haben, soll nun an einer
Beispielmessung demonstriert werden. 

Dazu wurde das in Kapitel \ref{Mess11} 
vorgestellte komplexe linearphasige System vermessen, wobei derselbe erregende
Prozess gew"ahlt wurde. Dem simulierten System wurde wieder eine 
St"orung "uberlagert, die in zwei Schritten erzeugt wurde. Zun"achst wurde 
ein komplexer wei"ser station"arer mittelwertfreier Prozess mit unkorrelierten 
Real- und Imagin"arteilprozessen erzeugt, wobei beide Prozessanteile eine Streuung von 
\mbox{$\sigma_{\Re\{\boldsymbol{n}\}}=\sigma_{\Im\{\boldsymbol{n}\}}=0,\!1$}
hatten. Danach wurde diesem komplexen Prozess der um einen Takt verz"ogerte
und konjugierte Prozess, sowie ein unabh"angiger station"arer komplexer Eintonst"orer
mit der Kreisfrequenz Eins und einer in \mbox{$[0;2\pi)$} gleichverteilten 
zuf"alligen Phase "uberlagert. Als Amplitude der Eintonst"orung wurde hier
abweichend von der Messung im letzten Unterkapitel 1/2 gew"ahlt. Der Anteil 
dieses Eintonst"orers im MLDS ist Null, w"ahrend der Anteil im LDS 
ein Dirac-Impuls der St"arke $\pi/2$ bei der Kreisfrequenz Eins ist.

In den theoretischen N"aherungen \mbox{$\Tilde{\Phi}_{\boldsymbol{n}}(\mu)$} 
erh"alt man als Anteil, der von dem Eintonst"orer verursacht wird, 
das im Raster \mbox{$\Omega=\mu\CdoT2\pi/M$} mit \mbox{$M\!=\!128$} 
abgetastete um Eins verschobene Betragsquadratspektrum 
der Fensterfolge, w"ahrend die theoretischen N"aherungen
\mbox{$\Tilde{\Psi}_{\boldsymbol{n}}(\mu)$} von dem Eintonst"orer
nicht beeinflusst werden. Der Anteil des
komplexen Gau"sprozesses in der theoretischen LDS-N"aherung
\mbox{$\Tilde{\Phi}_{\boldsymbol{n}}(\mu)$} ist konstant
\mbox{$4\cdot0,\!1^2$}. F"ur die theoretische MLDS-N"aherung
ergibt sich \mbox{$\Tilde{\Psi}_{\boldsymbol{n}}(\mu) = 
4\cdot0,\!1^2\cdot d(1)\cdot\cos(\mu\CdoT2\pi/M)$}, wobei \mbox{$d(k)$}
die nach Gleichung (\ref{2.21}) definierte Fenster-AKF ist.
Bei dem hier verwendeten Fenster, das nach Kapitel \ref{Algo} mit 
\mbox{$N\!=\!4$} berechnet wurde, ergibt sich f"ur den Wert der 
Fenster-AKF f"ur \mbox{$k\!=\!1$} ziemlich genau Eins.

Nach \mbox{$L\!=\!12$} Einzelmessungen wurde die Messung unterbrochen, und es wurden die
Mess\-werte \mbox{$\Hat{\Psi}_{\boldsymbol{n}}(\mu)$}, sowie deren
Messwert(ko)varianzen und die daraus bestimmten elliptischen
Konfidenzgebiete mit \mbox{$\alpha\!=\!0,\!1$} f"ur die diskreten
Frequenzen \mbox{$\mu=20\;(12)\;44$} berechnet. Nach weiteren
$1188$ Einzelmessungen wurden diese Messergebnisse erneut berechnet.
Diese Frequenzpunkte wurden ausgesucht, weil zum einen die Frequenz
\mbox{$20\CdoT2\pi/M\approx1$} unmittelbar neben der Frequenz des
periodischen St"orers liegt, weil zum zweiten bei der Frequenz
\mbox{$32\CdoT2\pi/M=\pi/2$} eine Nullstelle von
\mbox{$\Tilde{\Psi}_{\boldsymbol{n}}(\mu)$} vorliegt, und weil
es zum dritten sinnvoll ist, zum Vergleich den Wert zu verwenden,
bei dem \mbox{$\Tilde{\Psi}_{\boldsymbol{n}}(44) =
-\Tilde{\Psi}_{\boldsymbol{n}}(20)$} gilt. 

Die Messwerte
\mbox{$\Hat{\Psi}_{\boldsymbol{n}}(\mu)$}, sowie die theoretischen
Verl"aufe \mbox{$\Tilde{\Psi}_{\boldsymbol{n}}(\mu)$} sind in Bild
\ref{b5d} einerseits "uber der Frequenz getrennt nach Real-
und Imagin"arteil als auch darunter f"ur die drei ausgesuchten Frequenzpunkte
in der komplexen Ebene jeweils f"ur \mbox{$L\!=\!12$} und \mbox{$L\!=\!1200$}
dargestellt. Die Messwerte der drei Frequenzpunkte \mbox{$\mu=20\;(12)\;44$}
sind als Kreuzchen und die entsprechenden theoretischen Wert als kleine
Kreise eingetragen. In den Darstellungen in der komplexen Ebene
sind die Konfidenzellipsen mit ihren Halbachsen eingezeichnet.

Wenn man die unterschiedlichen Achsenskalierungen beachtet,
erkennt man, dass man nach \mbox{$L\!=\!12$} Einzelmessungen bei keinem der
drei Messwerte eine sinnvolle Aussage "uber die zu messenden
Gr"o"sen erh"alt, weil einerseits bei allen drei Messwerten
der Koordinatenursprung in oder zumindest in unmittelbarer
Umgebung der Konfidenzellipse liegt, und andererseits die
Messwertvarianz, die sich aus der Gr"o"se der Konfidenzellipsen
absch"atzen l"asst, im Vergleich zu den anderen Messwerten
noch viel zu gro"s ist. 

Nach \mbox{$L\!=\!1200$} Einzelmessungen hat sich daran
bei dem Messwert der Frequenz \mbox{$\mu\!=\!20$} noch nichts ge"andert.
Man kann daher bei diesem Messwert auch ohne die Kenntnis des theoretischen
Verlaufs sagen, dass die "`Peaks"', die man im gemessenen
Frequenzgang vorfindet, nicht etwa auf eine im Spektrum vorhandene
starke Korrelation zur"uckzuf"uhren sind, sondern lediglich
durch eine starke St"orung bei der Messung verursacht werden.

Bei dem Messwert der Frequenz \mbox{$\mu\!=\!32$} erkennt man, dass die Varianz
des Messwertes so klein ist, dass man zuverl"assig sagen kann,
dass bei dieser Frequenz offensichtlich keine im Vergleich zu
den anderen Messwerten nennenswerte Korrelation der Spektralwerte
bei positiver und bei negativer Frequenz vorhanden ist. Man kann daher
diesen Messwert in Gegensatz zu dem Messwert bei \mbox{$\mu\!=\!20$} als
zuverl"assig bezeichnen, obwohl bei beiden Messwerten der 
Koordinatenursprung innerhalb der Konfidenzellipse liegt.
Eine Aussage dar"uber, inwieweit die angegebene Konfidenzellipse
mit dem Konfidenzgebiet "ubereinstimmt, das sich mit der 
theoretischen Messwertverteilung ergeben w"urde, sollte man
dennoch auch bei dem Messwert mit \mbox{$\mu\!=\!32$} nicht machen. 

Bei dem Messwert
mit \mbox{$\mu\!=\!44$} hingegen ist zu erwarten, dass dessen Zuverl"assigkeit
mit dem Konfidenzgebiet gut abgesch"atzt werden kann, weil man annehmen
kann, dass die n"aherungsweise angenommene Normalverteilung
des Messwertes vorliegt, da dessen Konfidenzellipse sehr weit
von Koordinatenursprung entfernt zu finden ist.

\section{Beispielsystem: \mbox{$\Sigma\Delta$}-Wandler}\label{SD-DAC}

Zum Schluss sei noch ein Beispiel f"ur das RKM mit Fensterung zur 
Messung reellwertiger Systeme angef"uhrt, wobei an diesem Beispiel auch 
die Anwendbarkeit des von mir in Kapitel \ref{myChirp} vorgestellten 
Chirpsignals demonstriert werden soll. Es handelt sich bei dem Beispielsystem 
um den Teil eines Systems, bei dessen Messung mit dem RKM nach 
\cite{Sch/D} Probleme bei der Messung ohne Fensterung auftraten. Das 
gab den Anlass sich damit zu besch"aftigen, wie das RKM zu modifizieren 
ist, wenn das System nicht ad"aquat gemessen werden kann. 

Bei dem Gesamtsystem handelt es sich um einen digital gesteuerten 
Oszillator (\,DCO\,) der intern aus einem hybriden Phasenregelkreis 
(\,PLL\,) besteht. Die wesentlichen \mbox{Bauteile} dieses 
Systems\footnote{Dieser DCO wurde als Hardware auch aufgebaut. 
Von einem steuernden PC konnte "uber die serielle Schnittstelle die 
gew"unschte Frequenz des Oszillators, die bis zu ca. 20~MHz betragen durfte, 
in extrem engen Frequenzschritten (\,im mHz Bereich\,) verstellt werden. 
Damit konnte man die auf einen externen Referenztakt bezogene Phase des 
Oszillators f"ur jeden Zeitpunkt so einstellen, wie man dies am Steuerrechner 
vorgab.} sind neben einem Signalprozessor, der die digitalen 
Teile des hybriden PLL realisiert und die digitalen Steuerworte f"ur 
die Sollfrequenz "ubernimmt, noch ein Digital-Analog-Wandler (\,DAC\,), 
ein einfacher analog steuerbarer Oszillator, ein digital einstellbarer 
Frequenzteiler und ein Analog-Digital-Wandler (\,ADC\,). Bei dem DCO 
wird ein linearer Zusammenhang zwischen dem digitalen Signal am Eingang 
und der Momentanfrequenz des Oszillatorausgangssignals gew"unscht. 
Daher kann dieser Zusammenhang durch die "Ubertragungsfunktion 
eines linearen Systems beschrieben werden. 

Unter anderem verursachen die in den Wandlern auftretenden Quantisierungen 
eine St"orung in der Momentanfrequenz des Oszillators. Da sich herausstellte, 
dass zwischen zwei aufeinanderfolgenden DA-Wandlungen im Signalprozessor 
noch gen"ugend Rechenzeit zur Verf"ugung stand, wurde versucht, die 
Quantisierungsfehler in der Momentanfrequenz des Oszillators mit Hilfe eines 
\mbox{$\Sigma\Delta$}-Gliedes vor dem DAC im Bereich niedriger 
Frequenzen zu Lasten h"oherer Frequenzen zu verringern. Zun"achst 
wurde bei einer Simulation des DCO die "Ubertragungsfunktion und das 
LDS mit dem RKM nach \cite{Sch/D} gemessen. Dabei wurde beobachtet, 
dass sich beim LDS erhebliche Abweichungen von den theoretisch zu 
erwartenden Frequenzverl"aufen ergaben. Diese Abweichungen lie"sen sich 
durch eine Vergr"o"serung der Mittelungsanzahl $L$ nicht beeinflussen, 
und auch eine Ver"anderung des Parameters $M$ erbrachte nur eine 
geringf"ugige Verbesserung. Da bekannt ist, dass eine Fensterung 
bei der Spektralsch"atzung eine Messung mit einer h"oheren 
Frequenzselektivit"at erm"oglicht, wurde nun untersucht, wie eine 
Fensterung beim RKM eingesetzt werden kann, und welche Forderungen 
die dabei verwendete Fensterfolge zu erf"ullen hat. 

Bei dem im folgenden beschriebenen Beispielsystem handelt es sich 
um genau das in Bild \ref{b5x}
\begin{figure}[btp]
\begin{center}
{ 
\begin{picture}(450,265)
\put(0,225){\makebox(0,0)[lb]{\mbox{$\Sigma\Delta$}-Glied:}}
\put(25,200){\circle{6}}
\put(190,200){\circle{6}}
\put(152,200){\line(1,0){35}}
\put(110,200){\circle*{4}}
\put(45,160){\line(1,0){15}}
\put(45,195){\line(0,1){10}}
\put(110,155){\line(0,1){10}}
\put(170,160){\line(0,1){40}}
\put(28,200){\vector(1,0){12}}
\put(40,200){\vector(1,0){88}}
\put(45,160){\vector(0,1){35}}
\put(115,160){\vector(-1,0){25}}
\put(110,200){\vector(0,-1){35}}
\put(170,160){\vector(-1,0){55}}
\put(40,195){\framebox(10,10){}}
\put(105,155){\framebox(10,10){}}
\put(128,188){\framebox(24,24){}}
\put(130,190){\framebox(20,20){Q}}
\put(60,150){\framebox(30,20){$P(z)$}}
\put(20,200){\makebox(0,0)[r]{$\boldsymbol{v}(k)$}}
\put(197,200){\makebox(0,0)[l]{$\boldsymbol{y}(k)$}}
\put(108,204){\makebox(0,0)[b]{$\boldsymbol{v}_{\!\Delta\!}(k)$}}
\put(33,202){\makebox(0,0)[b]{${\scriptstyle +}$}}
\put(47,188){\makebox(0,0)[l]{${\scriptstyle +}$}}
\put(108,172){\makebox(0,0)[r]{${\scriptstyle +}$}}
\put(123,159){\makebox(0,0)[t]{${\scriptstyle -}$}}

\put(0,105){\makebox(0,0)[lb]{Ersatzschaltbild:}}
\put(25,80){\circle{6}}
\put(25,50){\circle{6}}
\put(190,80){\circle{6}}
\put(115,80){\line(1,0){72}}
\put(100,50){\line(1,0){20}}
\put(120,75){\line(0,1){10}}
\put(28,80){\vector(1,0){87}}
\put(28,50){\vector(1,0){22}}
\put(120,50){\vector(0,1){25}}
\put(115,75){\framebox(10,10){}}
\put(50,40){\framebox(50,20){$1\!-\!P(z)$}}
\put(20,80){\makebox(0,0)[r]{$\boldsymbol{v}(k)$}}
\put(20,50){\makebox(0,0)[r]{$\boldsymbol{n}(k)$}}
\put(197,80){\makebox(0,0)[l]{$\boldsymbol{y}(k)$}}
\put(108,82){\makebox(0,0)[b]{${\scriptstyle +}$}}
\put(122,68){\makebox(0,0)[l]{${\scriptstyle +}$}}

\put(210,250){\makebox(0,0)[lb]{Quantisierungskennlinie:}}
\put(210,120){\line(1,0){74}}
\put(286,120){\line(1,0){88}}
\put(210,20){\line(1,0){30}}
\put(240,40){\line(1,0){20}}
\put(260,60){\line(1,0){10}}
\put(300,90){\line(1,0){10}}
\put(310,110){\line(1,0){20}}
\put(330,130){\line(1,0){20}}
\put(350,150){\line(1,0){10}}
\put(390,180){\line(1,0){10}}
\put(400,200){\line(1,0){20}}
\put(420,220){\line(1,0){30}}
\put(327,20){\line(1,0){6}}
\put(327,220){\line(1,0){6}}
\put(330,0){\line(0,1){74}}
\put(330,76){\line(0,1){88}}
\put(240,20){\line(0,1){20}}
\put(260,40){\line(0,1){20}}
\put(310,90){\line(0,1){20}}
\put(329.7,110){\line(0,1){20}}
\put(330.2,110){\line(0,1){20}}
\put(350,130){\line(0,1){20}}
\put(400,180){\line(0,1){20}}
\put(420,200){\line(0,1){20}}
\put(230,117){\line(0,1){6}}
\put(430,117){\line(0,1){6}}
\put(350,117){\line(0,1){6}}
\put(376,120){\vector(1,0){74}}
\put(330,166){\vector(0,1){74}}
\multiput(370,160)(5,5){3}{\circle*{1}}
\multiput(280,70)(5,5){3}{\circle*{1}}
\qbezier(374,124)(372,122)(374,120)
\qbezier(374,120)(376,118)(374,116)
\qbezier(376,124)(374,122)(376,120)
\qbezier(376,120)(378,118)(376,116)
\qbezier(284,124)(282,122)(284,120)
\qbezier(284,120)(286,118)(284,116)
\qbezier(286,124)(284,122)(286,120)
\qbezier(286,120)(288,118)(286,116)
\qbezier(326,164)(328,166)(330,164)
\qbezier(330,164)(332,162)(334,164)
\qbezier(326,166)(328,168)(330,166)
\qbezier(330,166)(332,164)(334,166)
\qbezier(326,74)(328,76)(330,74)
\qbezier(330,74)(332,72)(334,74)
\qbezier(326,76)(328,78)(330,76)
\qbezier(330,76)(332,74)(334,76)
\put(450,125){\makebox(0,0)[br]{$\boldsymbol{v}_{\!\Delta\!}(k)$}}
\put(325,240){\makebox(0,0)[rt]{$\boldsymbol{y}(k)$}}
\put(230,115){\makebox(0,0)[t]{$\!\!-1$}}
\put(430,115){\makebox(0,0)[t]{$1$}}
\put(325,20){\makebox(0,0)[r]{$-1$}}
\put(325,220){\makebox(0,0)[r]{$1$}}
\put(332,118){\makebox(0,0)[lt]{$0$}}
\put(350,115){\makebox(0,0)[t]{$Q$}}

\end{picture}}
\end{center}\vspace{-10pt}
\setlength{\belowcaptionskip}{-6pt}
\caption{Beispielsystem: \mbox{$\Sigma\Delta$}-Glied}
\label{b5x}
\rule{\textwidth}{0.5pt}\vspace{-10pt}
\end{figure}
links oben gezeigte 
\mbox{$\Sigma\Delta$}-Glied, das im DCO eingesetzt werden sollte. 
F"ur die dabei simulierte gleichm"a"sige Quantisierung wurde 
die im linken Teilbild dargestellte punktsymmetrische Kennlinie 
gew"ahlt, bei der sich bei einer Wortl"ange von $12$ Bit eine 
Quantisierungsstufenh"ohe von \mbox{${\D Q=2/(2^{12}\!\!-\!1)}$} 
ergibt. Wenn man die Quantisierung durch eine additive, gleichverteilte, 
von \mbox{$\boldsymbol{v}(k)$} unabh"angige, wei"se, zuf"allige St"orung 
\mbox{$\boldsymbol{n}(k)$} mit der Varianz \mbox{${\D Q^2/12}$} ersetzt, 
erh"alt man das im linken unteren Teilbild dargestellte Ersatzschaltbild. 
Als R"uckkopplungsfilter \mbox{$P(z)$} f"ur den Quantisierungsfehler wurde 
ein FIR-Filter gew"ahlt, das im Ersatzschaltbild im Pfad der St"orung zu 
einem FIR-Filter mit der Z-Transformierten 
\begin{gather*}
1\!-\!P(z)\;=\;z^{-5}\Cdot\big(z\!-\!1\big)\cdot
\big(z^2\!-\!2\CdoT\cos(0,\!246)\cdot z\!+\!1\big)\cdot
\big(z^2\!-\!2\CdoT\cos(0,\!4)\cdot z\!+\!1\big)
\\*[-6pt plus12pt minus0pt]
\end{gather*}
f"uhrte, bei der am Einheitskreis im Bereich niedriger Frequenzen 
f"unf Nullstellen liegen, mit deren Hilfe die St"oranteile bei niedrigen 
Frequenzen abgesenkt werden sollten. 

Bei der Messung wurde das System mit dem in Kapitel \ref{myChirp} 
beschriebenen zuf"alligen Chirpsignal erregt. Dabei wurde ein 
maximaler Crest-Faktor von \mbox{$\text{Cr}_{\text{max}}\!=\!1,\!6$} 
zugelassen. Da am Eingang der Quantisierung nicht nur das Chirpsignal 
\mbox{$\boldsymbol{v}(k)$} sondern auch noch der gefilterte 
Quantisierungsfehler anliegt, wurde f"ur den bei allen Einzelmessungen 
maximal zul"assigen Spitzenwert des Betrags des Chirpsignals 
\mbox{$\max|v_{\lambda}(k)|$} nicht der Wert Eins sondern 
nur \mbox{$0,\!96$} gew"ahlt, um so eine "Ubersteuerung der 
Quantisierungskennlinie zu vermeiden. Der Effektivwert des 
Chirpsignals ist dann \mbox{$0,\!6$} und als Betrag aller 
Spektrallinien ergibt sich mit Gleichung (\ref{4.16}) der Wert 
\mbox{$V_C=0,\!6\CdoT\sqrt{M\,}$}. Die maximale Amplitude des 
zuf"alligen Phasenhubs des sinusf"ormigen Anteils an der Gesamtphase 
des Chirp-Spektrums berechnet sich mit Gleichung (\ref{4.18}), 
mit der man \mbox{$\varphi_{\text{max}}\;=\;7/32$} erh"alt, zu 
\mbox{$\varphi_{\text{max}}\CdoT M/\pi$}. Bei den Einzelmessungen 
wurden Phasenwerte \mbox{$\varphi_{\lambda}$} verwendet, 
die als Stichproben aus einer im Bereich nach Gleichung (\ref{4.17}) 
gleichverteilten Zufallsgr"o"se gewonnen wurden. Das Vorzeichen des 
Chirpsignals wurde bei den Einzelmessungen gleichwahrscheinlich 
zuf"allig gew"ahlt, wie dies am Ende von Kapitel \ref{myChirp} 
beschrieben ist. Um die Stationarit"at des Quantisierungsfehlers 
zu gew"ahrleisten, wurden die $M$ Elemente der Vektoren 
\mbox{$\Vec{v}_{\lambda}$}, die das bei der Einzelmessung 
gerade verwendete Chirpsignal vor der periodischen Fortsetzung 
enthalten, um eine zuf"allige gleichwahrscheinliche Anzahl von Elementen 
zyklisch rotiert. 

Die Zustandsgr"o"sen des R"uckkopplungsfilters im \mbox{$\Sigma\Delta$}-Glied 
wurden nur am Anfang der gesamten Messung mit Null initialisiert. 
Bei den folgenden Einzelmessungen wurden die Zustandsgr"o"sen 
nicht mehr zur"uckgesetzt. Um dennoch die Unabh"angigkeit der 
Einzelmessungen zu gew"ahrleisten wurde die Einschwingzeit\footnote{Am 
Ersatzschaltbild erkennt man, dass trotz der R"uckkopplung im realen 
\mbox{$\Sigma\Delta$}-Glied das lineare Teilsystem keine Einschwingzeit 
ben"otigt, da es die "Ubertragungsfunktion Eins aufweist. Die Messung 
best"atigt dies, da man praktisch dieselben Ergebnisse erh"alt, wenn man 
\mbox{$E\!=\!0$} einstellt.} auf \mbox{$E\!=\!1000$} gesetzt. 

Ganz oben in Bild~\ref{b5y}
\begin{figure}[btp]
\begin{center}
{ 
\begin{picture}(454,599)

\input{mbild5y1}
\put(35,560){\makebox(0,0)[r]{\small$-70$}}
\put(35,540){\makebox(0,0)[r]{\small$-90$}}
\put(35,520){\makebox(0,0)[r]{\small$-110$}}
\put(35,500){\makebox(0,0)[r]{\small$-130$}}
\put(42,467){\makebox(0,0)[lb]{\small$0$}}
\put(130,467){\makebox(0,0)[b]{\small$\pi/4$}}
\put(220,467){\makebox(0,0)[b]{\small$\pi/2$}}
\put(310,467){\makebox(0,0)[b]{\small$3\pi/4$}}
\put(400,467){\makebox(0,0)[b]{\small$\pi$}}
\put(455,475){\makebox(0,0)[tr]{${\T\Omega\!=\!\mu\CdoT\frac{2\pi}{M}}$}}
\put(5,590){\makebox(0,0)[l]{
$10\CdoT\log_{10}\bigl(\,\bigl|\Hat{\Phi}_{\boldsymbol{n}}(\mu)\bigr|\,\bigr)$}}
\put(120,590){\makebox(0,0)[l]{Rechteckfenster}}
\put(360,560){\makebox(0,0)[l]{${\D10\CdoT\log_{10}(Q^2/12)}$}}
\put(360,520){\makebox(0,0)[l]{$L\!=\!10$}}
\put(360,500){\makebox(0,0)[l]{$M\!=\!256$}}

\input{mbild5y2}
\put(35,410){\makebox(0,0)[r]{\small$-70$}}
\put(35,390){\makebox(0,0)[r]{\small$-90$}}
\put(35,370){\makebox(0,0)[r]{\small$-110$}}
\put(35,350){\makebox(0,0)[r]{\small$-130$}}
\put(42,317){\makebox(0,0)[lb]{\small$0$}}
\put(130,317){\makebox(0,0)[b]{\small$\pi/4$}}
\put(220,317){\makebox(0,0)[b]{\small$\pi/2$}}
\put(310,317){\makebox(0,0)[b]{\small$3\pi/4$}}
\put(400,317){\makebox(0,0)[b]{\small$\pi$}}
\put(455,325){\makebox(0,0)[tr]{${\T\Omega\!=\!\mu\CdoT\frac{2\pi}{M}}$}}
\put(5,440){\makebox(0,0)[l]{
$10\CdoT\log_{10}\bigl(\,\bigl|\Hat{\Phi}_{\boldsymbol{n}}(\mu)\bigr|\,\bigr)$}}
\put(120,440){\makebox(0,0)[l]{Rechteckfenster}}
\put(360,410){\makebox(0,0)[l]{${\D10\CdoT\log_{10}(Q^2/12)}$}}
\put(360,370){\makebox(0,0)[l]{$L\!=\!1000$}}
\put(360,350){\makebox(0,0)[l]{$M\!=\!256$}}

\input{mbild5y3}
\put(35,260){\makebox(0,0)[r]{\small$-70$}}
\put(35,240){\makebox(0,0)[r]{\small$-90$}}
\put(35,220){\makebox(0,0)[r]{\small$-110$}}
\put(35,200){\makebox(0,0)[r]{\small$-130$}}
\put(42,167){\makebox(0,0)[lb]{\small$0$}}
\put(130,167){\makebox(0,0)[b]{\small$\pi/4$}}
\put(220,167){\makebox(0,0)[b]{\small$\pi/2$}}
\put(310,167){\makebox(0,0)[b]{\small$3\pi/4$}}
\put(400,167){\makebox(0,0)[b]{\small$\pi$}}
\put(455,175){\makebox(0,0)[tr]{${\T\Omega\!=\!\mu\CdoT\frac{2\pi}{M}}$}}
\put(5,290){\makebox(0,0)[l]{
$10\CdoT\log_{10}\bigl(\,\bigl|\Hat{\Phi}_{\boldsymbol{n}}(\mu)\bigr|\,\bigr)$}}
\put(120,290){\makebox(0,0)[l]{Rechteckfenster}}
\put(360,260){\makebox(0,0)[l]{${\D10\CdoT\log_{10}(Q^2/12)}$}}
\put(360,220){\makebox(0,0)[l]{$L\!=\!10$}}
\put(360,200){\makebox(0,0)[l]{$M\!=\!4096$}}

\input{mbild5y4}
\put(35,110){\makebox(0,0)[r]{\small$-70$}}
\put(35,90){\makebox(0,0)[r]{\small$-90$}}
\put(35,70){\makebox(0,0)[r]{\small$-110$}}
\put(35,50){\makebox(0,0)[r]{\small$-130$}}
\put(42,17){\makebox(0,0)[lb]{\small$0$}}
\put(130,17){\makebox(0,0)[b]{\small$\pi/4$}}
\put(220,17){\makebox(0,0)[b]{\small$\pi/2$}}
\put(310,17){\makebox(0,0)[b]{\small$3\pi/4$}}
\put(400,17){\makebox(0,0)[b]{\small$\pi$}}
\put(455,25){\makebox(0,0)[tr]{${\T\Omega\!=\!\mu\CdoT\frac{2\pi}{M}}$}}
\put(5,140){\makebox(0,0)[l]{
$10\CdoT\log_{10}\bigl(\,\bigl|\Hat{\Phi}_{\boldsymbol{n}}(\mu)\bigr|\,\bigr)$}}
\put(120,140){\makebox(0,0)[l]{Fenster mit \mbox{$N\!=\!4$}}}
\put(360,110){\makebox(0,0)[l]{${\D10\CdoT\log_{10}(Q^2/12)}$}}
\put(360,70){\makebox(0,0)[l]{$L\!=\!10$}}
\put(360,50){\makebox(0,0)[l]{$M\!=\!256$}}

\end{picture}}
\end{center}\vspace{-30pt}
\caption[LDS des \mbox{$\Sigma\Delta$}-Gliedes.]{
LDS des \mbox{$\Sigma\Delta$}-Gliedes.\makebox[200pt]{}
\protect\\
Messung mit Chirpsignal und \mbox{$E\!=\!1000$}.}
\label{b5y}
\end{figure}
ist das Ergebnis 
der LDS-Messung f"ur \mbox{$M\!=\!256$} Frequenzpunkte bei einer 
Mittelung "uber zehn Einzelmessungen dargestellt, wobei keine 
Fensterung stattfand, was einer impliziten Fensterung mit einem 
Rechteckfenster der L"ange $M$ entspricht. Wider erwarten ergibt sich 
bei der Messung im unteren Frequenzbereich durch das Einf"ugen der 
Fehlerr"uckkopplung keine nennenswerte Absenkung im gemessenen LDS. 

Zum Vergleich ist auch das LDS eingezeichnet, das sich im Ersatzschaltbild
theoretisch ergibt, wenn man das mit \mbox{${\D Q^2/12}$} konstante LDS
des Quantisierungsfehlers \mbox{$\boldsymbol{n}(k)$} mit dem
Quadrat des Betrags der "Ubertragungsfunktion
\mbox{$1-P\big(e^{j\Omega}\big)$} im St"orpfad des Ersatzschaltbildes
gewichtet. Nun liegt es f"ur jemanden, der sich noch nicht allzu intensiv
mit der Theorie des RKM besch"aftigt hat, zun"achst nahe anzunehmen,
dass die Varianz der Messwerte einfach noch zu hoch sei, so dass man
in diesem Frequenzbereich lediglich die Betragsquadrate zuf"alliger
Spektralwerte misst. Daher wird man als n"achstes die Mittelungsanzahl
erh"ohen. F"ur \mbox{$L\!=\!1000$} sind die Messergebnisse im
darunterliegenden Teilbild dargestellt. Abgesehen davon, dass nun
die Messwertvarianz offensichtlich zur"uckgegangen ist, hat sich an dem
merkw"urdigen Verhalten des gemessenen Verlaufs im Bereich niedriger
Frequenzen nichts wesentliches ge"andert. Da man bei einer Erh"ohung der
Anzahl der gemessenen Frequenzpunkte auf \mbox{$M\!=\!4096$} im dritten
\pagebreak[2]Teilbild ein leicht abgesenktes, gemessenes LDS im niederfrequenten
Bereich erh"alt, kann man davon ausgehen, dass die Ursache f"ur dieses
Ph"anomen nicht bei dem tats"achlich vorhanden LDS, sondern bei dem
angewandten Messverfahren zu suchen ist. Wenn man nun grob extrapoliert,
wieviele Frequenzpunkte $M$ man messen m"usste, um in den Bereich zu kommen,
in dem sich das theoretische LDS des Ersatzschaltbildes befindet, so erkennt
man bereits ohne gro"se Rechnung, dass der f"ur $M$ notwendigen Wert
in einer Gr"o"senordnung liegen d"urfte, bei dem eine Berechnung in
akzeptabler Zeit unm"oglich w"are. Wie das unterste Teilbild deutlich
zeigt, kann durch die relativ aufwandsarme Erweiterung des Messverfahrens
um die Fensterung mit einer geeigneten Fensterfolge die gew"unschte
Messbarkeit des LDS im Bereich niedriger Frequenzen erreicht werden.

\chapter{Zusammenfassung und Ausblick}

Das Rauschklirrmessverfahren bietet bei einer Vielzahl von n"aherungsweise 
linearen Systemen eine effiziente M"oglichkeit zugleich das 
lineare "Ubertragungsverhalten zu messen, und eine Absch"atzung 
f"ur die spektrale Leistungsdichte der Abweichungen von diesem 
linearen Verhalten zu erhalten. Erstmalig wurde die ideale Modellierung 
des realen Systems durch ein lineares Modellsystem als L"osung 
eines theoretischen Regressionsproblems ausf"uhrlich beschrieben. 
Dabei wurde von Anfang an eine Erweiterung auf komplexwertige Systeme 
vorgenommen. Es stellte sich heraus, dass eine sinnvolle Art der 
Beschreibung der spektralen Eigenschaften des Regressionsfehlers 
nur dann m"oglich ist, wenn man auch schon bei der L"osung der 
theoretischen Regression eine geeignete Fensterung vornimmt. 
Es konnte dabei eine hinreichende Bedingung f"ur die dabei zu 
verwendende Fensterfolge angegeben werden, deren Einhaltung 
sicherstellt, dass die L"osung der theoretischen Regression 
durch die Fensterung nicht verf"alscht wird. Durch die zus"atzliche 
Angabe eines modifizierten Leistungsdichtespektrums wurde die Beschreibung 
der zweiten Momente des komplexen Regressionsfehlers vervollst"andigt. 

Die Aufgabe des RKM besteht nun darin, Sch"atzwerte f"ur die bis 
dahin hergeleiteten Regressionskoeffizienten messtechnisch zu 
bestimmen, und die theoretischen zweiten Momente des Spektrums des 
Regressionsfehlers abzusch"atzen. Das bisher "ubliche Messverfahren 
wurde so modifiziert und erweitert, dass bei komplexen Systemen 
zus"atzlich das modifizierte Leistungsdichtespektrum des Regressionsfehlers 
abgesch"atzt werden kann. Dabei stellte sich heraus, dass die nun 
dabei vorgenommene Fensterung nur eine minimale Erh"ohung des 
Rechenaufwands bewirkt, wenn man eine Fensterfolge verwendet, 
die den bei der L"osung der theoretischen Regression aufgestellten 
Forderungen gen"ugt. Es wird gezeigt, dass man mit dem so erweiterten 
Messverfahren erwartungstreue\footnote{Das bisher "ubliche RKM lieferte 
teils nur asymptotisch (\,\mbox{$L\!\to\!\infty$}\,) erwartungstreue 
Messergebnisse.} und konsistente Messwerte erh"alt. Erstmals werden 
nun auch Sch"atzwerte f"ur die Konfidenzintervalle reeller 
Messwerte und Konfidenzellipsen komplexer Messwerte angegeben. 

Da oft reellwertige Systeme zu messen sind, wurde gezeigt, welche 
Vereinfachungen und Besonderheiten sich bei dem Messverfahren in 
diesem Fall ergeben. Ebenso wurde die reine Spektralsch"atzung eines 
Zufallsprozesses als ein Sonderfall des RKM beschrieben. F"ur die 
Messung reellwertiger Systeme kann es unter Umst"anden erforderlich sein, 
andere Zufallssignale zur Erregung des zu messenden realen Systems zu verwenden, 
als bei komplexwertigen Systemen. Es ist bei komplexwertigen 
Systemen oft vorteilhaft, diese --- wenn m"oglich --- mit einem 
zuf"alligen Chirpsignal zu erregen. Da das komplexe zuf"allige 
Chirpsignal manchmal bei reellwertigen Systemen nicht eingesetzt werden 
kann, wurde hier erstmals ein geeignetes reelles zuf"alliges Chirpsignal 
angegeben, das "ahnlich gute Eigenschaften besitzt, wie sein komplexes 
Pendant.

Beim RKM l"asst sich nicht jede beliebige Fensterfolge einsetzen.
Unter den bisher bekannten Fensterfolgen, die f"ur die Spektralsch"atzung
"ublicherweise verwendet werden, fand sich keine, die alle Forderungen
zugleich und mit der gew"unschten Pr"azision erf"ullt. Es wurde
ein Prinzip f"ur die Konstruktion geeigneter Fensterfolgen 
entwickelt, und daraus ein Algorithmus abgeleitet, mit dessen 
Hilfe die Fensterfolgen berechnet werden k"onnen. Es wurde untersucht, 
welche Anforderungen der Algorithmus bez"uglich der Genauigkeit der 
Berechnung zu erf"ullen hat, wenn die damit berechneten Fensterfolgen 
die Forderungen hochgenau erf"ullen sollen, die f"ur die Einsetzbarkeit 
beim RKM notwendig sind. Ein gro"ser Vorteil des hier angegebenen 
Algorithmus ist es, dass dieser nicht iterativ arbeitet, und somit 
keine Probleme hinsichtlich der Konvergenz, des Startwertes, des 
Abbruchs oder der Genauigkeit der Berechnung auftreten. Anhand einiger
Beispiele wird unter anderem auch demonstriert, wie pr"azise die so 
berechneten Fensterfolgen die beim RKM notwendigen Forderungen erf"ullen. 
Weitere Beispiele demonstrieren einerseits die Vorteile einer 
geeigneten Fensterung bei der Messung mit dem RKM, und dienen 
andererseits der Erl"auterung einiger bei der Theorie des RKM 
dargestellter Sachverhalte. Der Anhang enth"alt neben einigen Details, 
die speziell f"ur die Beschreibung des RKM und der Fensterfolgen 
ben"otigt werden, auch einige recht allgemein hilfreiche Beweise 
und Anmerkungen. 

Bei der Beschreibung des RKM ergab sich an mehreren Stellen 
die Notwendigkeit Einschr"ankungen vorzunehmen. So wurde hier nur 
die Variante des RKM untersucht, die sich zur Messung zeitinvarianter 
Systeme mit station"arer und mittelwertfreier St"orung eignet. 
Um den Rahmen dieser Abhandlung nicht zu sprengen, wurden die RKM-Varianten 
zur Messung solcher Systeme, die sich durch periodisch zeitvariante 
Modellsysteme modellieren lassen, und die alle Kreuzkorrelationen des 
Systemein- und -ausgangs ber"ucksichtigen, und auch solcher Systeme, bei denen 
sich bei der Modellierung ein zyklostation"arer Approximationsfehlerprozess 
mit einem zeitabh"angigen Mittelwert ergibt, in einer gesonderten 
Abhandlung \cite{Erg} n"aher beschrieben. Dort ist auch angegeben, wie 
sich weitere Fensterfolgen mit einem erweiterten Algorithmus berechnen 
lassen, die ebenfalls f"ur den Einsatz beim RKM geeignet sind, und bei denen 
man weitere Freiheitsgrade nutzen kann, um bestimmte Eigenschaften der 
Fensterfolgen positiv zu beeinflussen. Wie man kontinuierliche 
Fensterfunktionen mit analogen Eigenschaften berechnen kann, wird auch 
dort gezeigt. Weitere praxisorientierte Betrachtungen zum Ablauf der 
RKM-Messung sowie zur Berechnung der Fensterfolgen sind ebenfalls in 
\cite{Erg} zu finden.

Andere Aspekte, die sich bei der Herleitung des RKM ergaben,
konnten bisher nicht n"aher untersucht werden. So w"are es
beispielsweise w"unschenswert, das RKM auch zur Messung von
Systemen im laufenden Betrieb verwenden zu k"onnen. In diesem
Fall ist es nicht m"oglich als Erregung periodische Testsignalsequenzen
zu verwenden, die man mit einem geeigneten Zufallsgenerator selbst
erzeugt. Da es jedoch bei dem hier untersuchten RKM erst durch die
bereichsweise Periodizit"at der Erregung m"oglich wird,
die Matrizen des beim RKM zu l"osenden Gleichungssystems
durch eine DFT zu diagonalisieren, und so eine erhebliche
Reduktion des Aufwands zu erzielen, w"are im Fall einer
vorgegebenen nicht periodischen Erregung das RKM zu modifizieren,
um mit einem vertretbaren Mehraufwand auch solche Messungen vornehmen
zu k"onnen. 

Ein m"oglicher Ansatz hierzu besteht im wesentlichen
aus zwei Ver"anderungen. Zum einen m"usste man die gleichwertige
Mittelung aller Messergebnisse durch eine dynamische Mittelung
ersetzen, bei der die jeweils letzten Einzelmessungen st"arker
gewichtet in die Mess"-er"-geb"-nisse eingehen (\,z.~B. durch eine rekursive
Tiefpassfilterung der Messergebnisse\,). Zum zweiten w"are bei jeder
Einzelmessung mit Hilfe der bis dahin gewonnenen Messergebnisse
f"ur das lineare Verhalten des Systems eine Sch"atzung f"ur die
Antwort des Systems auf diejenige Erregung vorzunehmen, die der
realen Erregung additiv "uberlagert eine periodische Erregung
ergibt. Wenn man nun die gesch"atzte Antwort des Systems auf diese
periodifizierende Erregung vom gemessenen Systemausgangssignal
subtrahiert, kann man bei der jeweiligen Einzelmessung dieselben
Berechnungen durchf"uhren, wie bei dem hier untersuchten RKM.
Die Fehler der Sch"atzung der subtrahierten Systemantwort w"urden
dann "ahnlich wie der eigentlich zu messende Approximationsfehler
in die Messergebnisse einflie"sen. Wenn nun die Sch"atzung
des linearen Systemverhaltens im Laufe der Messung mit der
dynamischen Mittelung der Messergebnisse immer besser wird
--- was zu vermuten ist, aber nicht pauschal feststeht ---,
werden die Sch"atzfehler der Antwort auf die periodifizierende
Erregung immer geringer und die Messwerte der spektralen
Eigenschaften des Approximationsfehlers werden bei einer
dynamischen Mittelung ebenfalls immer genauer. Es soll
zuk"unftigen Untersuchungen vorbehalten sein, diesen Ansatz,
das RKM zu modifizieren, auf Anwendbarkeit zu testen,
und die Theorie dazu n"aher zu ergr"unden.

Fensterfolgen wurden auch bisher schon f"ur andere Anwendungen als 
die Spektralsch"atzung eingesetzt. Es liegt nun nahe zu untersuchen, 
ob auch die hier vorgestellte Fensterfolge f"ur diese Zwecke 
verwendet werden kann. Ein Beispiel hierf"ur ist die Methode 
des Filterentwurfs mit Hilfe der Fenstermethode. Da wird der 
Wunschverlauf des Spektrums des zu entwerfenden Filters abgetastet und einer 
DFT unterworfen. Die so entstandene Filterimpulsantwort bildet 
meist noch keine sehr gute Approximation des gew"unschten Filters, 
da der Filterfrequenzgang in der Regel nicht tolerierbare Abweichungen 
vom gew"unschten Verlauf aufweist, die z.~B. durch das Gibbssche Ph"anomen 
verursacht werden. Eine Verbesserung kann meist dadurch erreicht werden, 
dass man die Filterimpulsantwort mit einer Fensterfolge gewichtet. 
Da jedoch die damit erzeugten Filter im Vergleich zu Filtern gleicher
L"ange, die mit anderen Methoden erzeugt worden sind, in der Regel
unterlegen sind, und da die von mir entwickelte Fensterfolge
zudem nicht linearphasig ist, wurde bisher nicht untersucht, ob
sie sich bei der Filterentwurfsmethode mit der Fensterung einsetzen
l"asst.

Es ist lediglich eine Frage der Interpretation, ob man eine endlich 
lange zeitdiskrete Folge als eine Fensterfolge oder als die 
Impulsantwort eines FIR-Filters betrachtet. Daher w"are es denkbar, 
die Tiefp"asse, deren Filterkoeffizienten die Werte der Fensterfolge 
sind, auch "uberall dort einzusetzen, wo Tiefp"asse mit "ahnlichen 
spektralen Eigenschaften ben"otigt werden. Beispielweise erscheint 
es mir erfolgversprechend, das Fenster als Prototyp-Filter einer 
DFT-Filterbank oder als Filter zur Sendeimpulsformung bei einer 
digitalen "Ubertragung einzusetzen. Diese Anwendungen wurden von 
mir jedoch nicht n"aher untersucht.

\renewcommand{\thechapter}{}
\chapter{Literaturverzeichnis}


\chapter{Anhang}
\renewcommand{\thechapter}{A}
\section[Zweites Moment des Abstands der Regressionsfl"ache der ersten Art 
zur Regressionshyperebene der zweiten Art]{Zweites Moment des Abstands der 
Regressions\-fl"ache der ersten Art zur Regressionshyperebene der zweiten Art}\label{Regress}

Beginnen wir mit einer Vorbemerkung. Der bedingte Erwartungswert
\mbox{$\text{E}\{\mbox{funktion}(\boldsymbol{y})\,\pmb{|}\,x\}$}
ist der Erwartungswert der Funktion der Zufallsgr"o"se \mbox{$\boldsymbol{y}$}, 
der sich abh"angig von der Bedingung $x$ ergibt. Die Erwartungswertbildung 
erfolgt dabei durch Integration\footnote{Gegebenenfalls ist im Stieltjesschen 
Sinne "uber die Verteilungfunktion zu integrieren.} der mit der bedingten
Verteilungsdichtefunktion \mbox{$p_{\boldsymbol{y}}(y\,\pmb{|}\,x)$} 
gewichteten Funktion von $y$ "uber den gesamten Ergebnisraum der Zufallsgr"o"se 
\mbox{$\boldsymbol{y}$}. Sowohl Verteilungsdichtefunktion als auch der 
Ergebnisraum k"onnen dabei vom Wert \mbox{$x$} abh"angen. Es handelt sich 
bei diesem bedingten Erwartungswert also um eine Funktion, die jedem Wert
des Definitionsbereichs von \mbox{$x$} einen konkreten nicht
zuf"alligen Wert zuordnet. Wenn wir nun f"ur die unabh"angige Variable
\mbox{$x$} eine Zufallsgr"o"se \mbox{$\boldsymbol{x}$} einsetzen, so ist
der Erwartungswert eine Funktion der Zufallsgr"o"se \mbox{$\boldsymbol{x}$} 
und somit selbst eine Zufallsgr"o"se. Wir schreiben daher in diesem
Fall sowohl die Bedingung, als auch den bedingten Erwartungswert
\mbox{$\text{\bf E}\{\mbox{funktion}(\boldsymbol{y})\,\pmb{|}\,\boldsymbol{x}\}$}
fettgedruckt, um damit zum Ausdruck zu bringen, dass es sich hierbei um 
Zufallsgr"o"sen handelt. Diese Betrachtung l"asst sich problemlos auf 
Zufallsvektoren erweitern. In diesem Kapitel des Anhangs wird dies im weiteren
der Zufallsvektor $\Vec{\boldsymbol{V}}$ sein, der als Elemente die Zufallsgr"o"sen 
\mbox{$\boldsymbol{V}\!(\mu)$} enth"alt.

Es werden nun f"ur die zuf"alligen Erwartungswerte einige Beziehungen 
angegeben, deren G"ultigkeit man leicht nachweisen kann, wenn man 
ber"ucksichtigt, dass bei der Berechnung der Erwartungswerte jeweils 
diejenigen Faktoren vor das Integral gezogen werden k"onnen, 
die nicht von der Integrationsvariablen abh"angen.
\begin{gather}
\text{\bf E}\big\{\,\boldsymbol{V}\!(\mu)\;\pmb{\big|}\;\Vec{\boldsymbol{V}}\,\big\}\;=\;\boldsymbol{V}\!(\mu)\label{A.1.1}\\*[6pt]
\text{\bf E}\big\{\,|\boldsymbol{V}\!(\mu)|^2\;\pmb{\big|}\;\Vec{\boldsymbol{V}}\,\big\}\;=\;|\boldsymbol{V}\!(\mu)|^2\label{A.1.2}\displaybreak[1]\\[6pt]
\text{\bf E}\big\{\,\boldsymbol{y}(k)^*\!\cdot\boldsymbol{V}\!(\mu)\;\pmb{\big|}\;\Vec{\boldsymbol{V}}\,\big\}\;=\;
\boldsymbol{V}\!(\mu)\cdot\text{\bf E}\big\{\,\boldsymbol{y}(k)^*\;\pmb{\big|}\;\Vec{\boldsymbol{V}}\,\big\}\label{A.1.3}\displaybreak[1]\\[6pt]
\text{\bf E}\Big\{\,\boldsymbol{y}(k)^*\cdot\text{\bf E}\big\{\,\boldsymbol{y}(k)\;\pmb{\big|}\;\Vec{\boldsymbol{V}}\,\big\}\;\pmb{\Big|}\;\Vec{\boldsymbol{V}}\,\Big\}\;=\;
\Big|\,\text{\bf E}\big\{\,\boldsymbol{y}(k)\;\pmb{\big|}\;\Vec{\boldsymbol{V}}\,\big\}\Big|^2\label{A.1.4}\\*[6pt]
\text{\bf E}\Big\{\,\big|\text{\bf E}\big\{\,\boldsymbol{y}(k)\;\pmb{\big|}\;\Vec{\boldsymbol{V}}\,\big\}\big|^2\pmb{\Big|}\;\Vec{\boldsymbol{V}}\,\Big\}\;=\;
\Big|\,\text{\bf E}\big\{\,\boldsymbol{y}(k)\;\pmb{\big|}\;\Vec{\boldsymbol{V}}\,\big\}\Big|^2\label{A.1.5}\displaybreak[1]\\[6pt]
\begin{flalign}
&\text{\bf E}\Bigg\{\;\boldsymbol{y}(k)^*\cdot\bigg(
\frac{1}{M}\cdoT\Sum{\mu=0}{M-1}\!H\!\big({\T\mu\CdoT\frac{2\pi}{M}}\big)\CdoT\boldsymbol{V}\!(\mu)\cdot e^{j\cdot\frac{2\pi}{M}\cdot\mu\cdot k}
\bigg)\;\pmb{\Bigg|}\;\Vec{\boldsymbol{V}}\,\Bigg\}\;=&&
\end{flalign}\label{A.1.6}\notag\\*[4pt]\begin{flalign*}
&&=\;\text{\bf E}\big\{\,\boldsymbol{y}(k)\;\pmb{\big|}\;\Vec{\boldsymbol{V}}\,\big\}^{\!*}\cdot\bigg(
\frac{1}{M}\cdoT\Sum{\mu=0}{M-1}H\!\big({\T\mu\CdoT\frac{2\pi}{M}}\big)\CdoT\boldsymbol{V}\!(\mu)\cdot e^{j\cdot\frac{2\pi}{M}\cdot\mu\cdot k}\bigg)&
\end{flalign*}\notag\\[10pt]
\text{\bf E}\Bigg\{\,\bigg|\,\frac{1}{M}\cdoT\Sum{\mu=0}{M-1}
H\!\big({\T\mu\CdoT\frac{2\pi}{M}}\big)\CdoT\boldsymbol{V}\!(\mu)\cdot e^{j\cdot\frac{2\pi}{M}\cdot\mu\cdot k}\:\bigg|^2
\;\pmb{\Bigg|}\;\Vec{\boldsymbol{V}}\,\Bigg\}\;=\;
\bigg|\,\frac{1}{M}\cdoT\Sum{\mu=0}{M-1}
H\!\big({\T\mu\CdoT\frac{2\pi}{M}}\big)\CdoT\boldsymbol{V}\!(\mu)\cdot
e^{j\cdot\frac{2\pi}{M}\cdot\mu\cdot k}\:\bigg|^2
\raisetag{-6pt}\label{A.1.7}\\[10pt]
\text{\bf E}\big\{\,|\boldsymbol{y}(k)|^2\;\pmb{\big|}\;\Vec{\boldsymbol{V}}\,\big\}\;=\;
\Big|\,\text{\bf E}\big\{\,\boldsymbol{y}(k)\;\pmb{\big|}\;\Vec{\boldsymbol{V}}\,\big\}\,\Big|^2+
\text{\bf E}\bigg\{\,\Big|\,\boldsymbol{y}(k)-
\text{\bf E}\big\{\,\boldsymbol{y}(k)\;\pmb{\big|}\;\Vec{\boldsymbol{V}}\,\big\}\,\Big|^2\;\pmb{\bigg|}\;\Vec{\boldsymbol{V}}\,\bigg\}
\label{A.1.8}
\end{gather}
Mit Hilfe dieser Gleichungen und mit der Gleichung (\ref{2.10})
k"onnen wir nun die zu minimierende Varianz des Approximationsfehlers 
in zwei Schritten berechnen. Zun"achst berechnen wir den bedingten Erwartungswert
\mbox{$\text{E}\big\{\,|\boldsymbol{n}(k)|^2\;\pmb{\big|}\;\Vec{V}\,\big\}$}, 
also die Varianz des Approxima\-tions\-fehlers unter der Bedingung $\Vec{V}$.
In einem zweiten Schritt ber"ucksichtigen wir dann, dass der Vektor $\Vec{V}$ 
zuf"allig aus dem Zufallsvektor $\Vec{\boldsymbol{V}}$ gewonnen wird.
F"ur die Varianz des Approximationsfehlers erhalten wir:
\begin{gather}
\text{E}\big\{\,|\boldsymbol{n}(k)|^2\big\}\;=\;
\text{E}\Big\{\,\text{\bf E}\big\{\,|\boldsymbol{n}(k)|^2
\;\pmb{\big|}\;\Vec{\boldsymbol{V}}\,\big\}\,\Big\}\;=
\label{A.1.9}\\*[6pt]
=\;\text{E}\Bigg\{\,\text{\bf E}\Bigg\{\,\bigg|\,\boldsymbol{y}(k)-
\frac{1}{M}\cdoT\Sum{\mu=0}{M-1}H\!\big({\T\mu\CdoT\frac{2\pi}{M}}\big)\CdoT
\boldsymbol{V}\!(\mu)\cdot e^{j\cdot\frac{2\pi}{M}\cdot\mu\cdot k}
\,\bigg|^2\;\pmb{\Bigg|}\;\Vec{\boldsymbol{V}}\,\Bigg\}\,\Bigg\}\;=\;
\notag\\[8pt]\begin{flalign*}
&=\;\text{E}\Bigg\{\;\bigg|\,\text{\bf E}\big\{\,\boldsymbol{y}(k)
\;\pmb{\big|}\;\Vec{\boldsymbol{V}}\,\big\}-
\frac{1}{M}\cdoT\Sum{\mu=0}{M-1}H\!\big({\T\mu\CdoT\frac{2\pi}{M}}\big)\cdot
\boldsymbol{V}\!(\mu)\cdot e^{j\cdot\frac{2\pi}{M}\cdot\mu\cdot k}
\:\bigg|^2\:\Bigg\}\:+\:{}&&
\end{flalign*}\notag\\*\begin{flalign*}
&&{}\:+\:\text{E}\Bigg\{\,\text{\bf E}\bigg\{\,\Big|\,\boldsymbol{y}(k)\!-\!
\text{\bf E}\big\{\,\boldsymbol{y}(k)\;\pmb{\big|}\;
\Vec{\boldsymbol{V}}\big\}\,\Big|^2\;\pmb{\bigg|}\;
\Vec{\boldsymbol{V}}\,\bigg\}\,\Bigg\}&
\end{flalign*}\notag
\end{gather}
Der erste Summand auf der rechten Seite der Gleichungskette (\ref{A.1.9}) 
ist das zweite Moment des Abstands der Regressionshyperebene der zweiten 
Art von der Regressionsfl"ache der ersten Art. Der zweite Summand 
ist der Erwartungswert der bedingten Varianz der Zufallsgr"o"se 
\mbox{$\boldsymbol{y}(k)$}. Dieser ist von den Regressionskoeffizienten 
\mbox{$H(\mu\CdoT2\pi/M)$} unabh"angig, und spielt daher bei der Minimierung 
der Varianz der Zufallsgr"o"se \mbox{$\boldsymbol{n}(k)$} bez"uglich dieser 
Koeffizienten keine Rolle. Da die Regressionskoeffizienten, welche die 
Minimierungsaufgaben (\ref{2.10}) l"osen, den Term auf der linken Seite 
der Gleichungskette (\ref{A.1.9}) minimieren, wird zugleich 
das zweite Moment des Abstands der Hyperebene der linearen 
N"aherung zu der Regressionsfl"ache der ersten Art minimal.

\section[Cauchy-Schwarzsche Ungleichung f"ur die zweiten Momente
zweier\\komplexer Zufallsgr"o"sen]{\markright{\small\sf\thesection.\ Cauchy-Schwarzsche Ungleichung f"ur die zweiten Momente
zweier komplexer Zufallsgr"o"sen}Cauchy-Schwarzsche Ungleichung f"ur die zweiten Momente
zweier komplexer Zufallsgr"o"sen}\label{Cauchy}

Gegeben seien zwei komplexe Zufallsgr"o"sen \mbox{$\boldsymbol{X}$} und 
\mbox{$\boldsymbol{Y}$}, von denen vorausgesetzt wird, dass die zweiten Momente
\mbox{$\text{E}\big\{\,|\boldsymbol{X}|^2\big\}$},
\mbox{$\text{E}\big\{\,|\boldsymbol{Y}|^2\big\}$} und
\mbox{$\text{E}\big\{\,\boldsymbol{X}\CdoT\boldsymbol{Y}\,\big\}$}
existieren. Mit einer beliebigen komplexen Zahl, die in
Polarkoordinatendarstellung als \mbox{$a\CdoT e^{j\cdot\alpha}$}
mit reellem nichtnegativem $a$ geschrieben wird, l"asst sich der
stets nichtnegative Erwartungswert
\begin{gather}
\label{A.2.1}
a^2\cdot\text{E}\big\{\,|\boldsymbol{X}|^2\big\}-
2\cdot a \cdot \Re\Big\{\,e^{j\cdot\alpha}\cdot
\text{E}\big\{\boldsymbol{X}\CdoT\boldsymbol{Y}\big\}\Big\}+
\text{E}\big\{\,|\boldsymbol{Y}|^2\big\}\;=\\[5pt]
=\;\text{E}\Big\{\,a^2\Cdot|\boldsymbol{X}|^2-
a\CdoT e^{j\cdot\alpha}\Cdot\boldsymbol{X}\CdoT\boldsymbol{Y}-
a\CdoT e^{-j\cdot\alpha}\Cdot\boldsymbol{X}^*\Cdot\boldsymbol{Y}^*\!+
|\boldsymbol{Y}|^2\,\Big\}\;=\notag\\[3pt]
=\;\text{E}\Big\{
\big(a\CdoT e^{j\cdot\alpha}\CdoT\boldsymbol{X}-
\boldsymbol{Y}^*\big)\CdoT
\big(a\CdoT e^{-j\cdot\alpha}\CdoT\boldsymbol{X}^*-
\boldsymbol{Y}\big)\Big\}\;=\;
\text{E}\Big\{\,\big|\,
a\CdoT e^{j\cdot\alpha}\CdoT\boldsymbol{X}-\boldsymbol{Y}^*\big|^2\,
\Big\}\;\ge\;0\notag
\end{gather}
angeben. Diese Ungleichung ist f"ur jeden Winkel $\alpha$, also auch f"ur
\begin{equation}
\alpha=-\winkel\big\{\text{E}\{
\boldsymbol{X}\CdoT\boldsymbol{Y}\}\big\}
\label{A.2.2}
\end{equation}
erf"ullt. Wenn wir diesen Winkel einsetzen, erhalten wir die Ungleichung
\begin{equation}
a^2\Cdot\text{E}\big\{\,|\boldsymbol{X}|^2\big\}-
2\cdot a\cdot\big|\,\text{E}\big\{
\boldsymbol{X}\CdoT\boldsymbol{Y}\big\}\big|+
\text{E}\big\{\,|\boldsymbol{Y}|^2\big\}\;\ge\;0,
\label{A.2.3}
\end{equation}
deren linke Seite ein quadratisches Polynom in $a$ mit reellen Koeffizienten
ist. Da der Koeffizient bei $a^2$ positiv ist, und der Koeffizient bei $a$
negativ ist, liegt das Minimum dieses quadratischen Polynoms immer im
Bereich \mbox{$a\ge0$}. Da die letzte Ungleichung f"ur alle Werte von $a$
immer erf"ullt ist, gilt sie auch f"ur das $a$, das den Minimalwert
des quadratischen Polynoms liefert. Nun setzt man diesen Wert
\begin{equation}
a\;=\;\frac{\;
\big|\text{E}\big\{\boldsymbol{X}\CdoT\boldsymbol{Y}\big\}\big|\;}{
\text{E}\big\{\,|\boldsymbol{X}|^2\big\}}
\label{A.2.4}
\end{equation}
ein und multipliziert die Ungleichung auf beiden Seiten mit dem stets 
positiven reellen Wert \mbox{$\text{E}\big\{\,|\boldsymbol{X}|^2\big\}$}
und man erh"alt die Cauchy-Schwarzsche Ungleichung f"ur die zweiten Momente
zweier komplexer Zufallsgr"o"sen:
\begin{equation}
\big|\,\text{E}\big\{\,\boldsymbol{X}\CdoT\boldsymbol{Y}\,\big\}\big|^2
\,\le\;
\text{E}\big\{\,|\boldsymbol{X}|^2\big\}\cdot
\text{E}\big\{\,|\boldsymbol{Y}|^2\big\}.
\label{A.2.5}
\end{equation}

\section[Zur Unabh"angigkeit der Zufallsgr"o"sen \mbox{$\boldsymbol{V}\!(\mu)$} und
\mbox{$\boldsymbol{N}_{\!\!f}(\mu)$}]{Zur Unabh"angigkeit der Zufallsgr"o"sen\\ 
\mbox{$\boldsymbol{V}\!(\mu)$} und \mbox{$\boldsymbol{N}_{\!\!f}(\mu)$}}\label{Unab}

Der zuf"allige Spektralwert \mbox{$\boldsymbol{N}_{\!\!f}(\mu)$} der gefensterten
St"orung des realen Systems bei einer Frequenz $\mu$ ist dann unabh"angig von dem 
zuf"alligen Spektralwert \mbox{$\boldsymbol{V}\!(\mu)$} der Erregung bei derselben 
Frequenz, wenn sich die zweidimensionale Verbundverteilung\footnote{Man beachte bei komplexen 
Zufallsgr"o"sen und -vektoren den Hinweis in Kapitel \ref{komplrel} auf Seite 
\pageref{komplrel} zur Interpretation der "`$<$"'-Relation im Argument 
der Verbundverteilungsfunktion.} des Zufallsgr"o"sentupels
\mbox{$\big[\boldsymbol{V}\!(\mu),\boldsymbol{N}_{\!\!f}(\mu)\big]^{\TT}$}
als das Produkt der Verteilungen der Zufallsgr"o"sen \mbox{$\boldsymbol{V}\!(\mu)$} 
und \mbox{$\boldsymbol{N}_{\!\!f}(\mu)$} schreiben l"asst.
\begin{gather}
P\Big(\boldsymbol{V}\!(\mu)\!<\!V\!(\mu)\wedge
\boldsymbol{N}_{\!\!f}(\mu)\!<\!N_{\!f}(\mu)\Big)\;\stackrel{!}{=}\;
P\Big(\boldsymbol{V}\!(\mu)\!<\!V\!(\mu)\Big)\cdot
P\Big(\boldsymbol{N}_{\!\!f}(\mu)\!<\!N_{\!f}(\mu)\Big)\notag\\*
\forall\qquad \mu=0\;(1)\;M\!-\!1
\label{A.3.1}
\end{gather}
Auch wenn diese Forderung nach Unabh"angigkeit der Spektralwerte 
\mbox{$\boldsymbol{N}_{\!\!f}(\mu)$} und \mbox{$\boldsymbol{V}\!(\mu)$} 
bei derselben Frequenz f"ur alle Frequenzen $\mu$ erf"ullt sein soll, 
bedeutet dies {\em nicht}, dass der Zufallsvektor
$\Vec{\boldsymbol{v}}$ der Erregung unabh"angig von dem
Approximationsfehlerprozess \mbox{$\boldsymbol{n}(k)$} sein muss, 
oder dass der Zufallsvektor $\Vec{\boldsymbol{V}}$ der L"ange $M$, 
der die Fouriertransformierten \mbox{$\boldsymbol{V}\!(\mu)$} aller 
Frequenzpunkte zu einem Vektor zusammenfasst, unabh"angig sein
muss von dem Zufallsvektor $\Vec{\boldsymbol{N}}_{\!\!f}$ der L"ange $M$,
der die Zufallsgr"o"sen \mbox{$\boldsymbol{N}_{\!\!f}(\mu)$} bei allen
diskreten Frequenzen enth"alt. Die beiden letztgenannten Vektoren sind
definitionsgem"a"s nur dann unabh"angig, wenn sich die gemeinsame
Verbundverteilung --- eine reelle Funktion, die im
\mbox{$2\CdoT M$-}dimensionalen komplexen Raum definiert ist --- aller
\mbox{$2\CdoT M$} Zufallsgr"o"sen \mbox{$\boldsymbol{V}\!(\mu)$} und
\mbox{$\boldsymbol{N}_{\!\!f}(\mu)$} mit \mbox{$\mu=0\;(1)\;M\!-\!1$} als das
Produkt der beiden Verbundverteilungen --- zwei reelle Funktionen, die
jeweils in einem \mbox{$M$-}dimensionalen komplexen Raum definiert sind ---
der beiden beteiligten Zufallsvektoren schreiben l"asst. Die Forderung
\begin{equation}
P\Big(\Vec{\boldsymbol{V}}<\Vec{V}\wedge\Vec{\boldsymbol{N}}_{\!\!f}<\Vec{N}_{\!f}\Big)
\;\stackrel{!}{=}\;
P\Big(\Vec{\boldsymbol{V}}<\Vec{V}\Big)
\cdot
P\Big(\Vec{\boldsymbol{N}}_{\!\!f}<\Vec{N}_{\!f}\Big),
\label{A.3.2}
\end{equation}
die ausgeschrieben 
\begin{gather*}
P\Big(\!\boldsymbol{V}\!(0)\!<\!V\!(0)\!\wedge\ldots\wedge\!
\boldsymbol{V}\!(M\!\!-\!\!1)\!<\!V\!(M\!\!-\!\!1)\!\wedge\!
\boldsymbol{N}_{\!\!f}(0)\!<\!N_{\!f}(0)\!\wedge\ldots\wedge\!
\boldsymbol{N}_{\!\!f}(M\!\!-\!\!1)\!<\!N_{\!f}(M\!\!-\!\!1)\!\Big)\!\stackrel{!}{=}\!\!{}
\\*[4pt]\begin{flalign*}
&=\;P\Big(\boldsymbol{V}\!(0)\!<\!V\!(0)\wedge\ldots\wedge
\boldsymbol{V}\!(M\!-\!1)\!<\!V\!(M\!-\!1)\Big)\cdot{}&&
\end{flalign*}\\*\begin{flalign*}
&&{}\cdot P\Big(\boldsymbol{N}_{\!\!f}(0)\!<\!N_{\!f}(0)\wedge\ldots\wedge
\boldsymbol{N}_{\!\!f}(M\!-\!1)<N_{\!f}(M\!-\!1)\Big)&
\end{flalign*}
\end{gather*}
lautet, muss also {\em nicht}\/ erf"ullt sein. Die geforderte Unabh"angigkeit
bezieht sich f"ur jeden einzelnen diskreten Frequenzpunkt $\mu$ lediglich 
auf die Verbundverteilung des Zufallsgr"o"sentupels
\mbox{$\big[\boldsymbol{V}\!(\mu),\boldsymbol{N}_{\!\!f}(\mu)\big]^{\Tt}\!\!$},
die eine Randverteilung der gemeinsamen Verbundverteilung aller
\mbox{$2\CdoT M$} Zufallsgr"o"sen ist, und sich daher als
\begin{gather}
P\Big(\boldsymbol{V}\!(\mu)<V\!(\mu)\wedge\boldsymbol{N}_{\!\!f}(\mu)<N_{\!f}(\mu)\Big)\;=
\label{A.3.3}\\*[3pt]
=\,P\Bigg(\!\big(\boldsymbol{V}\!(\mu)\!<\!V\!(\mu)\big)\wedge\!\!
\bigwedge_{\substack{\Tilde{\mu}=0\\\Tilde{\mu}\neq\mu}}^{M-1}\!\!
\big(\boldsymbol{V}\!(\Tilde{\mu})\!<\!\infty\!+\!j\infty\big)\wedge
\big(\boldsymbol{N}_{\!\!f}(\mu)\!<\!N_{\!f}(\mu)\big)\wedge\!\!
\bigwedge_{\substack{\Tilde{\mu}=0\\\Tilde{\mu}\neq\mu}}^{M-1}\!\!
\big(\boldsymbol{N}_{\!\!f}(\Tilde{\mu})\!<\!\infty\!+\!j\infty\big)\!\Bigg)
\notag\\*
\forall\qquad\mu=0\;(1)\;M\!-\!1\notag
\end{gather}
schreiben l"asst. W"ahrend die Forderung (\ref{A.3.2}) nur erf"ullt ist,
wenn sich die Verbundverteilung aller \mbox{$2\CdoT M$}
beteiligten Zufallsgr"o"sen f"ur {\em jeden}\/ Punkt des
\mbox{$2\CdoT M$-}dimensionalen komplexen Raumes als Produkt
zweier Faktoren schreiben l"asst, die jeweils nur von den
$M$ Werten \mbox{$V(\mu)$} bzw. den $M$ Werten \mbox{$N_{\!f}(\mu)$}
abh"angen, wird durch die Forderung (\ref{A.3.1}) lediglich die
Faktorisierbarkeit f"ur alle Punkte innerhalb des zweidimensionalen Randes,
der sich durch den Grenz"ubergang \mbox{$V(\Tilde{\mu})\to\infty\!+\!j\infty$} und
\mbox{$N_{\!f}(\Tilde{\mu})\to\infty\!+\!j\infty$} f"ur \mbox{$\Tilde{\mu}\neq\mu$}
ergibt, gefordert. Daher ist die Forderung (\ref{A.3.1}) bei weitem leichter
zu erf"ullen, als die Forderung (\ref{A.3.2}). Es ist daher auch bei
vielen Systemen, die gut ausgesteuerte\footnote{Die Streuung des Signals
am Eingang des Quantisierers sei wesentlich gr"o"ser als die
Quantisierungsstufenh"ohe} Quantisierungen enthalten, zu erwarten,
dass sie die im Hauptteil ben"otigte Forderung (\ref{A.3.1}) wenigstens
in guter N"aherung erf"ullen, obwohl sie die Forderung (\ref{A.3.2})
sicher markant verletzen. Die Tatsache, dass nur die Forderungen (\ref{A.3.1})
nach Faktorisierbarkeit der $M$ zweidimensionalen Randverbundverteilungen
und nicht die Forderung (\ref{A.3.2}) nach Faktorisierbarkeit der gesamten
\mbox{$2\CdoT M$-}dimensionalen Verbundverteilung erf"ullt sein muss,
bildet den wesentlichen Grund, warum das RKM auch auf
nahezu alle Systeme angewendet werden kann, bei denen ein deterministischer
Zusammenhang zwischen Systemeingangssignal und Sy"-stem"-aus"-gangs"-sig"-nal
besteht. Ob die hier geforderte paarweise Unabh"angigkeit
der Elemente der beteiligten Zufallsvektoren bei einem realen System
exakt gegeben ist, oder wenigstens n"aherungsweise in der Art erf"ullt
wird, dass die im weiteren hergeleiteten Ergebnisse als brauchbare
N"aherungen betrachtet werden k"onnen, ist entweder anhand einer
inhaltlichen "Uberlegung oder mit Hilfe eines Hypothesentests bei
dem konkret zu vermessenden System zu "uberpr"ufen.

{\small 
Anmerkung: Dass die paarweise Unabh"angigkeit der Zufallsgr"o"sen
der beiden Zufallsvektoren selbst dann gegeben sein kann, wenn ein
deterministischer Zusammenhang zwischen den Zufallsvektoren
$\Vec{\boldsymbol{V}}$ und $\Vec{\boldsymbol{N}}_{\!\!f}$ besteht,
weil diese Vektoren aus den Ein- und Ausgangssignalen eines
ungest"orten deterministischen nichtlinearen Systems gewonnen werden,
das mit Zufallssignalen erregt wird, sei nun an einem einfachen Beispiel
erl"autert. In diesem Beispiel sei $M=2$ und die Zufallsvektoren enthalten
jeweils zwei reelle Zufallsgr"o"sen. Die beiden Zufallsgr"o"sen
\mbox{$\boldsymbol{V}\!(0)$} und \mbox{$\boldsymbol{V}\!(1)$}
des Zufallsvektors $\Vec{\boldsymbol{V}}$ seien unabh"angig
und gleichverteilt im Intervall \mbox{$[-1;1]$}.
Ihre\linebreak[1] Verbundverteilungsdichte ist daher innerhalb des Gebietes
\mbox{$|V_0|\le 1\;\wedge\;|V_1|\le 1$} konstant \mbox{$1/4$} und sonst
Null. Die beiden Zufallsgr"o"sen \mbox{$\boldsymbol{N}_{\!\!f}(0)$} und
\mbox{$\boldsymbol{N}_{\!\!f}(1)$} des Zufallsvektors
$\Vec{\boldsymbol{N}}_{\!\!f}$
h"angen "uber den deterministischen Zusammenhang
\[
\begin{bmatrix}
\boldsymbol{N}_{\!\!f}(0)\\
\boldsymbol{N}_{\!\!f}(1)
\end{bmatrix}\;=\;
\begin{bmatrix}
\boldsymbol{V}\!(0)+\boldsymbol{V}\!(1)-
\text{sgn}\big(\boldsymbol{V}\!(0)\!+\!\boldsymbol{V}\!(1)\big)\\
\boldsymbol{V}\!(0)-\boldsymbol{V}\!(1)-
\text{sgn}\big(\boldsymbol{V}\!(0)\!-\!\boldsymbol{V}\!(1)\big)
\end{bmatrix}
\]
mit den Zufallsgr"o"sen \mbox{$\boldsymbol{V}\!(0)$} und
\mbox{$\boldsymbol{V}\!(1)$} zusammen. Dabei ist \mbox{sgn$(\ldots)$}
die Vorzeichenfunktion. Die Zufallsvektoren
\mbox{$\big[\boldsymbol{V}\!(0),\,\boldsymbol{V}\!(1),\,
\boldsymbol{N}_{\!\!f}(0)\big]^{\TT}$} und
\mbox{$\big[\boldsymbol{V}\!(0),\,\boldsymbol{V}\!(1),\,
\boldsymbol{N}_{\!\!f}(1)\big]^{\TT}$}
weisen jeweils eine dreidimensionale Verbundverteilungsdichte auf, die
sich mit Hilfe der Delta-Distribution als\vspace{0pt}
\[p_{[\boldsymbol{V}(0),\boldsymbol{V}(1),\boldsymbol{N}_{\!f}(0)]^{
\up{0.3}{\TT}}}
(V_0,V_1,N_0)\;=\;\begin{cases}
\frac{1}{4}\CdoT\delta_0(N_0\!-\!V_0\!-\!V_1\!+\!1)&
\text{ f"ur }\,V_0\!\le\!1\,\wedge\,V_1\!\le\!1\,\wedge\,V_1\!\ge\!-V_0\\
\frac{1}{4}\CdoT\delta_0(N_0\!-\!V_0\!-\!V_1\!-\!1)&
\text{ f"ur }\,V_0\!\ge\!-1\,\wedge\,V_1\!\ge\!-1\,\wedge\,V_1\!<\!-V_0\\
0&\text{ sonst }
\end{cases}\vspace{-6pt}\]
und\vspace{-6pt}
\[p_{[\boldsymbol{V}(0),\boldsymbol{V}(1),\boldsymbol{N}_{\!f}(1)]^{
\up{0.3}{\TT}}}
(V_0,V_1,N_1)\;=\;\begin{cases}
\frac{1}{4}\CdoT\delta_0(N_1\!-\!V_0\!+\!V_1\!+\!1)&
\text{ f"ur }\,V_0\!\le\!1\,\wedge\,V_1\!\ge\!-1\,\wedge\,V_1\!\le\!V_0\\
\frac{1}{4}\CdoT\delta_0(N_1\!-\!V_0\!+\!V_1\!-\!1)&
\text{ f"ur }\,V_0\!\ge\!-1\,\wedge\,V_1\!\le\!1\,\wedge\,V_1\!>\!V_0\\
0&\text{ sonst }
\end{cases}\]
angeben l"asst. Diese beiden Verbundverteilungsdichten lassen sich
jeweils {\em nicht}\/ als das Produkt zweier Faktoren schreiben,
\vadjust{\penalty-100}bei dem der eine Faktor nur von einer der
Variablen $N_0$ bzw. $N_1$ abh"angt w"ahrend der zweite Faktor
 nur von den beiden anderen Variablen $V_0$ und $V_1$ abh"angt.
Daher sind die beiden Zufallsvektoren $\Vec{\boldsymbol{V}}$ und
$\Vec{\boldsymbol{N}}_{\!\!f}$ {\em abh"angig}. Wie man durch Integration
dieser Verbundverteilungsdichten "uber die Variable $V_1$ bei der
ersten und $V_0$ bei der zweiten best"atigen kann, sind die beiden
zweidimensionalen Verbundverteilungsdichten der Zufallsgr"o"senpaare
\mbox{$\big[\boldsymbol{V}\!(0),\boldsymbol{N}_{\!\!f}(0)\big]^{\Tt}$} und
\mbox{$\big[\boldsymbol{V}\!(1),\boldsymbol{N}_{\!\!f}(1)\big]^{\Tt}$} immer
gleich der konstanten Gleichverteilung \mbox{$1/4$} innerhalb des Gebietes, in
dem beide Ver"anderlichen betraglich kleiner $1$ sind. Beide zweidimensionalen
Verbundverteilungsdichten lassen sich daher als das Produkt der
Randverteilungen (\,immer gleichverteilt mit \mbox{$1/2$} im Intervall
\mbox{$[-1;1]$}\,) schreiben. Die Zufallsgr"o"sen der beiden
Zufallsvektoren sind daher paarweise voneinander {\em unabh"angig}.
Der Erwartungswert einer Zufallsgr"o"se, die als Linearkombination der
Produkte der Zufallsgr"o"senpaare definiert ist, l"asst sich daher als
\begin{gather*}
\text{E}\big\{\,\alpha\CdoT\boldsymbol{V}\!(0)\CdoT\boldsymbol{N}_{\!\!f}(0)
+\beta\CdoT\boldsymbol{V}\!(1)\CdoT\boldsymbol{N}_{\!\!f}(1)\,\big\}\;=\;
\alpha\cdot
\text{E}\big\{\,\boldsymbol{V}\!(0)\CdoT\boldsymbol{N}_{\!\!f}(0)\,\big\}+
\beta \cdot
\text{E}\big\{\,\boldsymbol{V}\!(1)\CdoT\boldsymbol{N}_{\!\!f}(1)\,\big\}\;=
\\*[2pt]
=\;\alpha\cdot\text{E}\{\boldsymbol{V}\!(0)\}\cdot
\text{E}\{\boldsymbol{N}_{\!\!f}(0)\}+
\beta\cdot\text{E}\{\boldsymbol{V}\!(1)\}\cdot
\text{E}\{\boldsymbol{N}_{\!\!f}(1)\}
\end{gather*}
schreiben, obwohl die beiden Zufallsvektoren nicht unabh"angig sind.
An diesem Beispiel sieht man, dass die paarweise Unabh"angigkeit
der Zufallsgr"o"sen davon abh"angt, welche gemeinsame Verbundverteilung
die \mbox{$\boldsymbol{V}\!(0)$} und \mbox{$\boldsymbol{V}\!(1)$} aufweisen.
W"aren diese hier nicht verbundgleichverteilt, so w"are die paarweise
Unabh"angigkeit wohl nicht gegeben.}

\section{Zur Konditionierung der empirischen Kovarianzmatrix}\label{Komat}

Bei der Berechnung der Messwerte des RKM ist die empirisch gewonnene
Kovarianzmatrix einiger Werte des Spektrums der Erregung zu invertieren.
Da solch eine Matrix aus einer Stichprobe vom Umfang $L$ der Spektralwerte
der Erregung gewonnen wird, h"angt deren Konditionierung davon ab, welche
konkrete Stichprobe man gerade gezogen hat. Wir wollen daher eine obere
Grenze f"ur die Wahrscheinlichkeit herleiten, eine schlecht konditionierte 
Matrix zu erhalten, und zeigen, dass diese Grenze mit steigendem Umfang
$L$ der Stichprobe mindestens indirekt proportional abf"allt, wenn man die
zuf"alligen Spektralwerte so w"ahlt, dass deren theoretische
Kovarianzmatrix gut konditioniert ist. Es sei darauf hingewiesen,
dass die nun folgende Betrachtung auch f"ur die Varianz einer einzigen
Zufallsgr"o"se gilt, da es sich dabei um den Fall einer \mbox{$1\!\times\!1$}
Kovarianzmatrix handelt.

Gegeben seien $R$ komplexe, mittelwertfreie Zufallsgr"o"sen, die den Zufallsspaltenvektor
$\Vec{\boldsymbol{V}}$ bilden, und deren Momente bis zur vierten Ordnung alle
existieren sollen. Bei technisch relevanten Rauschprozessen findet
immer eine inh"arente Limitierung statt, so dass man davon ausgehen
kann, dass diese Momente existieren.
Ein Teil der zweiten Momente sind die Elemente der
theoretischen \mbox{$R\!\times\!R$} Kovarianzmatrix
\begin{equation}
\underline{C}_{\Vec{\boldsymbol{V}},\Vec{\boldsymbol{V}}}\;=\;
\text{E}\big\{\Vec{\boldsymbol{V}}\CdoT
\Vec{\boldsymbol{V}}^{\,\HH}\big\}\;=\;
\underline{C}_{\Vec{\boldsymbol{V}},\Vec{\boldsymbol{V}}}^{\,\Hh},
\label{A.4.1}
\end{equation}
die immer hermitesch und positiv semidefinit ist. Sie l"asst sich daher
unit"ar kongruent auf Diagonalform transformieren, wobei die Diagonalelemente
alle nichtnegativ reell sind, und mit dem Zeilenindex (\,=~Spaltenindex\,)
des Diagonalelements monoton fallen. Die Diagonalelemente sind die
Singul"arwerte $s_i$ mit \mbox{$i=1\;(1)\;R$} der Kovarianzmatrix,
die zugleich deren Eigenwerte sind. Es gilt \mbox{$s_i\ge s_{i+1}$}.
Die Spektralnorm \mbox{$\snorm{\ldots}$} der Kovarianzmatrix ist der gr"o"ste
Singul"arwert $s_1$, die Konditionszahl ist der Quotient des gr"o"sten und
des kleinsten Singul"arwertes \mbox{$K_{\Vec{\boldsymbol{V}}}=s_1/s_R$}.
Je n"aher die Konditionszahl bei Eins liegt, desto besser ist die
Kovarianzmatrix konditioniert, und desto genauer l"asst sich deren
Inverse berechnen. Jeder beliebige Vektor wird durch die Multiplikation
mit der Kovarianzmatrix auf einen Vektor abgebildet, dessen euklidische
Norm \mbox{$\snorm{\ldots}$} die Bedingung
\begin{equation}
s_R \cdot \snorm{\Vec{x}} \;\le\;
\snorm{\,\underline{C}_{\Vec{\boldsymbol{V}},\Vec{\boldsymbol{V}}} \CdoT \Vec{x}\,} \;\le\;
s_1 \cdot \snorm{\Vec{x}}
\label{A.4.2}
\end{equation}
erf"ullt. Da es sich bei der Spektralnorm um eine mit der euklidischen
Vektornorm kompatible Matrixnorm handelt, gilt au"serdem f"ur jeden
beliebigen Vektor:
\begin{equation}
\snorm{\,\underline{C}_{\Vec{\boldsymbol{V}},\Vec{\boldsymbol{V}}} \CdoT \Vec{x}\,} \;\le\;
\snorm{\underline{C}_{\Vec{\boldsymbol{V}},\Vec{\boldsymbol{V}}}} \cdot \snorm{\Vec{x}}.
\label{A.4.3}
\end{equation}

Die \mbox{$R\!\times\!L$} Matrix $\underline{V}$ enthalte eine
konkrete Stichprobe vom Umfang $L$ des Zufallsvektors $\Vec{\boldsymbol{V}}$,
d.~h. jede Spalte dieser Matrix ist ein Element der Stichprobe, also
eine konkrete Realisierung des Zufallsvektors $\Vec{\boldsymbol{V}}$, und es
wurden insgesamt $L$ konkrete Realisierungen dieses Zufallsvektors
zu einer Matrix zusammengefasst. Bei der Matrix handelt es sich um eine
konkrete Realisierung der Matrix $\underline{\boldsymbol{V}}$ der
mathematischen Stichprobe vom Umfang $L$ des Zufallsvektors
$\Vec{\boldsymbol{V}}$.
Dies ist der Fall, wenn die Stichprobenentnahme in der Art erfolgt ist,
dass jedes Element der Stichprobe --- also jede Spalte der Matrix
$\underline{\boldsymbol{V}}$ --- von jedem anderen Element unabh"angig
ist, und die gleiche Verbundverteilung besitzt, wie der Zufallsvektor
$\Vec{\boldsymbol{V}}$. Aus jeder konkreten Stichprobenmatrix
$\underline{V}$ kann man eine empirische Kovarianzmatrix berechnen.
\begin{equation}
\Hat{\underline{C}}_{\Vec{\boldsymbol{V}},\Vec{\boldsymbol{V}}}\;=\;
\frac{1}{L}\cdot\underline{V}\cdot\underline{V}^{\HH}
\label{A.4.4}
\end{equation}
Nun ben"otigen wir noch eine weitere Matrixnorm \mbox{$\fnorm{\ldots}$},
n"amlich die euklidische Matrixnorm, die auch Frobeniusnorm genannt wird,
und die die Wurzel aus der Summe der Betragsquadrate aller Matrixelemente
ist. Da auch diese Matrixnorm mit der euklidischen Vektornorm
kompatibel ist, gilt auch mit dieser Matrixnorm f"ur jeden
beliebigen Vektor:
\begin{equation}
\Snorm{\,\big(\underline{C}_{\Vec{\boldsymbol{V}},\Vec{\boldsymbol{V}}}\!-\!
\Hat{\underline{C}}_{\Vec{\boldsymbol{V}},\Vec{\boldsymbol{V}}}\big)\CdoT\Vec{x}\,} \,=\;
\Snorm{\,\big(\Hat{\underline{C}}_{\Vec{\boldsymbol{V}},\Vec{\boldsymbol{V}}}\!-\!
\underline{C}_{\Vec{\boldsymbol{V}},\Vec{\boldsymbol{V}}}\big)\CdoT\Vec{x}\,}\,\le\;
\fnorm{\,\Hat{\underline{C}}_{\Vec{\boldsymbol{V}},\Vec{\boldsymbol{V}}}\!-\!
\underline{C}_{\Vec{\boldsymbol{V}},\Vec{\boldsymbol{V}}}}
\cdot\snorm{\Vec{x}}.
\label{A.4.5}
\end{equation}
Zus"atzlich ben"otigen wir noch die Dreiecksungleichung
\begin{equation}
\snorm{\,\Vec{x}\!+\!\Vec{y}\,}\,\le\;\snorm{\Vec{x}}+\snorm{\Vec{y}}.
\label{A.4.6}
\end{equation}
Damit k"onnen wir absch"atzen, dass die Norm des Produkts eines
beliebigen Vektors mit der empirischen Kovarianzmatrix innerhalb eines
bestimmten Intervalls liegen muss.
\begin{gather}
\label{A.4.7}
\snorm{\,\Hat{\underline{C}}_{\Vec{\boldsymbol{V}},\Vec{\boldsymbol{V}}}\CdoT\Vec{x}\,}\,=\;
\Snorm{\,\underline{C}_{\Vec{\boldsymbol{V}},\Vec{\boldsymbol{V}}}\CdoT\Vec{x}+
\big(\Hat{\underline{C}}_{\Vec{\boldsymbol{V}},\Vec{\boldsymbol{V}}}\!-\!
\underline{C}_{\Vec{\boldsymbol{V}},\Vec{\boldsymbol{V}}}\big)
\CdoT\Vec{x}\,}\,\le\\*[4pt]
\le\;\snorm{\,\underline{C}_{\Vec{\boldsymbol{V}},\Vec{\boldsymbol{V}}}\CdoT\Vec{x}\,},+\;
\Snorm{\,\big(\Hat{\underline{C}}_{\Vec{\boldsymbol{V}},\Vec{\boldsymbol{V}}}\!-\!
\underline{C}_{\Vec{\boldsymbol{V}},\Vec{\boldsymbol{V}}}\big)
\CdoT\Vec{x}\,}\,\le\notag\\*[4pt]
\le\;\snorm{\underline{C}_{\Vec{\boldsymbol{V}},\Vec{\boldsymbol{V}}}} \Cdot \snorm{\Vec{x}}\,+\;
\fnorm{\,\Hat{\underline{C}}_{\Vec{\boldsymbol{V}},\Vec{\boldsymbol{V}}}\!-\!
\underline{C}_{\Vec{\boldsymbol{V}},\Vec{\boldsymbol{V}}}}\Cdot\snorm{\Vec{x}}\,=\;
\big(\,s_1 + \fnorm{\,\Hat{\underline{C}}_{\Vec{\boldsymbol{V}},\Vec{\boldsymbol{V}}}\!-\!
\underline{C}_{\Vec{\boldsymbol{V}},\Vec{\boldsymbol{V}}}}\,\big)
\cdot\snorm{\Vec{x}}\notag
\end{gather}
\begin{gather}
\label{A.4.8}
\snorm{\,\Hat{\underline{C}}_{\Vec{\boldsymbol{V}},\Vec{\boldsymbol{V}}}\CdoT\Vec{x}\,}\,=\;
\Snorm{\,\underline{C}_{\Vec{\boldsymbol{V}},\Vec{\boldsymbol{V}}}\CdoT\Vec{x}-
\big(\underline{C}_{\Vec{\boldsymbol{V}},\Vec{\boldsymbol{V}}}\!-\!
\Hat{\underline{C}}_{\Vec{\boldsymbol{V}},\Vec{\boldsymbol{V}}}\big)
\CdoT\Vec{x}\,}\,\ge\\*[4pt]
\ge\;\snorm{\,\underline{C}_{\Vec{\boldsymbol{V}},\Vec{\boldsymbol{V}}}\CdoT\Vec{x}\,}\,-\;
\Snorm{\,\big(\underline{C}_{\Vec{\boldsymbol{V}},\Vec{\boldsymbol{V}}}\!-\!
\Hat{\underline{C}}_{\Vec{\boldsymbol{V}},\Vec{\boldsymbol{V}}}\big)
\CdoT\Vec{x}\,}\,\ge\notag\\*[4pt]
\ge\;s_R\cdot\snorm{\Vec{x}}\,-\;
\fnorm{\,\Hat{\underline{C}}_{\Vec{\boldsymbol{V}},\Vec{\boldsymbol{V}}}\!-\!
\underline{C}_{\Vec{\boldsymbol{V}},\Vec{\boldsymbol{V}}}}\Cdot\snorm{\Vec{x}}\,=\;
\big(\,s_R - \fnorm{\,\Hat{\underline{C}}_{\Vec{\boldsymbol{V}},\Vec{\boldsymbol{V}}}\!-\!
\underline{C}_{\Vec{\boldsymbol{V}},\Vec{\boldsymbol{V}}}}\,\big)
\cdot\snorm{\Vec{x}}\notag
\end{gather}
Diese beiden Grenzen gelten f"ur beliebige Vektoren, also
auch f"ur Vektoren, die auf ihre euklidische Norm normiert worden sind
und daher die L"ange Eins haben. Der kleinste Singul"arwert 
\mbox{$\Hat{s}_R$} der empirischen Kovarianzmatrix ist die minimale
L"ange aller Vektoren, die durch eine Multiplikation mit der
empirischen Kovarianzmatrix aus allen Vektoren der L"ange Eins entstanden
sind. Die L"ange des l"angsten Bildvektors ist entsprechend der gr"o"ste
Singul"arwert \mbox{$\Hat{s}_1$} der empirischen
Kovarianzmatrix. Falls die Frobeniusnorm der Abweichung der empirischen
von der theoretischen Kovarianzmatrix kleiner als $s_R$ ist, kann
die Konditionszahl $\Hat{K}_{\Vec{\boldsymbol{V}}}$ der empirischen
Kovarianzmatrix mit den beiden letzten Ungleichungen abgesch"atzt
werden.
\begin{equation}
\Hat{K}_{\Vec{\boldsymbol{V}}}\;=\;\frac{\;\Hat{s}_1}{\;\Hat{s}_R}\;\le\;
\frac{\;s_1+\fnorm{\,\Hat{\underline{C}}_{\Vec{\boldsymbol{V}},\Vec{\boldsymbol{V}}}\!-\!
\underline{C}_{\Vec{\boldsymbol{V}},\Vec{\boldsymbol{V}}}}}
{\;s_R-\fnorm{\,\Hat{\underline{C}}_{\Vec{\boldsymbol{V}},\Vec{\boldsymbol{V}}}\!-\!
\underline{C}_{\Vec{\boldsymbol{V}},\Vec{\boldsymbol{V}}}}}
\label{A.4.9}
\end{equation}
Wenn die Frobeniusnorm der Abweichung der empirischen von der theoretischen
Kovarianzmatrix au"serdem noch kleiner als
\begin{equation}
\fnorm{\,\Hat{\underline{C}}_{\Vec{\boldsymbol{V}},\Vec{\boldsymbol{V}}}\!-\!
\underline{C}_{\Vec{\boldsymbol{V}},\Vec{\boldsymbol{V}}}}\,<\;
\frac{n-1}{\;n\!+\!K_{\Vec{\boldsymbol{V}}}^{-1}}\cdot s_R\;<\;s_R
\qquad\text{ mit }\quad n\in\mathbb{R}\quad\text{ und }\quad n>1
\label{A.4.10}
\end{equation}
ist, kann man sicher sein, dass die Konditionszahl der empirischen
Kovarianzmatrix h"ochstens $n$ mal so gro"s wie die Konditionszahl der
theoretischen Kovarianzmatrix ist:
\begin{equation}
\Hat{K}_{\Vec{\boldsymbol{V}}}\;\le\;
\frac{\;s_1+\fnorm{\,\Hat{\underline{C}}_{\Vec{\boldsymbol{V}},\Vec{\boldsymbol{V}}}\!-\!
\underline{C}_{\Vec{\boldsymbol{V}},\Vec{\boldsymbol{V}}}}}
{\;s_R-\fnorm{\,\Hat{\underline{C}}_{\Vec{\boldsymbol{V}},\Vec{\boldsymbol{V}}}\!-\!
\underline{C}_{\Vec{\boldsymbol{V}},\Vec{\boldsymbol{V}}}}}\;\le\;
\frac{\;s_1+\frac{n-1}{\;n+K_{\Vec{\boldsymbol{V}}}^{-1}}\cdot s_R}
{\;s_R-\frac{n-1}{\;n+K_{\Vec{\boldsymbol{V}}}^{-1}}\cdot s_R}\;=\;
n\CdoT K_{\Vec{\boldsymbol{V}}}.
\label{A.4.11}
\end{equation}
Falls nun die Wahrscheinlichkeit, dass die Frobeniusnorm der Matrixdifferenz 
gr"o"ser als die Schranke in der Ungleichung (\ref{A.4.10}) ist, mit steigendem 
$L$ gegen Null konvergiert, kann man durch eine Erh"ohung von $L$ erreichen, 
dass die Wahrscheinlichkeit, eine empirische Kovarianzmatrix zu erhalten, 
die $n$ mal schlechter als die theoretische Kovarianzmatrix konditioniert ist, 
unter einer beliebig kleinen tolerierbaren Schwelle bleibt. Wir wollen daher 
eine obere Schranke f"ur diese Wahrscheinlichkeit herleiten.

Dazu ben"otigen wir einen Satz, aus dem sich auch die Tschebyscheffsche
Ungleichung ableiten l"asst, und den wir aus \cite{Fisz} entnehmen.\\[6pt]
{\sl Satz: Nimmt eine zuf"allige Ver"anderliche $\boldsymbol{Y}$
nur nichtnegative Werte an, und besitzt sie einen endlichen Mittelwert
\text{E$\{\boldsymbol{Y}\}$}, so ist f"ur jede positive Zahl $K$
die Ungleichung
\begin{gather*}
P(\,\boldsymbol{Y}\!\ge\!K\,)\;\le\;\frac{\;\text{E}\{\boldsymbol{Y}\}\;}{K}
\\[-28pt]
\end{gather*}
erf"ullt.}\\[6pt]
Da das Quadrat der Frobeniusnorm einer zuf"alligen Matrix solch eine
zuf"allige Ver"anderliche ist, erhalten wir mit
\[
\boldsymbol{Y}\;=\;
\fnorm{\,\Hat{\underline{\boldsymbol{C}}}_{\Vec{\boldsymbol{V}},\Vec{\boldsymbol{V}}}\!-\!
\underline{C}_{\Vec{\boldsymbol{V}},\Vec{\boldsymbol{V}}}}^2
\qquad\qquad\text{ und mit }\qquad
K = \Big(\frac{n-1}{\;n\!+\!K_{\Vec{\boldsymbol{V}}}^{-1}}\cdot s_R\Big)^{\!2}
\]
die gesuchte obere Grenze\vspace{-8pt}
\begin{gather}
\label{A.4.12}
P\Big(\Hat{\boldsymbol{K}}_{\Vec{\boldsymbol{V}}}\ge
n\CdoT K_{\Vec{\boldsymbol{V}}}\Big)\;\le\\[8pt]
\le\;P\bigg(\,
\fnorm{\,\Hat{\underline{\boldsymbol{C}}}_{\Vec{\boldsymbol{V}},\Vec{\boldsymbol{V}}}\!-\!
\underline{C}_{\Vec{\boldsymbol{V}},\Vec{\boldsymbol{V}}}} \ge
\frac{n-1}{\;n\!+\!K_{\Vec{\boldsymbol{V}}}^{-1}}\cdot s_R \bigg)\;={}
\notag\\[8pt]
{}=\;P\bigg(\,\fnorm{\,\Hat{\underline{\boldsymbol{C}}}_{\Vec{\boldsymbol{V}},\Vec{\boldsymbol{V}}}\!-\!
\underline{C}_{\Vec{\boldsymbol{V}},\Vec{\boldsymbol{V}}}}^2\ge
\Big(\frac{n-1}{\;n\!+\!K_{\Vec{\boldsymbol{V}}}^{-1}}\cdot s_R\Big)^{\!2}
\,\bigg)\;\le
\notag\\[8pt]
\le\;\bigg(\!\frac{n\!+\!K_{\Vec{\boldsymbol{V}}}^{-1}}
{\;(n\!-\!1)\CdoT s_R\;}\!\bigg)^{\!\!2}\Cdot\text{E}\big\{\,
\fnorm{\,\Hat{\underline{\boldsymbol{C}}}_{\Vec{\boldsymbol{V}},\Vec{\boldsymbol{V}}}\!-\!
\underline{C}_{\Vec{\boldsymbol{V}},\Vec{\boldsymbol{V}}}}^2\,\big\}
\notag
\end{gather}
f"ur die Wahrscheinlichkeit, dass die Frobeniusnorm der
Differenz der empirischen und der theoretischen Kovarianzmatrix
oberhalb der zul"assigen Schwelle liegt, die gleichzeitig eine obere Grenze
f"ur die Wahrscheinlichkeit ist, dass die Konditionszahl der empirischen
Kovarianzmatrix h"ochstens $n$ mal so gro"s ist, wie die Konditionszahl
der theoretischen Kovarianzmatrix. Um diese Grenze angeben zu k"onnen,
m"ussen wir den Erwartungswert des Quadrats der Frobeniusnorm der
Differenz der empirischen und der theoretischen Kovarianzmatrix berechnen.
Nach der Definition der Frobeniusnorm ergibt sich diese als die Wurzel
aus der Summe der Betragsquadrate aller Matrixelemente. Der Erwartungswert
des Quadrats der Frobeniusnorm ist daher die Summe der Erwartungswerte
der Betragsquadrate aller zuf"alligen Elemente der Matrixdifferenz.

Mit \mbox{$\Vec{\boldsymbol{V}}_{\!\!i}$} sei die mathematische
Stichprobe vom Umfang $L$ der $i$-ten Zufallsgr"o"se des Zufallsvektors
$\Vec{\boldsymbol{V}}$, also die $i$-te Zeile der Zufallsmatrix
$\underline{\boldsymbol{V}}$ bezeichnet. Das Element
in der $i$-ten Zeile und $j$-ten Spalte der empirischen
Kovarianzmatrix ist
\begin{equation}
\Hat{\boldsymbol{C}}_{\boldsymbol{V}_{\!i},\boldsymbol{V}_{\!j}}\;=\;
\frac{1}{L}\cdot
\Vec{\boldsymbol{V}}_{\!\!i}\cdot
\Vec{\boldsymbol{V}}_{\!\!j}^{\hH}.
\label{A.4.13}
\end{equation}
Der Erwartungswert dieses Elements ist
\begin{equation}
\text{E}\big\{\,
\Hat{\boldsymbol{C}}_{\boldsymbol{V}_{\!i},\boldsymbol{V}_{\!j}}\big\}\;=\;
\text{E}\Big\{\;\frac{1}{L}\cdot
\Vec{\boldsymbol{V}}_{\!\!i}\cdot
\Vec{\boldsymbol{V}}_{\!\!j}^{\hH}\,\Big\}\;=\;
\text{E}\big\{\boldsymbol{V}_{\!\!i}\CdoT
\boldsymbol{V}_{\!\!j}^*\big\}\;=\;
C_{\boldsymbol{V}_{\!i},\boldsymbol{V}_{\!j}}
\label{A.4.14}
\end{equation}
und somit gleich dem entsprechenden Element der theoretischen
Kovarianzmatrix \mbox{$\underline{C}_{\Vec{\boldsymbol{V}},\Vec{\boldsymbol{V}}}$}.
Die Varianz des Elements in der $i$-te Zeile und $j$-ten Spalte 
der empirischen Kovarianzmatrix ist daher gleich dem Erwartungswert 
des Betragsquadrats des Elements in der $i$-te Zeile und $j$-ten 
Spalte der Differenz der empirischen und der theoretischen 
Kovarianzmatrix. Die Varianz dieses Matrixelements berechnet sich zu
\begin{gather}
\text{E}\Big\{\,\Big|
\Hat{\boldsymbol{C}}_{\boldsymbol{V}_{\!i},\boldsymbol{V}_{\!j}}-
\text{E}\{\Hat{\boldsymbol{C}}_{\boldsymbol{V}_{\!i},\boldsymbol{V}_{\!j}}\}
\Big|^2\Big\}\;=\;
\text{E}\Big\{\,\Big|
\Hat{\boldsymbol{C}}_{\boldsymbol{V}_{\!i},\boldsymbol{V}_{\!j}}-
C_{\boldsymbol{V}_{\!i},\boldsymbol{V}_{\!j}}\Big|^2\Big\}\;={}
\notag\displaybreak[1]\\[8pt]
{}=\;\text{E}\Big\{\,\Big|\frac{1}{L}\cdot\Vec{\boldsymbol{V}}_{\!\!i}\cdot
\Vec{\boldsymbol{V}}_{\!\!j}^{\hH}-
\text{E}\big\{\boldsymbol{V}_{\!\!i}\CdoT\boldsymbol{V}_{\!\!j}^*\big\}\Big|^2\Big\}\;=\;
\text{E}\Big\{\,\Big|\frac{1}{L}\cdot
\Sum{\lambda=1}{L}\boldsymbol{V}_{\!\!i,\lambda}\cdot
\boldsymbol{V}_{\!\!j,\lambda}^*-
\text{E}\big\{\boldsymbol{V}_{\!\!i}\CdoT\boldsymbol{V}_{\!\!j}^*\big\}\Big|^2\Big\}\;={}
\notag\\[8pt]
{}\!=\;\frac{1}{L^2}\cdoT\Sum{\lambda_1=1}{L}\;
\Sum{\substack{\lambda_2=1\;\;\\\lambda_2\neq\lambda_1}}{L}
\text{E}\big\{\boldsymbol{V}_{\!\!i,\lambda_1}\Cdot
\boldsymbol{V}_{\!\!j,\lambda_1}^*\big\}\cdot
\text{E}\big\{\boldsymbol{V}_{\!\!i,\lambda_2}^*\Cdot
\boldsymbol{V}_{\!\!j,\lambda_2}\big\}+
\frac{1}{L^2}\cdoT\Sum{\lambda=1}{L}
\text{E}\big\{\boldsymbol{V}_{\!\!i,\lambda}\Cdot
\boldsymbol{V}_{\!\!j,\lambda}^*\Cdot\boldsymbol{V}_{\!\!i,\lambda}^*\Cdot
\boldsymbol{V}_{\!\!j,\lambda}\big\}-{}
\notag\\*[4pt]
{}-\;\frac{1}{L}\cdoT\Sum{\lambda=1}{L}
\text{E}\big\{\boldsymbol{V}_{\!\!i,\lambda}\cdot
\boldsymbol{V}_{\!\!j,\lambda}^*\big\}\cdot
\text{E}\big\{\boldsymbol{V}_{\!\!i}^*\Cdot\boldsymbol{V}_{\!\!j}\big\}-
\frac{1}{L}\cdoT\Sum{\lambda=1}{L}
\text{E}\big\{\boldsymbol{V}_{\!\!i,\lambda}^*\Cdot
\boldsymbol{V}_{\!\!j,\lambda}\big\}\cdot
\text{E}\big\{\boldsymbol{V}_{\!\!i}\CdoT\boldsymbol{V}_{\!\!j}^*\big\}+{}
\notag\\*[4pt]
{}+\Big|\text{E}\big\{\boldsymbol{V}_{\!\!i}\CdoT\boldsymbol{V}_{\!\!j}^*\big\}\Big|^2\;=
\frac{1}{L}\cdot\text{E}\big\{|\boldsymbol{V}_{\!\!i}|^2\Cdot
|\boldsymbol{V}_{\!\!j}|^2\big\}-\frac{1}{L}\cdot
\big|\text{E}\big\{\boldsymbol{V}_{\!\!i}\CdoT\boldsymbol{V}_{\!\!j}^*\big\}\big|^2\;={}
\notag\\[8pt]
{}=\;\frac{1}{L}\cdot\text{E}\Big\{\Big|\boldsymbol{V}_{\!\!i}\cdot
\boldsymbol{V}_{\!\!j}^*-
\text{E}\big\{\boldsymbol{V}_{\!\!i}\CdoT\boldsymbol{V}_{\!\!j}^*\big\}\Big|^2\Big\}.
\label{A.4.15}
\end{gather}
Die Varianz der Elemente der empirischen Kovarianzmatrix nimmt 
also mit steigendem $L$ mit $\alpha/L$ ab, wobei $\alpha$ eine 
positive Konstante ist, die nur von den zweiten und vierten Momenten
der Zufallsgr"o"sen des Vektors $\Vec{\boldsymbol{V}}$ abh"angt. 
Da die Summe der Varianzen aller Matrixelemente der Erwartungswert 
des Quadrats der Frobeniusnorm der Differenz der empirischen und der 
theoretischen Kovarianzmatrix ist, und die Dimension dieser Matrix
mit \mbox{$R\!\times\!R$} von $L$ unabh"angig ist, nimmt auch
der Erwartungswert des Quadrats der Frobeniusnorm mindestens
mit der Ordnung $1/L$ ab. 
\begin{equation}
\text{E}\big\{\,
\fnorm{\,\Hat{\underline{\boldsymbol{C}}}_{\Vec{\boldsymbol{V}},\Vec{\boldsymbol{V}}}\!-\!
\underline{C}_{\Vec{\boldsymbol{V}},\Vec{\boldsymbol{V}}}}^2\,\big\}\;=\;
\Sum{i=1}{L}\Sum{j=1}{L}\text{E}\Big\{\,\Big|
\Hat{\boldsymbol{C}}_{\boldsymbol{V}_{\!i},\boldsymbol{V}_{\!j}}-
C_{\boldsymbol{V}_{\!i},\boldsymbol{V}_{\!j}}\Big|^2\Big\}\;\sim\;
\frac{1}{L}
\label{A.4.16}
\end{equation}
Setzt man dies in die Ungleichung (\ref{A.4.12}) ein, so sieht man,
dass auch die Wahrscheinlichkeit, eine Konditionszahl der empirischen
Kovarianzmatrix zu erhalten, die mehr als $n$ mal so gro"s ist wie die Konditionszahl
der theoretischen Kovarianzmatrix, wenigstens mit $1/L$ abnimmt.

Es sei noch angemerkt, dass dies nur eine {\em obere Schranke}\/
f"ur die Wahrscheinlichkeit ist, dass die empirische
Kovarianzmatrix eine mehr als $n$-fache Konditionszahl
der theoretischen Kovarianzmatrix aufweist. Da dieser oberen
Schranke eine Vielzahl von Ungleichungen zugrundeliegen,
ist anzunehmen, dass die wahre Wahrscheinlichkeit, deren
Berechnung --- wenn "uberhaupt --- nur bei Kenntnis der gemeinsamen
Verbundverteilung aller Elemente des Zufallsvektors $\Vec{\boldsymbol{V}}$
m"oglich w"are, wesentlich kleiner ist als die angegebene obere
Schranke. Die hier gemachte Herleitung hat den Vorteil, dass keine
Aussage "uber die genaue Verbundverteilung ben"otigt wird.
Es gen"ugt, wenn die theoretische Kovarianzmatrix gut
konditioniert ist, um zu gew"ahrleisten, dass auch die
empirische Kovarianzmatrix mit gro"ser Wahrscheinlichkeit brauchbar
konditioniert ist, wenn man $L$ nur gro"s genug w"ahlt.

\section{Zur Berechnung der Messwert(ko)varianzen}\label{4Mom}

Bei der Herleitung der zweiten Momente der Messwerte
\mbox{$\Hat{\boldsymbol{\Phi}}_{\boldsymbol{n}}(\mu)$} und
\mbox{$\Hat{\boldsymbol{\Psi}}_{\boldsymbol{n}}(\mu)$}
eines station"aren Approximationsfehlerprozesses in Kapitel \ref{Kova}
treten immer wieder die Erwartungswerte der Produkte zweier zuf"alliger bilinearer
Formen und eines zuf"alligen Faktors auf. Entsprechende Produkte
finden sich auch in \cite{Erg} bei der Berechnung der zweiten Momente
der Messwerte \mbox{$\Hat{\boldsymbol{\Phi}}_{\boldsymbol{n}}
\big(\mu,\mu\!+\!\Tilde{\mu}\CdoT\frac{M}{K_{\Phi}}\big)$}
und \mbox{$\Hat{\boldsymbol{\Psi}}_{\boldsymbol{n}}
\big(\mu,\mu\!+\!\Tilde{\mu}\CdoT\frac{M}{K_{\Phi}}\big)$}
eines zyklostation"aren Approximationsfehlerprozesses. 
Dabei sind sowohl die Elemente der zuf"alligen Matrizen der bilinearen 
Formen als auch der zuf"allige Faktor Funktionen, die nur von dem 
zuf"alligen Spektrum der Erregung abh"angen. Die Vektoren der 
bilinearen Formen sind ebenfalls zuf"allig und h"angen nur von den 
Spektralwerten des gefensterten Approximationsfehlerprozesses ab. 
Wegen der Gemeinsamkeit, die alle diese Erwartungswerte aufweisen,
werde ich in diesem Unterkapitel die Erwartungswerte
\begin{equation}
\text{E}\Big\{\;\boldsymbol{c}\cdot
\Vec{\boldsymbol{N}}_{\!1}\Cdot
\underline{\boldsymbol{A}}\cdot
\Vec{\boldsymbol{N}}_{\!2}^{\hH}\cdot\,
\Vec{\boldsymbol{N}}_{\!3}\Cdot
\underline{\boldsymbol{B}}\cdot
\Vec{\boldsymbol{N}}_{\!4}^{\hH}\,\Big\}
\label{A.5.1}
\end{equation}
der Produkte zweier bilinearer Formen und eines zuf"alligen Faktors mit den
allgemeineren Matrizen \mbox{$\underline{\boldsymbol{A}}$}
und \mbox{$\underline{\boldsymbol{B}}$}, dem allgemeineren
Faktor $\boldsymbol{c}$ und den allgemeineren Zufallsvektoren
\mbox{$\Vec{\boldsymbol{N}}_{\!1}$},
\mbox{$\Vec{\boldsymbol{N}}_{\!2}$},
\mbox{$\Vec{\boldsymbol{N}}_{\!3}$} und
\mbox{$\Vec{\boldsymbol{N}}_{\!4}$} berechnen, um so eine Vielzahl
von Berechnungen mit den im Einzelfall verwendeten Matrizen, Faktoren
und Vektoren zu vermeiden. Die Zufallsvektoren sind jedoch nicht 
beliebig w"ahlbar. Es muss sich bei den $L$ Spalten der Matrix, 
die die vier Zufallsvektoren
\mbox{$\Vec{\boldsymbol{N}}_{\!1}$},
\mbox{$\Vec{\boldsymbol{N}}_{\!2}$},
\mbox{$\Vec{\boldsymbol{N}}_{\!3}$} und
\mbox{$\Vec{\boldsymbol{N}}_{\!4}$} als Zeilenvektoren enth"alt, 
um die Elemente einer mathematischen Stichprobe vom Umfang $L$ 
des Zufallsspaltenvektors \mbox{$\big[\boldsymbol{N}_{\!1},\,
\boldsymbol{N}_{\!2},\,\boldsymbol{N}_{\!3},\,\boldsymbol{N}_{\!4}\big]^{\TT}$}
handeln. Somit sind alle $L$ Spaltenvektoren voneinander unabh"angig 
und haben dieselbe Verbundverteilung, n"amlich die Verbundverteilung 
des Zufallsspaltenvektors, aus dem die Stichprobe entnommen wurde. 
Die folgende Herleitung beschr"ankt sich auf den Fall, dass es sich
bei dieser Verbundverteilung um eine mittelwertfreie Normalverteilung 
handelt. Wenn mit \mbox{$\boldsymbol{N}_{\!i,\lambda}$} das $\lambda$-te 
Element des Stichprobenvektors \mbox{$\Vec{\boldsymbol{N}}_{\!i}$} 
bezeichnet ist, gilt f"ur die ersten und zweiten Momente somit 
\begin{gather}
\text{E}\big\{\boldsymbol{N}_{\!i,\lambda}\big\}\;=\;
\text{E}\big\{\boldsymbol{N}_{\!i}\big\}\;=\;0\qquad\text{ und }
\label{A.5.2}\\*[8pt]
\text{E}\Big\{\boldsymbol{N}_{\!i,\lambda_1}\Cdot
\boldsymbol{N}_{\!j,\lambda_2}\Big\}\;=\;
\begin{cases}
\;\text{E}\big\{\boldsymbol{N}_{\!i}\CdoT\boldsymbol{N}_{\!j}\big\}&
\text{ f"ur }\lambda_1 = \lambda_2 \\
\;\text{E}\big\{\boldsymbol{N}_{\!i}\big\}\cdot
\text{E}\big\{\boldsymbol{N}_{\!j}\big\}\quad&
\text{ f"ur }\lambda_1 \neq \lambda_2
\end{cases}\label{A.5.3}\\*[8pt]
\forall\qquad\qquad i,j = 1\;(1)\;4
\;;\qquad\lambda,\lambda_1,\lambda_2 = 1\;(1)\;L.\notag
\end{gather}
Entsprechendes gilt auch f"ur konjugierte Stichprobenelemente bez"uglich ihrer 
ersten und zweiten Momente und ihrer Unabh"angigkeit. Es sei darauf hingewiesen, 
dass die hier geforderte Unabh"angigkeit f"ur unterschiedliche Werte von $\lambda$
{\em nicht}\/ bedeutet, dass die Vektoren selbst voneinander unabh"angig sein 
m"ussen. Diese k"onnen z.~B. zueinander konjugiert oder sogar identisch sein. 
Auch f"ur die h"oheren Momente kann man eine Faktorisierung durchf"uhren. 
Wenn man den Erwartungswert eines Produktes bildet, deren Faktoren
Stichprobenelemente teils mit gleichem und teils mit unterschiedlichem
Index $\lambda$ sind, so gruppiert man zun"achst die Faktoren so um, dass
man ein Produkt von Teilprodukten erh"alt, bei dem jedes Teilprodukt
nur mehr Faktoren mit gleichem Index $\lambda$ enth"alt, w"ahrend die
Indizes aller Teilprodukte jeweils unterschiedlich sind. Der Erwartungswert
des gesamten Produktes ist dann das Produkt der Erwartungswerte der
Teilprodukte. Bei den Erwartungswerten der Teilprodukte kann dann der Index
weggelassen werden, weil das Teilprodukt als das Element mit der Nummer
$\lambda$ einer mathematischen Stichprobe vom Umfang $L$ der Zufallsgr"o"se
angesehen werden kann, die das Produkt der an dem Teilprodukt
beteiligten Zufallsgr"o"sen ist. Es gilt also beispielsweise folgendes:
\[
\text{E}\big\{\boldsymbol{N}_{\!1,2}\CdoT
\boldsymbol{N}_{\!1,3}^*\CdoT
\boldsymbol{N}_{\!4,2}\CdoT
\boldsymbol{N}_{\!4,3}\big\}\,=\,
\text{E}\big\{\boldsymbol{N}_{\!1,2}\CdoT
\boldsymbol{N}_{\!4,2}\big\}\CdoT
\text{E}\big\{\boldsymbol{N}_{\!1,3}^*\CdoT
\boldsymbol{N}_{\!4,3}\big\}\,=\,
\text{E}\big\{\boldsymbol{N}_{\!1}\CdoT\boldsymbol{N}_{\!4}\big\}\CdoT
\text{E}\big\{\boldsymbol{N}_{\!1}^*\CdoT\boldsymbol{N}_{\!4}\big\}.
\]
Damit die weitere Herleitung ihre G"ultigkeit beh"alt, m"ussen 
die Elemente der mit dem zuf"alligen Faktor $\boldsymbol{c}$ 
multiplizierten Matrizen \mbox{$\underline{\boldsymbol{A}}$} und 
\mbox{$\underline{\boldsymbol{B}}$} 
von den Elementen der Stichprobenvektoren
\mbox{$\Vec{\boldsymbol{N}}_{\!1}$},
\mbox{$\Vec{\boldsymbol{N}}_{\!2}$},
\mbox{$\Vec{\boldsymbol{N}}_{\!3}$} und
\mbox{$\Vec{\boldsymbol{N}}_{\!4}$} unabh"angig sein. 

Das Produkt der bilinearen Formen in Gleichung (\ref{A.5.1})
kann man als Vierfachsumme mit den Lauf"|indizes
\mbox{$\lambda_1,\,\lambda_2,\,\lambda_3,\,\lambda_4= 1\;(1)\;L$}
schreiben, wobei die Lauf"|indizes von links nach rechts den vier
Zufallsvektoren der bilinearen Form zugeordnet werden. Der
Faktor $\Hat{\boldsymbol{c}}$ kann in die Vierfachsumme gezogen werden. Der
Erwartungswert dieser Summe ist die Summe der Erwartungswerte der
einzelnen Summanden. 
\begin{equation}
\Sum{\lambda_1=1}{L}\Sum{\lambda_2=1}{L}\Sum{\lambda_3=1}{L}\Sum{\lambda_4=1}{L}
\text{E}\big\{\,\boldsymbol{N}_{\!1,\lambda_1}\Cdot
\boldsymbol{N}_{\!2,\lambda_2}^*\Cdot
\boldsymbol{N}_{\!3,\lambda_3}\Cdot
\boldsymbol{N}_{\!4,\lambda_4}^*\cdot\,
\boldsymbol{c}\cdot
\boldsymbol{A}_{\lambda_1,\lambda_2}\Cdot
\boldsymbol{B}_{\lambda_3,\lambda_4}\,\big\}
\label{A.5.4}
\end{equation}
Jeder Summand dieser Vierfachsumme besteht aus einem Produkt von 
sieben Faktoren. Dies sind zum einen je ein Element jedes der 
vier beteiligten Zufallsvektoren und zum anderen je ein Element 
jeder der zwei zuf"alligen Matrizen und der Faktor $\boldsymbol{c}$.
\mbox{$\boldsymbol{A}_{\lambda_1,\lambda_2}$} ist dabei das 
zuf"allige Element der Matrix \mbox{$\underline{\boldsymbol{A}}$}
in der Zeile \mbox{$\lambda_1$} und in der Spalte \mbox{$\lambda_2$}.
Entsprechendes gilt f"ur \mbox{$\boldsymbol{B}_{\lambda_3,\lambda_4}$}.
Unter der obengenannten Voraussetzung, kann man den Erwartungswert
jedes Produktes der sieben Faktoren als das Produkt von je zwei
Erwartungswerten schreiben. Dabei ist der eine Faktor der Erwartungswert
des Produkts der vier Elemente der Zufallsvektoren, w"ahrend der andere
Faktor der Erwartungswert des Produkts der zwei zuf"alligen Matrixelemente
und des zuf"alligen Faktors $\boldsymbol{c}$ ist. 
\begin{equation}
\Sum{\lambda_1=1}{L}\Sum{\lambda_2=1}{L}\Sum{\lambda_3=1}{L}\Sum{\lambda_4=1}{L}
\text{E}\big\{\,\boldsymbol{N}_{\!1,\lambda_1}\Cdot
\boldsymbol{N}_{\!2,\lambda_2}^*\Cdot
\boldsymbol{N}_{\!3,\lambda_3}\Cdot
\boldsymbol{N}_{\!4,\lambda_4}^*\big\}\cdot
\text{E}\big\{\,\boldsymbol{c}\cdot
\boldsymbol{A}_{\lambda_1,\lambda_2}\Cdot
\boldsymbol{B}_{\lambda_3,\lambda_4}\,\big\}
\label{A.5.5}
\end{equation}
\begin{table}[btp]
\[
{\renewcommand{\arraystretch}{1.41}
\begin{array}{||c|c|c||}
\hline
\hline
\text{ Zeile }&\text{ Bedingung f"ur die Indizes }&
\text{ gemeinsamer Faktor }\\
\hline
\hline
1&\lambda_1=\lambda_2=\lambda_3=\lambda_4&
\text{E}\{\boldsymbol{N}_{\!1}  \CdoT
          \boldsymbol{N}_{\!2}^*\CdoT
          \boldsymbol{N}_{\!3}  \CdoT
          \boldsymbol{N}_{\!4}^*     \}        \\
\hline
2&\lambda_1\neq\lambda_2=\lambda_3=\lambda_4&
\text{E}\{\boldsymbol{N}_{\!1}       \}        \cdot
\text{E}\{\boldsymbol{N}_{\!2}^*\CdoT
          \boldsymbol{N}_{\!3}  \CdoT
          \boldsymbol{N}_{\!4}^*     \}        \\
3&\lambda_2\neq\lambda_1=\lambda_3=\lambda_4&
\text{E}\{\boldsymbol{N}_{\!2}       \}^{\!\Kk}\Cdot
\text{E}\{\boldsymbol{N}_{\!1}  \CdoT
          \boldsymbol{N}_{\!3}  \CdoT
          \boldsymbol{N}_{\!4}^*     \}        \\
4&\lambda_3\neq\lambda_1=\lambda_2=\lambda_4&
\text{E}\{\boldsymbol{N}_{\!3}       \}        \cdot
\text{E}\{\boldsymbol{N}_{\!1}  \CdoT
          \boldsymbol{N}_{\!2}^*\CdoT
          \boldsymbol{N}_{\!4}^*     \}        \\
5&\lambda_4\neq\lambda_1=\lambda_2=\lambda_3&
\text{E}\{\boldsymbol{N}_{\!4}       \}^{\!\Kk}\Cdot
\text{E}\{\boldsymbol{N}_{\!1}  \CdoT
          \boldsymbol{N}_{\!2}^*\CdoT
          \boldsymbol{N}_{\!3}       \}        \\
\hline
6&\lambda_1=\lambda_2\neq\lambda_3=\lambda_4&
\text{E}\{\boldsymbol{N}_{\!1}  \CdoT
          \boldsymbol{N}_{\!2}^*     \}        \cdot
\text{E}\{\boldsymbol{N}_{\!3}  \CdoT
          \boldsymbol{N}_{\!4}^*     \}        \\
7&\lambda_1=\lambda_3\neq\lambda_2=\lambda_4&
\text{E}\{\boldsymbol{N}_{\!1}  \CdoT
          \boldsymbol{N}_{\!3}       \}        \cdot
\text{E}\{\boldsymbol{N}_{\!2}  \CdoT
          \boldsymbol{N}_{\!4}       \}^{\!\Kk}\\
8&\lambda_1=\lambda_4\neq\lambda_2=\lambda_3&
\text{E}\{\boldsymbol{N}_{\!1}  \CdoT
          \boldsymbol{N}_{\!4}^*     \}        \cdot
\text{E}\{\boldsymbol{N}_{\!2}  \CdoT
          \boldsymbol{N}_{\!3}^*     \}^{\!\Kk}\\
\hline
9&\lambda_1=\lambda_2\neq\lambda_3\neq\lambda_4\neq\lambda_1&
\text{E}\{\boldsymbol{N}_{\!3}       \}        \cdot
\text{E}\{\boldsymbol{N}_{\!4}       \}^{\!\Kk}\Cdot
\text{E}\{\boldsymbol{N}_{\!1}  \CdoT
          \boldsymbol{N}_{\!2}^*     \}        \\
10&\lambda_1=\lambda_3\neq\lambda_2\neq\lambda_4\neq\lambda_1&
\text{E}\{\boldsymbol{N}_{\!2}       \}^{\!\Kk}\Cdot
\text{E}\{\boldsymbol{N}_{\!4}       \}^{\!\Kk}\Cdot
\text{E}\{\boldsymbol{N}_{\!1}  \CdoT
          \boldsymbol{N}_{\!3}       \}        \\
11&\lambda_1=\lambda_4\neq\lambda_2\neq\lambda_3\neq\lambda_1&
\text{E}\{\boldsymbol{N}_{\!2}       \}^{\!\Kk}\Cdot
\text{E}\{\boldsymbol{N}_{\!3}       \}        \cdot
\text{E}\{\boldsymbol{N}_{\!1}  \CdoT
          \boldsymbol{N}_{\!4}^*     \}        \\
12&\lambda_2=\lambda_3\neq\lambda_1\neq\lambda_4\neq\lambda_2&
\text{E}\{\boldsymbol{N}_{\!1}       \}        \cdot
\text{E}\{\boldsymbol{N}_{\!4}       \}^{\!\Kk}\Cdot
\text{E}\{\boldsymbol{N}_{\!2}  \CdoT
          \boldsymbol{N}_{\!3}^*     \}^{\!\Kk}\\
13&\lambda_2=\lambda_4\neq\lambda_1\neq\lambda_3\neq\lambda_2&
\text{E}\{\boldsymbol{N}_{\!1}       \}        \cdot
\text{E}\{\boldsymbol{N}_{\!3}       \}        \cdot
\text{E}\{\boldsymbol{N}_{\!2}  \CdoT
          \boldsymbol{N}_{\!4}       \}^{\!\Kk}\\
14&\lambda_3=\lambda_4\neq\lambda_1\neq\lambda_2\neq\lambda_3&
\text{E}\{\boldsymbol{N}_{\!1}       \}        \cdot
\text{E}\{\boldsymbol{N}_{\!2}       \}^{\!\Kk}\Cdot
\text{E}\{\boldsymbol{N}_{\!3}  \CdoT
          \boldsymbol{N}_{\!4}^*     \}        \\
\hline
15&\lambda_1\!\neq\!\lambda_2\!\neq\!\lambda_3\!\neq\!\lambda_4\!\neq\!
\lambda_1\!\neq\!\lambda_3\!\neq\!\lambda_2\!\neq\!\lambda_4&
\text{E}\{\boldsymbol{N}_{\!1}       \}        \cdot
\text{E}\{\boldsymbol{N}_{\!2}       \}^{\!\Kk}\Cdot
\text{E}\{\boldsymbol{N}_{\!3}       \}        \cdot
\text{E}\{\boldsymbol{N}_{\!4}       \}^{\!\Kk}\\
\hline
\hline
\end{array}}
\]\vspace{-17pt}
\setlength{\belowcaptionskip}{-4pt}
\caption{Aufspaltung der Vierfachsumme des Produktes zweier bilinearer Formen
nach Gleichung (\ref{A.5.5}).}
\rule{\textwidth}{0.5pt}\vspace{-20pt}
\label{TA.1}
\end{table}
Nun fasst man die Summanden der Vierfachsumme zu disjunkten Gruppen zusammen, 
deren Indizes den Bedingungen gen"ugen, die in der Tabelle \ref{TA.1} aufgelistet 
sind. Wenn man ber"ucksichtigt, dass die einzelnen Messungen mit
unterschiedlichem Index unabh"angig sind, und dass die einzelnen Elemente
der Stichproben die gleichen Erwartungswerte besitzen, wie die
entsprechenden Zufallsgr"o"sen, aus denen die Stichproben entnommen wurden,
kann man aus jeder dieser Teilsummen einen gemeinsamen Faktor
ausklammern. Auch diese Faktoren sind in der Tabelle \ref{TA.1} f"ur alle
Teilsummen zusammengestellt. Da die gemeinsamen Faktoren der Teilsummen der 
Zeilen zwei bis f"unf sowie neun bis 15 jeweils den Erwartungswert einer 
einzelnen Zufalls"|gr"o"se wenigstens einmal als Faktor enthalten, und da dieser
aufgrund der vorausgesetzten Mittelwertfreiheit der Zufallsgr"o"sen
\mbox{$\boldsymbol{N}_{\!1}$},
\mbox{$\boldsymbol{N}_{\!2}$},
\mbox{$\boldsymbol{N}_{\!3}$} und
\mbox{$\boldsymbol{N}_{\!4}$} immer Null ist, liefern die Teilsummen dieser 
Zeilen keinen Beitrag zur gesamten Vierfachsumme \ref{A.5.5}. 
Die einzigen Teilsummen, die einen Beitrag liefern, sind die
Teilsummen in den Zeilen eins, sechs, sieben und acht der Tabelle \ref{TA.1}.
In Zeile eins ergibt sich die $\boldsymbol{c}$-fache Einfachsumme der 
Produkte der Hauptdiagonalelemente der Matrix \mbox{$\underline{\boldsymbol{A}}$} 
und der Hauptdiagonalelemente der Matrix \mbox{$\underline{\boldsymbol{B}}$}.
Diese Summe sei im weiteren mit
\begin{equation}
\boldsymbol{S}\;=\;\boldsymbol{c}\cdot
\Sum{\lambda=1}{L}
\boldsymbol{A}_{\lambda,\lambda}\cdot
\boldsymbol{B}_{\lambda,\lambda}
\label{A.5.6}
\end{equation}
bezeichnet. Bei den Doppelsummen der Zeilen sechs, sieben und acht erg"anzt man 
nun diese Summe $\boldsymbol{S}$ einmal mit positivem und einmal mit negativem
Vorzeichen, und berechnet die mit dem positiven Vorzeichen erg"anzte
Doppelsumme entweder als Produkt der beiden Matrixspuren oder als Spur eines
Matrixproduktes. Man erh"alt so die Werte der verbleibenden Teilsummen,
"uber die der Erwartungswert zu bilden ist.
\begin{equation}
{\renewcommand{\arraystretch}{1.3}
\begin{array}{||c|c||}
\hline
\hline
\text{ Zeile }&\text{ Wert der Teilsumme }\\
\hline
\hline
1&\boldsymbol{S}\\
\hline
6&\boldsymbol{c}\cdot
\text{spur}\big(\underline{\boldsymbol{A}}\big)\cdot
\text{spur}\big(\underline{\boldsymbol{B}}\big)
-\boldsymbol{S}\\
7&\boldsymbol{c}\cdot
\text{spur}\big(\,\underline{\boldsymbol{A}}^{\TT}\!\Cdot
\underline{\boldsymbol{B}}\,\big)-\boldsymbol{S}\\
8&\boldsymbol{c}\cdot
\text{spur}\big(\,\underline{\boldsymbol{A}}\Cdot
\underline{\boldsymbol{B}}\,\big)-\boldsymbol{S}\\
\hline
\hline
\end{array}}
\label{A.5.7}
\end{equation}
Die Einfachsumme, die der ersten Zeile der Tabelle \ref{TA.1} 
entspricht, enth"alt als gemeinsamen Faktor eines der vierten 
Momente der vier beteiligten Zufallsgr"o"sen. Im Kapitel \ref{Gauss} 
des Anhangs wird gezeigt, dass sich dieses vierte Moment im Fall, dass die
gemeinsame Verbundverteilung der daran beteiligten Zufallsgr"o"sen eine
Normalverteilung ist, durch einen Teil ihrer zweiten Momente
ausdr"ucken l"asst:
\begin{align}
\text{E}\{\boldsymbol{N}_{\!1}  \CdoT
          \boldsymbol{N}_{\!2}^*\CdoT
          \boldsymbol{N}_{\!3}  \CdoT
          \boldsymbol{N}_{\!4}^*     \}\;&{}=\,
\text{E}\{\boldsymbol{N}_{\!1}  \CdoT
          \boldsymbol{N}_{\!2}^*     \}\cdot
\text{E}\{\boldsymbol{N}_{\!3}  \CdoT
          \boldsymbol{N}_{\!4}^*     \}\;+{}\notag\\*[2pt]&{}+\;
\label{A.5.8}
\text{E}\{\boldsymbol{N}_{\!1}  \CdoT
          \boldsymbol{N}_{\!3}^{\phantom{*}}\}\cdot
\text{E}\{\boldsymbol{N}_{\!2}  \CdoT
          \boldsymbol{N}_{\!4}^{\phantom{*}}\}^{\!\Kk}+{}\\*[2pt]&{}+\;
\text{E}\{\boldsymbol{N}_{\!1}  \CdoT
          \boldsymbol{N}_{\!4}^*     \}\cdot
\text{E}\{\boldsymbol{N}_{\!2}  \CdoT
          \boldsymbol{N}_{\!3}^*     \}^{\!\Kk}\!.\notag
\end{align}
Dabei spielt es keine Rolle, ob einige der Zufallsgr"o"sen
\mbox{$\boldsymbol{N}_{\!1}$},
\mbox{$\boldsymbol{N}_{\!2}$},
\mbox{$\boldsymbol{N}_{\!3}$} und
\mbox{$\boldsymbol{N}_{\!4}$} zueinander konjugiert oder sogar identisch sind,
weil sie im Hauptteil dieser Abhandlung demselben Spektralwert des 
gefensterten Approximationsfehlerprozesses zugeordnet sind.
Vergleicht man diese zweiten Momente mit den Momenten in den Zeilen
sechs, sieben und acht der Tabelle~\ref{TA.1}, so stellt man fest, dass sich
die Einfachsumme in Zeile eins mit den Anteilen der Teilsummen der Zeilen
sechs, sieben und acht, die die Einfachsumme $-\boldsymbol{S}$ \pagebreak[2]
als Faktor aufweisen, gerade aufheben. Man erh"alt daher als
Erwartungswert der gesamten bilinearen Form:
\begin{align}
\label{A.5.9}
\text{E}\Big\{\;\boldsymbol{c}\cdot
\Vec{\boldsymbol{N}}_{\!1}\Cdot
\underline{\boldsymbol{A}}\cdot
\Vec{\boldsymbol{N}}_{\!2}^{\hH}\cdot\,
\Vec{\boldsymbol{N}}_{\!3}\Cdot
\underline{\boldsymbol{B}}\cdot
\Vec{\boldsymbol{N}}_{\!4}^{\hH}\,\Big\}
\;=\qquad\quad&\\*[3pt]
{}=\;
\text{E}\Big\{\boldsymbol{c}\cdot
\text{spur}\big(\underline{\boldsymbol{A}}\big)\cdot
\text{spur}\big(\underline{\boldsymbol{B}}\big)\Big\}&{}\cdot
\text{E}\{\boldsymbol{N}_{\!1}  \CdoT
          \boldsymbol{N}_{\!2}^*     \}\cdot
\text{E}\{\boldsymbol{N}_{\!3}  \CdoT
          \boldsymbol{N}_{\!4}^*     \}\;+{}\notag\\*
{}+\;
\text{E}\Big\{\boldsymbol{c}\cdot
\text{spur}\big(\,\underline{\boldsymbol{A}}^{\tT}\!\Cdot
\underline{\boldsymbol{B}}\,\big)\Big\}&{}\cdot
\text{E}\{\boldsymbol{N}_{\!1}  \CdoT
          \boldsymbol{N}_{\!3}^{\phantom{*}}\}\cdot
\text{E}\{\boldsymbol{N}_{\!2}  \CdoT
          \boldsymbol{N}_{\!4}^{\phantom{*}}\}^{\!\Kk}+{}\notag\\*
{}+\;
\text{E}\Big\{\boldsymbol{c}\cdot
\text{spur}\big(\,\underline{\boldsymbol{A}}\Cdot
\underline{\boldsymbol{B}}\,\big)\Big\}&{}\cdot
\text{E}\{\boldsymbol{N}_{\!1}  \CdoT
          \boldsymbol{N}_{\!4}^*     \}\cdot
\text{E}\{\boldsymbol{N}_{\!2}  \CdoT
          \boldsymbol{N}_{\!3}^*     \}^{\!\Kk}\!.\notag
\end{align}
Zur Berechnung der zweiten Momente der Messwerte braucht man
in diese Gleichung nur mehr den entsprechenden Faktor
\mbox{$\boldsymbol{c}$} und die Matrizen \mbox{$\underline{\boldsymbol{A}}$} 
und \mbox{$\underline{\boldsymbol{B}}$}, und f"ur die vom
Approximationsfehler abh"angigen Erwartungswerte die theoretischen Gr"o"sen
\mbox{$\Tilde{\boldsymbol{\Phi}}_{\boldsymbol{n}}(\mu)$} bzw.
\mbox{$\Tilde{\boldsymbol{\Psi}}_{\boldsymbol{n}}(\mu)$} einzusetzen.

\section[Vierte Momente eines komplexen mehrdimensionalen
Gau"sprozesses]{Vierte Momente eines komplexen\\mehrdimensionalen
Gau"sprozesses}\label{Gauss}

Die charakteristische Funktion eines reellen
Zufallsspaltenvektors $\Vec{\Tilde{\boldsymbol{n}}}$ der Dimension
\mbox{$2\CdoT\!R$} ist allgemein der Erwartungswert der skalaren Gr"o"se
\mbox{$e^{j\cdot\Vec{\Tilde{s}}\cdot\Vec{\Tilde{\boldsymbol{n}}}}$}
und ist somit eine Funk\-tion (\,komplex und skalar\,), die von
dem reellen Zeilenvektor $\Vec{\Tilde{s}}$ abh"angt, der genauso\-viele
Elemente aufweist wie der Zufallsvektor $\Vec{\Tilde{\boldsymbol{n}}}$.
Im Fall eines reellen \mbox{$2\CdoT\!R$}-dimensionalen, mittel\-wertfreien
und normalverteilten Zufallsvektors $\Vec{\Tilde{\boldsymbol{n}}}$ ergibt sich
nach \cite{Fisz} die "uber dem gesamten \mbox{$2\CdoT\!R$}-dimensionalen Raum
\mbox{${\D\Vec{\Tilde{s}}\in\mathbb{R}^{2\cdot R}}$}
definierte charakteristische Funktion:
\begin{equation}
\text{E}\big\{e^{j\cdot
\Vec{\Tilde{s}}\cdot\Vec{\Tilde{\boldsymbol{n}}}}\big\}\;=\;
e^{-\frac{1}{2}\cdot\Vec{\Tilde{s}}\cdot
\underline{C}_{\Vec{\Tilde{\boldsymbol{n}}},\Vec{\Tilde{\boldsymbol{n}}}}\cdot
\Vec{\Tilde{s}}^{\,\:\TT}}\!\!.
\label{A.6.1}
\end{equation}
Dabei ist $\underline{C}_{\Vec{\Tilde{\boldsymbol{n}}},\Vec{\Tilde{\boldsymbol{n}}}}$ 
die symmetrische, reelle Kovarianzmatrix
\begin{equation}
\underline{C}_{\Vec{\Tilde{\boldsymbol{n}}},\Vec{\Tilde{\boldsymbol{n}}}}\;=\;
\text{E}\big\{\,\Vec{\Tilde{\boldsymbol{n}}}\cdot
                \Vec{\Tilde{\boldsymbol{n}}}^{\uP{-0.5}{\;\TT}}\big\}.
\label{A.6.2}
\end{equation}
Bei einem mehrdimensionalen, reellen und normalverteilten Zufallsvektor
existiert die charakteristische Funktion immer, auch wenn zwischen
den Elementen des Zufallsvektors lineare Abh"angigkeiten existieren,
so dass die Zahl der reellen Freiheitsgrade kleiner als \mbox{$2\CdoT\!R$} ist.
Im Fall linearer Abh"angigkeit der beteiligten Zufallsgr"o"sen existiert
die Verbundverteilungsdichtefunktion des Zufallsvektor --- zumindest im
klassischen Sinne --- nicht. Dennoch existiert die charakteristische
Funktion. Sie wird dann entweder durch Integration im Stieltjesschen
Sinne aus der Verbundverteilung berechnet, oder nach Elimination der
linear abh"angigen Variablen durch Integration einer
Verbundverteilungsdichtefunktion geringerer Dimension im Riemannschen Sinne.

Der Zufallsvektor $\Vec{\Tilde{\boldsymbol{n}}}$ kann als
ein komplexer Zufallsvektor betrachtet werden, der mit der Wahrscheinlichkeit
Eins in dem Gebiet der reellen Vektoren des komplexen linearen Vektorraums
\mbox{${\D\mathbb{C}^{2\cdot R}}$} liegt. Es sei darauf hingewiesen,
dass das Gebiet der reellen Vektoren kein linearer Unterraum ist, da
z.~B. die Multiplikation eines reellen Vektors mit der Konstanten $j$
einen rein imagin"aren Vektor liefert, der nicht im Gebiet der reellen
Vektoren liegt, so dass die Abgeschlossenheit nicht gegeben ist.
Nun bilden wir den komplexen Zufallsvektor\vspace{-6pt}
\begin{equation}
\Vec{\boldsymbol{n}}\;=\;
\underline{T}\cdot\Vec{\Tilde{\boldsymbol{n}}},
\label{A.6.3}
\end{equation}
den wir durch eine lineare Abbildung mit der bis auf den
konstanten Faktor \mbox{$\sqrt{2\,}$} unit"aren Matrix\vspace{-6pt}
\begin{equation}
\underline{T}\;=\;
\begin{bmatrix}
\;\underline{E}\;&\;\;\,j\CdoT\underline{E}\; \\
\;\underline{E}\;&     -j\CdoT\underline{E}\;
\end{bmatrix}
\label{A.6.4}
\end{equation}
aus dem reellen Zufallsspaltenvektor $\Vec{\Tilde{\boldsymbol{n}}}$ erhalten.
Dabei ist die Matrix $\underline{E}$ die Einheitsmatrix der
Dimension \mbox{$R\!\times\!R$}.
Diese eineindeutige lineare Abbildung, erzeugt uns einen \mbox{komplexen}\linebreak
\mbox{$2\CdoT\!R$-dimensionalen}, mittelwertfreien und normalverteilten
Zufallsspaltenvektor, der in \linebreak den zweiten \mbox{$R$}
Zeilen genau die konjugierten Elemente der ersten \mbox{$R$} Zeilen enth"alt.
Der Zufallsvektor $\Vec{\boldsymbol{n}}$ liegt mit der Wahrscheinlichkeit
Eins in dem Gebiet des \mbox{${\D\mathbb{C}^{2\cdot R}}$}, das durch
die Umkehrabbildung\vspace{-4pt}
\begin{equation}
\Vec{\Tilde{\boldsymbol{n}}}\;=\;\frac{1}{2}\cdot
\underline{T}^{\,\HH}\Cdot\Vec{\boldsymbol{n}}
\label{A.6.5}
\end{equation}
auf das Gebiet der reellen Vektoren abgebildet wird. Mit
dem $R$-dimensionalen komplexen Zufallsspaltenvektor\vspace{-6pt}
\begin{equation}
\Vec{\Tilde{\boldsymbol{N}}}\;=\;
\big[\;\underline{E}\;,\;\;j\CdoT\underline{E}\;\big]\cdot
\Vec{\Tilde{\boldsymbol{n}}},
\label{A.6.6}
\end{equation}
der im gesamten Vektorraum \mbox{$\mathbb{C}^R$} definiert ist, kann man auch
\begin{equation}
\Vec{\boldsymbol{n}}\;=\;\begin{bmatrix}
\Vec{\Tilde{\boldsymbol{N}}}\;\\
\Vec{\Tilde{\boldsymbol{N}}}^{\kK}
\end{bmatrix}
\label{A.6.7}
\end{equation}
schreiben. Ersetzen wir den Vektor $\Vec{\Tilde{\boldsymbol{n}}}$, so ergibt
sich f"ur die symmetrische, reelle Kovarianzmatrix
\begin{equation}
\makebox[\displaywidth]{$\displaystyle 
\underline{C}_{\Vec{\Tilde{\boldsymbol{n}}},\Vec{\Tilde{\boldsymbol{n}}}}\,=\,
\text{E}\big\{\Vec{\Tilde{\boldsymbol{n}}}\CdoT
               \Vec{\Tilde{\boldsymbol{n}}}^{\uP{-0.5}{\,\TT}}\big\}\,=\,
\text{E}\big\{\Vec{\Tilde{\boldsymbol{n}}}\CdoT
              \Vec{\Tilde{\boldsymbol{n}}}^{\uP{-0.5}{\,\HH}}\big\}\,=\,
\text{E}\big\{\,\tfrac{1}{2}\cdot\underline{T}^{\,\HH}\!\Cdot
                                \Vec{\boldsymbol{n}}\cdot
                                \Vec{\boldsymbol{n}}^{\HH}\!\Cdot
                                \underline{T}\cdot\tfrac{1}{2}\,\big\}\,=\,
\tfrac{1}{4}\cdot\underline{T}^{\,\HH}\!\Cdot
\text{E}\big\{\Vec{\boldsymbol{n}}\cdot\Vec{\boldsymbol{n}}^{\HH}\big\}
\Cdot\underline{T}.$}
\label{A.6.8}
\end{equation}
Sie l"asst sich durch eine unit"are kongruente
"Ahnlichkeitstransformation einer Matrix berechnen, die sich
aus vier \mbox{$R\!\times\!R$} Bl"ocken zusammensetzt.
Die vier B"ocke sind dabei die beiden Kovarianzmatrizen
\mbox{$\text{E}\big\{\,\Vec{\Tilde{\boldsymbol{N}}}\CdoT
\Vec{\Tilde{\boldsymbol{N}}}^{\hH}\,\big\}$} und 
\mbox{$\text{E}\big\{\,\Vec{\Tilde{\boldsymbol{N}}}\CdoT
\Vec{\Tilde{\boldsymbol{N}}}^{\tT}\,\big\}$} 
des $R$-dimensionalen komplexen Zufallsvektors $\Vec{\Tilde{\boldsymbol{N}}}$ 
und deren Konjugierte. Setzen wir die Matrix nach Gleichung (\ref{A.6.8})
in die charakteristische Funktion nach Gleichung (\ref{A.6.1}) ein,
so erhalten wir mit
\begin{equation}
\text{E}\big\{e^{j\cdot
\Vec{\Tilde{s}}\cdot\Vec{\Tilde{\boldsymbol{n}}}}\big\}\,=\,
\text{E}\big\{e^{\frac{j}{2}\cdot
\Vec{\Tilde{s}}\cdot\underline{T}^{\HH}\Cdot\Vec{\boldsymbol{n}}}\big\}\,=\,
e^{-\frac{1}{2}\cdot\Vec{\Tilde{s}}\cdot
\underline{C}_{\Vec{\Tilde{\boldsymbol{n}}},\Vec{\Tilde{\boldsymbol{n}}}}\cdot
\Vec{\Tilde{s}}^{\,\:\TT}}=\,
e^{-\frac{1}{2}\cdot\Vec{\Tilde{s}}\cdot
\underline{C}_{\Vec{\Tilde{\boldsymbol{n}}},\Vec{\Tilde{\boldsymbol{n}}}}\cdot
\Vec{\Tilde{s}}^{\;\HH}}=\,
e^{-\frac{1}{8}\cdot\Vec{\Tilde{s}}\cdot\underline{T}^{\HH}\Cdot
\text{E}\{\Vec{\boldsymbol{n}}\cdot\Vec{\boldsymbol{n}}^{\HH}\}\cdot
\underline{T}\cdot\Vec{\Tilde{s}}^{\;\HH}}
\label{A.6.9}
\end{equation}
eine Funktion der Variablen des reellen Vektors \mbox{$\Vec{\Tilde{s}}$}.
Wir k"onnen die charakteristische Funktion auch als eine Funktion
von \mbox{$2\CdoT\!R$} komplexen Variablen, den Elementen eines
komplexen Vektors, auf"|fassen, wobei der Definitionsbereich der
Funktion gleich dem Gebiet der reellen Vektoren ist. Mit den
linear abgebildeten Vektoren
\begin{equation}
\Vec{s}\;=\;\Vec{\Tilde{s}}\cdot\underline{T}^{\,\HH}
\label{A.6.10}
\end{equation}
erhalten wir die charakteristische Funktion
\begin{equation}
\text{E}\big\{e^{\frac{j}{2}\cdot
\Vec{s}\cdot\Vec{\boldsymbol{n}}}\big\}\;=\;
e^{-\frac{1}{8}\cdot\Vec{s}\cdot\text{E}\{\Vec{\boldsymbol{n}}\cdot
\Vec{\boldsymbol{n}}^{\,\HH}\}\cdot\Vec{s}^{\,\HH}}\!\!,
\label{A.6.11}
\end{equation}
die in dem Gebiet definiert ist, das durch die Abbildung aus dem Gebiet der
reellen Vektoren hervorgeht. Die auf beiden Seiten der Gleichung
auftretende Exponentialfunktion l"asst sich im gesamten Vektorraum
\mbox{${\D\mathbb{C}^{2\cdot R}}$} --- also auch im Gebiet der
reellen Vektoren --- in eine Reihe entwickeln. Die Erwartungswertbildung
ziehen wir in die so entstandene Summe hinein.
\begin{equation}
\Sum{\nu=0}{\infty}\,
\frac{j^{\nu}}{\;2^{\nu}\Cdot\nu!\;}\cdot\text{E}\big\{
(\,\Vec{s}\cdot\Vec{\boldsymbol{n}}\,)^{\nu}\big\}\;=\;
\Sum{\nu=0}{\infty}\;
\frac{(-1)^{\nu}}{\;8^{\nu}\Cdot\nu!\;}\cdot\Big(\,
\Vec{s}\cdot\text{E}\big\{\Vec{\boldsymbol{n}}\cdot
\Vec{\boldsymbol{n}}^{\HH}\big\}\cdot\Vec{s}^{\,\HH}\,\Big)^{\!\nu}
\label{A.6.12}
\end{equation}
Nun f"uhren wir einen Koeffizientenvergleich bei dem Term vierter
Ordnung in $\Vec{s}$ durch. Dazu nehmen wir bei der Summe auf der
linken Seite der Gleichung den Summanden mit dem Index \mbox{$\nu=4$}
und bei der Summe auf der rechten Seite der Gleichung
den Summanden mit dem Index \mbox{$\nu=2$}. Man erh"alt so bei
dem Term vierter Ordnung in $\Vec{s}$:
\begin{equation}
\text{E}\big\{(\,\Vec{s}\cdot\Vec{\boldsymbol{n}}\,)^4\big\}\;=\;
3\cdot\Big(\,\Vec{s}\cdot\text{E}\big\{\Vec{\boldsymbol{n}}\cdot
\Vec{\boldsymbol{n}}^{\Hh}\big\}\cdot\Vec{s}^{\,\HH}\Big)^{\!2}
\label{A.6.13}
\end{equation}
In dem untersuchten Gebiet mit
\mbox{${\D\Vec{s}\cdot\underline{T}\in\mathbb{R}^{2\cdot R}}$}
des \mbox{${\D\mathbb{C}^{2\cdot R}}$} gilt
\begin{equation}
\makebox[\displaywidth]{$\displaystyle 
\Vec{\boldsymbol{n}}^{\HH}\!\Cdot\Vec{s}^{\,\HH}\,=\;
\Vec{\Tilde{\boldsymbol{n}}}^{\uP{-0.5}{\,\HH}}\!\!\CdoT
\underline{T}^{\,\HH}\!\CdoT
\underline{T}\CdoT\Vec{\Tilde{s}}^{\:\HH}=\;
2\CdoT\Vec{\Tilde{\boldsymbol{n}}}^{\uP{-0.5}{\,\HH}}\!\!\Cdot
\Vec{\Tilde{s}}^{\:\HH}=\;
2\CdoT\Vec{\Tilde{\boldsymbol{n}}}^{\uP{-0.5}{\,\TT}}\!\Cdot
\Vec{\Tilde{s}}^{\;\TT}=\;
2\CdoT\tfrac{1}{2}\CdoT\Vec{\boldsymbol{n}}^{\,\TT}\!\!\Cdot
\underline{T}^{\Kk}\!\CdoT
\tfrac{1}{2}\CdoT\underline{T}^{\:\TT}\!\Cdot\Vec{s}^{\;\TT}=\;
\Vec{\boldsymbol{n}}^{\,\TT}\!\Cdot\Vec{s}^{\;\TT}\!,$}
\label{A.6.14}
\end{equation}
so dass wir f"ur Gleichung (\ref{A.6.13}) auch
\begin{equation}
\text{E}\Big\{\big(\,\Vec{s}\cdot\Vec{\boldsymbol{n}}\cdot
       \Vec{\boldsymbol{n}}^{\,\TT}\!\Cdot\Vec{s}^{\;\TT}\,\big)^2\Big\}\;=\;
3\cdot\Big(\,\Vec{s}\cdot\text{E}\big\{\Vec{\boldsymbol{n}}\cdot
\Vec{\boldsymbol{n}}^{\,\TT}\big\}\cdot\Vec{s}^{\;\TT}\,\Big)^{\!2}
\label{A.6.15}
\end{equation}
schreiben k"onnen. Jede der in dieser Gleichung auftretenden bilinearen Formen 
l"asst sich als Doppelsumme schreiben. Da jeweils das Produkt zweier bilinearer 
Formen gebildet wird, erh"alt man schlie"slich zwei Vierfachsummen. Die
Erwartungwertbildung "uber die Vierfachsumme, die auf der linken Seite der
Gleichung steht, kann mit der Summation vertauscht werden. Ebenso kann
das Produkt der Elemente des Vektors $\Vec{s}$ aus der Erwartungwertbildung
herausgezogen werden, da es sich dabei nicht um Zufallsgr"o"sen, sondern um
die unabh"angigen Variablen der charakteristischen Funktion handelt.
Es seien die Indizes der vier Einzelsummen mit $i$, $j$, $k$ und $l$
bezeichnet. Man erh"alt so
\begin{equation}
\begin{aligned}
{}&
\Sum{i=1}{2\cdot R}\;
\Sum{j=1}{2\cdot R}\;
\Sum{k=1}{2\cdot R}\;
\Sum{l=1}{2\cdot R}\;
\text{E}\{\,\boldsymbol{n}_i\CdoT\boldsymbol{n}_j\CdoT
\boldsymbol{n}_k\CdoT\boldsymbol{n}_l\,\}\cdot
s_i\CdoT s_j\CdoT s_k\CdoT s_l\;=\\[8pt]
=\;3\cdot{}&
\Sum{i=1}{2\cdot R}\;
\Sum{j=1}{2\cdot R}\;
\Sum{k=1}{2\cdot R}\;
\Sum{l=1}{2\cdot R}\;
\text{E}\{\boldsymbol{n}_i\CdoT\boldsymbol{n}_j\}\cdot
\text{E}\{\boldsymbol{n}_k\CdoT\boldsymbol{n}_l\}\cdot
s_i\CdoT s_j\CdoT s_k\CdoT s_l
\end{aligned}
\label{A.6.16}
\end{equation}
Dabei sei $s_i$ das $i$-te Element des Vektors $\Vec{s}$ und $\boldsymbol{n}_i$
die $i$-te Zufallsgr"o"se des Zufallsvektors~$\Vec{\boldsymbol{n}}$. Bei
den Vierfachsummen m"ussen nun alle Summanden, die zu derselben Kombination
der Elemente von $\Vec{s}$ f"uhren, addiert, und deren Koeffizienten
verglichen werden. Dies sei bei der Vierfachsumme auf der rechten Seite
der Gleichung (\ref{A.6.16}) am Beispiel des Summanden mit dem Indexquadrupel
\mbox{$i\!=\!1\,\wedge\,j\!=\!2\,\wedge\,k\!=\!3\,\wedge\,l\!=\!4$}
erl"autert. F"ur diesen erh"alt man den Ausdruck
\[
\qquad\text{E}\{\boldsymbol{n}_1\CdoT\boldsymbol{n}_2\}\cdot
\text{E}\{\boldsymbol{n}_3\CdoT\boldsymbol{n}_4\}\cdot
s_1\CdoT s_2\CdoT s_3\CdoT s_4,
\]
der dieselbe Kombination der Elemente des Vektors $\Vec{s}$ aufweist, wie
der Summand mit dem Indexquadrupel
$i\!=\!3\,\wedge\,j\!=\!2\,\wedge\,k\!=\!1\,\wedge\,l\!=\!4$, f"ur den sich
\[
\qquad\text{E}\{\boldsymbol{n}_3\CdoT\boldsymbol{n}_2\}\cdot
\text{E}\{\boldsymbol{n}_1\CdoT\boldsymbol{n}_4\}\cdot
s_3\CdoT s_2\CdoT s_1\CdoT s_4,
\]
ergibt. Da jedoch die Erwartungswerte, deren Produkte als Koeffizienten
vor gleichen Produkten in $s$ auftreten, bei der Vierfachsumme
auf der rechten Seite der Gleichung~(\ref{A.6.16}) i.~Allg. --- wie auch
in unserem Beispiel --- verschieden sind, m"ussen vor dem
Koeffizientenvergleich alle m"oglichen Permutationen der
Indizes, die zu gleichen Produkten in~$s$ f"uhren, addiert werden.
Da die Anzahl dieser Permutationen, die sich mit Hilfe der
Polynomialkoeffizienten berechnen l"asst, davon abh"angt, ob paarweise gleiche
Indizes auftreten, oder nicht, wird dieses Verfahren wegen der dann notwendigen
Fallunterscheidung recht un"ubersichtlich. Die Fallunterscheidungen
kann man jedoch umgehen. Dazu schreibt man sich die Gleichung (\ref{A.6.16})
$4!\!=\!24$ mal untereinander. Jede dieser Gleichungen enth"alt wieder zwei
Vierfachsummen. Bei der Zuteilung der Indizes $i$, $j$, $k$ und $l$ zu den
einzelnen Summen (\,z.~B. $k$ sei der Index der "au"sersten Summe und $i$ der
der innersten Summe etc.\,) w"ahlt man bei jeder der Gleichungen und bei jeder
der Vierfachsummen jeweils eine andere Permutation. Als Permutationen w"ahlt
man dabei jeweils alle $4!$ m"oglichen verschiedenen Permutationen, die alle
das Produkt \mbox{$s_i\CdoT s_j\CdoT s_k\CdoT s_l$}
ergeben. Beispielsweise erg"abe sich bei der ersten Gleichung mit der
Permutation $ikjl$ auf der rechten Seite die Summe 
\[
\qquad\Sum{i=1}{2R}\;
\Sum{k=1}{2R}\;
\Sum{j=1}{2R}\;
\Sum{l=1}{2R}\;
\text{E}\{\boldsymbol{n}_i\CdoT\boldsymbol{n}_k\}\cdot
\text{E}\{\boldsymbol{n}_j\CdoT\boldsymbol{n}_l\}\cdot
s_i\CdoT s_k\CdoT s_j\CdoT s_l
\]
die sich durch Vertauschung der Reihenfolge der Summation und der Faktoren
der Produkte auch als\vspace{-12pt}
\[
\qquad\Sum{i=1}{2R}\;
\Sum{j=1}{2R}\;
\Sum{k=1}{2R}\;
\Sum{l=1}{2R}\;
\text{E}\{\boldsymbol{n}_i\CdoT\boldsymbol{n}_k\}\cdot
\text{E}\{\boldsymbol{n}_j\CdoT\boldsymbol{n}_l\}\cdot
s_i\CdoT s_j\CdoT s_k\CdoT s_l
\]
schreiben l"asst. Bei der Vierfachsumme auf der linken Seite der
Gleichung (\ref{A.6.16}), bewirkt die Vertauschung der Indizes
von $i$, $j$, $k$ und $l$ keine "Anderung des Erwartungwertes,
da eine Vertauschung der Reihenfolge der Faktoren innerhalb der
Erwartungswertbildung die Permutationen wieder ineinander "uberf"uhrt.
Vertauscht man auch auf den linken Seiten aller $4!$ Gleichungen
die Reihenfolge der Summation in der Art, dass die Summenzeichen
wieder in der Reihenfolge $i$, $j$, $k$, $l$ stehen, so erkennt man,
dass man auf den linken Seiten immer exakt dasselbe stehen hat.
Bei der Vierfachsumme auf der rechten Seite lassen sich
von den $24$ Gleichungen jeweils acht Gleichungen durch Vertauschung der
Reihenfolge der Faktoren innerhalb der Erwartungswertbildungen sowie durch
Vertauschung der Reihenfolge der Erwartungswerte ineinander
"uberf"uhren. Es entstehen so acht Gleichungen mit den Erwartungswertprodukten
\mbox{$\text{E}\{\boldsymbol{n}_i\CdoT\boldsymbol{n}_j\}\cdot
       \text{E}\{\boldsymbol{n}_k\CdoT\boldsymbol{n}_l\}$},
acht Gleichungen mit
\mbox{$\text{E}\{\boldsymbol{n}_i\CdoT\boldsymbol{n}_k\}\cdot
       \text{E}\{\boldsymbol{n}_j\CdoT\boldsymbol{n}_l\}$}
und weitere acht Gleichungen mit
\mbox{$\text{E}\{\boldsymbol{n}_i\CdoT\boldsymbol{n}_l\}\cdot
       \text{E}\{\boldsymbol{n}_j\CdoT\boldsymbol{n}_k\}$}.
Nun werden die $24$ Gleichungen addiert. Da alle Gleichungen lediglich
unterschiedliche Darstellungen derselben Gleichung (\ref{A.6.16}) sind,
entspricht das der Multiplikation der Gleichung (\ref{A.6.16}) mit dem Faktor
$24$ und damit einer "aquivalenten Umformung. Beachtet man, dass vor der
Vierfachsumme auf der rechten Seite der Faktor $3$ steht, so erkennt man,
dass sich der Faktor $24$ k"urzen l"asst und man somit die Gleichung
\begin{gather}
\Sum{i=1}{2\cdot R}\;
\Sum{j=1}{2\cdot R}\;
\Sum{k=1}{2\cdot R}\;
\Sum{l=1}{2\cdot R}
\text{E}\{\,\boldsymbol{n}_i\CdoT\boldsymbol{n}_j\CdoT
                \boldsymbol{n}_k\CdoT\boldsymbol{n}_l\,\}\cdot
s_i\CdoT s_j\CdoT s_k\CdoT s_l\;=
\label{A.6.17}\\*[8pt]
=\!
\Sum{i=1}{2\cdoT R}\!
\Sum{j=1}{2\cdoT R}\!
\Sum{k=1}{2\cdoT R}\!
\Sum{l=1}{2\cdoT R}\!
\Big(\!\text{E}\{\!\boldsymbol{n}_i\CdoT\boldsymbol{n}_j\!\}\!\CdoT\!
       \text{E}\{\!\boldsymbol{n}_k\CdoT\boldsymbol{n}_l\!\}\!+\!
       \text{E}\{\!\boldsymbol{n}_i\CdoT\boldsymbol{n}_k\!\}\!\CdoT\!
       \text{E}\{\!\boldsymbol{n}_j\CdoT\boldsymbol{n}_l\!\}\!+\!
       \text{E}\{\!\boldsymbol{n}_i\CdoT\boldsymbol{n}_l\!\}\!\CdoT\!
       \text{E}\{\!\boldsymbol{n}_j\CdoT\boldsymbol{n}_k\!\}\!\Big)\CdoT
s_i\CdoT s_j\CdoT s_k\CdoT s_l\notag
\end{gather}
erh"alt. In dieser Gleichung kann nun jede beliebige Permutation eines
Indexquadrupels eingesetzt werden, ohne dass sich dadurch der Koeffizient
vor dem Produkt der Elemente des Vektors $\Vec{s}$ ver"andert.
Unabh"angig davon, wieviele Permutationen f"ur ein konkretes
Indexquadrupel existieren\footnote{Beispielsweise
gibt es nur einen Summanden mit \mbox{$s_2^4$},
aber vier Summanden mit \mbox{$s_1^3\Cdot s_3$}}\!, ist deren Anzahl auf
beiden Seiten der Gleichung immer gleich. Ein Koeffizientenvergleich f"ur
ein konkretes Indexquadrupel liefert daher immer auf beiden Seiten
der Gleichung denselben konstanten Faktor, n"amlich den
Polynomialkoeffizienten dieser Anzahl der Summanden. Da dieser Faktor aber
auf beiden Seiten der Gleichung steht, kann er immer gek"urzt werden. Es
gen"ugt daher, sich aus allen m"oglichen Permutationen eines Indexquadrupels,
die zu derselben Kombination der Elemente von $\Vec{s}$ f"uhren, einen
beliebigen herauszugreifen, und damit einen Koeffizientenvergleich
durchzuf"uhren. Da dies gerade die einzelnen Summanden
der letzten Gleichung sind, kann der Koeffizientenvergleich mit jedem
einzelnen Summanden der letzten Gleichung getrennt durchgef"uhrt werden.
Man erh"alt daher:
\begin{equation}
\text{E}\{\boldsymbol{n}_i\CdoT\boldsymbol{n}_j\CdoT
          \boldsymbol{n}_k\CdoT\boldsymbol{n}_l\}\,=\,
\text{E}\{\boldsymbol{n}_i\CdoT\boldsymbol{n}_j\}\CdoT
\text{E}\{\boldsymbol{n}_k\CdoT\boldsymbol{n}_l\} +
\text{E}\{\boldsymbol{n}_i\CdoT\boldsymbol{n}_k\}\CdoT
\text{E}\{\boldsymbol{n}_j\CdoT\boldsymbol{n}_l\} +
\text{E}\{\boldsymbol{n}_i\CdoT\boldsymbol{n}_l\}\CdoT
\text{E}\{\boldsymbol{n}_j\CdoT\boldsymbol{n}_k\}.
\raisetag{38pt}\label{A.6.18}
\end{equation}
Damit ist das vierte Momente des \mbox{$2\CdoT\!R$}-dimensionalen,
komplexen, mittelwertfreien und normalverteilten Zufallsvektors
$\Vec{\boldsymbol{n}}$, der alle vier beteiligten Zufallsgr"o"sen
unkonjugiert enth"alt, auf die zweiten Momente zur"uckgef"uhrt.
Voraussetzung bei dieser Herleitung war, dass der komplexe
Zufallsvektor $\Vec{\boldsymbol{n}}$ mit der Wahrscheinlichkeit
Eins in dem Gebiet des \mbox{${\D\mathbb{C}^{2\cdot R}}$} liegt,
das die Vektoren enth"alt, die sich mit Gleichung (\ref{A.6.7}) als
die Aneinanderkettung des beliebigen komplexen Zufallsvektors 
\mbox{$\Vec{\Tilde{\boldsymbol{N}}}\in\mathbb{C}^R$}
und seines Konjugierten $\Vec{\Tilde{\boldsymbol{N}}}^{\uP{0.5}{*}}$
darstellen lassen. Nun kann man die Elemente von $\boldsymbol{n}$
in Gleichung (\ref{A.6.18}) durch die Elemente des Zufallsvektors
$\Vec{\Tilde{\boldsymbol{N}}}$ substituieren. Je nachdem in welchem
Bereich die Indizes $i$, $j$, $k$ und $l$ liegen, ergibt sich die Substitution
\begin{equation}
\boldsymbol{n}_m\;=\;\begin{cases}
\;\Tilde{\boldsymbol{N}}_m&\text{ f"ur }\;0<m\le R\\
\;\Tilde{\boldsymbol{N}}_{m-R}^*\quad&\text{ f"ur }\;R<m\le 2\cdoT R
\end{cases}\qquad\quad
\text{ mit }\quad m\in\{\,i,\,j,\,k,\,l\,\}.\qquad
\label{A.6.19}
\end{equation}
Durch entsprechende Wahl der Indizes erh"alt man aus Gleichung
(\ref{A.6.18}) alle vierten Momente des $R$-dimensionalen, komplexen
Zufallsvektors $\Vec{\Tilde{\boldsymbol{N}}}$. Mit \mbox{$R\!=\!4$},
\mbox{$\Tilde{\boldsymbol{N}}_{\!\nu}\!=\!\boldsymbol{N}_{\!\nu}$},
wobei \mbox{$\nu = 1\;(1)\;4$} zu setzen ist, und dem Indexquadrupel
\mbox{$[\,i,\,j,\,k,\,l\,] = [\,1,\,6,\,3,\,8\,]$} erh"alt man
die Beziehung (\ref{A.5.8}) f"ur eines der vierten Momente.
Die entsprechenden anderen vierten Momente erh"alt man indem man
andere Indexquadrupel w"ahlt. Rein formal entspricht das in 
Gleichung (\ref{A.5.8}) dem eventuellen Konjugieren der Zufallsvektoren
und dem anschlie"senden Ersetzten der Indizes durch die modulo $R$
berechneten Indizes des Indexquadrupels \mbox{$[\,i,\,j,\,k,\,l\,]$}.
Hingewiesen sein noch darauf, dass ebenso wie der reelle
Zufallsvektor $\Vec{\Tilde{\boldsymbol{n}}}$ auch der Zufallsvektor
$\Vec{\boldsymbol{N}}$ deterministisch abh"angige komplexe Zufallsgr"o"sen
enthalten darf. Insbesondere darf der Zufallsvektor $\Vec{\boldsymbol{N}}$ 
neben einer komplexen Zufallsgr"o"se auch deren Konjugierte enthalten, 
da diese Art der deterministischen Abh"angigkeit einer explizit zugelassenen 
linearen Abh"angigkeit im reellen Zufallsvektor $\Vec{\Tilde{\boldsymbol{n}}}$ 
entspricht. Wir k"onnen somit mit Hilfe der Gleichung (\ref{A.5.8}) alle vierten
Momente eines komplexen, mehrdimensionalen, mittelwertfreien
und normalverteilten Zufallsvektors berechnen.
Hierzu ein Beispiel: Gegeben sei die komplexe, mittelwertfreie
normalverteilte Zufallsgr"o"se $\boldsymbol{Z}$. Wenn wir in
Gleichung (\ref{A.5.8}) \mbox{$\boldsymbol{N}_{\!1}\!=\!\boldsymbol{N}_{\!2}\!=\!
\boldsymbol{N}_{\!3}\!=\!\boldsymbol{N}_{\!4}\!=\!\boldsymbol{Z}$} einsetzen, 
erhalten wir das vierte Moment
\mbox{$\text{E}\big\{\,|\boldsymbol{Z}|^4\big\}$}.
Dies entspricht mit $R\!=\!1$ und mit \mbox{$\Vec{\Tilde{\boldsymbol{N}}}\!=\!
\boldsymbol{Z}$} in Gleichung (\ref{A.6.7}) dem Indexquadrupel
\mbox{$[\,i,\,j,\,k,\,l\,] = [\,1,\,2,\,1,\,2\,]$}
oder jedem permutierten dieses Indexquadrupels in Gleichung (\ref{A.6.18}).
Wenn wir das vierte Moment \mbox{$\text{E}\big\{\boldsymbol{Z}^4\big\}$}
berechnen wollen, brauchen wir in Gleichung (\ref{A.5.8}) lediglich die beiden
Zufallsgr"o"sen \mbox{$\boldsymbol{N}_{\!2}$} und
\mbox{$\boldsymbol{N}_{\!4}$} durch ihre Konjugierten zu ersetzen
und k"onnen mit der so entstandenen Formel,
die in Gleichung (\ref{A.6.18}) dem Indexquadrupel
\mbox{$[\,i,\,j,\,k,\,l\,] = [\,1,\,2,\,3,\,4\,]$} mit $R\!=\!4$ entspricht,
das gesuchte vierte Moment berechnen, indem wir wieder
\mbox{$\boldsymbol{N}_{\!1}\!=\!\boldsymbol{N}_{\!2}\!=\!
\boldsymbol{N}_{\!3}\!=\!\boldsymbol{N}_{\!4}\!=\!\boldsymbol{Z}$} einsetzen.
Andererseits k"onnen wir auch mit $R\!=\!4$ und mit\vspace{-4pt}
\[
\Vec{\Tilde{\boldsymbol{N}}}\;=\;
\begin{bmatrix}
\boldsymbol{Z}\;\,\\
\boldsymbol{Z}^*\\
\boldsymbol{Z}\;\,\\
\boldsymbol{Z}^*
\end{bmatrix}\;=\;
\begin{bmatrix}
\boldsymbol{N}_{\!1}\\
\boldsymbol{N}_{\!2}\\
\boldsymbol{N}_{\!3}\\
\boldsymbol{N}_{\!4}
\end{bmatrix}
\]
einen vierdimensionalen Zufallsvektor schaffen, der deterministisch abh"angige
Zufallsgr"o"sen enth"alt. Setzen wir dies in Gleichung (\ref{A.5.8}) ein
(\,\mbox{$\boldsymbol{N}_{\!2}\!=\!\boldsymbol{N}_{\!4}\!=\!\boldsymbol{Z}^*$}
und
\mbox{$\boldsymbol{N}_{\!1}\!=\!\boldsymbol{N}_{\!3}\!=\!\boldsymbol{Z}$}\,), 
so erhalten wir ebenfalls den gleichen Ausdruck f"ur das gesuchte vierte Moment
\mbox{$\text{E}\big\{\boldsymbol{Z}^4\big\}$}, den wir auch durch die
formale Ersetzung der beiden Zufallsgr"o"sen \mbox{$\boldsymbol{N}_{\!2}$} und
\mbox{$\boldsymbol{N}_{\!4}$} durch ihre Konjugierten und die anschlie"sende
Substitution aller vier Zufallsgr"o"sen durch $\boldsymbol{Z}$ erhalten haben.

\section{Zu den Quantisierungsfehlern bei der Berechnung des
Logarithmus}\label{log}

Hier soll nun untersucht werden, wann der Einfluss welcher der 
beiden Hauptquellen f"ur Quantisierungsfehler bei der Berechnung 
des Logarithmus "uberwiegt.

Gegeben sei die positiv reelle Zahl $x$. Wenn wir nun annehmen,
dass diese Zahl nicht extrem klein oder gro"s ist, l"asst sie sich
an einem Rechner mit Flie"skommaarithmetik mit einem relativen Fehler
in der Gr"o"senordnung von $\varepsilon$ darstellen. Wir modellieren dies,
indem wir die quantisierte, im Rechner verwendete Zahl $x_Q$ als das
Produkt \mbox{$x_Q=x\cdot(1\!+\!\varepsilon\CdoT\boldsymbol{q}_x)$}
schreiben, wobei $\boldsymbol{q}_x$ die Zufallsgr"o"se des
Quantisierungsfehlers ist, der die quantisierte Zahl $x_Q$ entstehen l"asst.
Die Streuung dieser Zufallsgr"o"se liegt in der Gr"o"senordnung
von Eins. Mit\vspace{-15pt}
\begin{gather*}
\log_B(x_Q)-\log_B(x)\;=\;
\log_B\big(x\cdot(1\!+\!\varepsilon\CdoT\boldsymbol{q}_x)\big)-\log_B(x)\;=\\
=\;\log_B(1\!+\!\varepsilon\CdoT\boldsymbol{q}_x)\;=\;
\frac{\ln(1\!+\!\varepsilon\CdoT\boldsymbol{q}_x)}{\ln(B)}\;\approx\;
\frac{\varepsilon}{\ln(B)}\cdot\boldsymbol{q}_x
\end{gather*}
kann der absolute Fehler des exakten theoretischen Logarithmus zur Basis $B$
abgesch"atzt werden. Die Streuung dieses absoluten Fehlers ist also
mit $\varepsilon/\ln(B)$ von $x$ unabh"angig. 

Wenn wir davon ausgehen,
dass ein Algorithmus zur Berechnung des Logarithmus verwendet wird, der
immer diejenige am Rechner darstellbare Zahl als Ergebnis liefert, die dem
theoretischen Wert \mbox{$\log_B(x_Q)$} der quantisierten Gr"o"se $x_Q$ am
n"achsten liegt, erhalten wir einen zus"atzlichen relativen Fehler in der
Gr"o"senordnung von $\varepsilon$, also einen absoluten Fehler, den wir
durch \mbox{$\big|\log_B(x_Q)\big|\CdoT\varepsilon\CdoT\boldsymbol{q}_y$}
modellieren k"onnen, wobei die Zufallsgr"o"se $\boldsymbol{q}_y$
den Quantisierungsfehler des Ergebnisses modelliert, und eine Streuung 
in der Gr"o"senordnung von Eins aufweist. Die Streuung dieses Anteils 
des absoluten Fehlers steigt also mit dem Logarithmus von
\mbox{$x_Q\!\approx\!x$} monoton an. Wenn wir davon ausgehen, dass
die beiden Quantisierungen vor und nach der Berechnung des Logarithmus
unkorrelierte Fehler bewirken, addieren sich die Varianzen dieser beiden
Fehlerquellen. Die Streuung des Gesamtfehlers wird dann als die Wurzel
aus der Summe der beiden Varianzen im wesentlichen durch die Varianz
der Fehlerquelle mit der gr"o"seren Streuung bestimmt. Um die Grenze
f"ur die Werte von $x$ zu bestimmen, ab der die Varianz des zweiten
absoluten Fehlers "uberwiegt, werden nun die Streuungen der beiden
absoluten Fehler gleichgesetzt. Mit
\[
\frac{\varepsilon}{\ln(B)}\;\stackrel{!}{=}\;
\big|\log_B(x_Q)\big|\CdoT\varepsilon\;=\;
\frac{\varepsilon}{\ln(B)}\cdot\big|\ln(x_Q)\big|\;\approx\;
\frac{\varepsilon}{\ln(B)}\cdot\big|\ln(x)|
\]
erhalten wir die Grenze \mbox{$\big|\ln(x)\big|\approx1$}, die von der
Basis $B$ des Logarithmus unabh"angig ist. F"ur Werte von $e^{-1}\ll x\ll e$ 
kann man also die Quantisierungsfehler des Ergebnisses des Logarithmus
vernachl"assigen. Da in unserem Fall die Streuung des absoluten Fehlers 
m"oglichst f"ur alle Frequenzwerte, f"ur die der Logarithmus zu berechnen 
ist, m"oglichst klein sein soll, werden diese Werte vor der Berechnung 
des Logarithmus in der Art normiert, dass der Maximalwert gleich
dem Reziproken des Minimalwertes ist. 

\section[{\tt MATLAB}-Programmrumpf zur Berechnung der
Fensterfolge]{{\tt MATLAB}-Programmrumpf zur Berechnung der\\
Fensterfolge}\label{MatFen1}

F"ur den im Kapitel \ref{Algo} vorgestellten Algorithmus zu 
Berechnung der Fensterfolge sollen nun die entscheidenden 
Zeilen eines Programms in der Interpretersprache {\tt MATLAB}
angegeben werden. Dabei wird darauf verzichtet, die Teile des Programms
abzudrucken, die nicht f"ur die eigentliche Berechnung der Fensterfolge
ben"otigt werden, die aber bei einem guten Programm immer vorhanden
sein sollten, wie zum Beispiel eine "Uberpr"ufung der Eingabeparameter
oder ein ad"aquate Behandlung von Spezial- und Ausnahmef"allen.
Zun"achst werden die Programmzeilen in \verb|Schreibmaschinenschrift|
aufgelistet, wobei diese durchnummeriert sind, um im folgenden
Kommentar auf die Zeilen Bezug nehmen zu k"onnen. Auch ist dieser
Programmauszug nur stellenweise f"ur eine besonders schnelle
Berechnung der Fensterfolge optimiert. Die im Hauptteil der
Abhandlung durchgef"uhrten Betrachtungen zur Genauigkeit der Berechnung
sind hier alle ber"ucksichtigt. Auch wurde weitgehend versucht, bei der
Art der Berechnung die in Kapitel \ref{Algo} beschriebene
Vorgehensweise beizubehalten, so dass dieses Programm auch dazu dienen
soll, dem Leser den konkreten Ablauf zu zeigen, und zu demonstrieren,
dass die dort f"ur die theoretische Herleitung angegebenen Formeln
praktisch unver"andert zur Berechnung der Fensterfolge in ein Programm
"ubernommen werden k"onnen. Der hier angegebene Programmrumpf ben"otigt
die zwei Parameter $N$ und $M$ als skalare Gr"o"sen (\,\mbox{$1\!\times\!1$}
Matrizen\,) \verb|M| und \verb|N|, die beim Programmaufruf "ubergeben werden.
Von diesen wird angenommen, das es sich dabei um ganze Zahlen gr"o"ser Eins
handelt. Au"serdem m"ussen die relative Rechnergenauigkeit $\varepsilon$,
also die Differenz zwischen der Zahl Eins und der n"achstgr"o"seren am
Rechner darstellbaren Zahl, sowie die Zahl $\pi$ als globale Variablen
\verb|eps| und \verb|pi| vorhanden sein. Beim Aufruf des Programms
{\tt MATLAB} werden \verb|eps| und \verb|pi| automatisch generiert,
so dass diese Variablen normalerweise nicht extra bestimmt und an das
Programm "ubergeben werden m"ussen. Als Ergebnis werden von diesem
Programm die Werte der Fensterfolge auf dem Vektor \verb|f_k| sowie die
Fourierreihenkoeffizienten f"ur \mbox{$\nu=0\;(1)\;N\!-\!1$}
auf dem Vektor \verb|F_nu| zur"uckgegeben.
{\renewcommand{\labelenumi}{{\footnotesize\arabic{enumi}:}}
\begin{enumerate}
\setlength{\itemsep}{-6pt plus1pt minus0pt}
\setlength{\parsep}{0pt plus1pt minus0pt}
\item\verb|function [ f_k, F_nu ] = fenster( N, M )|\label{P1Z0}\begin{npb}
\item\verb|F = N * M|\label{P1Z1}\end{npb}
\item\verb|c = 2 / ( 1 + (N/2)^(M/3/(1-M)) / tan(pi/2/M) )|\label{P1Z2}\begin{npb}
\item\verb|Ms = -log(eps)/log(2) * 2^log(N/3) * 3.6^(1/M)|\label{P1Z3}
\item\verb|Ms = 2^ceil(log(Ms)/log(2))|\label{P1Z4}\end{npb}
\item\verb|eta = 0:Ms/2|\label{P1Z5}\begin{npb}
\item\verb|F_eta = zeros(1,Ms/2+1)|\label{P1Z6}\end{npb}
\item\verb|NF = 1 / ( c^2 + 4 * (1-c) * sin( pi/F*(N-1) )^2 )|\label{P1Z7}
\item\verb|for nu_1 = (1-N)/2:(N-1)/2|\label{P1Z8}\begin{npb}
\item\verb|  F_cumprod = ones(1,Ms/2+1)|\label{P1Z9}
\item\verb|  for nu_2 = [(1-N):(nu_1-(N+1)/2),(nu_1+(N+1)/2):(N-1)]|\label{P1Z10}
\item\verb|    K = NF * ( c^2  + 4 * (1-c) * sin( pi/F * nu_2 )^2 )|\label{P1Z11}
\item\verb|    Psi = atan( ( (1-c) * sin(2*pi/F*nu_2) ) / ...|\label{P1Z12}
\item[]\verb|                  ( c + 2 * (1-c) * sin(pi/F*nu_2)^2 ) )|
\item\verb|    F_cumprod = F_cumprod .* K .* sin( pi/Ms*eta-pi/F*nu_2-Psi ).^2|\label{P1Z13}
\item\verb|  end|\label{P1Z14}
\item\verb|  F_eta = F_eta + F_cumprod|\label{P1Z15}
\item\verb|end|\label{P1Z16}\end{npb}
\item\verb|NF = 1 / sqrt( max(F_eta) * min(F_eta) )|\label{P1Z17}\begin{npb}
\item\verb|F_eta = NF * F_eta|\label{P1Z18}\end{npb}
\item\verb|L_eta = log(F_eta)|\label{P1Z19}
\item\verb|Ceps_2 = ifft( [L_eta,L_eta(Ms/2:-1:2)] )|\label{P1Z20}\begin{npb}
\item\verb|Ceps_2 = real( Ceps_2 )|\label{P1Z21}
\item\verb|Ceps_2 = ( Ceps_2 + Ceps_2([1,Ms:-1:2]) ) / 2|\label{P1Z22}\end{npb}\vadjust{\penalty-100}
\item\verb|nu = 1:N-1|\label{P1Z23}
\item\verb|Omega = 2*pi/F * nu|\label{P1Z24}
\item\verb|Omega_s = Omega + 2 * atan( ( (1-c) * sin(Omega) ) ./ ...|\label{P1Z25}\begin{npb}
\item[]\verb|                            ( c + 2 * (1-c) * sin(Omega/2).^2 ) )|\end{npb}
\item\verb|phi = zeros(1,N-1)|\label{P1Z26}
\item\verb|for nu_i = nu|\label{P1Z27}\begin{npb}
\item\verb|  phi(nu_i) = sin(Omega_s(nu_i)*[Ms/2-1:-1:1]) * Ceps_2(Ms/2:-1:2).'|\label{P1Z28}
\item\verb|end|\label{P1Z29}\end{npb}
\item\verb|phi = phi - ( N - 1 ) * Omega_s / 2 + ( F - N ) * Omega / 2|\label{P1Z30}\vadjust{\penalty-200}
\item\verb|nu = 0:N-1|\label{P1Z31}\begin{npb}
\item\verb|F_nu = zeros(1,N)|\label{P1Z32}\end{npb}
\item\verb|NF = exp( 2*log(8/pi*F) - sum( log(5:2:4*N) )/(N-1) )|\label{P1Z33}
\item\verb|for nu_1 = (1-N)/2:(N-1)/2|\label{P1Z34}\begin{npb}
\item\verb|  F_cumprod = ( ( nu < nu_1+N/2 ) & ( nu > nu_1-N/2) )|\label{P1Z35}
\item\verb|  for nu_2 = (1-N)/2:(N-3)/2|\label{P1Z36}
\item\verb|    nu_3 = nu - nu_1 - nu_2 - ( nu <= nu_1 + nu_2 )|\label{P1Z37}
\item\verb|    F_cumprod = F_cumprod ./ ( NF .* sin( pi/F*nu_3 ).^2 )|\label{P1Z38}
\item\verb|  end|\label{P1Z39}
\item\verb|  F_nu = F_nu + F_cumprod|\label{P1Z40}
\item\verb|end|\label{P1Z41}\end{npb}
\item\verb|F_nu = sqrt(F_nu)|\label{P1Z42}\begin{npb}
\item\verb|F_nu = F_nu / ( N * F_nu(1) )|\label{P1Z43}\end{npb}
\item\verb|F_nu(2:N) =  F_nu(2:N) .* exp(-j*phi)|\label{P1Z44}\vadjust{\penalty-200}
\item\verb|f_k = zeros(1,F)|\label{P1Z45}\begin{npb}
\item\verb|si_k = [ 0:2:(F+0.5)/2,...|\label{P1Z46}
\item[]\verb|         F-[2*floor((F+0.5)/4)+2:2:(3*F+0.5)/2],...|
\item[]\verb|         [2*floor((3*F+0.5)/4)+2:2:2*F-1]-2*F      ]|
\item\verb|si_k = sin( (pi/F) * si_k )|\label{P1Z47}\end{npb}
\item\verb|co_k = [ F-[0:4:2*F-1], [4*floor((2*F-1)/4)+4:4:4*F-2]-3*F ]|\label{P1Z48}\begin{npb}
\item\verb|co_k = sin( (pi/(2*F)) * co_k )|\label{P1Z49}\end{npb}
\item\verb|for nu = N-1:-1:1|\label{P1Z50}\begin{npb}
\item\verb|  k_nu = rem( [0:F-1]*nu, F ) + 1|\label{P1Z51}
\item\verb|  f_k = f_k + 2 * real( F_nu(nu+1) ) * co_k(k_nu) - ...|\label{P1Z52}
\item[]\verb|              2 * imag( F_nu(nu+1) ) * si_k(k_nu)|
\item\verb|end|\label{P1Z53}
\item\verb|f_k = f_k + 1/N|\label{P1Z54}\end{npb}
\end{enumerate}}
In den Zeilen~\ref{P1Z1} bis \ref{P1Z3} werden die Fensterl"ange $F$ mit
Gleichung (\ref{6.2}), der Bilineartransformationsparameter $c$ nach Gleichung
(\ref{6.30}) und die L"ange $\widetilde{M}$ der FFT f"ur die Bestimmung des
Cepstrums nach Gleichung (\ref{6.31}) berechnet. $\widetilde{M}$
wird in Zeile~\ref{P1Z4} auf die n"achsth"ohere Zweierpotenz aufgerundet. 
Die Berechnung der Phase des minimalphasigen Anteils
von \mbox{$\widetilde{D}_N\big(e^{j\widetilde{\Omega}}\big)$} erfolgt
nach Gleichung (\ref{6.29}), wobei dies f"ur alle diskreten Frequenzen
\mbox{$\widetilde{\Omega}=\eta\CdoT2\pi/\widetilde{M}$} mit
\mbox{$\eta=0\;(1)\;\widetilde{M}/2$}, die in einem Vektor
zusammengefasst sind, der in Zeile~\ref{P1Z5} definiert wird,
vektoriell geschieht, so dass hierf"ur keine \verb|for|-Schleife
ben"otigt wird. F"ur die anderen Frequenzwerte mit
\mbox{$\eta=\widetilde{M}/2\!+\!1\;(1)\;\widetilde{M}\!-\!1$} braucht
\mbox{$\widetilde{D}_N\big(e^{j\widetilde{\Omega}}\big)$} wegen der
Symmetrie nicht berechnet zu werden. Die Summe "uber $\nu_1$ in
Gleichung (\ref{6.29}) wird dadurch realisiert, dass man
zun"achst in Zeile~\ref{P1Z6} den Vektor \verb|F_eta| bereitstellt,
den man zu Null initialisiert, und dass man zu diesem Vektor nach und nach
in einer \verb|for|-Schleife mit dem Schleifenindex \verb|nu_1|,
die in Zeile~\ref{P1Z8} beginnt und in Zeile~\ref{P1Z16} endet, die
einzelnen Summanden in Zeile~\ref{P1Z15} f"ur alle Werte von $\eta$
zugleich addiert. Entsprechend wird das kumulative Produkt in
Gleichung (\ref{6.29}) realisiert, indem man in Zeile~\ref{P1Z9} den
Vektor \verb|F_cumprod| auf Eins initialisiert, und diesen in der 
\verb|for|-Schleife, die in Zeile~\ref{P1Z10} beginnt und in 
Zeile~\ref{P1Z14} endet, in Zeile~\ref{P1Z13} nach und nach mit den 
einzelnen Faktoren f"ur alle Werte von $\eta$ zugleich multipliziert. 
Der Schleifenindex \verb|nu_2| nimmt dabei die im Anschluss an 
Gleichung (\ref{6.19}) beschriebenen Werte an. Die in der Summe
der Gleichung (\ref{6.29}) auftretenden Faktoren \mbox{$K_{\nu_2}$}
nach Gleichung (\ref{6.25}), k"onnen f"ur gro"se Werte von $M$
und dementsprechend kleine Werte von $c$ sehr klein werden. 
Um zu vermeiden, dass die Endwerte der kumulativen Produkte
in Gleichung (\ref{6.29}) den darstellbaren Zahlenbereich verlassen,
werden in Zeile~\ref{P1Z11} alle von $\eta$ unabh"angigen Faktoren
\mbox{$K_{\nu_2}$} mit einem gemeinsamen, konstanten Normierungsfaktor
\verb|NF| multipliziert, der in Zeile~\ref{P1Z7} berechnet wird, und der
so gew"ahlt wurde, dass der maximal auftretende normierte Faktor \verb|K|,
der in Zeile~\ref{P1Z11} berechnet wird, zu Eins wird. Da der 
Schleifenindex \verb|nu_2| maximal den Wert \mbox{$(N\!-\!1)$}
annimmt, wird n"amlich der Faktor \mbox{$K_{\nu_2}$} nach Gleichung 
(\ref{6.25}) maximal gleich dem Reziprokwert des in Zeile~\ref{P1Z7} 
berechneten Normierungsfaktors \verb|NF|. Der ebenfalls von $\eta$ 
unabh"angige Anteil $\Tilde{\psi}_{\nu_2}$ des halben Winkels der 
Nullstellenrotation nach Gleichung (\ref{6.26}) wird in Zeile~\ref{P1Z12} 
berechnet. Mit der Differenz aus dem halben Winkel der
bilinear transformierten und um \mbox{$e^{j\cdot\nu_2\CdoT2\pi/F}$}
rotierten Nullstelle und dem halben Winkel des Punktes
\mbox{$\Tilde{z}\!=\!e^{j\cdot\eta\cdot2\pi/\widetilde{M}}$}, f"ur den
das Spektrum \mbox{$\widetilde{D}_N(\Tilde{z})$} zu berechnen ist, kann
ein Faktor des kumulativen Produktes berechnet werden. F"ur alle Werte
von $\eta$ kann dieser Faktor nun in Zeile~\ref{P1Z13} mit
dem bisher berechneten kumulativen Produkt auf dem Vektor
\verb|F_cumprod| multipliziert werden. Das vollst"andig berechnete
kumulative Produkt in Gleichung (\ref{6.29}) wird in Zeile~\ref{P1Z15}
zu der bisher berechneten Summe auf dem Vektor \verb|F_eta| addiert. 
F"ur die so berechneten Spektralwerte
\mbox{$\widetilde{D}_N\big(e^{j\widetilde{\Omega}}\big)$} 
mit \mbox{$\widetilde{\Omega}=\eta\CdoT2\pi/\widetilde{M}$} 
wird in Zeile~\ref{P1Z17} der Normierungsfaktor berechnet, der 
daf"ur sorgt, dass der Maximalwert gleich dem Reziproken des 
Minimalwertes wird. Die Normierung erfolgt in Zeile~\ref{P1Z18}. 
In Zeile~\ref{P1Z19} wird davon der Logarithmus berechnet. 
Eine inverse FFT der durch Spiegelung f"ur die nicht berechneten 
$\eta$-Werte erg"anzten, normierten und logarithmierten Spektralwerte 
\mbox{$\widetilde{D}_N\big(e^{j\widetilde{\Omega}}\big)$} 
liefert uns in Zeile~\ref{P1Z20} das doppelte Cepstrum 
des minimalphasigen Anteils von \mbox{$\widetilde{D}_N(\Tilde{z})$}. 
Die Realteilbildung in Zeile~\ref{P1Z21} unterdr"uckt dabei den 
Imagin"arteil, der nur durch Quantisierungsfehler bei der inversen 
FFT entsteht. Auch die Unsymmetrie des doppelten Cepstrums, die aus 
demselben Grund entsteht, wird in Zeile~\ref{P1Z22} beseitigt. 
Die Werte des doppelten Cepstrums sind f"ur \mbox{$k\!>\!0$} die
Fouriersinusreihenkoeffizienten der Phase des minimalphasigen Anteils von
\mbox{$\widetilde{D}_N\big(e^{j\widetilde{\Omega}}\big)$}. Diese
wird nun f"ur die vor der Bilineartransformation "aquidistanten
Frequenzen \mbox{$\Omega=\nu\CdoT2\pi/F$} mit \mbox{$\nu=1\;(1)\;N\!-\!1$}
berechnet. Diese Frequenzen werden in Zeile~\ref{P1Z24} auf dem Vektor
\verb|Omega| aus den Werten \verb|nu| berechnet, die zuvor in
Zeile~\ref{P1Z23} auf einem Vektor abgelegt wurden. F"ur 
\mbox{$\nu\!=\!0$} braucht die Phase nicht berechnet zu werden, da diese
immer Null ist. Nach Gleichung (\ref{6.22}) entsprechen die Frequenzen
des Vektors \verb|Omega| nach der Bilineartransformation den
Frequenzen $\widetilde{\Omega}$, die in Zeile~\ref{P1Z25}
dem Vektor \verb|Omega_s| zugewiesen werden. Die Summe 
der Sinusreihe wird f"ur jede Frequenz des Vektors \verb|Omega_s| 
in der \verb|for|-Schleife mit dem Schleifenindex \verb|nu_i|, 
die in Zeile~\ref{P1Z27} beginnt und in Zeile~\ref{P1Z29} endet, 
in Zeile~\ref{P1Z28} als Skalarprodukt zweier Vektoren berechnet. Der eine
Vektor enth"alt dabei die Koeffizienten der Sinusreihe und der andere die
Werte der Sinusfunktion der $k$-fachen Grundkreisfrequenz --- die hier Eins
ist, weil die Phase eine in $2\pi$ periodische Funktion ist --- bei der zu
berechnenden Frequenz $\widetilde{\Omega}$. Die Reihenfolge der 
Elemente der Vektoren wird dabei umgekehrt, um eine h"ohere 
Genauigkeit zu erzielen\footnote{Es wurde "uberpr"uft, dass {\tt MATLAB} 
die Reihenfolge nicht vertauscht, wenn es die Zeile~\ref{P1Z28}
in Maschinencode umsetzt.}. Die Phase des minimalphasigen Anteils 
von \mbox{$D_P\big(e^{j\cdot\frac{2\pi}{F}\cdot\nu}\big)$} ist das 
\mbox{$N\!-\!1$}-fache des in Gleichung (\ref{6.28}) angegebenen Terms.
Wie ein Vergleich der Gleichungen (\ref{6.22}) und (\ref{6.28}) zeigt,
sind die halben negativen Frequenzen \mbox{$\widetilde{\Omega}/2$} um
\mbox{$\Omega/2$} kleiner als der Beitrag eines Faktors
\mbox{$\big(z\!-\!(1\!-\!c)\big)/z$} zum minimalphasigen Anteil
von \mbox{$D_P(z)$}. Da der Phasenbeitrag des minimalphasigen
Anteils von \mbox{$D_E(z)$} gleich
\mbox{$(F\!-\!2\CdoT\!N\!+\!1)\CdoT\Omega/2$} ist, kann man den Fehler,
den man macht, wenn man \mbox{$(N\!-\!1)\CdoT\widetilde{\Omega}/2$}
statt \mbox{$(N\!-\!1)\CdoT(\widetilde{\Omega}\!-\!\Omega)/2$}
als Phasenanteil des minimalphasigen Anteils von \mbox{$D_P(z)$} subtrahiert,
einfach dadurch kompensieren, dass man den Phasenbeitrag des
minimalphasigen Anteils von \mbox{$D_E(z)$} entsprechend um
\mbox{$(N\!-\!1)\CdoT\Omega/2$} erh"oht. Die Gesamtphase berechnet
sich dann wie dies in Zeile~\ref{P1Z30} angegeben ist.

Nachdem in den Zeilen~\ref{P1Z31} bis \ref{P1Z41} die Betragsquadrate 
der gesuchten Spektralwerte mit Gleichung (\ref{6.18}) bis auf eine 
bei allen Frequenzpunkten gleiche Normierungskonstante berechnet 
worden sind, werden in Zeile~\ref{P1Z42} daraus die Betr"age durch
Wurzelziehen bestimmt. Da die Spektralwerte bei der Berechnung der
Fourierreihe nach Gleichung (\ref{6.16}) mit $F$ zu dividieren sind,
werden die bis dahin berechneten Betr"age am Ende nicht so normiert, dass
sich die Abtastwerte des Spektrums mit \mbox{$F(0)\!=\!M$} ergeben, sondern
gleich so, dass man die Betr"age der Fourierreihenkoeffizienten mit
\mbox{${\tt F\_nu(1)}=F(0)/F=1/N$} \mbox{erh"alt.} Da bei der Berechnung der
Betragsquadrate der gesuchten Spektralwerte bei \mbox{allen} Frequenzen eine
gemeinsame Normierungskonstante \verb|NF| verwendet wird, kann die
abschlie"sende Normierung in Zeile~\ref{P1Z43} am einfachsten dadurch
erfolgen, dass man auch den Betrag des Spektralwertes bei der Frequenz
Null mit derselben vorl"aufigen Normierung berechnet, und dann alle in
derselben Art normierten Spektralwertbetr"age durch das $N$-fache des
vorl"aufig normierten Betrags des Spektralwertes bei der Frequenz Null
dividiert. F"ur \mbox{alle} Frequenzen \mbox{$\Omega=\nu\CdoT2\pi/F$} mit
dem Frequenzparameter \mbox{$\nu=0\;(1)\;N\!-\!1$}, dessen Werte in
Zeile~\ref{P1Z31} auf einem Vektor abgelegt werden, k"onnen die
Betragsquadrate der gesuchten Spektralwerte nach Gleichung (\ref{6.18})
vektoriell berechnet werden, so dass hierf"ur keine \verb|for|-Schleife
ben"otigt wird. Die Summe "uber $\nu_1$ in Gleichung (\ref{6.18})
wird dadurch realisiert, dass man zun"achst in Zeile~\ref{P1Z32} den
Vektor \verb|F_nu| bereitstellt, den man zu Null initialisiert, und
zu dem man nach und nach in einer \verb|for|-Schleife mit dem Schleifenindex
\verb|nu_1|, die in Zeile~\ref{P1Z34} beginnt und in Zeile~\ref{P1Z41}
endet, die einzelnen Summanden in Zeile~\ref{P1Z40} f"ur 
\mbox{alle} Werte von $\nu$ zugleich addiert. In den Grenzen der
Summe in Gleichung (\ref{6.18}) wird ber"ucksichtigt, dass
bei einer festen Frequenz \mbox{$\nu\CdoT2\pi/F$} die Summanden
f"ur einige Werte des Verschiebungsparameters $\nu_1$ Null sind,
weil das um \mbox{$\nu_1\CdoT2\pi/F$} verschobene Betragsquadratspektrum
der Basisfensterfolge bei dieser Frequenz \mbox{$\nu\CdoT2\pi/F$} eine
doppelte Nullstelle aufweist, die nicht durch eine doppelte Polstelle
kompensiert wird. Da die Berechnung im Programm f"ur alle $\nu$-Werte
zugleich erfolgt, kann dieser Fakt nicht in den Summationsgrenzen in
Zeile~\ref{P1Z34} ber"ucksichtigt werden. Daher wird stattdessen in
Zeile~\ref{P1Z35} der Vektor \verb|F_cumprod|, mit dessen Hilfe das
kumulative Produkt in Gleichung (\ref{6.18}) berechnet wird, nur f"ur
die Frequenzwerte $\nu$ mit Eins initialisiert, f"ur die das um
\mbox{$\nu_1\CdoT2\pi/F$} verschobene Betragsquadratspektrum der
Basisfensterfolge keine doppelte Nullstelle aufweist. 
F"ur die anderen Frequenzwerte $\nu$ wird \verb|F_cumprod|
mit Null initialisiert, so dass auch das kumulative Produkt dieser
Frequenzwerte $\nu$ bei dem Summanden, der einer Verschiebung um
\mbox{$\nu_1\CdoT2\pi/F$} entspricht, die in Gleichung (\ref{6.18})
durch die Summationsgrenzen ausgeschlossen wurde, Null wird.
Die Faktoren des kumulativen Produkts in Gleichung (\ref{6.18})
weisen alle den negativen Exponenten $-2$ auf. Daher kann das kumulative
Produkt als \verb|for|-Schleife, die in Zeile~\ref{P1Z36} beginnt und in
Zeile~\ref{P1Z39} endet, realisiert werden, indem der bereits geeignet
initialisierte Vektor \verb|F_cumprod| in Zeile~\ref{P1Z38} nach und
nach durch den Vektor der normierten Quadrate der Sinusfunktion
dividiert wird. Im Argument der Sinusfunktion tritt der von $\nu$
abh"angige Term \mbox{$\nu\!-\!\nu_2\!-\!\nu_1$} auf. Abgesehen
von dem im folgenden erkl"arten Detail werden die Werte dieses Terms
f"ur alle Werte von $\nu$ als Elemente des Vektors \verb|nu_3| verwendet. 

In Gleichung (\ref{6.18}) bewirken die Grenzen des
Summenindex $\nu_1$, dass der beim kumulativen Produkt explizit
ausgenommene Faktor mit \mbox{$\nu_2\neq\nu\!-\!\nu_1$}, bei
dem eine doppelte Nullstelle gegen eine doppelte Polstelle gek"urzt 
wurde, bei dem kumulativen Produkt jedes Summanden genau einmal 
vorkommen w"urde, wenn man ihn nicht ausgeschlossen h"atte.
Bei der im Programm verwendeten Grenze des Schleifenindex \verb|nu_1|
in Zeile~\ref{P1Z34} kann es vorkommen, dass der f"ur $\nu_2$ explizit
ausgenommene Wert $\nu\!-\!\nu_1$ sowieso nicht innerhalb der f"ur $\nu_2$
in Gleichung (\ref{6.18}) angegebenen Grenzen liegt. Ist dieser Fall
gegeben, so ist das entsprechende Elemente des Vektors \verb|F_cumprod|
zu Null initialisiert worden. Daher kann man diese Elemente durch
beliebige Werte au"ser durch Null dividieren, ohne Fehler zu machen.
Diese Tatsache kann genutzt werden, den Ausschluss des Falls
\mbox{$\nu_2=\nu\!-\!\nu_1$} einfach zu realisieren. Einerseits
wird in Zeile~\ref{P1Z36} der Endwert f"ur den Schleifenindex
\verb|nu_2| gegen"uber dem Endwert f"ur $\nu_2$ in Gleichung
(\ref{6.18}) um Eins verringert. Andererseits wird das der gek"urzten
doppelten Polstelle entsprechende Argument der Sinusfunktion
"ubersprungen. Das "Uberspringen der gek"urzten doppelten Polstelle
wird realisiert, indem man bei der Berechnung des Terms im Argument
der Sinusfunktion in Zeile~\ref{P1Z37} f"ur die Faktoren des kumulativen
Produkts mit \mbox{$\nu_2<\nu\!-\!\nu_1$} den Wert \verb|nu_3 = nu-nu_1-nu_2|
verwendet, und f"ur die Faktoren mit \mbox{$\nu_2>\nu\!-\!\nu_1$}
den Wert \verb|nu_3 = nu-nu_1-(nu_2+1)|. Der Wert \mbox{${\tt nu\_3}\!=\!0$}
ist dadurch f"ur alle Werte von \verb|nu_2| ausgeschlossen. Bei welchem
Schleifendurchlauf das "Uberspringen der gek"urzten doppelten Polstelle
passiert, h"angt von $\nu$ ab, und ist somit bei den Elementen des Vektors
\verb|F_cumprod| unterschiedlich. Nur bei den Summanden mit dem Index
\verb|nu_1|, die im Programm vorhanden sind, aber nicht in Gleichung
(\ref{6.18}), wird durch die eben beschriebene Indexarithmetik ein
Faktor zu wenig berechnet. Da dann aber das entsprechende Element
des Vektors \verb|F_cumprod| zu Null initialisiert ist, und im Argument
der Sinusfunktion nie der Wert Null auftritt, bleibt dieser Wert
korrekterweise Null.

Da auf die Normierungskonstante \verb|NF| im Hauptteil des Textes nicht
n"aher eingegangen wurde, sei nun erl"autert, warum dieser so gew"ahlt wurde,
wie dies in Zeile~\ref{P1Z33} angegeben ist. Ziel der Normierung ist
es, zu erreichen, dass der Spektralwert \verb|F_nu(1)| f"ur die Frequenz
\mbox{$\Omega\!=\!0$} vor der abschlie"senden Normierung etwa in der
Gr"o"senordnung von Eins liegt, so dass keine Gefahr besteht, dass es
w"ahrend der Berechnung zu einem "Uberlauf kommt, weil die einzelnen
Faktoren des kumulativen Produkts alle zu klein oder zu gro"s sind,
so dass das kumulative Produkt insgesamt kleiner bzw. gr"o"ser als die
kleinste bzw. gr"o"ste am Rechner darstellbare Zahl wird. 
Wir wollen zun"achst annehmen, dass $N$ eine ungerade nat"urliche Zahl 
ist. Um auf die in Zeile~\ref{P1Z33} berechnete Normierungskonstante 
zu kommen, werden drei N"aherungen gemacht. 
Bei der Frequenz \mbox{$\Omega\!=\!0$} ist f"ur ungerades $N$
der Summand mit \mbox{$\nu_1\!=\!0$}, also der Summand, bei dem das
Betragsquadratspektrum der Basisfensterfolge unverschoben auftritt,
am gr"o"sten, weil dann das Produkt der Quadrate der Sinusfunktionen,
die mit einem negativen Exponenten auftreten, am kleinsten wird. Wenn wir
nun als erste N"aherung die anderen, kleineren Summanden vernachl"assigen,
sch"atzen wir die Summe der Produkte der Quadrate der Sinusfunktionen
schlimmstenfalls um den Faktor $1/N$ zu klein ab, weil insgesamt genau $N$
Summanden in Gleichung (\ref{6.18}) vorhanden sind. Da $N$ im Vergleich
zu der gr"o"sten am Rechner darstellbaren Zahl extrem klein ist,
wird es auch dann nicht zu einem "Uberlauf bei der Berechnung kommen,
wenn sich aufgrund dieser groben Absch"atzung ein Spektralwert \verb|F_nu(1)|
ergibt, der nicht exakt $1$ ist, sondern der zwischen $1$ und $N$ liegt. 
Die zweite N"aherung ergibt sich aus folgender "Uberlegung. 
Die Gefahr eines "Uberlaufs tritt eigentlich nur dann auf, wenn 
alle Faktoren des kumulativen Produkts gro"s sind. Dies ist der Fall,
wenn $M$ gro"s ist, und dadurch das Argument der Sinusfunktion
klein wird. In diesem Fall kann man die Sinusfunktion in guter
N"aherung durch eine Gerade mit der Steigung Eins ann"ahern. Zum dritten 
n"ahern wir noch die vierte Potenz \mbox{${\D(4\CdoT\nu_2)^4}$} durch 
das Produkt \mbox{$(4\CdoT\nu_2\!-1\!,\!5)\CdoT(4\CdoT\nu_2\!-\!0,\!5)\CdoT
(4\CdoT\nu_2\!+\!0,\!5)\CdoT(4\CdoT\nu_2\!+\!1,\!5)$}, um zu der Gleichung 
zu kommen, mit deren Hilfe wir die Normierungskonstante \verb|NF| 
bestimmen k"onnen.\vspace{-2pt} 
\begin{gather*}\begin{flalign*} 
&\quad{\tt F\_nu(1)}\;=\; 
\Sum{\nu_1=-\frac{N-1}{2}}{\frac{N-1}{2}}\;\; 
\Prod{\substack{\nu_2=\nu_1-\frac{N-1}{2}\\\nu_2\neq0}} 
{\nu_1+\frac{N-1}{2}}\!\!\!\Big({\tt NF}\cdot 
\sin\!\big({\T\nu_2\CdoT\frac{\pi}{F}}\big)^2\Big)^{\!\!-1}\;\approx&&
\end{flalign*}\\[6pt]
\approx\;\Prod{\substack{\nu_2=-\frac{N-1}{2}\\\nu_2\neq0}}
{\frac{N-1}{2}}\!\!\!\Big({\tt NF}\cdot
\sin\!\big({\T\nu_2\CdoT\frac{\pi}{F}}\big)^2\Big)^{\!\!-1}\;=\;
\Prod{\nu_2=1}{\frac{N-1}{2}}\!\Big({\tt NF}\cdot
\sin\!\big({\T\nu_2\CdoT\frac{\pi}{F}}\big)^2\Big)^{\!\!-2}\;\approx\qquad
\\\begin{flalign*}
&&\approx\;\Prod{\nu_2=1}{\frac{N-1}{2}}\!\Big({\tt NF}^{-2}\CdoT
\big({\T4\CdoT\nu_2\CdoT\frac{\pi}{4\cdot F}}\big)^{\!-4}\Big)\;\approx\;
{\tt NF}^{1-N}\cdoT\Prod{\nu_2=2,5}{2\cdot N-0,5}\!\!\!
\big({\T\nu_2\CdoT\frac{\pi}{4\cdot F}}\big)^{\!-1}\;\stackrel{!}{=}\;1&
\end{flalign*}\\[10pt]
\Longrightarrow\qquad\qquad
{\tt NF}\;=\;\Big(\frac{\pi}{8\CdoT F}\Big)^{\!\!-2}\CdoT
\Bigg(\Prod{\nu_2=2,5}{2\cdot N-0,5}\!\!\!
(2\CdoT\nu_2)\Bigg)^{\!\!\frac{-1}{N-1}}\qquad\qquad\qquad
\end{gather*}
Die zuletzt gemachte N"aherung bewirkt, dass man diese Normierungskonstante
sowohl f"ur gerades als auch f"ur ungerades $N$ problemlos verwenden kann.
Da das Produkt bei der Berechnung der Normierungskonstante im wesentlichen
fakultativ mit $N$ ansteigt, besteht die Gefahr, dass es bei der
Berechnung der Normierungskonstante zu einem "Uberlauf kommt, bevor die
\mbox{$N\!-\!1$}-te Wurzel berechnet wird. Daher wird in Zeile~\ref{P1Z33}
zun"achst der Logarithmus der Normierungskonstante berechnet,
bei dem das Radizieren einer Division durch \mbox{$N\!-\!1$} entspricht. 
Erst danach wird die Exponentialfunktion dieses Wertes berechnet.
Dadurch wird die Normierungskonstante zwar ungenau --- was nicht weiter
st"ort, weil \verb|NF| bei allen Faktoren und allen Frequenzen gleich
ist ---, aber ohne die Gefahr eines "Uberlaufs bei gro"sen Werten von
$N$ berechnet.

Nachdem die Betr"age der gesuchten Fourierreihenkoeffizienten der 
Fensterfolge berechnet sind, werden diese in Zeile~\ref{P1Z44} noch 
mit der Exponentialfunktion der Phase multipliziert, und es kann mit 
der Berechnung der Fensterfolge durch Auswertung einer Sinus- und einer 
Kosinusreihe begonnen werden. Es wird keine DFT verwendet, da diese 
erstens zu gro"se Fehler verursacht, und zweitens nur $N$ Werte des 
Spektrums von Null verschieden sind. Da die Werte der Sinus- und der 
Kosinusfunktion mehrfach ben"otigt werden, werden diese zun"achst mit 
gr"o"stm"oglicher Genauigkeit auf zwei Vektoren berechnet. Zuerst 
werden in Zeile~\ref{P1Z46} die $F/\pi$-fachen Argumentwerte der 
Sinusfunktion berechnet, die sich am Rechner exakt darstellen lassen, 
da sie ganzzahlig sind. Dabei werden die trigonometrischen Formeln 
\mbox{$\sin(x)\!=\!\sin(\pi\!-\!x)$} und \mbox{$\sin(x)\!=\!\sin(x\!-\!2\pi)$} 
verwendet, so dass die exakt darstellbaren auf $\pi/F$ normierten 
Argumente der Sinusfunktion betraglich nie gr"o"ser als $F/2$ und somit 
die Argumente der Sinusfunktion betraglich nie gr"o"ser als $\pi/2$ 
werden. Auf diese Weise wird bei der Berechnung der Abtastwerte der 
Sinusfunktion in Zeile~\ref{P1Z47} die h"ochstm"ogliche Genauigkeit 
erreicht. Analog erh"alt man mit den trigonometrischen Formeln 
\mbox{$\cos(x)\!=\!\sin(\pi/2\!-\!x)$} und \mbox{$\cos(x)\!=\!\sin(x\!-\!3\pi/2)$} 
in der Zeile~\ref{P1Z49} die Werte der Kosinusfunktion m"oglichst genau, 
wenn man in die Sinusfunktion das $\pi/F$-fache der in Zeile~\ref{P1Z48} 
exakt berechneten Werte einsetzt. Die Summe der Sinus- und Kosinusreihe 
wird dadurch realisiert, dass man zun"achst in Zeile~\ref{P1Z45} den 
Vektor \verb|f_k| bereitstellt, den man zu Null initialisiert, und 
zu dem man nach und nach in einer \verb|for|-Schleife mit dem Schleifenindex 
\verb|nu|, die in Zeile~\ref{P1Z50} beginnt und in Zeile~\ref{P1Z53} 
endet, die einzelnen Reihenglieder in Zeile~\ref{P1Z52} 
f"ur alle Werte von \mbox{$k=0\;(1)\;F\!-\!1$} zugleich addiert. 
Mit der in Zeile~\ref{P1Z51} angegebenen Modulo-Arithmetik kann man 
sich die bei dem Reihenglied jeweils ben"otigten Werte der Sinus- und 
Kosinusfunktion herauspicken. Um die Fehler bei der Auswertung der 
Sinus- und Kosinusreihe klein zu halten, erfolgt die Berechnung der 
Sinus- und Kosinusreihen in umgekehrter Reihenfolge, was man am 
Schleifenindex in Zeile~\ref{P1Z50} erkennen kann. Abschlie"send 
wird in Zeile~\ref{P1Z54} der Gleichanteil $1/N$ addiert. Damit 
befindet sich die Fensterfolge \mbox{$f(k)$} auf dem Vektor \verb|f_k|,
der in Zeile~\ref{P1Z0} als Ergebnis der Funktion \verb|fenster(N,M)|
definiert wurde.

\renewcommand{\thechapter}{}
\renewcommand{\chaptermark}[1]{\markboth{\small\sf\thechapter\ #1}{\small\sf\thechapter\ #1}}
\renewcommand{\sectionmark}[1]{\markright{\small\sf\thesection\ #1}}
\addtolength{\abovechaptervspace}{-20pt}
\chapter{Liste der verwendeten Abk"urzungen und Formelzeichen}
{\bf Abk"urzungen}\\
Neben den allgemein gebr"auchlichen Abk"urzungen, die im Duden aufgef"uhrt
sind, werden im gesamten Text die Abk"urzungen\vspace{-8pt}
\begin{list}{}{\setlength{\itemsep}{0.2ex plus0.2ex}
\setlength{\labelwidth}{4em}\setlength{\labelsep}{1em}
\setlength{\leftmargin}{5em}}
\item[AKF\hfill] Autokorrelationsfolge 
\item[MAKF\hfill] modifizierte Autokorrelationsfolge 
\item[DFT\hfill] diskrete Fouriertransformation 
\item[FIR\hfill] $\underline{\text{f}}$inite $\underline{\text{i}}$mpule 
                 $\underline{\text{r}}$esponse = System mit endlicher 
                 Impulsantwort =\\
                 nichtrekursives Filter. 
\item[FFT\hfill] schnelle diskrete Fouriertransformation 
\item[LDS\hfill] Leistungsdichtespektrum 
\item[MLDS\hfill] modifiziertes Leistungsdichtespektrum 
\item[RKM\hfill] Rauschklirrmessverfahren\vspace{-8pt}
\end{list}
verwendet. Einige wenige weitere Abk"urzungen werden nur lokal in eng
begrenztem Zusammenhang gebraucht und sind dort erkl"art, wo sie zum
ersten Mal auftreten.\vspace{6pt}

{\bf Konstanten}\vspace{-8pt}
\begin{list}{}{\setlength{\itemsep}{0.2ex plus0.2ex}
\setlength{\labelwidth}{4em}\setlength{\labelsep}{1em}
\setlength{\leftmargin}{5em}}
\item[$e$\hfill] $=\;2,718282\ldots$\begin{npb}
\item[$j$\hfill] imagin"are Einheit
\item[$\varepsilon$\hfill] \mbox{$=\;2^{1-\text{Mantissenwortl"ange}}$}.
     Relative Rechnergenauigkeit bei
     Gleitkommazahlendarstellung\footnote{Unter der Mantissenwortl"ange sei
     hier die Anzahl der signifikanten Bits verstanden. Diese ist
     beim IEEE Standard 754 sowohl bei "`single precision"' als auch bei
     "`double precision"' um Eins gr"o"ser als die Anzahl der Bits, die
     zur Abspeicherung der Mantisse ben"otigt werden, da das "`most
     significant bit"' bei der verwendeten, normierten Zahlendarstellung
     immer Eins ist, und daher nicht abgespeichert wird.} =
     Differenz zwischen $1$ und der n"achstgr"o"seren am Rechner
     darstellbaren Zahl.
\item[$\pi$\hfill] $=\;3,141593\ldots$\end{npb}
\end{list}\pagebreak[2]

{\bf Allgemeine Formelzeichen und Funktionen}
\begin{list}{}{\setlength{\itemsep}{0.2ex plus 0.2ex minus 0.3ex}
\setlength{\labelwidth}{4em}\setlength{\labelsep}{1em}
\setlength{\leftmargin}{5em}}
\item[$\forall$\hfill] f"ur jedes $\ldots$
\item[$\exists$\hfill] es gibt ein $\ldots$
\item[$\in$\hfill] ist Element der Menge $\ldots$
\item[$\mathbb{C}$\hfill] Menge der komplexen Zahlen
\item[$\mathbb{N}$\hfill] Menge der nat"urlichen Zahlen ( ohne $0$ )
\item[$\mathbb{R}$\hfill] Menge der reellen Zahlen
\item[$\mathbb{Z}$\hfill] Menge der ganzen Zahlen
\item[$\times$\hfill] Trennung bei der Angabe der Dimensionen einen Matrix
\item[$\ast$\hfill] Faltung
\item[$\cdot$\hfill] ( Matrix- ) Multiplikation
\item[$(...)^{\Kk}$\hfill] konjugiert komplexe Zahl oder Matrix
\item[$(...)^{\Hh}$\hfill] transponierte konjugiert komplexe Matrix
\item[$(...)^{\Tt}$\hfill] transponierte Matrix
\item[$(...)^{-1}$\hfill] Reziprokwert einer Zahl, invertierte Matrix oder Umkehrfunktion.
\item[$\Vec{0}$\hfill] Zeilenvektor, der nur Nullen enth"alt.
\item[$\underline{0}$\hfill] Matrix, die nur Nullen enth"alt.
\item[$\underline{E}$\hfill] Einheitsmatrix
\item[$||...||_2$\hfill] euklidische Vektornorm; Spektralnorm einer Matrix
\item[$||...||_F$\hfill] euklidische Matrixnorm = Frobeniusnorm
\item[$[a;b)$\hfill] Intervall von $a$ bis $b$. Die untere Intervallgrenze
     geh"ort zum Intervall, w"ahrend die obere Intervallgrenze {\em nicht }\/
     zum Intervall geh"ort.
\item[$x=a\;(b)\;c$\hfill] Inkrementoperator \mbox{$x\in
     \big\{\,x=a\!+\!\mu\CdoT b\;\;|\;\;a\!\le\!x\!\le\!c\;\wedge\;
     a,b,c\in\mathbb{R}\;\wedge\;\mu\in\mathbb{Z}\;\big\}$}
\item[$\text{P}(...)$\hfill] Wahrscheinlichkeit eines Ereignisses
\item[$\text{P}(\Vec{\boldsymbol{x}}\!<\!\Vec{x})$\hfill] Verbundverteilung
     der Real- und Imagin"arteile der Zufallsgr"o"sen, die die Elemente des
     Zufallsvektors $\Vec{\boldsymbol{x}}$ sind. Es handelt sich dabei um
     die symbolische Kurzschreibweise f"ur die Wahrscheinlichkeit,
     dass ein mehrdimensionales Experiment ein Ergebnis liefert, das
     in dem Unterraum der Ergebnismenge liegt, der nur die Ergebnisse
     enth"alt, denen die komplexen Vektoren zugeordnet sind, deren
     Elemente einen Realteil aufweisen,\pagebreak[2] der bei allen Elementen kleiner
     ist als der Realteil der Elemente des Vektors $\Vec{x}$ und deren
     Elemente einen Imagin"arteil aufweisen, der bei allen Elementen kleiner
     ist als der Imagin"arteil der Elemente des Vektors $\Vec{x}$.
\item[$\text{p}_{\Vec{\boldsymbol{x}}}(\Vec{x})$\hfill] mehrdimensionale
     Verbundverteilungsdichte eines Vektors von Zufallsgr"o"sen.
\item[$\text{E}\Left\{...\right\}$\hfill] Erwartungswertbildung. 
     Dieser Erwartungswert wird von einer Zufallsgr"o"se gebildet,  
     und ist selbst eine deterministische Gr"o"se. Bei Vektoren 
     und Matrizen erfolgt die Erwartungswertbildung elementweise. 
\item[$\text{E}\{\boldsymbol{y}\;\pmb{|}\;x\}$\hfill] bedingter
     Erwartungswert = Erwartungswert des Zufallsprozesses $\boldsymbol{y}$
     unter der Bedingung des deterministischen Parameters $x$. Dieser
     Erwartungswert ist ebenfalls eine deterministische Gr"o"se.
\item[$\text{\bf E}\{\boldsymbol{y}\;\pmb{|}\;\boldsymbol{x}\}$\hfill]
     bedingter Erwartungswert = Erwartungswert des Zufallsprozesses
     $\boldsymbol{y}$ unter der Bedingung des Zufallsprozesses
     $\boldsymbol{x}$. Dieser Erwartungswert ist selbst eine Zufallsgr"o"se. 
\item[$\Re\{...\}$\hfill] Realteilbildung
\item[$\Im\{...\}$\hfill] Imagin"arteilbildung
\item[$|...|$\hfill] Betrag einer komplexen Zahl
\item[$\winkel\{...\}$\hfill] Winkel einer komplexen Zahl
\item[$\ln(...)$\hfill] nat"urlicher Logarithmus. Bei komplexem Argument
     ist gegebenenfalls zum Hauptwert des Logarithmus ein geeignetes
     Vielfaches von \mbox{$j\CdoT2\pi$} zu addieren.
\item[$\log_B(...)$\hfill] Logarithmus zur Basis $B$
\item[max$(...)$\hfill] Maximalwert
\item[sgn$(...)$\hfill] Signumfunktion
\item[si$(...)$\hfill] \mbox{${\D = \frac{\sin(\ldots)}{(\ldots)}=}$}
     si-Funktion
\item[erfc$(...)$\hfill] komplement"are Fehlerfunktion
     nach Gleichung (\ref{3.71}).
\item[$a!$\hfill] \mbox{$1\CdoT2\Cdot\ldots\cdoT(a\!-\!1)\CdoT a=$}
     Fakult"at von $a$.
\item[$\tbinom{a}{b}$\hfill] \mbox{$\frac{a!}{b!\cdot(a-b)!}=$}
     Binomialkoeffizient "`$a$ "uber $b$"'.
\item[$\delta_0(t)$\hfill] Dirac-Impuls. Distribution, die jeder
     Funktion den Wert der Funktion an der Stelle Null zuordnet.
\item[$\gamma_0(k)$\hfill] Folge die f"ur \mbox{$k=0$} Eins ist,
     und sonst Null.
\item[{\raisebox{0pt}[19pt][1pt]{$\Sum{\mu=a}{b}$}}/{\raisebox{0pt}[1pt][1pt]{$\Prod{\mu=a}{b}$}}\hfill] Summe / Produkt "uber alle
     Terme mit dem Laufindex $\mu$ mit \mbox{$\mu=a\;(1)\;b$}. Man beachte,
     dass bei dieser Definition $a$ {\em keine}\/ ganze Zahl sein muss,
     w"ahrend das Inkrement immer $1$ ist. Wenn f"ur die untere Intervallgrenze
     \mbox{$\mu=-\infty$} eingesetzt ist, wird wie "ublich \mbox{$\mu\in\mathbb{Z}$} angenommen.\pagebreak[2]
\end{list}

{\bf Spezielle Formelzeichen}\\[4pt]
Folgende Konvention f"ur die Schreibweise von Formelzeichen wird
weitgehend einheitlich im gesamten Text befolgt. Vektoren werden mit einen
Pfeil $\Vec{\phantom{x}}$ versehen (~z.~B. $\Vec{v}$~). Bei Matrizen
wird das Formelzeichen unterstrichen (~z.~B. $\underline{M}$~).
Zufallsgr"o"sen, Zufallsvektoren, Zufallsmatrizen und Zufallsprozesse
werden fettgedruckt dargestellt (~z.~B. $\boldsymbol{Y}$~). Konkrete
Stichproben werden {\em nicht}\/ fettgedruckt und mit dem Index
${\vphantom{x}_{\lambda}}$ gekennzeichnet. Daraus gewonnene Sch"atzwerte
werden mit einem Dach $\Hat{\phantom{x}}$ auf dem Symbol markiert
(~z.~B. $\Hat{H}$~).
Die "uber den Symbolen stehenden Markierungen $\Tilde{\phantom{x}}$,
$\Breve{\phantom{x}}$, $\Check{\phantom{x}}$, $\Bar{\phantom{x}}$ und
$\Hat{\phantom{x}}$ werden auch zur Unterscheidung von Argumenten und
Indizes "ahnlicher Bedeutung (~z.~B. die diskrete Zeit in $v(k\!-\!\Tilde{k})$~)
verwendet. Zu diesem Zweck werden auch nat"urliche Zahlen als Subindizes
verwendet, wenn damit eine Reihenfolge verdeutlicht werden soll. Z.~B.
ist bei einer Vierfachsumme der Laufindex der "au"sersten Summe
${\vphantom{x}_{\lambda_1}}$, w"ahrend der Laufindex der innersten Summe
${\vphantom{x}_{\lambda_4}}$ ist. Elemente von Vektoren und Matrizen
werden entweder mit der Indexschreibweise (~z.~B. $\Hat{\Vec{H}}_n$~) bzw.
(~z.~B. $\underline{M}_{\text{Zeile,Spalte}}$~) angesprochen, oder, wenn es
sich bei den Elementen um die Werte einer Sequenz einer diskreten Folge
handelt, auch "uber die Argumentschreibweise (~z.~B. $v(k)$ sind die
Elemente des Vektors $\Vec{v}$~).
Gr"o"sen, die im Index ein $f$ aufweisen, sind mit Hilfe einer
Fensterung entstanden. Formelzeichen, die sich aus den im
folgenden angegeben Formelzeichen gem"a"s dieser Regeln unzweifelhaft
herleiten lassen, sind hier teilweise nicht extra
aufgef"uhrt. Auch werden einige weitere Formelzeichen nur lokal in eng
begrenztem Zusammenhang gebraucht und sind dort erkl"art, wo sie zum
ersten Mal auftreten.\vspace{-3pt}
\begin{list}{}{\setlength{\itemsep}{0.2ex plus 0.2ex minus 0.3ex}
\setlength{\labelwidth}{4em}\setlength{\labelsep}{1em}
\setlength{\leftmargin}{5em}}
\item[$\Hat{A}_{\Phi}(\mu)$\hfill] Sch"atzwert f"ur die halbe Breite des
     Konfidenzintervalls des Messwertes
     \mbox{$\Hat{\Phi}_{\boldsymbol{n}}(\mu)$}.
\item[$A_{1,H}(\mu)$\hfill] komplexer Zeiger der l"angeren Halbachse der
     Konfidenzellipse des Messwertes \mbox{$\Hat{H}(\mu)$}.
\item[$A_{2,H}(\mu)$\hfill] komplexer Zeiger der k"urzeren Halbachse der
     Konfidenzellipse des Messwertes \mbox{$\Hat{H}(\mu)$}.
\item[$B$\hfill] Parameter eines Beispiels eines LDS. Dieses hat eine
     \mbox{$2\CdoT\!B$}-fache Nullstelle bei \mbox{$\Omega\!=\!0$}.
\item[$c$\hfill] Parameter der Bilineartransformation bei der Konstruktion
     der diskreten Fensterfolge (\,\mbox{$0\!<\!c\!<\!2$}\,).
\item[$c_f$\hfill] Konstanter Faktor, mit dem die Fensterfolge so skaliert 
     wird, dass deren Spektrum bei \mbox{$\Omega\!=\!0$} den Wert $M$ annimmt.
\item[$c_{\Phi}(\mu)$\hfill]Normierungsfaktor bei der Berechnung des LDS, 
     der sich mit Hilfe der Stichprobenvektoren \mbox{$\Vec{V}(\mu)$} und 
     \mbox{$\Vec{V}(\!-\mu)$} berechnen l"asst.
\item[$c_{\Psi}(\mu)$\hfill]Normierungsfaktor bei der Berechnung des MLDS, 
     der sich mit Hilfe der Stichprobenvektoren \mbox{$\Vec{V}(\mu)$} und 
     \mbox{$\Vec{V}(\!-\mu)$} berechnen l"asst.\pagebreak[2]
\item[$C_{\boldsymbol{A},\boldsymbol{B}}$\hfill]
     \mbox{$=\;\text{E}\{\boldsymbol{A}\CdoT\boldsymbol{B}^*\}\;=\;$}
     Kovarianz von $\boldsymbol{A}$ und $\boldsymbol{B}$
     (~$\boldsymbol{B}$ wird dabei konjugiert!~).
\item[$\underline{C}_{\Vec{\boldsymbol{A}},\Vec{\boldsymbol{B}}}$\hfill]
     \mbox{$=\;\text{E}\big\{\Vec{\boldsymbol{A}}\CdoT\Vec{\boldsymbol{B}}^{\HH}\big\}\;=\;$}
     Kovarianzmatrix der Elemente der Zufallsspaltenvektoren 
     \mbox{$\Vec{\boldsymbol{A}}$} und \mbox{$\Vec{\boldsymbol{B}}$}.
\item[$C(k)$\hfill] Cepstrum des Anteils \mbox{$\widetilde{D}_N(\Tilde{z})$}, 
     der die Nullstellen in der bilineartransformierten $\Tilde{z}$-Ebene enth"alt. 
\item[$d(k)$\hfill]
     \mbox{$=\;\frac{1}{M}\cdot\big(\,f(k)^{\Kk}\!\ast\!f(\!-k)\,\big)\;=\;$}
     diskrete Fenster-AKF
\item[$D(z)$\hfill] Z-Transformierte der Fenster-AKF
     \mbox{$d(k)$}.
\item[$D_E(z)$\hfill] Anteil der Z-Transformierten \mbox{$D(z)$}
     der Fenster-AKF, der nur die Nullstellen am Einheitskreis enth"alt.
\item[$D_{\overline{E}}(z)$\hfill] Anteil der Z-Transformierten \mbox{$D(z)$}
     der Fenster-AKF ohne die Nullstellen am Einheitskreis.
\item[$\widetilde{D}_{\overline{E}}(\Tilde{z})$\hfill] Polynom in $\Tilde{z}$,
     das durch Bilineartransformation aus \mbox{$D_{\overline{E}}(z)$}
     entsteht.
\item[$\widetilde{D}_N(\Tilde{z})$\hfill] Anteil von
     \mbox{$\widetilde{D}_{\overline{E}}(\Tilde{z})$}, der die
     Nullstellen in der bilineartransformierten $\Tilde{z}$-Ebene enth"alt, und
     dessen minimalphasiger Anteil einen nichtlinearen Phasenbeitrag liefert,
     der mit Hilfe des Cepstrums berechnet wird.
\item[$\widetilde{D}_P(\Tilde{z})$\hfill] Anteil von
     \mbox{$\widetilde{D}_{\overline{E}}(\Tilde{z})$}, der die
     Polstellen in der bilineartransformierten $\Tilde{z}$-Ebene enth"alt,
     und dessen minimalphasiger Anteil nach der bilinearen R"ucktransformation
     einen nichtlinearen Phasenbeitrag liefert, der geschlossen berechnet
     werden kann.
\item[$D_P(z)$\hfill] Bilinear R"ucktransformierte von
     \mbox{$\widetilde{D}_P(\Tilde{z})$}
\item[$E$\hfill] Einschwingzeit des Systems
\item[$f(k)$\hfill] diskrete Fensterfolge
\item[$F$\hfill] L"ange der Fensterfolge. Bei der in Kapitel
     \ref{Algo} vorgestellten Fensterfolge gilt \mbox{$F=M\CdoT N$}.
\item[$F(\Omega)$\hfill] Fouriertransformierte der Fensterfolge \mbox{$f(k)$}.
\item[$g(k)$\hfill] Basisfensterfolge, die zur theoretischen Herleitung der
     Konstruktion der diskreten Fensterfolge \mbox{$f(k)$} ben"otigt wird.
\item[$g_Q(k)$\hfill] \mbox{$=\;g(k)\ast g(\!-k)\;=\;M$}-fache diskrete
     Basisfenster-AKF
\item[$G(z)$\hfill] Z-Transformierte der Basisfensterfolge \mbox{$g(k)$}\pagebreak[2]
\item[$h(k)$\hfill] zeitinvariante Impulsantwort des linearen
     zeitinvarianten Modellsystems ${\cal S}_{lin}$
\item[$h_{\kappa}(k)$\hfill] zeitvariante Impulsantwort des linearen
     Modellsystems ${\cal S}_{lin}$
\item[$H(\Omega)$\hfill] "Ubertragungsfunktion des Modellsystems
     ${\cal S}_{lin}$ = wahrer Wert der zu messenden "Ubertragungsfunktion
     des linearen Teilsystems, das sich bei der theoreti\-schen Aufspaltung des
     gest"orten, nichtlinearen, realen Systems ${\cal S}$ mit Hilfe der \linebreak[2]wahren
     i.~Allg. unbekannten Erwartungswerte der anliegenden Prozesse ergibt.
\item[$\Hat{H}(\mu)$\hfill] Messwerte der "Ubertragungsfunktion des
     Modellsystems ${\cal S}_{lin}$ f"ur die Frequenzen
     \mbox{$\Omega=\mu\CdoT2\pi/M$}.
\item[$\Hat{\Vec{H}}$\hfill] Spaltenvektor, dessen Elemente die $M$ 
     Messwerte der "Ubertragungsfunktion \mbox{$\Hat{H}(\mu)$} sind.
\item[$\boldsymbol{\Delta}\Hat{\boldsymbol{H}}(\mu)$\hfill] Zufallsprozess der 
     Abweichung des Messwertes der "Ubertragungsfunktion von dessen theoretisch 
     exaktem Wert.
\item[$k$\hfill] diskrete Zeitvariable. Bei kontinuierlichen
     abgetasteten Signalen wird die kontinuierliche Zeit $t$ auf die
     Abtastperiode normiert.
\item[$\boldsymbol{\Delta}\boldsymbol{k}$\hfill] Zuf"allige Zeitverschiebung bei 
     Erregung des zu messenden Systems mit einem Chirpsignal.
\item[$K_{\nu_2}$\hfill] konstanter Faktor der $\nu_2$-ten Nullstelle des
     bilineartransformierten Polynoms mit den Nullstellen am Einheitskreis.
\item[$L$\hfill] Mittelungsanzahl beim RKM = Anzahl der Einzelmessungen
     am realen System.
\item[$M$\hfill] L"ange der DFT beim RKM = Anzahl der zu berechnenden
     Frequenzwerte.
\item[$\widetilde{M}$\hfill] L"ange der FFT bei der Berechnung der Phase des
     Spektrums der Fensterfolge "uber das Cepstrum.
\item[$\boldsymbol{n}(k)$\hfill] additiver Rauschprozess am Ausgang des
     Modellsystems ${\cal S}_{lin}$, der sich aus St"orungen
     zusammensetzt, die einerseits durch externe St"oreinfl"usse
     \mbox{$\boldsymbol{n}_{Ext}(k)$} und andererseits durch interne
     Nichtlinearit"aten des realen Systems ${\cal S}$ verursacht werden.
     Bei Erregung mit dem Zufallsprozess \mbox{$\boldsymbol{v}(k)$} ergibt
     sich der Zufallsprozess \mbox{$\boldsymbol{n}(k)$} aus dem stochastischen
     Signal \mbox{$\boldsymbol{x}(k)$} am Ausgang des Modellsystems
     ${\cal S}_{lin}$ und dem Zufallsprozess \mbox{$\boldsymbol{y}(k)$}
     am Ausgang des realen Systems ${\cal S}$.
\item[$\boldsymbol{n}_{Ext}(k)$\hfill] Rauschprozess, der von au"sen in das
     zu messende System ${\cal S}$ einstreut.
\item[$\boldsymbol{n}_f(k)$\hfill]Zufallsprozess der L"ange $M$,
     der sich durch Fensterung und blockweise "Uberlagerung aus dem Zufallsprozess
     \mbox{$\boldsymbol{n}(k)$} ergibt.\pagebreak[2]
\item[$N$\hfill] Faktor zwischen der DFT-L"ange $M$ beim RKM und
     Fensterl"ange \mbox{$F=M\CdoT N$}.
\item[$\Tilde{N}$\hfill] Anzahl der Glieder der Kosinusreihe bei Fensterfolgen,
     die durch Abtastung einer Periode eines Signals entstehen, das sich als
     endliche Kosinusreihe darstellen l"asst.
\item[$\boldsymbol{N}_{\!\!f}(\mu)$\hfill] Zufallsprozess der L"ange $M$,
     der sich durch eine DFT aus dem Zufallsprozess
     \mbox{$\boldsymbol{n}_f(k)$} ergibt.
\item[$\Hat{\Vec{N}}_{\!f}(\mu)$\hfill] Stichprobenspaltenvektor,
     der durch Projektion des Stichprobenvektors \mbox{$\Vec{Y}_{\!f}(\mu)$}
     mit der Matrix \mbox{$\underline{V}_{\bot}\!(\mu)$} auf den Orthogonalraum 
     der Stichprobe des Erregungsspektrums entsteht.
\item[$s$\hfill] komplexe Frequenzvariable der Laplacetransformierten
     einer kontinuierlichen Funktion.
\item[${\cal S}$\hfill] zu messendes, reales, nichtlineares und
     gest"ortes System.
\item[${\cal S}_{lin}$\hfill] lineares Teilsystem. Dieses Modellsystem ergibt
     sich bei der theoretischen Aufspaltung des gest"orten, nichtlinearen
     realen Systems ${\cal S}$ mit Hilfe der wahren, i.~Allg.
     unbekannten Erwartungswerte der anliegenden Prozesse.
\item[$t$\hfill] kontinuierliche Zeitvariable
\item[$\boldsymbol{v}(k)$\hfill] Zufallsprozess am Eingang des zu messenden
     Systems ${\cal S}$.
\item[$\Vec{\boldsymbol{v}}$\hfill] Zufallsspaltenvektor mit den $M$ Elementen
     \mbox{$\boldsymbol{v}(k)$} f"ur \mbox{$k=0\;(1)\;M\!-\!1$}.
\item[$\boldsymbol{V}(\mu)$\hfill] Zufallsprozess, der durch eine
     DFT der L"ange $M$ aus einer Periode (\,=~Ausschnitt der L"ange $M$\,)
     des Zufallsprozesses \mbox{$\boldsymbol{v}(k)$} entsteht.
\item[$\Vec{\boldsymbol{V}}$\hfill] Zufallsspaltenvektor mit den $M$ Elementen
     \mbox{$\boldsymbol{V}(\mu)$}.
\item[$\Hat{\Vec{\boldsymbol{V}}}(\mu)$\hfill] Zufallsspaltenvektor mit den
     $2$ Elementen \mbox{$\boldsymbol{V}(\mu)$} und \mbox{$\boldsymbol{V}(\!-\mu)^{\Kk}$}. 
\item[$\underline{V}_{\bot}\!(\mu)$\hfill] Matrix, deren Spaltenvektoren den 
     Orthogonalraum zu den beiden Stichprobenvektoren, die in der Matrix 
     \mbox{$\Hat{\underline{V}}(\mu)$} zusammengefasst sind, aufspannen.
\item[$\boldsymbol{x}(k)$\hfill] Zufallsprozess am Ausgang des Modellsystems
     ${\cal S}_{lin}$, der bei Erregung des Modellsystems mit dem
     Zufallsprozess \mbox{$\boldsymbol{v}(k)$} entsteht.
\item[$\boldsymbol{y}(k)$\hfill] Zufallsprozess, der sich am Ausgang des
     realen Systems bei Erregung mit dem Zufallsprozess
     \mbox{$\boldsymbol{v}(k)$} ergibt.
\item[$\boldsymbol{y}_{\!f}(k)$\hfill]Zufallsprozess der L"ange $M$,
     der sich durch Fensterung und blockweise "Uberlagerung aus dem Zufallsprozess
     \mbox{$\boldsymbol{y}(k)$} ergibt.\pagebreak[2]
\item[$\boldsymbol{Y}(\mu)$\hfill] Zufallsprozess, der durch eine
     DFT der L"ange $M$ aus einem Ausschnitt der L"ange $M$
     des Zufallsprozesses \mbox{$\boldsymbol{y}(k)$} entsteht.
\item[$\boldsymbol{Y}_{\!\!\!f}(\mu)$\hfill] Zufallsprozess der L"ange $M$,
     der sich durch eine DFT aus dem Zufallsprozess
     \mbox{$\boldsymbol{y}_{\!f}(k)$} ergibt.
\item[$\Vec{\boldsymbol{Y}}_{\!\!\!f}$\hfill] Zufallsspaltenvektor mit den $M$ Elementen
     \mbox{$\boldsymbol{Y}_{\!\!\!f}(\mu)$}.
\item[$z$\hfill]  komplexe Frequenzvariable der Z-transformierten einer
     diskreten Folge (\,vor der Bilineartransformation\,).
\item[$\Tilde{z}$\hfill]  komplexe Frequenzvariable der Z-transformierten
     einer diskreten Folge nach der Bilineartransformation.
\item[$\Tilde{z}_{\nu_2}$\hfill] $\nu_2$-te Nullstelle am Einheitskreis
     des einen Polynomfaktors von \mbox{$\widetilde{D}_N(\Tilde{z})$}
     nach der Bilineartransformation.
\item[$\zu$\hfill] unbekannte Nullstelle der Z-Transformierten,
     deren Cepstrum berechnet wird.
\item[$\alpha$\hfill] Wahrscheinlichkeit, dass das Konfidenzintervall den 
     wahren zu messenden Wert nicht beinhaltet. \mbox{$1\!-\!\alpha=$} 
     Konfidenzniveau. In Kapitel \ref{Andere} hat $\alpha$ eine andere Bedeutung!
\item[$\alpha$\hfill] Parameter der kontinuierlichen Fensterfunktionen. 
     Je nach Fensterfunktion, die in Kapitel \ref{Andere} auf ihre Tauglichkeit 
     zum Einsatz beim RKM untersucht wird, hat $\alpha$ eine andere Bedeutung!
\item[$\eta$\hfill] diskrete Frequenzvariable f"ur die nach der
     Bilineartransformation "aquidistanten Frequenzen
     \mbox{$\widetilde{\Omega}=\eta\CdoT2\pi/\widetilde{M}$}.
\item[$\kappa$\hfill] diskrete Zeitvariable.
\item[$\lambda$\hfill] Laufvariable der Mittelung beim RKM.
\item[$\mu$\hfill] diskrete Frequenzvariable f"ur die "aquidistanten
     normierten Kreisfrequenzen \mbox{$\Omega=\mu\CdoT2\pi/M$}. Bei der
     Berechnung der Fensterfolge handelt es sich um die diskrete
     Frequenzvariable vor der Bilineartransformation.
\item[$\nu$\hfill] diskrete Frequenzvariable f"ur die vor der
     Bilineartransformation "aquidistanten Frequenzen
     \mbox{$\Omega=\nu\CdoT2\pi/F$}.
\item[$\sigma_{\boldsymbol{n}}$\hfill] Streuung des Prozesses 
     \mbox{$\boldsymbol{n}(k)$}.
      Bei komplexen Zufallsgr"o"sen handelt es sich um die Wurzel der
      Summe der Real- und Imagin"arteilvarianzen
      \mbox{$ = \sqrt{C_{\boldsymbol{n}(k),\boldsymbol{n}(k)}\;}$} .
\item[$\boldsymbol{\varphi}(\mu)$\hfill] Zufallsphase bei Erregung des zu messenden 
     Systems mit einem Mehrtonsignal.
\item[$\boldsymbol{\varphi}$\hfill] Zufallsphase bei Erregung des zu messenden 
     Systems mit einem Chirpsignal.
\item[$\phi(\nu)$\hfill] Anteil der Phase der Fourierreihenkoeffizienten der 
     Fensterfolge, der "uber das Cepstrum berechnet wird.
\item[$\phi_{\boldsymbol{n}}(\kappa)$\hfill] AKF des
      Prozesses \mbox{$\boldsymbol{n}(k)$} gem"a"s Gleichung (\ref{1.2}).
\item[$\Phi_{\boldsymbol{n}}(\Omega)$\hfill] Leistungsdichtespektrum des
     Prozesses \mbox{$\boldsymbol{n}(k)$} gem"a"s Gleichung (\ref{1.3}).
\item[$\Bar{\Phi}_{\boldsymbol{n}}(\mu)$\hfill] Stufenapproximation des
     Leistungsdichtespektrums des Prozesses \mbox{$\boldsymbol{n}(k)$}.
\item[$\Tilde{\Phi}_{\boldsymbol{n}}(\mu)$\hfill] N"aherungswerte der
     Stufenapproximation des Leistungsdichtespektrums des Prozesses
     \mbox{$\boldsymbol{n}(k)$}, die man mit einer endlich langen
     Fensterfolge gewinnt.
\item[$\Hat{\Phi}_{\boldsymbol{n}}(\mu)$\hfill] Messwerte des
     Leistungsdichtespektrums des Prozesses \mbox{$\boldsymbol{n}(k)$}.
\item[$\psi_{\boldsymbol{n}}(\kappa)$\hfill] Modifizierte Autokorrelationsfolge des
      Prozesses \mbox{$\boldsymbol{n}(k)$} gem"a"s Gleichung (\ref{1.4}).
\item[$\Tilde{\psi}_{\nu_2}$\hfill] Ein Teil des Winkels der Nullstelle
     \mbox{$\Tilde{z}_{\nu_2}$} des Polynoms
     \mbox{$\widetilde{D}_N(\Tilde{z})$}.
\item[$\Psi_{\boldsymbol{n}}(\Omega)$\hfill] Modifiziertes Leistungsdichtespektrum des
     Prozesses \mbox{$\boldsymbol{n}(k)$} gem"a"s Gleichung (\ref{1.5}).
\item[$\Bar{\Psi}_{\boldsymbol{n}}(\mu)$\hfill] Stufenapproximation des
     modifizierten Leistungsdichtespektrums des Prozesses \mbox{$\boldsymbol{n}(k)$}.
\item[$\Tilde{\Psi}_{\boldsymbol{n}}(\mu)$\hfill] N"aherungswerte der
     Stufenapproximation des modifizierten Leistungsdichtespektrums des Prozesses
     \mbox{$\boldsymbol{n}(k)$}, die man mit einer endlich langen
     Fensterfolge gewinnt.
\item[$\Hat{\Psi}_{\boldsymbol{n}}(\mu)$\hfill] Messwerte des
     modifizierten Leistungsdichtespektrums des Prozesses \mbox{$\boldsymbol{n}(k)$}.
\item[$\omega$\hfill] Kreisfrequenz.
\item[$\Omega$\hfill] normierte Kreisfrequenz. Bei kontinuierlichen
     abgetasteten Signalen wird auf die Abtastfrequenz normiert. Bei
     der Berechnung der Fensterfolge handelt es sich um die Kreisfrequenz
     vor der Bilineartransformation.
\item[$\widetilde{\Omega}$\hfill] normierte Kreisfrequenz nach der
     Bilineartransformation.
\item[$\widetilde{\Omega}_\nu$\hfill]Kreisfrequenzen, die durch die 
     Bilineartransformation aus den diskreten Frequenzen \mbox{$\Omega=\nu\CdoT2\pi/F$} entstanden sind.
\end{list}
\newpage
\thispagestyle{empty}
\rule{0pt}{0pt}
\vfill
{\tiny .}


\end{document}